\documentclass[english, 9pt]{article}
\usepackage{jcappub}
\usepackage{lmodern}
\usepackage{graphicx}
\usepackage{esint}
\usepackage[toc,page]{appendix}
\usepackage{array}
\usepackage{longtable}
\usepackage{booktabs}
\usepackage{titlesec}
\usepackage{slashed}
\usepackage[utf8]{inputenc}
\usepackage{ifthen}
\usepackage{enumitem}
\usepackage{adjustbox}
\setcitestyle{square}

\begin{document}
\title{Machian Gravity: Modeling Rotation Curves and Radial Acceleration in the SPARC Galaxy Sample}
\author{Santanu Das}
\affiliation{Oridigm Inc., 43575 Mission Blvd \# 716, Fremont, CA 94539, USA}
\emailAdd{santanu@cosmocommunity.in}
\date{\today}

\abstract{
This paper investigates the potential of Machian Gravity (MG), a five-dimensional theory of gravity, to explain the acceleration law governing rotationally bound systems, in particular spiral galaxies. MG was proposed as a framework capable of accounting for a range of astrophysical and cosmological phenomena—including galactic rotation curves, mass distributions in galaxy clusters, and cosmic expansion—without invoking additional dark components. In this study, we apply the MG acceleration law to a large sample of galaxies drawn from the SPARC database. Through a detailed analysis, we determine the optimal MG parameters for each individual galaxy, successfully fitting their observed rotation curves. Similar to Modified Newtonian Dynamics (MOND), our results indicate the existence of a characteristic acceleration scale associated with galactic dynamics, which regulates rotational behavior in the outer regions. Notably, this acceleration scale varies from galaxy to galaxy, but typically remains of order $10^{-8} {\rm cm/s^2}$.
}

\maketitle

\section{Introduction}

The behavior of rotation curves in galaxies, as predicted by Newton's gravitational theory, should exhibit a Keplerian fall-off in the orbital rotational speed $v$ at the outer edge of the galaxy, given by $v^2 \propto M(r)/r$, where $M(r)$ represents the mass enclosed within the radius $r$. However, observations tell a different story. Instead of the anticipated decline, rotation curves tend to flatten out~\cite{rubin1980rotational,persic1996universal,borriello2001dark,binney2011galactic}. Contrary to expectations, velocities often increase towards the center of galaxies and then stabilize at around $v \sim 200 \-- 300 \, \text{km/s}$. On occasion, velocity might fall off or increase at a large radius while not complying with the Newtonian predictions. This inconsistency leads to a dynamic mass for galaxies that significantly surpasses their luminous mass.

This behavior of galactic rotation curves can be accounted for by introducing additional invisible matter or dark matter. This dark matter is presumed to exist in a spherical halo enveloping galaxies. The general consensus is that it consists of a cold, pressureless fluid. Their interaction with baryonic matter is extremely small but expected to be nonzero. 
Numerous candidates have been proposed to explain dark matter, with supersymmetric particles being a prominent contender~\cite{MARTIN_1998}. However, the lack of experimental confirmation of such particles, especially from the Large Hadron Collider (LHC), strengthens alternative propositions such as axions and ultra-light scalar field dark matter, etc~\cite{mielke2002nontopological, mbelek2004modelling, hernandez2004scalar, boehmer2007can, bertacca2008halos, lee2009galaxies, cervantes2009spherical, lee2010minimum, harko2010galactic, PhysRevLett.40.279,PhysRevD.53.2236}.

While dark matter may solve some of the velocity profiles, the evidence for dark matter relies on indirect observations. The observations only say that the baryonic matter that is present in a galaxy in the form of stars, neutral and high-temperature gas, etc., is not enough to provide sufficient gravity to explain the observed acceleration in these systems provided the calculations are made using the Newtonian law.  Nevertheless, it's plausible that the Newtonian mechanics that hold in laboratories and the solar system might not be directly applicable to galaxies.  Adjusting the equations could potentially account for observed accelerations without invoking dark matter. 

In recent years, a range of theories has emerged in attempts to explain dark matter. Empirical theories like Modified Newtonian Dynamics (MOND) have successfully matched galactic velocity profiles, though it violates the momentum conservation laws~\citep{Milgrim1983,Milgrim1983a,Milgrim1983b,Milgrom2011}. 
Consequently, a mathematically sound theory that can replicate the empirical achievements of MOND, may offer a plausible explanation for dark matter. Bekenstein proposed AQUAL to provide a mathematical foundation to MOND~\cite{Bekenstein1984,Bekenstein2009,Milgrom1986}. Other theories, such as Modified Gravity STVG~\cite{Bekenstein2005}, Tensor-Vector-Scalar (TeVeS) theory~\cite{Moffat2005,Brownstein2005,Brownstein2005a,Moffat2005a}, and Massive Gravity~\cite{Dam1970,Zakharov1970,Babichev2010,Babichev2013}, have also emerged to reproduce galactic velocity profiles without requiring dark matter. Higher-dimensional concepts like Induced Matter Theory has also been proposed by researchers~\cite{Overduin1998,Ponce1993,Wesson1992,de2010schwarzschild,moraes2016cosmic}.

However, all of these theories are proposed to fit the observed data rather than being derived from a foundational logical principle like general relativity.  A robust theory should ideally be grounded in a strong logical or mathematical foundation. The inception of the general theory of relativity (GR) was initially intended to provide a mathematical formulation for Mach's hypothesis. However, it became evident that GR did not align with Mach's principle. Despite this, as GR successfully accounted for various observations on the scale of the solar system, Einstein did not further attempt to reconcile this discrepancy or incorporate Mach's argument into the theory.

In my previous work,  I proposed a theory based on Mach's principle to address the shortcomings of GR and provide an explanation for the inertia of objects. The theory, named as Machian Gravity (MG) is a five-dimensional theory. It can be derived from an action principle, thereby ensuring compliance with conservation principles. The fifth dimension, is intricately linked to the distribution of distant matter across the universe and plays a role in the inertia of particles. It has been demonstrated in~\cite{das2023aspects} that MG converges towards GR in the context of the solar system, assuming all other matter is significantly distant. However, it deviates from GR on the galactic scale. Our earlier studies also illustrated how MG can provide insights into spiral galactic velocity profiles, galaxy cluster mass, and cosmic expansion history without necessitating extra dark components in the universe~\cite{das2023aspects}.

In MG the Yukawa like term arise from the background mass distribution in a completely natural way, without much assumptions. There are a huge amount of literature available on testing the Yukawa term against galaxy velocity profile, galaxy cluster mass profile, binary system and so on. Therefore, the theory allow us to use those literature for our purpose. For instance, in~\cite{benisty2022testing,edwards2017yukawa,berge2018interpretation,de2018analysis} Yukawa potential has been extensively studied in various 2 body problems and change in the orbital precession rate etc are extensively studied. Galactic velocity profile for the thin disk under Yukawa potential has been mathematically studied in~\cite{de2007solution}. 

In this paper, I investigate a range of rotationally supported systems, focusing primarily on spiral galaxies, within the framework of MG. I compare the resulting fits with those obtained from alternative models, including Modified Newtonian Dynamics (MOND) and the standard dark matter paradigm with a Navarro–Frenk–White (NFW) halo profile. The analysis primarily employs observational data from the Spitzer Photometry and Accurate Rotation Curves (SPARC) dataset, which contains detailed rotation curves and associated properties for 175 spiral galaxies. Using this dataset, I demonstrate that MG alone can accurately reproduce galactic rotation profiles, without invoking any additional non-baryonic dark matter. Most importantly, I show that on galactic scales MG exhibits behavior similar to MOND, thereby providing a theoretical framework for MOND-like dynamics.

The structure of this paper is organized as follows. Section~\ref{sec:section2} reviews the various models that have been proposed over the years to address the missing-mass problem in galaxies. Section~\ref{sec:section3} introduces the MG model and outlines its applicability to the calculation of galactic rotation velocities. In Section~\ref{sec:section4}, I provide a brief overview of the SPARC dataset. Section~\ref{sec:section5} presents the application of MG to the SPARC data and compares the resulting fits with those obtained from MOND and the standard dark matter model with an NFW halo profile. In Section~\ref{sec:section6}, I discuss why the MG model provides a better explanation of the observational data and highlight its advantages over competing models. Finally, Section~\ref{sec:section7} summarizes the main conclusions of this work.

\section{\label{sec:section2}Existing approaches to missing mass and their limitations}

Since the discovery of the discrepancy between luminous matter and the observed galactic rotation curves, numerous theoretical frameworks have been developed to account for the missing mass problem. According to the standard model of cosmology, galaxies are formed inside the  the virialised halos of cold dark matter, which are massive, gravitationally bound structures that grow from the collapse of small density fluctuations in the early universe~\cite{white1978core}. These perturbations are subsequently stretched to macroscopic scales by inflation. Since dark matter behaves as a collisionless and pressure-less fluid, it can gravitationally collapse to form the first potential wells into which baryonic matter later cools. 

The  most widely used galaxy dark matter halo model is given by the NFW profile.It was originally based on simulation runs using the P3M code developed by Efstathiou et al~\cite{navarro1997universal,efstathiou1985numerical}. The evolution of the DM fluid is modeled through numerical simulations~\cite{davis1985evolution}. Navarro et al.~\cite{navarro1997universal} found that the results of N-body simulations of cold dark matter halos can be well described by a density profile of the form  

\begin{equation}
\rho_{\mathrm{NFW}}(r) = \frac{\rho_s}{\left(r / r_s\right)\left(1 + r / r_s\right)^2},
\end{equation}

\noindent where $\rho_s$ and $r_s$ denote the characteristic density and scale radius of the halo, respectively. Although these two parameters are, in principle, independent, later studies  have demonstrated a correlation between them~\cite{bullock2001profiles, wechsler2002concentrations}. Consequently, the NFW profile can effectively be described as a one-parameter family characterized by the virial mass $M_{\mathrm{vir}}$.  

The relations connecting $M_{\mathrm{vir}}$ with the concentration parameter $c = r_{\mathrm{vir}} / r_s$, as well as with $r_s$ and $\rho_s$, can be expressed at redshift $z = 0$ for a cosmology with $\Lambda = 0.7$ and $\Omega_0 = 0.3$. We can define  
\begin{equation}
M_{\mathrm{vir}} \equiv \frac{4}{3}\pi \Delta_{\mathrm{vir}} \rho_c r_{\mathrm{vir}}^3,
\end{equation}

\noindent where $\Delta_{\mathrm{vir}}$ is the virial over-density ($\sim 337$ at $z = 0$), $\rho_c$ is the critical density of the Universe, and $r_{\mathrm{vir}}$ is the virial radius. The corresponding empirical relations are:  

\begin{equation}
c \simeq 20\left(\frac{M_{\mathrm{vir}}}{10^{11}\,M_\odot}\right)^{-0.13}, \qquad
r_s \simeq 5.7\left(\frac{M_{\mathrm{vir}}}{10^{11}\,M_\odot}\right)^{0.46}\,\mathrm{kpc}, \qquad
\rho_s \simeq \frac{101}{3}\,\frac{c^3}{\ln(1 + c) - \frac{c}{1 + c}}\,\rho_c.
\end{equation}

\noindent The NFW halo thus exhibits a central cusp, with $\rho_{\mathrm{NFW}}(r) \propto r^{-1}$ as $r \rightarrow 0$, and its overall amplitude and shape are determined by the single free parameter $M_{\mathrm{vir}}$~\cite{gentile2004cored}.  

Although the NFW profile has been widely used to describe dark matter halos, later studies proposed alternative forms that better reproduce the inner regions. Navarro et al.~\cite{navarro2004inner,maccio2008concentration} showed that the \textit{Einasto profile} describe the distribution of stellar light and mass in galaxies---provides a superior fit. Unlike the NFW model, the Einasto profile does not feature a central cusp; instead, its logarithmic slope follows $\mathrm{d}\ln\rho / \mathrm{d}\ln r \propto -r^{1/n}$, approaching zero near the centre. Its density profile is given by  

\begin{equation}
\rho_{\mathrm{E}}(r) = \rho_{-2}\,\exp\!\left[-2n\left(\left(\frac{r}{r_{-2}}\right)^{1/n} - 1\right)\right],
\end{equation}

\noindent where $r_{-2}$ is the radius at which the logarithmic slope equals $-2$, $\rho_{-2}$ is the density at that radius, and $n$ is the Einasto index controlling the overall shape of the profile. Increasing $n$ (at fixed $r_{-2}$ and $\rho_{-2}$) steepens the inner density slope, making the profile more cuspy, with a central density $\rho_0 = \rho_{-2}\exp(2n)$. Halos of Milky Way--like or smaller mass in dark matter only simulations typically exhibit $n \sim 6$, indicating a mildly cuspy structure, whereas smaller $n$ values (e.g., $n \sim 1$) correspond to more cored profiles, with $r_{-2}$ serving as a good proxy for the core size. While Einasto profiles are simple its very difficult to calculate any of the physical quantities even mass within a given radius using analytical method are extremely complex. This profile has been studied by various researchers~\cite{baes2022einasto, ghari2019dark, acharyya2024modelling}. However, as shown by~\cite{chemin2011improved}, fitting Einasto profiles to observed galaxy rotation curves often yields values of $n$ that are inconsistent with dark matter only simulations.

The cold dark matter halos typically predicts a central cusp. However, observational data often reveal large, constant-density cores in the central regions of many, rotationally supported galaxies~\cite{mcgaugh2001high,gentile2004cored}. This discrepancy between simulated cusps and observed cores becomes even more striking when considering that galaxies residing in halos with similar maximum circular velocities exhibit a wide variety of rotation curve shapes in their central regions~\cite{oman2015unexpected}. This diversity contrasts sharply with the remarkable uniformity found in the empirical relation between baryonic gravitational acceleration and total gravitational acceleration by various authors. 

To explain this correlation between accelerations Milgrom proposed MOND~\cite{milgrom1983modification}, where he hypothesized that the force on a body is not proportional to the acceleration, instead to $\mu\left(\frac{a}{a_0}\right)a$, where $\mu(x)$ is some interpolating function and $a_0$ is an acceleration scale. Several authors have tested MOND~\cite{janz2016mass, tiret2009mond} and proposed various empirical formula for $\mu(x)$. For instance  McGaugh and Lelli used the empirical form 

\begin{equation}
    a_{\text {obs }}=\frac{a_{\text {N }}}{1-e^{-\sqrt{a_{\text {N }} / a_{0}}}}
    \label{MONDLelli}
\end{equation}

\noindent where  $a_{0} = 1.20 \pm 0.02
 \text{(random)} \pm 0.24 \text{(systematic)} \times 10^{-10} m/s^2$.

\begin{equation}
a_{\mathrm{N}}=\left|\frac{\partial \Phi_{\mathrm{bar}}}{\partial R}\right|   \,,\qquad\qquad
a_{\text {obs }}=\frac{V^2(R)}{R}\,.
\end{equation}
They studied galaxies with very different morphologies, masses, sizes, and gas fractions. The correlation persists even when dark matter dominates. Consequently, the dark matter contribution is fully specified by that of the baryons~\cite{mcgaugh2016radial,lelli2017one,li2018fitting}. Various other authors also tested the mass discrepancy relation for galaxy of various morphology against the acceleration and find striking correlations~\cite{desmond2016statistical,sanders1990mass}. A comparative study between the MOND and the NFW dark matter profile for SPARC galaxies is shown in~\cite{khelashvili2024sparc}.

Several alternative interpolating functions within MOND have also been studied. Two of the most well-known MOND interpolating functions are the ``simple" one~\cite{famaey2005modified}:

\begin{equation}
a=a_{\mathrm{N}}\left(\frac{1}{2}+\sqrt{\frac{a_0}{a_{\mathrm{N}}}+\frac{1}{4}}\right),
\end{equation}

\noindent and ``standard" one~\cite{mcgaugh2008milky}

\begin{equation}
a=a_{\mathrm{N}} \sqrt{\frac{1}{2}+\sqrt{\frac{a_0^2}{a_{\mathrm{N}}^2}+\frac{1}{4}}} .
\end{equation}

A detailed account of MOND is provided in~\cite{milgrom2008mond}. While MOND successfully explains galactic rotation curves, it fails to account for the missing mass in galaxy clusters. In addition, MOND faces challenges in explaining gravitational lensing observations~\cite{clowe2004weak}. The Bullet Cluster, in particular, has been presented as strong evidence for the existence of dark matter and a major challenge for MOND~\cite{angus2006can}.

Another class of models explored by several researchers involves modifying Newtonian gravity. Since galaxies can be approximated within the weak-field regime of general relativity, most work in this area introduces a Yukawa-like correction to the Newtonian potential. In these models, the modified Newtonian potential is taken as

\begin{equation}
    \Phi = \Phi_\text{N} + \Phi_\text{Yu}, \qquad \text{where} \qquad \Phi_\text{N} = \frac{GM}{r}, \qquad \Phi_\text{Yu} = \frac{GM\beta}{r}e^{-\lambda r}\,.
\end{equation}

\noindent Here $\beta$ is a coupling constant and $\lambda$ is some inverse length scale. This form of potential was originally proposed by Sanders~\cite{sanders1984anti}, and since then it has been explored theoretically and applied by many authors to fit a wide range of gravitationally bound systems~\cite{Moffat2005,de2018galaxy,drummond2001bimetric,stabile2011rotation,mannheim2012fitting,pizzuti2017clash,hassan2025milky,d2024testing}. Although these models can reproduce galactic rotation curves with excellent precision, their main criticism is that they relay on two adjustable parameters that vary from one galaxy or cluster to another. This reduces their predictive power compared to MOND, where single universal constant $a_0$ is hypothesized to
successfully explain the detailed shapes and scaling relations of hundreds of galaxies. Even though in~\cite{chang2019there} authors have shown that a global value of $a_0$ may not be able to give good fit for the galaxies. 

Although these models have been extensively tested, it remains difficult to favor one over the others. This suggests that we need a more fundamental theory that is capable of explaining why such diverse models reproduce galactic rotation curves so well. In other words, there should exist a final theory that should reduce to several of these existing models in appropriate limits.

\section{\label{sec:section3}A brief overview of Machian Gravity}

The general theory of relativity was proposed based on the philosophy that ``the laws of physics should remain unchanged regardless of the choice of the coordinate system". However, it does not fully adhere to this principle.

To better understand the conceptual challenge posed by the general theory of relativity, let's consider a thought experiment. The velocity and acceleration of a particle are relative quantities, necessitating a reference frame from which to measure them. Imagine there are only two point masses in the universe, with one being significantly heavier than the other. Let's denote the heavier mass as A and the smaller mass as B, orbiting A in a circular path. We'll examine two coordinate systems, both centered at A, with the z-axis perpendicular to B's orbital plane. Suppose one of these coordinate systems is rotating with respect to the other at an angular velocity $\omega_z$. Since there's no distant object in the universe to establish the inertial coordinate system, we face ambiguity in distinguishing between the inertial and non-inertial coordinate systems. If we assume that the gravitational force is balanced by B's centripetal acceleration, we need to know the angular velocity of B, which differs between the two reference frames. Consequently, the centripetal acceleration of B varies between these frames, rendering the measurement of gravitational acceleration impossible as there is no way to know the centrifugal acceleration.

In fact, if we align the x-axis of a reference frame with particle B, B will have no centripetal acceleration in that reference frame, leading to zero/undefined gravitational force on B in that frame. Since there are no distant matter ( stars, or galaxies ) in the universe except these two particles, the reference frames are equivalent, unless we consider that there is some absolute space-time. The laws of physics should remain independent of the chosen reference frame, and the absence of centrifugal force in one frame should imply its absence in the other.

Einstein's general theory of relativity explains acceleration using the curvature of the of the space-time. In cases like this, it becomes impossible to discern the acceleration or predict which coordinate will experience curvature and to what extent. At a deeper level, this issue arises because, in GR or in Newtonian mechanics, the concept of inertia — or the definition of inertial reference frames – must be introduced apriori~\cite{sciama1953origin}. To address this limitation, we require a theoretical framework in which inertia emerges naturally from the dynamics of the theory, rather than being imposed externally. The Earnest Mach's solution was that the inertia must originate due to the  motion of distance objects in the universe. Therefore, in the above thought experiment, the very concept of inertia is not well defined, as the universe is otherwise empty. This gives rise to the problem discussed earlier.

A literature survey can identify that there are three primary effects that are  introduced by Mach’s hypothesis. First, distant objects (such as stars and galaxies) generate a gravitational potential acting on a body. While, the gravitational potential from a single object may be small, but the combined gravitational potential from all the other objects of the universe is significant, in fact $\frac{\varphi}{c^2} \sim \mathcal{O}(1)$, where $\varphi$ is the combined gravitational potential on the body from all other object in the universe. On a static or on an object moving at a constant velocity the combined gravitational field will be $0$. However, in the reference frame of an accelerating object, all distant masses appear to accelerate. This gives rise to inertial forces such as the centrifugal, Coriolis, and Euler forces etc. This idea was first formulated mathematically by Sciama and subsequently explored by various authors~\cite{sciama1953origin, berman2008machian, sciama1964physical}.

Second, the inertial mass of a particle may depend on the distribution of distant matter, since, according to Mach’s hypothesis, inertia arises from the interaction with distant masses. As a result, identical objects may exhibit different inertial properties depending on their location in the universe, thereby violating the Strong Equivalence Principle.~\cite{brans1962mach, benedetto2024equivalence, darabi2013flat, singleton2016global}. Third, the Newtonian gravitational constant itself may vary due to the motion of distant matter, particularly as a consequence of the expansion of the universe. This effect was originally studied in detail by Brans and Dicke, who, while acknowledging the first two effects, primarily focused on the variation of the gravitational field implied by Mach’s principle~\cite{Brans1961}. This becomes the basis of several Scalar Tensor theories.

Earlier, we demonstrated that all these three effects can be unified within a five-dimensional framework. The detailed derivation is presented in~\cite{das2023aspects}; here we take a detour and summarize the essential idea. In a local Minkowski frame, the special-relativistic dispersion relation $E^2 - \vec{p}^2 = {m_0}^2$ should hold. However, since the inertial mass $m_0$ may vary, we bring it to the left-hand side and postulate that the five-dimensional line element

\begin{equation}
    ds^2 = dt^2 -dx^2 -dy^2 -dz^2 -d\zeta^2
\end{equation}

\noindent remains invariant. Here $\zeta$ is the 5th dimension and this actually provides the variation in the inertial mass. Provided the gravitational and inertial masses coincide, we must have $\frac{d\zeta}{ds} = 1$. However, in situations where the inertial mass varies across spacetime, the strong equivalence principle will get violated and this quantity need not remain unity.

The fifth dimension, referred to as the background dimension, is associated with the origin of particle inertia. In curved spacetime, the five-dimensional line element $ds^{2}=\widetilde{g}_{AB}dx^{A}dx^{B}$ replaces the flat form, rather than the four-dimensional metric. Here, the indices $A$, $B$, etc., refer to the coordinates in the five-dimensional system. The tilde $(\widetilde{..})$ signifies quantities in the five-dimensional context.
In our earlier work, we derived this metric starting from the above three assumptions and demonstrated that, under appropriate limits, the resulting five-dimensional framework can reproduce general relativity, Sciama’s theory, or Brans–Dicke–type theories.

In the present article, we focus on the gravitational field equations for a static, spherically symmetric vacuum solution of the Machian gravity framework under weak gravitational field and fit galactic rotation curve data.

\subsection{Static, Spherically symmetric, Vacuum solution for weak gravitation field}

When considering a vacuum, the field equation for MG becomes $\widetilde{G}_{AB}=0$, which, after some algebraic manipulation, can be written as $\widetilde{R}_{AB}=0$, where $\widetilde{R}_{AB}$ denotes the Ricci tensor. In the context of a weak gravitational field, the metric can be approximated as a perturbation over the Minkowski metric, written as $\widetilde{g}_{AB}=\widetilde{\eta}_{AB}+\widetilde{\gamma}_{AB}$. Here, $\widetilde{\eta}_{AB}$ is the Minkowski metric defined by $\widetilde{\eta}_{AB}={\rm diag}(c^2, -1, -1, -1, -1)$, and $\widetilde{\gamma}_{AB}$ represents the metric perturbation. It should be emphasized that by vacuum we refer to the vanishing of the stress–energy tensor. In a universe entirely without background matter, the concept of inertia would lose its meaning within this theoretical framework. Equivalently, one may regard the gravitational mass at a given point as contributing to the stress–energy tensor, while the remainder of the universe serves as the background.
 
For weak gravitational fields, the dominant component is the time component. Consequently, the ${00}$ component of the Ricci tensor, $\widetilde{R}_{00}=\widetilde{R}^C_{0 C 0}$, simplifies to a form involving the Riemann tensor.  The Riemann tensor can be expressed as:

\begin{equation}
\widetilde{R}_{0 A 0}^B=\partial_A \widetilde{\Gamma}_{00}^B-\partial_0 \widetilde{\Gamma}_{A 0}^B+\widetilde{\Gamma}_{A C}^B \widetilde{\Gamma}_{00}^C-\widetilde{\Gamma}_{0 C}^B \widetilde{\Gamma}_{A 0}^C \;.
\end{equation}

\noindent The second term here is a time derivative, which vanishes for static fields. The third and fourth terms are of the form $(\widetilde{\Gamma})^2$, and since $\widetilde{\Gamma}$ is first-order in the metric perturbation, these contribute only at second order and can be neglected, giving 

\begin{equation}
\widetilde{R}_{00} = \widetilde{R}_{0 A 0}^A = \partial_A\left(\frac{1}{2} \widetilde{g}^{A C}\left(\partial_0 \widetilde{g}_{C 0}+\partial_0 \widetilde{g}_{0 C}-\partial_C \widetilde{g}_{00}\right)\right)  =-\frac{1}{2} \widetilde{g}^{A B} \partial_A \partial_B \widetilde{\gamma}_{00} \;.
\end{equation}

\noindent For the static solution, the time derivative also vanishes, and the equation becomes

\begin{equation}
\partial_{\zeta}^{2}\widetilde{\gamma}_{00}+\partial_{x}^{2}\widetilde{\gamma}_{00}+\partial_{y}^{2}\widetilde{\gamma}_{00}+\partial_{z}^{2}\widetilde{\gamma}_{00}=0\,.\label{eq:laplace equation}
\end{equation}

\noindent Here $\zeta$ is the fifth dimension, which we also sometimes refer to as the background dimension. It is somehow related to the matter distribution in the entire universe and is responsible for the inertial properties of matter~\cite{das2023aspects}. Under the assumption of spherical symmetry of the special part, it can be written as

\begin{equation}
\partial_{\zeta}^{2}(r\widetilde{\gamma}_{00})+\partial_{r}^{2}(r\widetilde{\gamma}_{00})=0\,.\label{eq:waveequation}
\end{equation}

\noindent Using `separation of variables' and considering $(r\widetilde{\gamma}_{00})=R(r)\chi(\zeta)$, we get 

\begin{equation}
\frac{1}{R}\frac{\partial^{2}R}{\partial r^{2}}=-\frac{1}{\chi}\frac{\partial^{2}\chi}{\partial\zeta^{2}}=\pm \lambda^{2}\,, \label{eq:speOfVar}
\end{equation}

\noindent where, $\lambda$ is a real constant. There are distinct cases, depending on the sign of the quantity on the right-hand side. It is essential to treat these cases separately, as they lead to different solutions. 
\paragraph{Solution 1:}
When the sign is positive, the particular integrals for $R$ and $\chi$ are given by

\begin{equation}
R_{PI}=P_{1}e^{\lambda r}+P_{2}e^{-\lambda r}\,, \qquad\qquad 
\chi_{PI}=Q_{1}\cos(\lambda\zeta)+Q_{2}\sin(\lambda\zeta)\,,
\end{equation}

\noindent The complimentary functions will be given by 
\begin{equation}
R_{CF}=S_1 + S_2 r\,, \qquad\qquad 
\chi_{CF}=S_3 + S_4 \zeta\,,
\end{equation}

\noindent where, $P_{1}$, $P_{2}$, $Q_{1}$, $Q_{2}$, $S_{1}$, $S_{2}$, $S_{3}$, and $S_{4}$ are constants. Combining all these we get

\begin{equation}
(r\tilde{\gamma}_{00})=\left( S_1 + S_2 r +  P_{1}e^{\lambda r}+P_{2}e^{-\lambda r}\right)\left(S_3 + S_4 \zeta + Q_{1}\cos(\lambda\zeta)+Q_{2}\sin(\lambda\zeta)\right)\,.\label{eq:potential-equation}
\end{equation}

\noindent According to our definition, $\frac{d\zeta}{ds} = \frac{m_{0}}{m_{g}} \sim 1$. As long as the weak equivalence principle holds, this ratio should remain unity. However, on galactic scales, the background mass distribution may vary across different regions of the galaxy and, consequently, this ratio may deviate from unity, although it is expected to remain of order one. Thus, $\zeta$ can be regarded as a monotonically increasing coordinate, similar to time.

Since we do not want the potential to vary with time, we must set $S_{4} = 0$ and assume $Q_{1}, Q_{2} \ll S_{3}$. 
Therefore, the expression in the brackets on the right-hand side effectively reduces to $\approx S_{3}$. Without loss of generality, we can set $S_{3} = 1$, as it is merely a constant whose value can be absorbed into the constants appearing in the radial part.

Now let us consider the radial part. Clearly, we do not want the potential to grow exponentially with radius. The term $\tilde{\gamma}_{00}$, under the weak-field approximation, represents a first-order perturbation and therefore cannot increase exponentially with distance. Consequently, we must set $P_{1} = 0$ and obtain

\begin{equation}
\tilde{\gamma}_{00}=\frac{S_1}{r} + S_2  +  \frac{P_{2}}{r}e^{-\lambda r}\,.
\end{equation}

\noindent The equation for a geodesic path is given by

\begin{equation}
\frac{d^2 x^A}{d s^2}-\widetilde{\Gamma}_{B C}^A \frac{d x^B}{d s} \frac{d x^C}{d s}=0\,.
\end{equation}

Under the above assumptions and within the weak-field limit, it is straightforward to show that
$\frac{d^2 x^A}{dt^2} = \frac{1}{2} \partial_A \gamma_{00}$ (same 
as in general relativity). This allows us to relate $\gamma_{00}$ to the Newtonian gravitational potential, giving $\tilde{\gamma}_{00} = 2\Phi$, where $\Phi$ denotes the Newtonian potential of the gravitational field. The term $S_2$ merely contributes a constant to the potential and can therefore be ignored. To recover the Newtonian limit at the solar system scale, we set
\begin{equation}
P_2 = -2KGM, \qquad S_1 = 2(1+K)GM \,,
\end{equation}

\noindent and, substituting $\tilde{\gamma}_{00} = 2\Phi$, we obtain the gravitational potential as

\begin{equation}
\Phi=\frac{GM}{r}\left[1+K\left(1-e^{-\lambda r}\right)\right]\,.\label{eq:potential}
\end{equation}

\noindent Here, $M$ denotes the mass at the center, and $G$ is Newton's gravitational constant. The quantities $\lambda$ and $K$ are background-dependent; they may depend on the mass $M$ but are independent of $r$. 

\paragraph{Solution 2:} We now turn to the other solution of Eq.~\ref{eq:speOfVar}, corresponding to a negative signature of $\lambda^2$. In this case, using a similar calculation, we can show that

\begin{equation}
(r\tilde{\gamma}_{00})=\left( S_1 + S_2 r +  P_{1}\sin{\lambda r}+P_{2}\cos{\lambda r}\right)\left(S_3 + S_4 \zeta + Q_{1}e^{-\lambda\zeta}+Q_{2}e^{\lambda\zeta}\right)\,.\label{eq:potential-equation1}
\end{equation}

In this case, we must set $S_4 = Q_2 = 0$; otherwise, the potential would grow with time, which is not physically feasible. 
The decaying term in $\zeta$ vanishes, reducing the second bracket to the constant $S_3$. 
As before, without loss of generality, we can set $S_3 = 1$. 

Turning to the radial part, one could in principle obtain an oscillating solution. 
Such a solution may be of interest and could be explored in future work; however, in this article, we do not consider it. 
Since such oscillatory behavior is not observed in nature, we must assume $P_1, P_2 \ll S_1$. 
Under this assumption, the solution naturally reduces to the standard Newtonian potential.

\subsection{Calculation of Radial Velocity and the Tully–Fisher Relation}

Returning to Eq.~\ref{eq:potential}, observations of galactic rotation curves suggest that $\lambda^{-1}$ typically lies on the order of a few kiloparsecs. For small values of $r$, the exponential term $e^{-\lambda r}$ approaches unity, and consequently, the potential $\Phi$ reduces to the Newtonian form, $\Phi = \frac{GM}{r}$. This alignment with the Newtonian potential is particularly significant on the solar system scale. 

In the asymptotic limit of $r\rightarrow\infty$, the
exponential term goes to $0$. Hence, for large values of $r$, it becomes $(1+K)$ times that of Newtonian potential and can provide additional gravitational force in large gravitationally balanced systems, such as galaxies, galaxy clusters, etc. A similar form of potential has previously been used 
by other groups to explain the galactic velocity profile correctly \cite{Moffat2009,Moffat2005,Brownstein2005,Brownstein2005a,Moffat2005a}. 

\begin{figure}
    \centering
    \includegraphics[trim=0cm 0cm 0cm 0cm, clip=true,width=0.48\columnwidth]{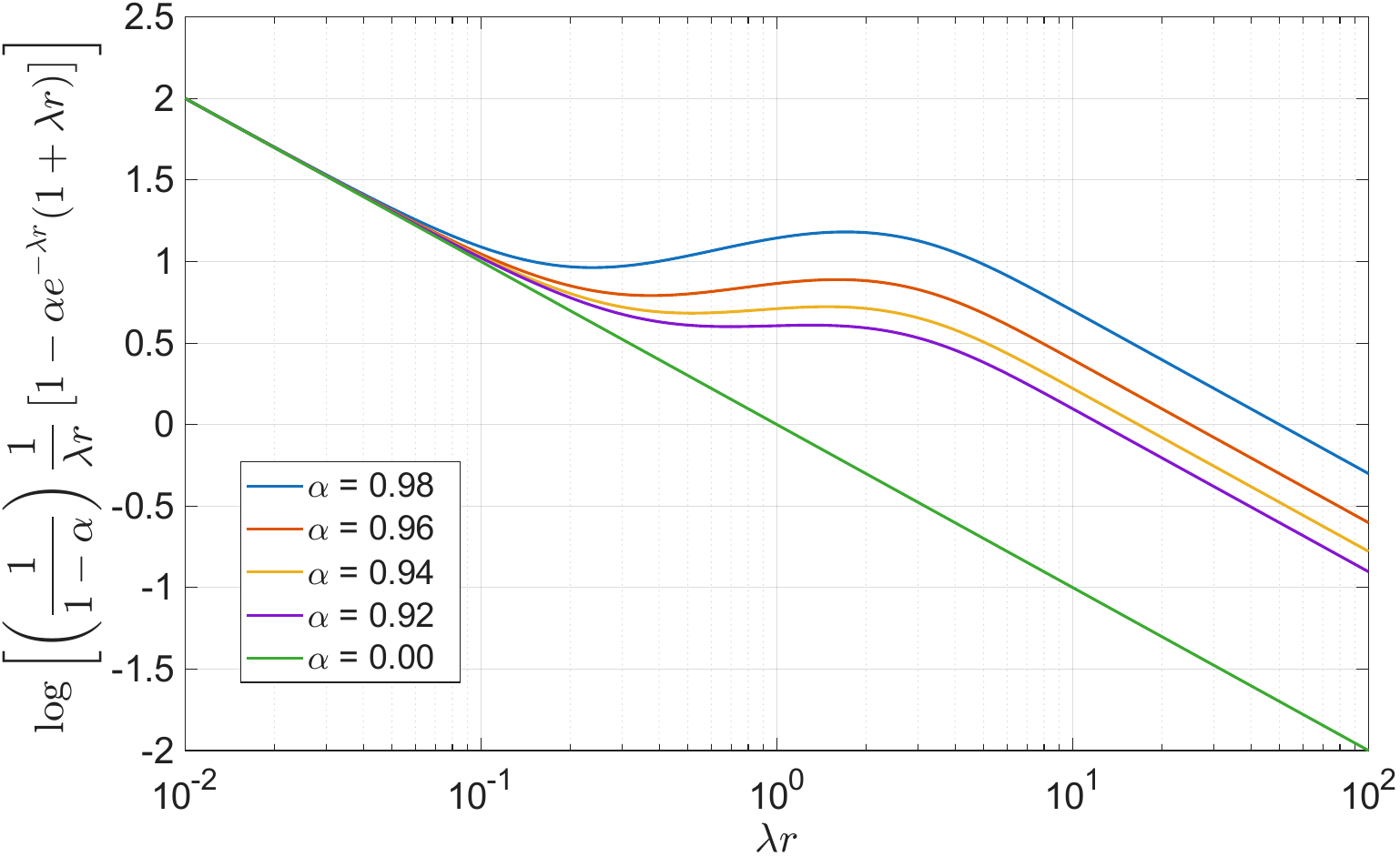}
    \includegraphics[trim=0cm 0cm 0cm 0cm, clip=true,width=0.51\columnwidth]{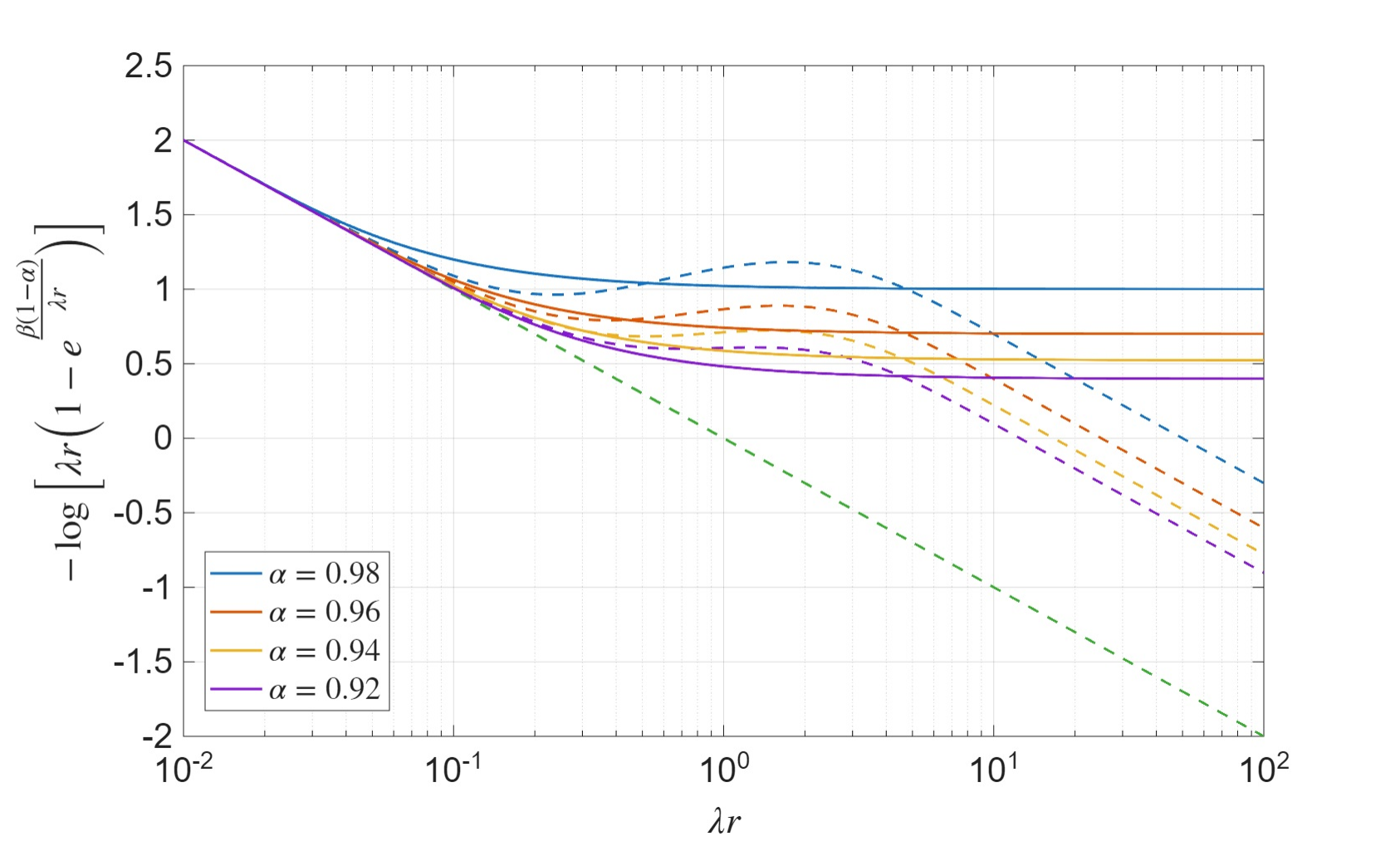}
    \caption{In the left panel, I plot a quantity derived from Eq.~\ref{eq:velocity} within the MG framework. For different values of $\alpha$, there exists a range of $\lambda r$ over which the curves approach a flat behavior. In the right panel, I show the corresponding quantity for MOND, represented by the solid curves. For comparison, the Machian Gravity results (identical to those shown in the left panel) are overplotted using dotted curves. It is evident that the two sets of curves agree well for $\lambda r \lesssim 6$. However, at larger values of $\lambda r$, the Machian Gravity curves deviate and tend toward a Newtonian behavior, whereas the MOND curves remain asymptotically flat.}
    \label{fig:alpha_values1}
\end{figure}

It is important to emphasize that both $K$ and $\lambda$ should remain independent of the mass of the gravitating objects, as any such dependence would lead to a violation of symmetry. To illustrate this point, consider a scenario where two particles mutually exert gravitational attraction on each other. If $K$ were to rely on the mass of just one of these particles, the equation governing gravitational energy would become asymmetric with respect to the masses of both particles. Consequently, in order to preserve symmetry, $K$ and $\lambda$ are not influenced by the masses of the interacting entities. Instead, $K$ and $\lambda$ are shaped by the combined mass distribution of all particles that exist nearby. Specifically. if there were only two particles in the universe, then $K$ should be $0$. However, when a third particle is introduced, it affects the gravitational interaction between the initial two particles through the  $K$ and $\lambda$ term.

This also aligns with the fundamental concepts of Mach's principle, which proposes that local physical phenomena are interconnected with the distribution of matter throughout the universe. In essence, the influence of neighboring masses is coming into the gravitational energy through the  $K$ and $\lambda$.

As the potential due to a static spherically symmetric gravitational 
field is given by Eq.(\ref{eq:potential}), we can calculate the acceleration due to the gravitational field as $\frac{\partial\Phi}{\partial r}$. If a particle  orbits mass $M$ in a circular orbit of radius $r$, and its orbital velocity is $v$, then we can calculate $v$ by equating the centripetal force with the gravitational field, giving

\begin{equation}
v^2=\frac{GM}{r}\left[1+K\left(1-e^{-\lambda r}\left(1+\lambda r\right)\right)\right]=\frac{GM(1+K)\lambda}{\lambda r}\left[1- \alpha e^{-\lambda r}\left(1+\lambda r\right)\right] \,.
\label{eq:velocity}
\end{equation}

\noindent where $\alpha = \frac{K}{1+K}$. The velocity has some interesting property. For $\alpha\in(0.92,0.95)$ and $\lambda r \in (0.4,2.5)$ the velocity becomes almost independent of $r$. This can be seen in Fig.~\ref{fig:alpha_values1}. From the range of $\alpha$ we can derive the range of $K$ to be $\in (11,19)$. 

The velocity in the outer part of the spiral galaxies (rotationally bounded system) does not decrease with increasing radius, as suggested by Keplarian velocity. In fact, it is almost independent of radius $r$.  Therefore, Eq.~\ref{eq:velocity} can be used to explain the velocity profile of spiral galaxies. This was first explained in \cite{sanders1984anti, sanders2002modified}.  

For this particular range of $r$, the velocity of the test particle behaves as $v^2 \sim GM(1+K)\lambda$. However, according to the Tully–Fisher relation, the mass of a spiral galaxy is linked to its asymptotic velocity as $M \sim v^\gamma$, where $\gamma \in (3.5,4)$. If we assume that $M\sim v^4$, then we can take 

\begin{equation}
    (1+K) \propto\frac{1}{\sqrt{M}} \qquad\implies\qquad K=\sqrt{\frac{M_c}{M}} - 1
    \label{eq:KSpiralGalaxy}
\end{equation}

\noindent Here $M_c$ is some constant mass. Putting everything together, the expression for the final velocity becomes

\begin{equation}
v^2=\frac{GM}{r}\left[1+\left(\sqrt{\frac{M_c}{M}} - 1\right)\left(1-e^{-\lambda r}\left(1+\lambda r\right)\right)\right] \,.
\label{eq:velocityFinal}
\end{equation}

This equation follows Newtonian velocity for a particle in orbit for $\lambda r \ll 1$. For $\lambda r \in (0.4, 2.5)$, velocity becomes constant and follows the Tully Fisher relation i.e., $v^4\sim M$. Finally for $\lambda r\gg 2.5$, it behaves as $v^2\sim\frac{\sqrt{M}}{r}$.  At this point, we should also like to point out that $K$ is a background-dependent quantity. For spiral galaxy $K$ follows the empirical relation Eq.~\ref{eq:KSpiralGalaxy} does not imply that $K$ should follow similar expressions for any kind of mass distribution. For other kinds of mass distribution, the form of $K$ may vary as we don't have any theoretical expression for $K$.  

\begin{figure}
    \centering
    \includegraphics[trim=1cm 1cm 2cm 1cm, clip=true,width=0.7\columnwidth]{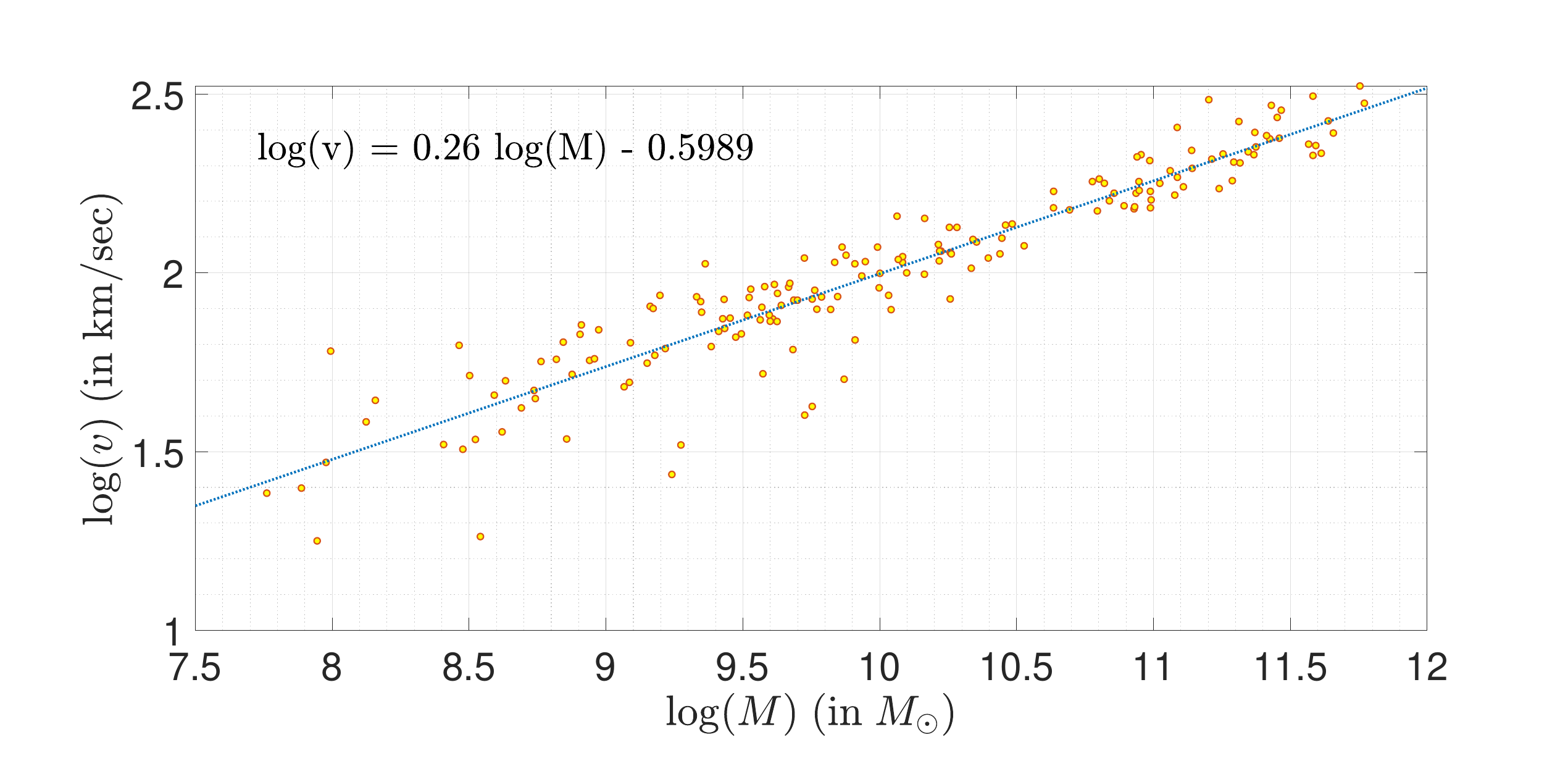}         
    \caption{The graph illustrates the measured velocities at the outer edges of 175 SPARC galaxies against the masses of these galaxies. Additionally, a blue straight line represents the optimal fit through these data points, accompanied by the equation characterizing this linear fit. The graph provides $M\sim v^{3.85}$, aligning with the expectations of the Tully Fisher relation.}
    \label{fig:tully-fisher}
\end{figure}

\subsection{Approximation and comparison with MOND}

MOND introduces an intrinsic acceleration scale, whereas no such fundamental acceleration scale exists in Machian Gravity (MG). However, as can be seen from Eq.~\ref{eq:velocityFinal}, the resulting velocity becomes independent of radius in the limit $\lambda r \rightarrow 1$. Therefore, expanding the velocity in a Taylor series centered around $\lambda r \rightarrow 1$ we can get:

\begin{eqnarray}
v^2
&=& \frac{GM}{r}\left[1+\left(\sqrt{\frac{M_c}{M}} - 1\right)\left(1-e^{-1}\left(3-\lambda r \right)\right)\right] \nonumber\\
&=& \frac{GM}{r}\left[3e^{-1}+\left(\sqrt{\frac{M_c}{M}}\lambda r e^{-1} - \lambda r e^{-1}\right) + \left( 1-3e^{-1} \right)\sqrt{\frac{M_c}{M}} \right] \nonumber\\ 
\frac{v^2}{r}&\approx& \frac{GM}{r^2}\left[1.1+0.37 \sqrt{\frac{\frac{GM_c}{\lambda^{-2}}}{\frac{GM}{r^2}}}  - 0.37 \lambda r  -0.1\sqrt{\frac{M_c}{M}} \right]
\label{Eq:MONDBreaking}
\end{eqnarray}

\noindent Now provided $\lambda r > 0.27$ the second term will dominate over the 4th term. Also, if we assume that $\frac{M_c}{M} \sim 10 - 50$, then the third term is also small. Therefore, if we ignore these terms, we can write the equation as 

\begin{equation}
a \approx a_\text{N}\left[\kappa+0.37 \sqrt\frac{a_0^\text{MG}}{a_\text{N}}  \right] 
= a_\text{N}\mu\left(\frac{a_0^\text{MG}}{a_\text{N}}  \right)\qquad\qquad\text{where, }\kappa =1.1  - 0.37 \lambda r  -0.1\sqrt{\frac{M_c}{M}}  \,.
\end{equation}

\noindent Here, $a_\text{N} = \frac{GM}{r^2}$ is the Newtonian acceleration, while $a^\text{MG}_0 = \frac{GM_c}{\lambda^{-2}}$ is a characteristic acceleration. $\mu$ is a function. In the equation $\kappa$ is much smaller than the second term as discussed before. Consequently, near the galaxy's periphery or when $\lambda r \rightarrow 1$, our equation behaves as MOND. Therefore, the mass discrepancy is predominantly linked to acceleration, with a slight dependence on $\lambda r$. This $\lambda r \rightarrow 1$ dependence is consistent with the findings in Fig.~\ref{fig:alpha_values1}.
As the mass discrepancy is influenced by $a^\text{MG}_0 = \frac{GM_c}{\lambda^{-2}}$,  it also shows that $M_c$ and $\lambda$ are not entirely independent. Instead, we should expect a strong correlation between $M_c$ and $\lambda^{-2}$. 

Here one should note that I have adopted the choice $K = \sqrt{M_c/M} - 1$. However, the Tully--Fisher relation implies that the baryonic mass scales with the asymptotic rotational velocity as $M \sim v^{\gamma}$, with $\gamma \in (3.5,4)$. If instead one chooses a general form of $K$ as $K = (M_c/M)^{1-2/\gamma} - 1$, then the resulting mass discrepancy acquires a dependence on the combination $M/\lambda^{\gamma/(\gamma-2)}$. I will discuss the implications of this choice in the context of SPARC galaxies in a later section.

In contrast, within the MOND framework, Lelli’s relation (Eq.~\ref{MONDLelli}) can be expanded in the low-acceleration regime $a_{\mathrm{N}} < a_0^{\mathrm{MOND}}$, yielding

\begin{equation}
    a_{\text{obs}} = a_{\text{N}} \frac{1}{  
    \left(\sqrt{ \frac{a_{\text{N}}}{a_0^{\text{MOND}}}}\right) - \frac{1}{2}\left(\sqrt{ \frac{a_{\text{N}}}{a_0^{\text{MOND}}}}\right)^2 + \ldots  } \approx \sqrt{a_{\text{N}}a^{\text{MOND}}_0} \,.
\end{equation}

\noindent Therefore, comparing with the previous result, we find that $a_0^{\mathrm{MG}} \sim 7.3\, a_0^{\mathrm{MOND}}$. The MOND acceleration scale may also be written as $a_0^{\mathrm{MOND}} = G M_c / \lambda^{-2}$ for appropriate choices of the characteristic mass $M_c$ and length scale $\lambda^{-1}$. 
The right panel of Fig.~\ref{fig:alpha_values1} shows the corresponding rotation curves for MOND dynamics. To facilitate a direct comparison with the Machian Gravity model, I introduce the parameter 
$\alpha = K/(1+K)$, where $K = \sqrt{M_c/M} - 1$. Although this definition has no intrinsic significance within MOND, it provides a convenient means of comparing the effective accelerations in MOND and Machian Gravity.

In this figure, the solid curves represent $-\log\!\left[\lambda r \left(1 - \exp\!\left(\frac{\beta(1-\alpha)}{\lambda r}\right)\right)\right]$ plotted as a function of $\lambda r$ for different values of $\alpha$. These are directly comparable to the quantity 
$\log\!\left[\left(\frac{1}{1-\alpha}\right)\frac{1}{\lambda r}\left(1 - \alpha e^{-\lambda r}(1+\lambda r)\right)\right]$ arising in the Machian Gravity model, which is shown by the dotted curves. 
For illustrative purposes, I adopt $\beta = 6$. Both sets of curves exhibit similar behavior for $\lambda r \lesssim 6$. However, at larger values of $\lambda r$, the MOND curves asymptotically flatten, whereas the Machian Gravity curves decay. Consequently, while both models provide comparable descriptions for systems with $\lambda r \lesssim 6$, such as galaxies, their predictions are expected to diverge for larger systems, including galaxy clusters.

\section{\label{sec:section4}A brief overview of SPARC galaxy database}

An ideal galaxy sample should include all the galaxies within a sufficiently extensive volume of the universe~\cite{eckert2015resolve}. However, in practice, such a sample can never exist as there will be limitations on the minimum luminosity and surface brightness of galaxies below which the galaxies can not be detected. Therefore, the best thing is to sample randomly across the mass function~\cite{mcgaugh2019dynamical}. 

SPARC (Spitzer Photometry and Accurate Rotation Curves) database provides a sample of 175 nearby galaxies with new surface photometry at $3.6 {\rm\mu m}$ and high-quality rotation curves from previous ${\rm HI/H\alpha}$ studies~\cite{Lelli_2016}. SPARC rotation curves are drawn from multiple papers. The rotation curves are generally smooth but can exhibit large-scale features that correspond directly to variations in the surface brightness profile. Miraculously, this behavior follows Renzo’s rule: ``For any feature in the luminosity profile, there is a corresponding feature in the rotation curve and vice versa"~\cite{sancisi2003visible,chan2022mysterious}. This implies that if dark matter is an independent substance in galaxies then it may not be able to explain this feature. However, the further details of these will be discussed in a later section. 

\begin{figure}[t]
    \centering
    \includegraphics[trim=0cm 0cm 1cm 1cm, clip=true,width=0.49\columnwidth]{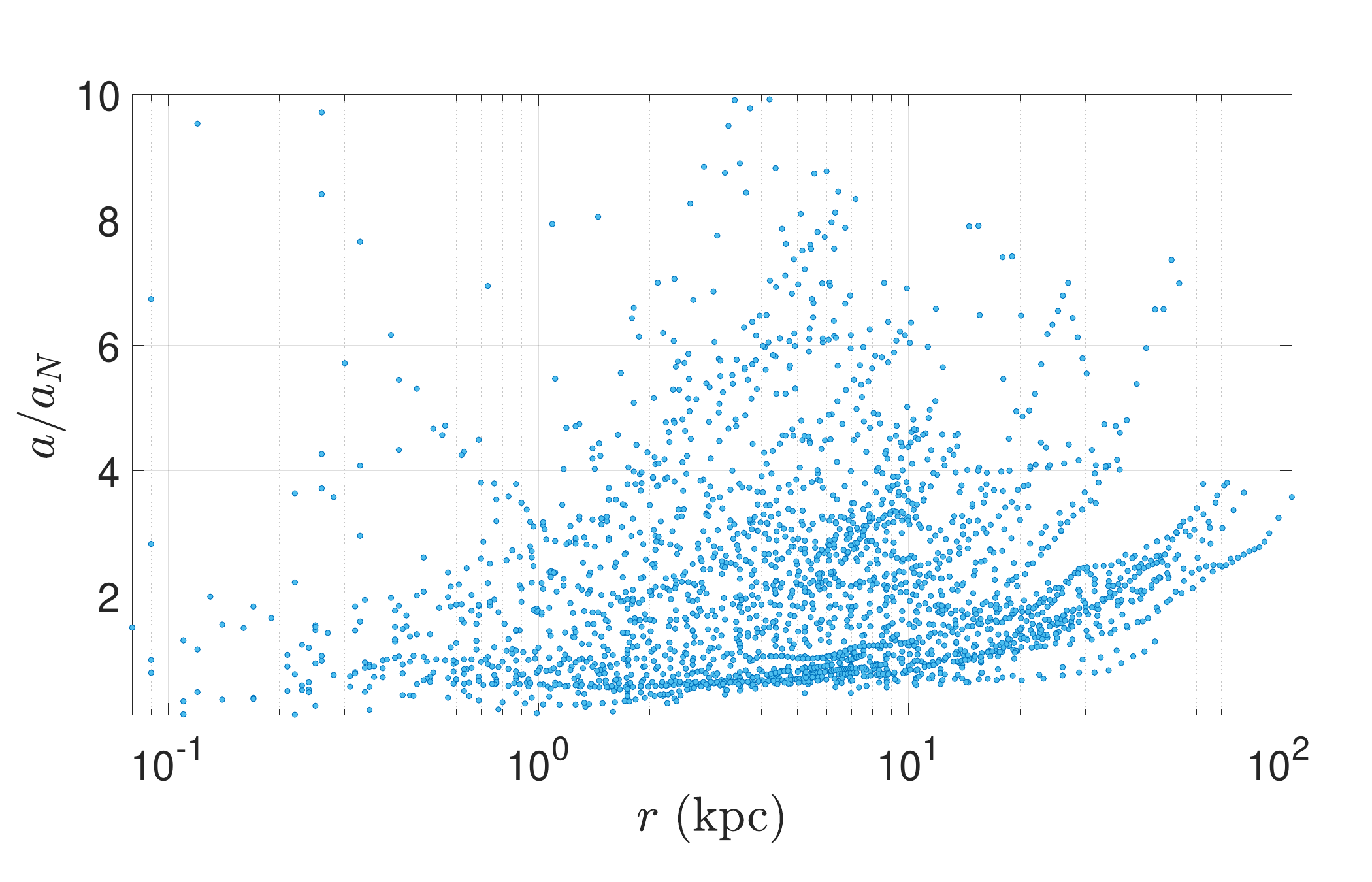}  
    \includegraphics[trim=0cm 0cm 1cm 1cm, clip=true,width=0.49\columnwidth]{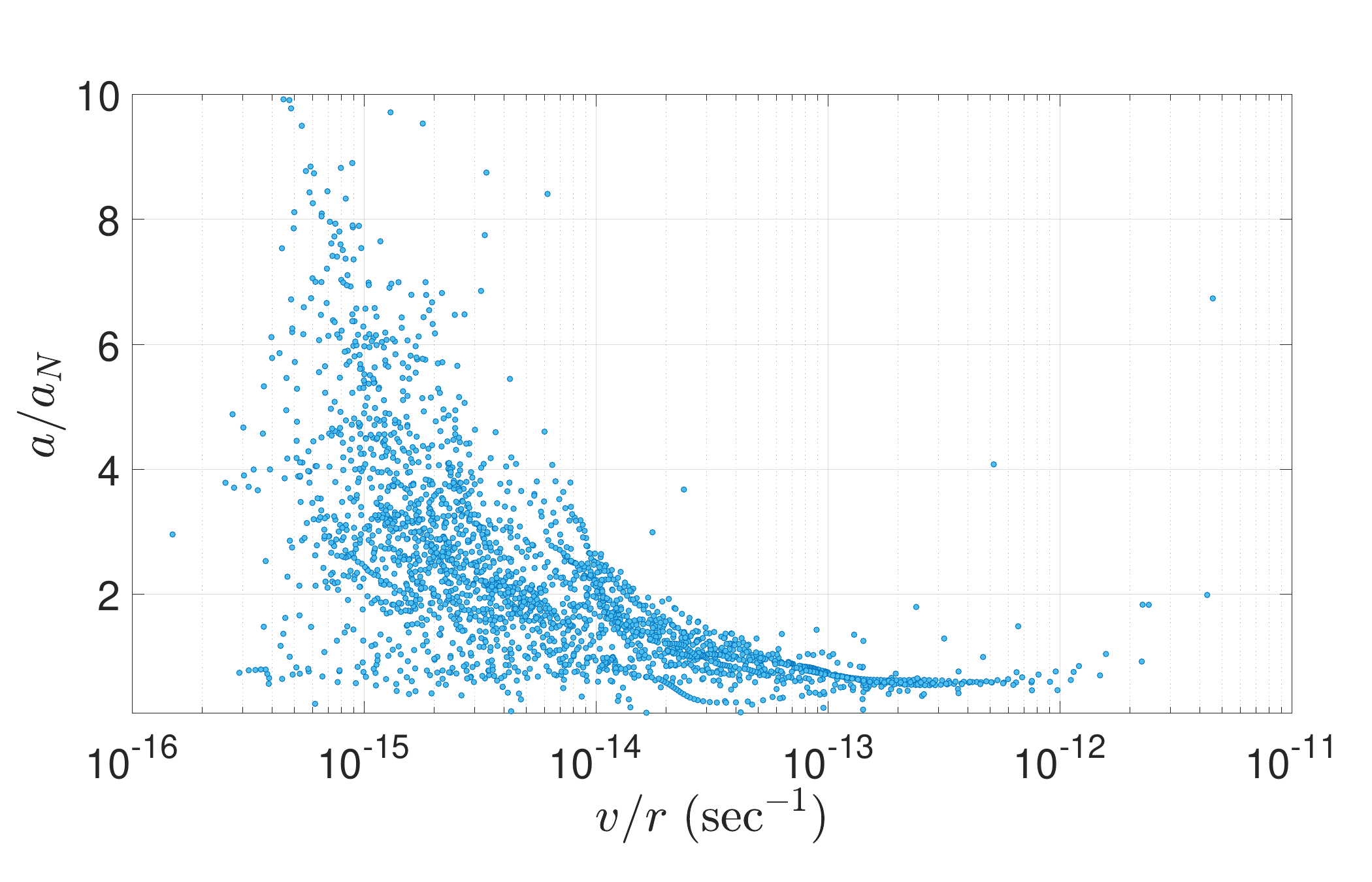}   
    \includegraphics[trim=0cm 0cm 1cm 1cm, clip=true,width=0.49\columnwidth]{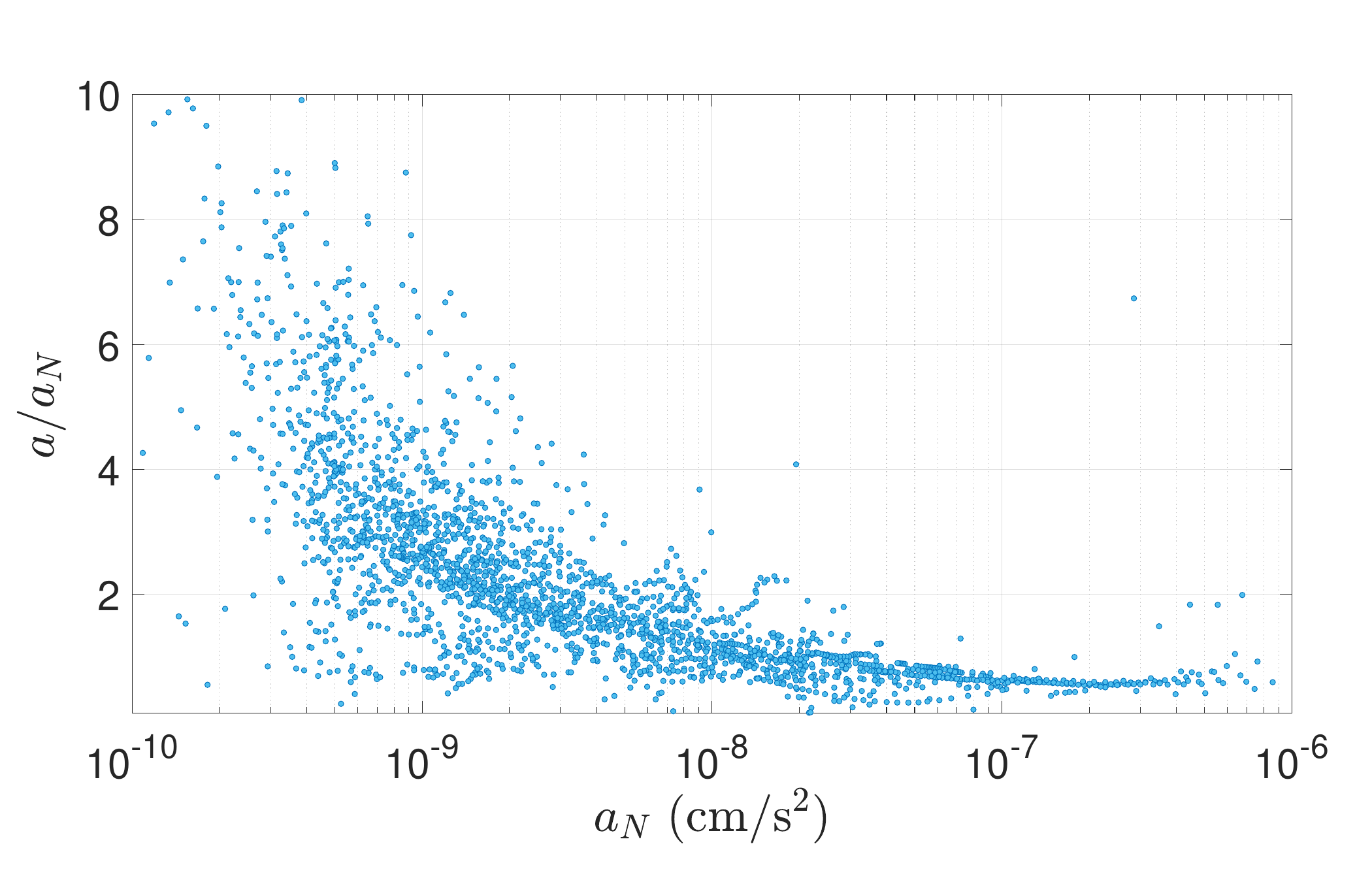}  
    \includegraphics[trim=0cm 0cm 1cm 1cm, clip=true,width=0.49\columnwidth]{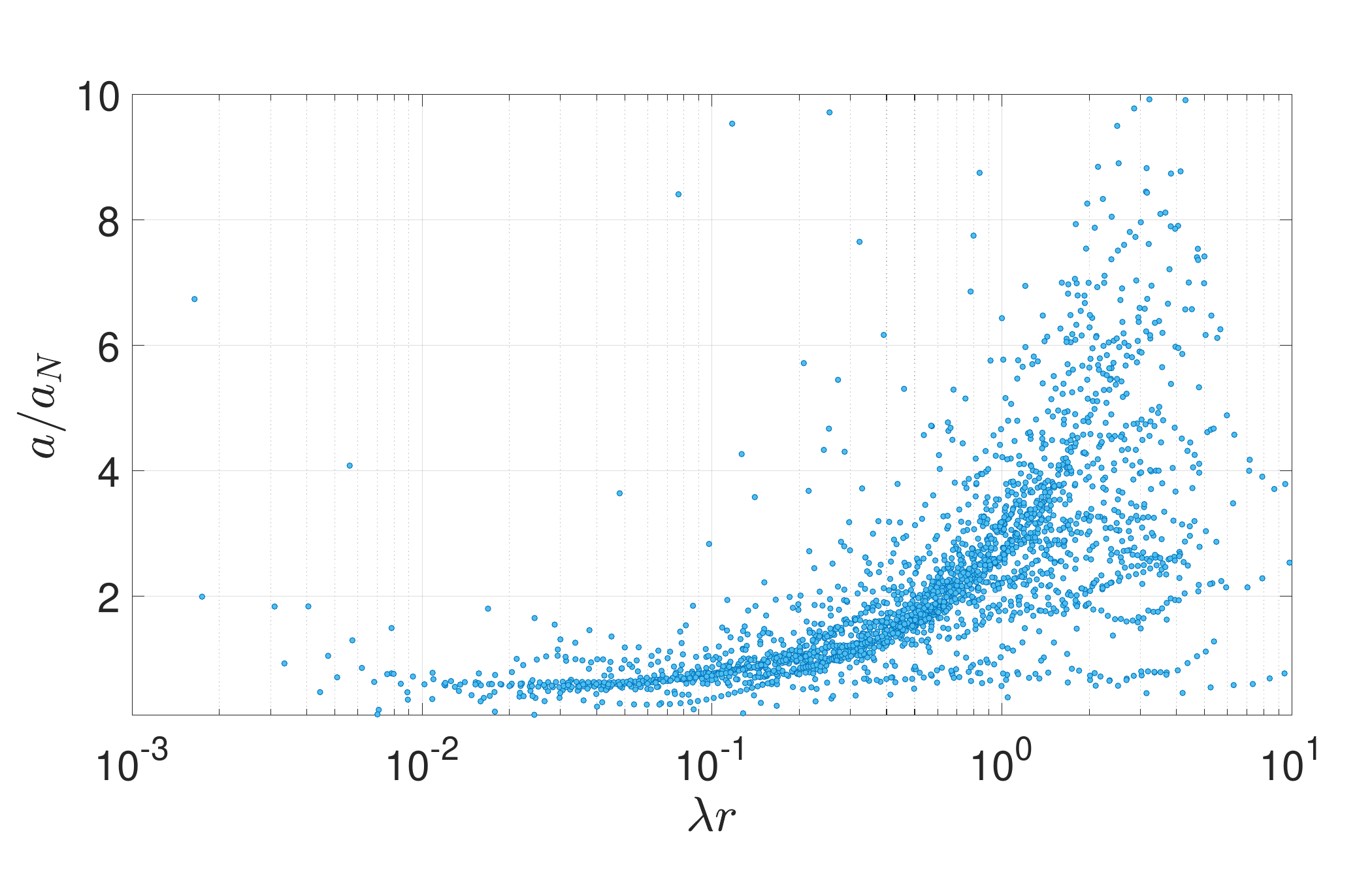} 
    \caption{The figure illustrates the relationship between mass discrepancy  to four variables: $r$, $v/r$, $a_N$, and $\lambda r$. These plots are based on a dataset comprising 2385 data points from the SPARC galaxies. None of the plots exhibit a very prominent correlation. Specifically, the top two plots display relatively weaker correlations and appear more dispersed, whereas the bottom two plots show relatively stronger correlations. }
    \label{fig:alpha_values}
\end{figure}

In the SPARC dataset, the baryonic velocity is divided into three components, the gas velocity $v_\text{gas}$, disk velocity $v_\text{disc}$, and the bulge velocity $v_\text{bul}$. The total baryonic velocity is defined as 

\begin{equation}
    v_\text{bar} = \sqrt{\epsilon_\text{gas}v_\text{gas}^2 + \epsilon_\text{disc}\gamma_\text{disc} v_\text{disc}^2 + \epsilon_\text{bul} \gamma_\text{bul} v_\text{bul}^2} \;.
\end{equation}

\noindent $\gamma_\text{disc}$ and $\gamma_\text{bul}$ are the mass to light ratio of the stars in the disc and the bulge. $\epsilon_{\ldots}$ represents the signature of different components of the velocities, which in some cases can be negative. Most importantly $V_\text{gas}$ can sometimes be negative in the innermost regions. This occurs when
the gas distribution has a significant central depression and the material in the outer regions exerts a stronger gravitational force than that in the inner parts~\cite{Lelli_2016}.

In the SPARC dataset, the mass-to-light ratios for the disk and bulge components are fixed to $\gamma_{\mathrm{disc}} = 0.5$ and $\gamma_{\mathrm{bul}} = 0.7$, respectively. These constant values are adopted in our calculations. Ideally, one expects $v_{\mathrm{obs}} > v_{\mathrm{bar}}$. However, in the SPARC dataset there exist several galaxies for which $v_{\mathrm{bar}}/v_{\mathrm{obs}} > 1$, particularly at small radii. While this behavior is observed in multiple systems, the discrepancy becomes significant for approximately 21 galaxies. In addition, for many galaxies the assumption of a constant mass-to-light ratio may not be appropriate as we discussed in later section and noted inin Ref.~\cite{Lelli_2016}.

For our calculations, we have used $v_\text{bar}$ given in the SPARC data set and then we obtain $M(r)$ inside a radius $r$ using $M(r)=v_\text{bar}^2 r/G$. In Fig.~\ref{fig:tully-fisher},  we have plotted the velocity at the outermost data point of all 175 galaxies against mass of the galaxies inside that radius. We can see that there is a very strong correlation between the mass of the galaxy and to velocity in the outermost data-points, in the log-log plot. The best linear fit is given by

\begin{equation}
    \log(v) = 0.26 \log (M) - 0.5989\,.
\end{equation}

\noindent This is equivalent to $M\sim v^{3.85}$, which is in perfect agreement with the Tully-Fisher relation~\cite{tully1977new}.

\begin{figure}
    \centering
    \raisebox{2pt}{\includegraphics[width=0.47\linewidth]{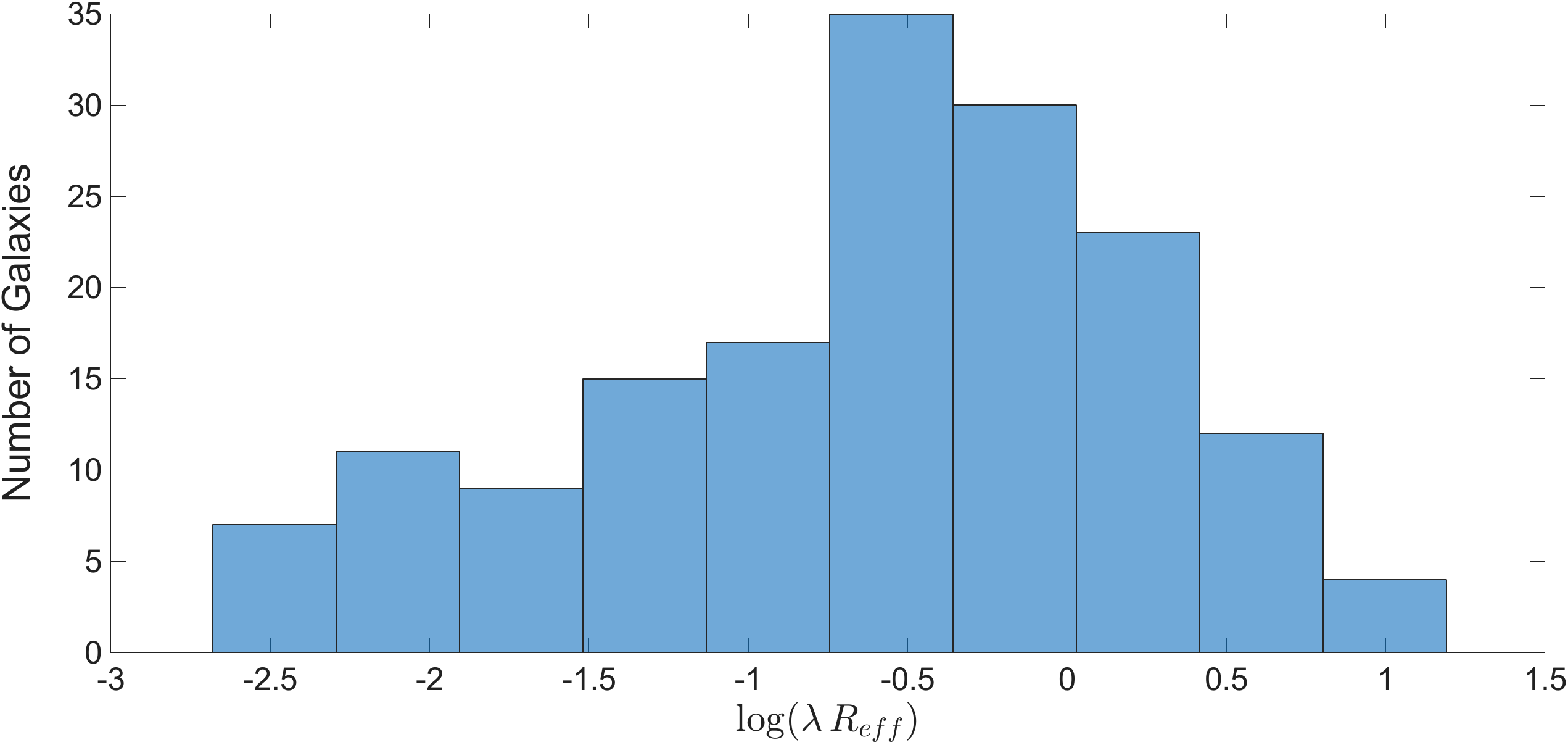}}
    \raisebox{0pt}{\includegraphics[width=0.47\linewidth]{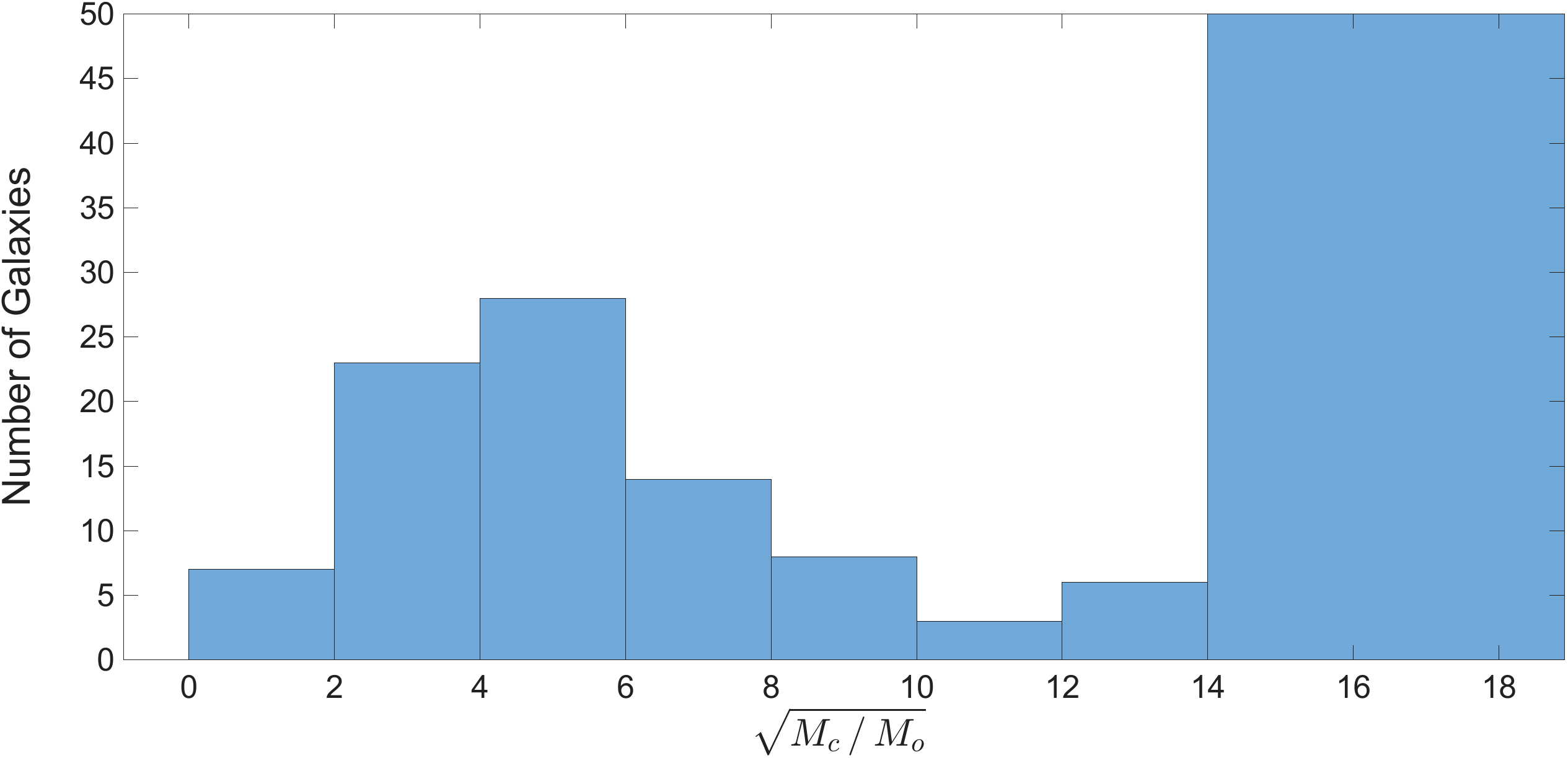}}
    \caption{Here $R_{\text{eff}}$ is the effective radius of the galaxy encompassing
 half of the total luminosity. Except 6 galaxies (not included in the plot), for all the galaxies the $\lambda R_{\text{eff}}$ is within $15$. If I remove another 8 galaxies then for all other galaxies $\lambda R_{\text{eff}}$ is within $3.7$. Its very interesting to see that within this  range our formula provides same result as MOND equation. In the second plot $M_o$ is the baryonic mass of the galaxy at the outer most data point. In the histogram of $\sqrt{M_c / M_o}$ I have listed all the quantities $> 13$ in a single bin. This is because for most of the things the $\lambda R_\text{eff}$ is very small and therefore the second quantity in Eq.~\ref{eq:velocityFinal} is almost $0$ making the second part irrelevant.}
    \label{fig:acc_descripency}
\end{figure}

\begin{figure}[t]
    \centering
    \includegraphics[trim=4cm 0cm 4cm 0cm, clip=true,width=0.96\columnwidth]{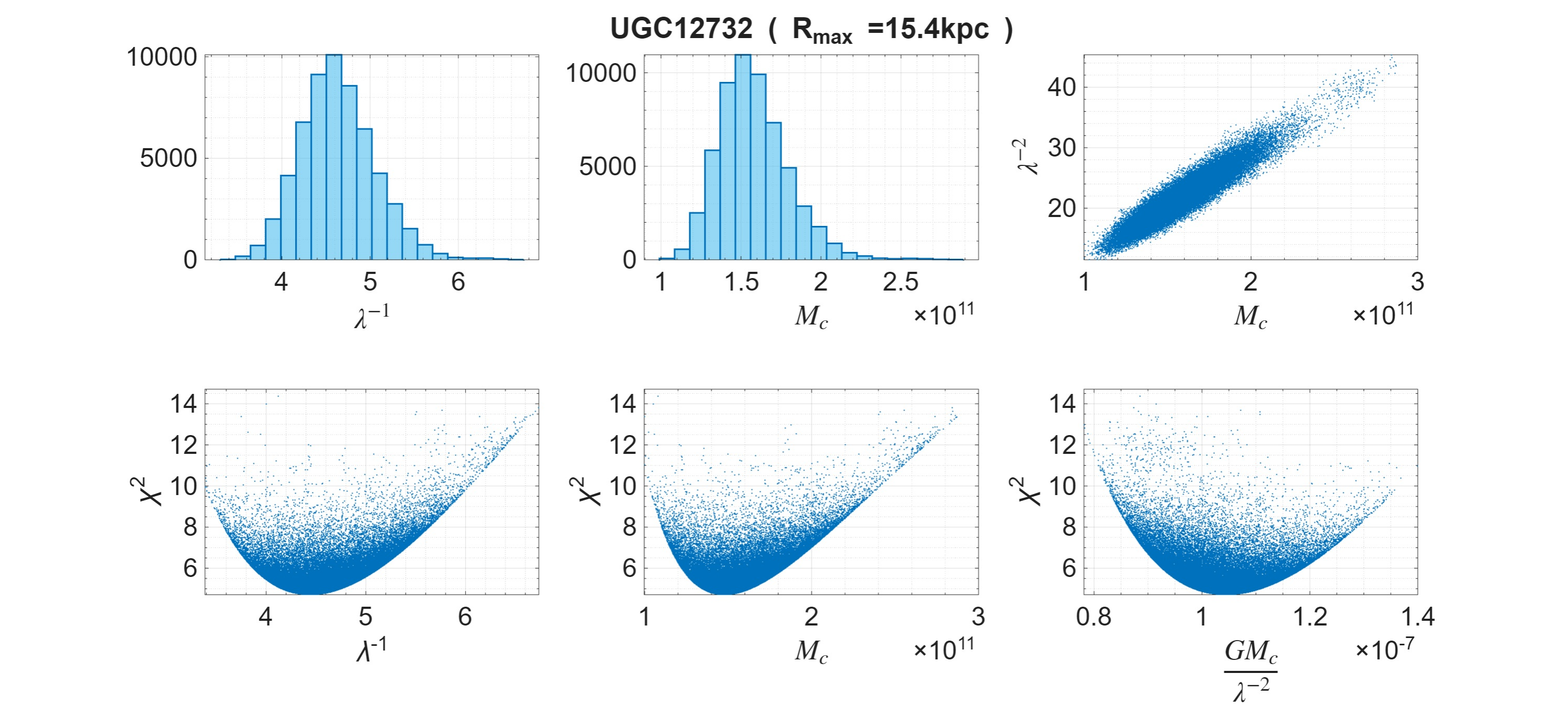}
    \includegraphics[trim=4cm 0cm 4cm 0cm, clip=true,width=0.96\columnwidth]{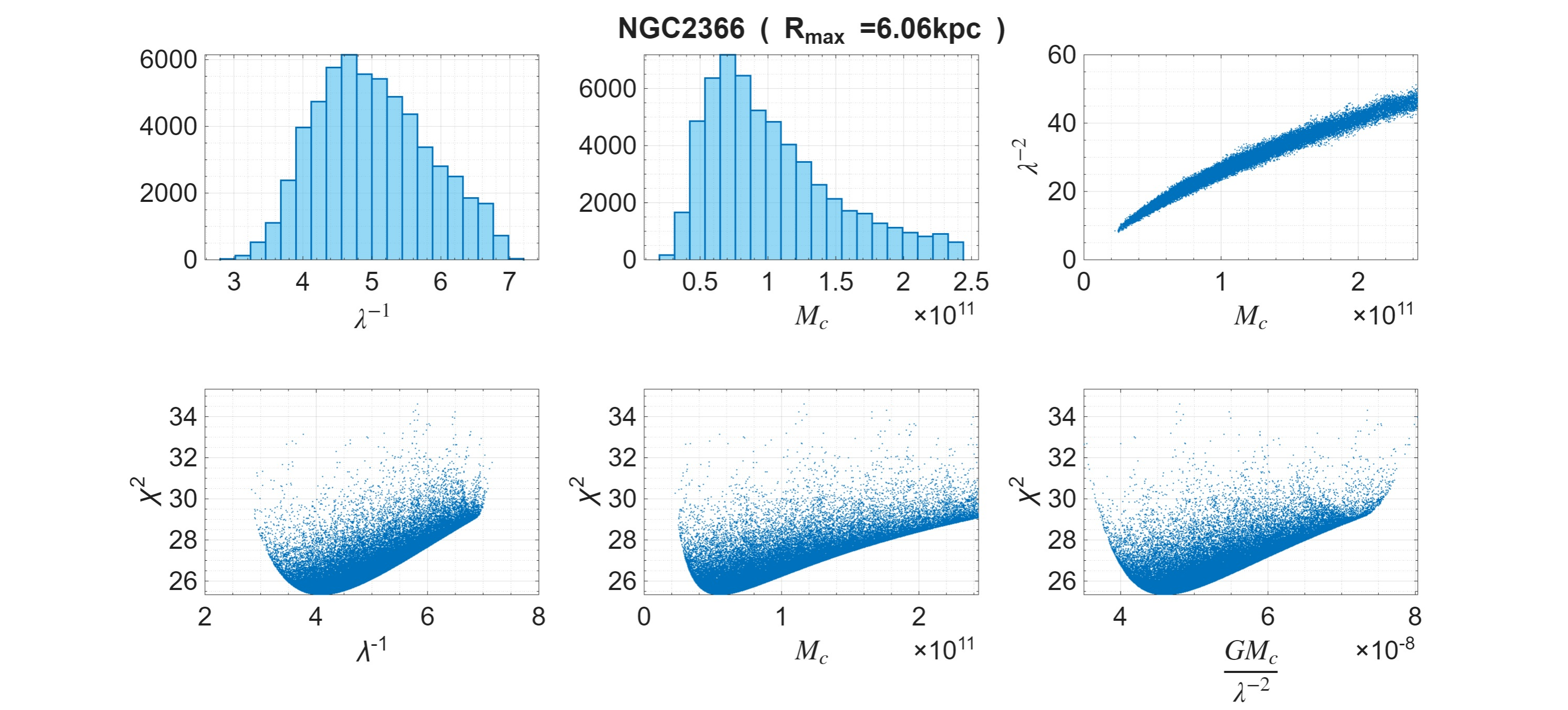}
    \caption{The illustration portrays the distributions of $\lambda^{-1}$ and $M_c$, their correlation, and the relationship of $\chi^2$ with these parameters, alongside $a_0 = \frac{GM_c}{\lambda^{-2}}$. In the figure, $\lambda^{-1}$ is scaled in kiloparsecs ($\rm kpc$) and $M_c$ is scaled in solar masses ($M_\odot$). Notably, the value of $a_0$ for the best-fit $\chi^2$ is approximately on the order of $10^{-8} \, \rm cm/s^2$. The value of $R_{\text{max}}$ corresponds to the radius of the last velocity data point for each galaxy. Importantly, $\lambda^{-1}$ aligns roughly with the order of $R_\text{max}$. Analysis of other galaxies can be found \href{https://machiangravity.github.io/Galactic-Velocity-Profile-/two_param_Mc_o_M(r)/}{here}.}
    \label{fig:mcmc}
\end{figure}

\begin{figure}
    \centering
    \includegraphics[trim=3cm 0cm 4cm 1cm, clip=true, width=0.24\linewidth]{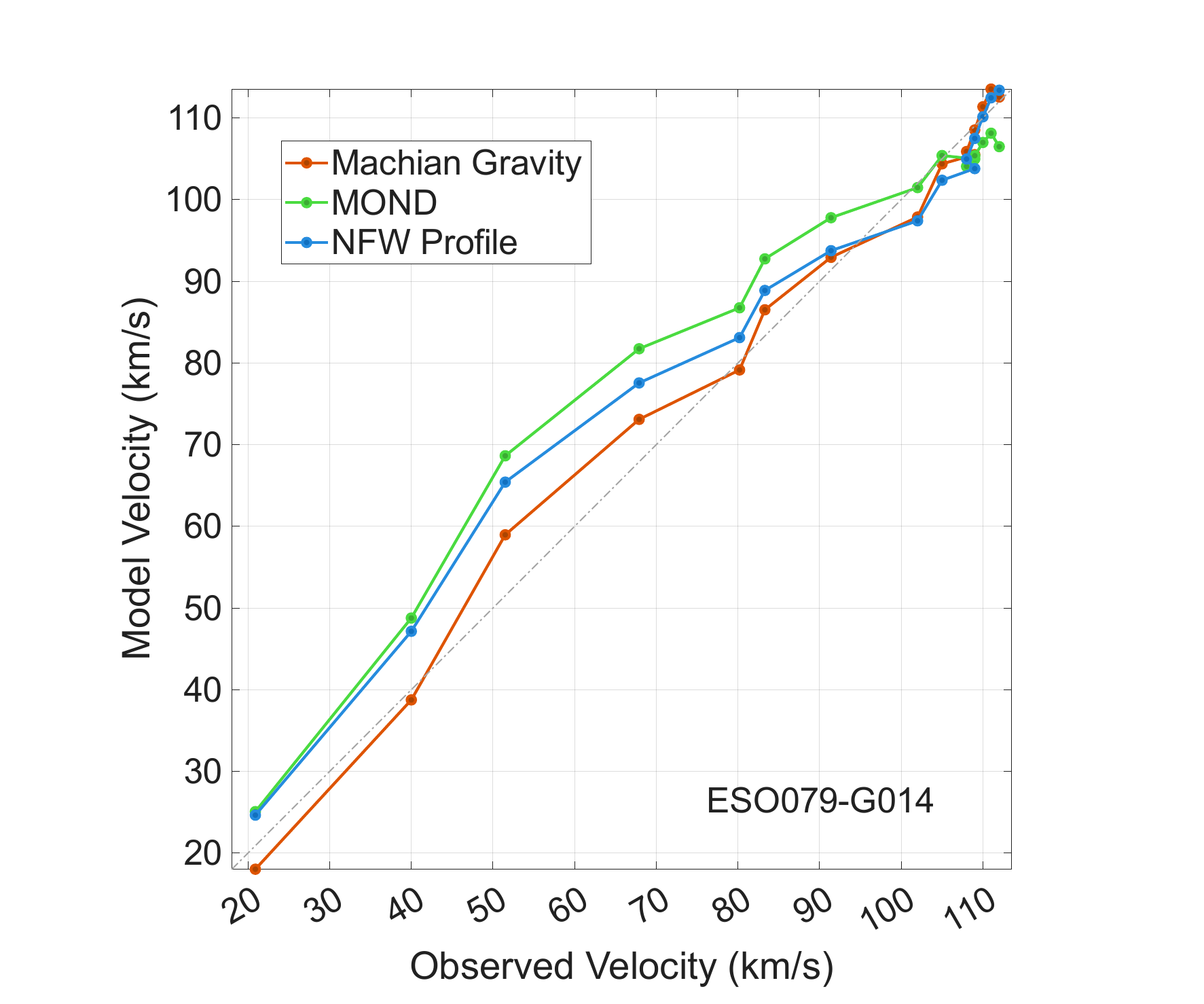}
    \includegraphics[trim=3cm 0cm 4cm 1cm, clip=true, width=0.24\linewidth]{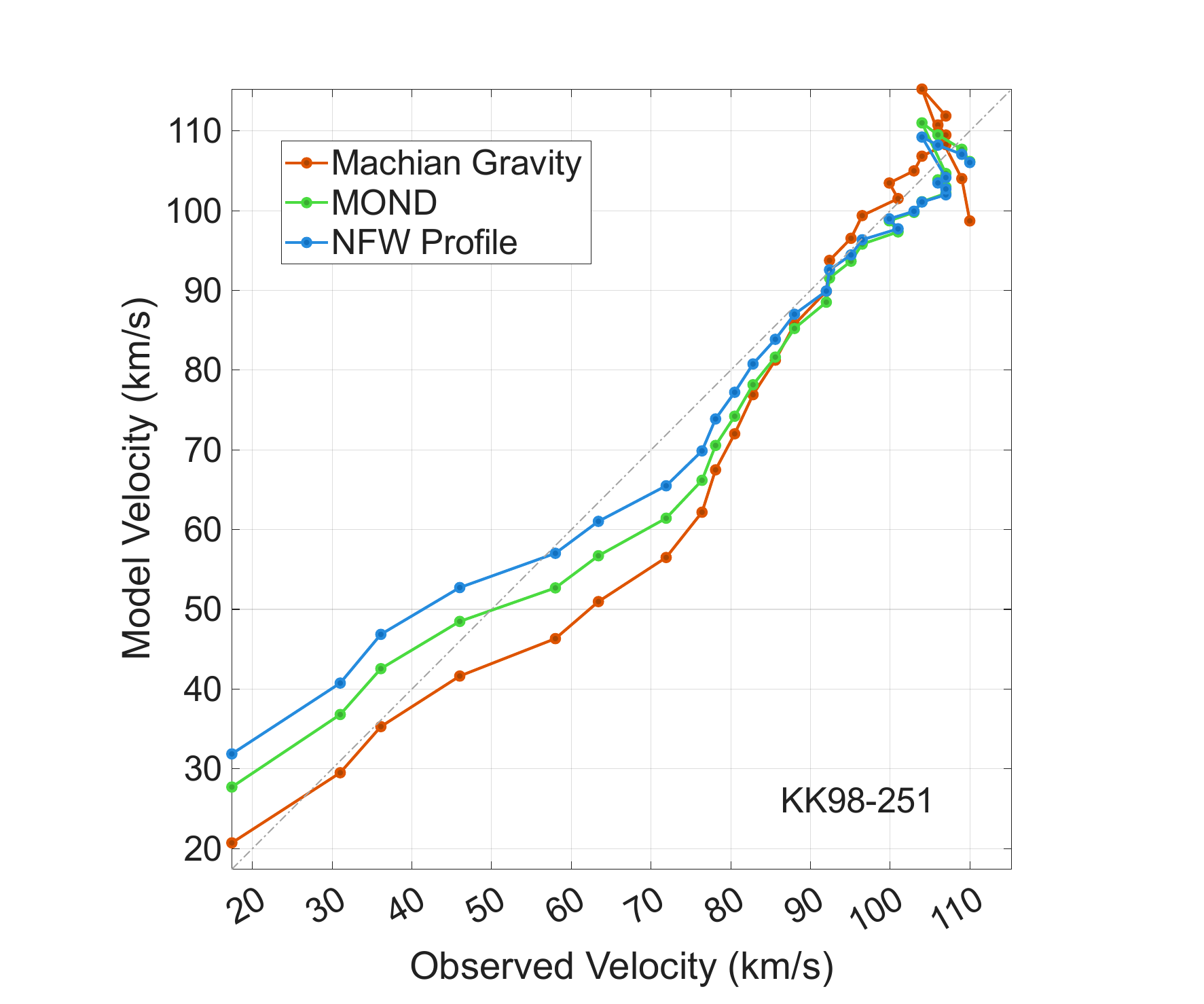}
    \includegraphics[trim=3cm 0cm 4cm 1cm, clip=true, width=0.24\linewidth]{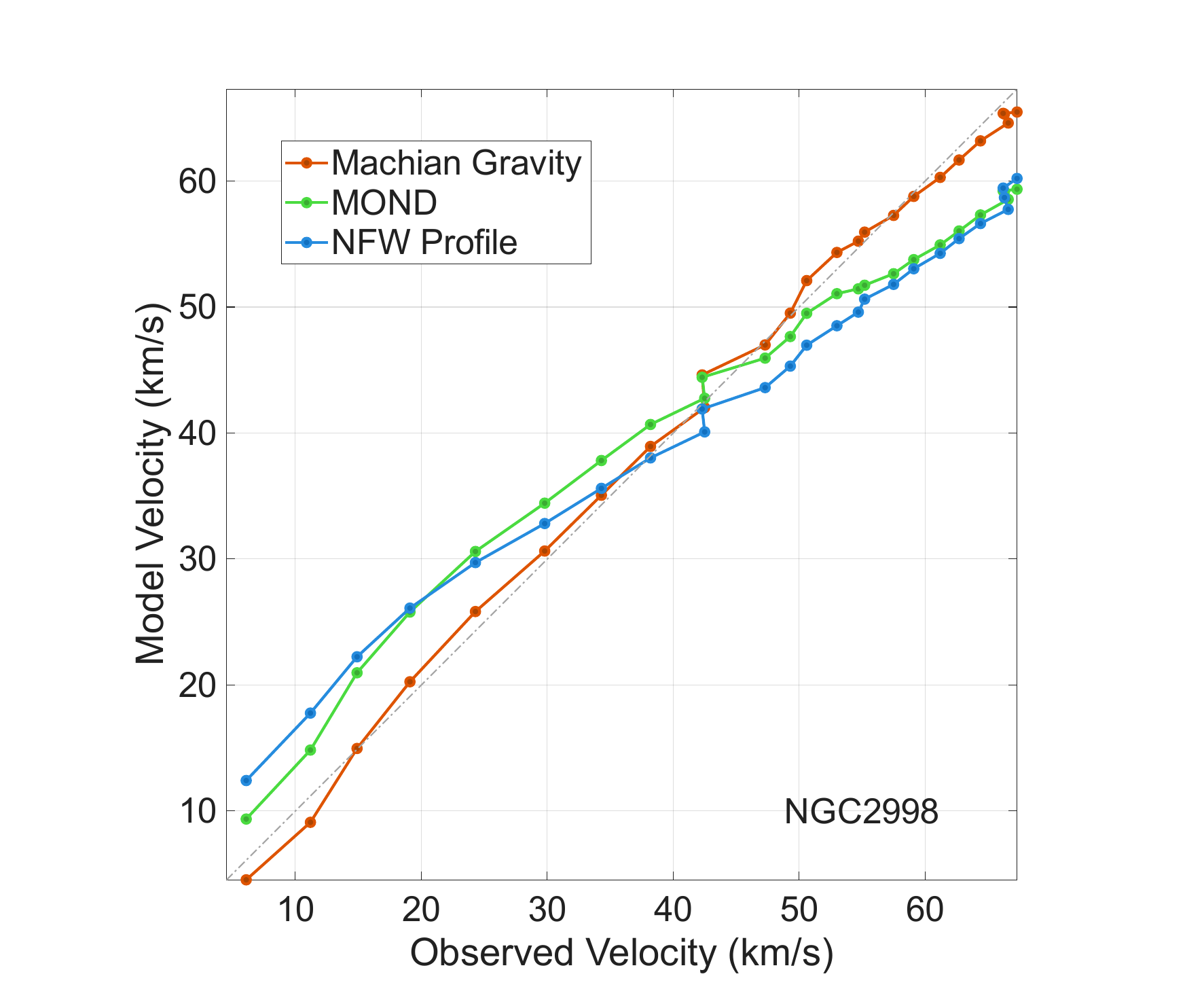}
    \includegraphics[trim=3cm 0cm 4cm 1cm, clip=true, width=0.24\linewidth]{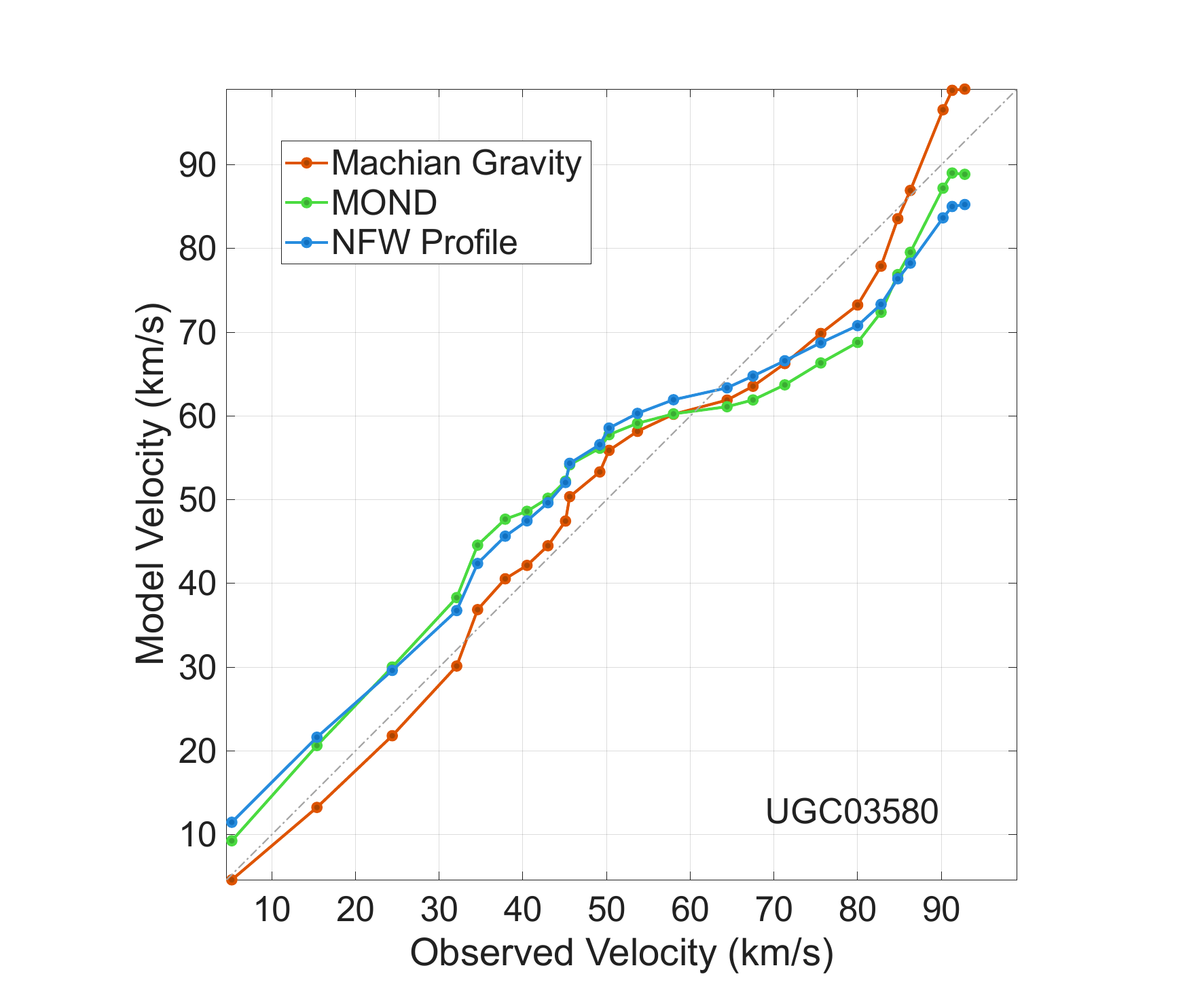}
    \includegraphics[trim=3cm 0cm 4cm 1cm, clip=true, width=0.24\linewidth]{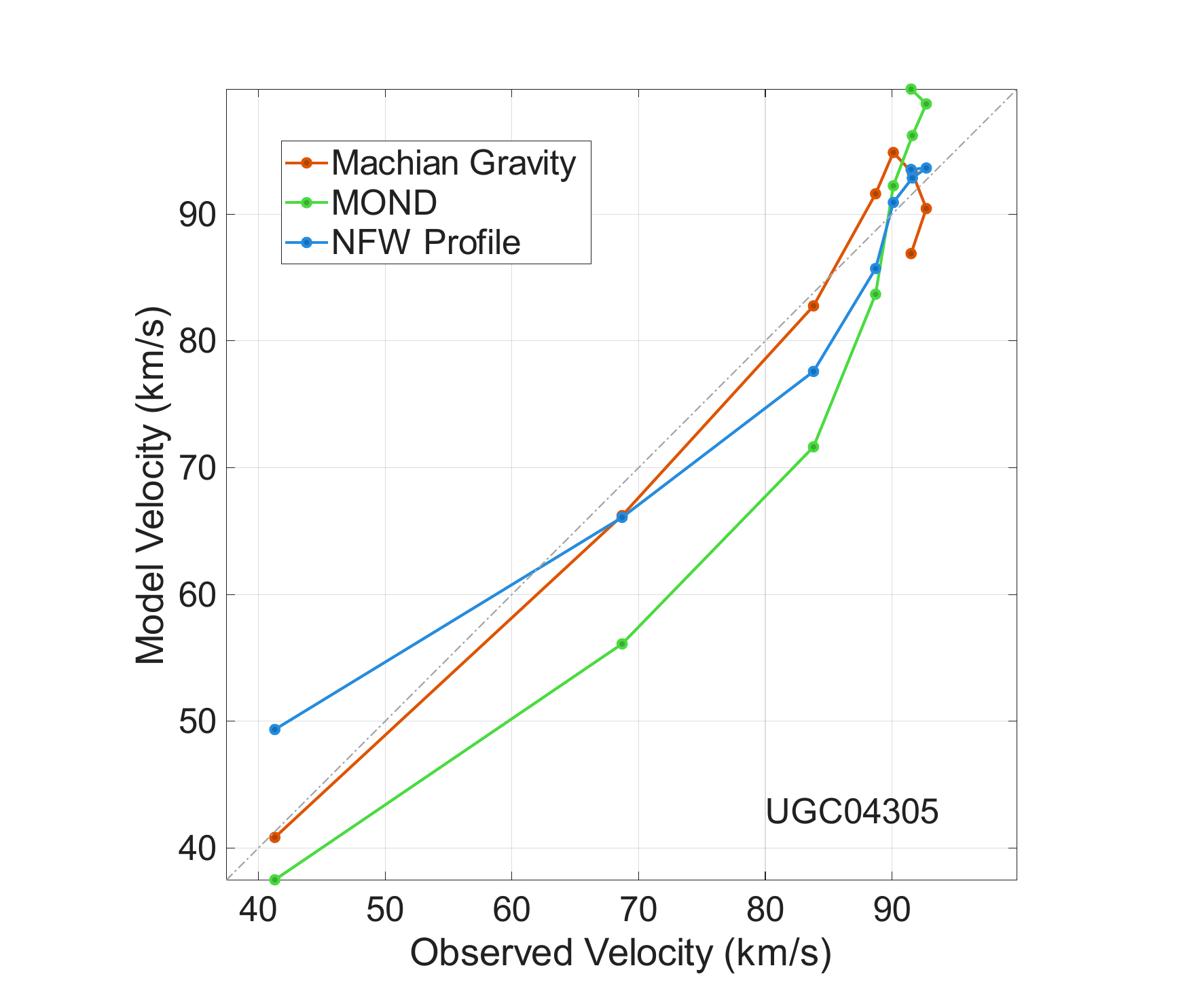}
    \includegraphics[trim=3cm 0cm 4cm 1cm, clip=true, width=0.24\linewidth]{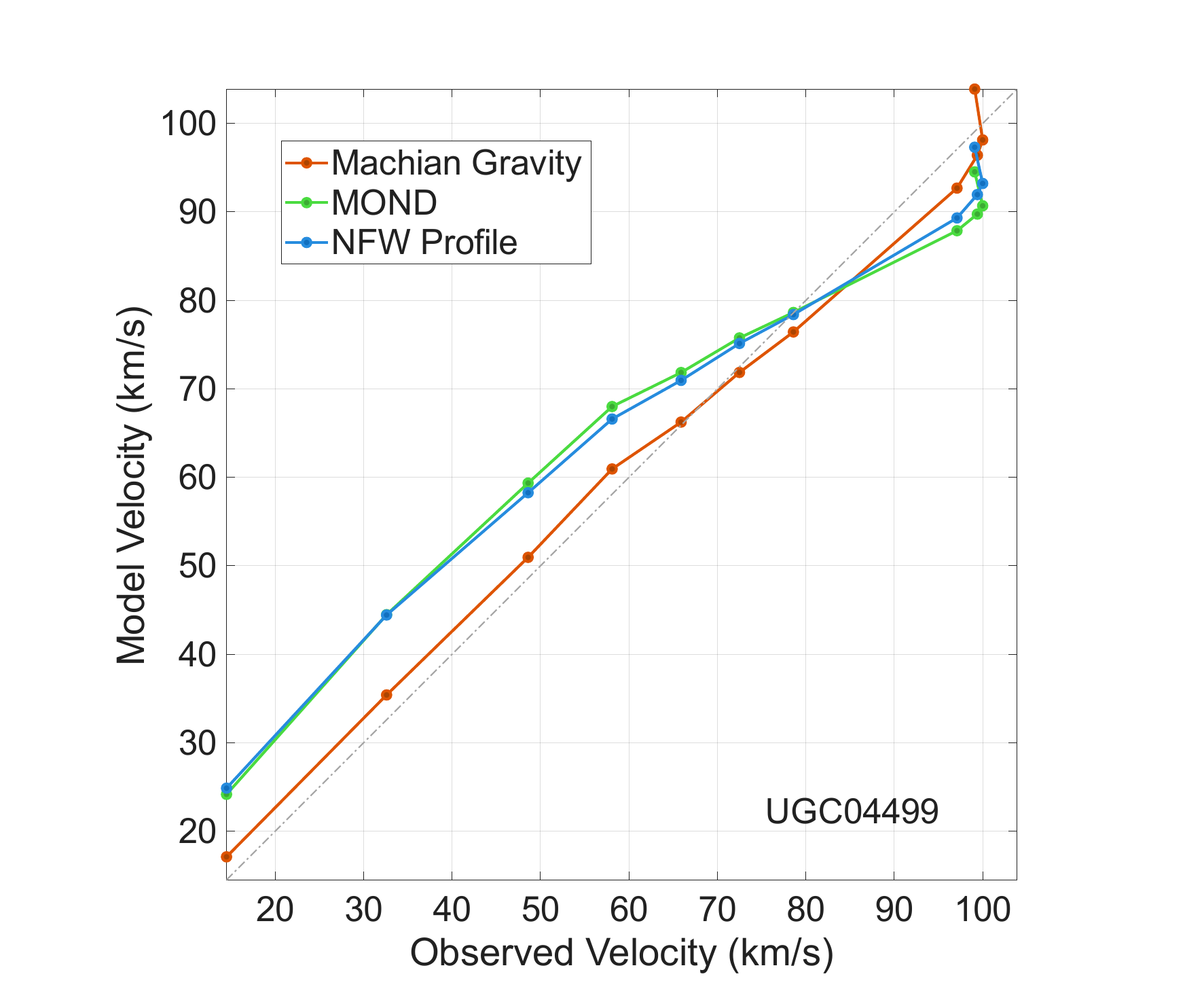}
    \includegraphics[trim=3cm 0cm 4cm 1cm, clip=true, width=0.24\linewidth]{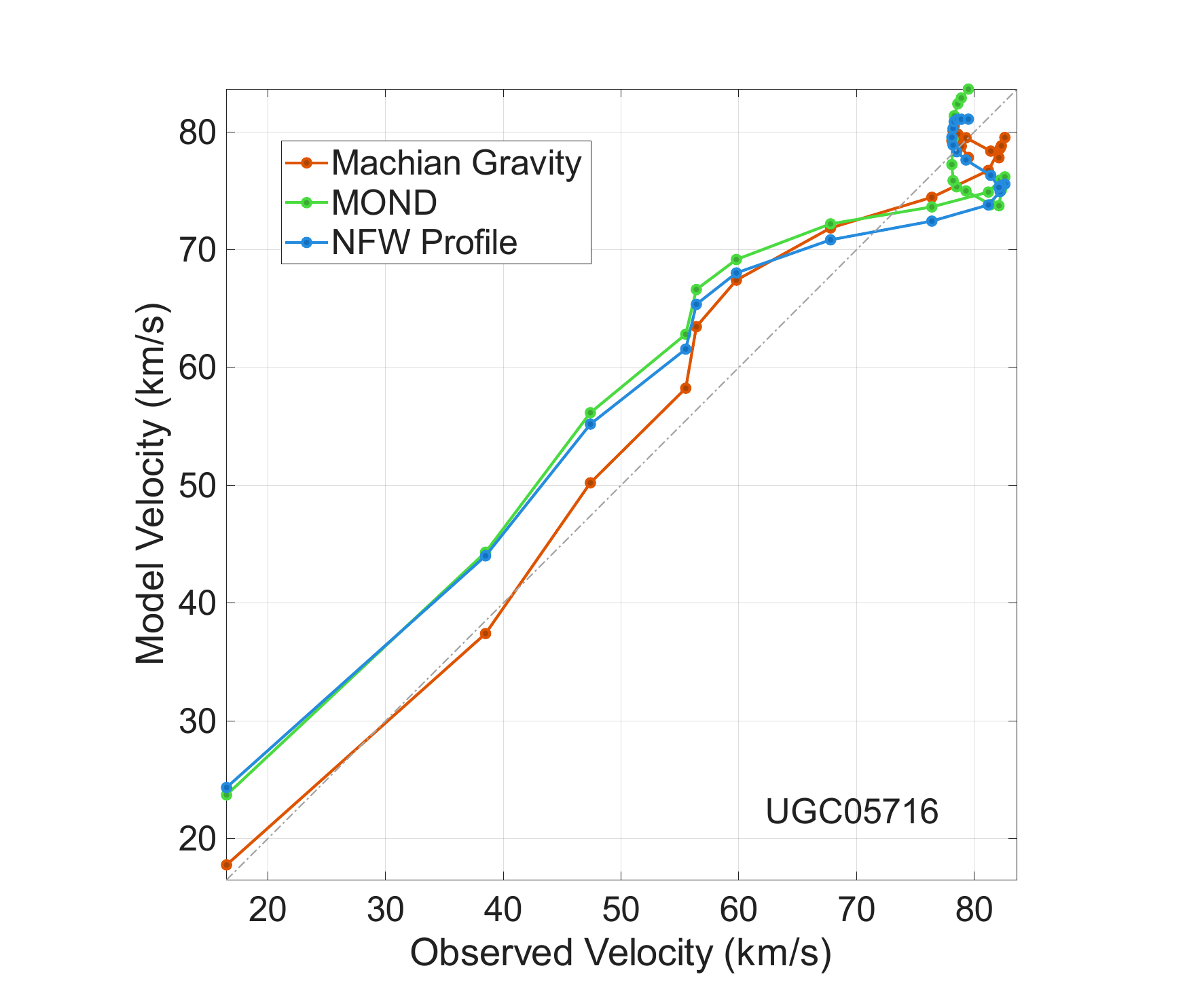}
    \includegraphics[trim=3cm 0cm 4cm 1cm, clip=true, width=0.24\linewidth]{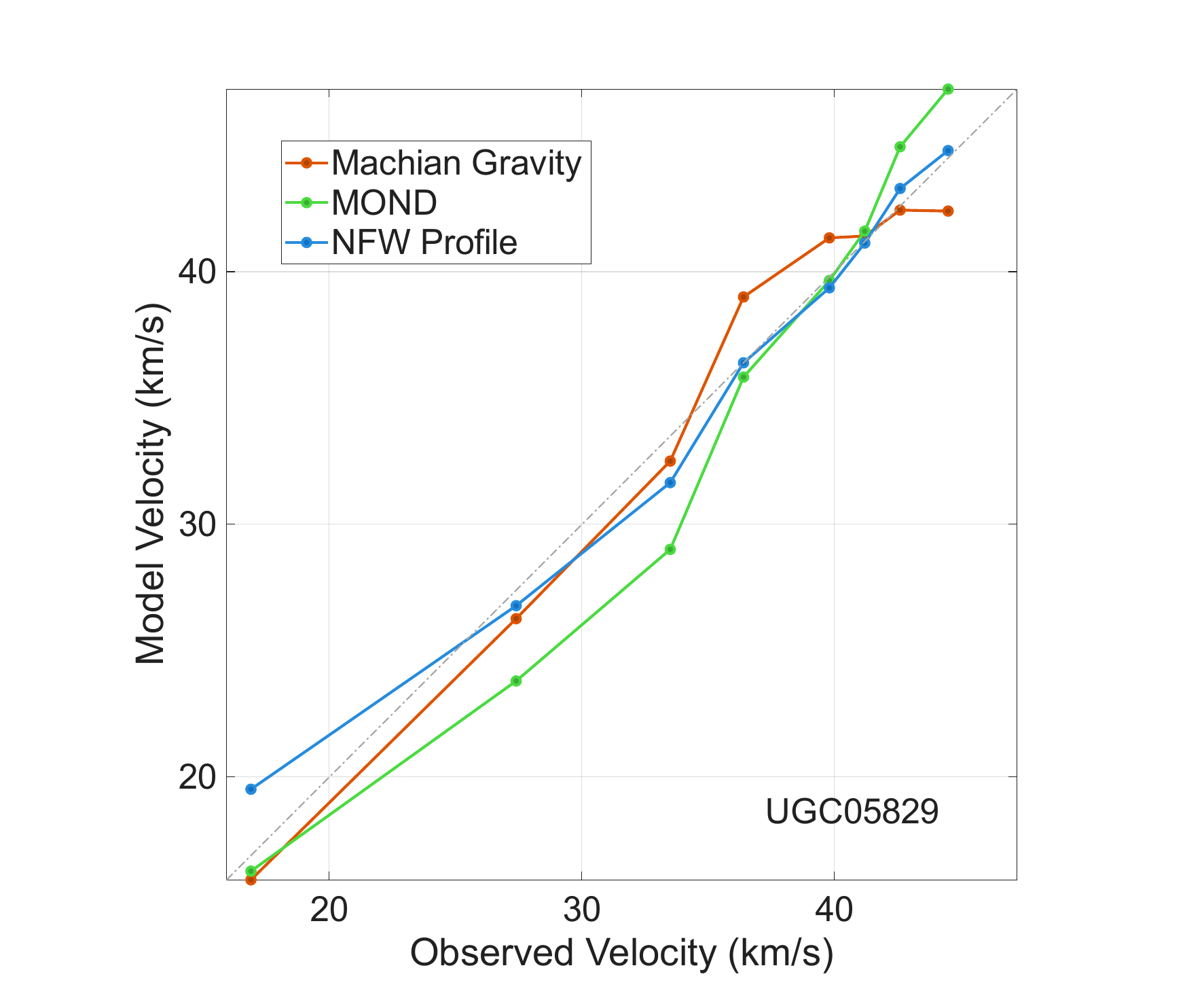}
    \includegraphics[trim=3cm 0cm 4cm 1cm, clip=true, width=0.24\linewidth]{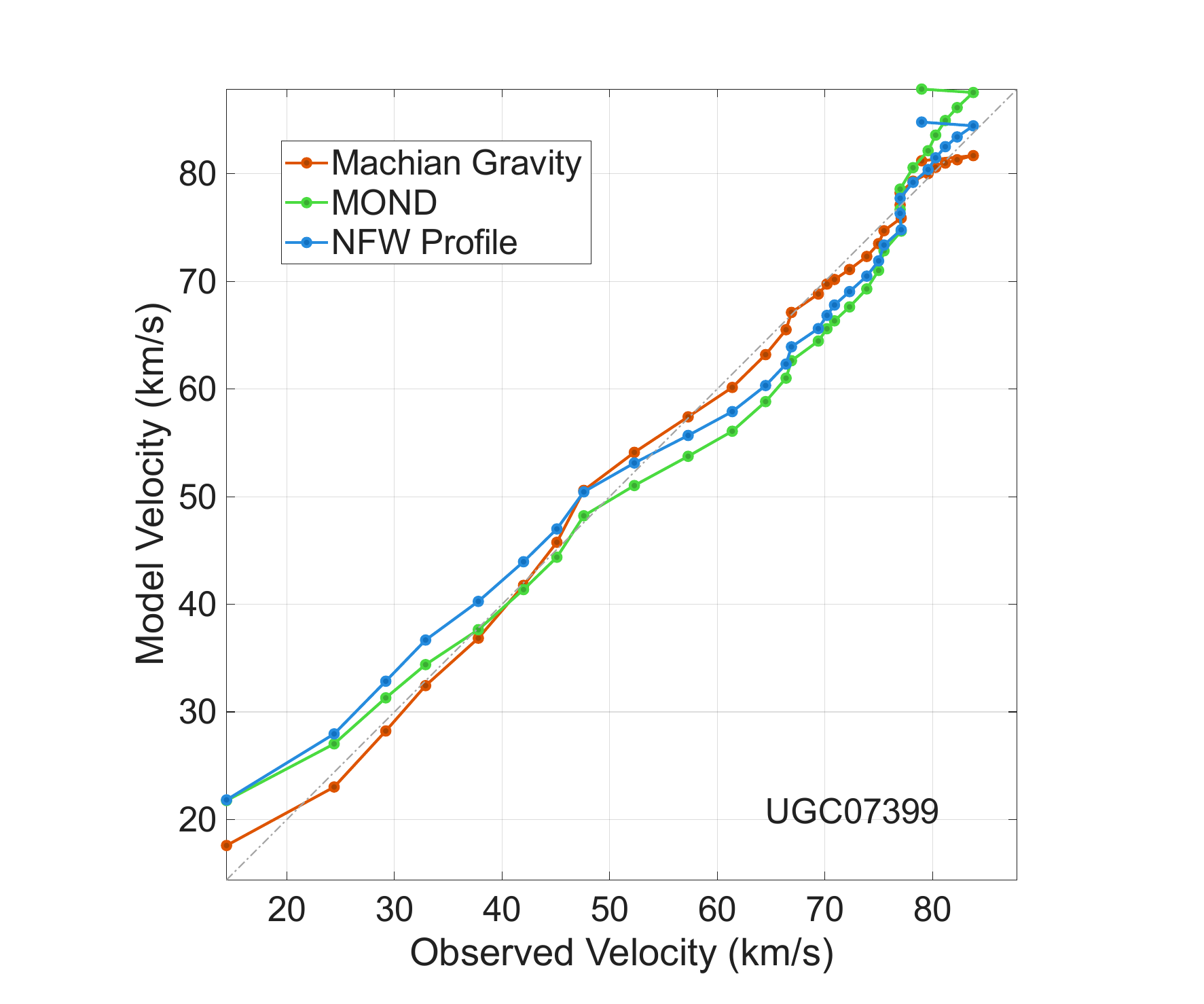}
    \includegraphics[trim=3cm 0cm 4cm 1cm, clip=true, width=0.24\linewidth]{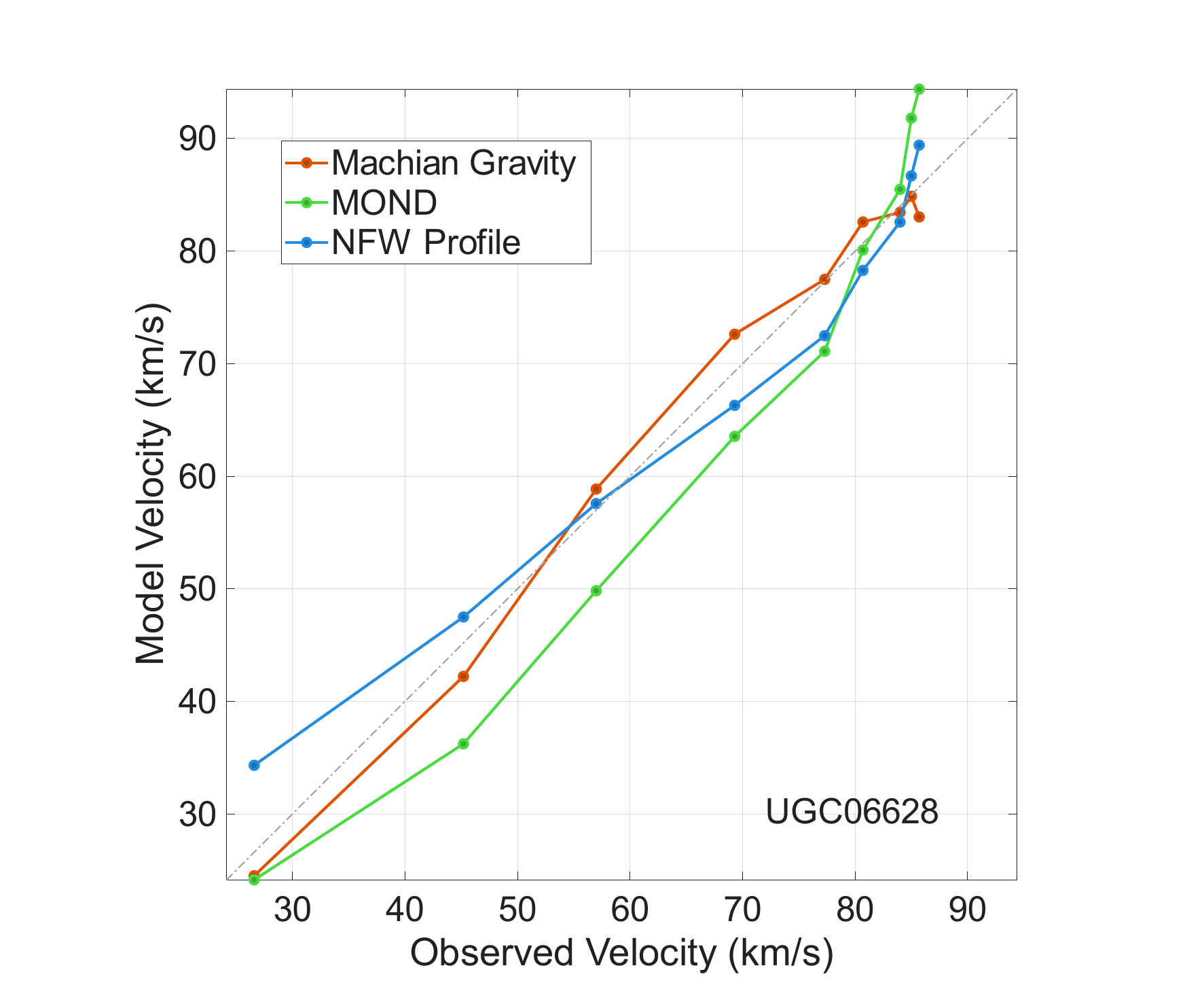}
    \includegraphics[trim=3cm 0cm 4cm 1cm, clip=true, width=0.24\linewidth]{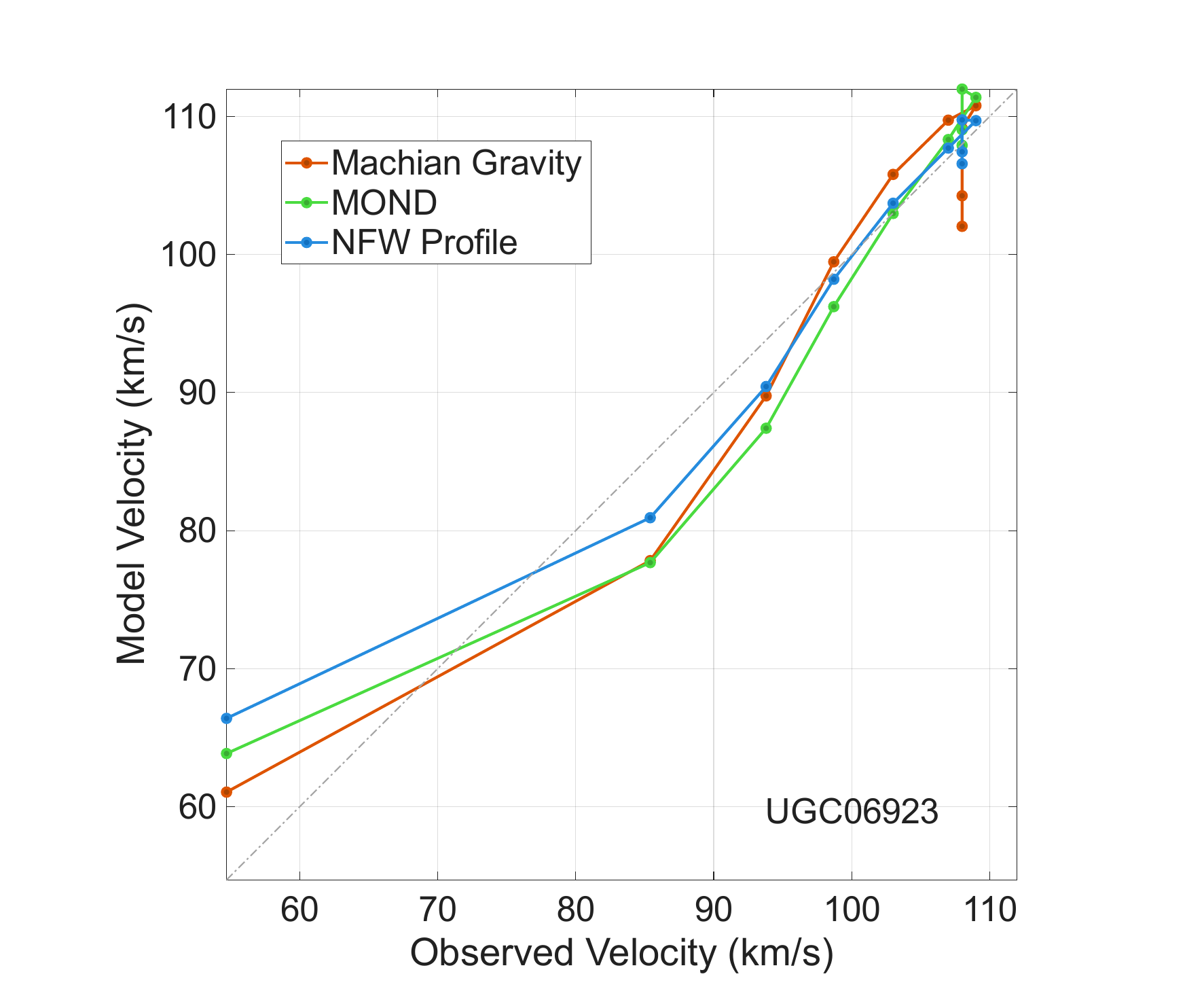}
    \includegraphics[trim=3cm 0cm 4cm 1cm, clip=true, width=0.24\linewidth]{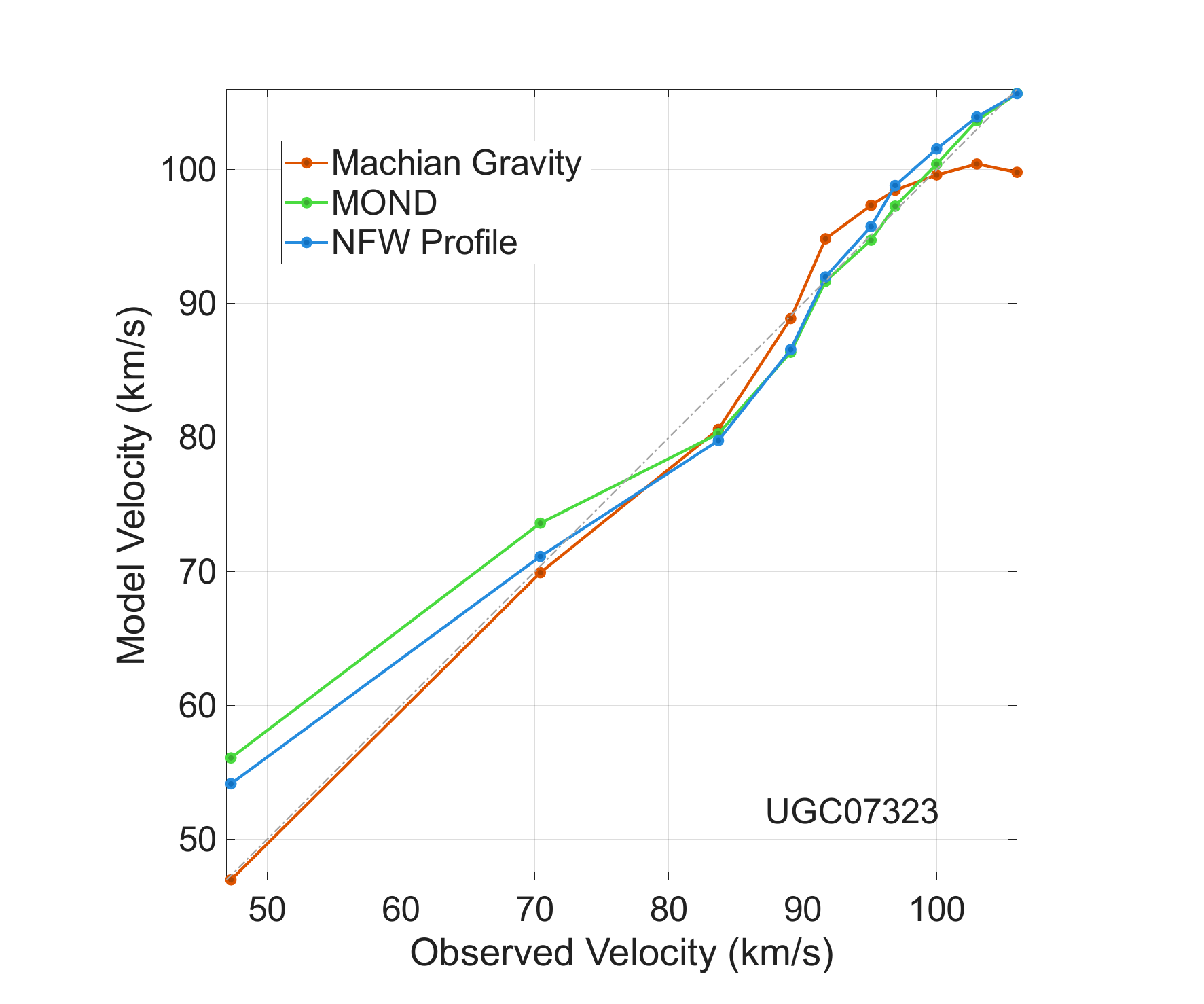}    
    \caption{Along the X-axis I have plotted the observed velocities, and along the Y-axis I have shown the velocities predicted by the different models. If a model predicts the data accurately, its points should lie on the diagonal dotted reference line. In several cases, the curves deviate significantly from this line. However, for most galaxies, the Machian gravity predictions lie much closer to the diagonal compared to the other two models. NFW and Machian gravity each have two free parameters, while MOND uses only one. I also observe that at smaller radii, both the NFW profile and MOND slightly tend to overpredict the velocities, whereas the Machian gravity results remain closer to the diagonal reference line. Full analysis can be found \href{https://machiangravity.github.io/Galactic-Velocity-Profile-/Velocity_Velocity_Comparison/}{here}.}
    \label{fig:velocityVelocity}
\end{figure}

\section{\label{sec:section5}Testing the theory against observations}

The core objective of this article is to investigate whether Machian gravity can offer an explanation for observed galactic velocity profile. Fig.~\ref{fig:alpha_values} displays the mass discrepancy of the galaxies, given by $\frac{a}{a_{N}}$ against various parameters,where $a$ is the observed acceleration of a star at radius $r$ and $a_N$ Newtonian acceleration calculated using the observed mass of a galaxy. This discrepancy  can be mathematically expressed as~\cite{mcgaugh2004mass}:

\begin{equation}
    \frac{a}{a_\text{N}} = \frac{v_\text{obs}^2}{v_\text{bar}^2}\;.
\end{equation}

In our investigation, we exclude 21 galaxies with $v_\text{bar}/v_\text{obs}>1$, as these are likely a consequence of fixed mass-to-light ratio selections, as discussed in~\cite{Lelli_2016}. Among the remaining  154  galaxies, several data points exhibit $v_\text{bar}/v_\text{obs}>1$ at smaller radii, although the errors may not be severe. Overall, we plot a total of $2385$ data points.

Different galaxies have distinct rotation curves. Given the substantial dataset encompassing numerous galaxies, it is expected that the resulting plot would exhibit a scattered pattern. This is indeed evident in the first plot (top-left) of Fig.~\ref{fig:alpha_values}, which illustrates mass discrepancies in relation to radius.  We can see that there are galaxies where the mass discrepancy is not apparent at low radii while there are other galaxies where the mass discrepancy kicks in even at a very small radius.   The  plot looks significantly scattered and no obvious relation between the mass discrepancy and the radius is seen in the plots. 

A similar trend is observed in the subsequent (top-right) panel, which shows the mass discrepancy as a function of orbital angular velocity. Although this plot exhibits a slightly higher degree of correlation with the mass discrepancy, the corresponding scatter indicates that the correlation is not particularly strong. In contrast, the lower-left panel, which displays the mass discrepancy as a function of the Newtonian acceleration, reveals a much clearer correlation. In general, the mass discrepancy becomes significant below an acceleration scale of $a \sim 10^{-8}\,\mathrm{cm\,s^{-2}}$, thereby establishing a direct connection between the Newtonian acceleration and the observed mass discrepancy. This relationship has been extensively explored in the literature (see, e.g., Refs.~\cite{mcgaugh2004mass,milgrom2001monda,Banik_2022}).

From the plot, it is evident that the relationship between the mass discrepancy and $a_N$ is not linear. To quantify this correlation, I therefore employ the Spearman rank correlation coefficient. The resulting value, $\rho = -0.8377$, indicates a very strong correlation. I further find that if the mass discrepancy is instead compared with the quantity $GM/r^{2.25}$, the correlation becomes slightly stronger, with $\rho = -0.8484$ for the SPARC dataset. However, this improvement is modest and may not be statistically significant.

The last (bottom-right) plot depicts the mass deficiency against $\lambda r$, where the best-fit $\lambda$ value has been individually computed for each galaxy through MCMC analysis. In this case also we don't see any strong correlation between mass discrepancy and $\lambda r$. The value of the Spearman rank corelation coefficient is $\rho = 0.57$, which indicate a weak correlation. However, we can notice that the mass discrepancy becomes apparent for $\lambda r>0.3$, and this pattern of discrepancy is prevalent in nearly all galaxies within the range of $\lambda r \in (0.3,5)$. This range aligns with our earlier expectations from the preceding section. This observation bears significant implications, particularly concerning theories such as MoG and TeVeS, which introduce additional massive scalar and vector fields. However, the fixed mass of these fields makes them unable to account for this observed relationship.

\begin{figure}
    \centering
    \includegraphics[width=1\linewidth]{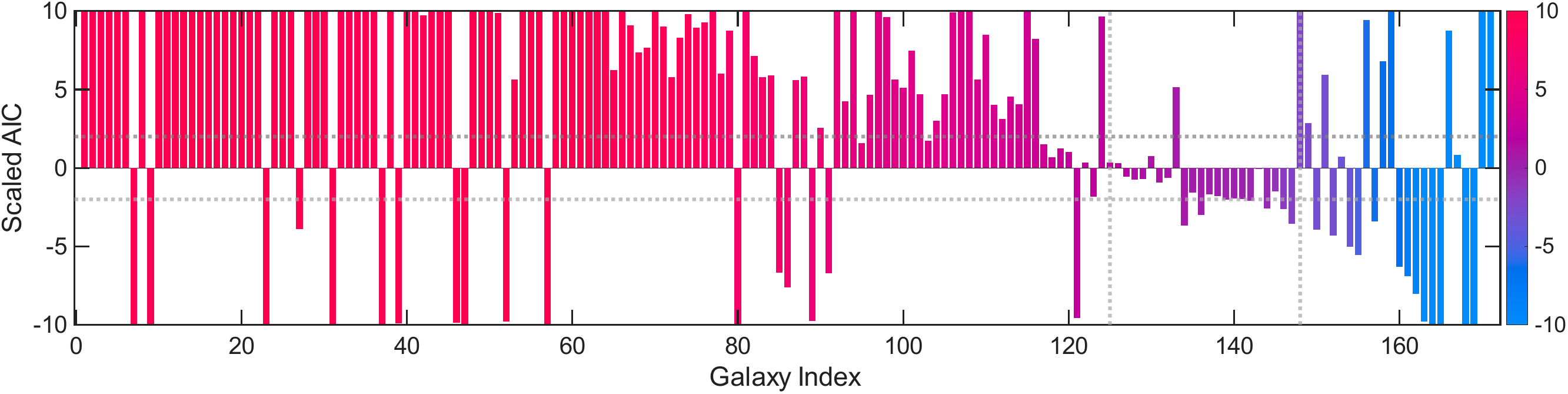}
    \caption{
    This figure compares three models— MG, MOND, and the NFW dark matter profile—using the Akaike Information Criterion (AIC), which balances goodness of fit and model complexity. MG and the NFW profile each have two free parameters, while MOND has one. The color of each bar represents $\mathrm{AIC}_{\mathrm{MOND}}-\mathrm{AIC}_{\mathrm{MG}}$, and the bar height represents $\mathrm{AIC}_{\mathrm{NFW}}-\mathrm{AIC}_{\mathrm{MG}}$. Values are smoothly truncated within $\pm 10$ using a $\tanh$ function, as differences larger than $\sim 7$–$8$ indicate a decisive preference. Galaxies are sorted by $\mathrm{AIC}_{\mathrm{MOND}}-\mathrm{AIC}_{\mathrm{MG}}$ to produce a gradual color variation. Red indicates a better fit for MG relative to MOND, while blue indicates the opposite. Similarly, positive bar heights indicate a better fit for MG relative to the NFW profile. The $\pm 2$ lines mark the threshold below which models are considered statistically equivalent.}
    \label{fig:AICPlot}
\end{figure}

\subsection{Results from MCMC analysis}

I fit the rotation curves of each galaxy independently using the Machian Gravity model. For each galaxy, I optimize two free parameters, $M_c$ and $\lambda^{-1}$, adopting flat priors for both. The overall results of this analysis are summarized in Fig.~\ref{fig:acc_descripency}.

The left panel shows the distribution of galaxies as a function of $\lambda R_{\mathrm{eff}}$, where $\lambda^{-1}$ is the best-fit scale length for a given galaxy and $R_{\mathrm{eff}}$ is the effective radius enclosing half of the total luminosity. Interestingly, with the exception of about 14 galaxies, the majority of the sample (161 out of 175 galaxies) satisfies $\lambda R_{\mathrm{eff}} < 3.7$. Among the remaining systems, four galaxies have $3.7 < \lambda R_{\mathrm{eff}} < 7$, while another four fall in the range $10 < \lambda R_{\mathrm{eff}} < 15$. For approximately 65 galaxies, $\lambda R_{\mathrm{eff}} < 0.2$.

In this small-$\lambda R_{\mathrm{eff}}$ regime, a strong degeneracy between $M_c$ and $\lambda^{-1}$ is observed. This arises because, in this limit, the velocity profile reduces to $v^2 \sim GM/r$, and the contribution of the second term in Eq.~\ref{eq:velocityFinal} becomes negligible. As a result, the MCMC sampler effectively attempts to keep this subdominant term fixed, leading to poor constraints on both $M_c$ and $\lambda^{-1}$. It decreases $\lambda$ and increases $M_c$, while keeping $M_c/\lambda^{-2}$ almost fixed. This behavior manifests as the characteristic elongated structures seen in the two-dimensional posterior distributions for these galaxies.

For most galaxies, however, the parameter range $0.2 < \lambda R_{\mathrm{eff}} < 3.7$ applies. In this regime, the relevant function can be well approximated by a linear form, as evident from the figure, and the resulting dynamics closely resemble the MOND acceleration law. In contrast, in the limit of very small $\lambda R_{\mathrm{eff}}$, the model naturally approaches the Newtonian regime. Consequently, the Machian Gravity model reproduces MOND-like behavior over an intermediate range of scales while recovering Newtonian dynamics at small radii.

The right panel shows the distribution of $\sqrt{M_c/M_0}$, where $M_c$ is the best-fit value obtained from the MCMC analysis and $M_0$ is the baryonic mass of the galaxy evaluated at the outermost observed radius. Although the outermost measured radius does not necessarily correspond to a well-defined physical boundary of the galaxy and thus may not provide an optimal normalization scale—unlike $R_{\mathrm{eff}}$—it nonetheless offers a useful reference mass scale.

From the histogram, it is evident that for most galaxies $\sqrt{M_c/M_0} < 12$, with the distribution peaking between 4 and 6. For roughly 50 galaxies, this ratio exceeds 15. However, for all such systems, $\lambda^{-1} R_{\mathrm{eff}}$ is very small, which renders the term $1 - e^{-\lambda r}(1+\lambda r)$ nearly zero. Consequently, variations in $\sqrt{M_c/M_0}$ have a negligible impact on the predicted velocities, as can be seen from Eq.~\ref{eq:velocityFinal}. In this regime, the precise value of $M_c$ is therefore not physically significant, and the rotation curves are effectively governed by the Newtonian contribution from the baryonic matter alone.

In Fig.~\ref{fig:mcmc}, I present the posterior distributions obtained from the MCMC analysis for two representative galaxies, UGC12732 and NGC2366. The first two panels in the top row show the marginalized distributions of $\lambda^{-1}$ and $M_c$, respectively, while the third panel illustrates their joint posterior distribution. A strong correlation between $\lambda^{-2}$ and $M_c$ is clearly visible, reflecting the role of the characteristic acceleration scale $a_0^{\mathrm{MG}} = GM_c/\lambda^{-2}$ in determining the gravitational field. The second row shows the $\chi^2$ distributions as functions of $\lambda^{-1}$, $M_c$, and $a_0^{\mathrm{MG}}$, respectively.

Two distinct behaviors are evident from these plots. In the first case, i.e. the case for UGC12732, exemplified by the top-right joint posterior, both parameters are well constrained. In the second case, i.e. for NGC2366, the two-dimensional posterior distribution is significantly elongated, indicating weak constraints on the parameters. These two behaviors are consistently observed across the full galaxy sample. The first type typically occurs for galaxies with a large mass discrepancy, where the MG contribution is significant and both parameters are tightly constrained. In contrast, the second type arises for galaxies with a small mass discrepancy. In such systems, $\lambda^{-1}$ becomes large, rendering the second term in Eq.~\ref{eq:velocityFinal} negligible. As a consequence, $M_c$ can take very large values without significantly affecting the predicted velocity, leading to the observed degeneracy between $M_c$ and $\lambda^{-1}$.

Although the $\chi^2$ distributions shown in Fig.~\ref{fig:mcmc} exhibit a single global minimum for all parameters, there exist many galaxies in the sample for which multiple local minima are present. Notably, galaxies belonging to the first category—where the parameters are well constrained—consistently display a single, well-defined minimum in $\chi^2$, while for the second category multiple local minimas are seen.

\begin{figure}
    \centering
    \includegraphics[width=1\linewidth]{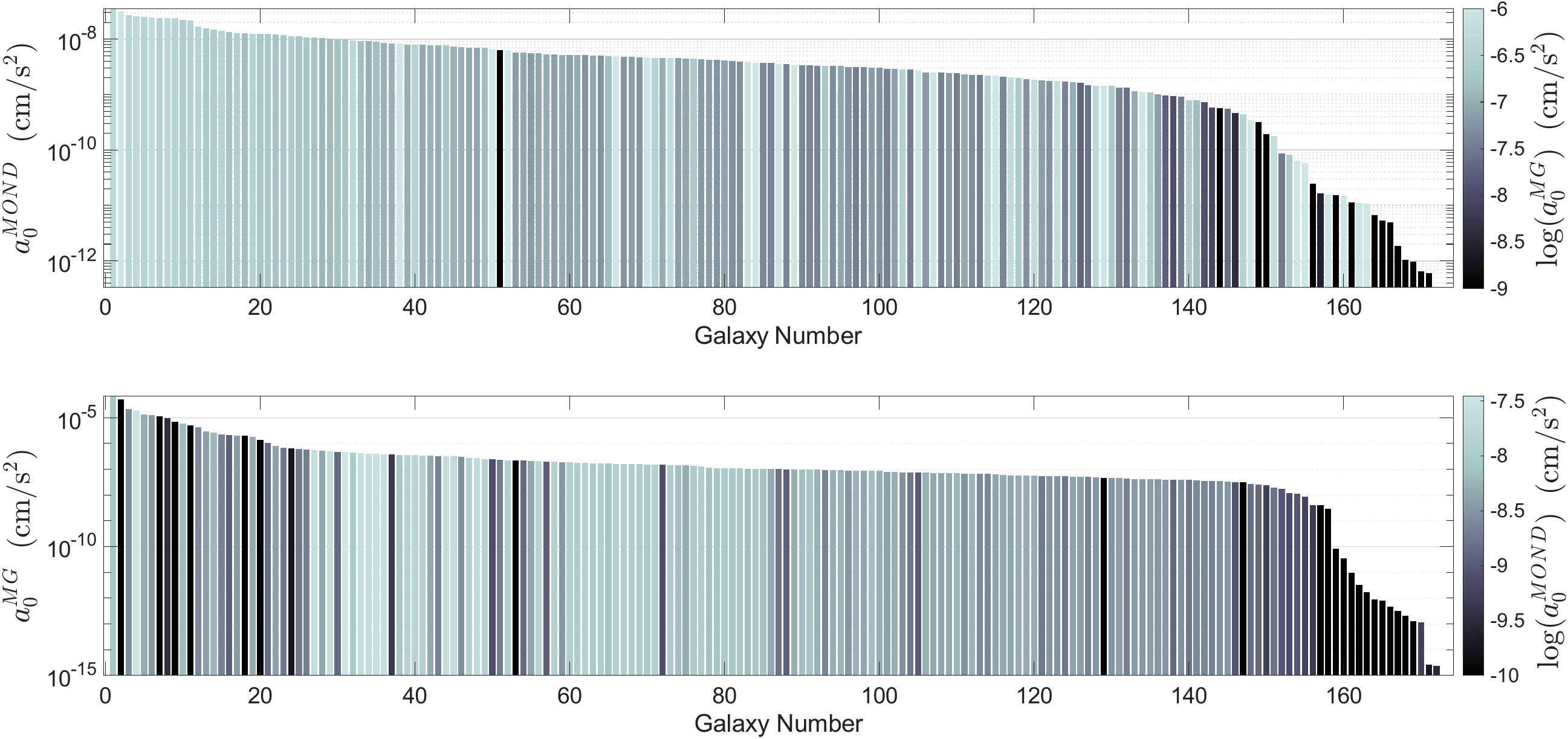}
    \caption{In the top panel, I plot the MOND acceleration scale and sort the galaxies by their corresponding value of $a_0^\text{MOND}$. Aside from a few galaxies at the very beginning and end of the sequence, the acceleration scale is nearly constant across the sample, although it does show a gradual variation. The bars are color-coded according to $a_0^\text{MG} = \frac{GM_c}{\lambda^{-2}}$. This illustrates that the acceleration scales predicted by the two models do not necessarily match for every galaxy. That is, the $n$‑th galaxy in the MOND-sorted list does not, in general, lie near the $n$‑th position when ranked by the Machian gravity acceleration scale. The figure indicates that, for many galaxies, these rankings can differ substantially. In the bottom panel, I present the analogous comparison but with the bar height representing the Machian gravity acceleration scale and the color encoding the MOND acceleration scale. Computing the median accelerations, we find that the median value of $a^\text{MOND}$ is roughly an order of magnitude smaller than the corresponding median of $a^\text{MG}$, in agreement with Eq.~\ref{Eq:MONDBreaking}. }
    \label{fig:a0_MG_MOND}
\end{figure}

\subsection{Compering between models}

We consider three different models, \-- MG, MOND, and the NFW dark matter profile—to fit the galactic rotation curves. For the MG model, I determine the best-fit values of the two free parameters, $M_c$ and $\lambda^{-1}$, for each galaxy using an MCMC analysis. For MOND, instead of adopting a fixed value of the acceleration scale $a^\text{MOND}_0$, I fit $a^\text{MOND}_0$ independently for each galaxy. Similarly, for the NFW dark matter profile, I obtain the best-fit values of the halo parameters $\rho_c$ and $r_s$ separately for each galaxy. A comparative analysis is essential to assess which model best reproduces the observational data. In Fig.~\ref{fig:velocityVelocity}, I plot the predicted velocities from each model against the corresponding observed velocities for $12$ galaxies. A model that provides a good fit is expected to produce points lying close to the diagonal line.

The red curve corresponds to the MG model, while the blue and green curves represent the NFW dark matter profile and MOND, respectively. From the figure, it is evident that the red curve follows the diagonal line more closely than the other two, indicating that MG provides a better overall fit to the data. 
Another interesting point to note is that at low velocities (which typically correspond to small radii, since galactic rotation curves generally increase monotonically with radius), the velocities predicted by MOND and the NFW profile are, in most cases, higher than the observed velocities. This suggests that both MOND and the NFW model systematically overpredict velocities in the inner regions of galaxies.

While the previous plot provides valuable visual insight, a quantitative statistical comparison is necessary for a robust model assessment. Therefore, I compare the MG, MOND, and the NFW dark matter profile using the Akaike Information Criterion (AIC). Both Machian Gravity and the NFW profile involve two free parameters, whereas MOND contains a single free parameter. Moreover, none of these models is a limiting case of another. Therefore, the AIC offers an appropriate framework for comparing their relative performance while accounting for model complexity. The AIC is defined as $2k + \chi^2$, where $k$ is the number of free parameters. This formulation penalizes models with a larger number of parameters. A lower AIC value show that a particular model is preferred over other. If the difference in AIC between two models is less than 2, the models may be regarded as statistically equivalent in terms of fit quality.

I compute the AIC values for all three models for each galaxy. In Fig.~\ref{fig:AICPlot}, the color of each bar represents $\mathrm{AIC}_{\mathrm{MOND}} - \mathrm{AIC}_{\mathrm{MG}}$, while the bar height denotes $\mathrm{AIC}_{\mathrm{NFW}} - \mathrm{AIC}_{\mathrm{MG}}$. To prevent extreme values from dominating the visualization, I smoothly truncate the AIC differences to the range $\pm 10$ using a $\tanh$ function, since values larger than about 7--8 already indicate a decisive preference for one model. The galaxies are sorted by $\mathrm{AIC}_{\mathrm{MOND}} - \mathrm{AIC}_{\mathrm{MG}}$, producing a gradual transition in color across the plot. Red colors indicate galaxies for which Machian Gravity fits better than MOND, while blue colors indicate a better MOND fit. Similarly, positive bar heights correspond to cases where Machian Gravity outperforms the NFW profile.

I have marked the $\pm 2$ levels along the Y-axis to indicate where the bar length exceeds this value. Similarly, wherever the color index crosses $\pm 2$ along the X-axis, I have indicated this with dotted lines. We find that for approximately 139 galaxies, Machian gravity provides a better fit than MOND. Likewise, for about 123 galaxies, Machian gravity yields a better fit than the NFW profile. This indicates a clear preference for Machian gravity in the SPARC data. However, we should note that, in this analysis, the mass-to-light ratios for the disk and bulge have been kept fixed at constant values, which may influence some of the results.

\subsection{Comparison of Acceleration Scales in MOND and MG}

In Fig.~\ref{fig:a0_MG_MOND}, I have shown a comparative study of the acceleration scales in MOND and MG. The top panel shows the MOND acceleration scale, with galaxies sorted by $a_0^{\mathrm{MOND}}$. Apart from a small number of galaxies at the beginning and end of the distribution, the acceleration scale remains nearly constant, although a gradual variation is still present. The bars are color-coded according to the corresponding values of $a_0^{\mathrm{MG}}$, illustrating that the acceleration scales inferred from the two models do not exhibit a one-to-one correspondence. In particular, a galaxy occupying a given position in the MOND-sorted sequence does not necessarily lie in the same relative position when ordered by the MG acceleration scale.

The bottom panel displays the same information but with the roles interchanged: the bar height now corresponds to $a_0^{\mathrm{MG}}$, while the color scale encodes $a_0^{\mathrm{MOND}}$. In the MG case, the characteristic acceleration varies substantially across different galaxies, as expected, since in the MG model the two parameters $M_c$ and $\lambda^{-1}$ are fitted independently with flat priors. (Note that this can give a misleading visual impression that the MG acceleration scale is uniformly distributed; however, the $y$-axis ranges are different in the two panels.) Comparing the median values, we find that $a_0^{\mathrm{MOND}}$ is roughly an order of magnitude smaller than $a_0^{\mathrm{MG}}$, consistent with Eq.~\ref{Eq:MONDBreaking} (though it does not exactly reproduce the numerical value given there). It is also worth pointing out that, in both panels, if we disregard a few bars at the extreme left and right, the accelerations vary within about 1–1.5 orders of magnitude, which is somewhat noteworthy given that, in the MG framework, the relevant acceleration scale is not being fitted independently.

\begin{figure}
    \centering
    \includegraphics[trim=3cm 0cm 4cm 1cm, clip=true, width=0.19\linewidth]{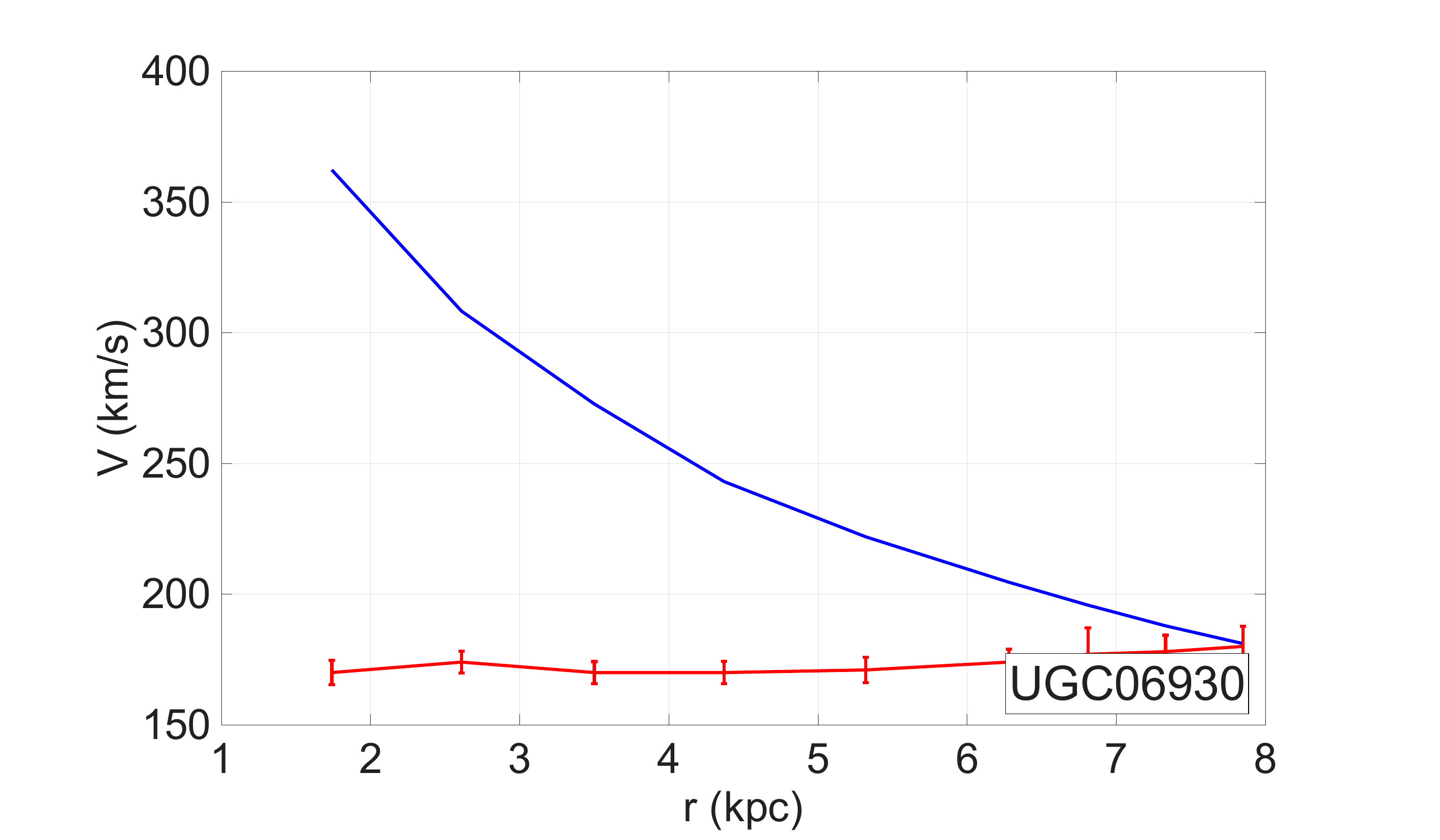}
    \includegraphics[trim=3cm 0cm 4cm 1cm, clip=true, width=0.19\linewidth]{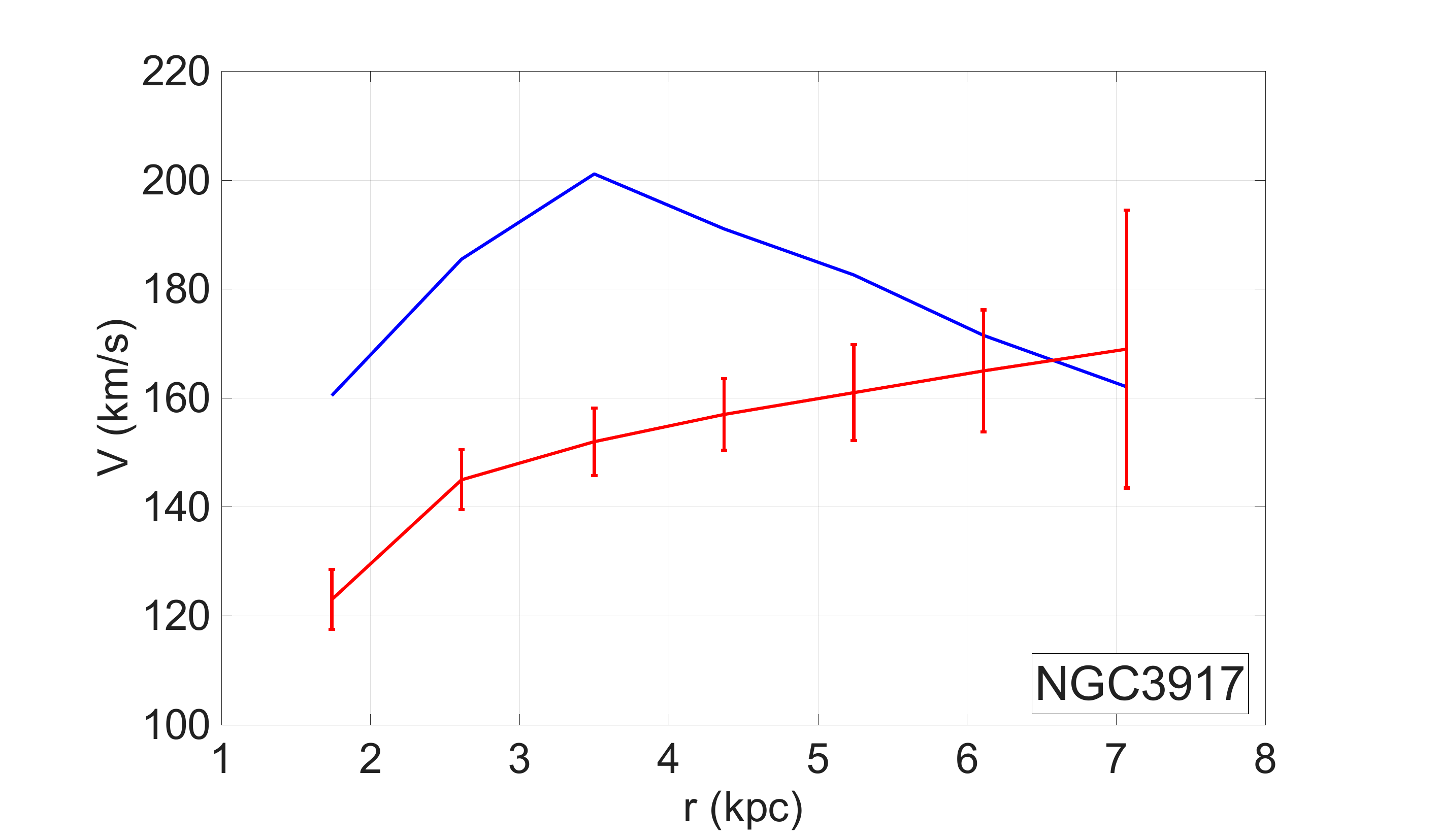}
    \includegraphics[trim=3cm 0cm 4cm 1cm, clip=true, width=0.19\linewidth]{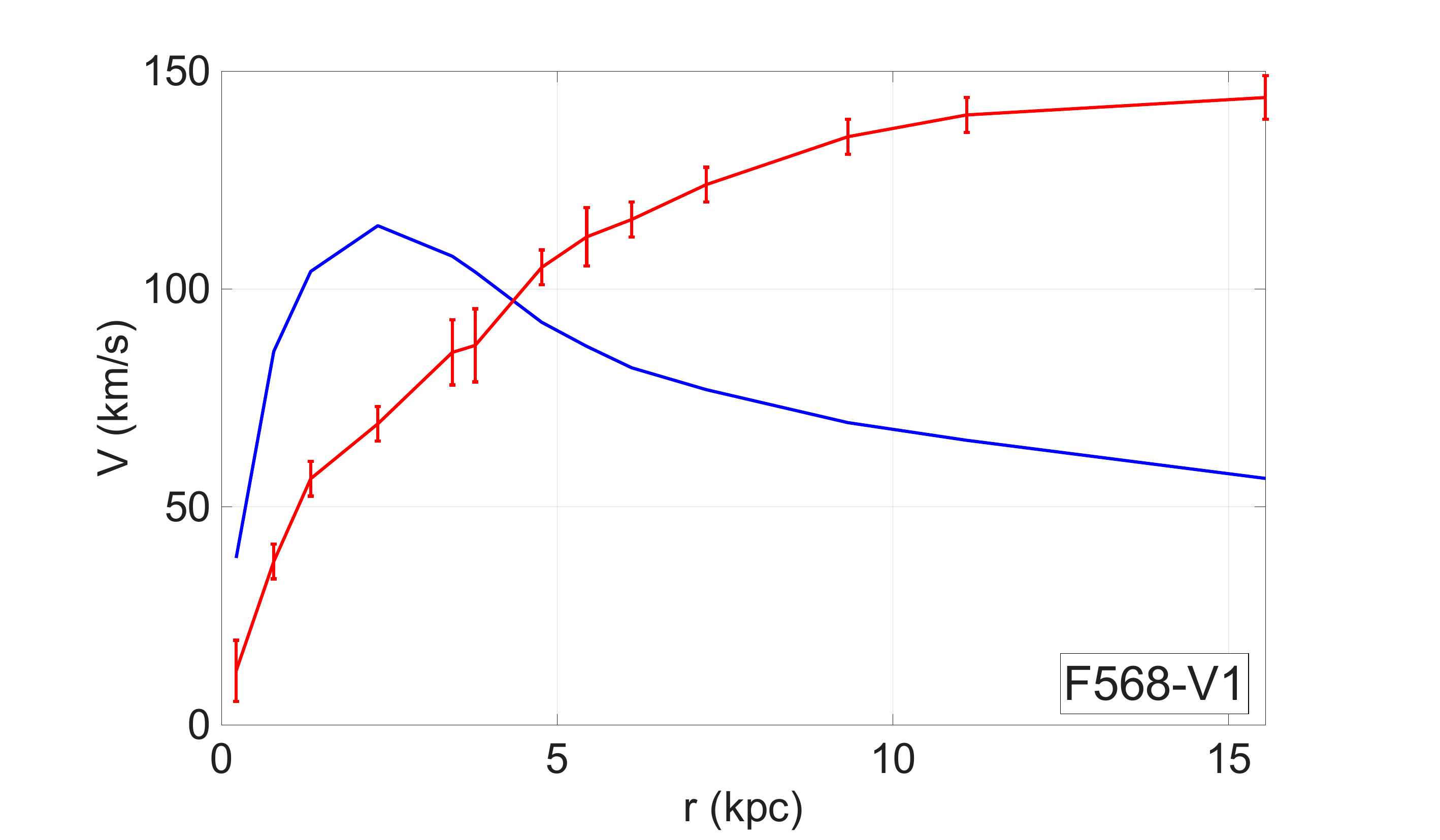}   
    \includegraphics[trim=3cm 0cm 4cm 1cm, clip=true, width=0.19\linewidth]{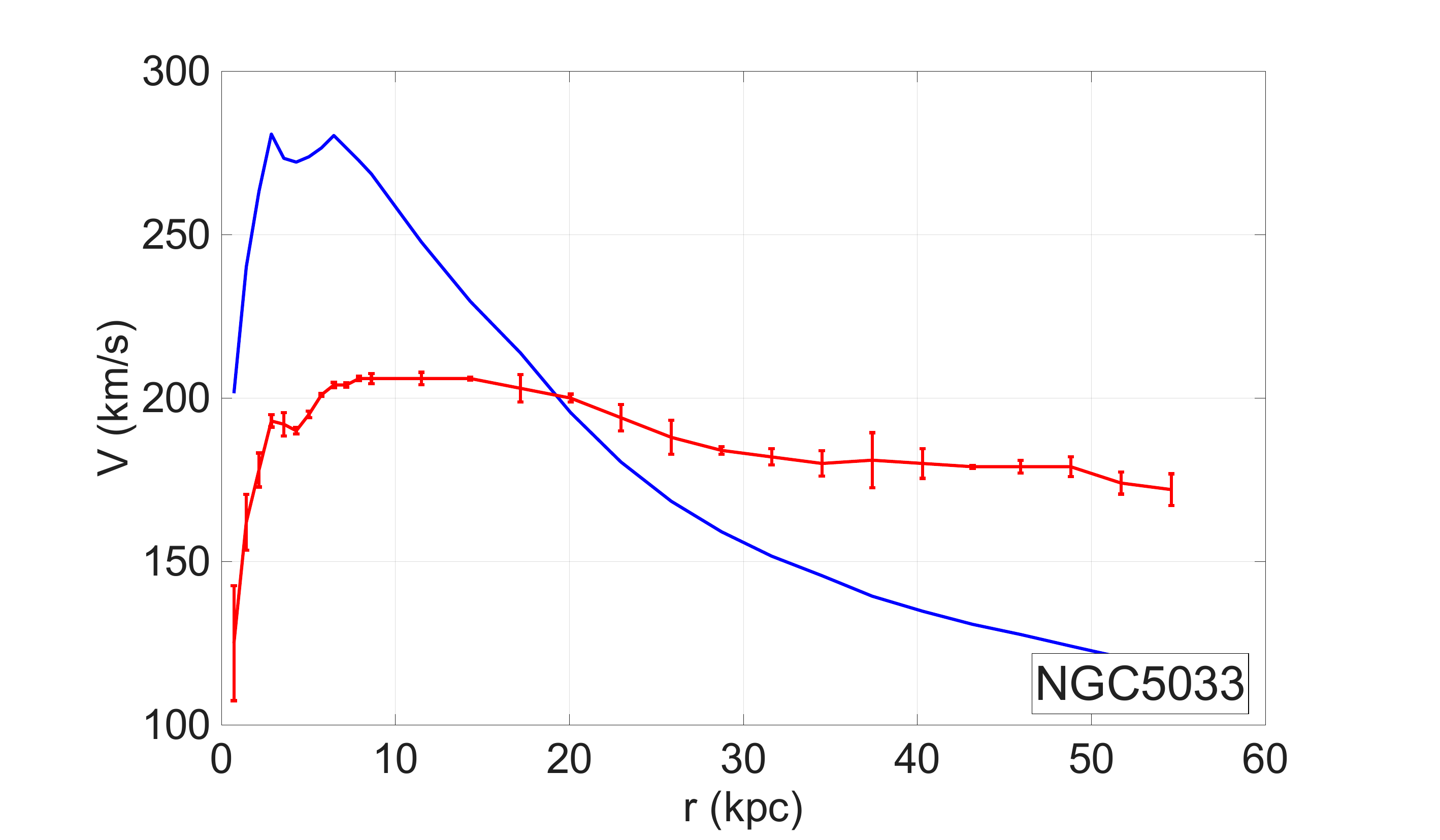}
    \includegraphics[trim=3cm 0cm 4cm 1cm, clip=true, width=0.19\linewidth]{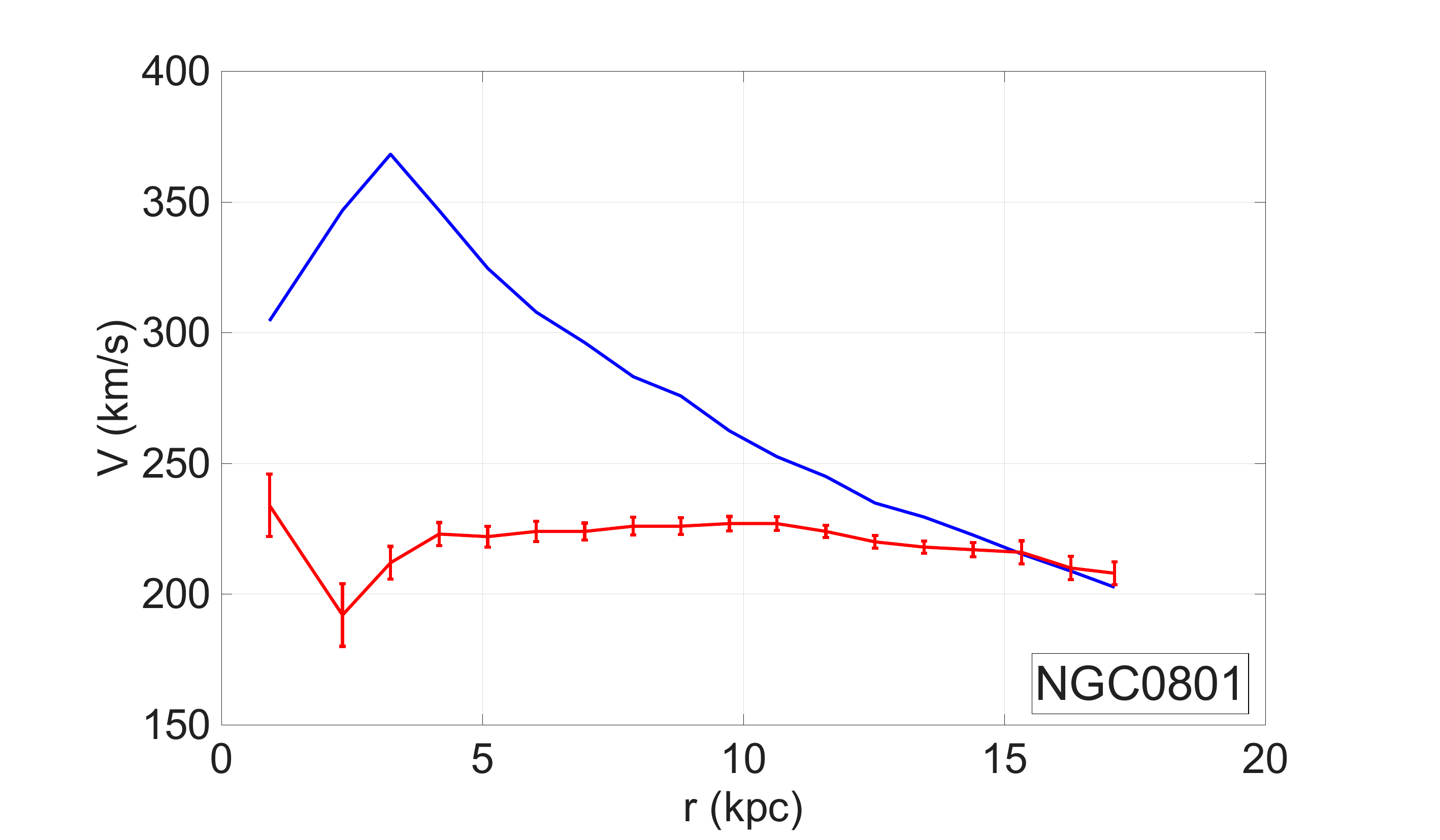}    
    \includegraphics[trim=3cm 0cm 4cm 1cm, clip=true, width=0.19\linewidth]{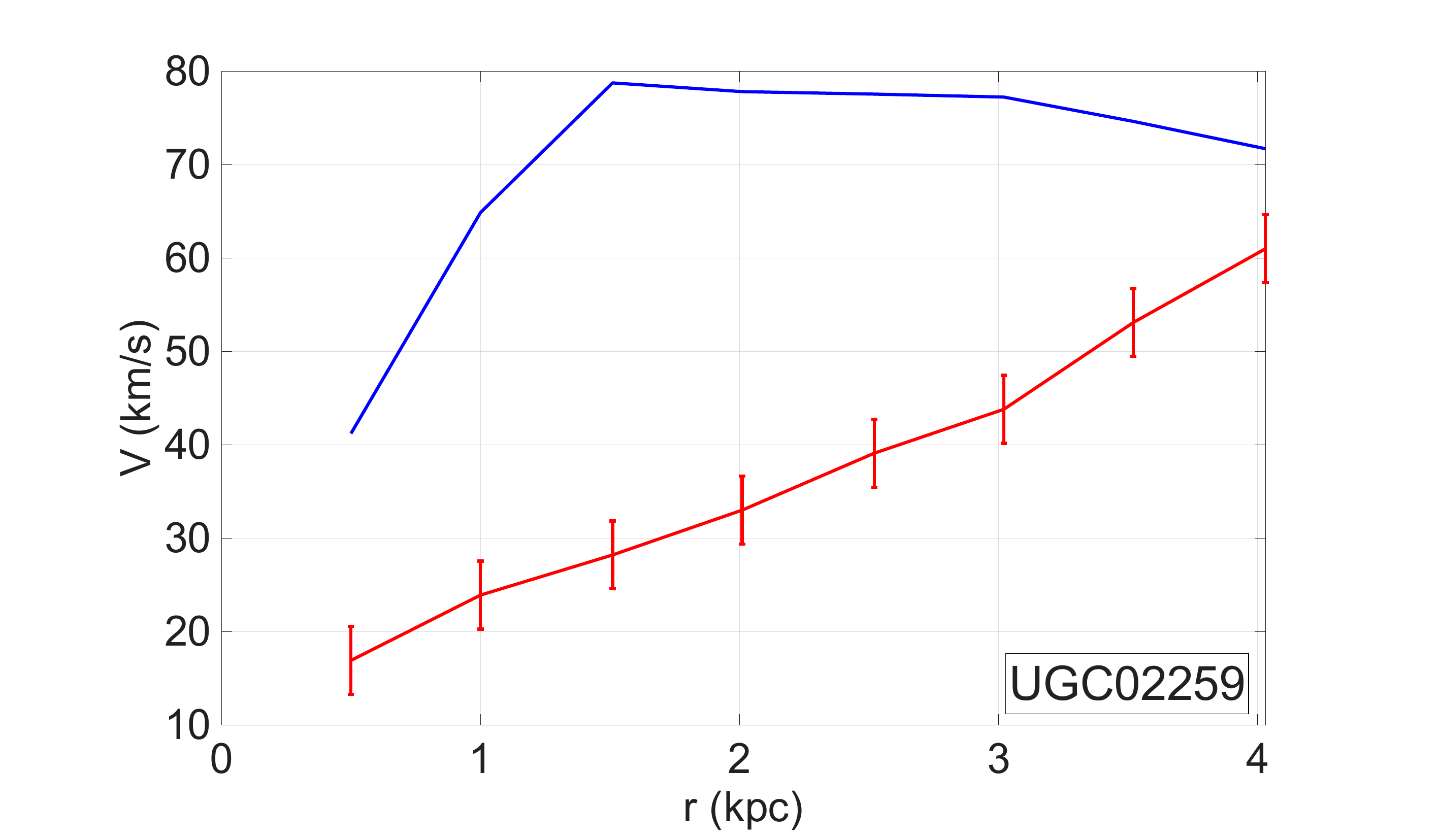}   
    \includegraphics[trim=3cm 0cm 4cm 1cm, clip=true, width=0.19\linewidth]{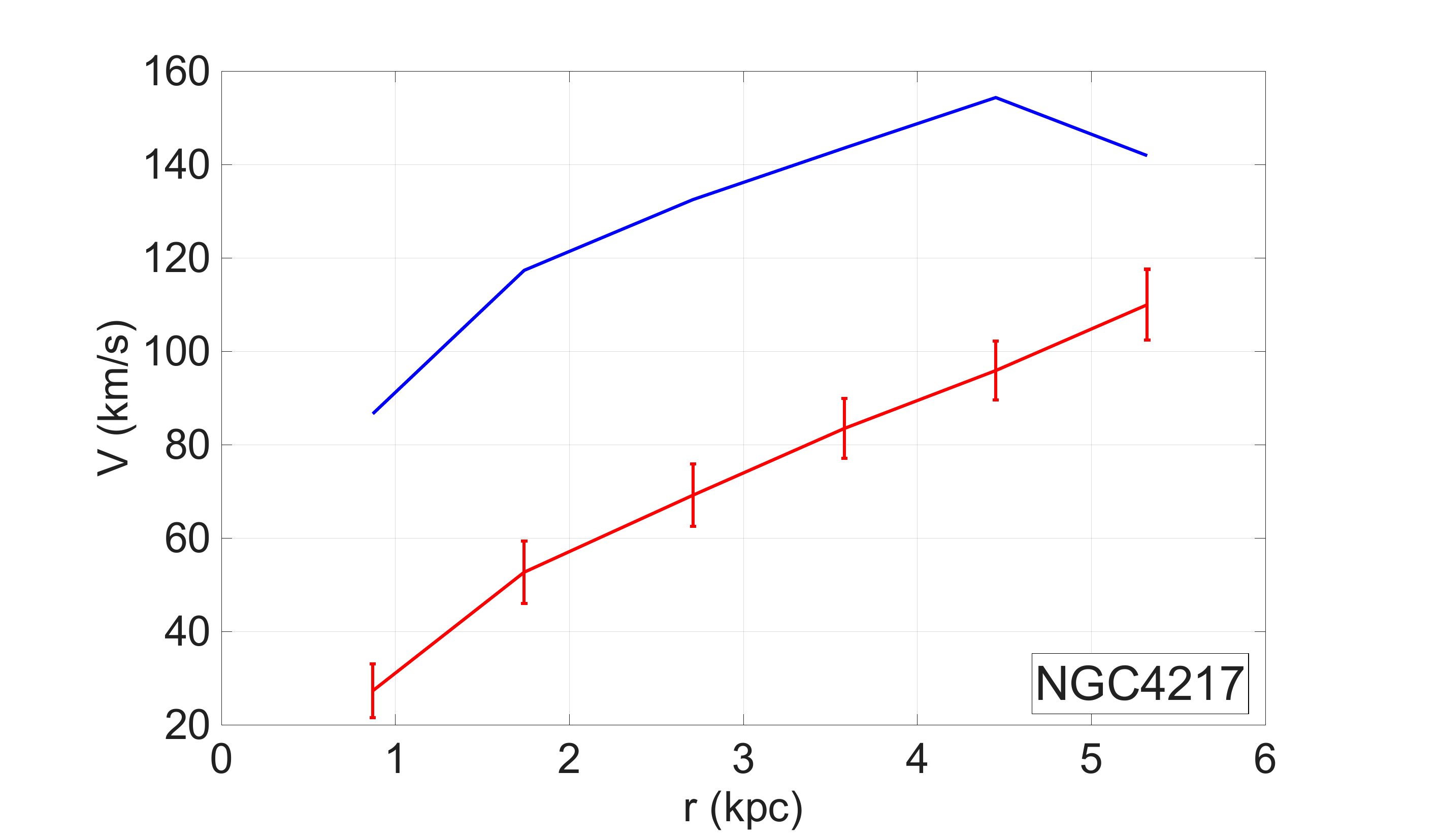}    
    \includegraphics[trim=3cm 0cm 4cm 1cm, clip=true, width=0.19\linewidth]{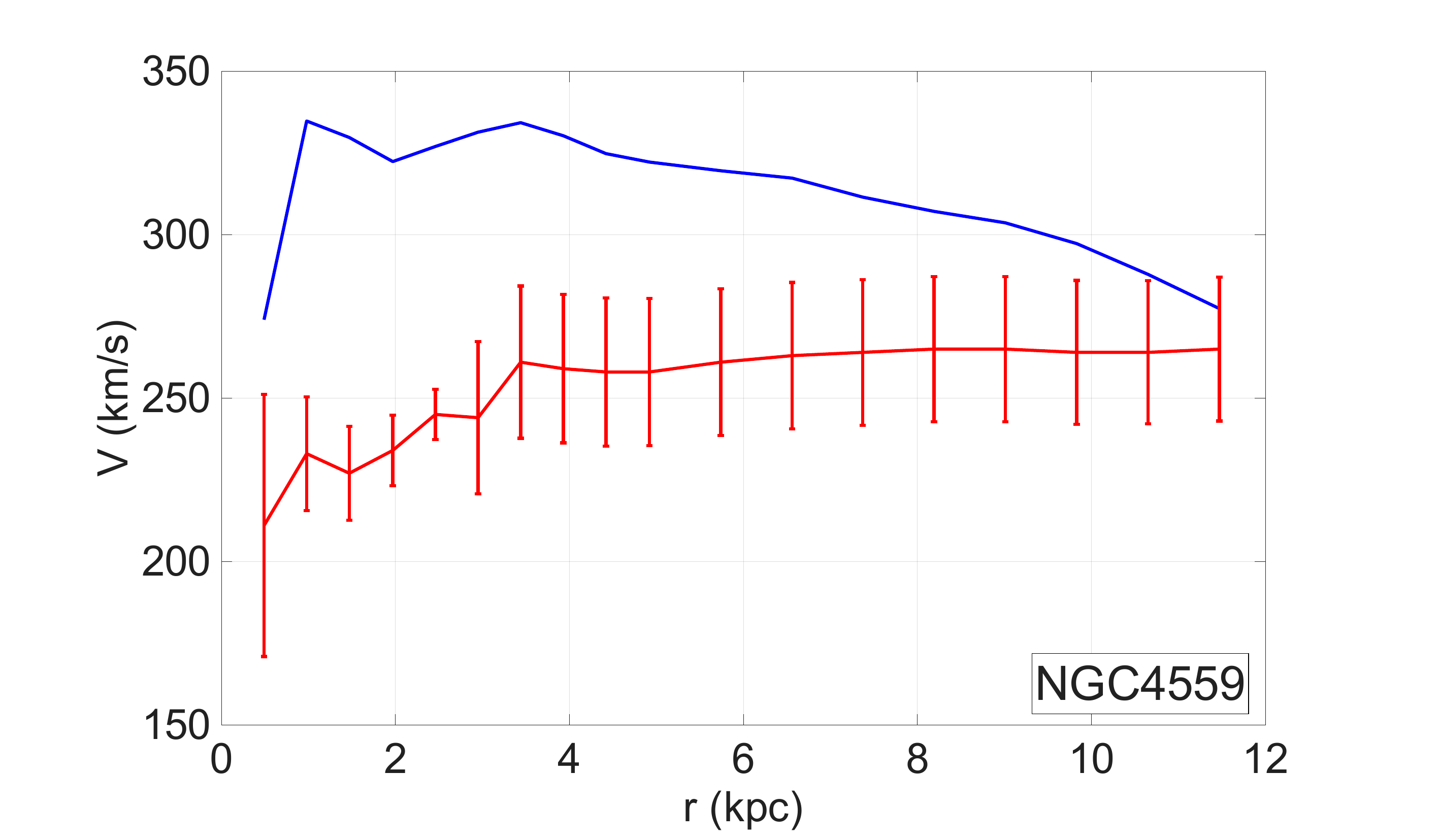}        
    \includegraphics[trim=3cm 0cm 4cm 1cm, clip=true, width=0.19\linewidth]{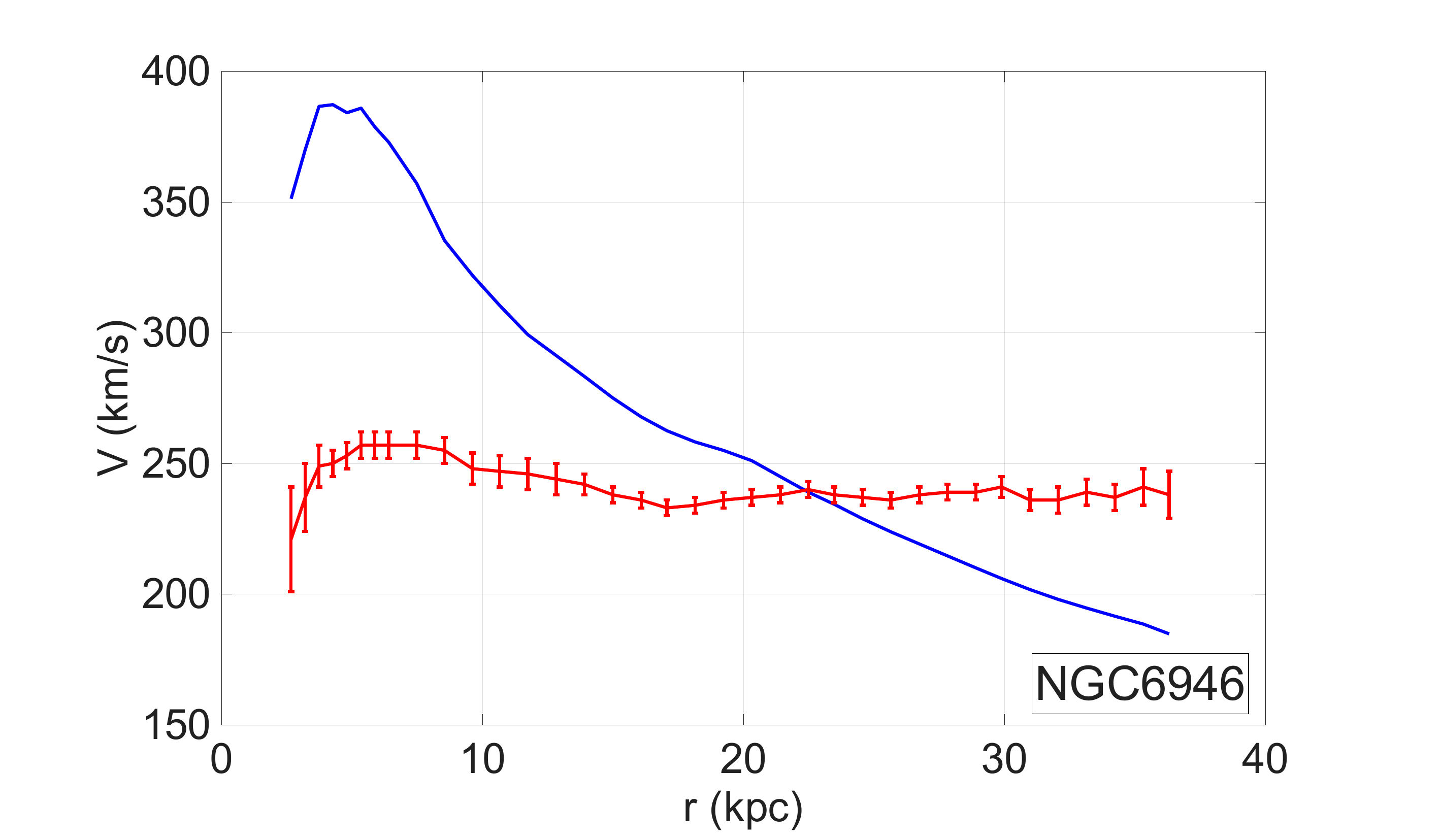} 
    \includegraphics[trim=3cm 0cm 4cm 1cm, clip=true, width=0.19\linewidth]{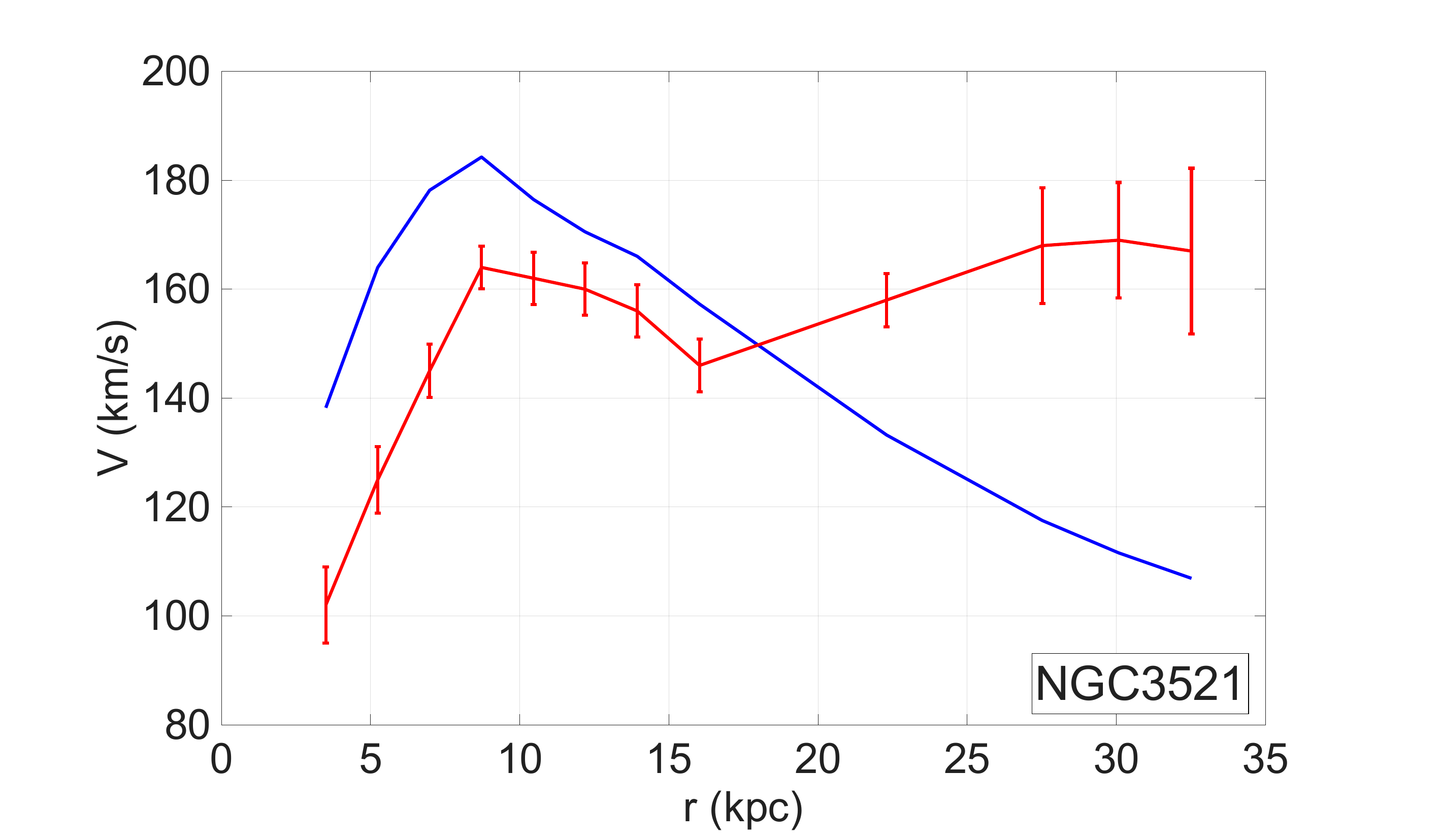}     
    \includegraphics[trim=3cm 0cm 4cm 1cm, clip=true, width=0.19\linewidth]{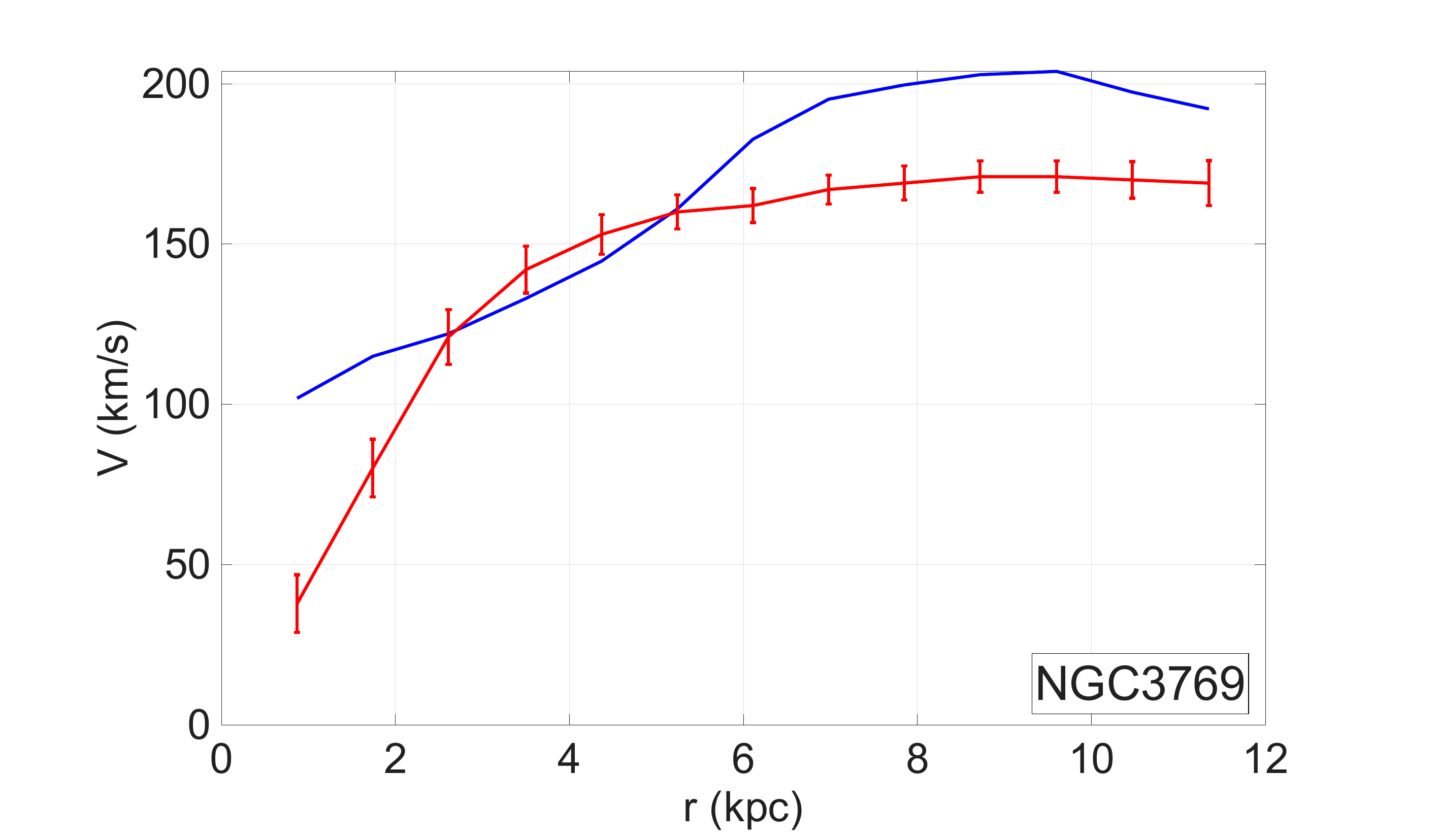}     
    \includegraphics[trim=3cm 0cm 4cm 1cm, clip=true, width=0.19\linewidth]{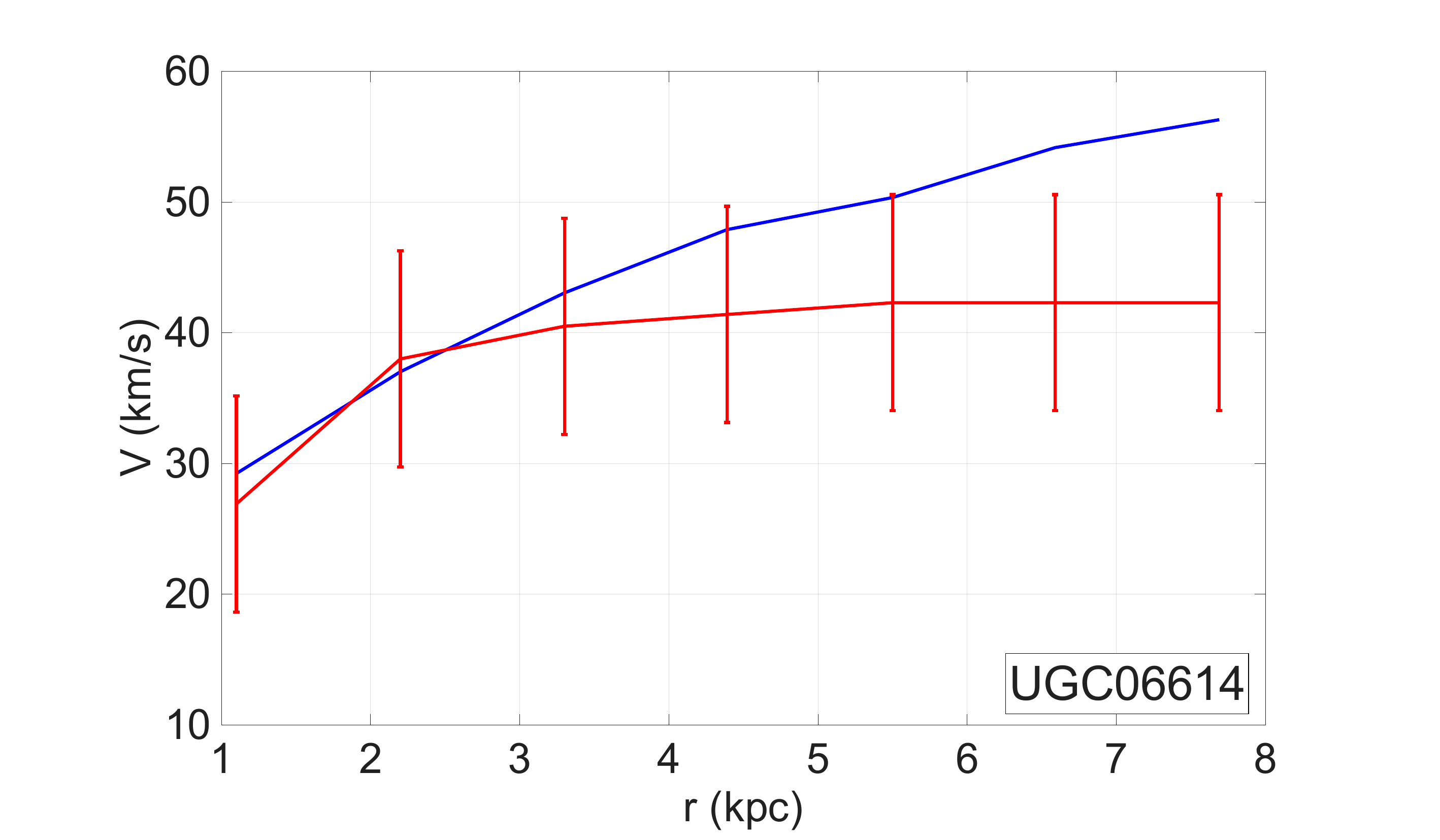}         
    \includegraphics[trim=3cm 0cm 4cm 1cm, clip=true, width=0.19\linewidth]{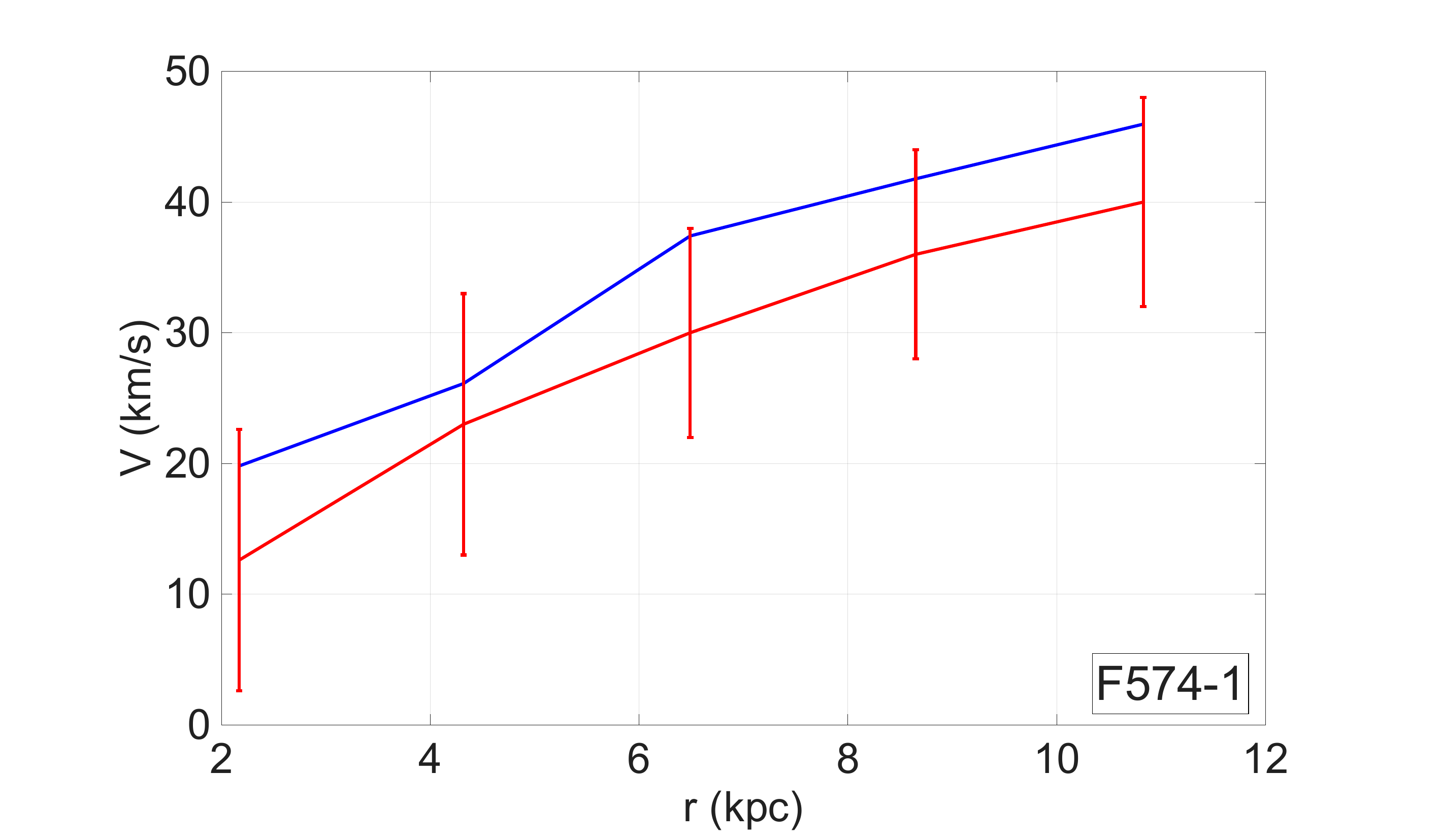}
    \includegraphics[trim=3cm 0cm 4cm 1cm, clip=true, width=0.19\linewidth]{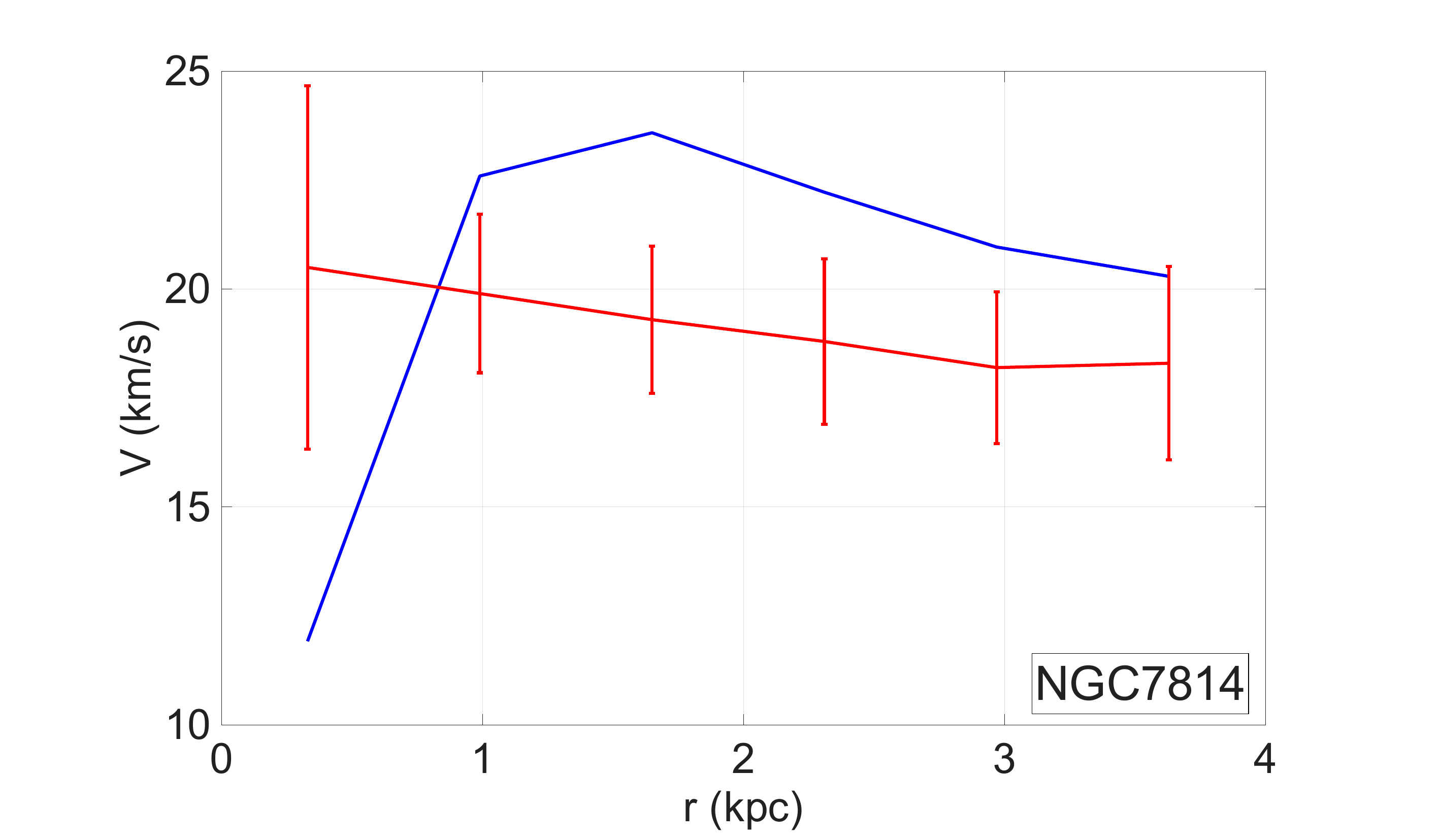}
    \includegraphics[trim=3cm 0cm 4cm 1cm, clip=true, width=0.19\linewidth]{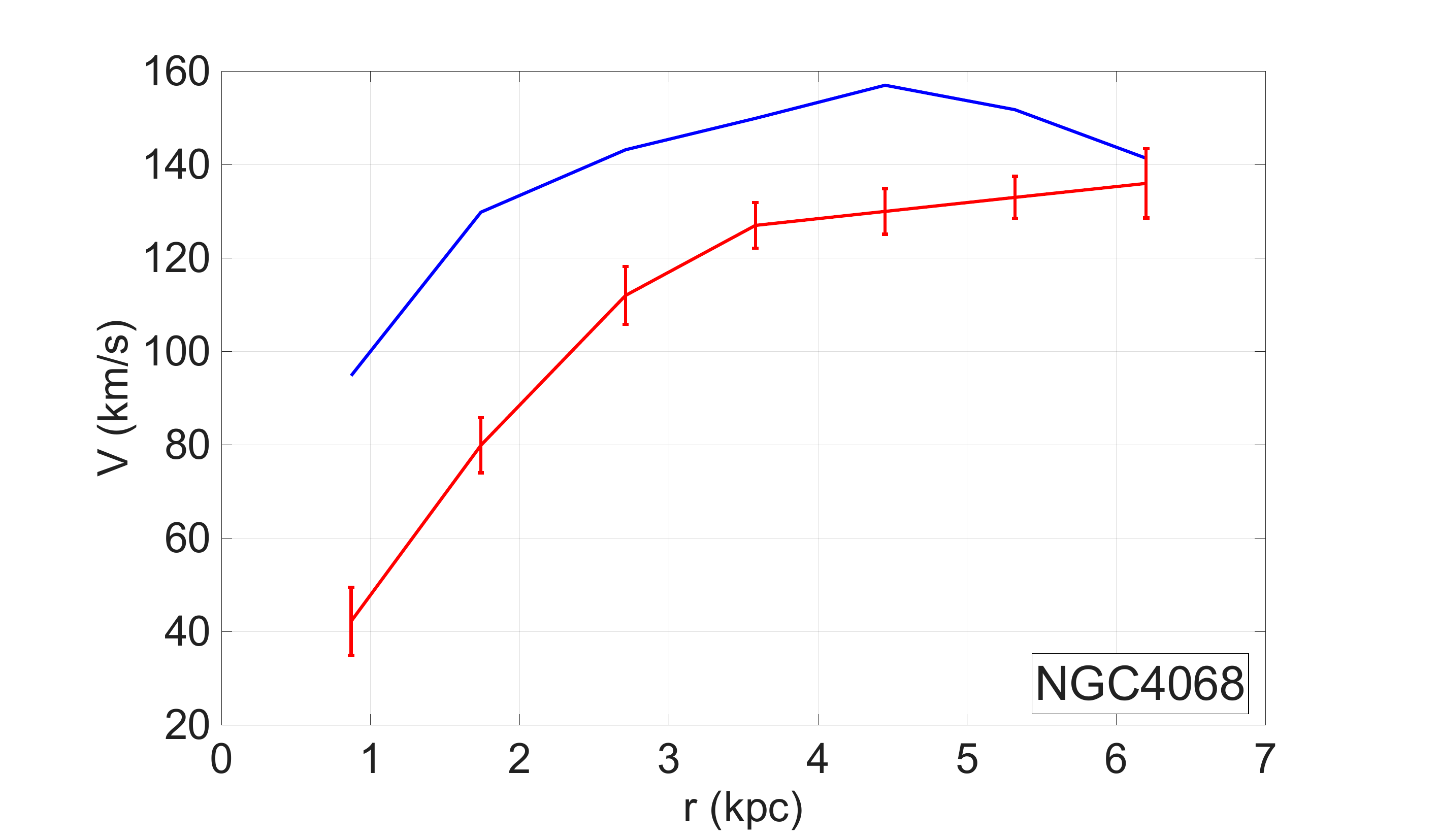}   
    \includegraphics[trim=3cm 0cm 4cm 1cm, clip=true, width=0.19\linewidth]{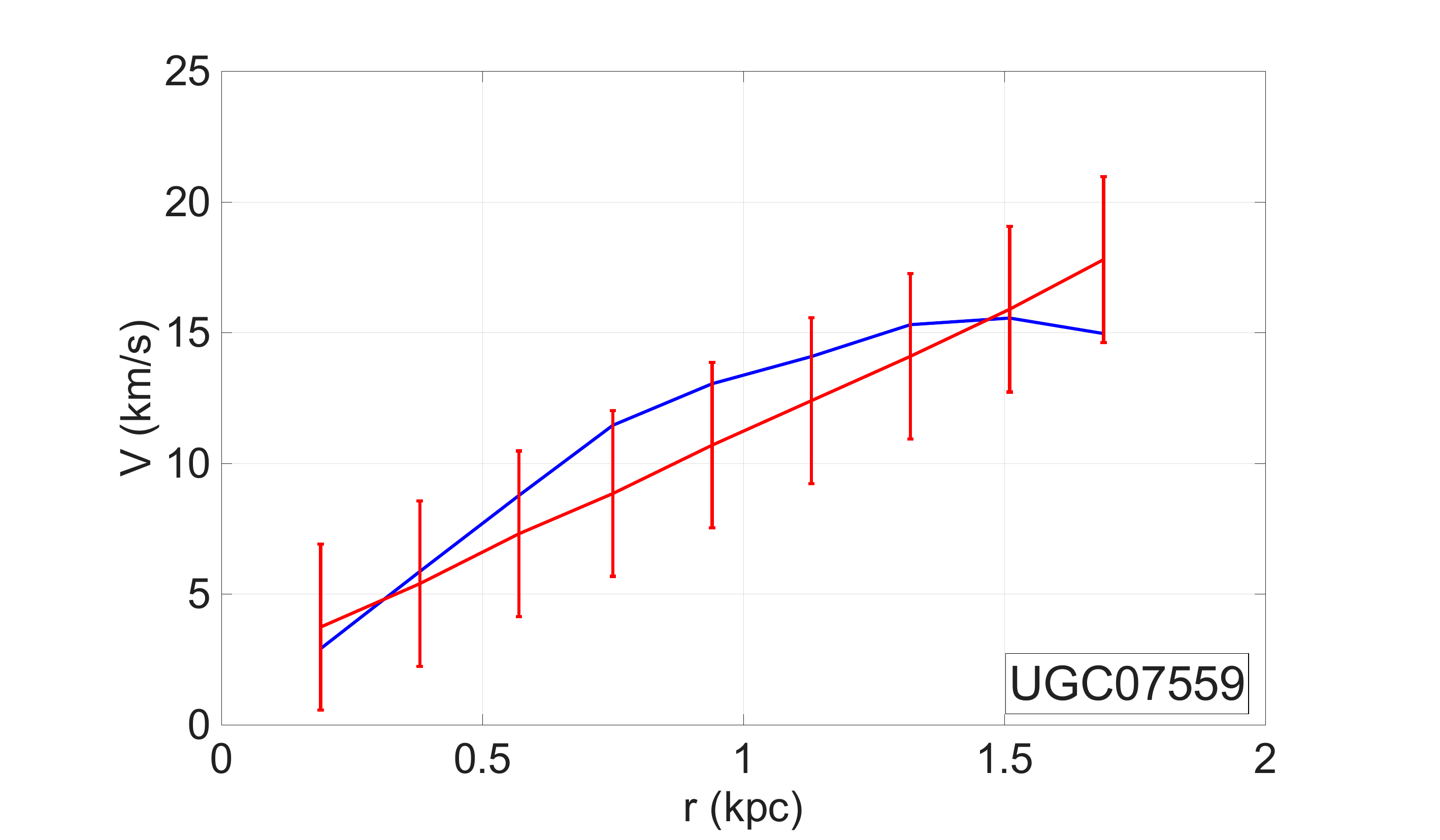}
    \includegraphics[trim=3cm 0cm 4cm 1cm, clip=true, width=0.19\linewidth]{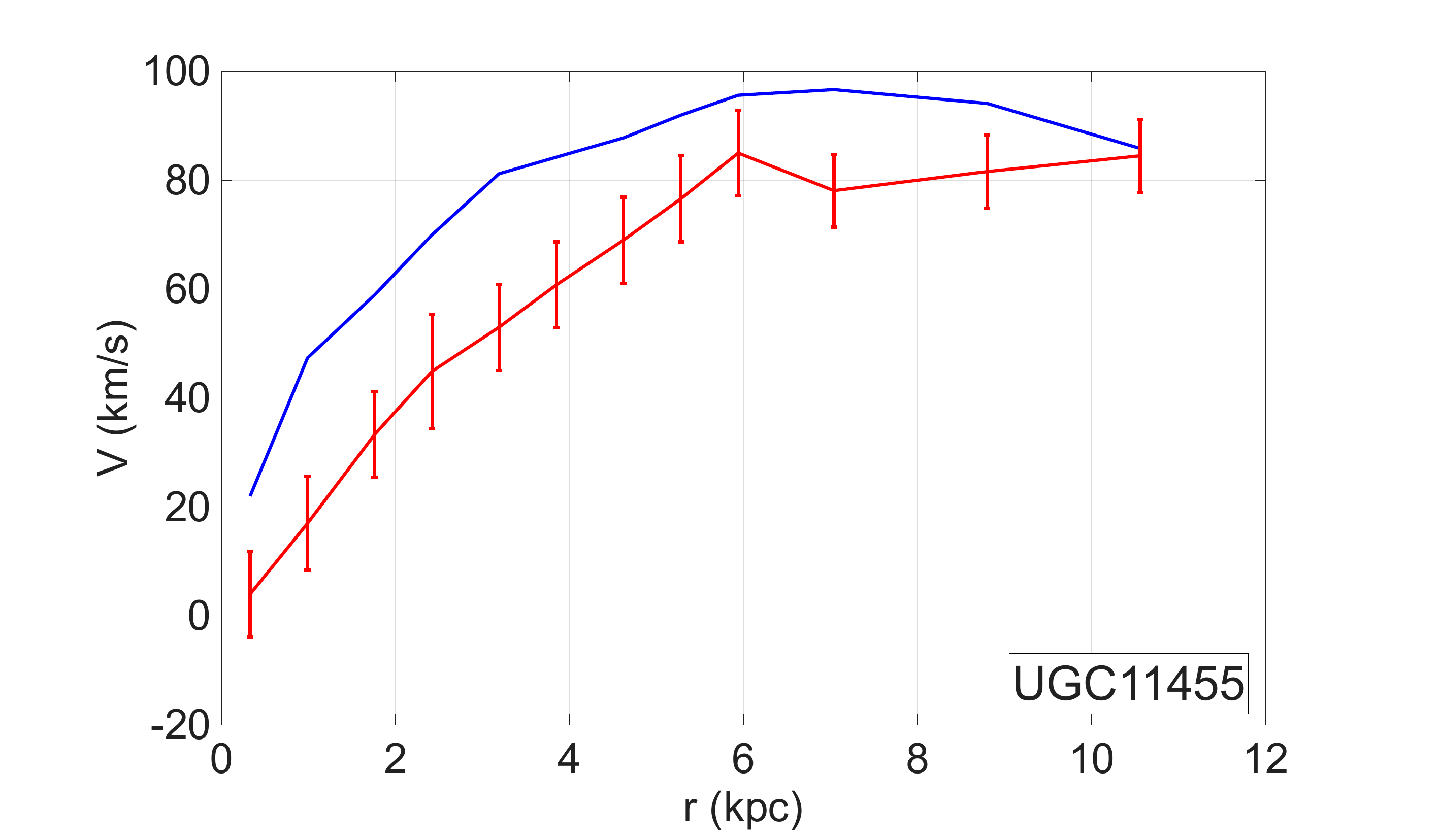}    
    \includegraphics[trim=3cm 0cm 4cm 1cm, clip=true, width=0.19\linewidth]{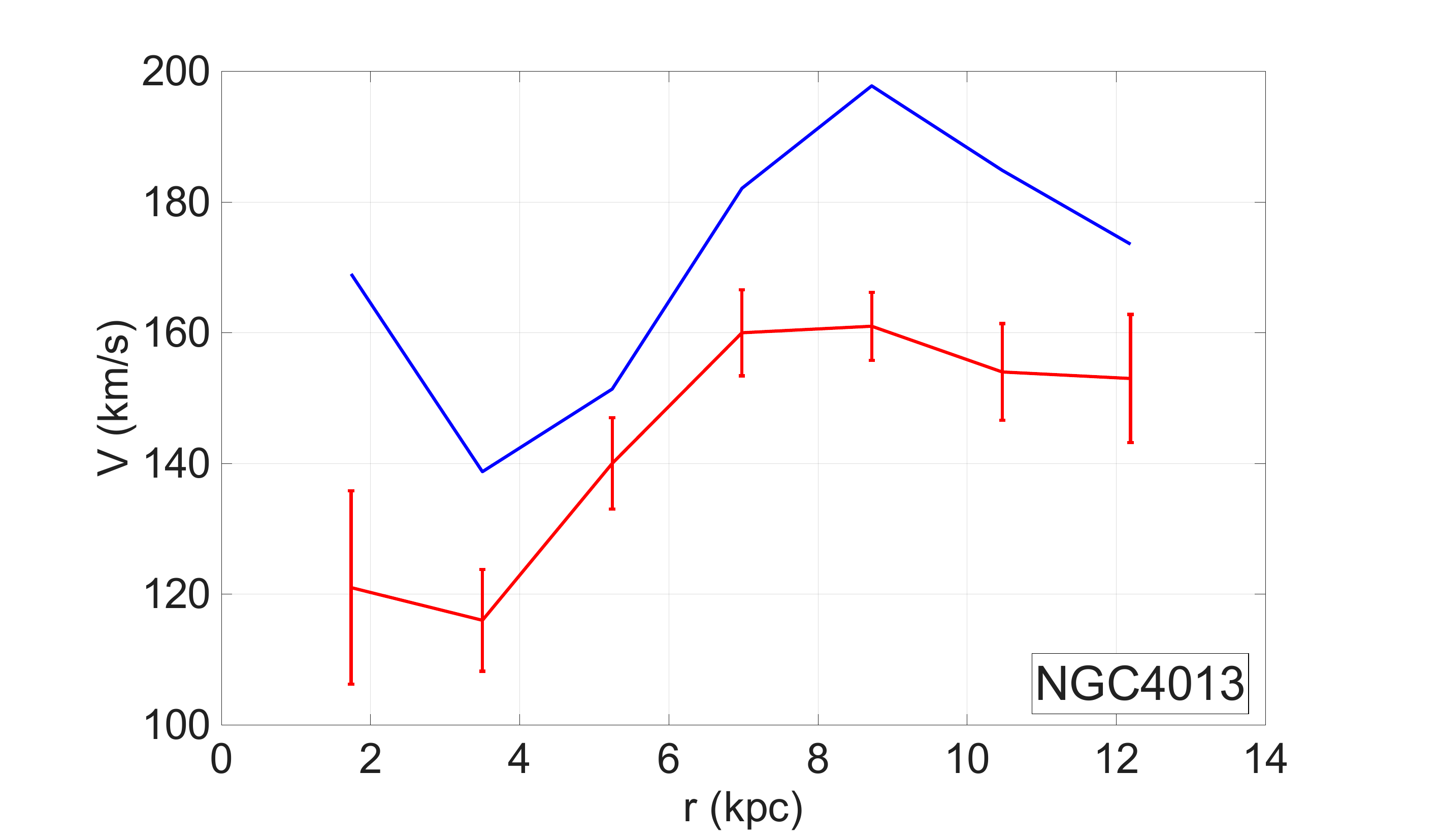}   
    \includegraphics[trim=3cm 0cm 4cm 1cm, clip=true, width=0.19\linewidth]{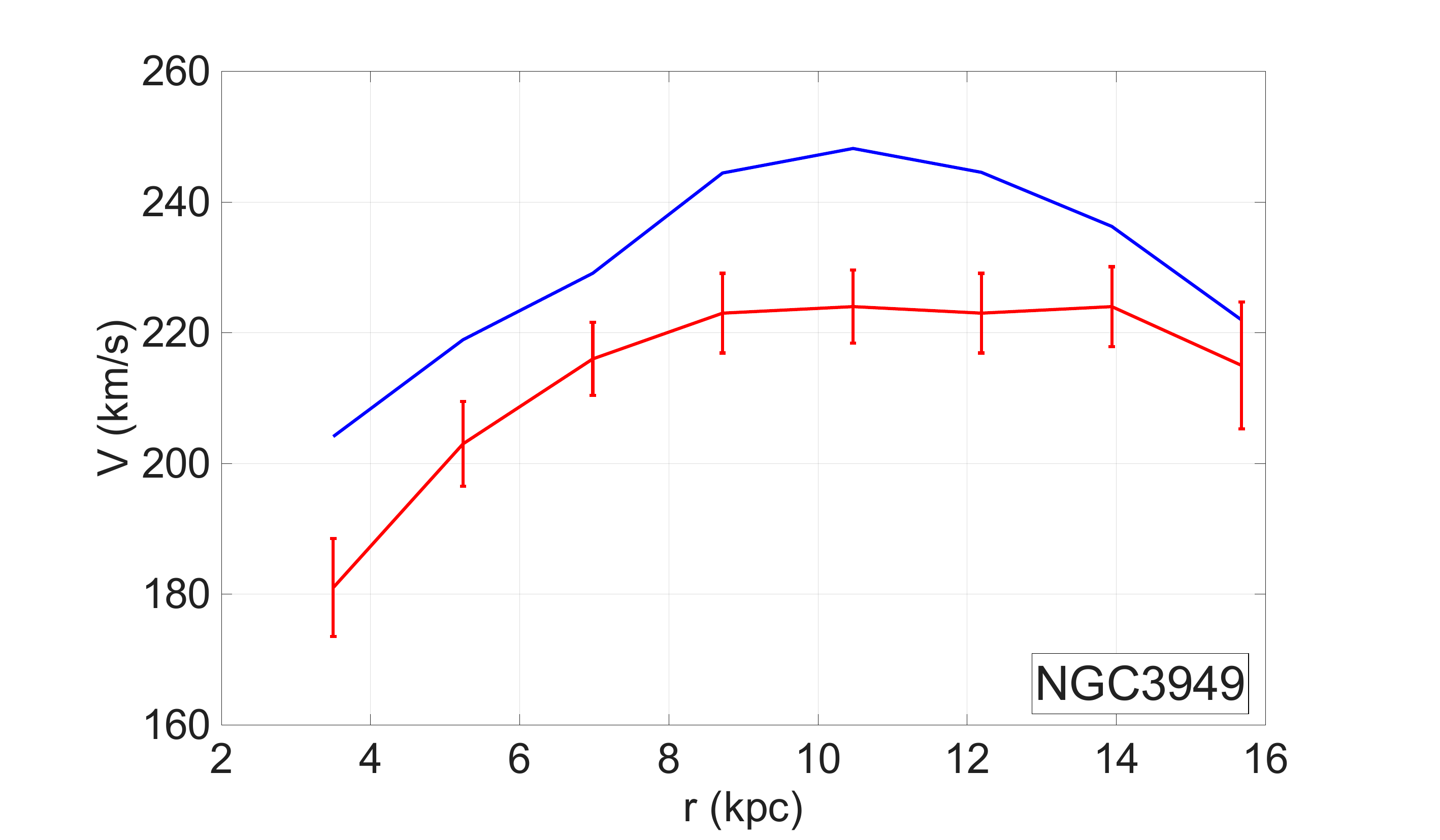}    
    \includegraphics[trim=3cm 0cm 4cm 1cm, clip=true, width=0.19\linewidth]{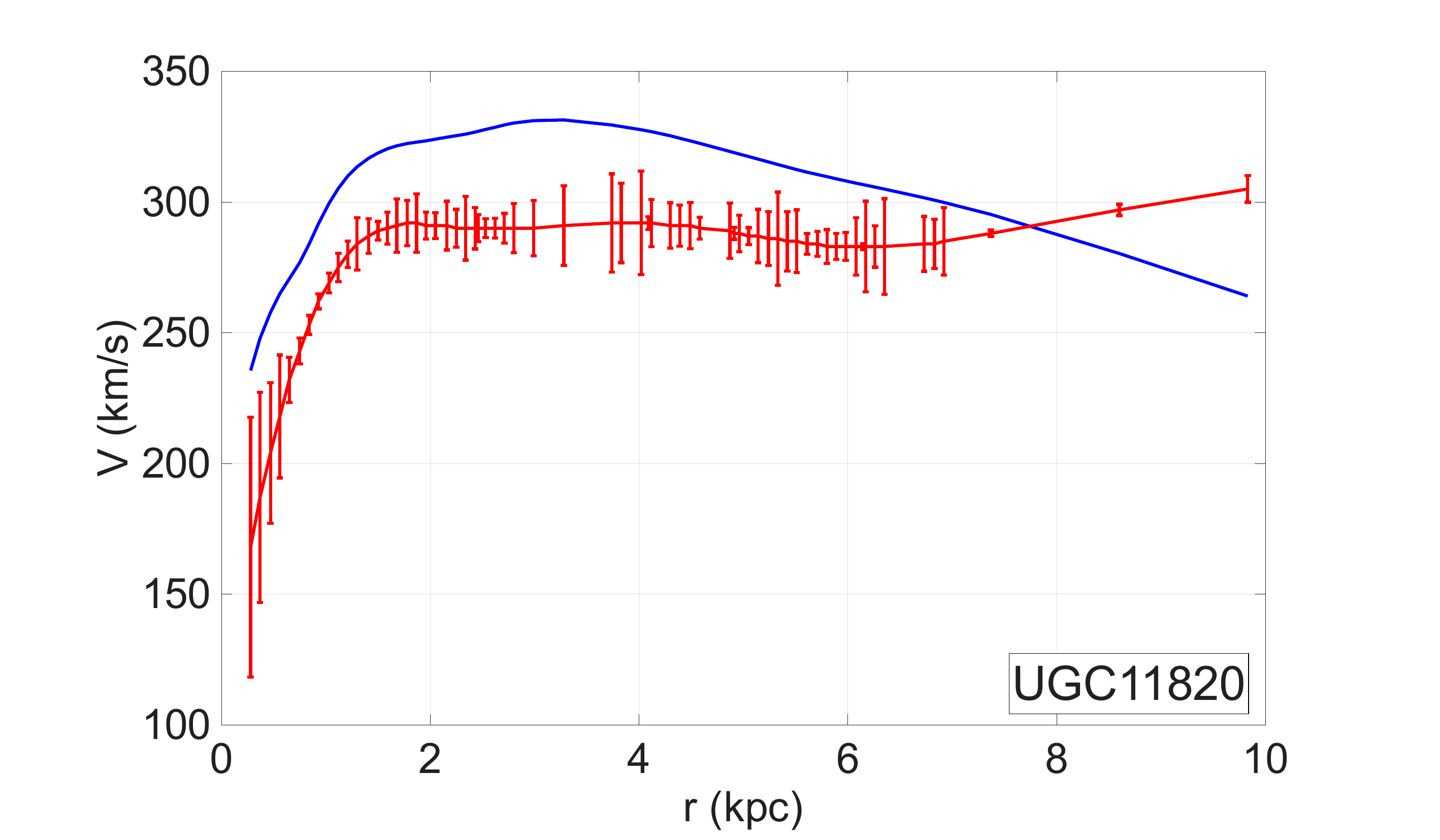}        
    \includegraphics[trim=3cm 0cm 4cm 1cm, clip=true, width=0.19\linewidth]{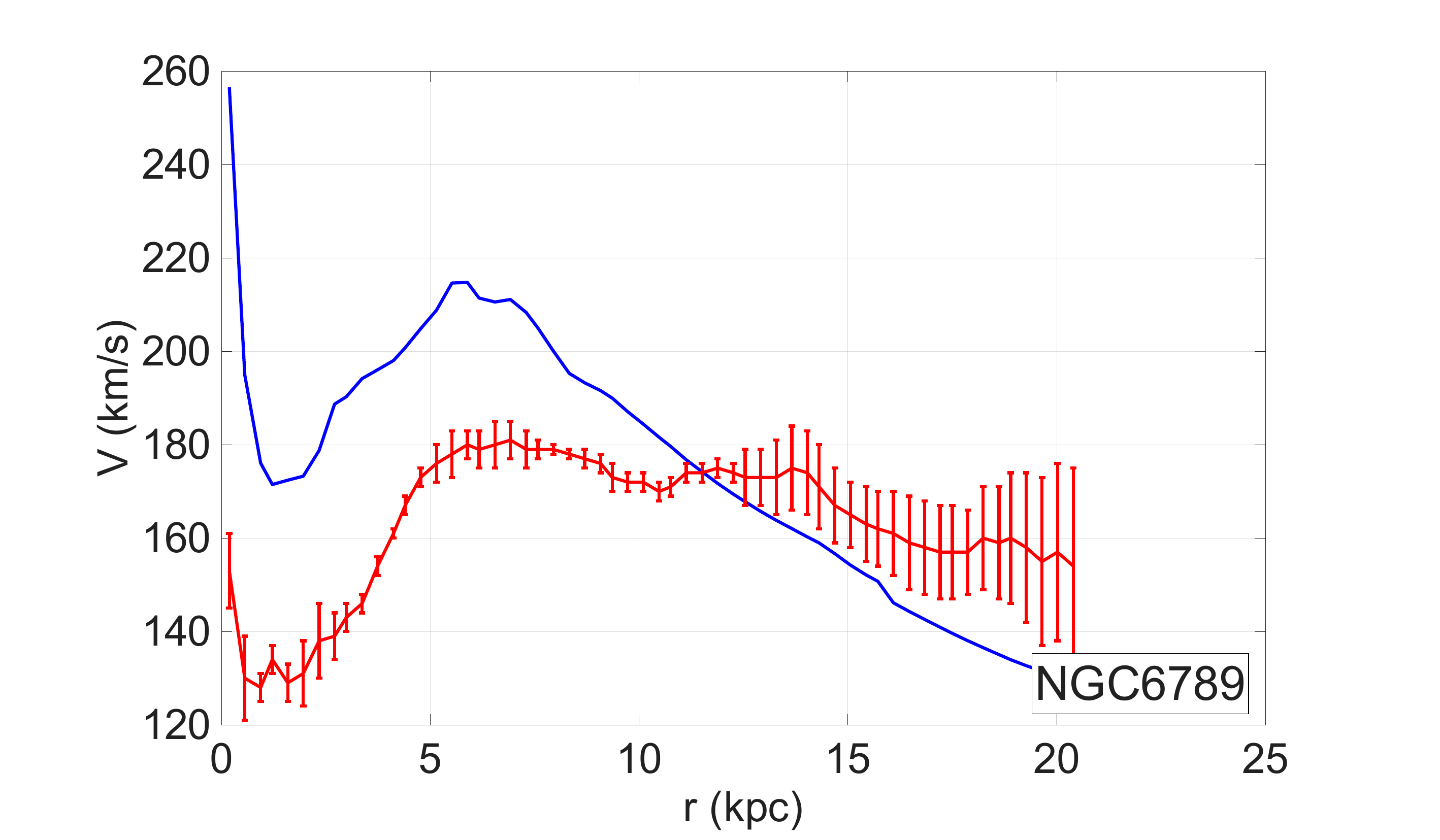} 
    \includegraphics[trim=3cm 0cm 4cm 1cm, clip=true, width=0.19\linewidth]{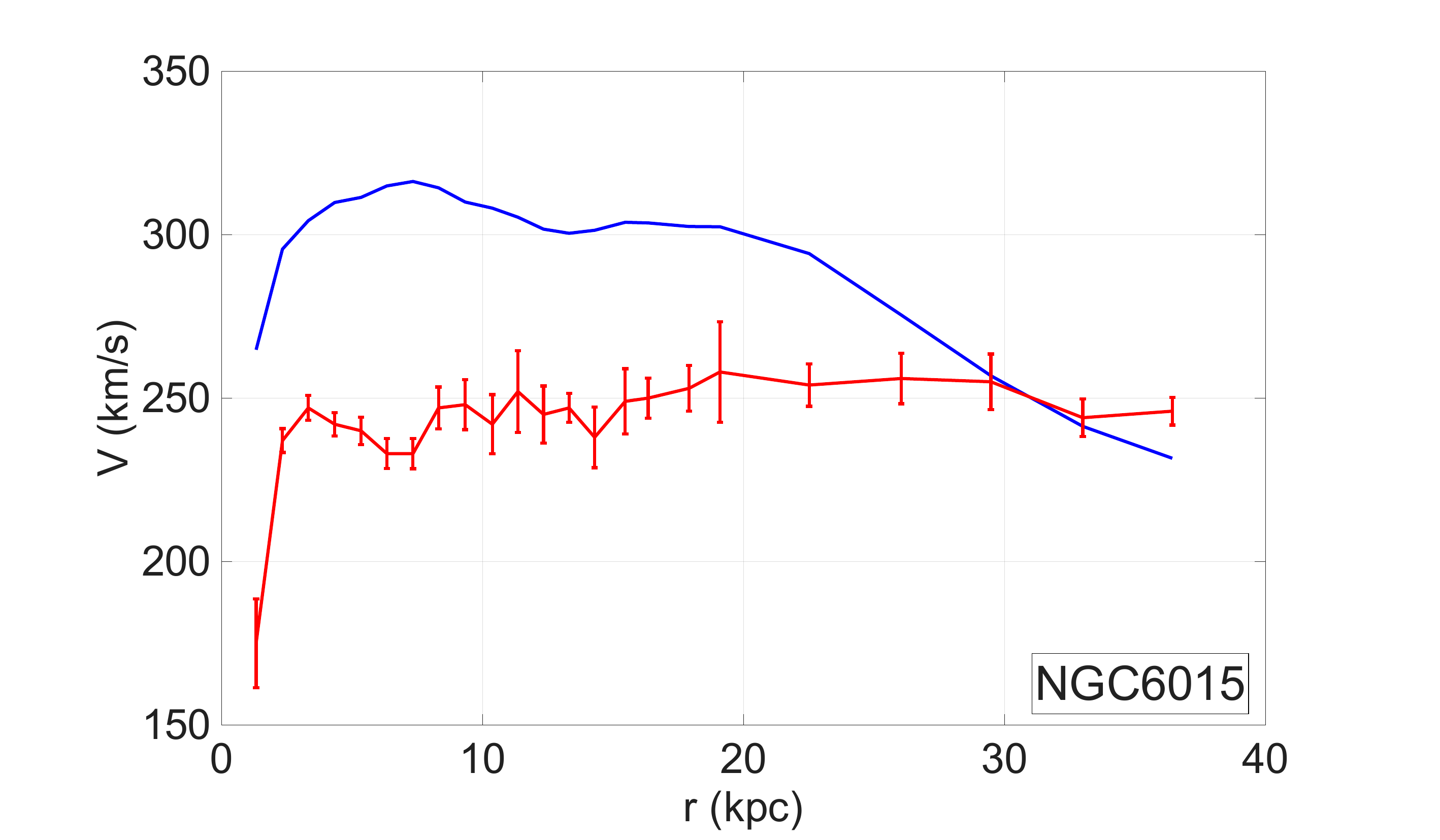}     
    \includegraphics[trim=3cm 0cm 4cm 1cm, clip=true, width=0.19\linewidth]{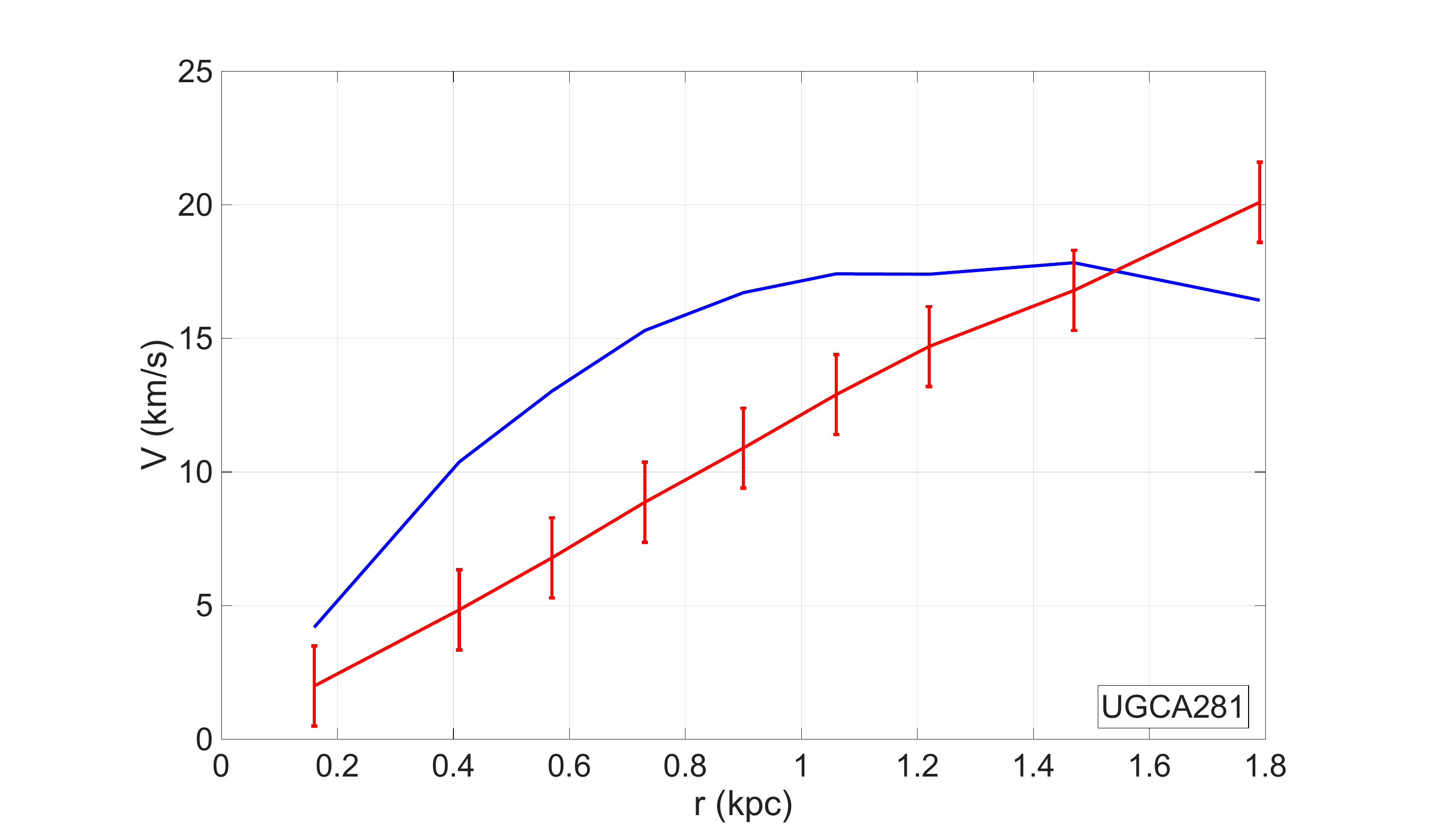}     
    \includegraphics[trim=3cm 0cm 4cm 1cm, clip=true, width=0.19\linewidth]{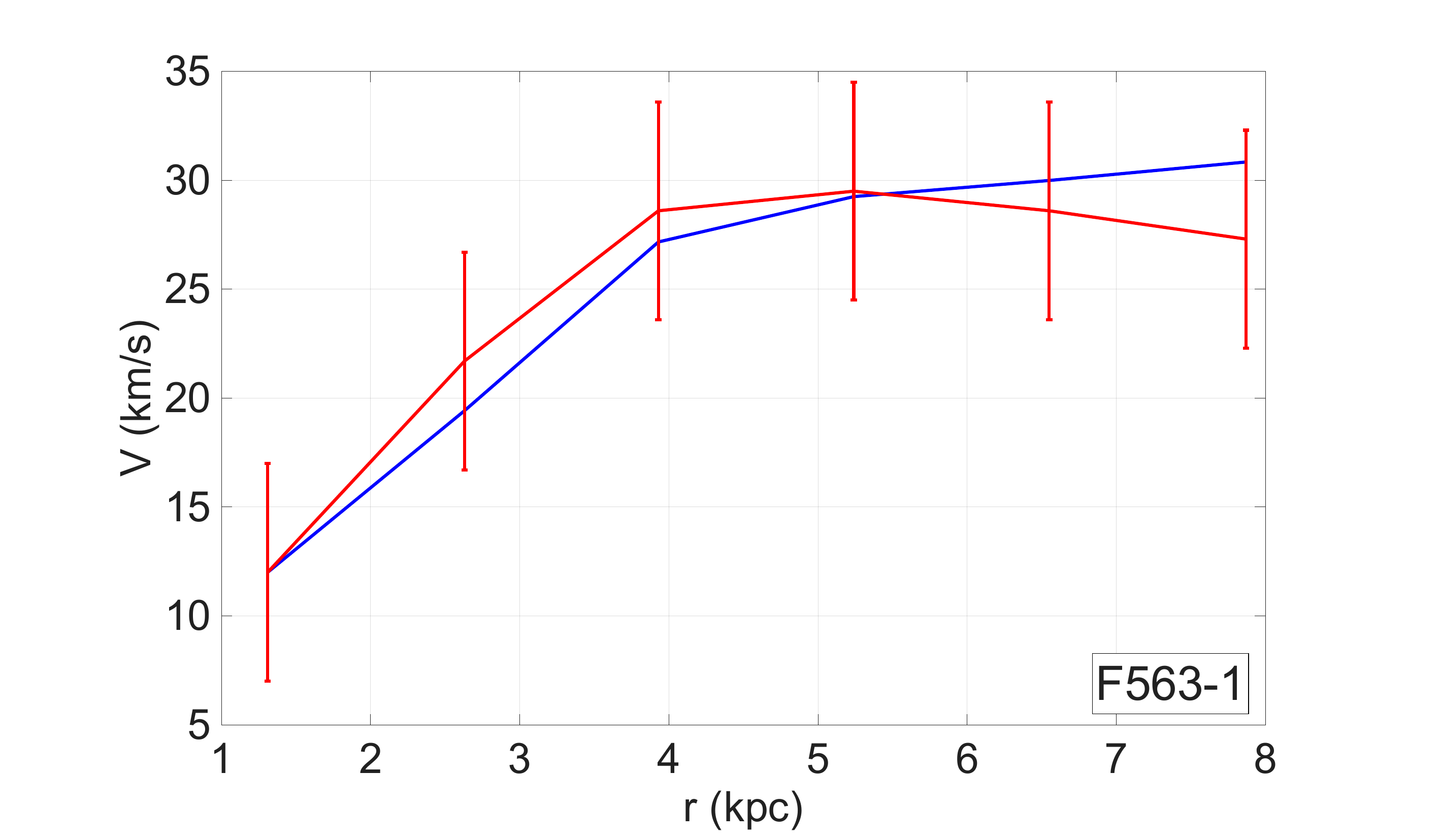}   
    \includegraphics[trim=3cm 0cm 4cm 1cm, clip=true, width=0.19\linewidth]{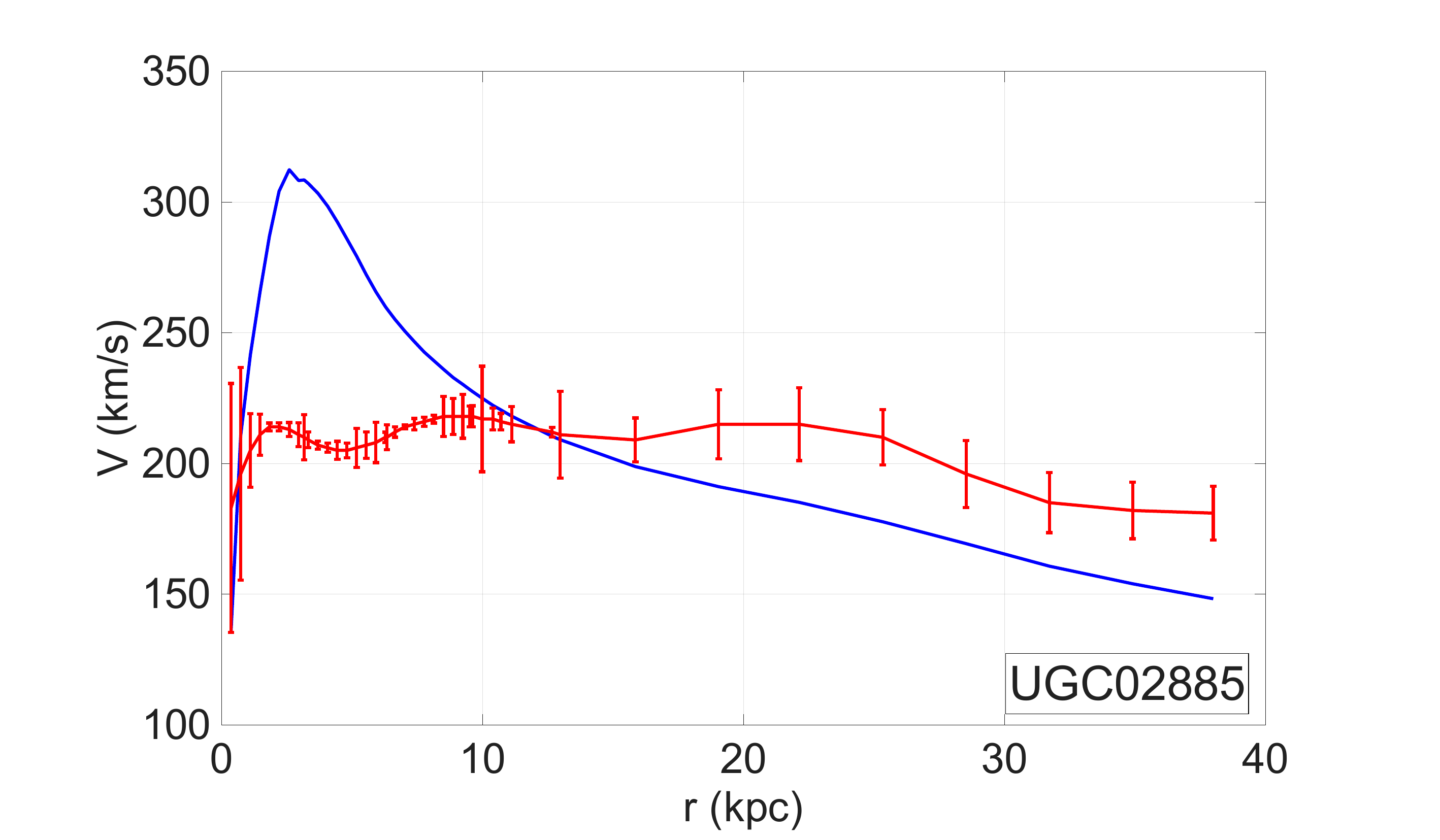}   
    \caption{Some galaxies located at the lower and upper extremes of the Machian Gravity acceleration scale in Fig.~\ref{fig:a0_MG_MOND} display noticeable irregularities in their baryonic velocity profiles. In this figure, we show the observed rotation velocities (red error bars) together with the baryonic velocities computed using Newtonian mechanics (blue curves) for 25 such galaxies. For most of these systems, the baryonic velocities exceed the observed velocities, suggesting that the fixed mass-to-light ratios adopted in the SPARC database may not yield accurate results for these galaxies. }
    \label{fig:abnormalVelocity}
\end{figure}

Approximately 20 bars at the beginning of the plot (corresponding to $a_0^{\mathrm{MG}}$) and about 15–20 bars at the end exhibit anomalous behavior in the inferred acceleration scales. To investigate the origin of these discrepancies, I plot the baryonic and observed velocity profiles for 25 of these galaxies in Fig.~\ref{fig:abnormalVelocity}. From these plots, it is evident that, for all such galaxies, the velocities computed from the baryonic mass distribution exceed the observed velocities, indicating potential issues with the assumed mass-to-light ratios. In particular, the use of constant mass-to-light ratios in the SPARC analysis may not be appropriate for these systems. This suggests that the mass-to-light ratios for these galaxies may need to be reassessed using independent observations or treated as free parameters in the MCMC analysis.Therefore, the unusual acceleration values derived for these galaxies should not be viewed as a failure or limitation of the MG model itself.

It is true that Machian gravity predicts an approximately constant value of $a_0^\text{MG}$, lying within about one to one and a half orders of magnitude. However, this does not mean that if we simply fix $a_0^\text{MG}$ and then determine a single pair of parameters $(M_c,\lambda^{-1})$, that pair will account for all galactic rotation curves. If that were the case, we would effectively obtain a universal acceleration scale, as proposed in MOND. The fact that no such universal pair exists instead indicates that we require a specific, fixed pair of $M_c$ and $\lambda^{-1}$ for each system. Here, $\lambda^{-1}$ sets the characteristic length scale of the mass distribution. When this length scale varies, it alters the inertia of the object, in accordance with the Machian gravity framework.

\subsection{Comparison of Machian Gravity with the NFW Dark Matter Profile}

\begin{figure}
    \centering
    \includegraphics[trim=3cm 3cm 4cm 3cm, clip=true, width=0.32\linewidth]{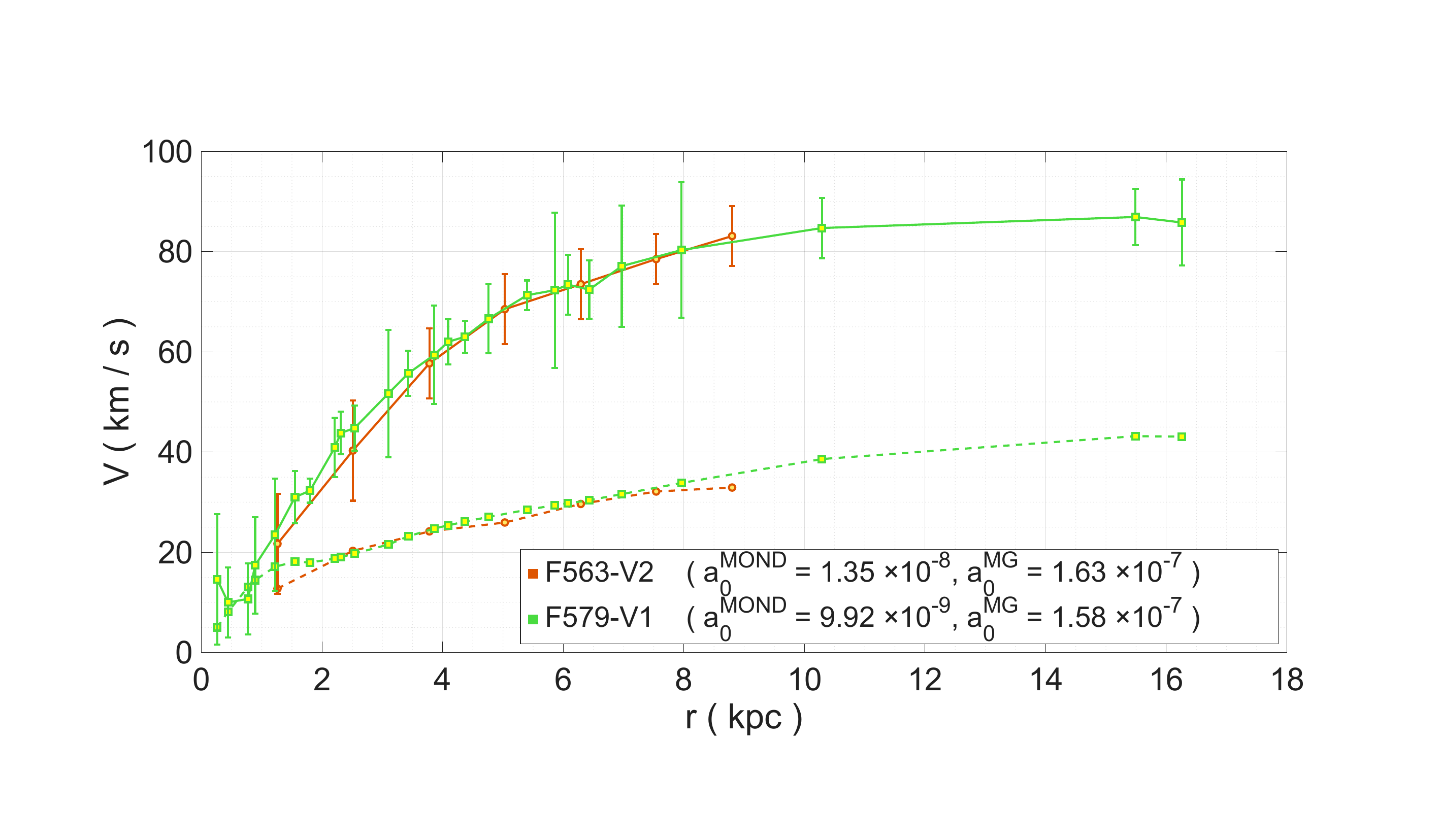}
    \includegraphics[trim=3cm 3cm 4cm 3cm, clip=true, width=0.32\linewidth]{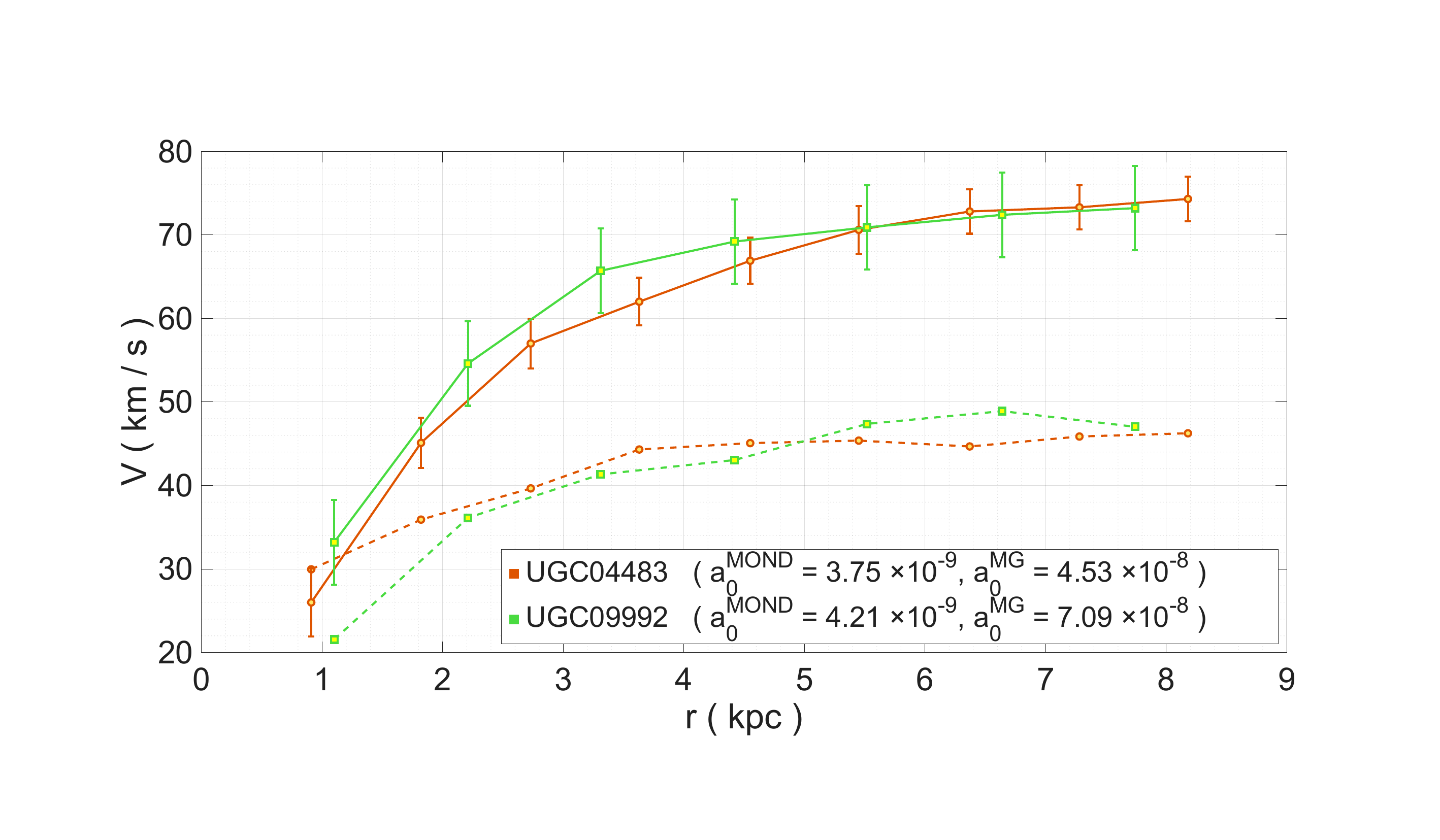}
    \includegraphics[trim=3cm 3cm 4cm 3cm, clip=true, width=0.32\linewidth]{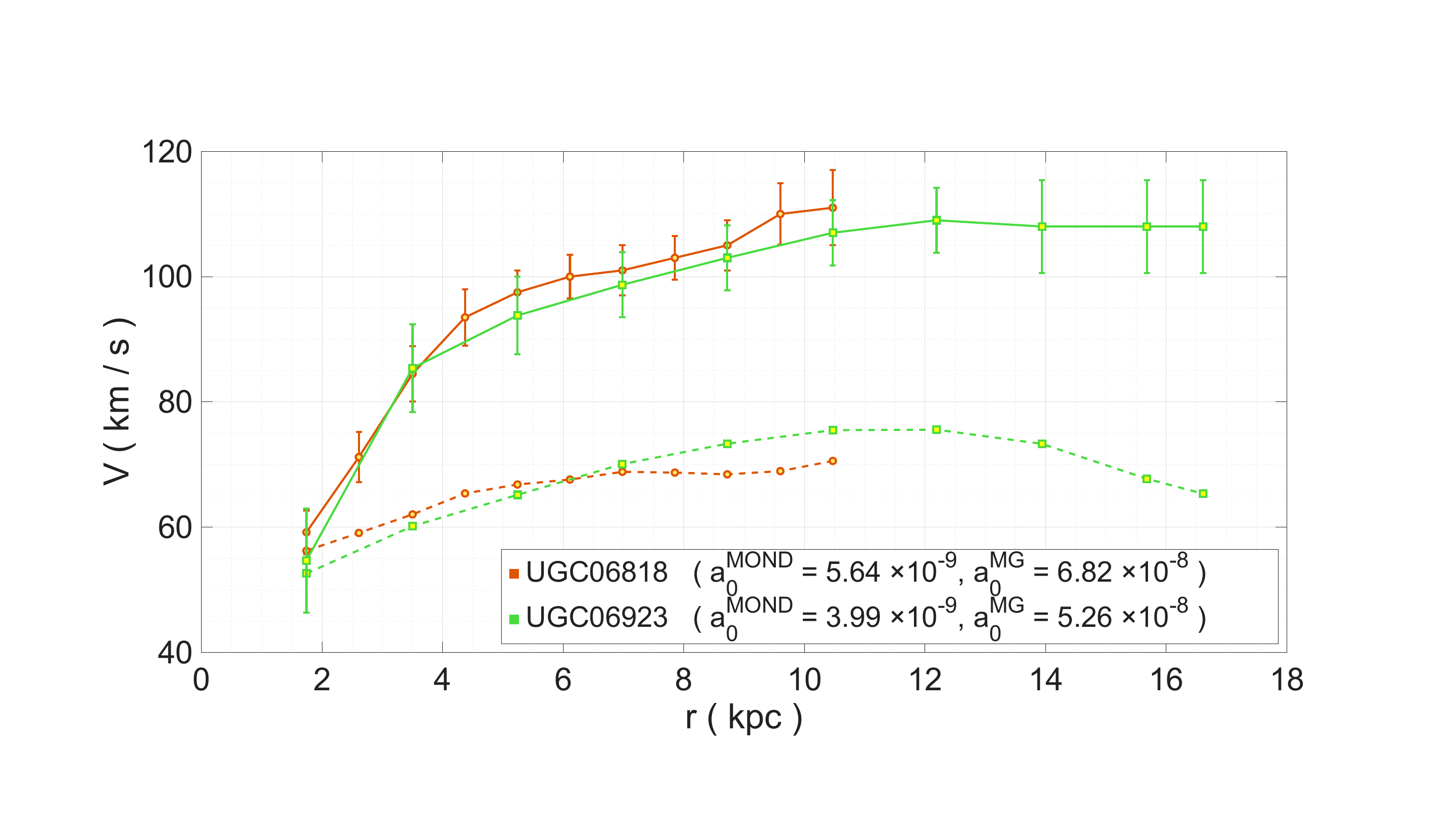}   
    \includegraphics[trim=3cm 3cm 4cm 3cm, clip=true, width=0.32\linewidth]{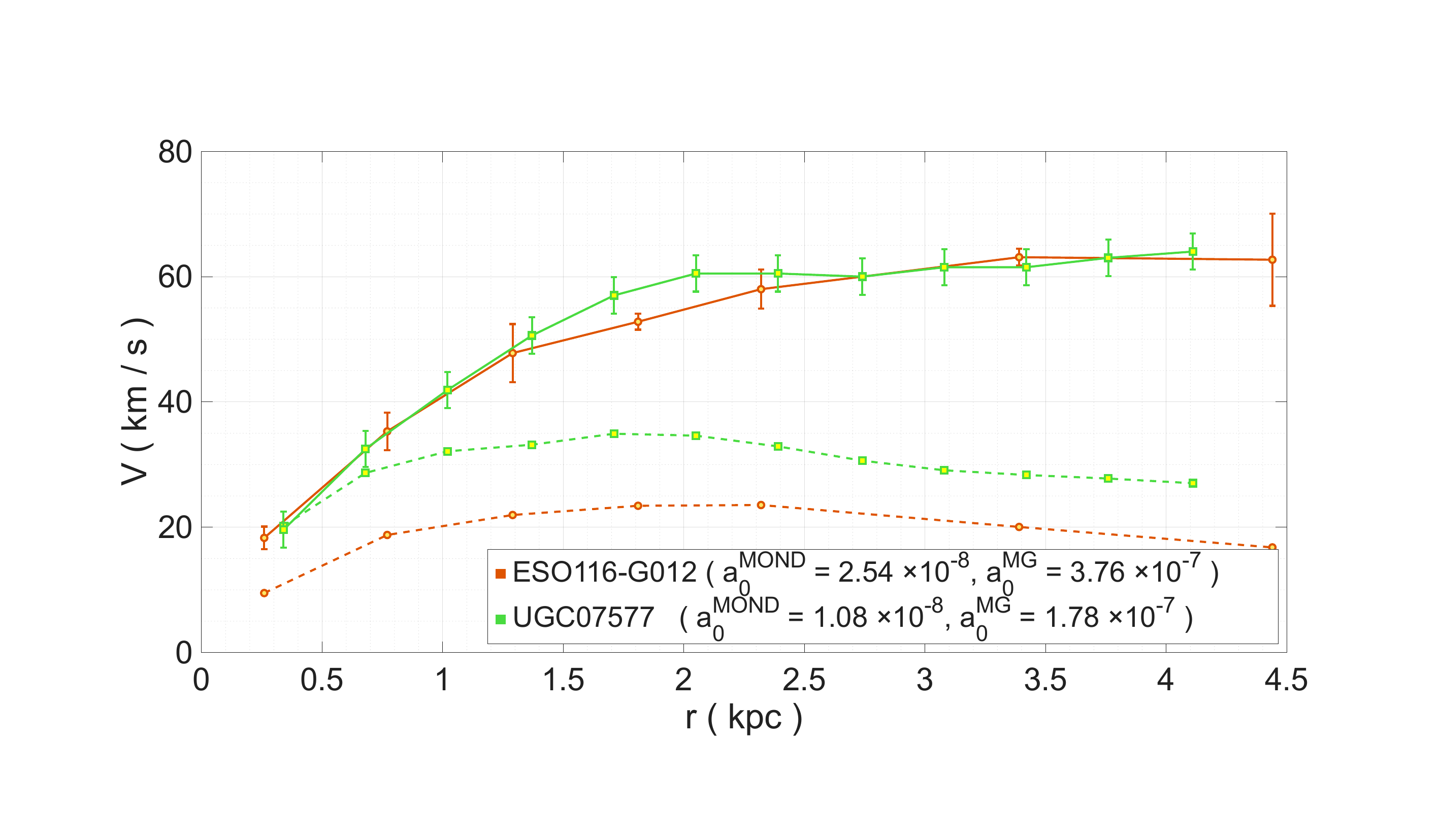}
    \includegraphics[trim=3cm 3cm 4cm 3cm, clip=true, width=0.32\linewidth]{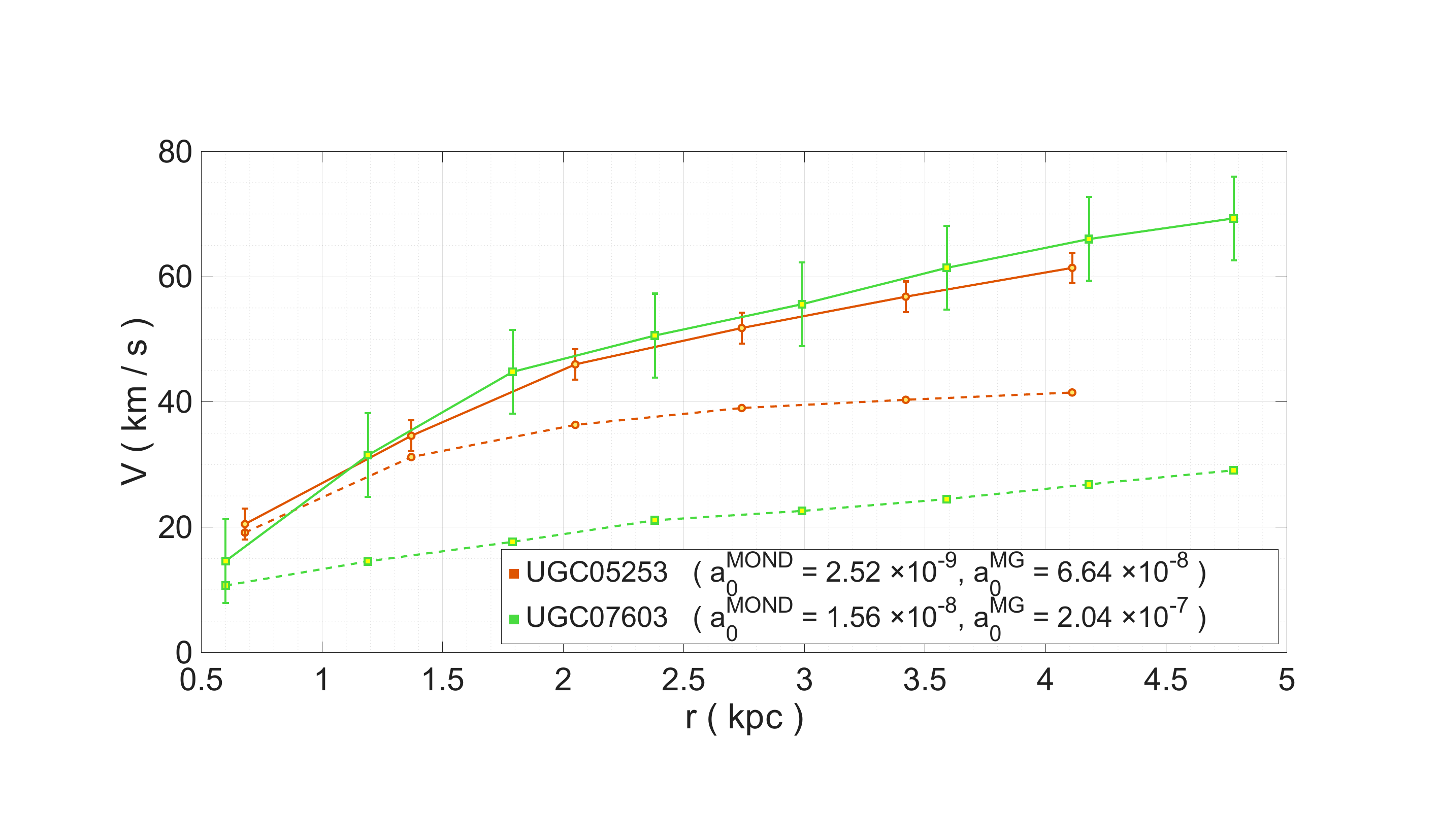}    
    \includegraphics[trim=3cm 3cm 4cm 3cm, clip=true, width=0.32\linewidth]{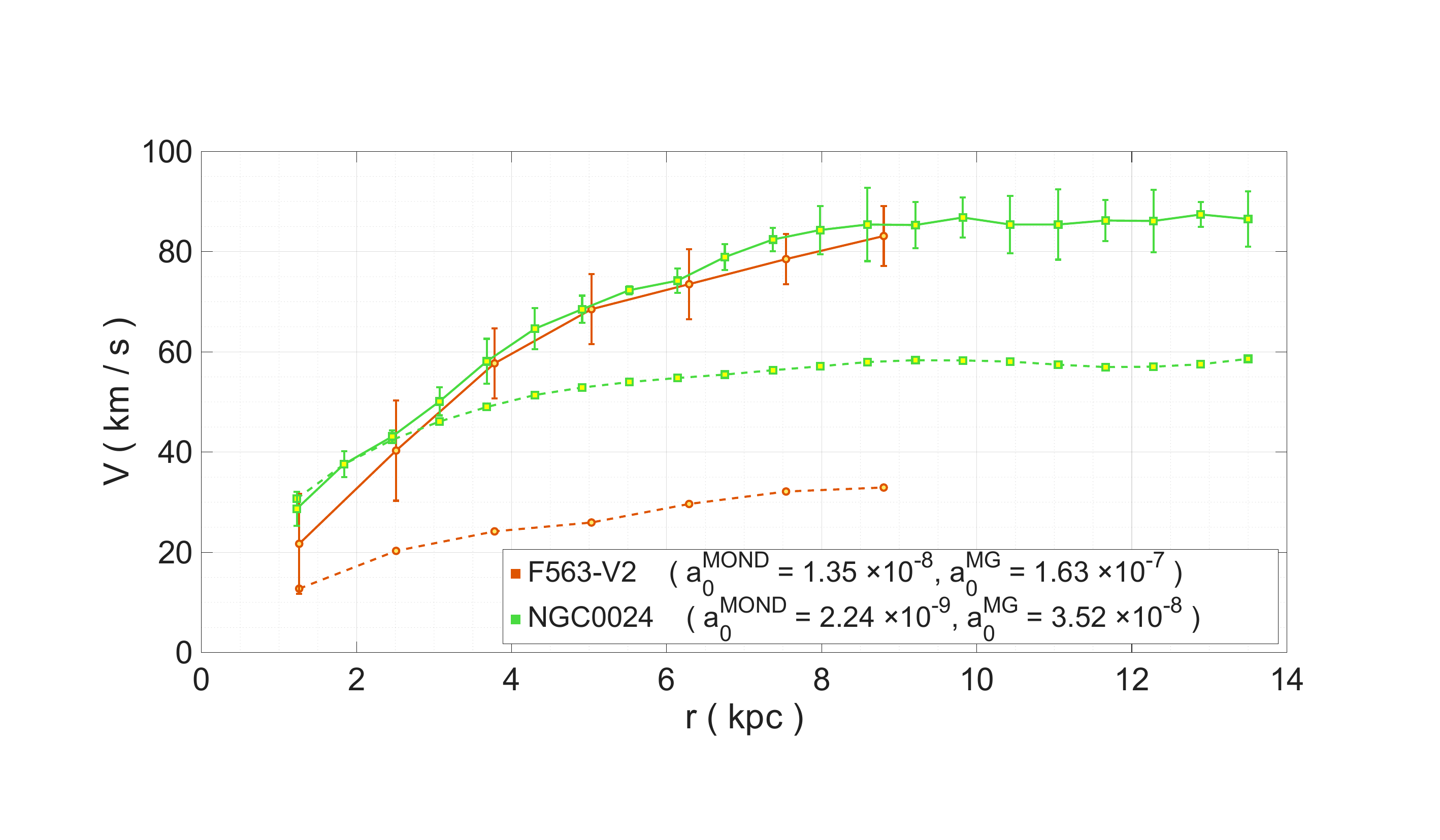}   
    \includegraphics[trim=3cm 3cm 4cm 3cm, clip=true, width=0.32\linewidth]{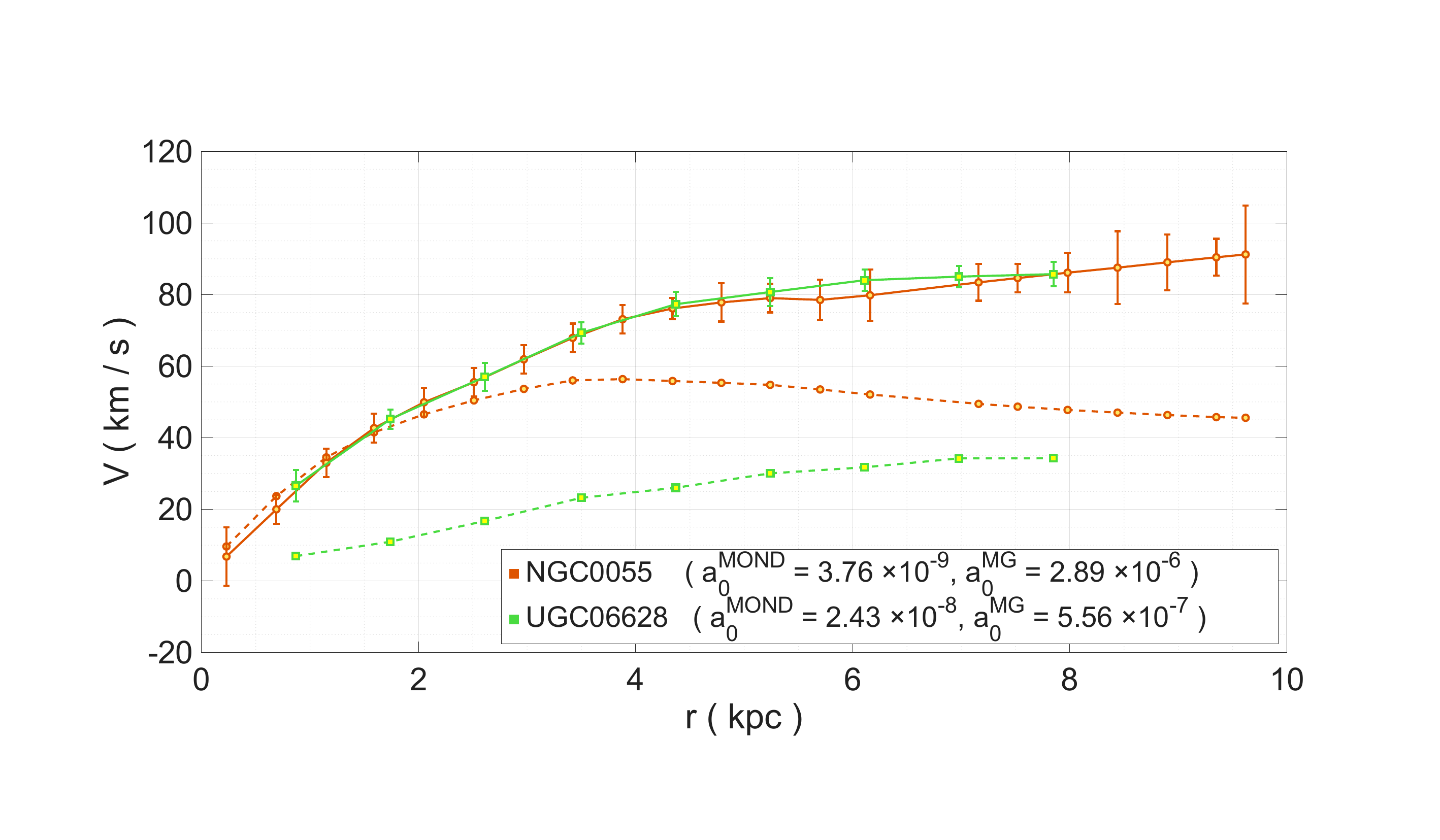}    
    \includegraphics[trim=3cm 3cm 4cm 3cm, clip=true, width=0.32\linewidth]{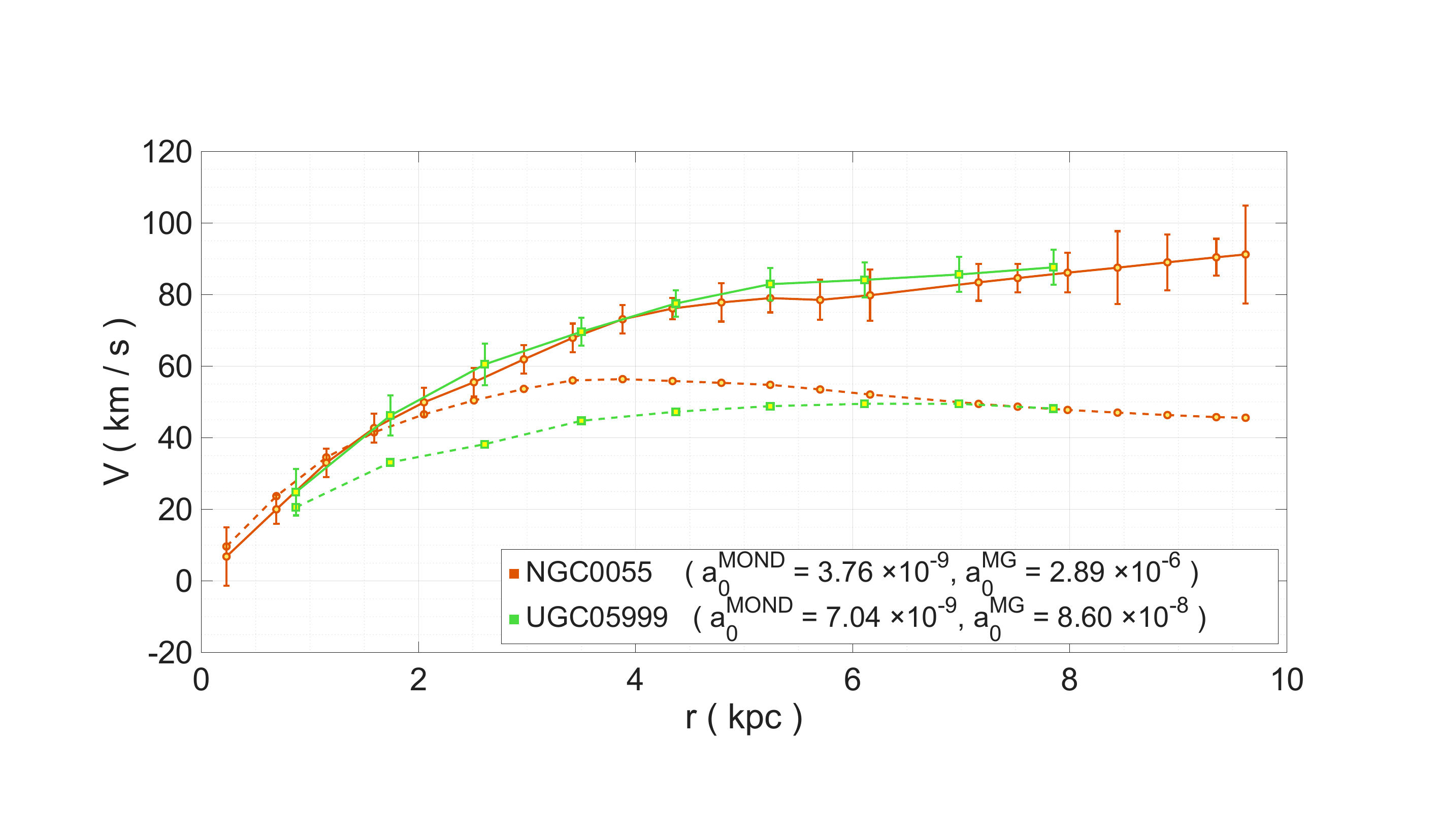}        
    \includegraphics[trim=3cm 3cm 4cm 3cm, clip=true, width=0.32\linewidth]{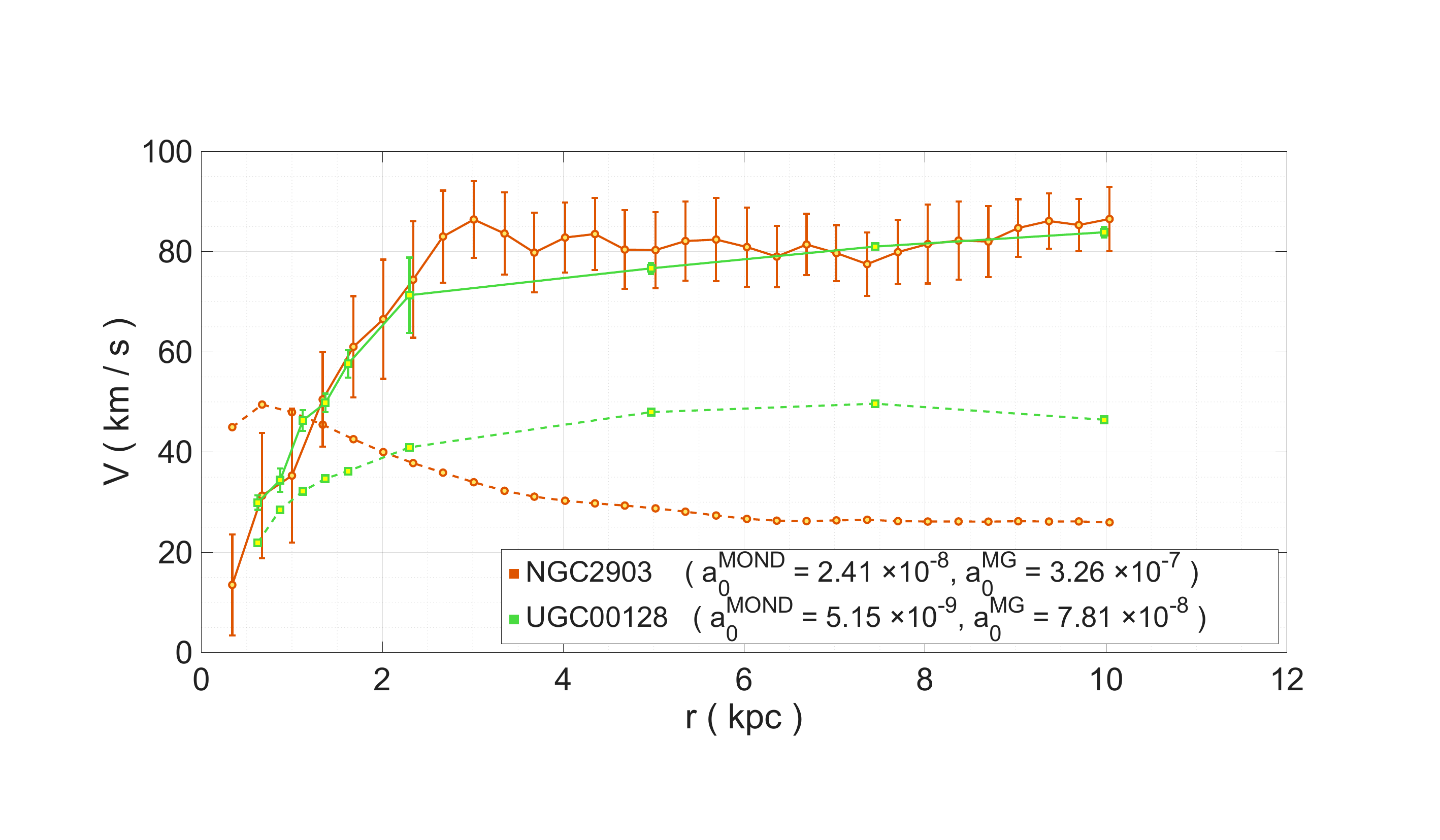} 
    \includegraphics[trim=3cm 3cm 4cm 3cm, clip=true, width=0.32\linewidth]{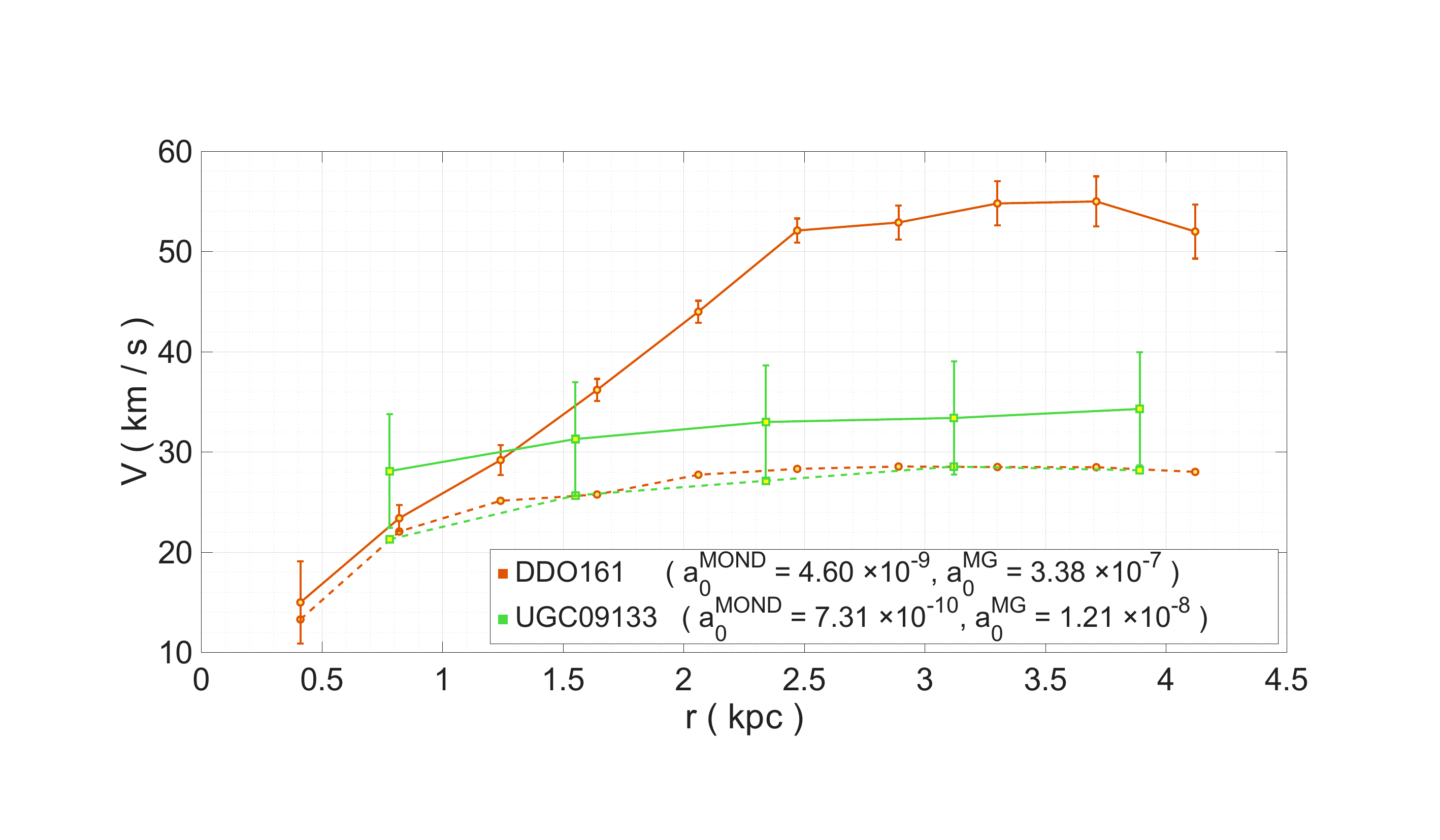}     
    \includegraphics[trim=3cm 3cm 4cm 3cm, clip=true, width=0.32\linewidth]{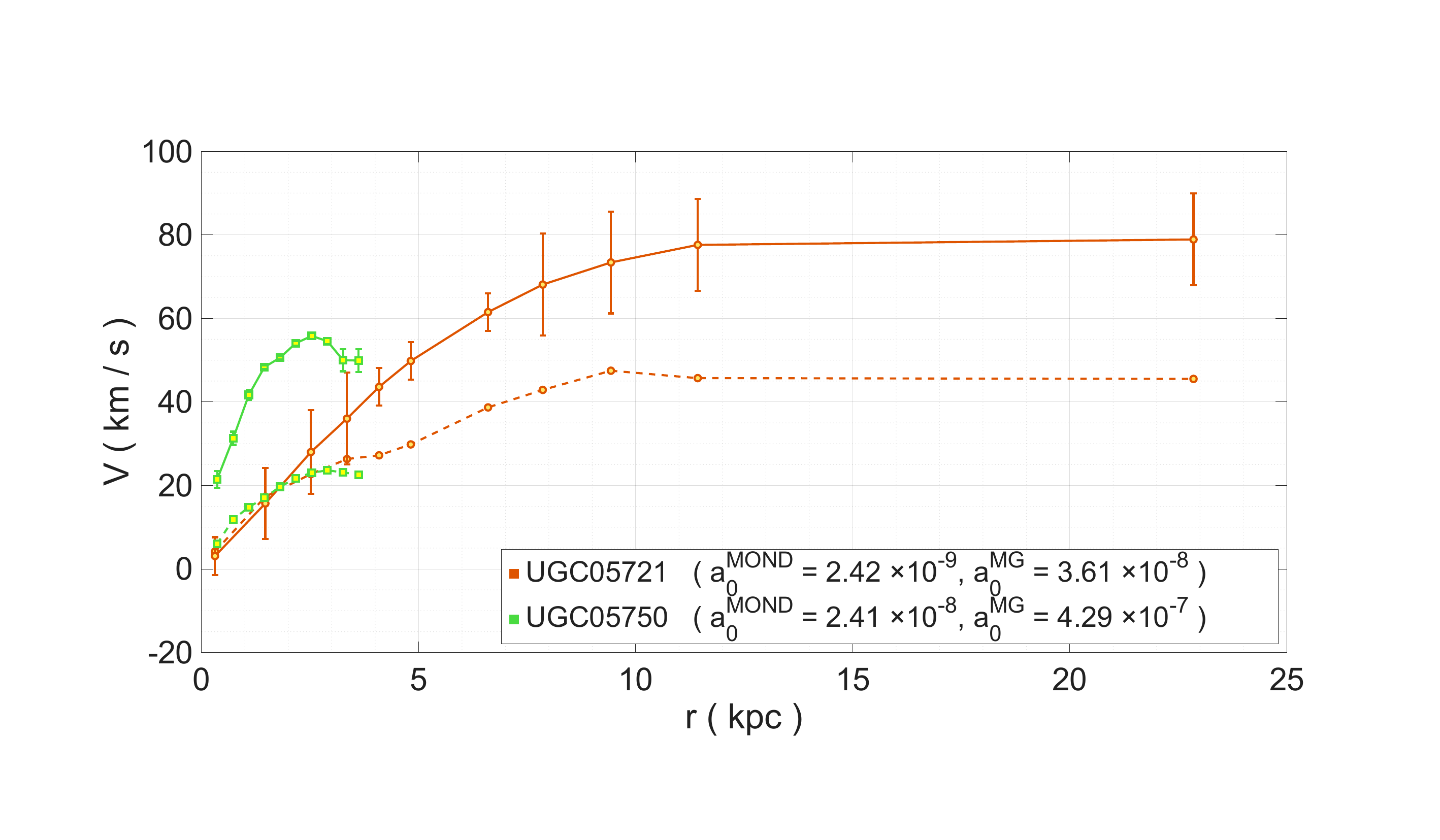}     
    \includegraphics[trim=3cm 3cm 4cm 3cm, clip=true, width=0.32\linewidth]{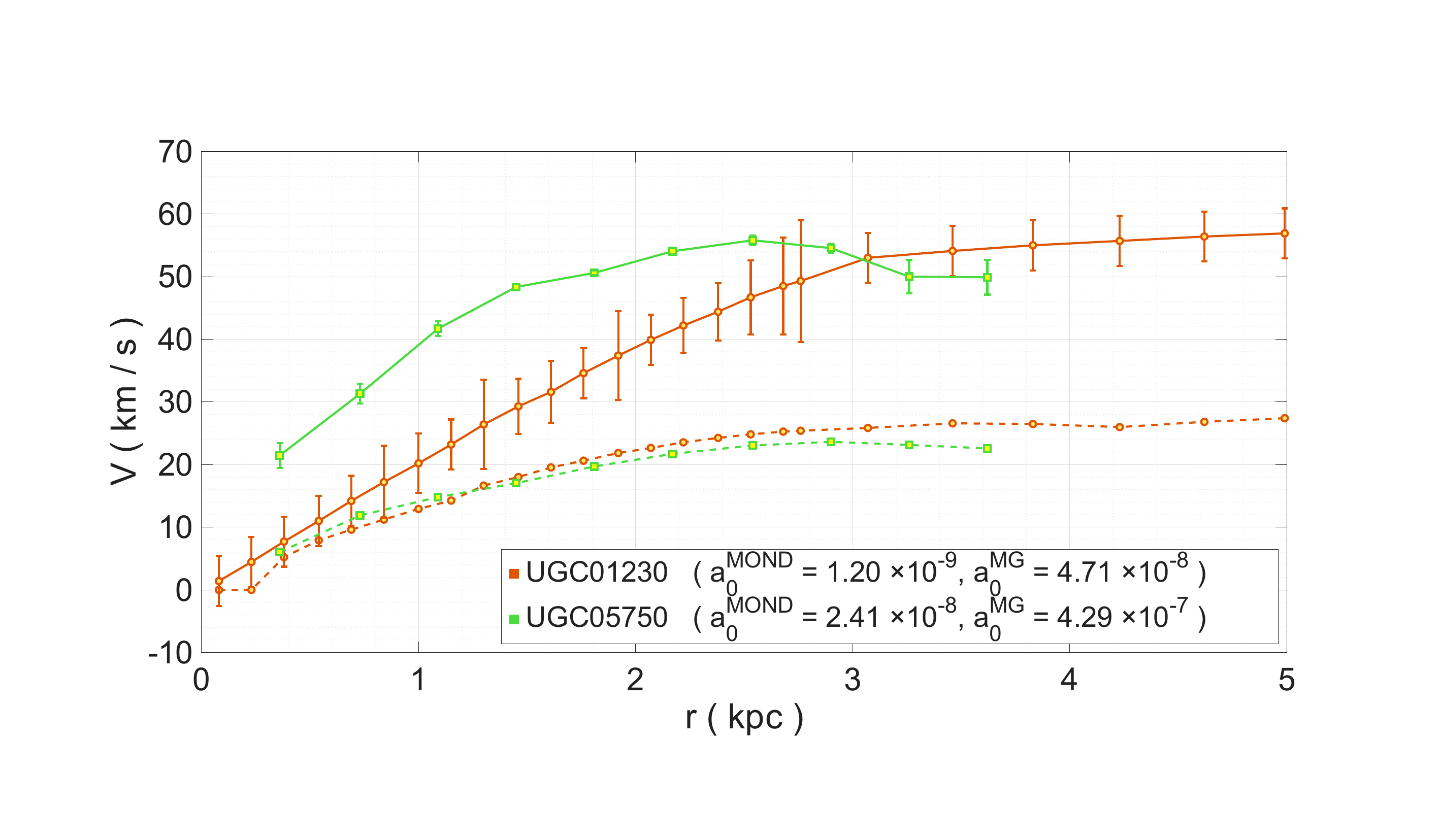}         
    \caption{This figure presents 12 pairs of galaxies whose baryonic mass profiles or rotation (velocity) curves are similar. In the first row, I show 3 galaxy pairs for which the velocities predicted from the baryonic mass distributions agree well with the observed rotation curves. In the second row, I show 3 galaxy pairs where the observed velocity profiles match, and the baryon-based velocity curves also look similar but do not coincide exactly. However, these plots suggest that the baryonic profiles could also be made to match by adjusting the mass-to-light ratios. In the third row, I show three additional galaxy pairs where the velocities computed from the baryons alone display slightly different shapes. It seems that taking a somewhat lower bulge mass-to-light ratio for the red curves could reconcile the remaining discrepancies between the mass profiles of the paired galaxies. Overall, in all these cases, MOND and MG appear to provide a more satisfactory description than standard dark matter, since in a model where dark matter is a completely independent component, there is no clear reason for such close agreement between baryonic and total mass distributions. In the last row, I show some galaxy pairs where the baryon-based velocity curves coincide, but the shapes of the observed rotation curves differ. Such galaxies are relatively rare and may be disregarded, for instance due to potential observational issues. For other galaxies check \href{https://machiangravity.github.io/Galactic-Velocity-Profile-/Velocity_Pair/}{Velocity Pair} and \href{https://machiangravity.github.io/Galactic-Velocity-Profile-/Baryon_Pair/}{Baryon Pair}.}
    \label{fig:pairVelocity}
\end{figure}

In the standard dark matter scenario, the dark matter distribution is assumed to be independent of the baryonic matter. Consequently, there is no fundamental reason for the dark matter distributions in different galaxies to closely trace their baryonic distributions. To examine this issue, Fig.~\ref{fig:pairVelocity} presents several pairs of galaxies with similar rotation curves.

In the first row, I show three pairs of galaxies whose observed velocity profiles closely match. Notably, the velocities computed from their baryonic mass distributions also agree remarkably well. If dark matter were truly independent of the baryonic distribution, there would be no a priori reason for both the baryonic and total mass distributions of two different galaxies to align so closely. Nevertheless, several such galaxy pairs are observed, suggesting a strong linkage between baryonic matter and the resulting velocity profiles.

In the second row, I present three additional galaxy pairs whose rotation curves nearly overlap, while their inferred baryonic distributions show slight differences. A simple visual inspection shows that the overall shapes of the baryonic distributions are similar, indicating that the discrepancies may arise from uncertainties in the adopted mass-to-light ratios. If consistent mass-to-light ratios were chosen for both the disk and bulge components, these baryonic distributions would likely coincide more closely.

The third row shows three further galaxy pairs with nearly identical rotation curves. In these cases, adjusting the mass-to-light ratios of the disk and bulge components appropriately can reconcile the apparent differences in the baryonic mass distributions. In most instances, the discrepancies appear to originate from an overestimation of the bulge mass-to-light ratio, leading to artificially distinct baryonic profiles. Otherwise, when the observed velocity profiles match, the velocities computed from the baryonic mass distributions also match, a behavior that is difficult to explain if dark matter is an independent component.

However, I should also point out that there are some galaxies as shown in the bottom row of the plot, for which the baryonic distributions are the same, yet the velocity profiles differ significantly. If the baryonic distribution is identical, one would normally expect similar velocity profiles. Therefore, in these particular cases, one might argue that there could be galaxies for which dark matter is required to account for the observed velocities. 

However, in the second and third examples, the radii of the galaxies are not the same, so it is possible that the characteristic length scales, $\lambda^{-1}$, differ as well. Moreover, these systems represent only a small fraction of the total galaxy sample and can potentially be disregarded, assuming that observational uncertainties or other effects are responsible for these anomalies.

\begin{figure}
\centering
\includegraphics[trim=4cm 3cm 5cm 4cm, clip=true, width=0.325\columnwidth]{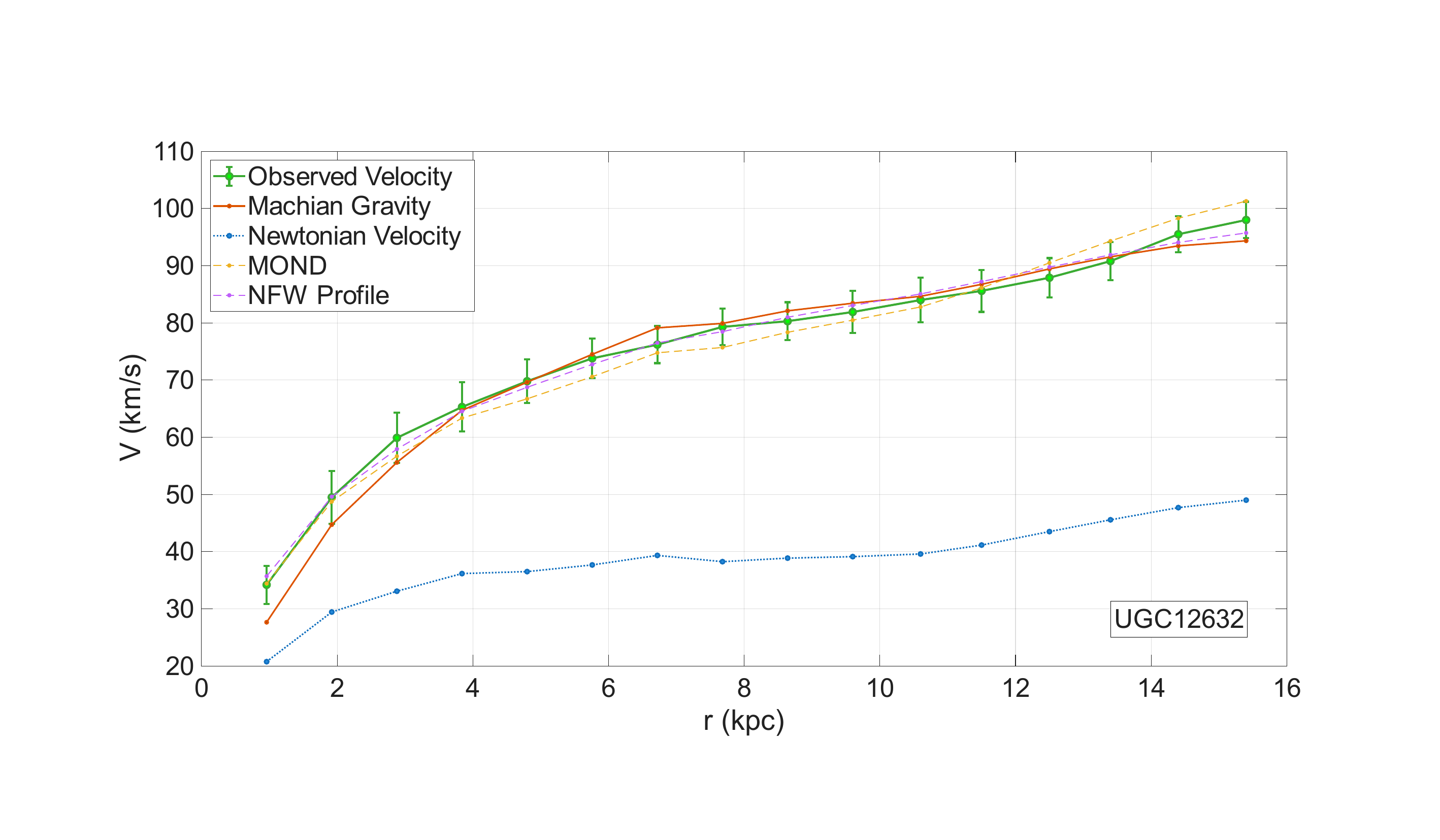}
\includegraphics[trim=4cm 3cm 5cm 4cm, clip=true, width=0.325\columnwidth]{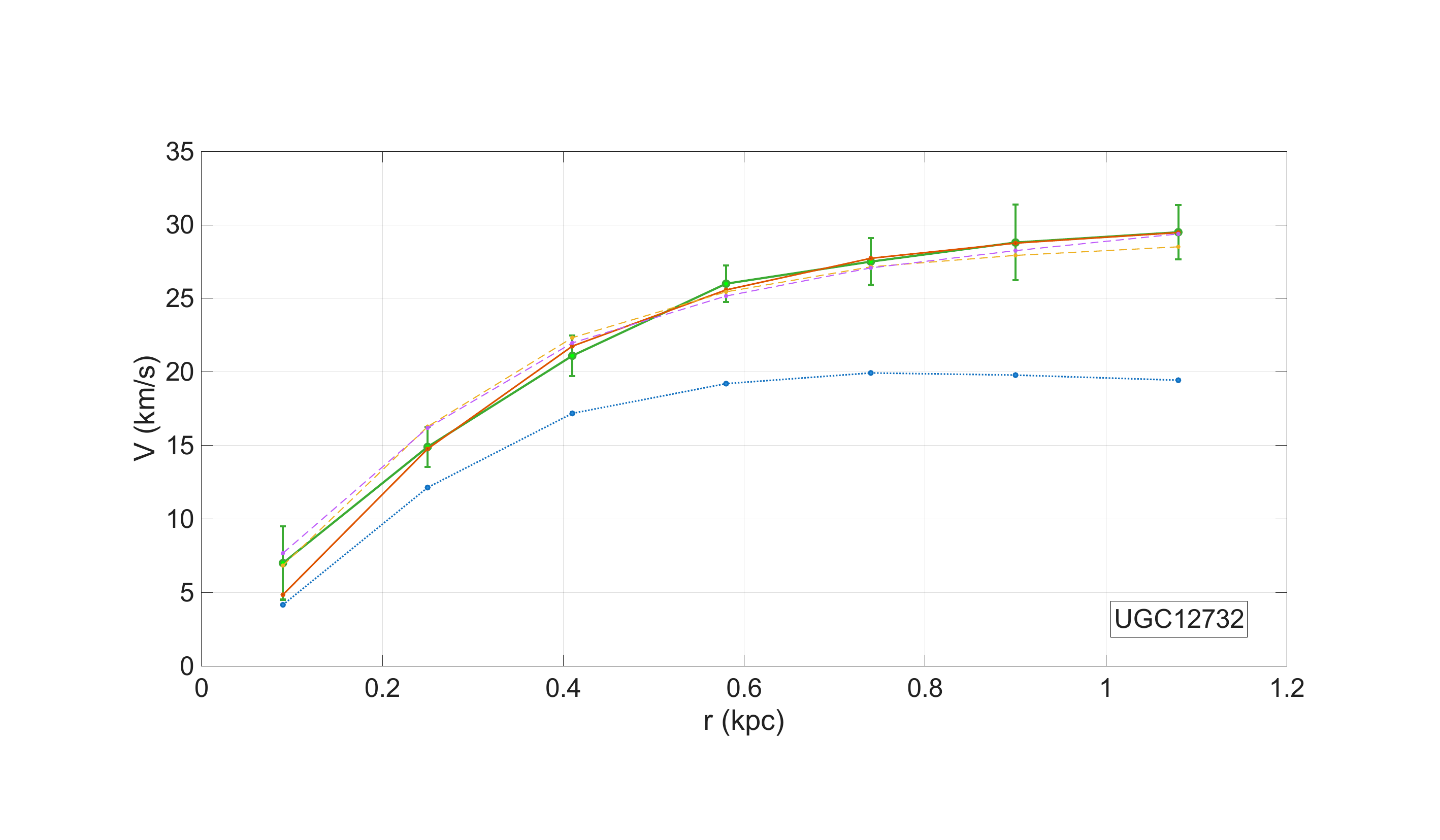}
\includegraphics[trim=4cm 3cm 5cm 4cm, clip=true, width=0.325\columnwidth]{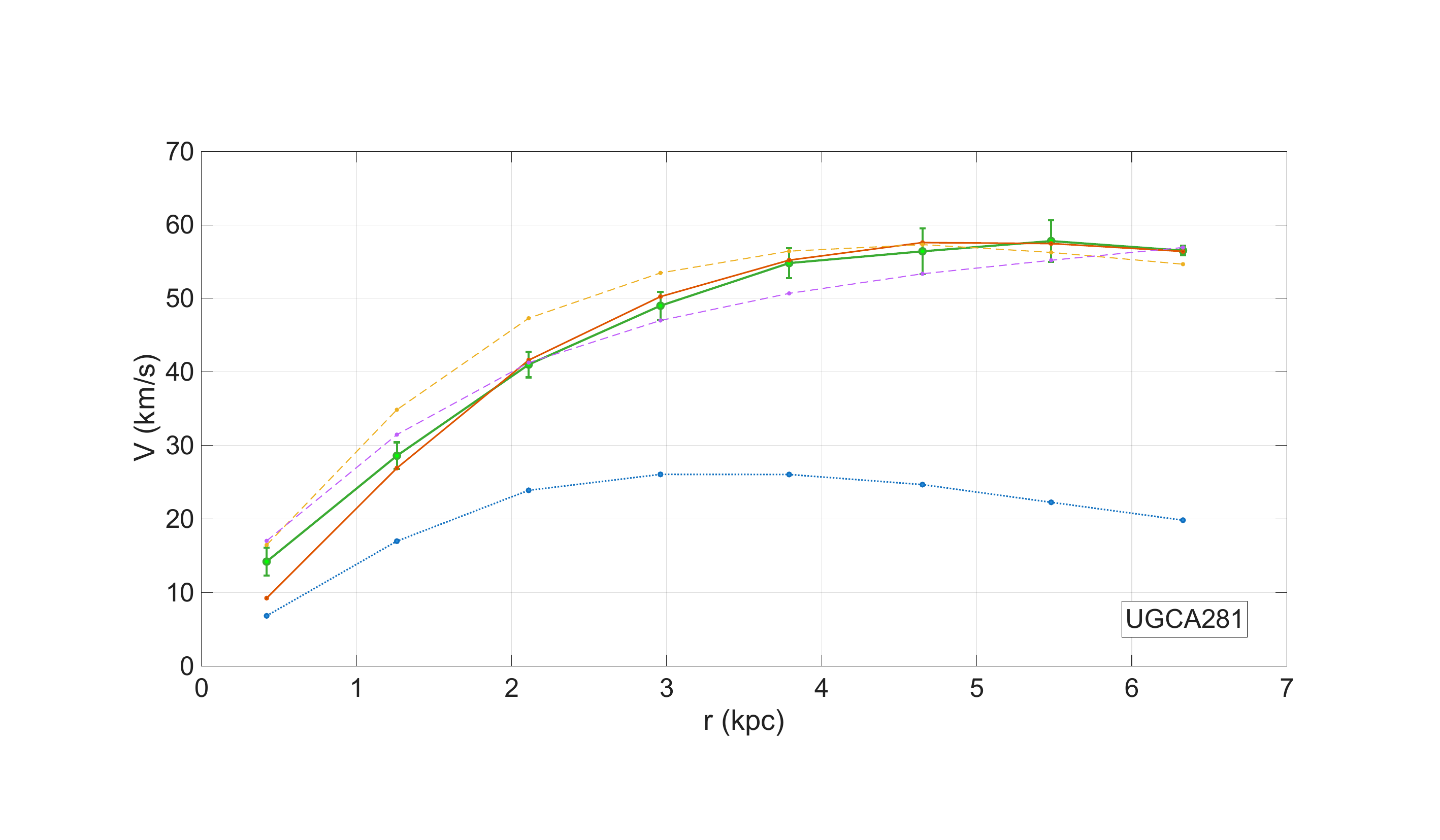}
\includegraphics[trim=4cm 3cm 5cm 4cm, clip=true, width=0.325\columnwidth]{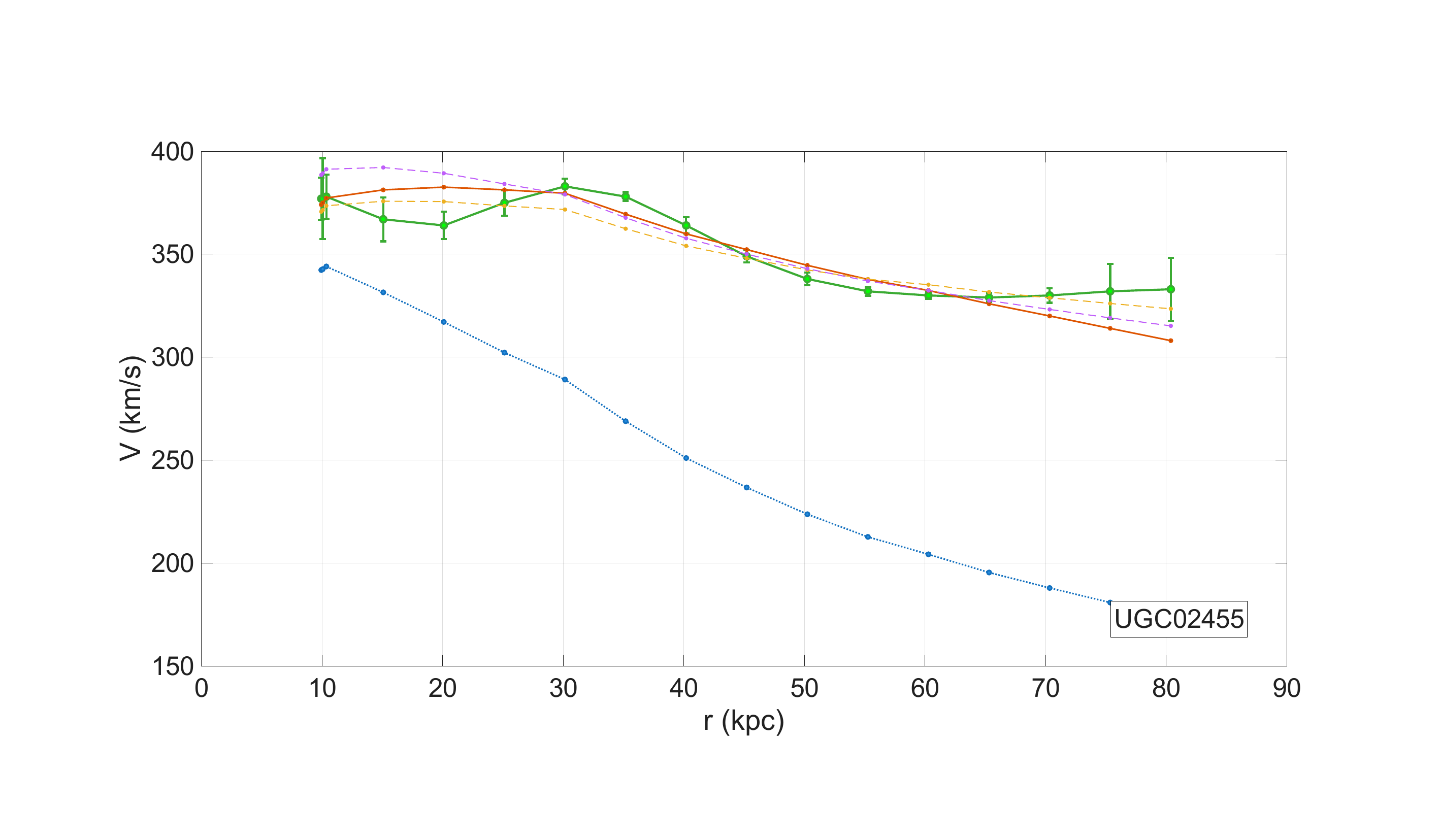}
\includegraphics[trim=4cm 3cm 5cm 4cm, clip=true, width=0.325\columnwidth]{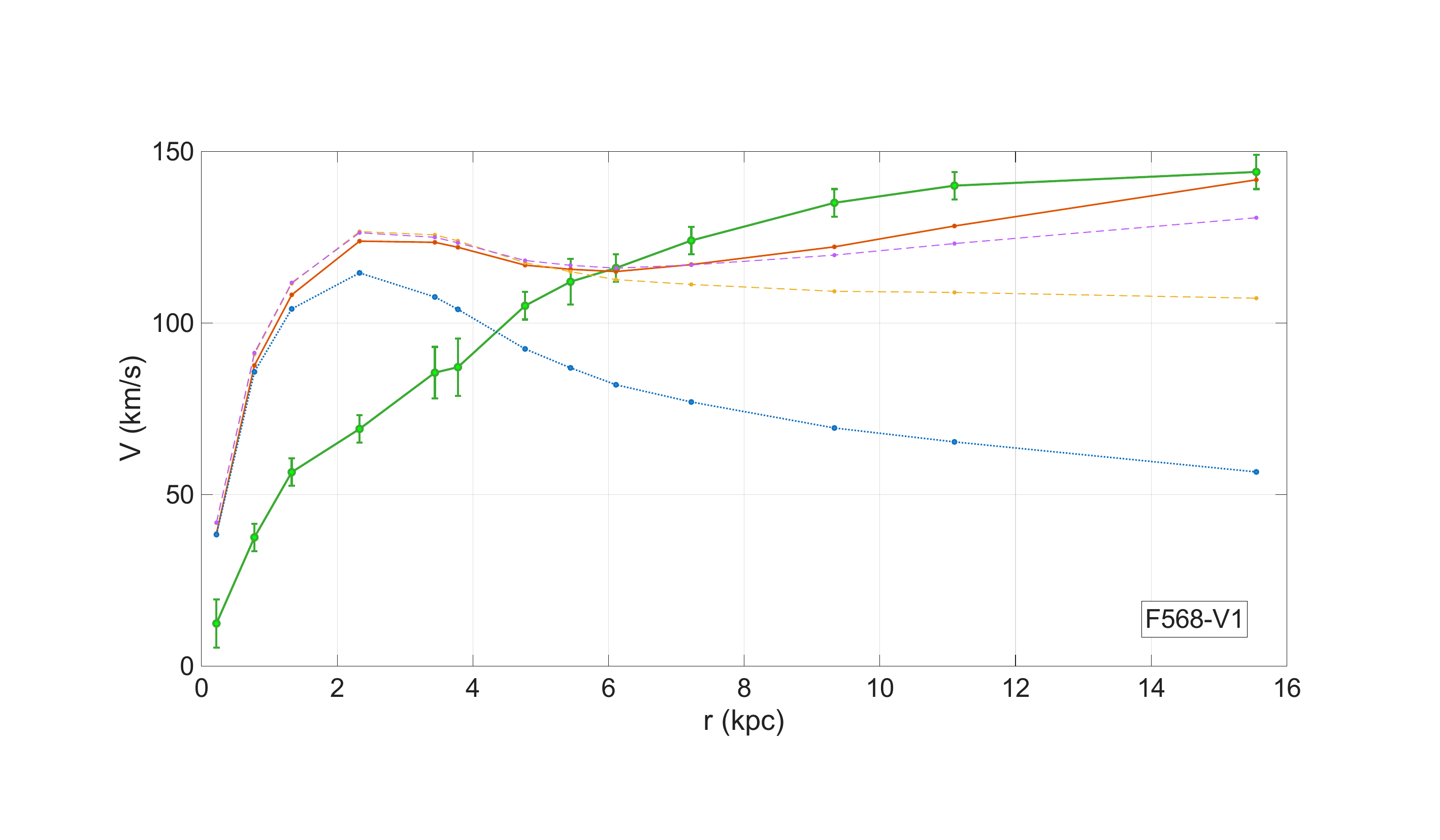}
\includegraphics[trim=4cm 3cm 5cm 4cm, clip=true, width=0.325\columnwidth]{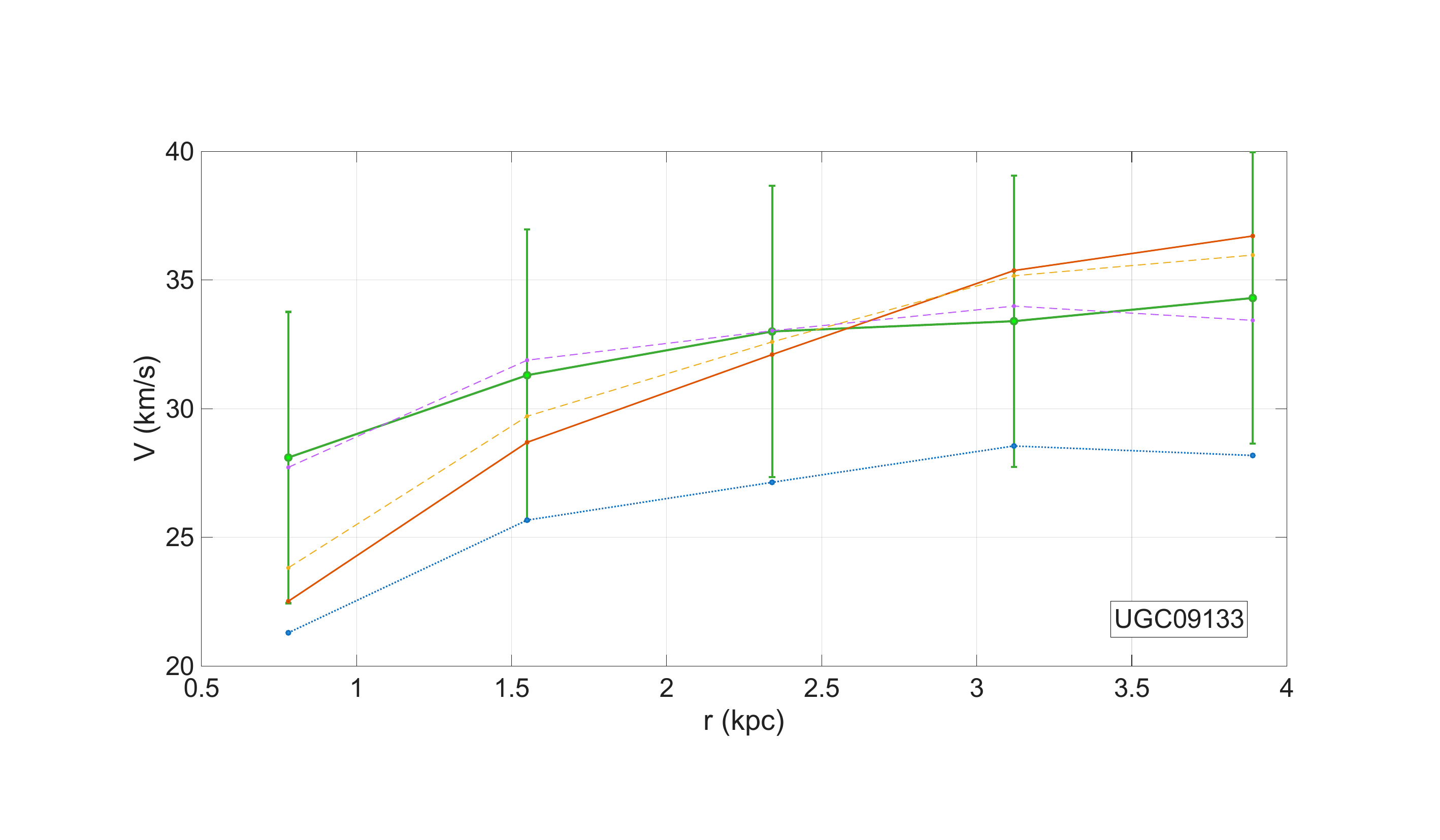}
\caption{The plots present the MG fits applied to the SPARC galaxy dataset. The green error bars denote the observed rotation velocities of the SPARC galaxies. The blue dotted curves represent the velocities calculated using Newtonian mechanics based solely on the baryonic mass, while the solid red curves show the velocities obtained from the MG model. For comparison, the best-fitting MOND and NFW dark matter profile fits are also shown as yellow and magenta dotted curves, respectively.}
\label{SPARC149}
\end{figure}

In Fig.~\ref{SPARC149}, we present rotation curves for 149 galaxies. The blue lines with points represent the velocities calculated from the baryonic matter in each galaxy using Newtonian mechanics. The green curves with error bars depict the observed rotation velocity profiles. The solid red curve corresponds to the Machian gravity fit. The yellow and purple dotted lines indicate the MOND and NFW dark matter profile fits, respectively. From this figure, we see that the fits are generally very good for most galaxies, with only a few showing noticeable deviations. Overall, the profiles look broadly similar across the sample. Fig.~\ref{SPARC26} displays several SPARC galaxies for which the fits are comparatively poor. However, the data for these systems suggest that there may be issues with the observations themselves—for example, an incorrect mass-to-light ratio or other observational uncertainties. Table~\ref{clusterProperties} summarizes our results, listing the best-fit values of $M_c$, $\lambda^{-1}$, and $a_0$ for the different galaxies, along with additional galaxy-specific properties such as baryonic mass and characteristic radius.

\section{\label{sec:section6}Discussion}

MG has several direct and indirect effects. Below, I outline some of the key consequences for galaxies.

1. Several massive gravity models have been proposed to explain galactic rotation curves. Readers can check~\cite{de2018galaxy,clifton2008parametrized,stabile2011rotation,rahvar2014observational} and the references therein. However, a key difficulty in such approaches is that introducing an additional massive field coupled to gravity typically requires the specification of a fixed mass term. Observational analyses of galaxy rotation curves indicate that a single, fixed mass scale is insufficient to reproduce the velocity profiles of all galaxies. Moreover, the associated coupling constants would also need to vary from galaxy to galaxy, making it problematic to describe all systems using a universal set of parameters. In theories like Scalar-Tensor-Vector gravity etc. the issue is addressed by promoting the mass and the coupling constants etc as new scalar fields. However, while such constructions may be mathematically correct, they remain largely ad hoc from a conceptual standpoint.

In contrast, our analysis shows that while the combination $G M_c / \lambda^{-2}$ remains bounded within a relatively narrow range (roughly within 1 to 1.5 order of magnitude), the individual parameters $M_c$ and $\lambda$ can vary significantly across different galaxies. Furthermore, the effective coupling in the model appears to depend on the mass distribution in the galaxy itself. These features suggest that standard massive gravity models with fixed parameters are not well suited to explain galactic dynamics. In MG, however, these parameters naturally depend on the mass distribution within each galaxy, making the framework better suited to account for the observed dynamics.

Secondly, in massive gravity models where the graviton has a physical or effective mass, the propagation speed of gravity would generally differ from that of light, a feature that can be tested using multi-messenger observations of merger events and through the frequency dependence of gravitational-wave propagation. However, current observations show that gravitational waves and electromagnetic signals propagate at the same speed within experimental uncertainties. This problem does not arise in the Machian Gravity (MG) framework, where the derivative terms effectively behave as dark components of the Universe and reproduce dark-matter-like dynamics, ensuring that the propagation speed of gravity remains identical to that of light (see Appendix~E of~\cite{das2023aspects}).

2. All of the theories introduce a characteristic mass scale and a characteristic length scale. In the NFW dark matter profile, the distribution of dark matter depends on radius, whereas in the other two cases it depends on acceleration. In NFW profile

\begin{equation}
M_\text{DM} = 4 \pi \rho_s r_s^3 \left(\log(1 + r/r_s) - \frac{r}{r_s + r}\right)\,.
\end{equation}

\noindent Here, the factor outside the brackets can be interpreted as a mass scale, while the expression inside the brackets depends only on $r$, with $r_s$ serving as a characteristic length scale. Earlier studies have reported that, for SPARC galaxies, dark matter profiles fit the data better than MOND. However, from the analysis in Section~\ref{sec:section4}, we found that the mass discrepancy is primarily a function of the acceleration. Although there is some correlation between the mass discrepancy and the scaled radius, this correlation is not particularly strong. Dark matter alone cannot account for this behavior, nor can it naturally explain the $M \sim v^4$ relation. In contrast, when we adopt a Machian gravity framework, we find that the effective additional mass is given by

\begin{equation}
M_\text{effectiveDM} = M\left(\sqrt{\frac{M_c}{M}} -1\right)(1-e^{-\lambda r} (1+ \lambda r))   
\end{equation}

\noindent In this case, as we can see from the Taylor expansion in Eq.~\ref{Eq:MONDBreaking}, the dependence is primarily on the acceleration. There is also some dependence on the scaled radius and scaled mass, but these effects are much weaker than the acceleration dependence, especially near the edge of the galaxy. Consequently, this framework can account for the observational results more effectively.

Of course, in the MOND framework the mass discrepancy of a galaxy depends solely on the acceleration $a$, and in that sense it accounts for the observations. However, the weak correlation between the mass discrepancy and the scaled radius as shown in section~\ref{sec:section5}, remains unexplained in this context. Moreover, our results indicate that the MOND acceleration scale $a_0$ varies significantly from one galaxy to another. The assumption that a single universal value of $a_0$ can fit the rotation curves of all galaxies has already been strongly criticized by several authors~\cite{rodrigues2018absence,li2021cautionary}. In contrast, we find that Machian gravity yields a more natural scaling for $a^\text{MG}_0$ than MOND and no need to assume any fundamental acceleration scale.

3. MOND is not successful for larger systems such as galaxy clusters. As evident from Fig.~\ref{fig:alpha_values1}, the velocity there behaves approximately as $v^2 \sim G M \lambda$. Consequently, MOND cannot account for cluster-scale dynamics, where the characteristic length scale is much greater. In contrast, Machian gravity reproduces the Newtonian $1/r^2$ behavior at large radii, and it also yields the scaling $v^4 \sim M$, which is observed in galaxy clusters through the Faber–Jackson relation. Thus, while Machian gravity exhibits MOND-like behavior on galactic scales, it departs from MOND at larger, cluster scales. MOND itself is a phenomenological framework, whereas Machian gravity provides a theoretical basis that can reproduce MOND-like effects at galactic scales.

4. One of the predication of MOND is that there exists a critical value of the surface density

\begin{equation}
\Sigma_m \approx a^\text{MOND}_0 / G .
\end{equation}

\noindent If a system, such as a spiral galaxy has a surface density of matter greater than $\Sigma_m$, then the internal accelerations are greater than $a_0^\text{MOND}$, so the system is in the Newtonian regime. In systems with $\Sigma \geq \Sigma_m$ (HSB galaxies) there should be a small discrepancy between the visible and classical Newtonian dynamical mass within the optical disk. But in LSB galaxies $\left(\Sigma \ll \Sigma_m\right)$ there is a low internal acceleration, so the discrepancy between the visible and dynamical mass would be large. Milgrom predicted, before the actual discovery of LSB galaxies, that there should be a serious discrepancy between the observable and dynamical mass within the luminous disk of such systems- should they exist. In Machian gravity there is no such global acceleration scale. However, we have an order of acceleration scale $a^\text{MG}_0$, which matches with the MOND scale. Therefore, in MG also the MOND-like predictions hold. However in standard dark matter profile this phenomenon can not be explain.

\section{\label{sec:section7}Conclusion}

This study shows how the Machian gravity (MG) model, as proposed in~\cite{das2023aspects,Mach4}, can explain spiral galactic velocity profiles using the SPARC database. Spiral galaxies, being rotationally bound systems, require dark matter to account for their velocity profiles. Over time, various modified gravity theories have been put forth to explain these velocity profiles in spiral galaxies empirically. Our investigation demonstrates how Machian gravity aligns with these empirical formulations.

The velocity profile for spiral galaxies in MG is characterized by Eq.~\ref{eq:velocityFinal}. It has two parameters: a characteristic mass scale $M_c$ and a length scale $\lambda^{-1}$. Our analysis reveals that the length scale $\lambda^{-1}$ is comparable with the effective radius of the galaxy and varies across different galaxies. Consequently, the conventional approach of adopting a fixed $\lambda$, as postulated by some prior researchers~\cite{Brownstein2005}, may not comprehensively capture the intricacies of galactic velocity profiles.

Furthermore, we establish that near the edge of a galaxy, where $\lambda r \rightarrow 1$, an acceleration scale emerges, $a_0=\frac{GM_c}{\lambda^{-2}}$. In this region, the stellar acceleration tends to a MOND-like behavior dictated by this acceleration scale. However, our analysis also shows that this acceleration scale varies for different galaxies, depending on their individual structures. Although these accelerations are typically on the order of $10^{-8} \, {\rm cm/sec^{2}}$, their variation from one galaxy to another indicates that a single characteristic acceleration scale may not fully describe the entire dataset. Our results further demonstrate that, in addition to acceleration, a galaxy’s mass discrepancy also depends on $\lambda r$, a parameter that has not been examined in prior studies. It should be emphasized, however, that the present analysis assumes a uniform mass-to-light ratio for all galaxies, which may introduce some degree of bias. Consequently, future datasets incorporating galaxy-specific mass-to-light ratio measurements could modify some of these conclusions, though the overall statistical trends are unlikely to change substantially.

Finally and most importantly, while several other modified gravity has been designed to explain observational phenomena, the MG model stands apart from these theories as it emerges from a purely mathematical quest to formulate Mach's principle. Notably, this model adeptly explains the galactic velocity profiles of  154  SPARC galaxies, demonstrating remarkable agreement. For the remaining 21 galaxies, discrepancies in the mass model are apparent, suggesting potential inaccuracies in those data. We are confident that refining the mass model for these galaxies through parameter adjustments, such as varying the mass-to-light ratio, could potentially align their velocities with the predictions of the MG model.

\section{Data Availability}
The data used in this analysis are publicly available on the project’s \href{https://machiangravity.github.io/Galactic-Velocity-Profile-/}{GitHub} page. 

\bibliographystyle{JHEP}
\bibliography{reference1}

\begin{figure}
\centering
\includegraphics[trim=4cm 3cm 5cm 4cm, clip=true, width=0.325\columnwidth]{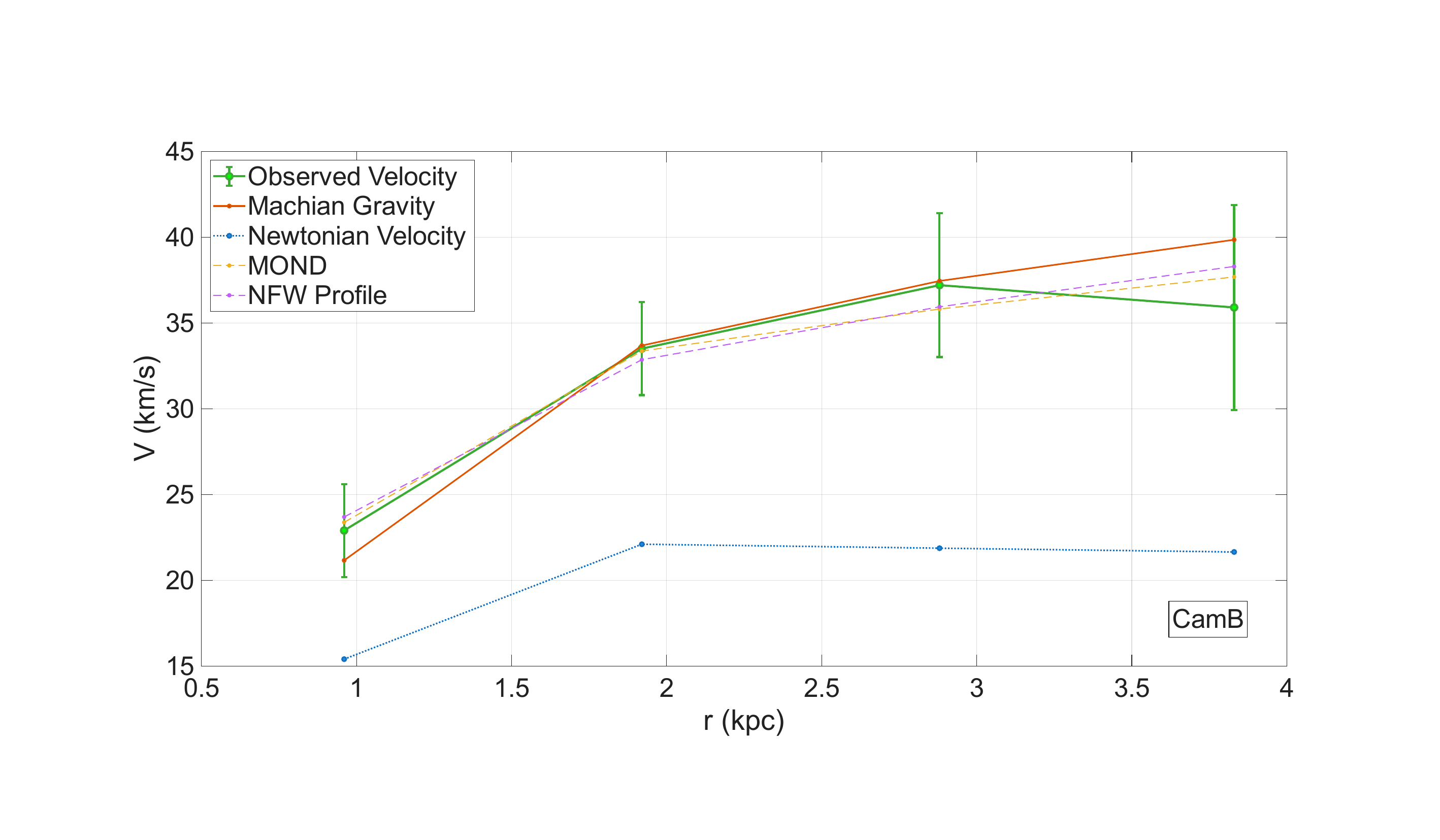}
\includegraphics[trim=4cm 3cm 5cm 4cm, clip=true, width=0.325\columnwidth]{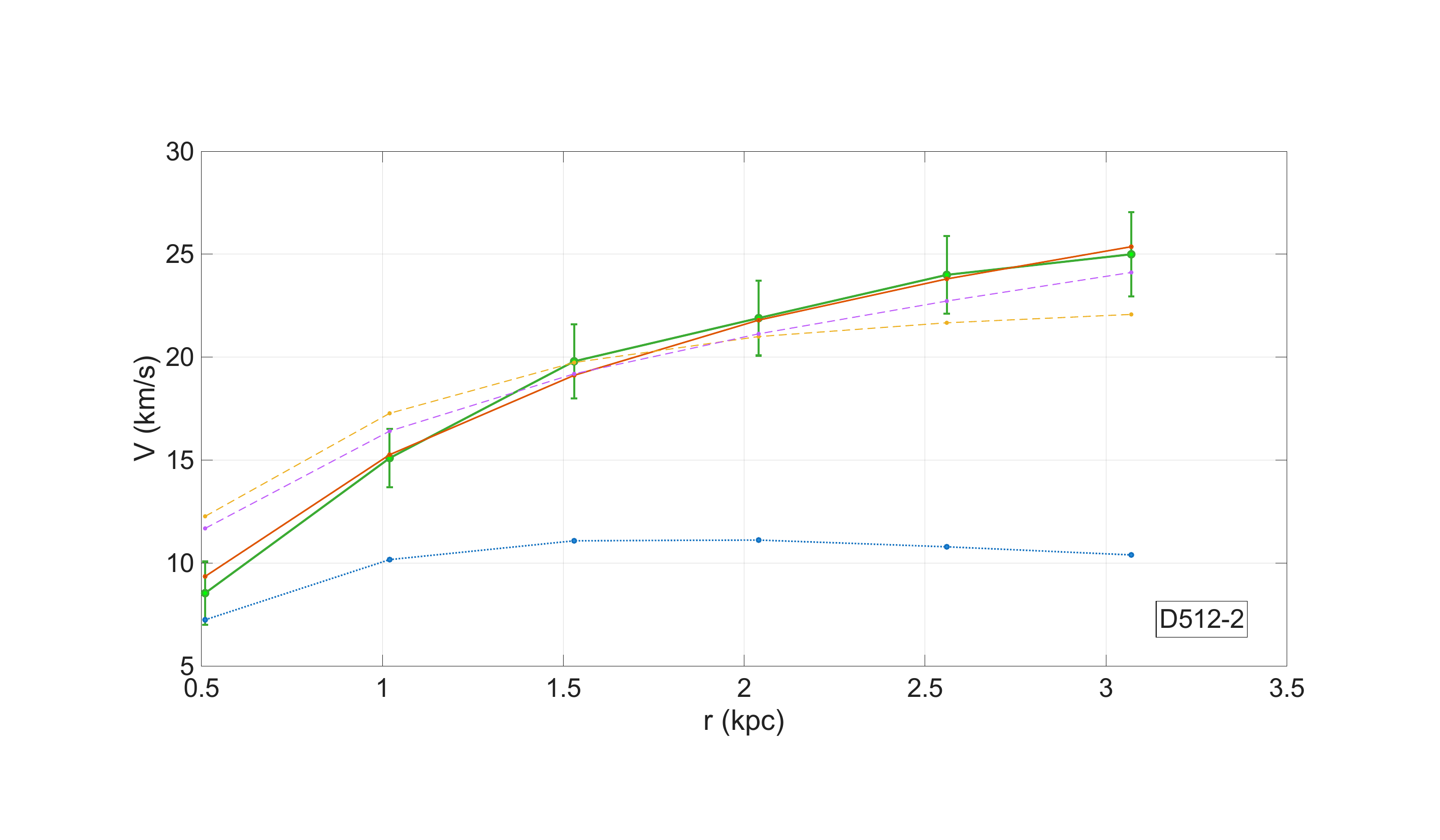}
\includegraphics[trim=4cm 3cm 5cm 4cm, clip=true, width=0.325\columnwidth]{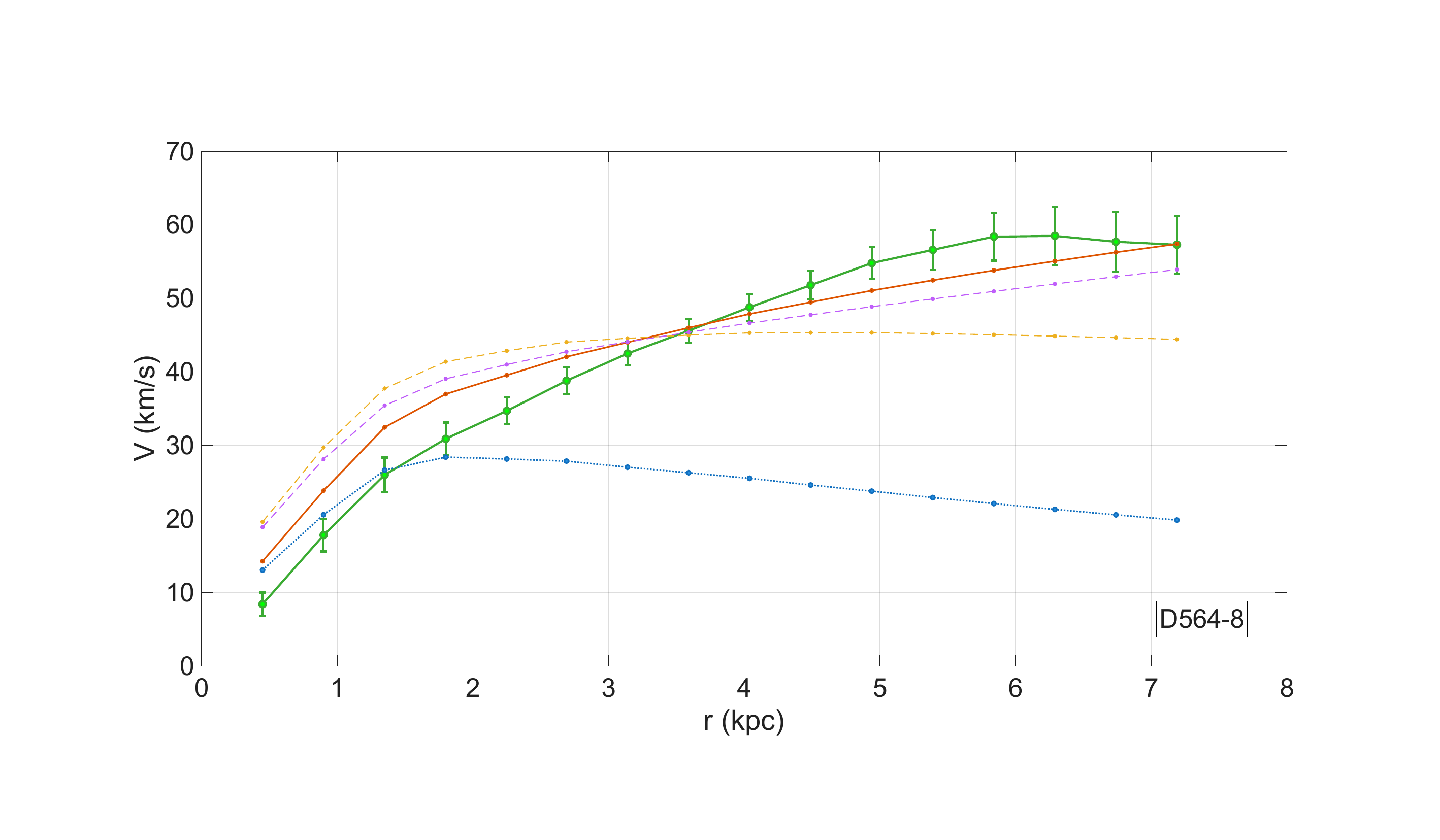}
\includegraphics[trim=4cm 3cm 5cm 4cm, clip=true, width=0.325\columnwidth]{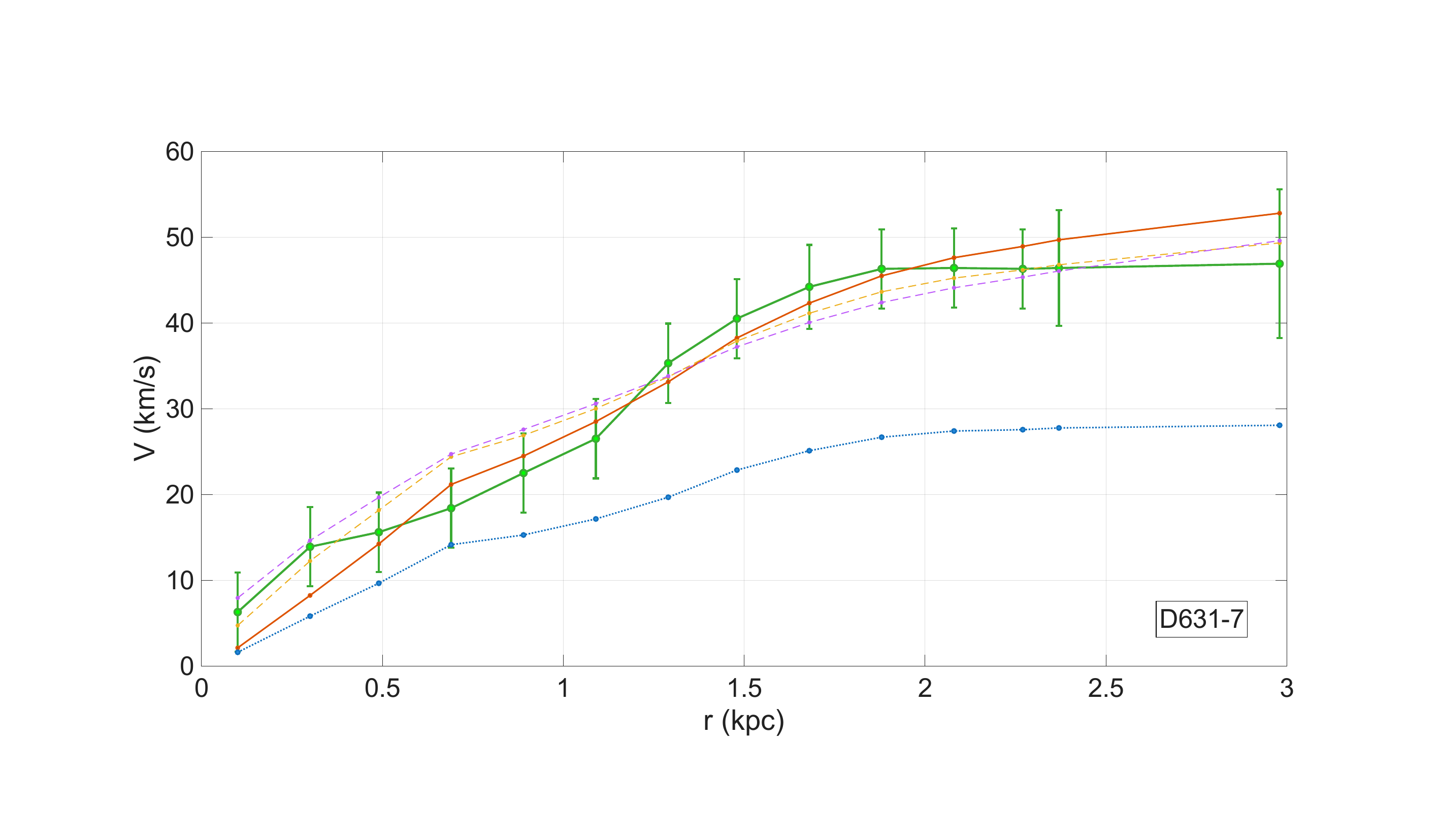}
\includegraphics[trim=4cm 3cm 5cm 4cm, clip=true, width=0.325\columnwidth]{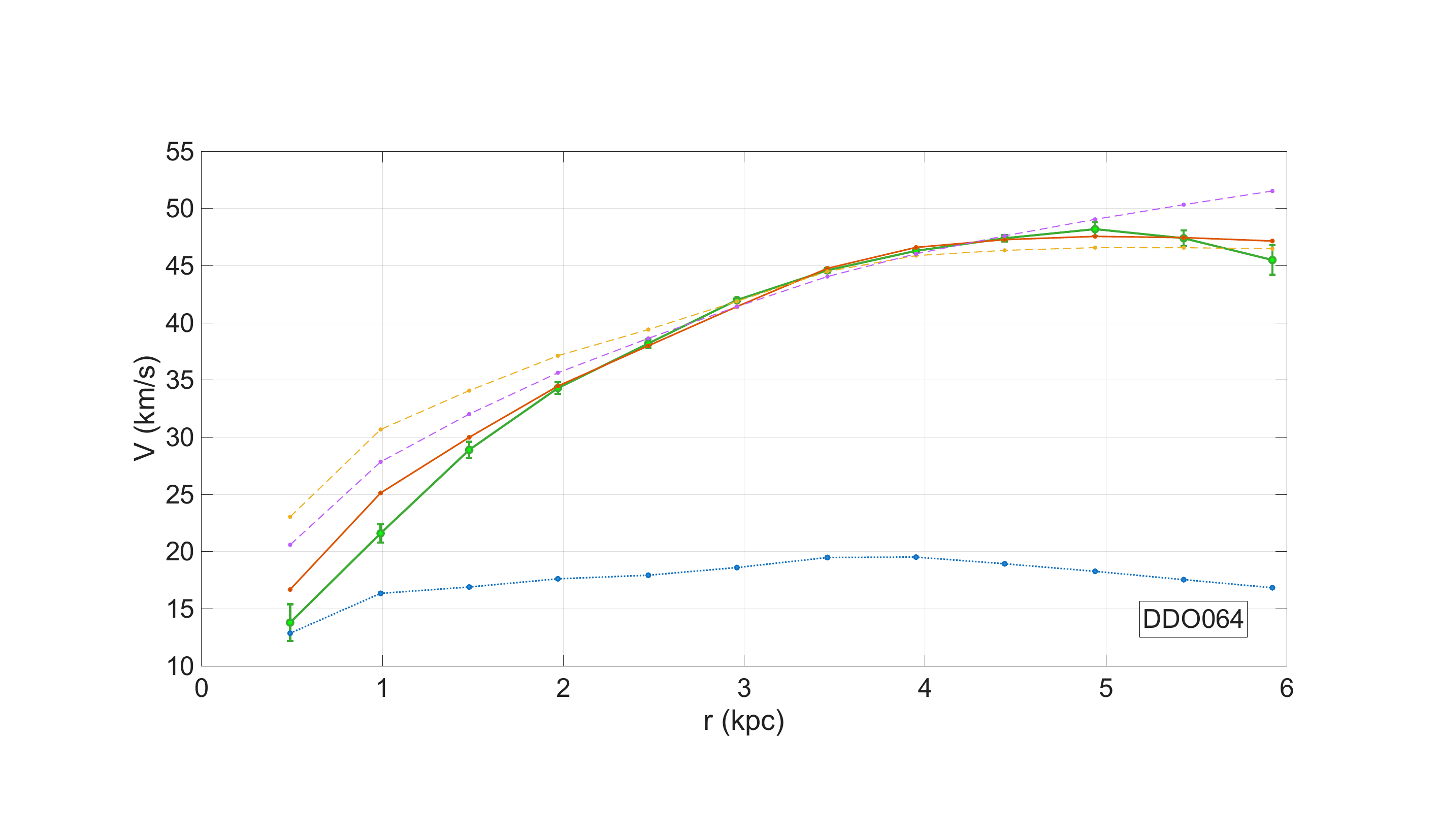}
\includegraphics[trim=4cm 3cm 5cm 4cm, clip=true, width=0.325\columnwidth]{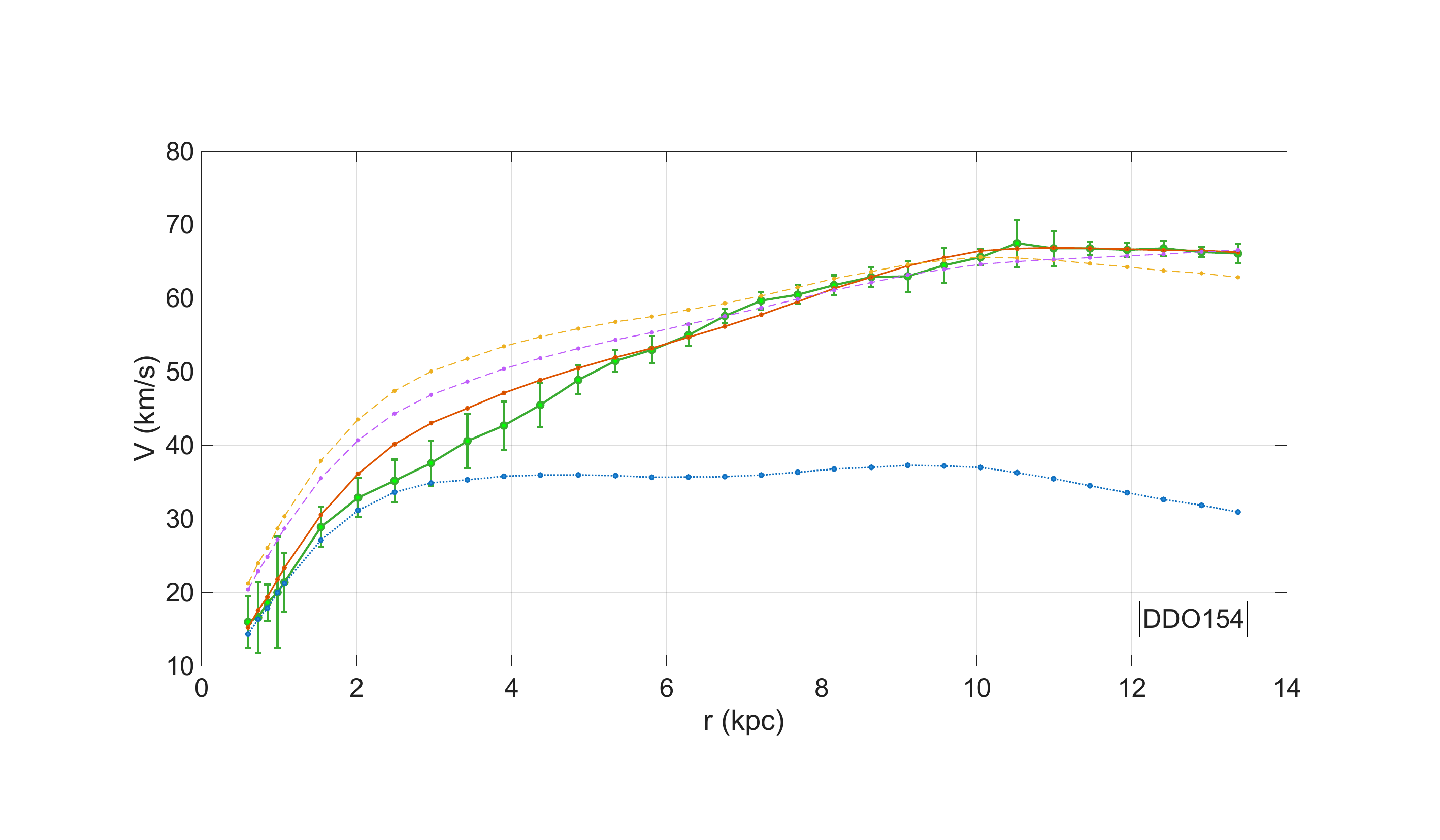}
\includegraphics[trim=4cm 3cm 5cm 4cm, clip=true, width=0.325\columnwidth]{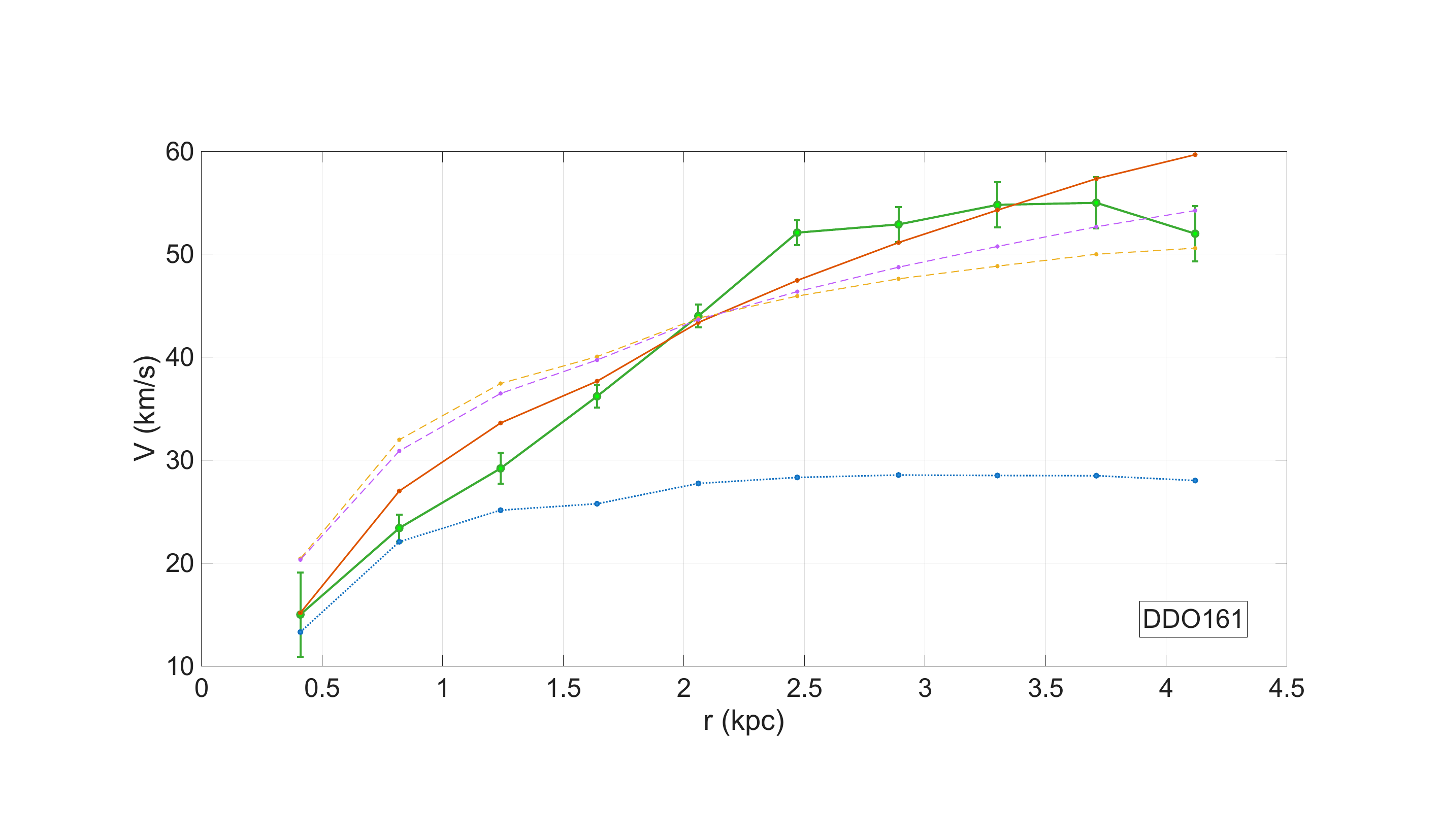}
\includegraphics[trim=4cm 3cm 5cm 4cm, clip=true, width=0.325\columnwidth]{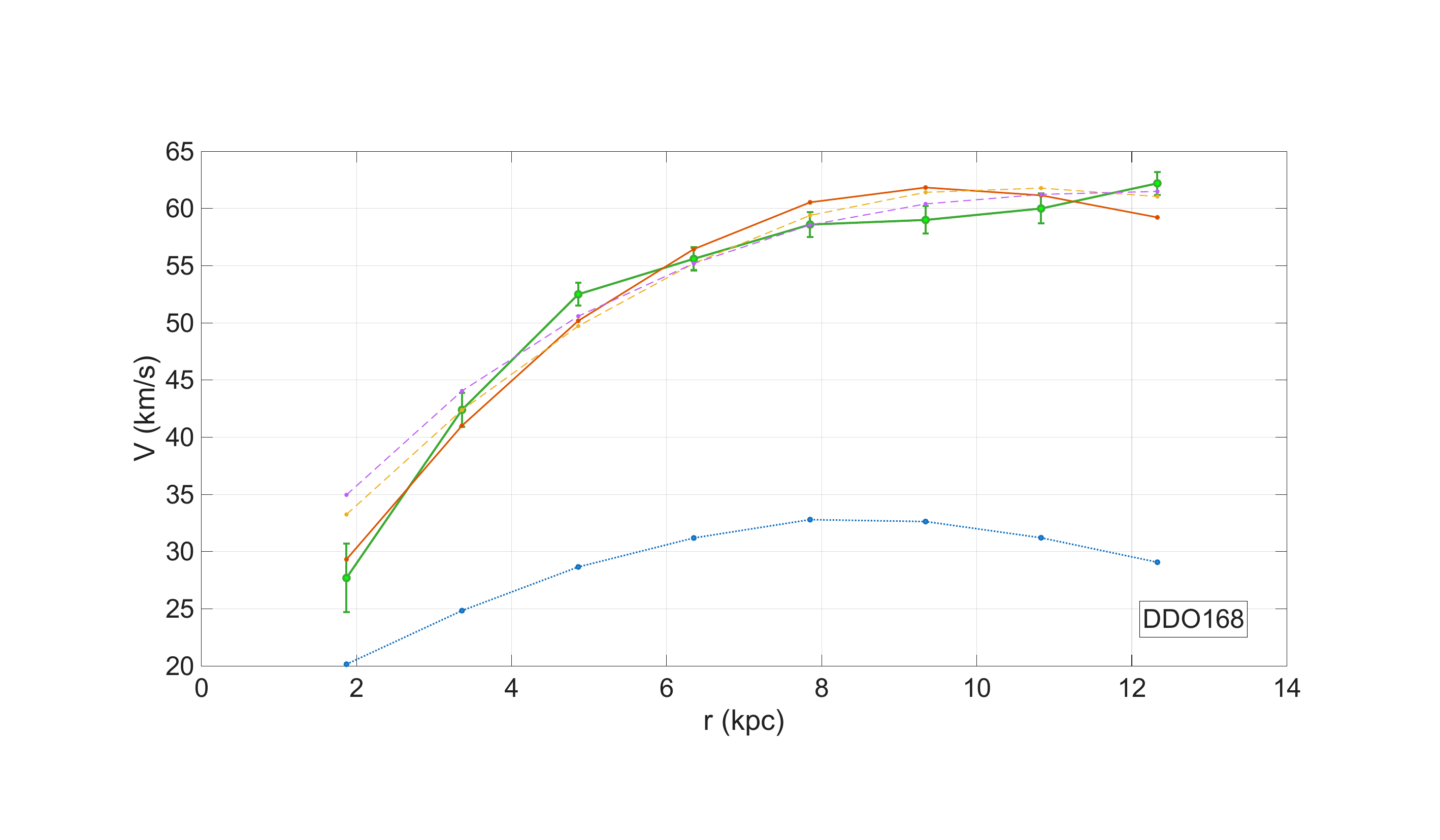}
\includegraphics[trim=4cm 3cm 5cm 4cm, clip=true, width=0.325\columnwidth]{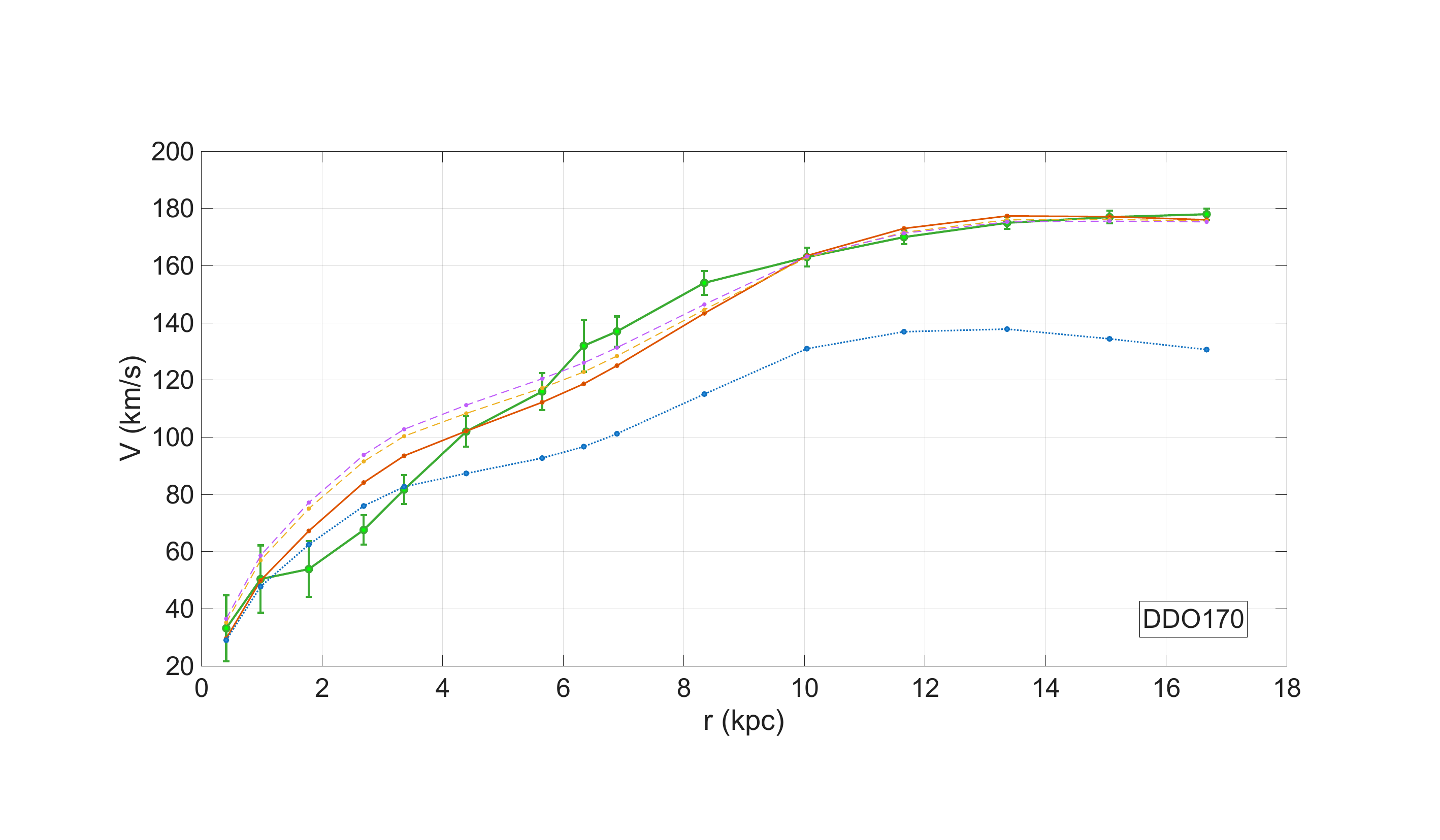}
\includegraphics[trim=4cm 3cm 5cm 4cm, clip=true, width=0.325\columnwidth]{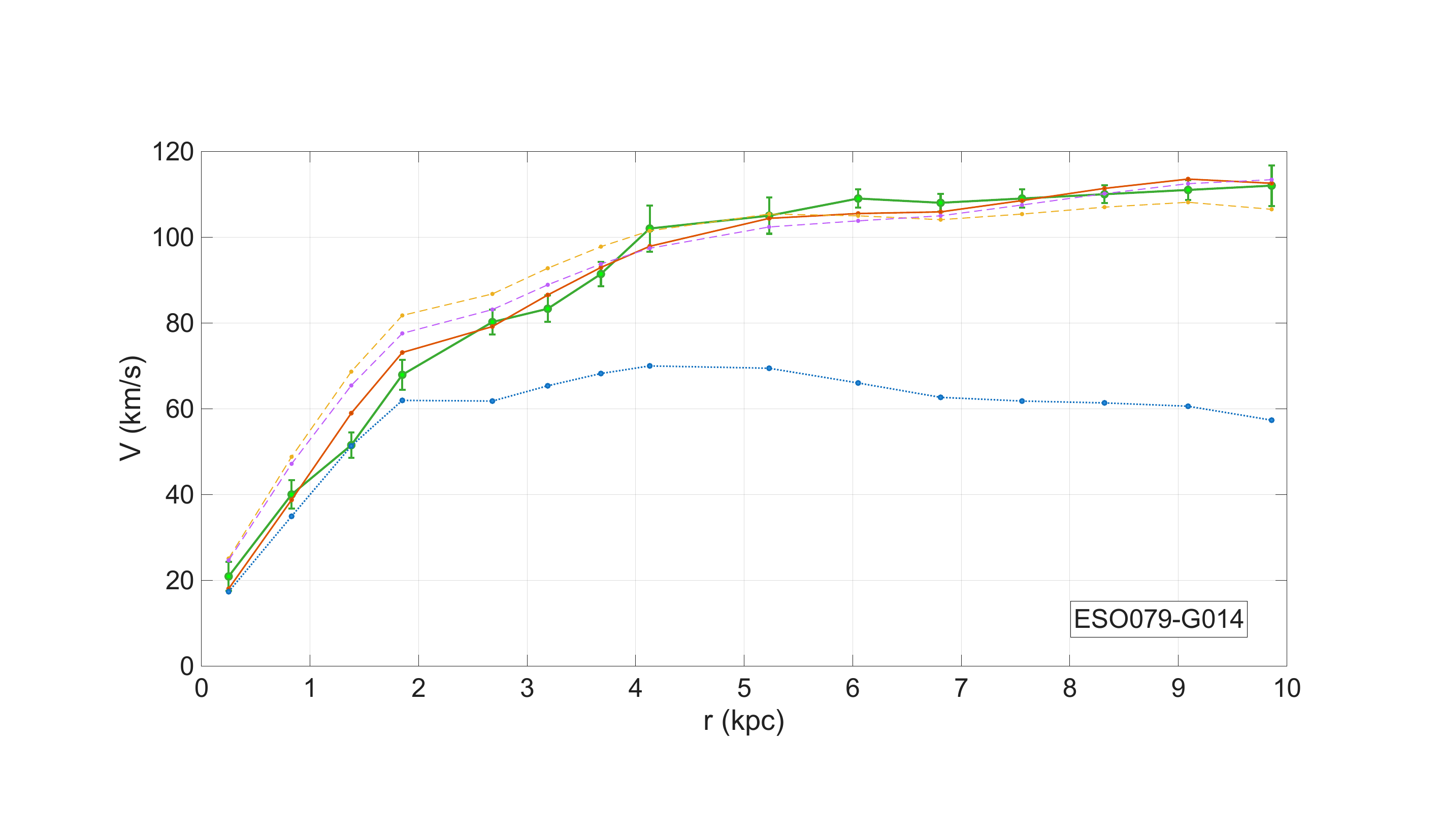}
\includegraphics[trim=4cm 3cm 5cm 4cm, clip=true, width=0.325\columnwidth]{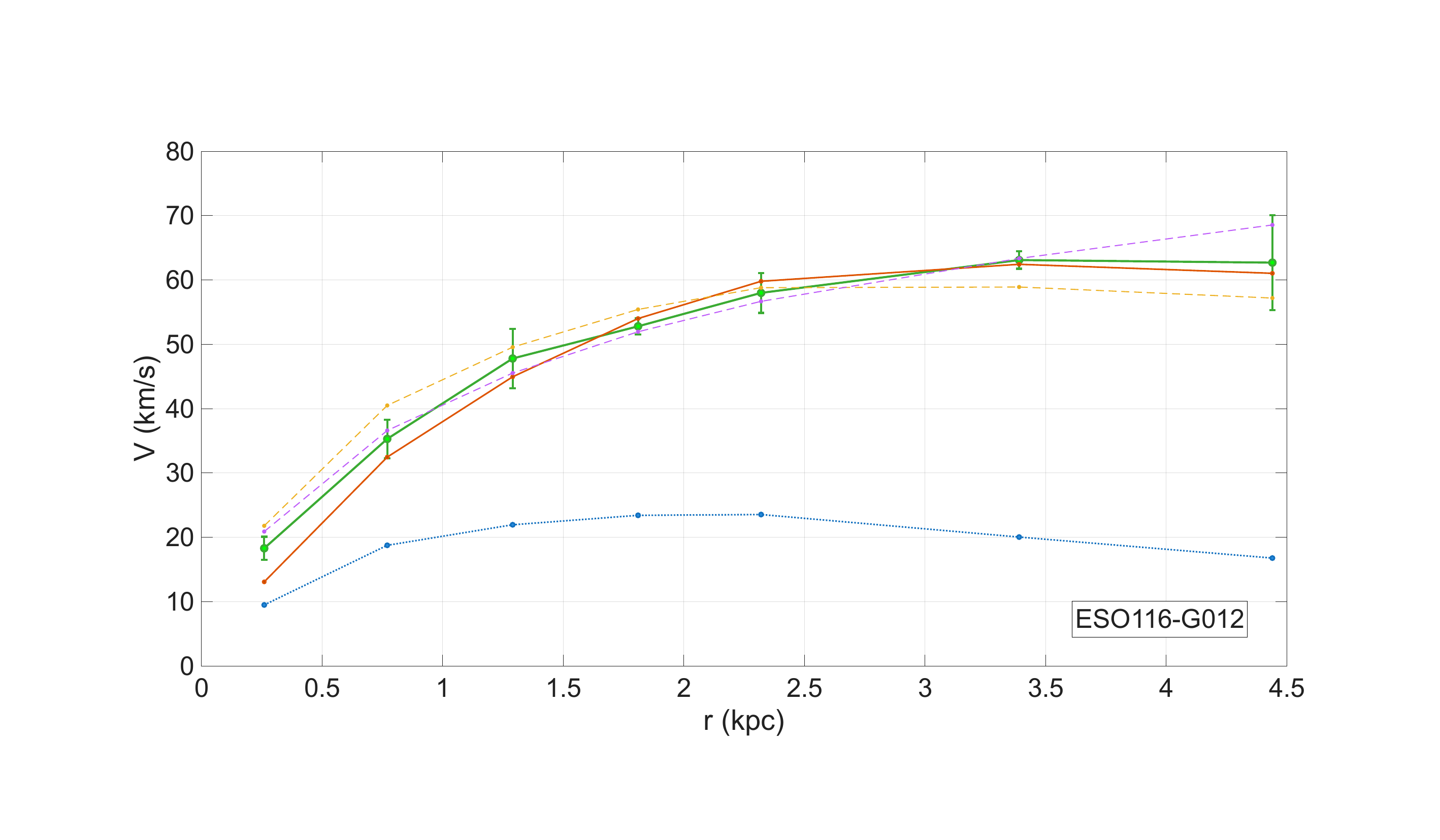}
\includegraphics[trim=4cm 3cm 5cm 4cm, clip=true, width=0.325\columnwidth]{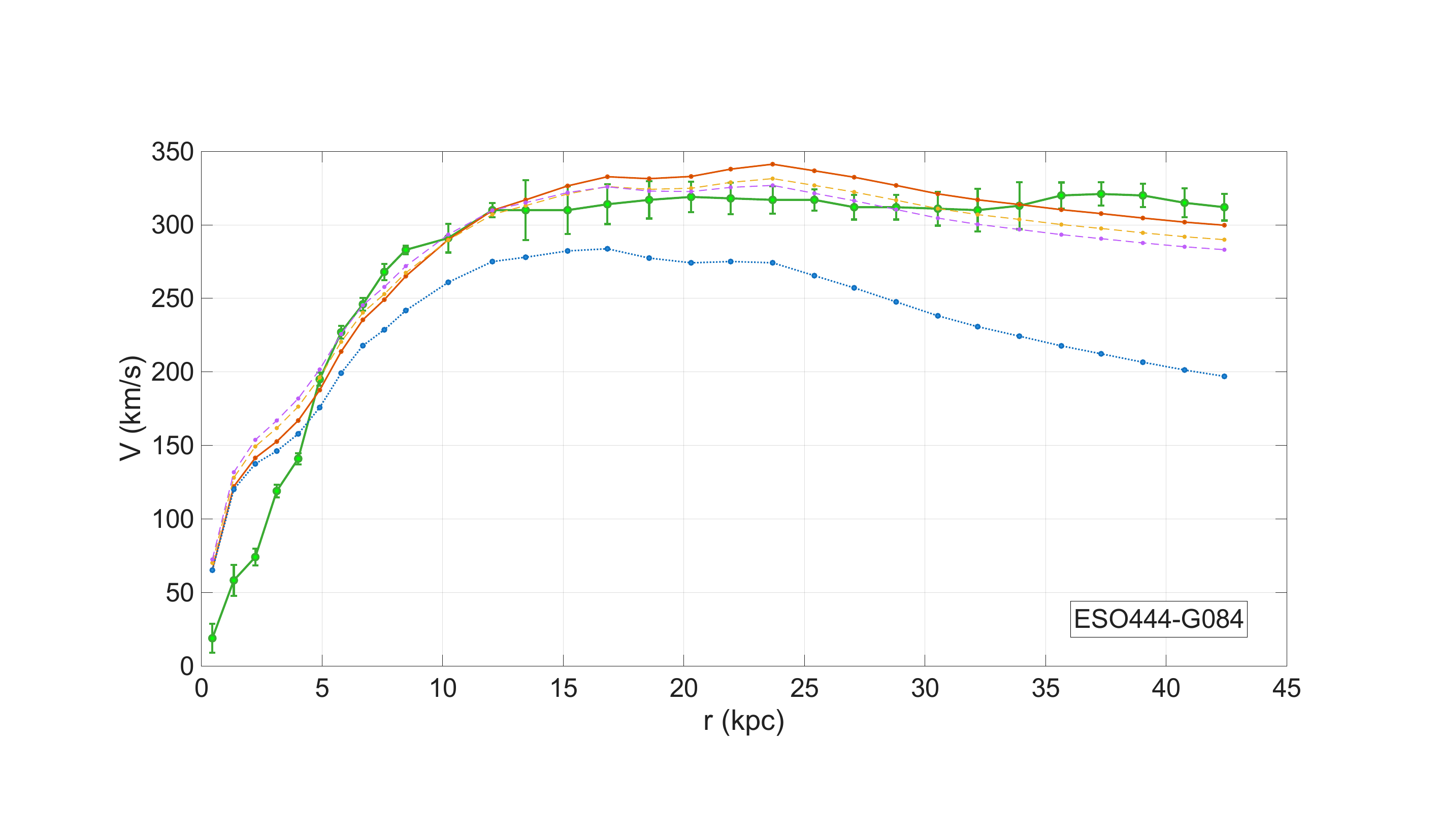}
\includegraphics[trim=4cm 3cm 5cm 4cm, clip=true, width=0.325\columnwidth]{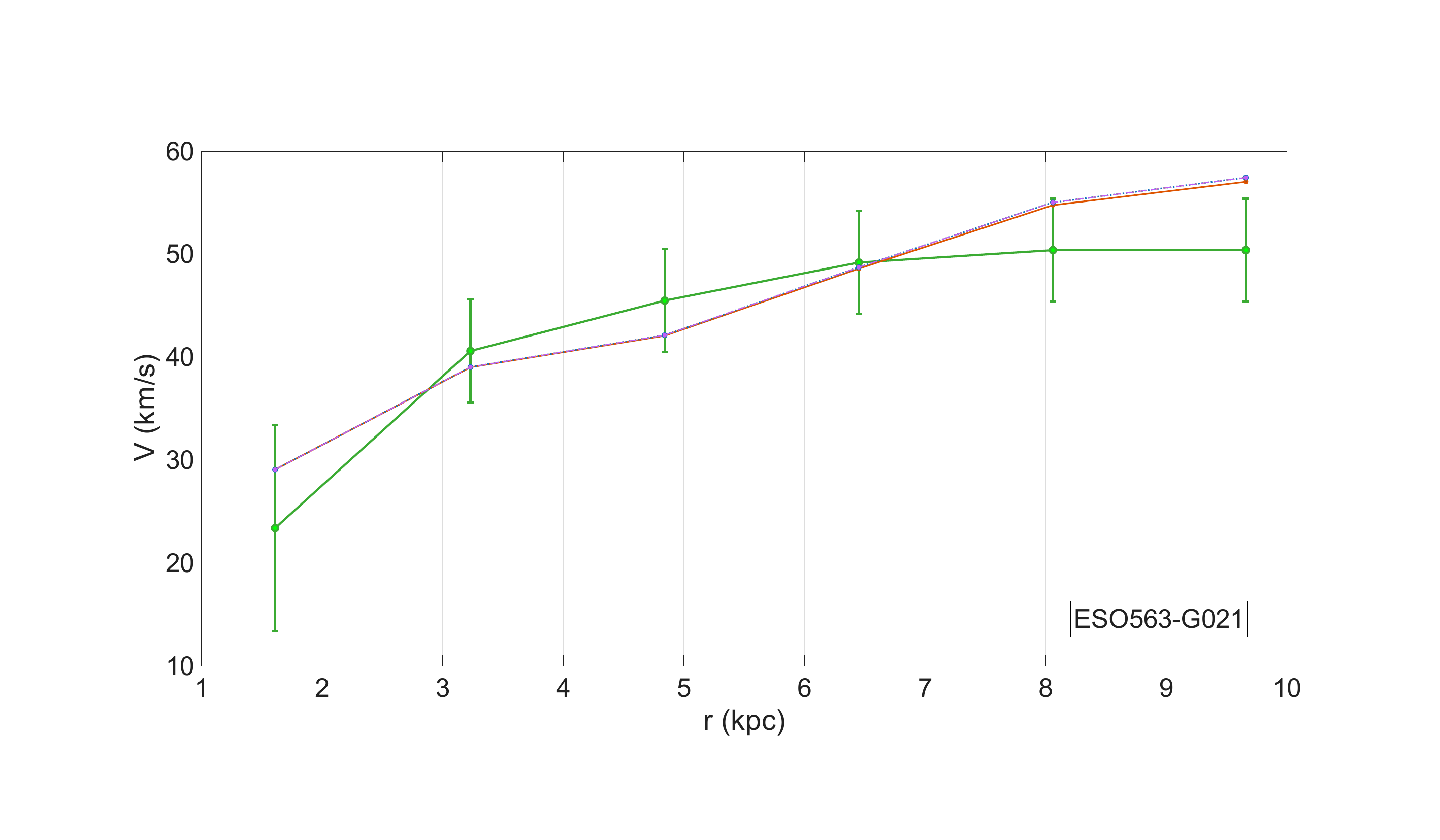}
\includegraphics[trim=4cm 3cm 5cm 4cm, clip=true, width=0.325\columnwidth]{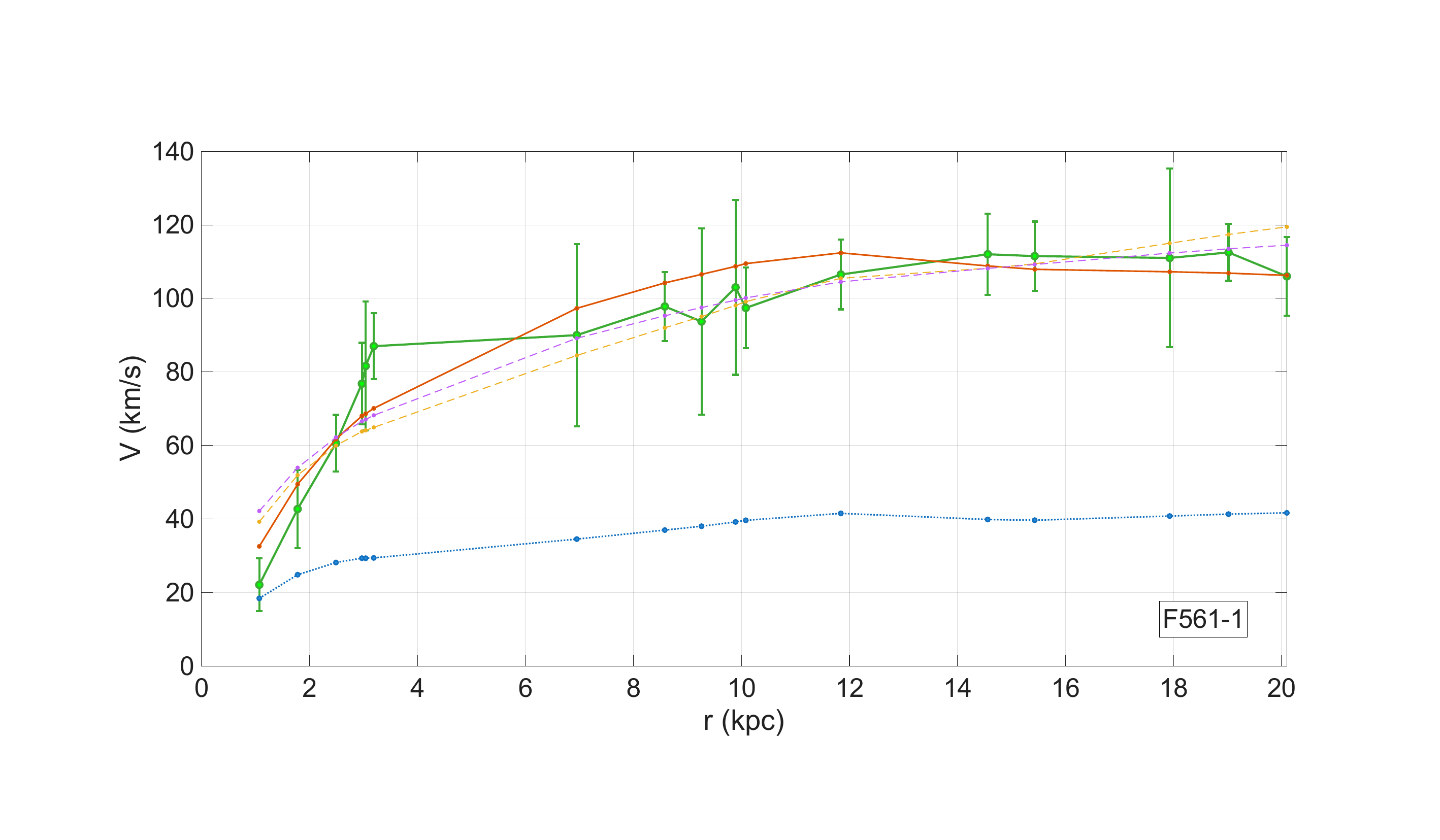}
\includegraphics[trim=4cm 3cm 5cm 4cm, clip=true, width=0.325\columnwidth]{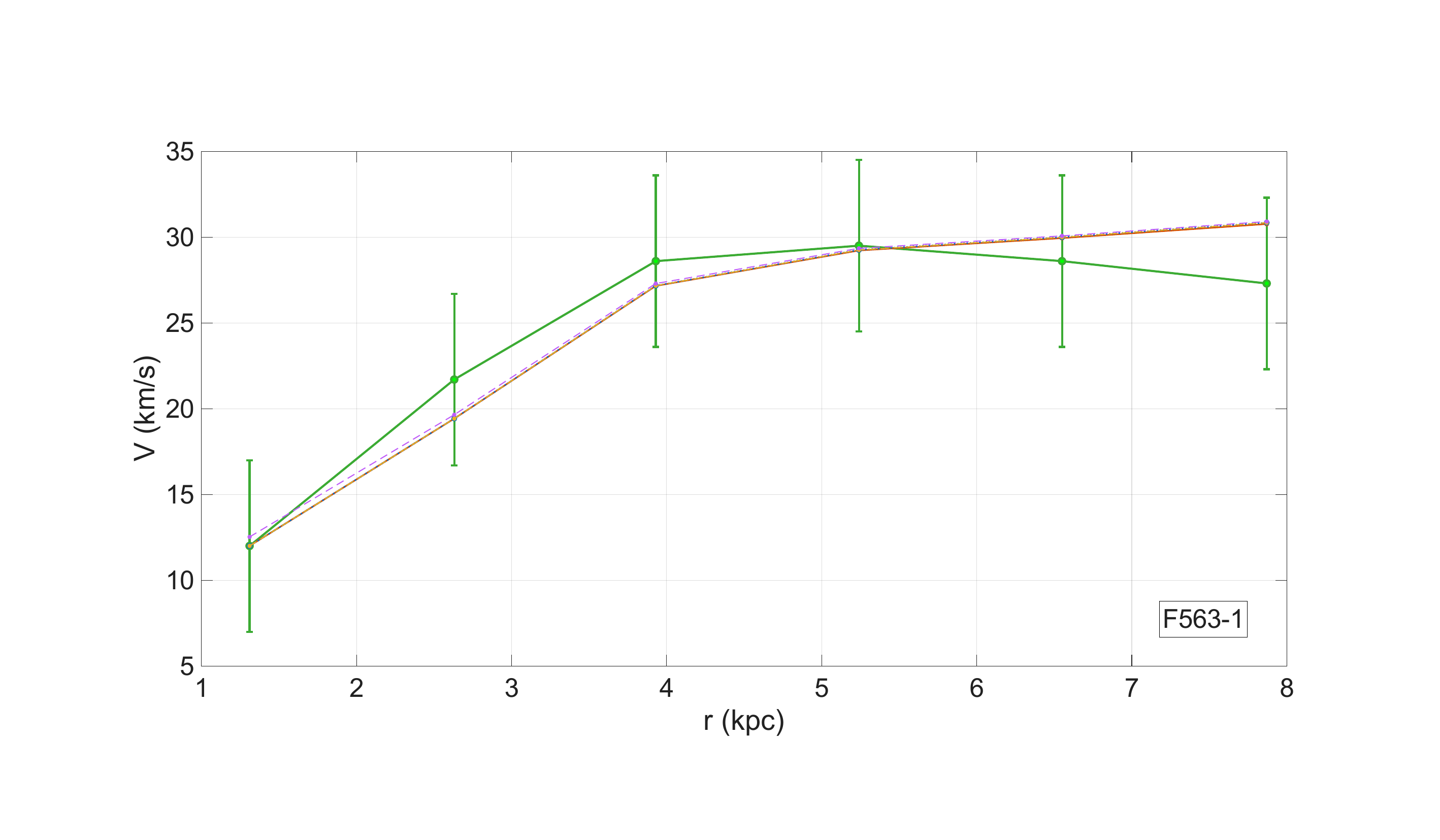}
\includegraphics[trim=4cm 3cm 5cm 4cm, clip=true, width=0.325\columnwidth]{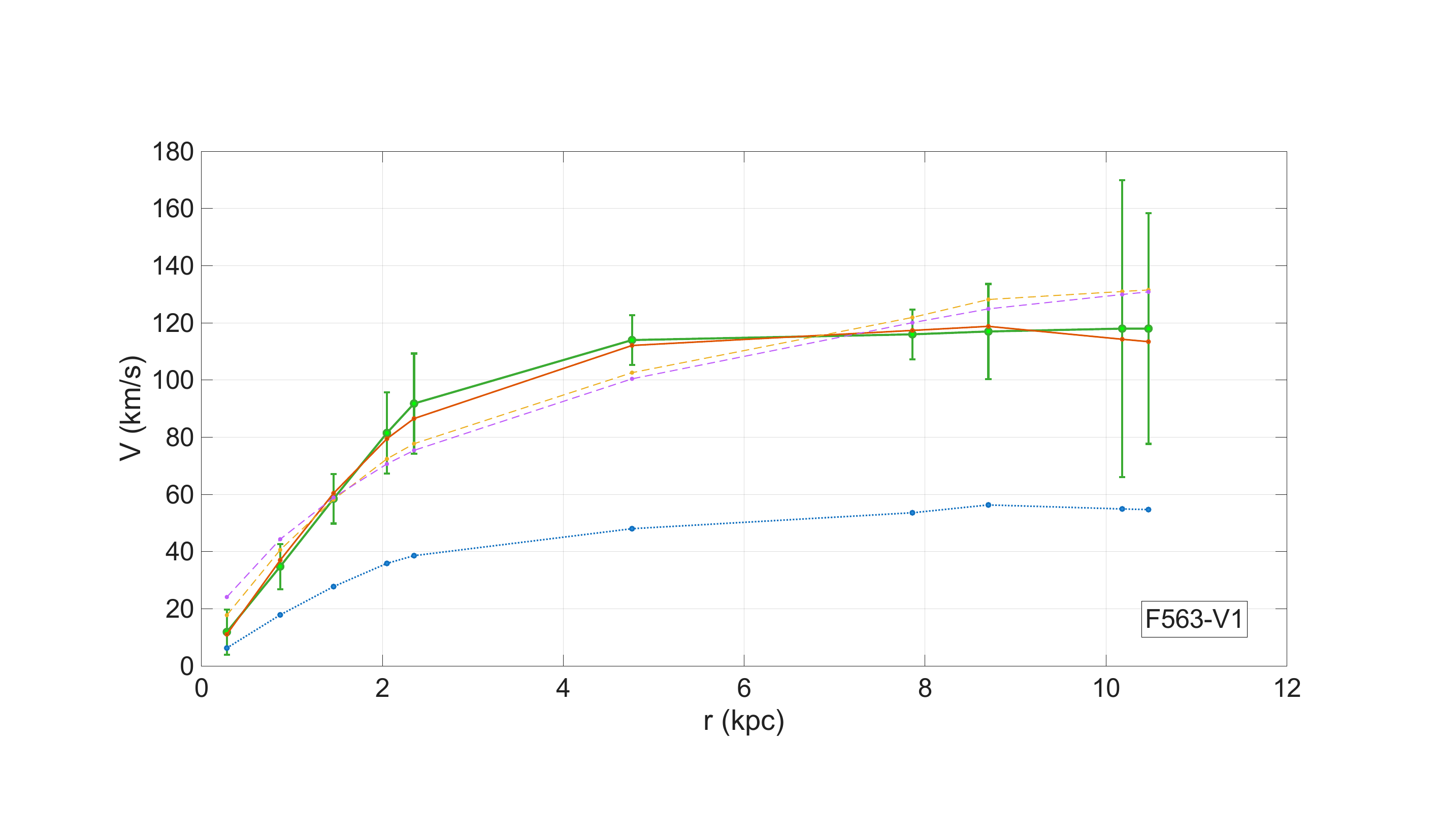}
\includegraphics[trim=4cm 3cm 5cm 4cm, clip=true, width=0.325\columnwidth]{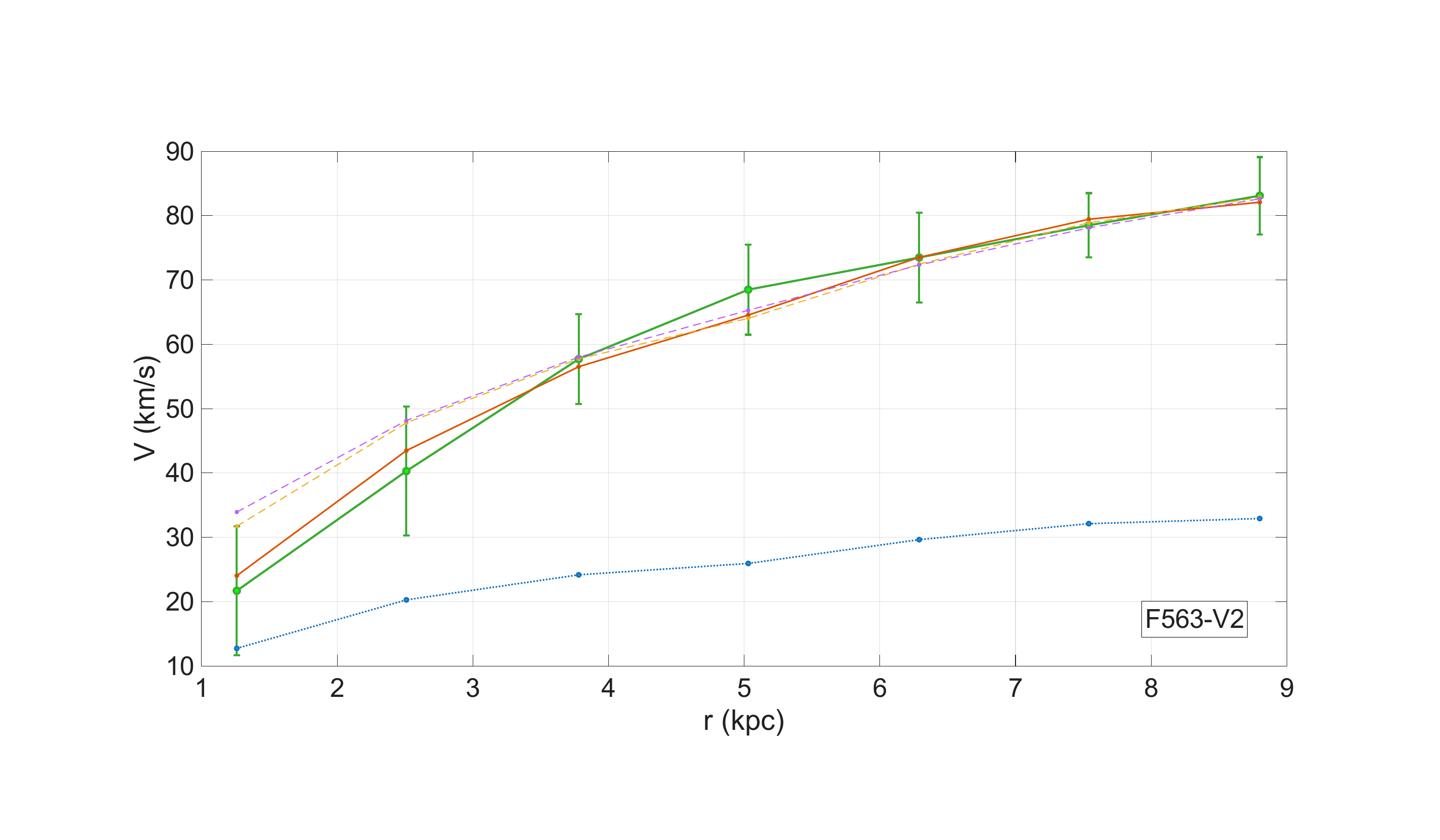}
\includegraphics[trim=4cm 3cm 5cm 4cm, clip=true, width=0.325\columnwidth]{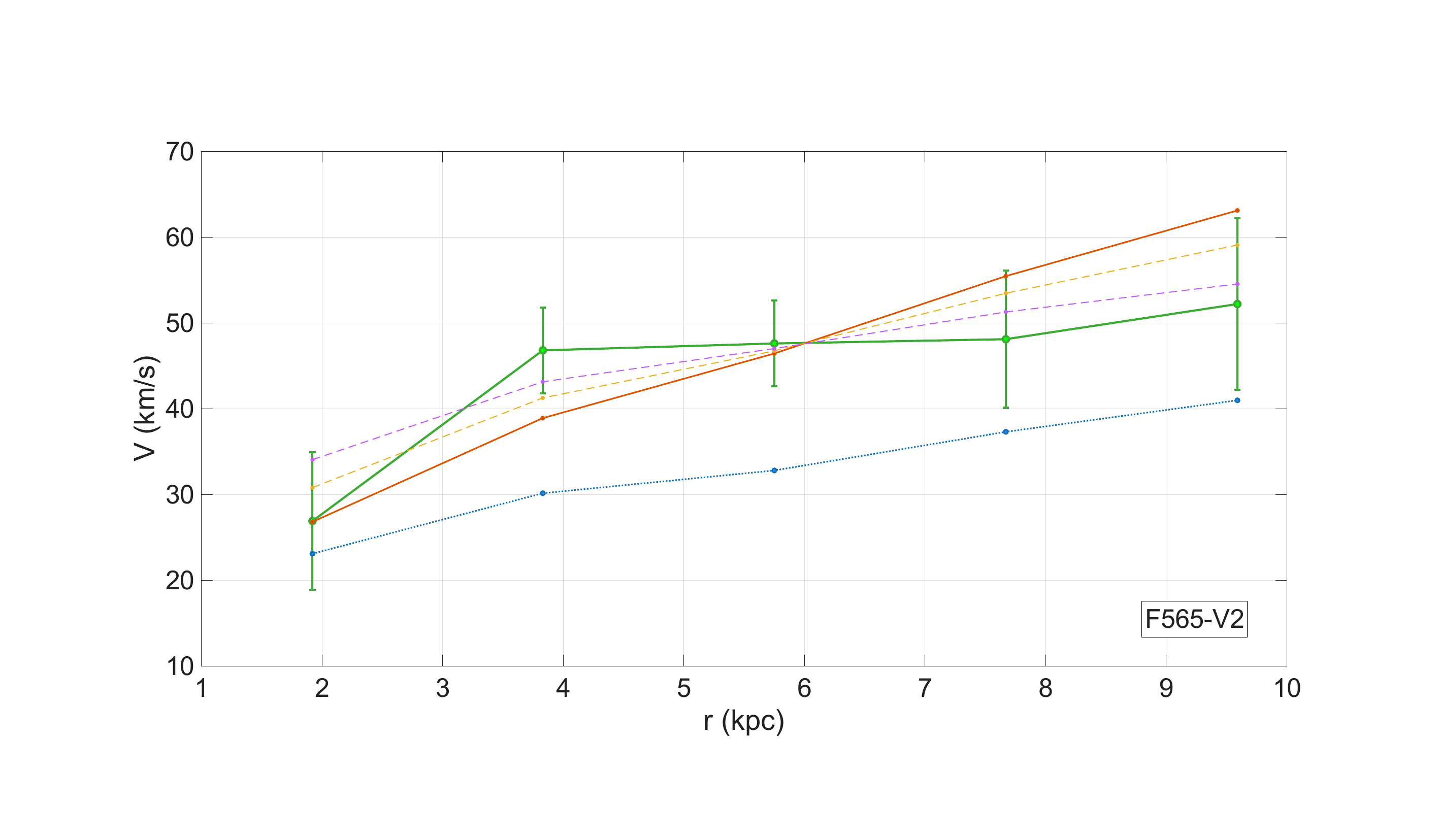}
\includegraphics[trim=4cm 3cm 5cm 4cm, clip=true, width=0.325\columnwidth]{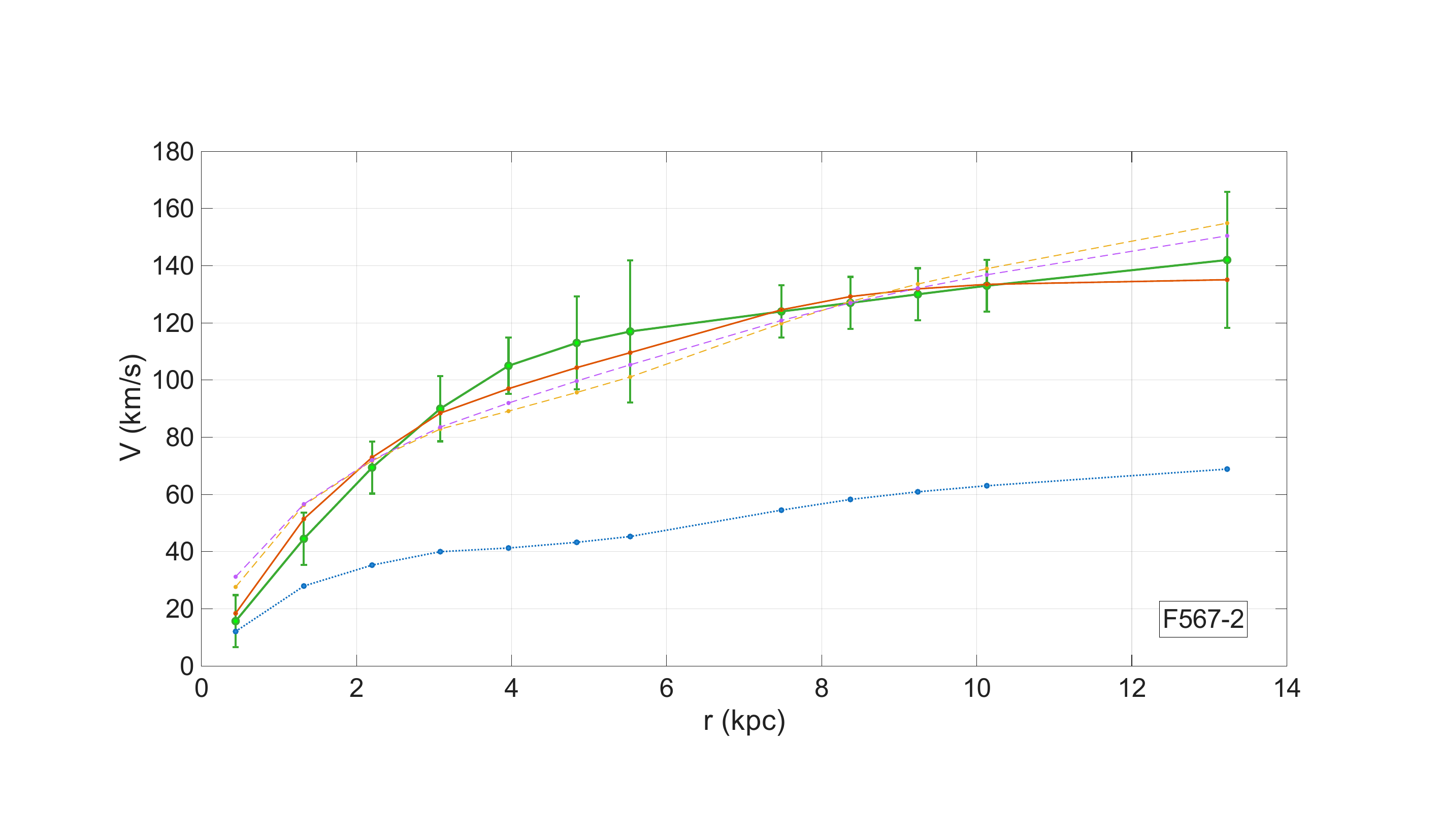}
\includegraphics[trim=4cm 3cm 5cm 4cm, clip=true, width=0.325\columnwidth]{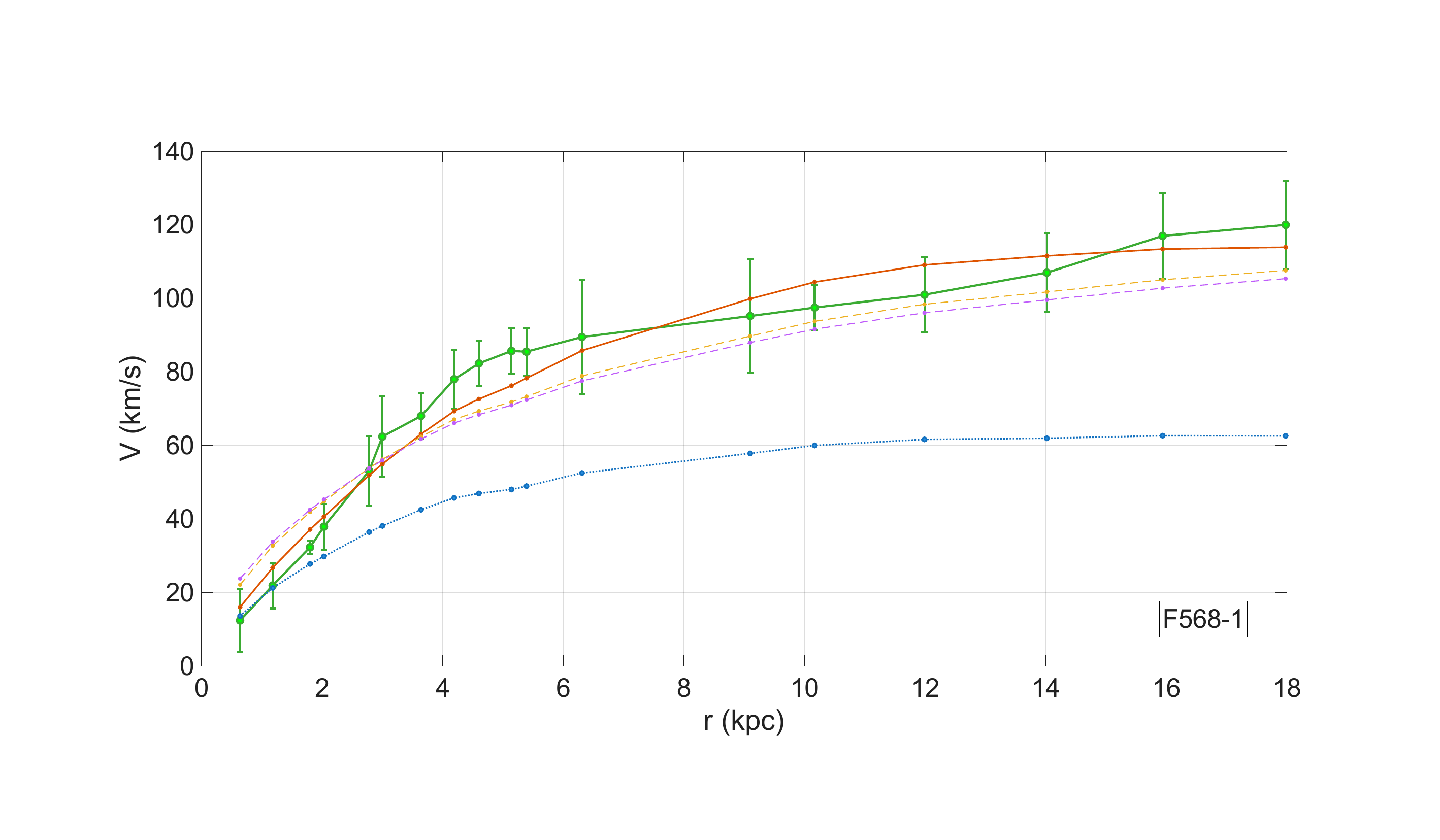}
\includegraphics[trim=4cm 3cm 5cm 4cm, clip=true, width=0.325\columnwidth]{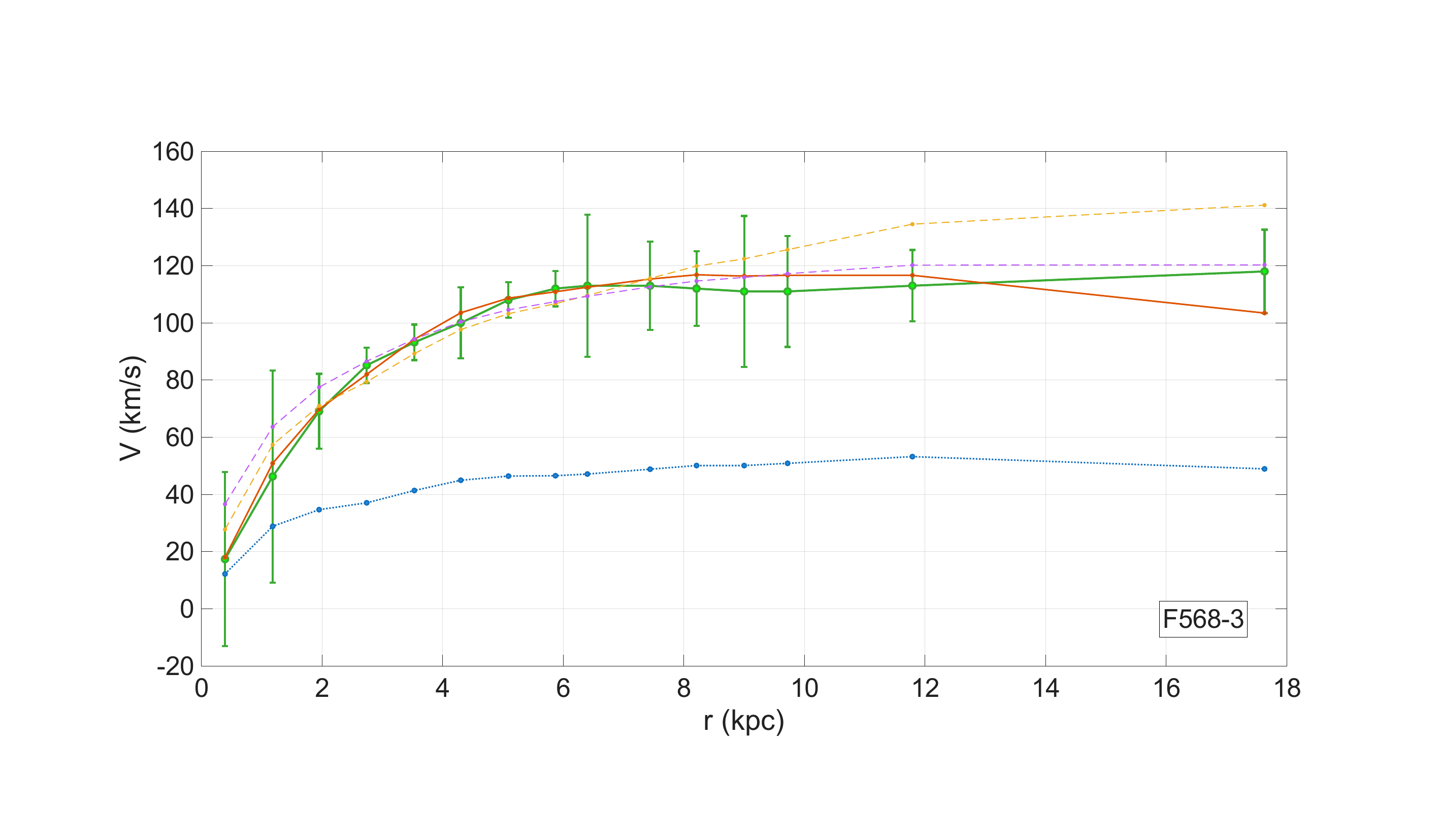}
\includegraphics[trim=4cm 3cm 5cm 4cm, clip=true, width=0.325\columnwidth]{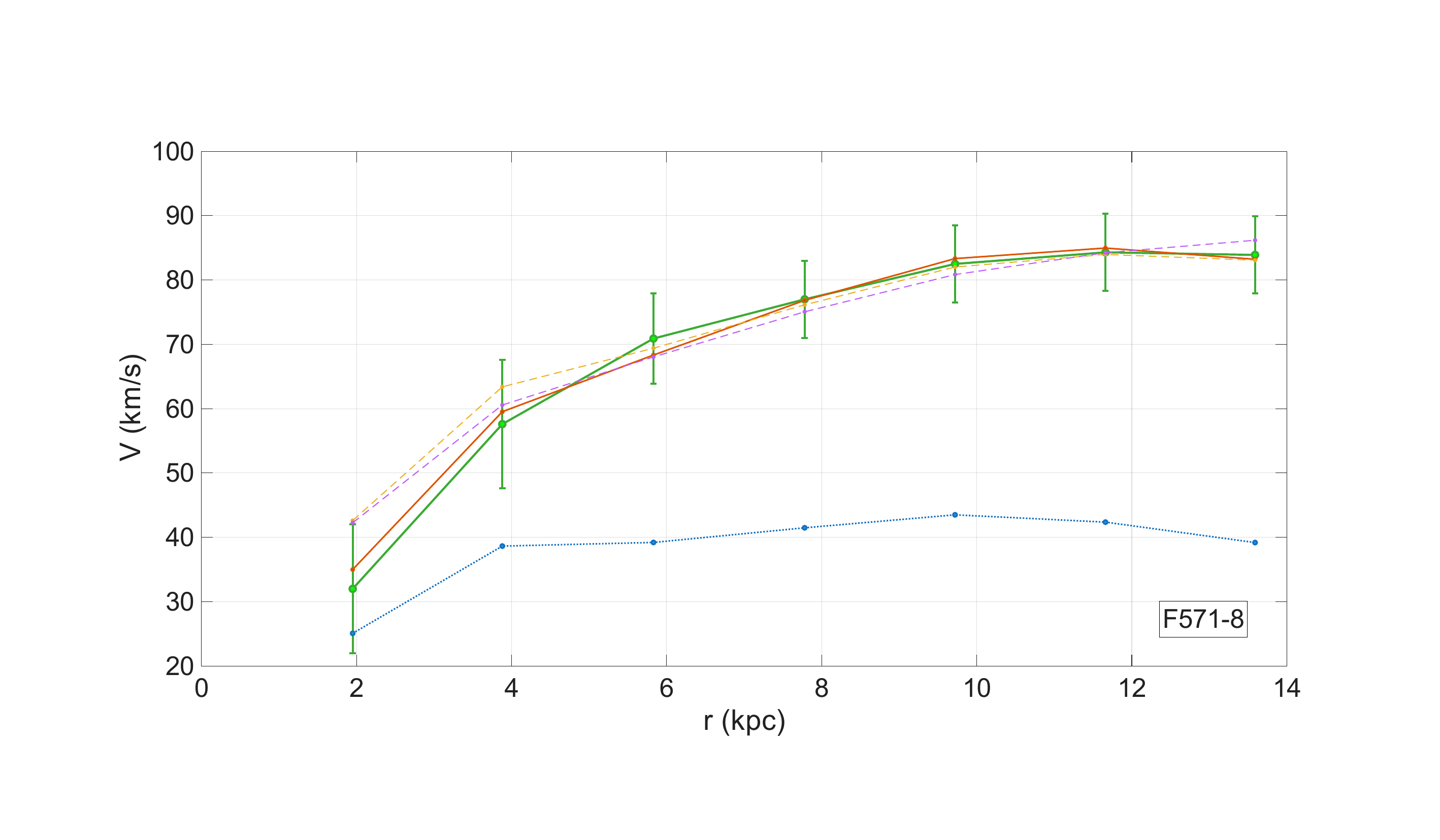}
\includegraphics[trim=4cm 3cm 5cm 4cm, clip=true, width=0.325\columnwidth]{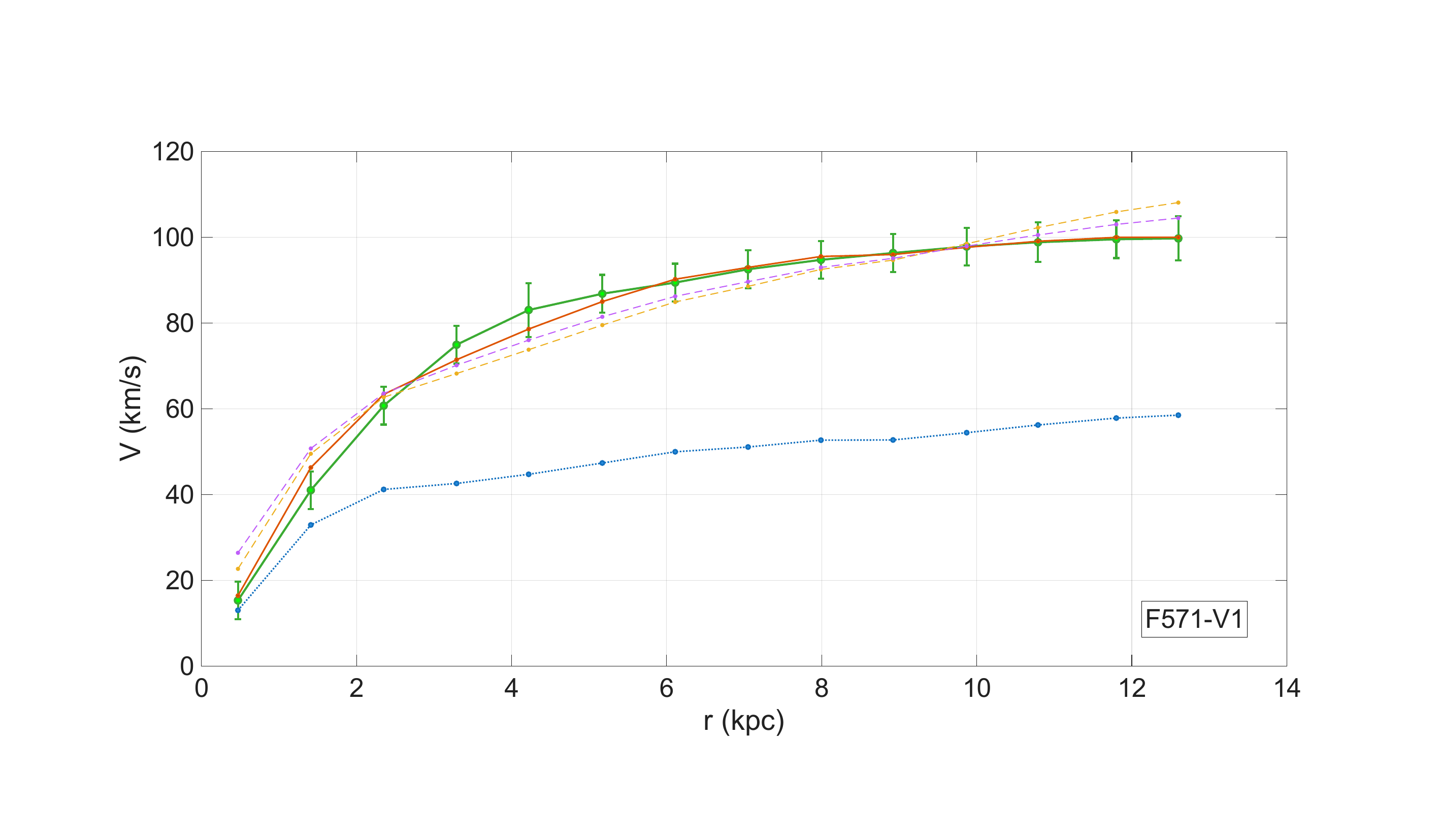}
\includegraphics[trim=4cm 3cm 5cm 4cm, clip=true, width=0.325\columnwidth]{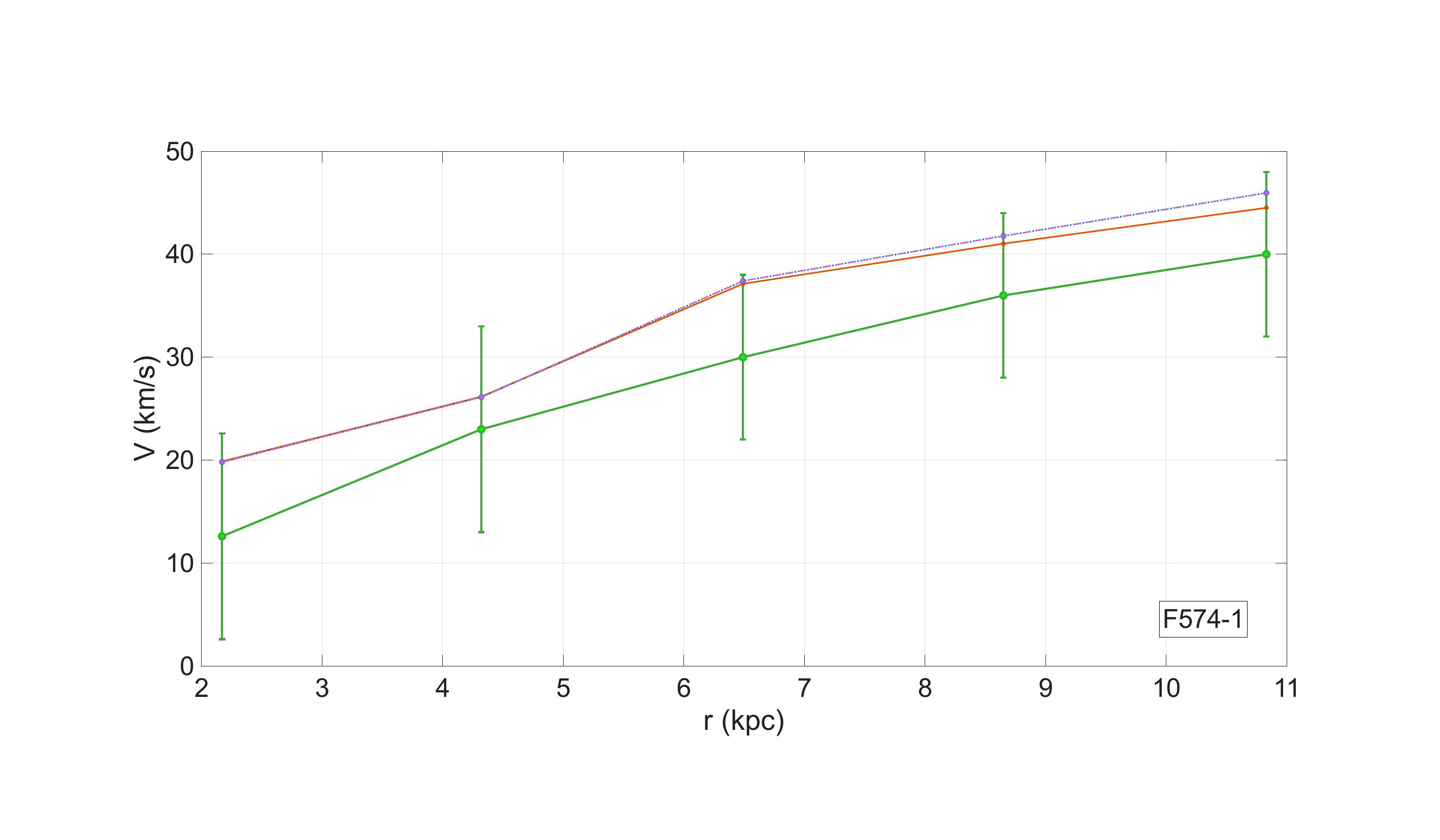}
\end{figure}
\begin{figure}
\centering
\includegraphics[trim=4cm 3cm 5cm 4cm, clip=true, width=0.325\columnwidth]{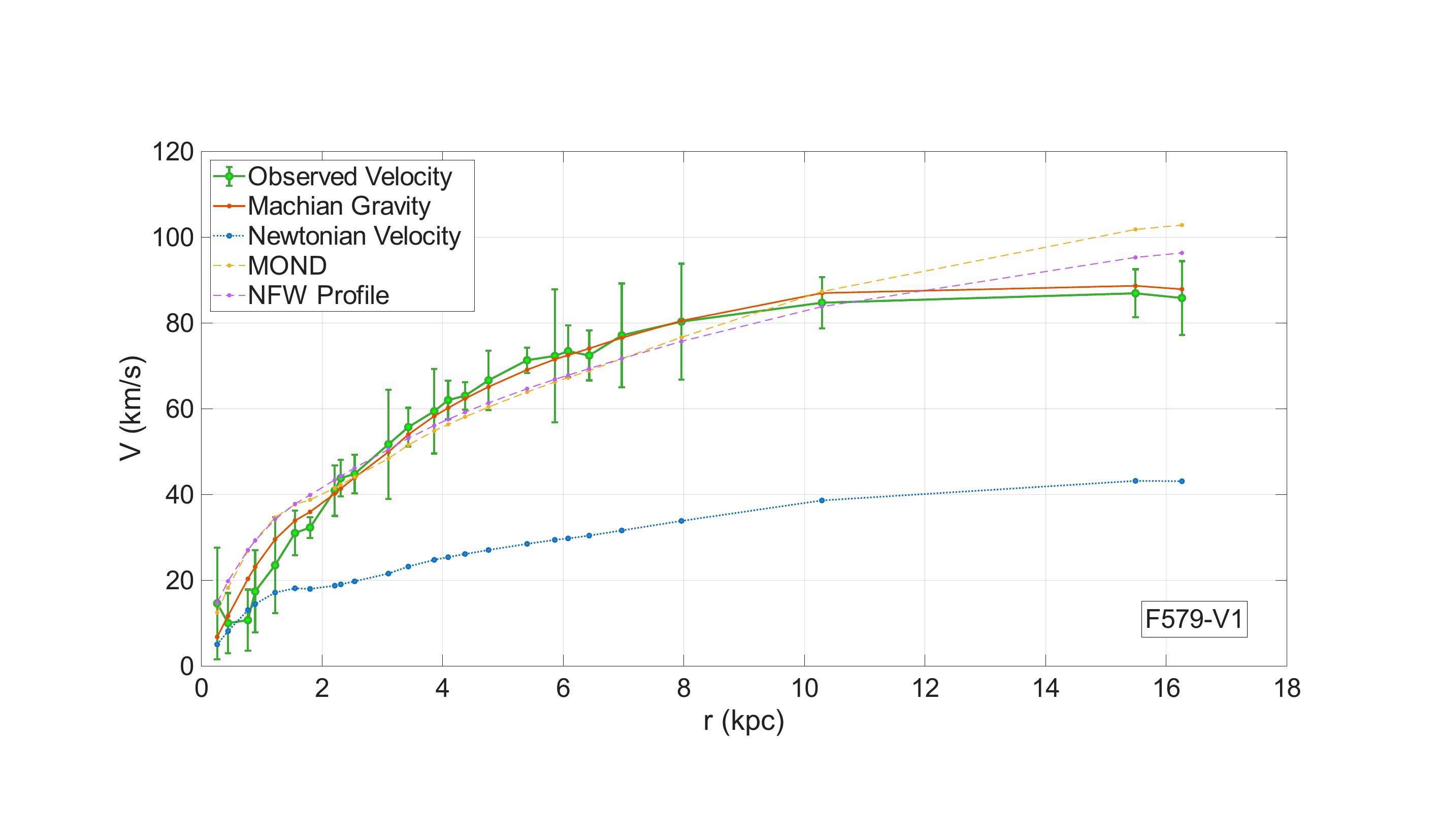}
\includegraphics[trim=4cm 3cm 5cm 4cm, clip=true, width=0.325\columnwidth]{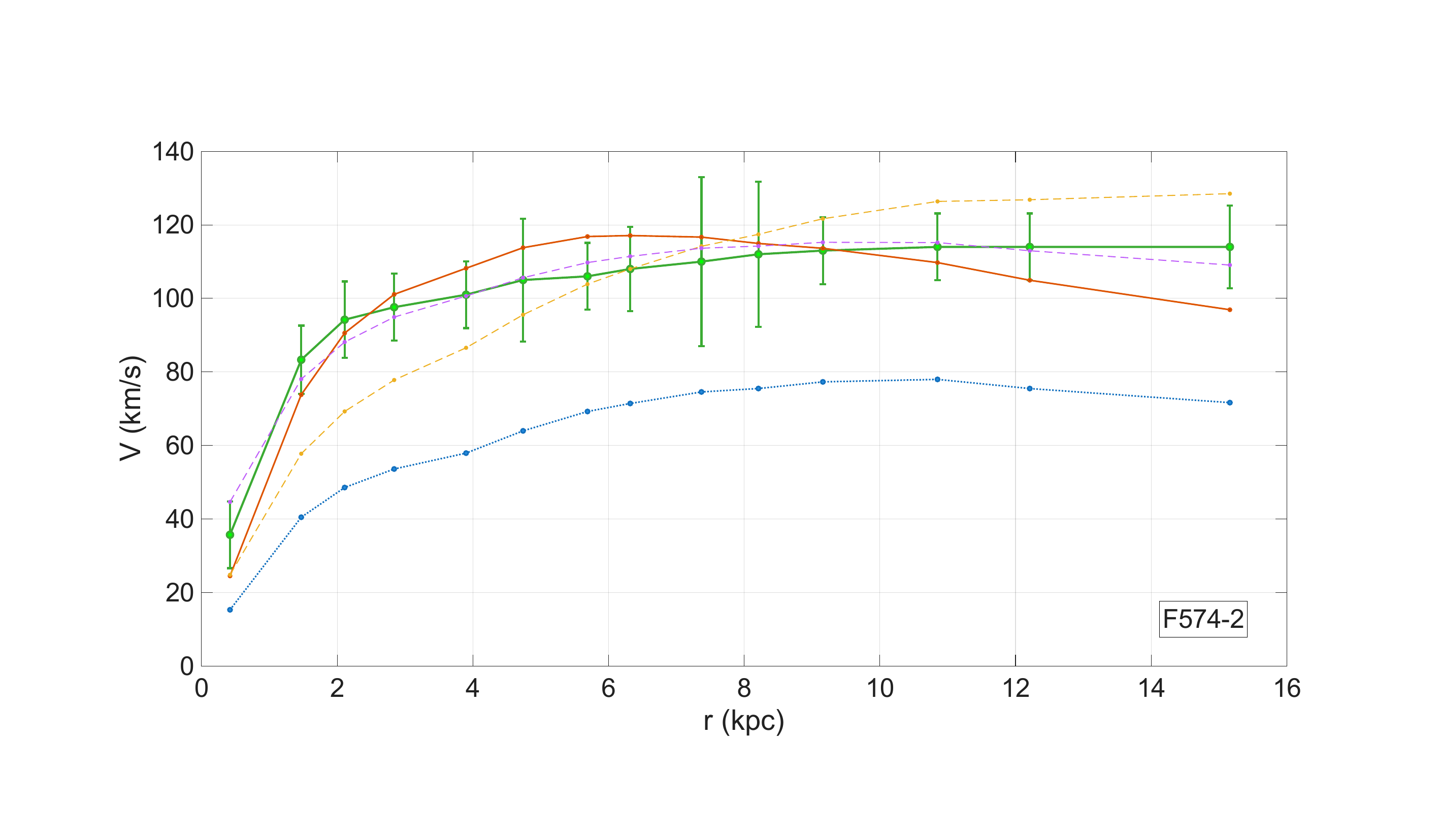}
\includegraphics[trim=4cm 3cm 5cm 4cm, clip=true, width=0.325\columnwidth]{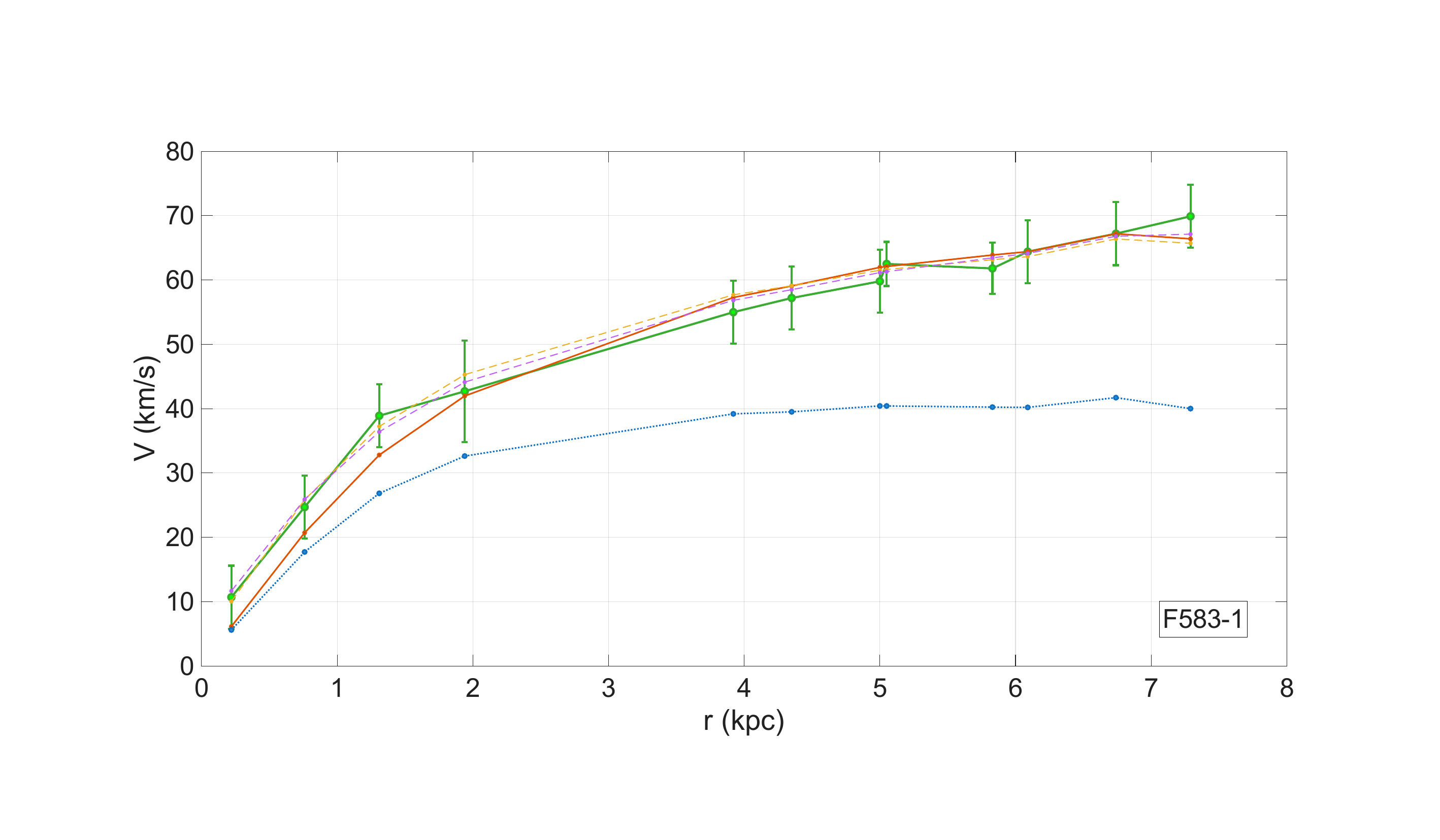}
\includegraphics[trim=4cm 3cm 5cm 4cm, clip=true, width=0.325\columnwidth]{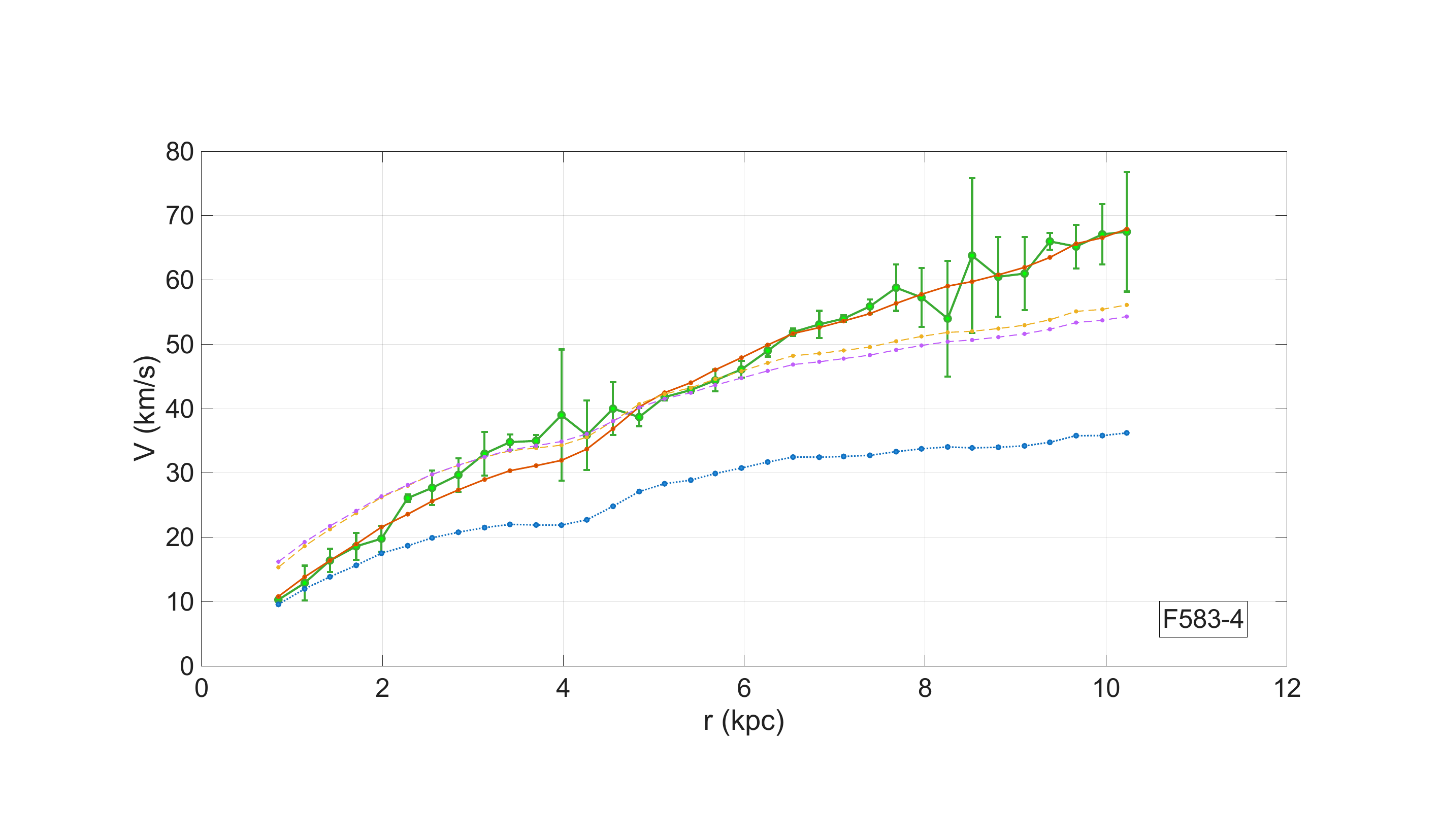}
\includegraphics[trim=4cm 3cm 5cm 4cm, clip=true, width=0.325\columnwidth]{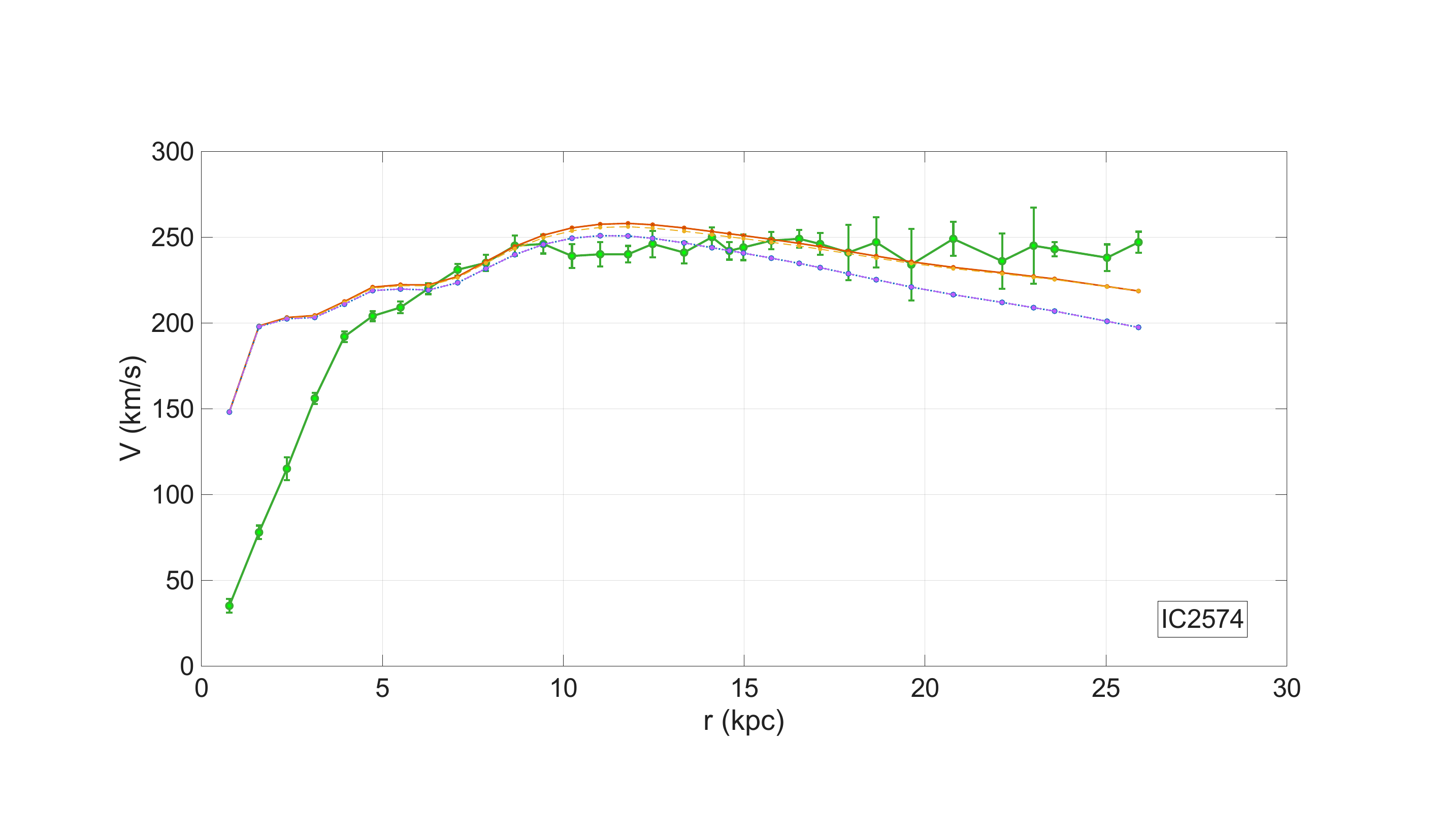}
\includegraphics[trim=4cm 3cm 5cm 4cm, clip=true, width=0.325\columnwidth]{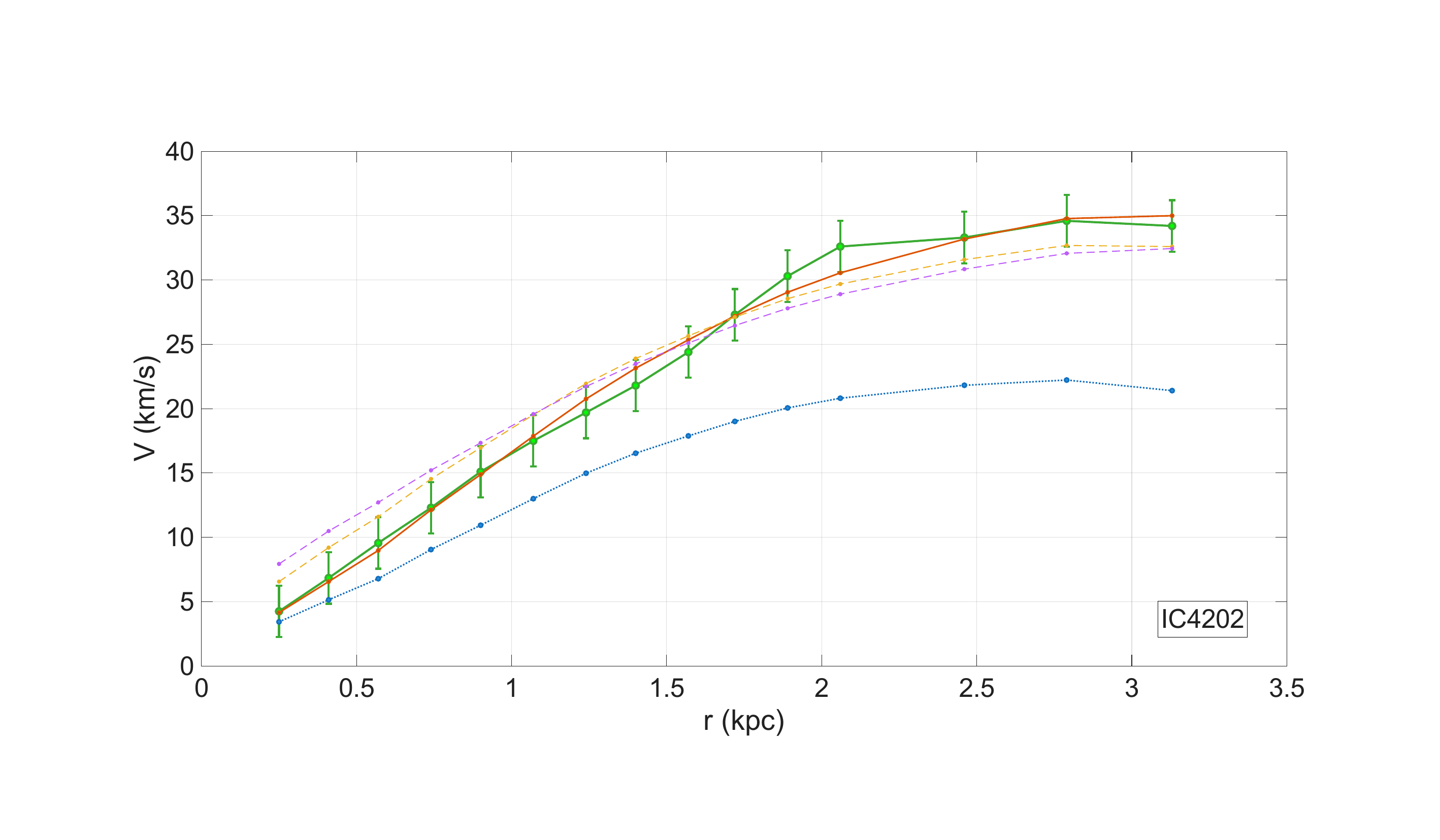}
\includegraphics[trim=4cm 3cm 5cm 4cm, clip=true, width=0.325\columnwidth]{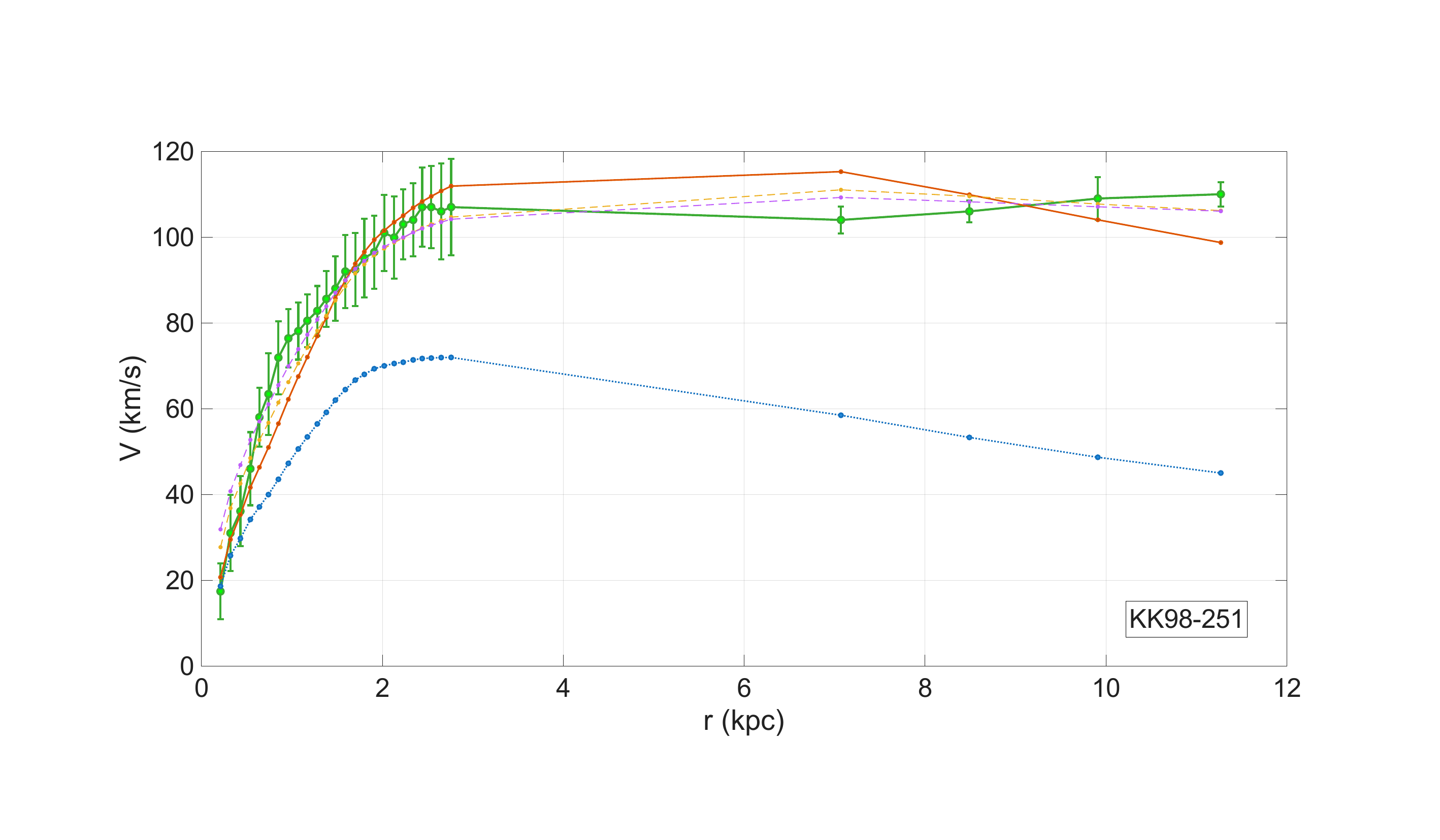}
\includegraphics[trim=4cm 3cm 5cm 4cm, clip=true, width=0.325\columnwidth]{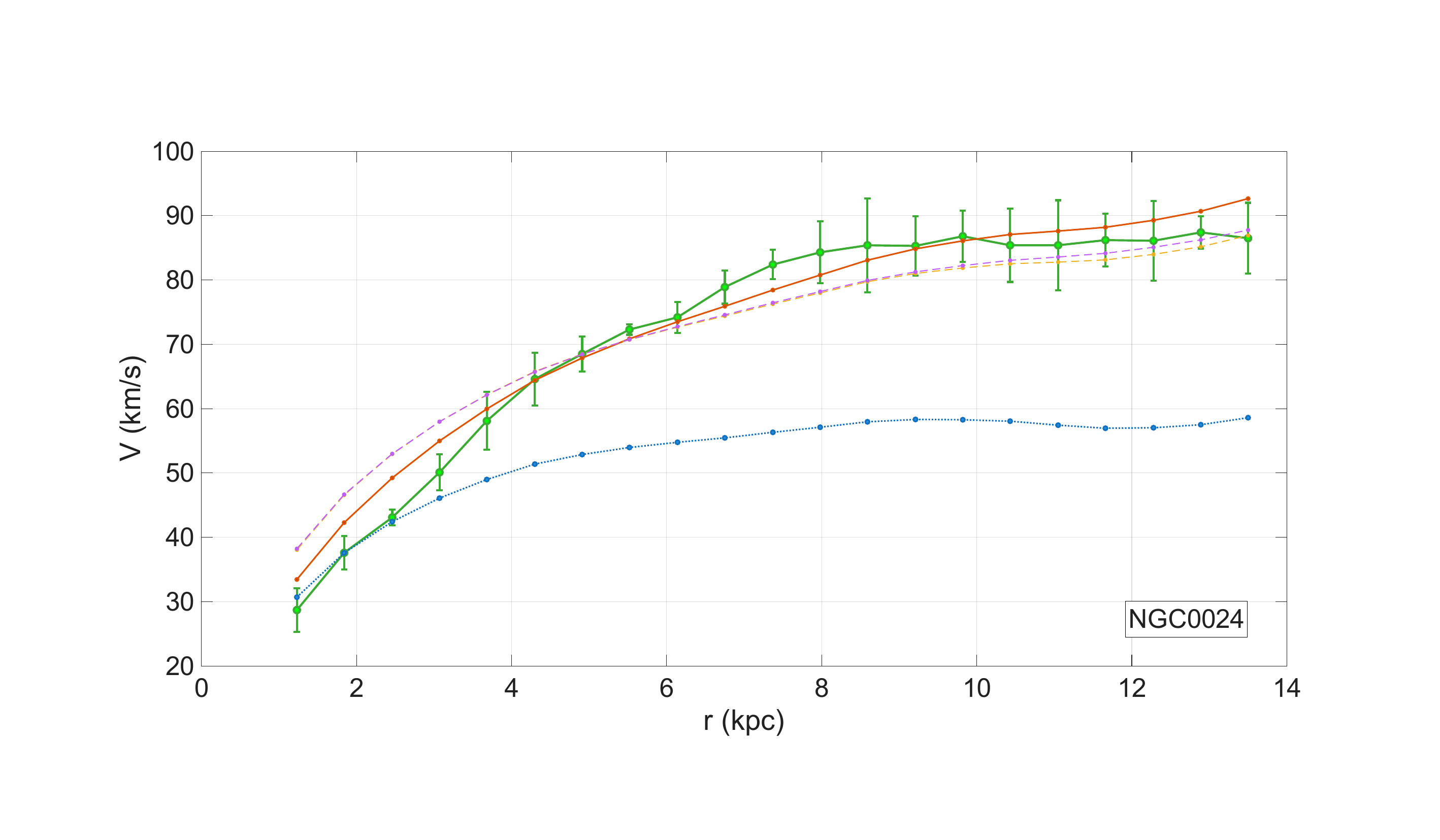}
\includegraphics[trim=4cm 3cm 5cm 4cm, clip=true, width=0.325\columnwidth]{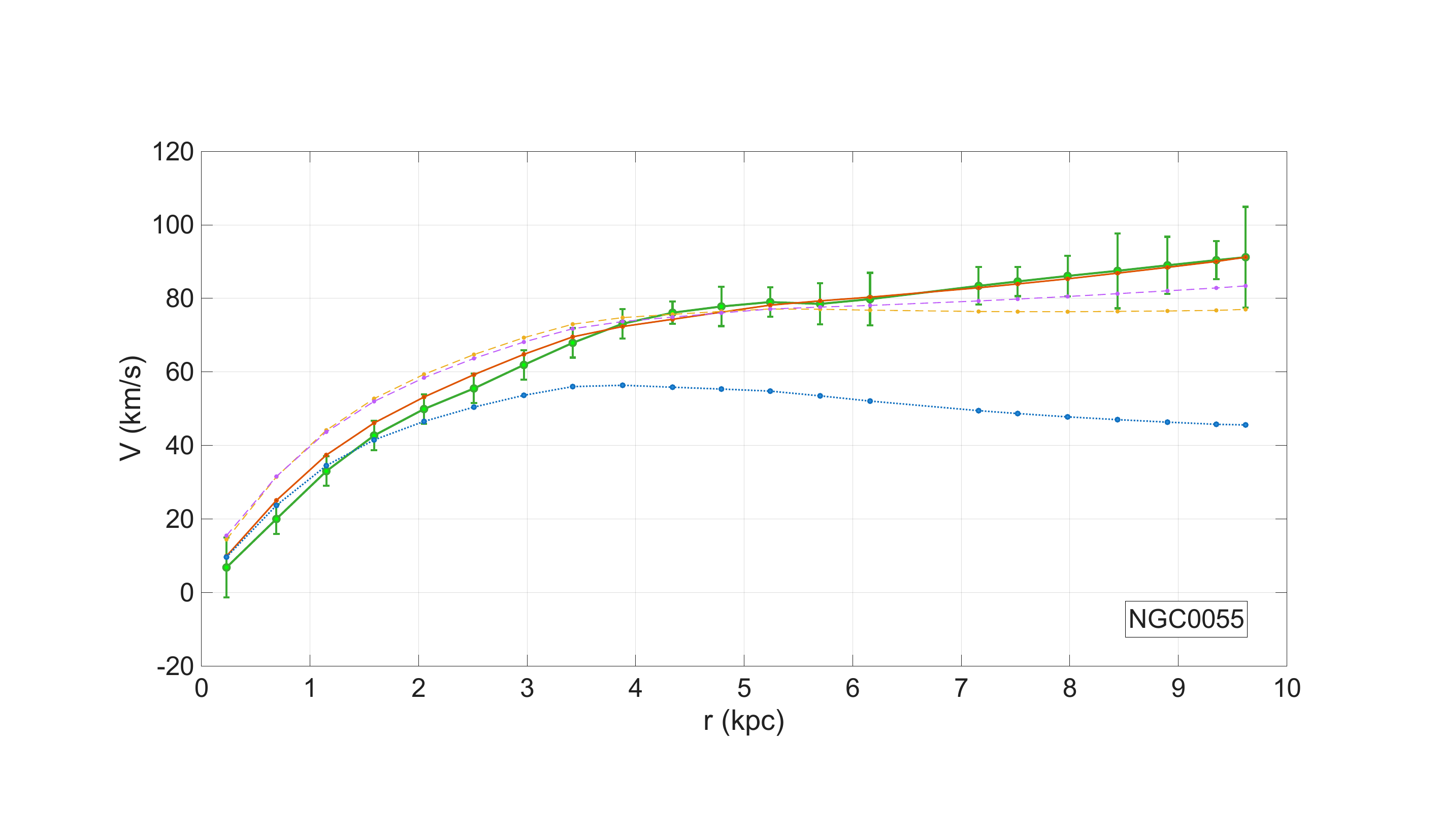}
\includegraphics[trim=4cm 3cm 5cm 4cm, clip=true, width=0.325\columnwidth]{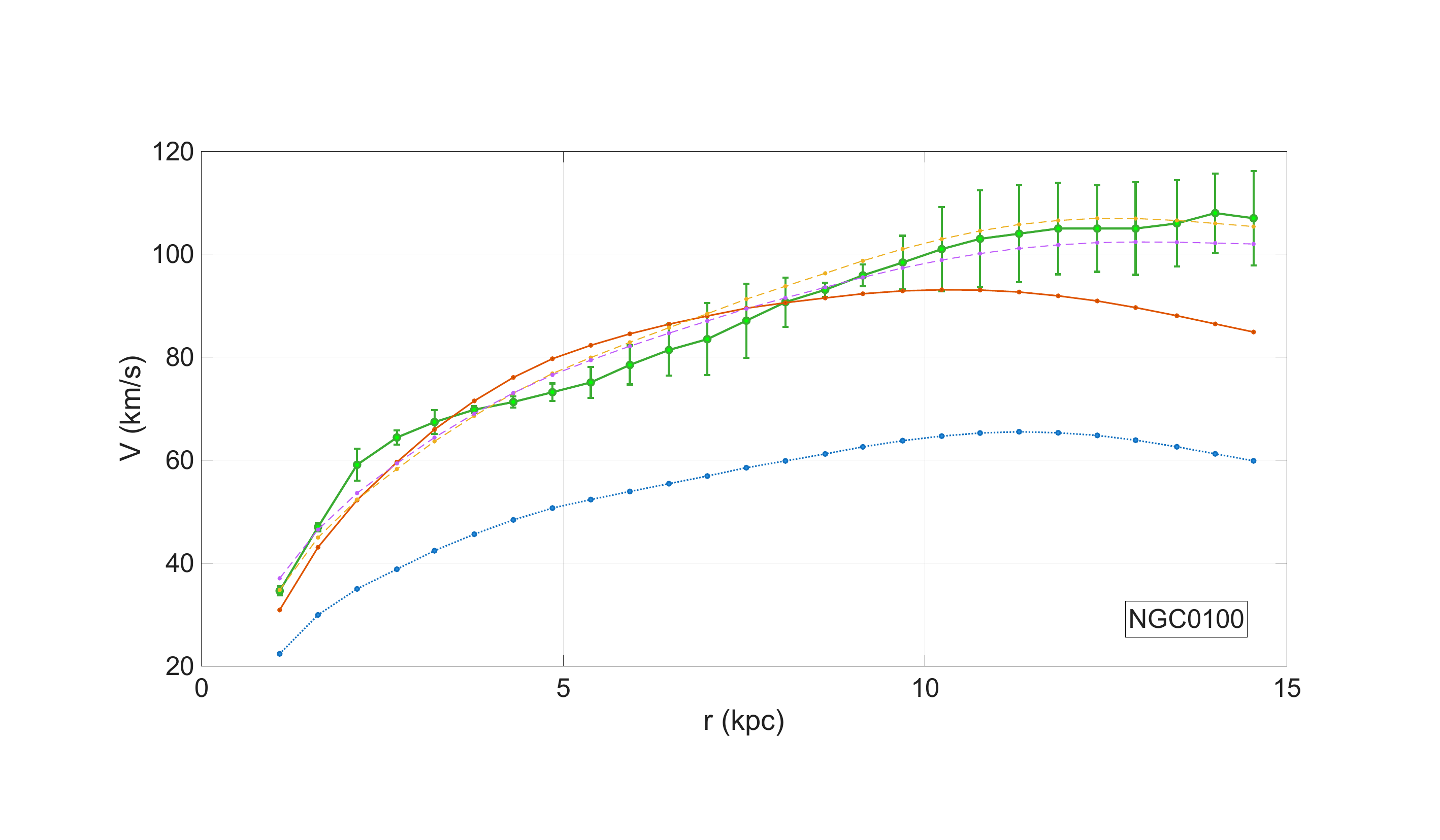}
\includegraphics[trim=4cm 3cm 5cm 4cm, clip=true, width=0.325\columnwidth]{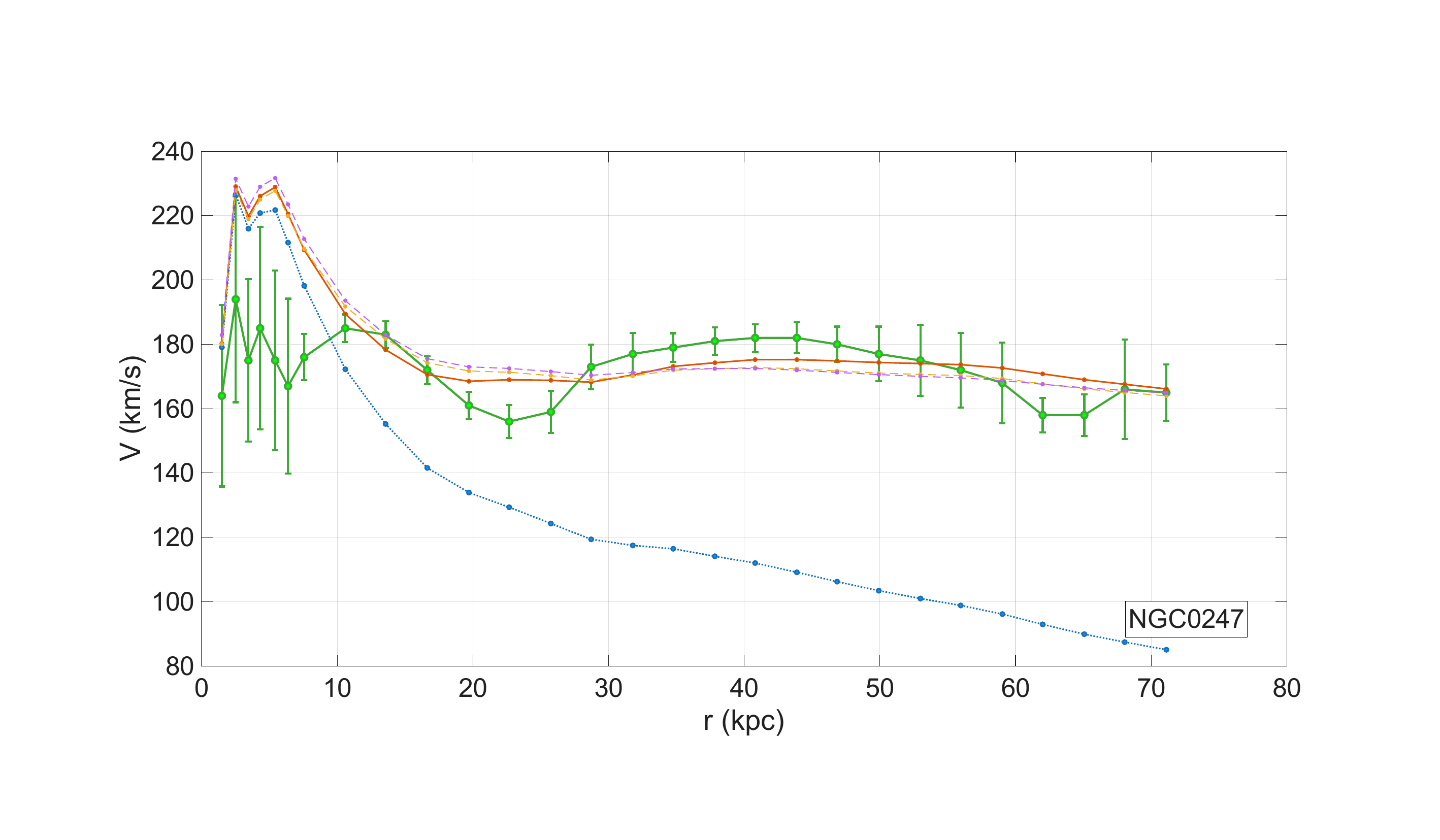}
\includegraphics[trim=4cm 3cm 5cm 4cm, clip=true, width=0.325\columnwidth]{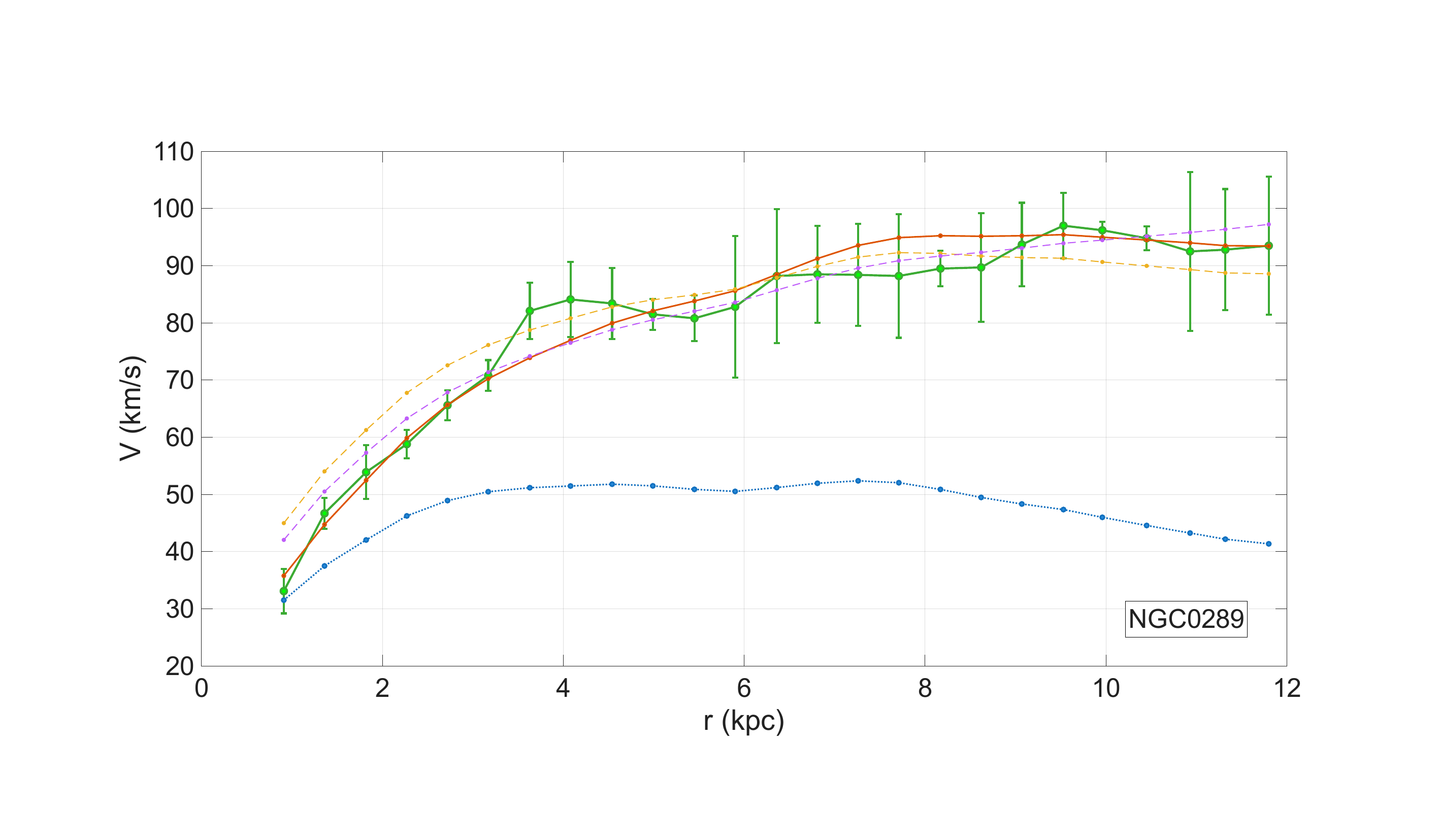}
\includegraphics[trim=4cm 3cm 5cm 4cm, clip=true, width=0.325\columnwidth]{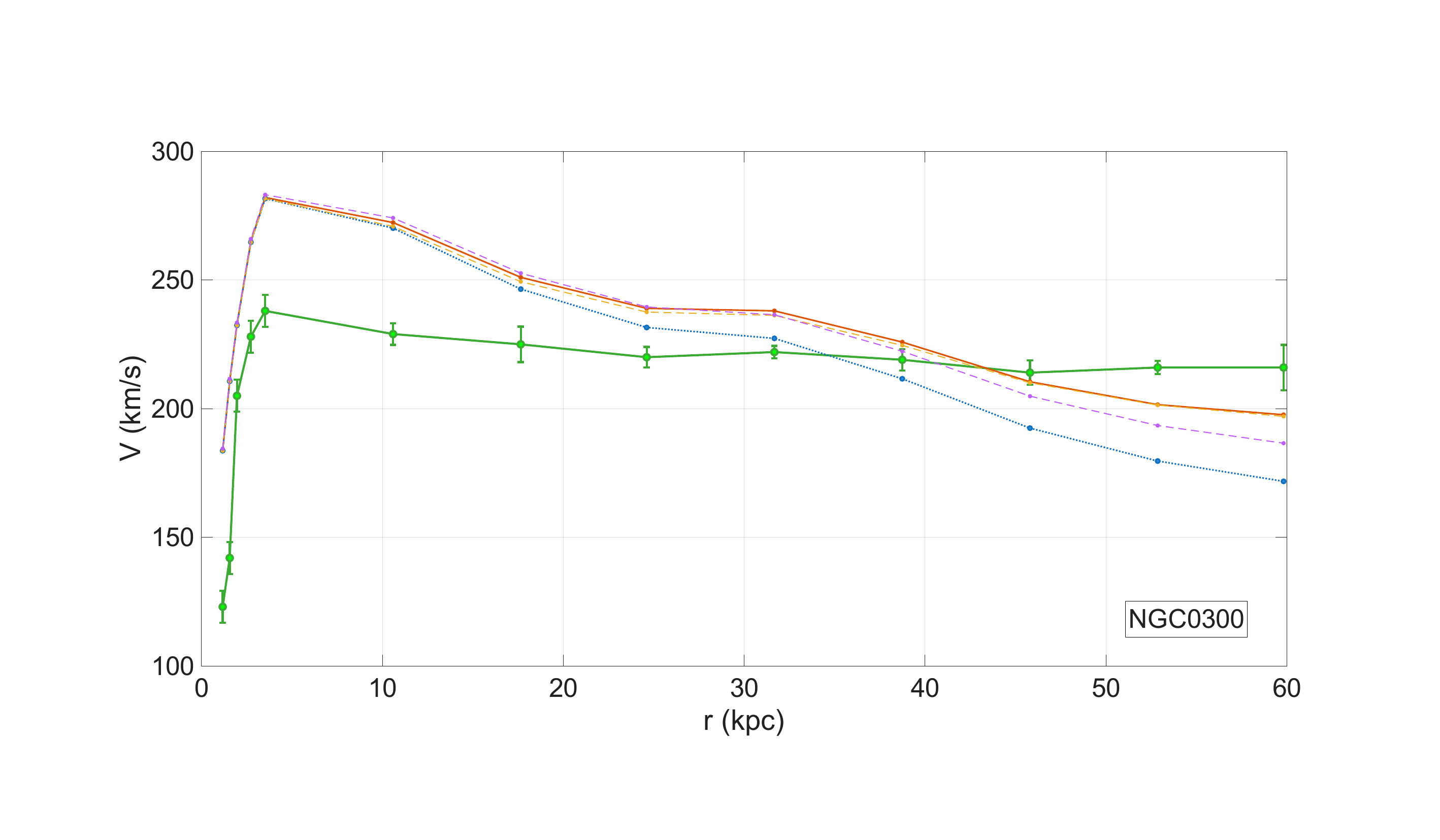}
\includegraphics[trim=4cm 3cm 5cm 4cm, clip=true, width=0.325\columnwidth]{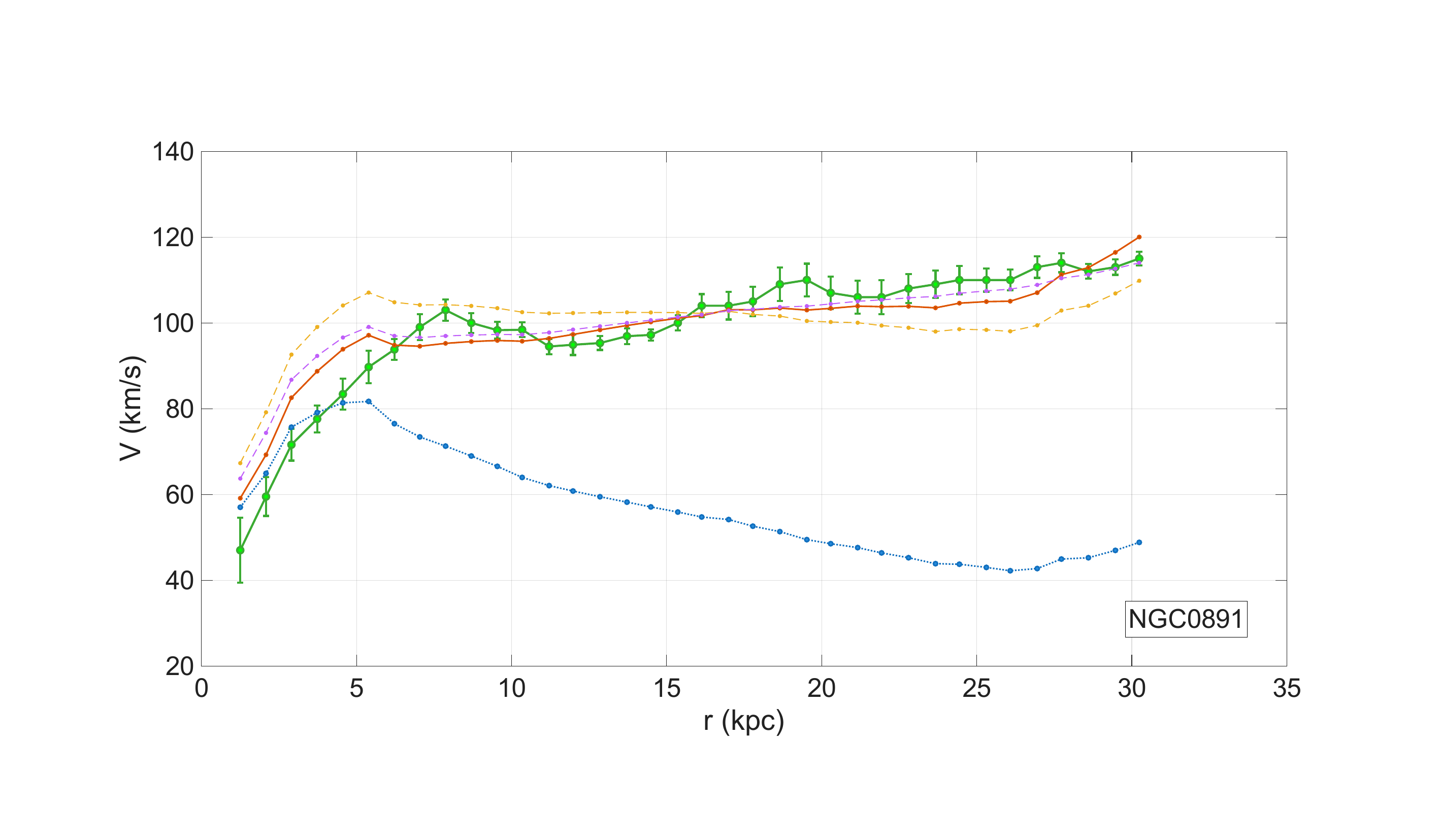}
\includegraphics[trim=4cm 3cm 5cm 4cm, clip=true, width=0.325\columnwidth]{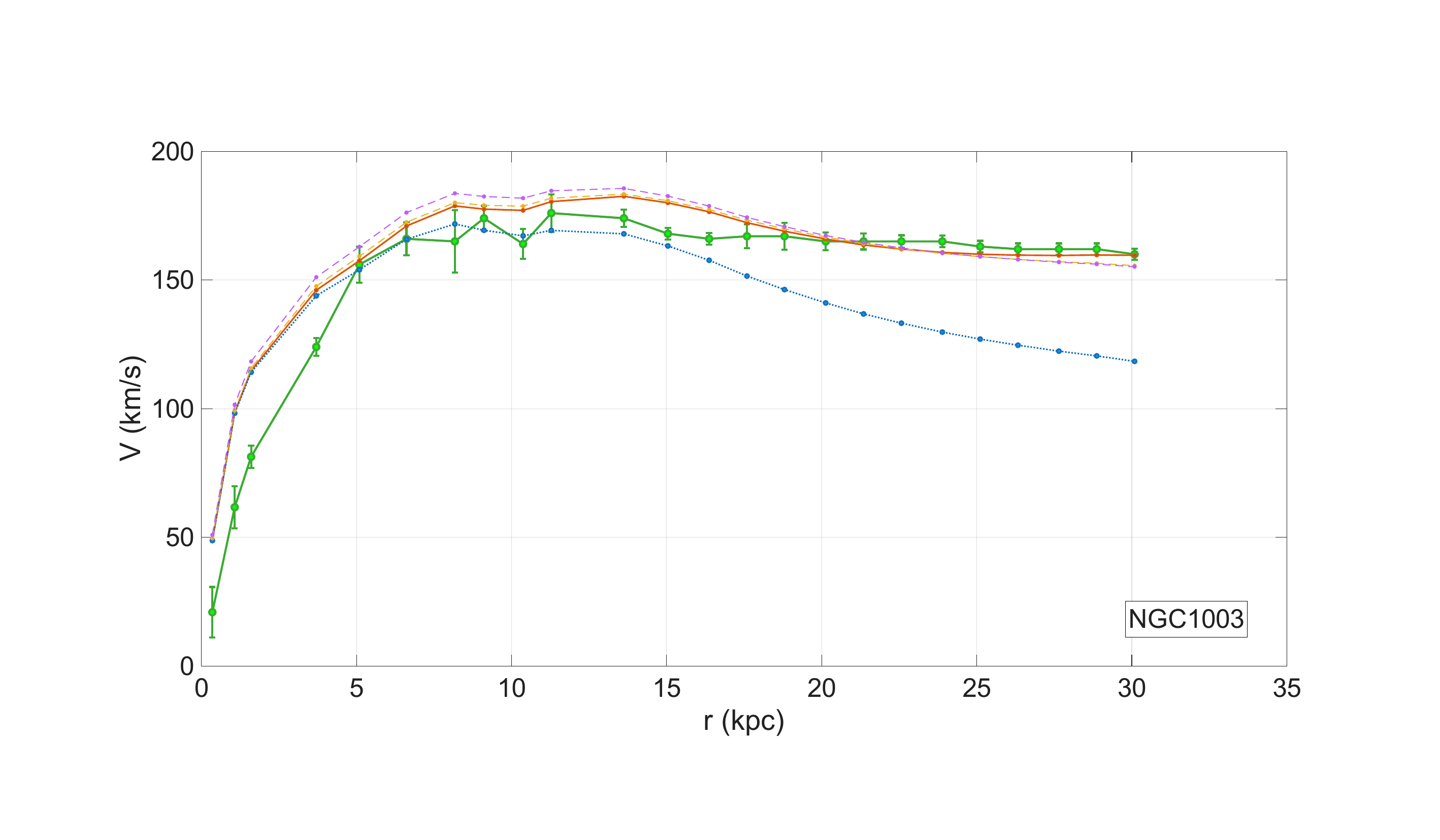}
\includegraphics[trim=4cm 3cm 5cm 4cm, clip=true, width=0.325\columnwidth]{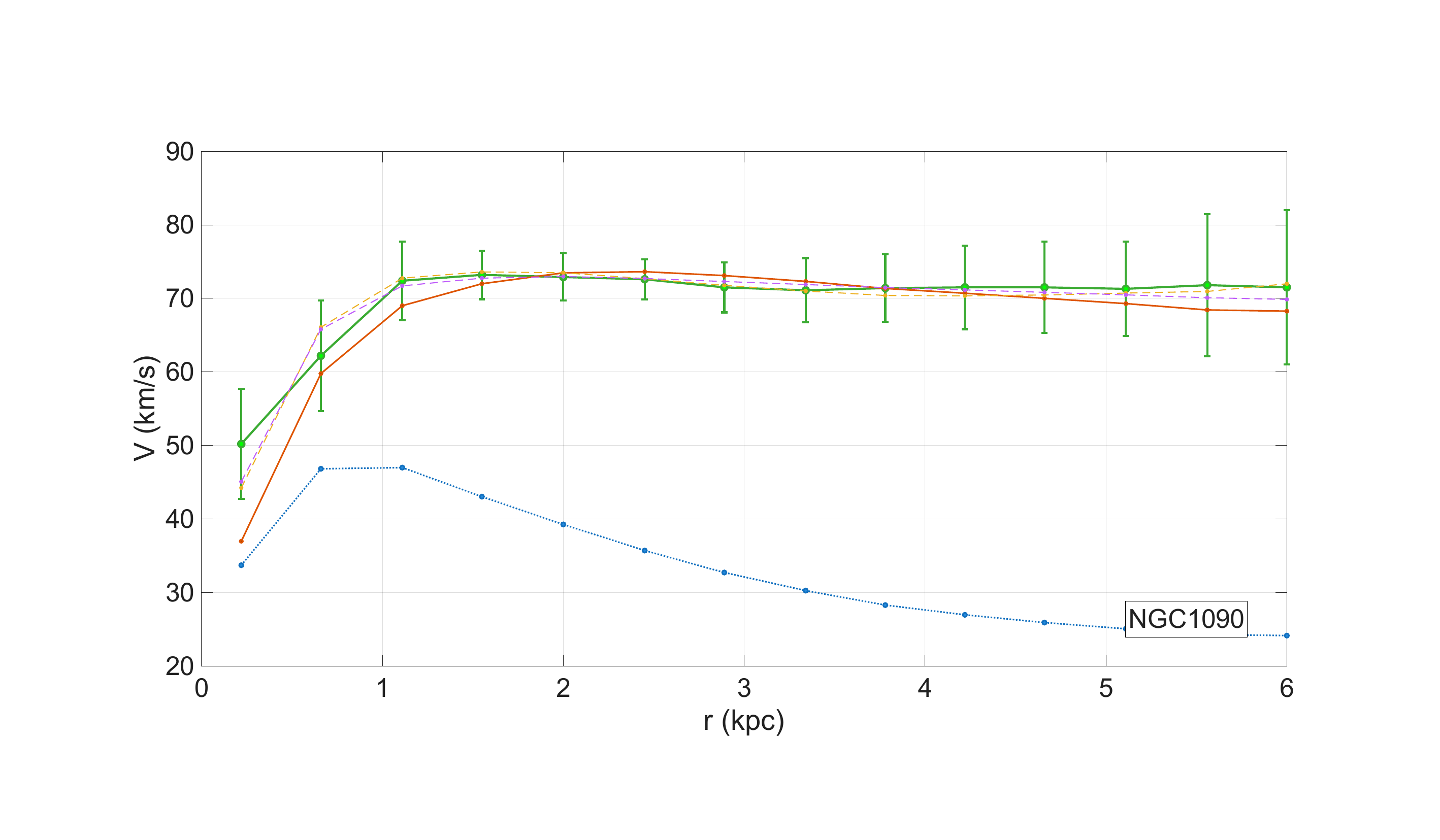}
\includegraphics[trim=4cm 3cm 5cm 4cm, clip=true, width=0.325\columnwidth]{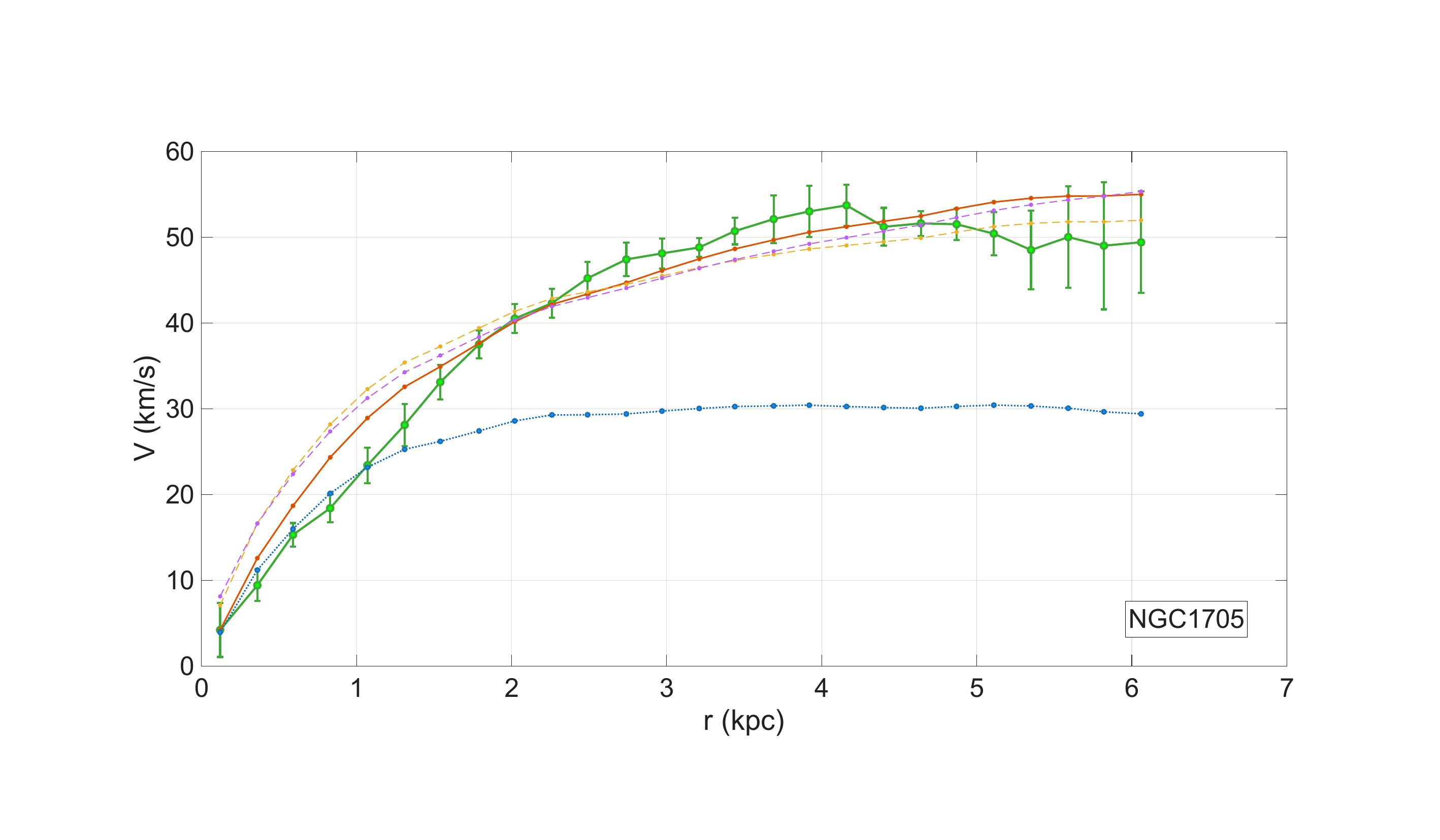}
\includegraphics[trim=4cm 3cm 5cm 4cm, clip=true, width=0.325\columnwidth]{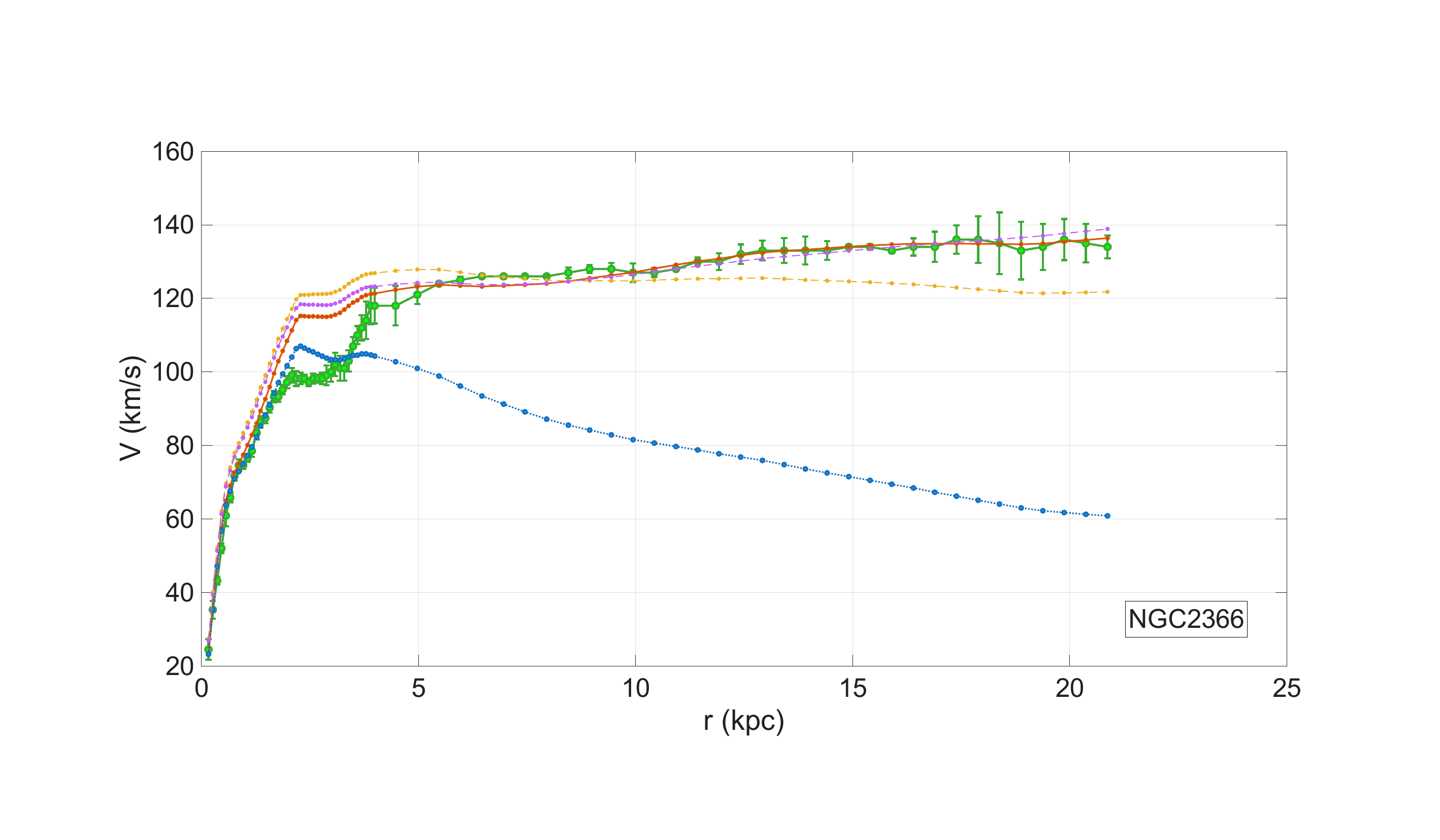}
\includegraphics[trim=4cm 3cm 5cm 4cm, clip=true, width=0.325\columnwidth]{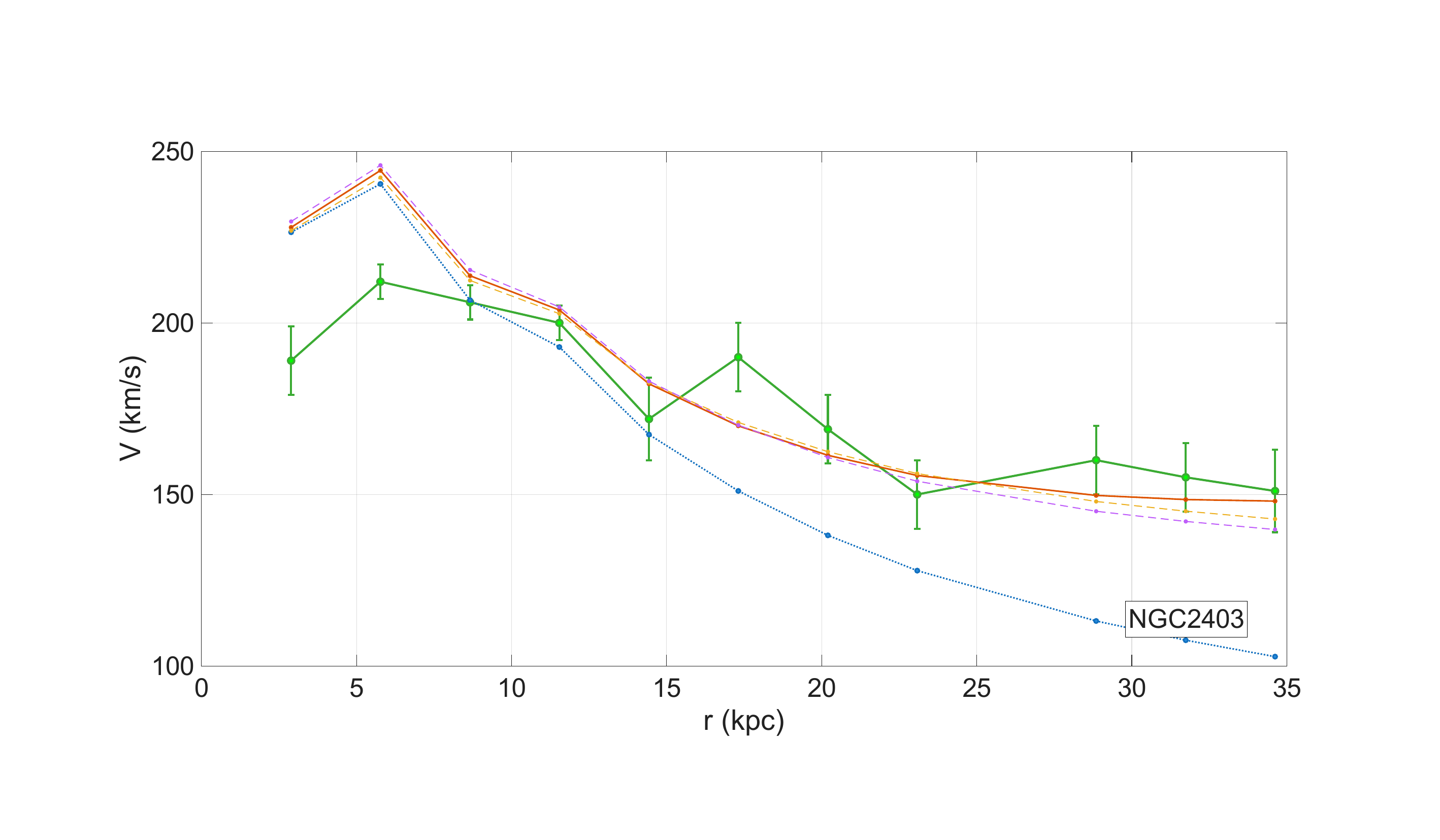}
\includegraphics[trim=4cm 3cm 5cm 4cm, clip=true, width=0.325\columnwidth]{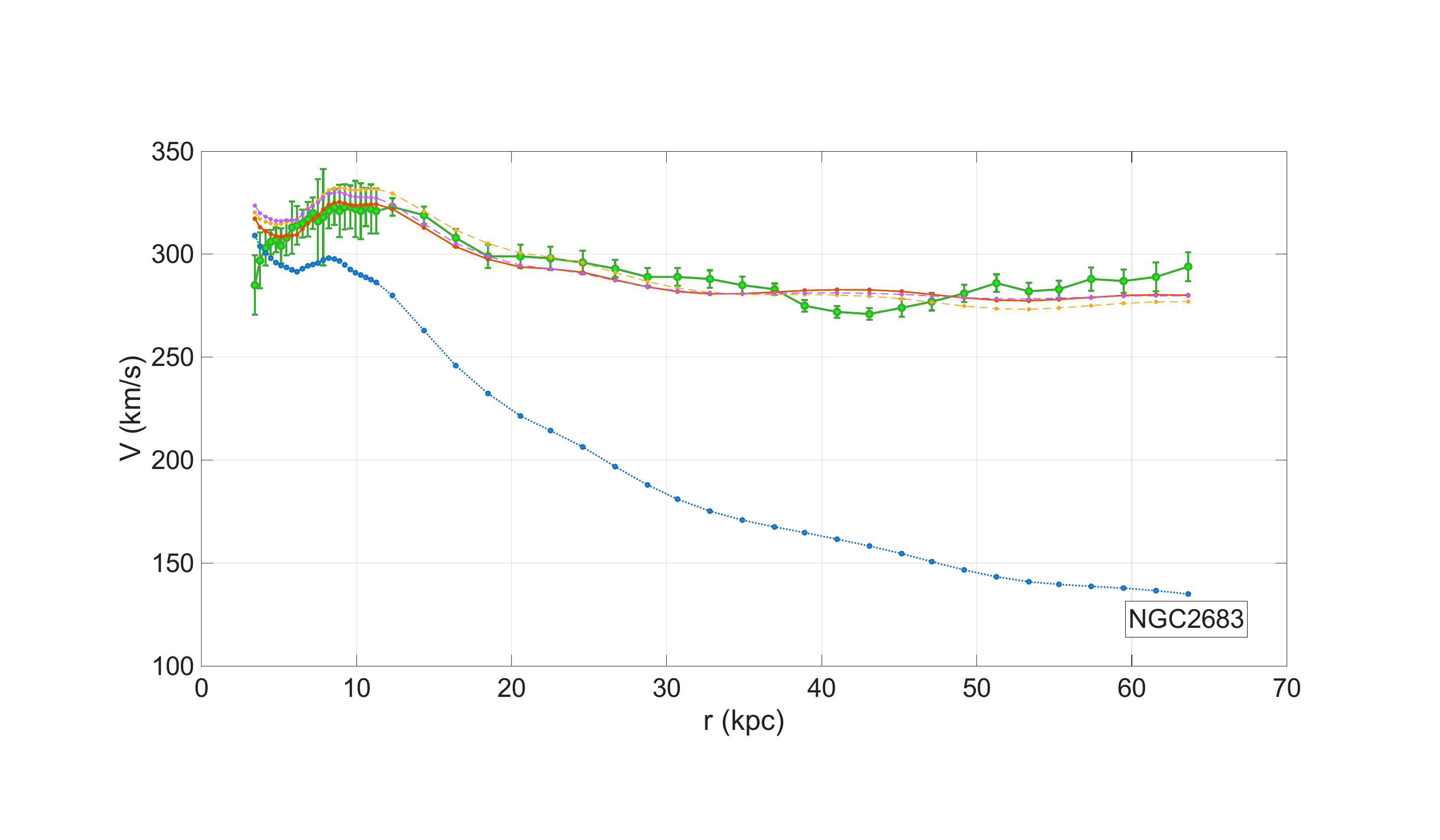}
\includegraphics[trim=4cm 3cm 5cm 4cm, clip=true, width=0.325\columnwidth]{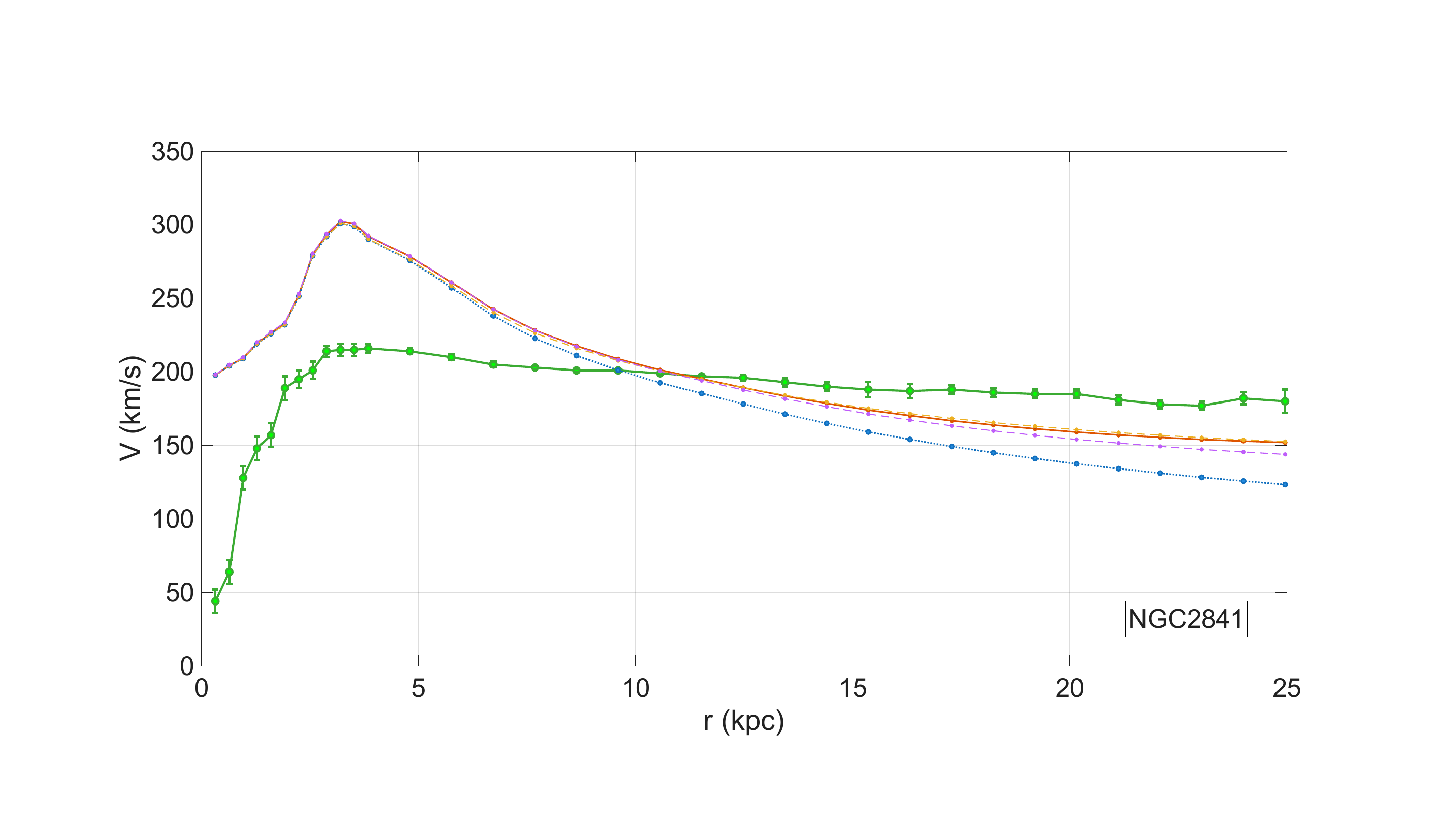}
\includegraphics[trim=4cm 3cm 5cm 4cm, clip=true, width=0.325\columnwidth]{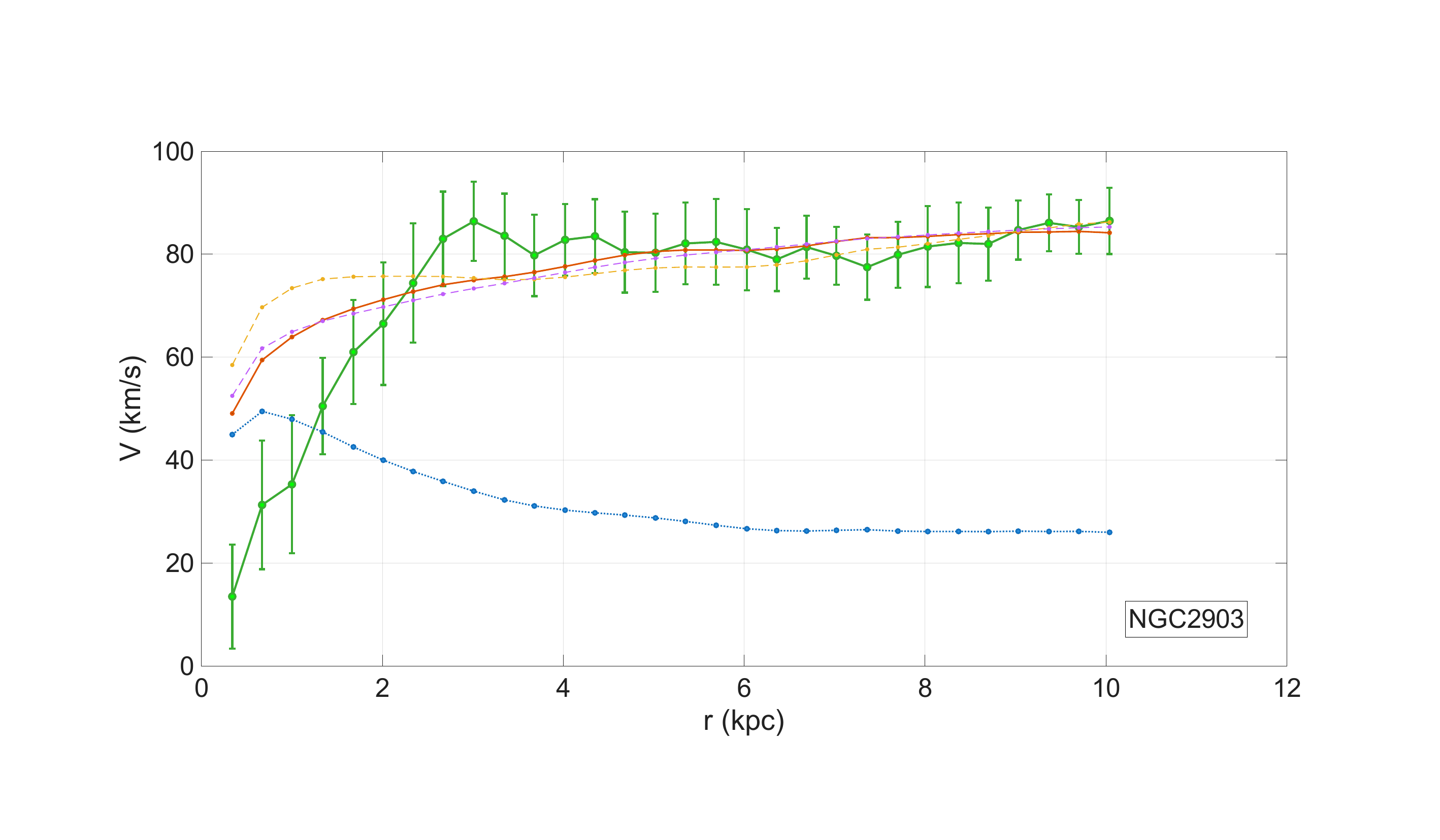}
\includegraphics[trim=4cm 3cm 5cm 4cm, clip=true, width=0.325\columnwidth]{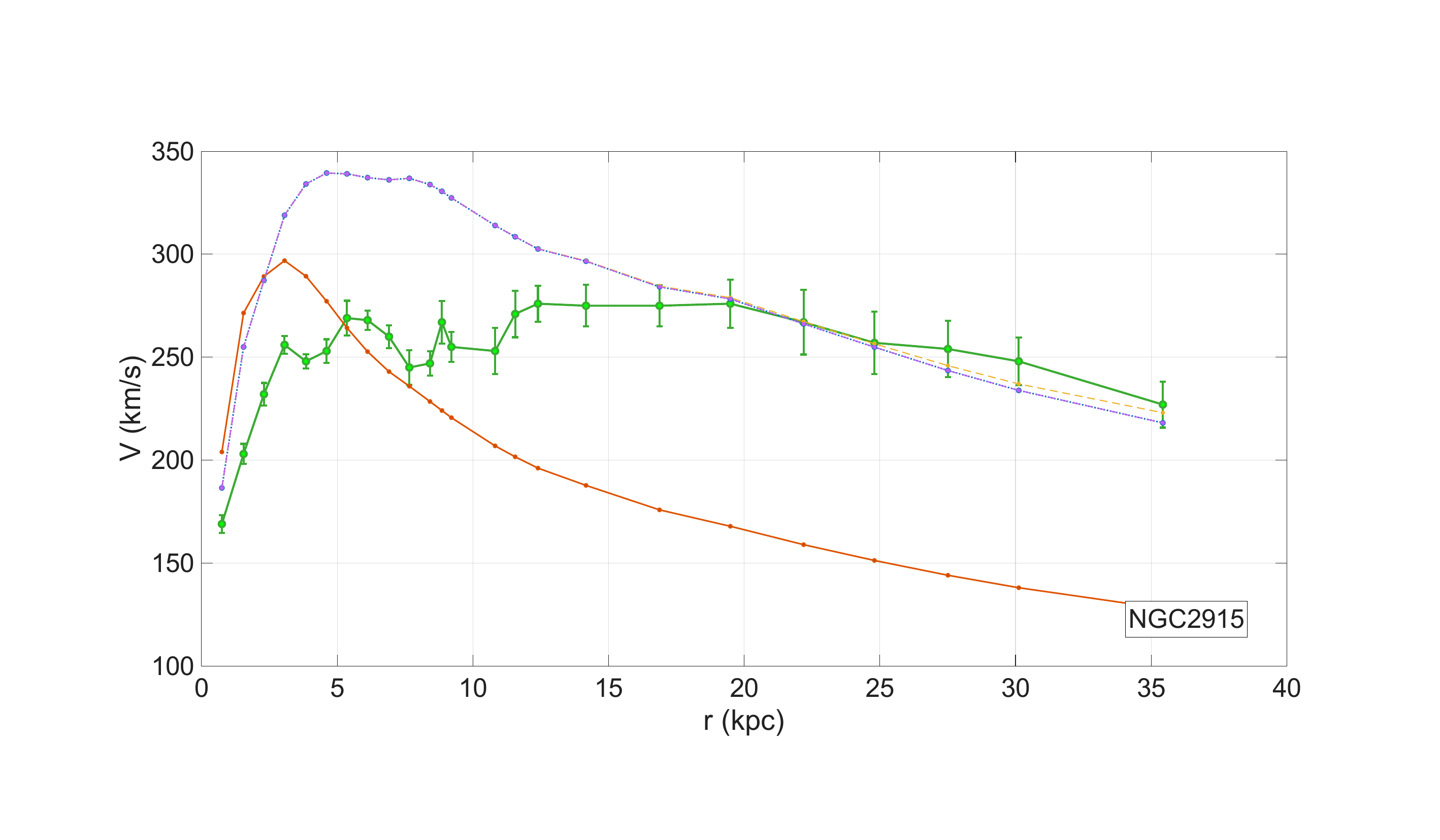}
\includegraphics[trim=4cm 3cm 5cm 4cm, clip=true, width=0.325\columnwidth]{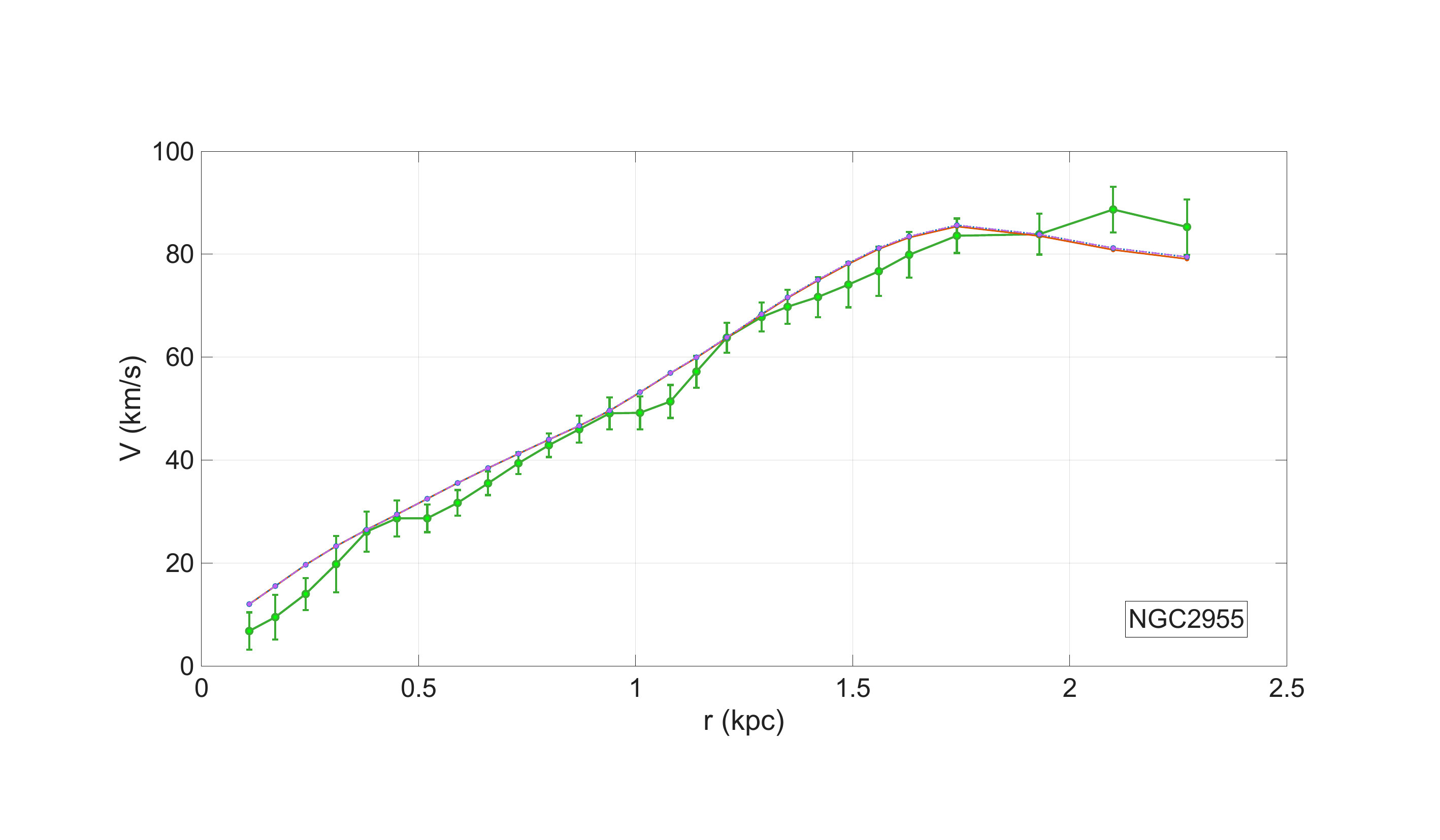}
\end{figure}
\begin{figure}
\centering
\includegraphics[trim=4cm 3cm 5cm 4cm, clip=true, width=0.325\columnwidth]{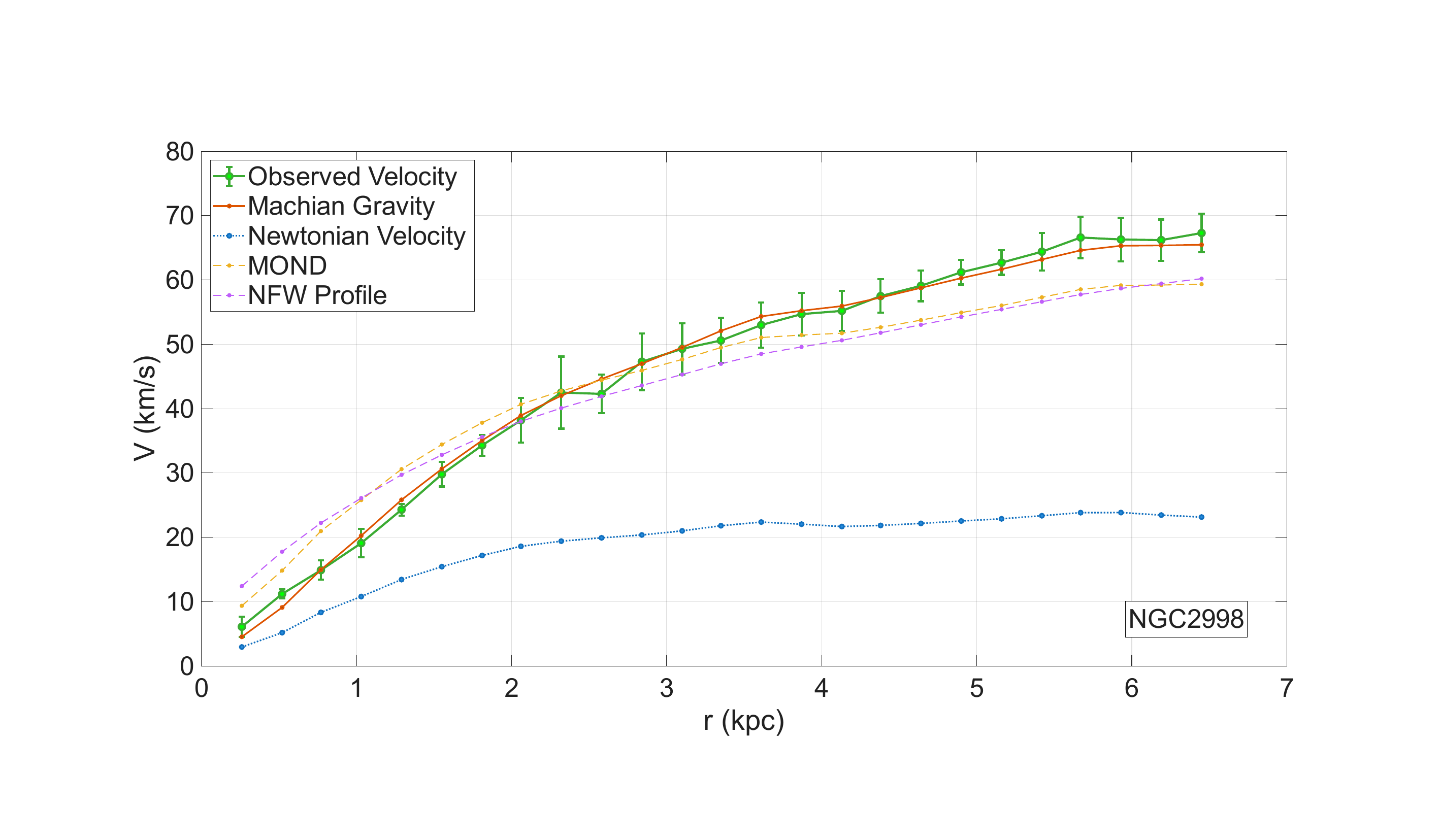}
\includegraphics[trim=4cm 3cm 5cm 4cm, clip=true, width=0.325\columnwidth]{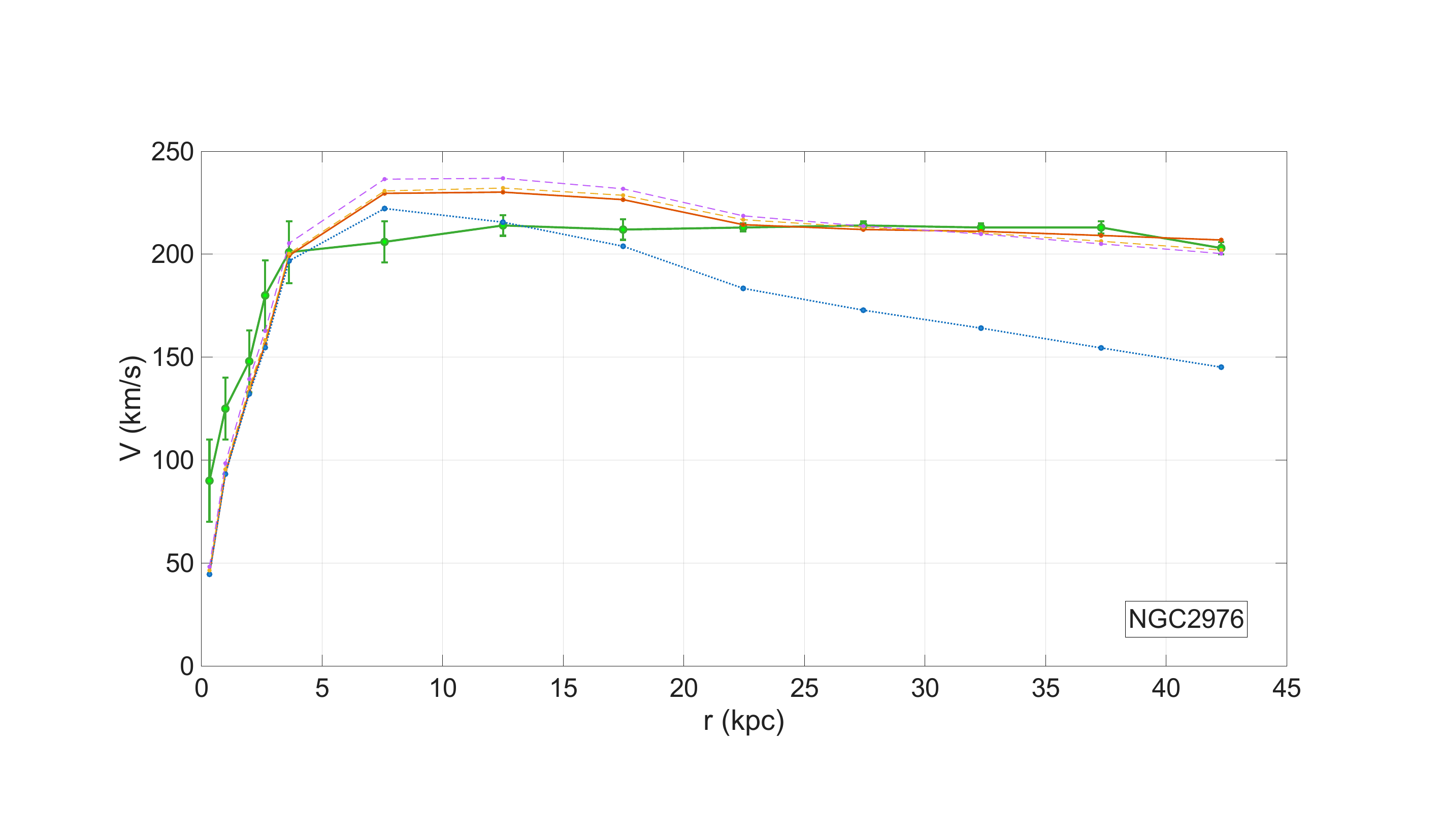}
\includegraphics[trim=4cm 3cm 5cm 4cm, clip=true, width=0.325\columnwidth]{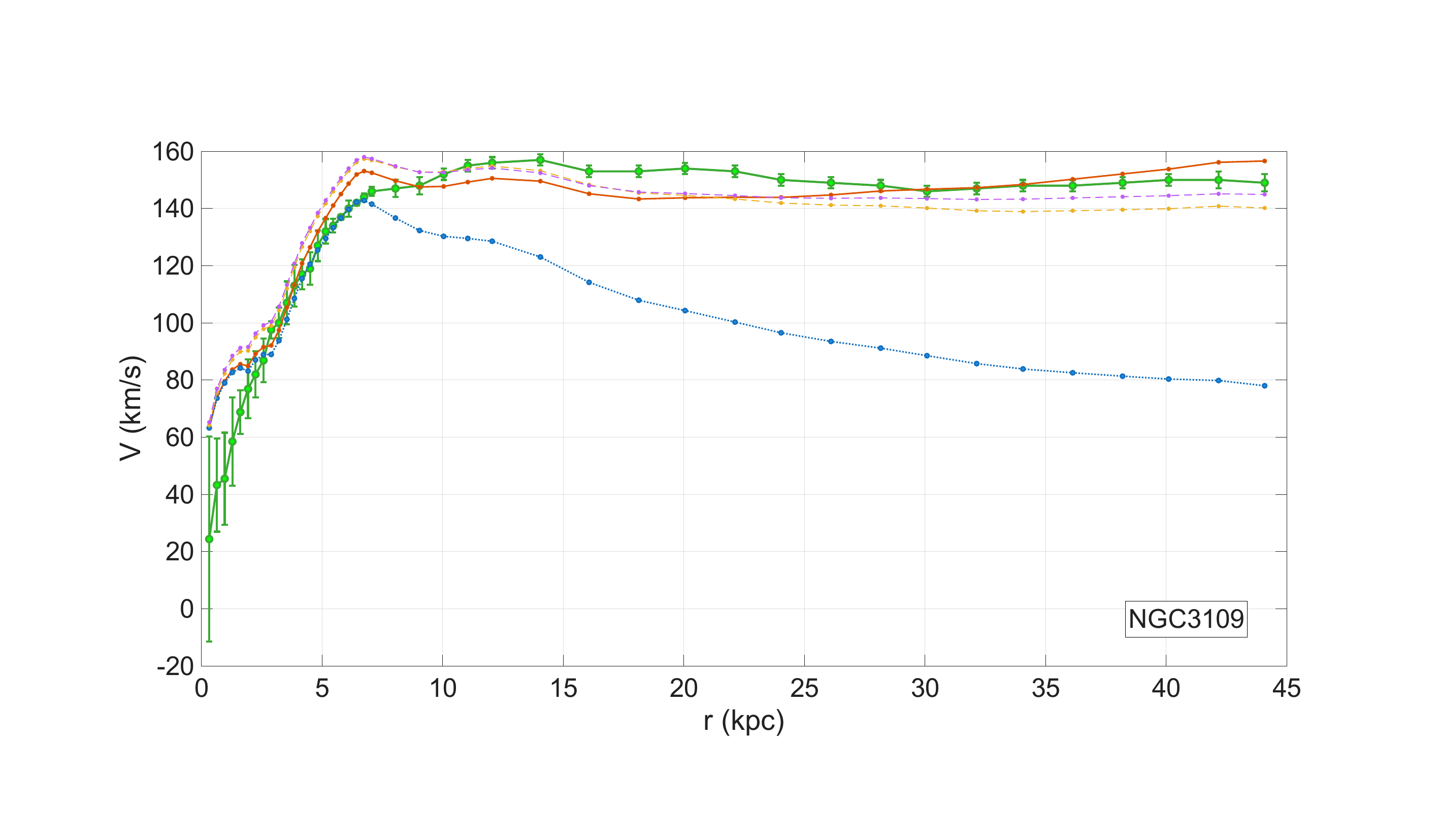}
\includegraphics[trim=4cm 3cm 5cm 4cm, clip=true, width=0.325\columnwidth]{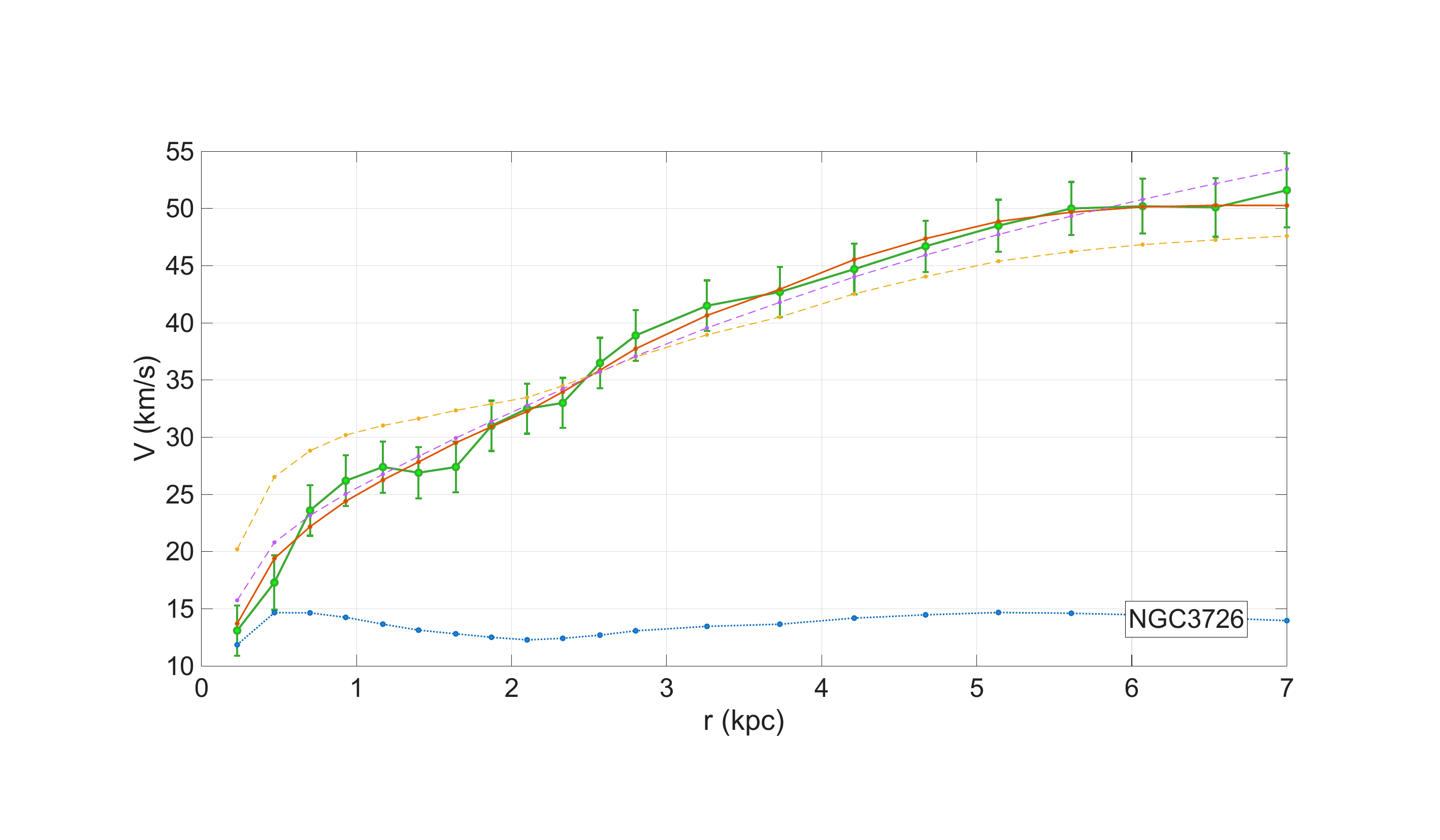}
\includegraphics[trim=4cm 3cm 5cm 4cm, clip=true, width=0.325\columnwidth]{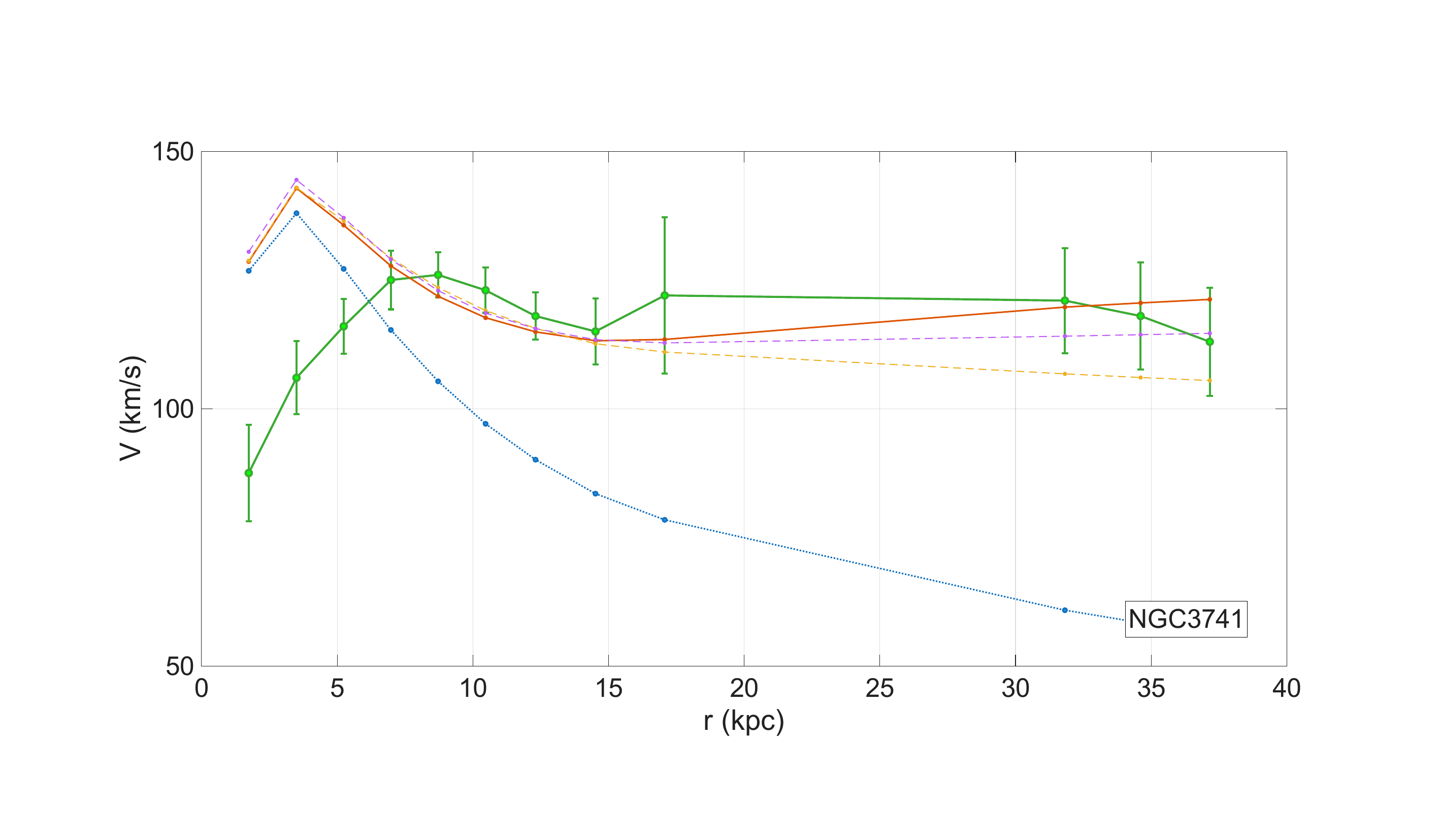}
\includegraphics[trim=4cm 3cm 5cm 4cm, clip=true, width=0.325\columnwidth]{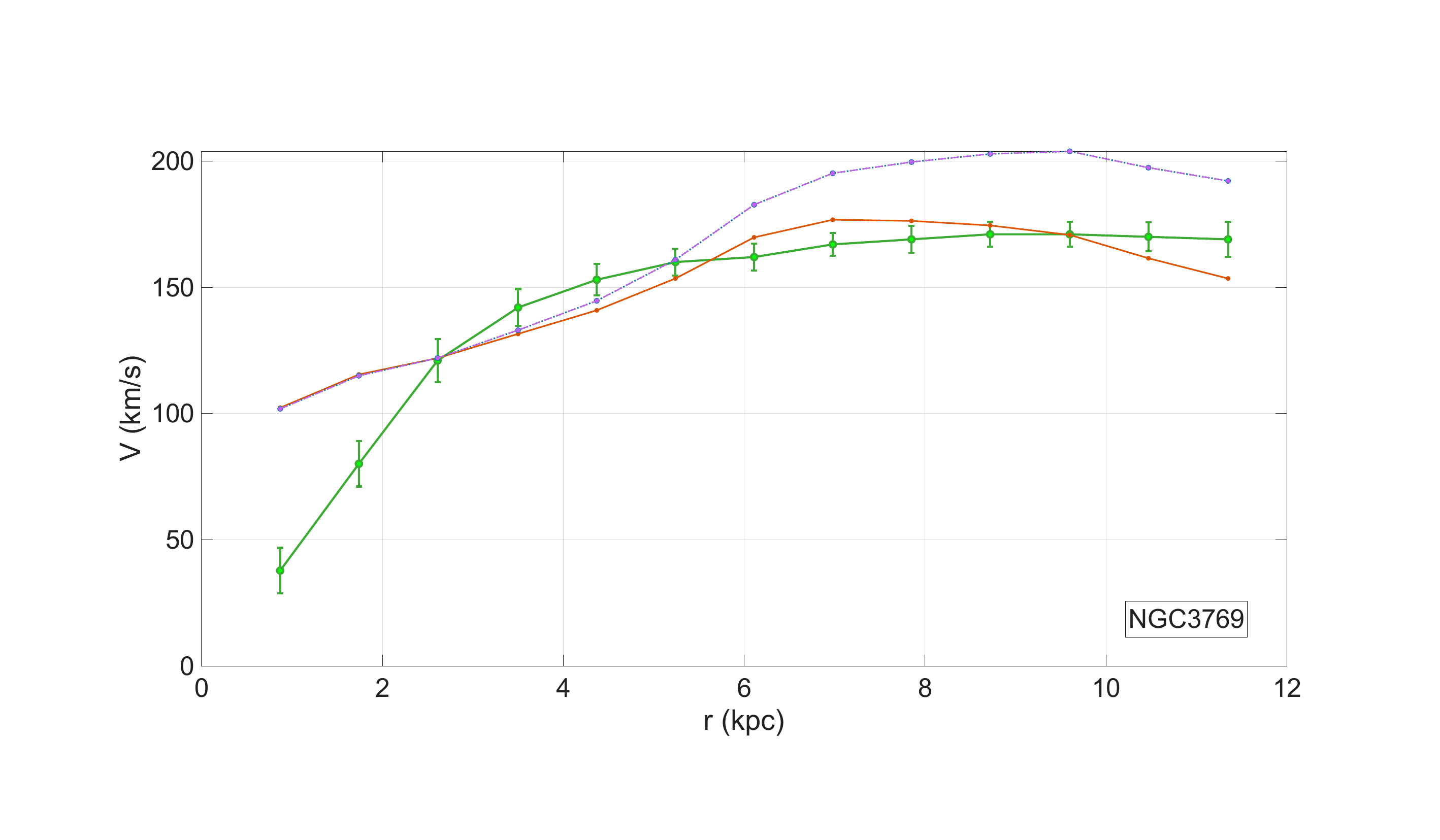}
\includegraphics[trim=4cm 3cm 5cm 4cm, clip=true, width=0.325\columnwidth]{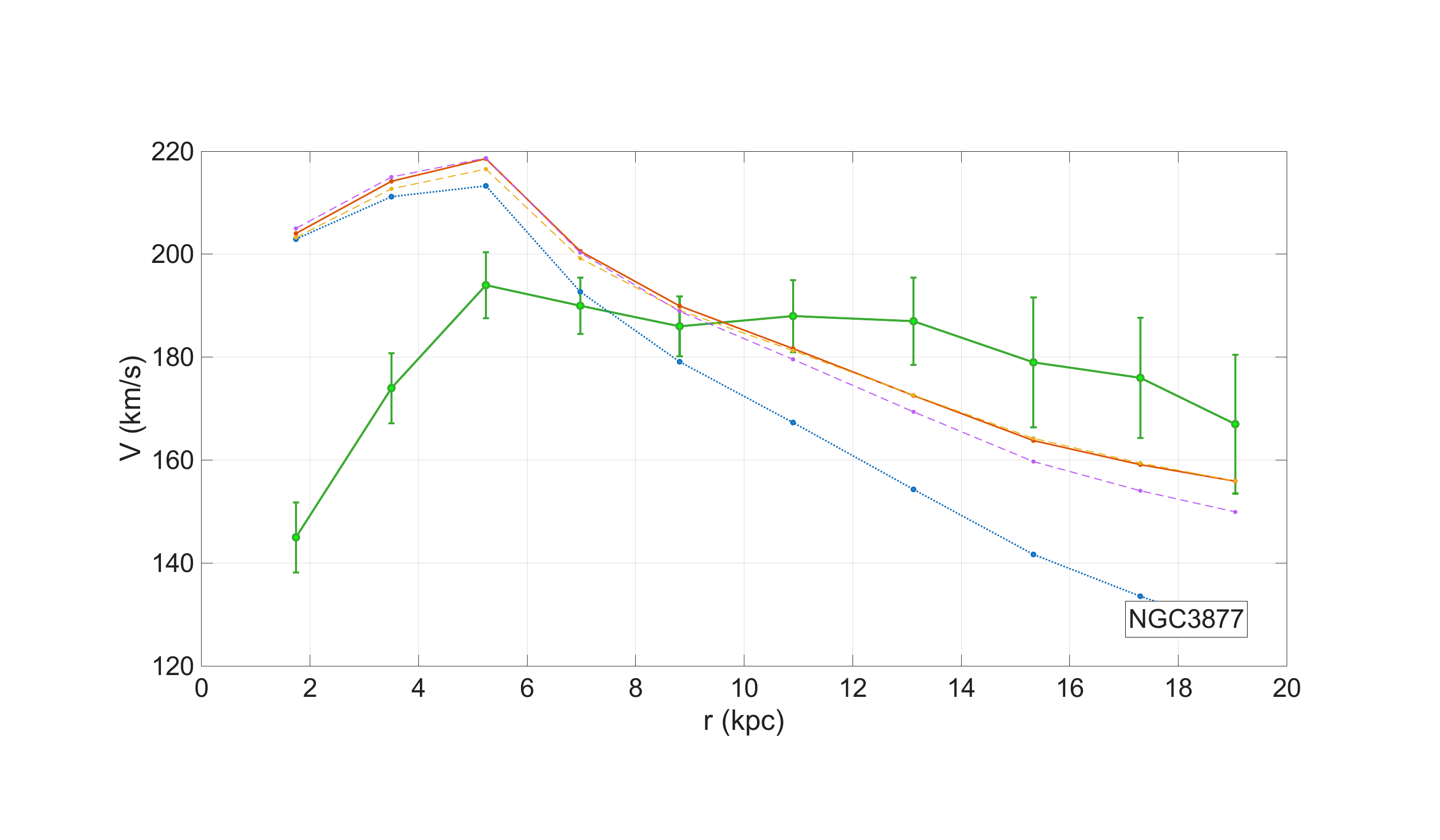}
\includegraphics[trim=4cm 3cm 5cm 4cm, clip=true, width=0.325\columnwidth]{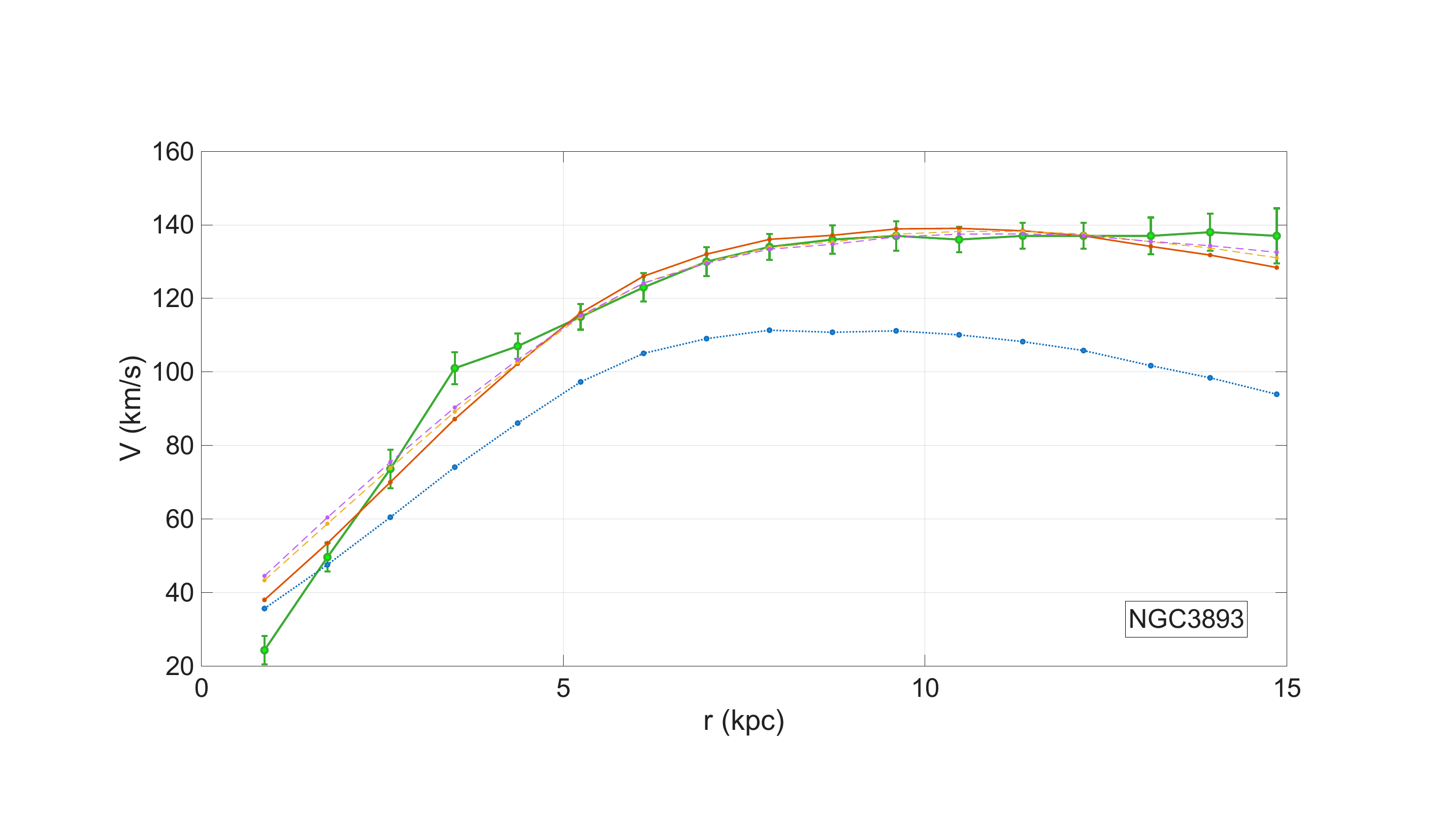}
\includegraphics[trim=4cm 3cm 5cm 4cm, clip=true, width=0.325\columnwidth]{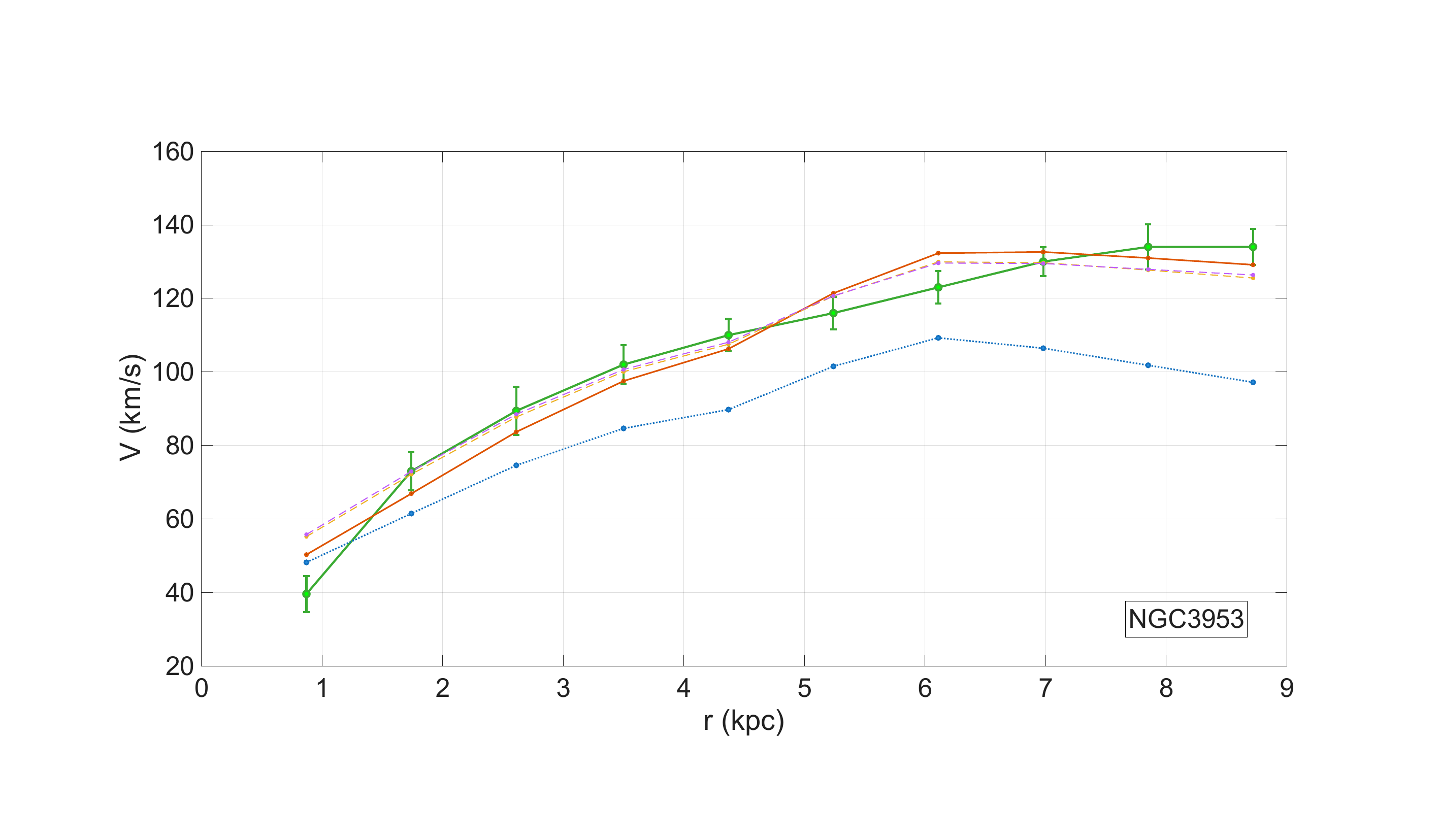}
\includegraphics[trim=4cm 3cm 5cm 4cm, clip=true, width=0.325\columnwidth]{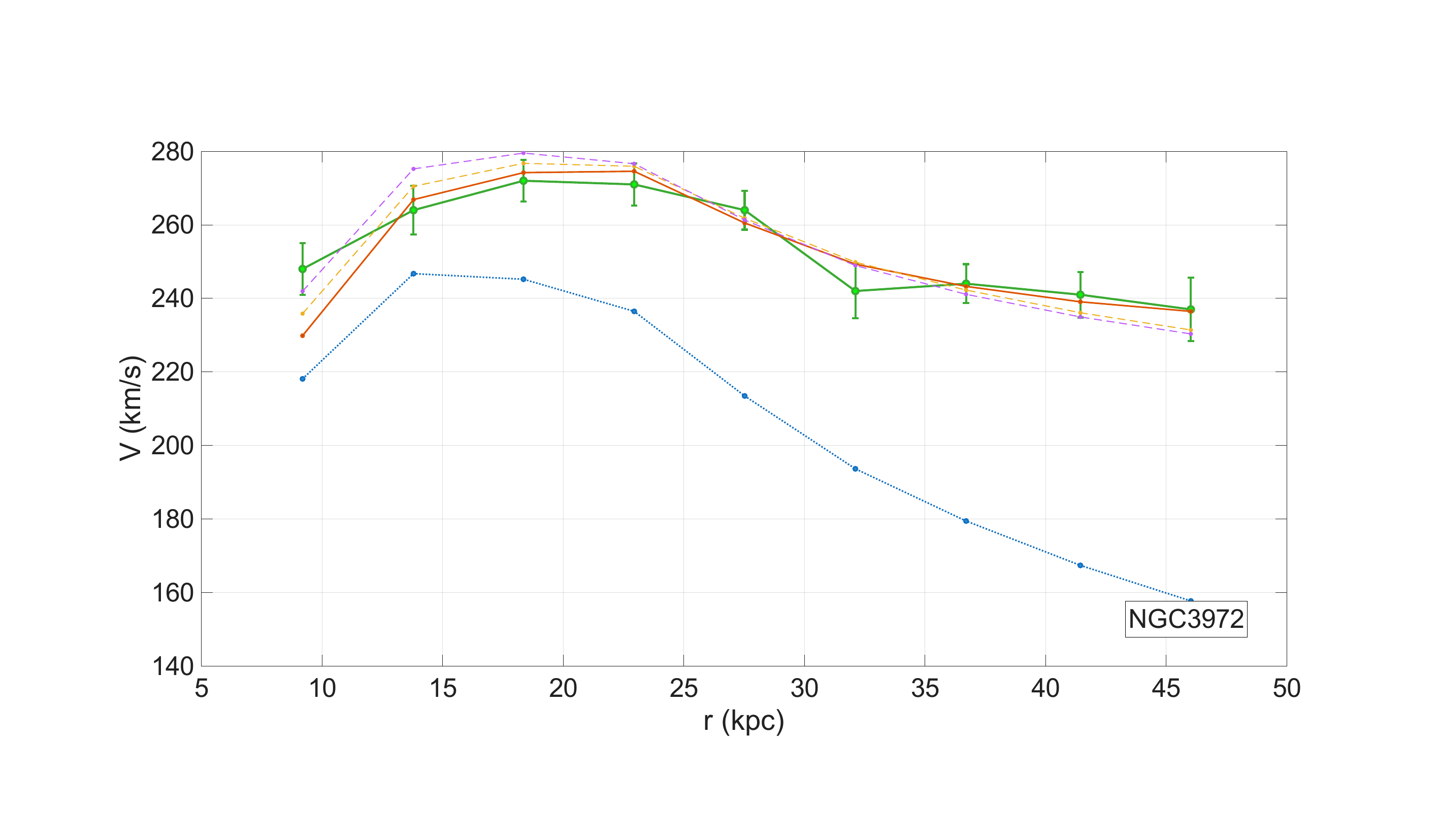}
\includegraphics[trim=4cm 3cm 5cm 4cm, clip=true, width=0.325\columnwidth]{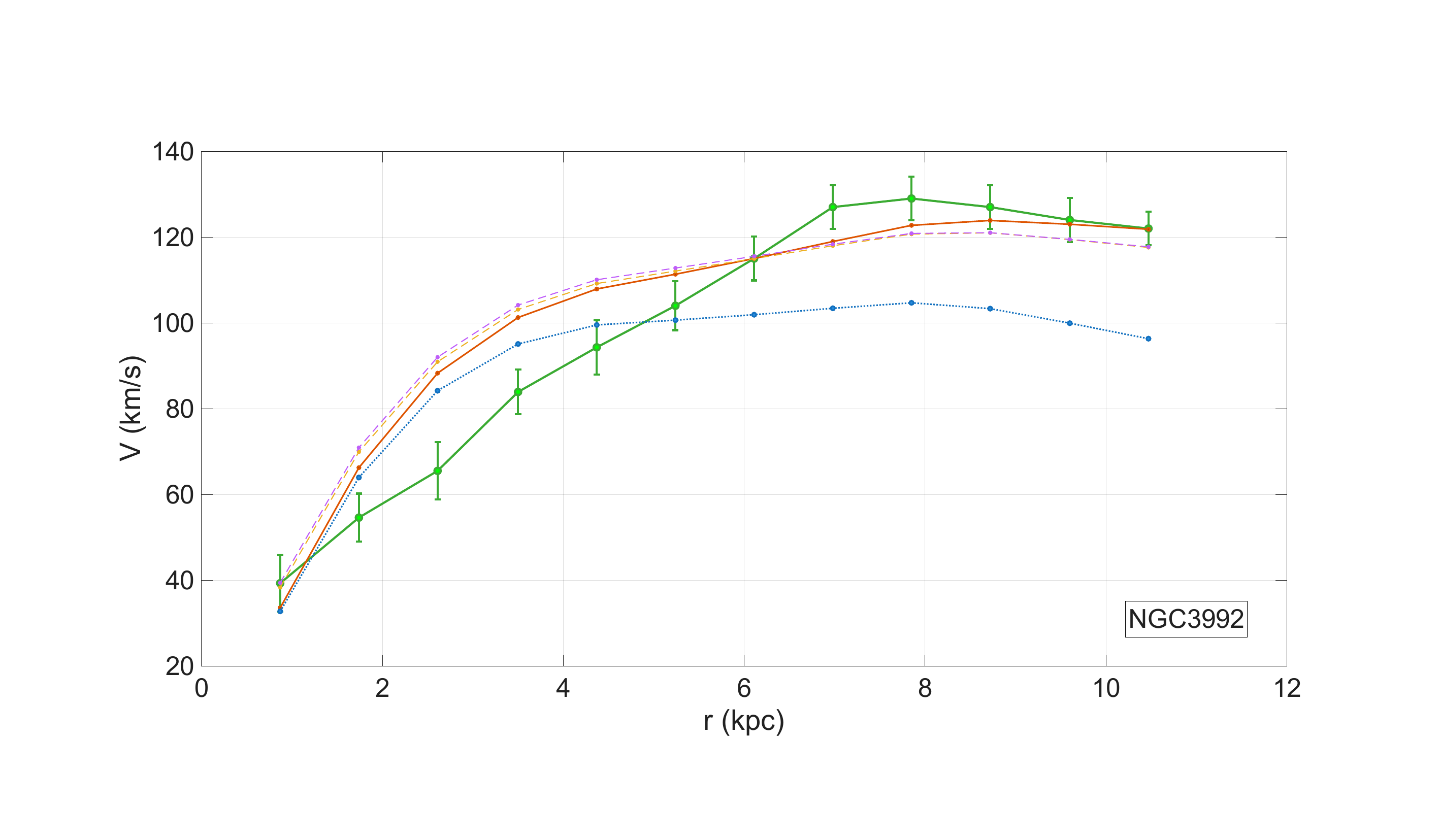}
\includegraphics[trim=4cm 3cm 5cm 4cm, clip=true, width=0.325\columnwidth]{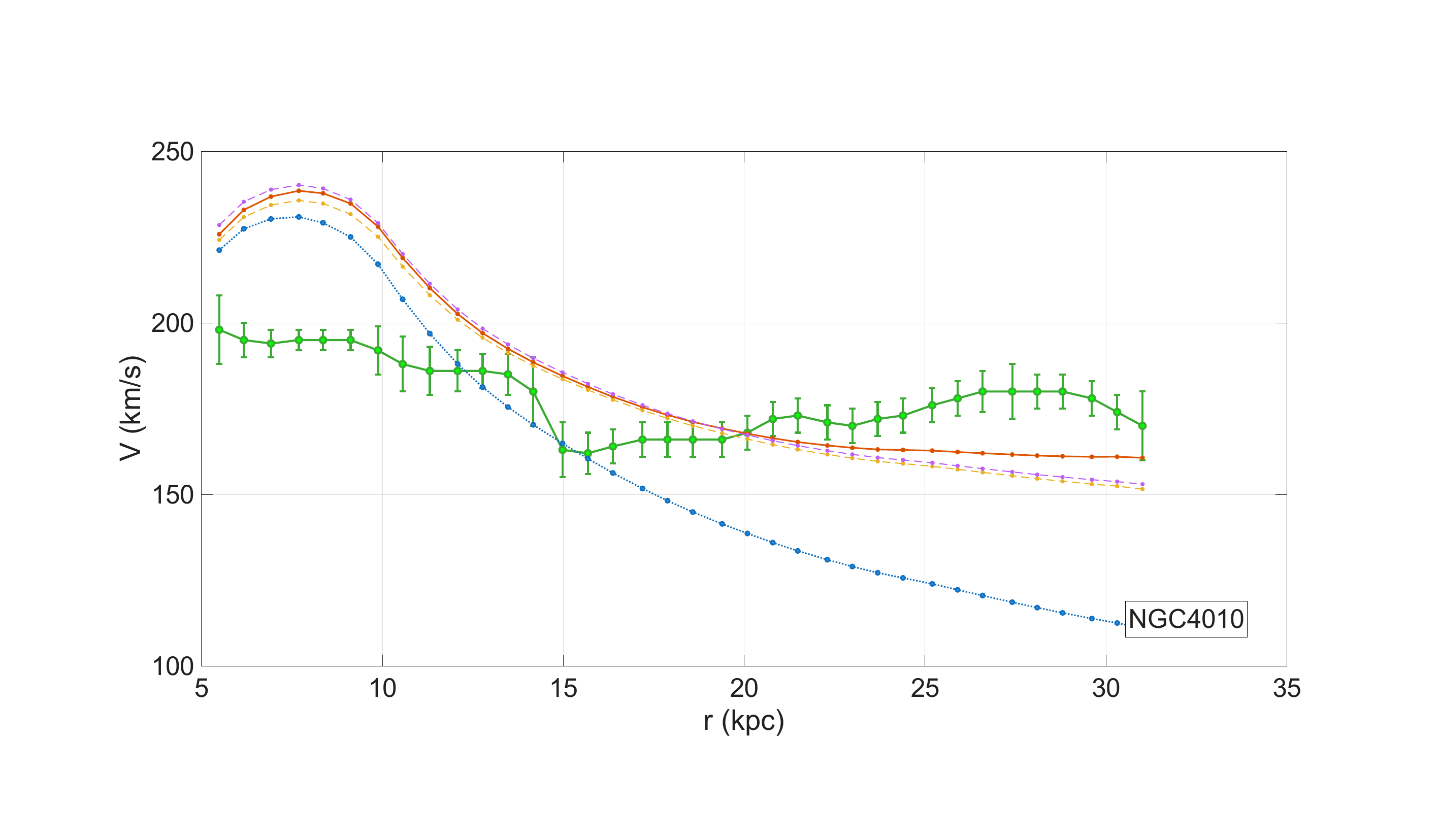}
\includegraphics[trim=4cm 3cm 5cm 4cm, clip=true, width=0.325\columnwidth]{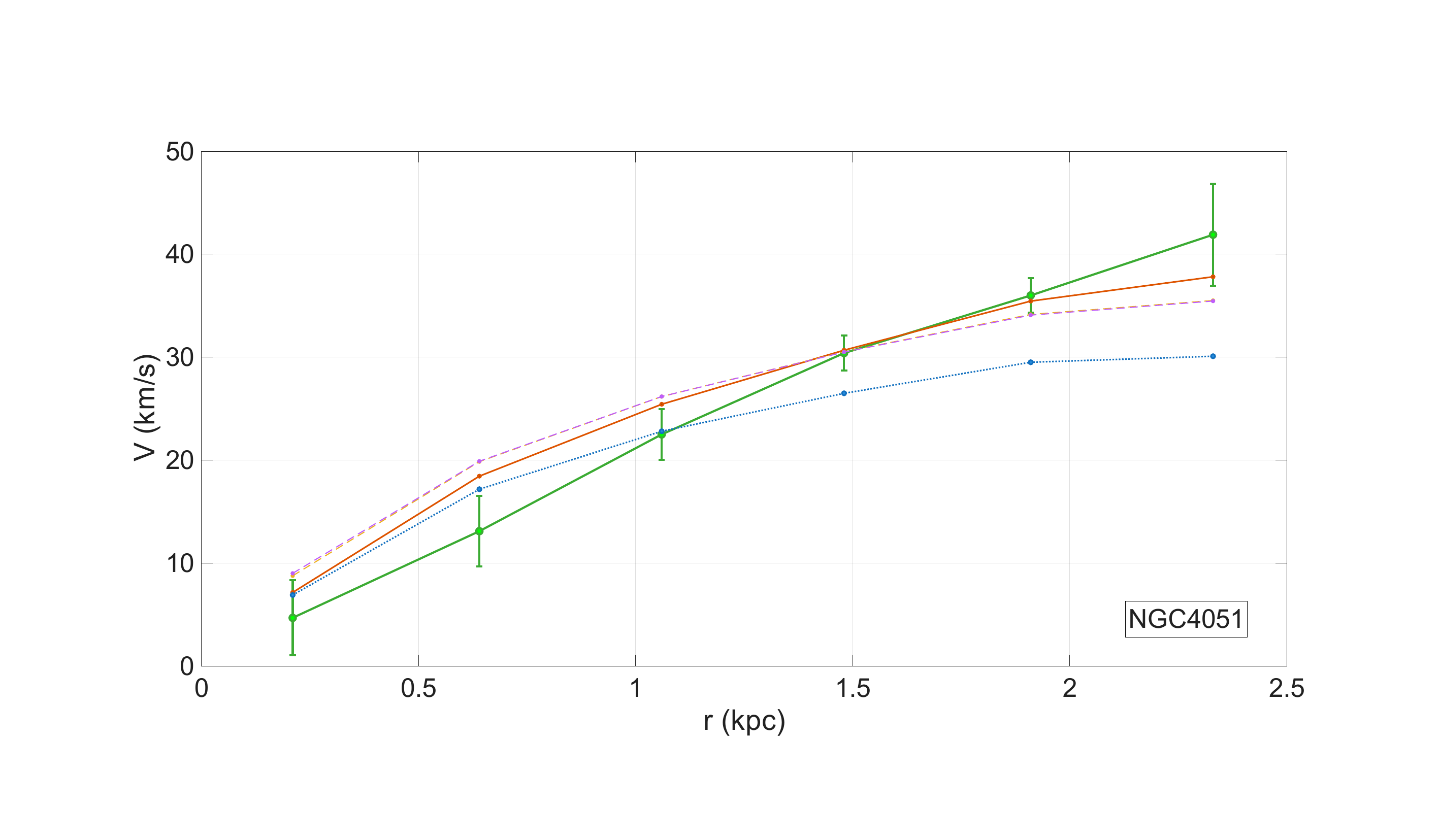}
\includegraphics[trim=4cm 3cm 5cm 4cm, clip=true, width=0.325\columnwidth]{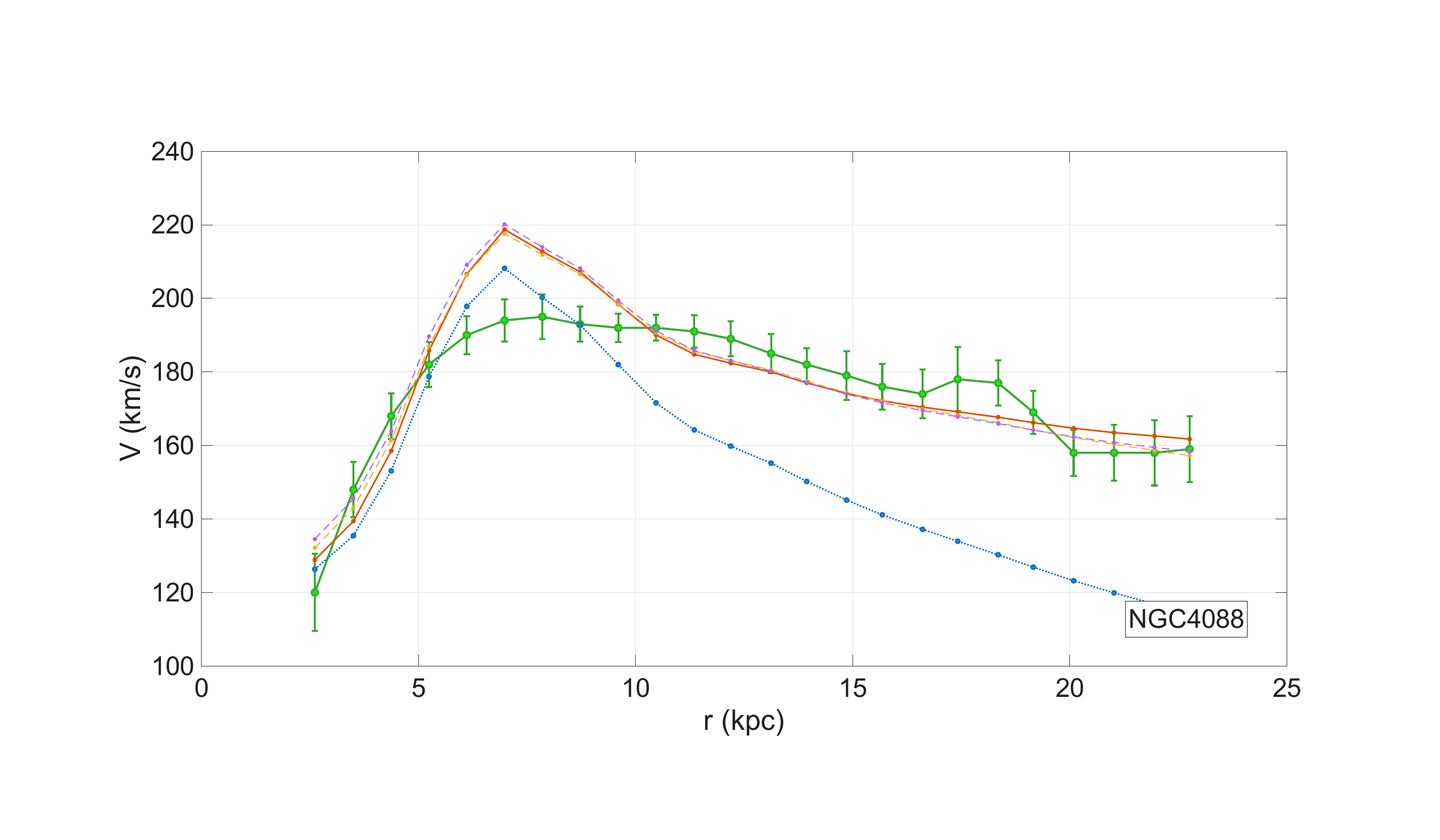}
\includegraphics[trim=4cm 3cm 5cm 4cm, clip=true, width=0.325\columnwidth]{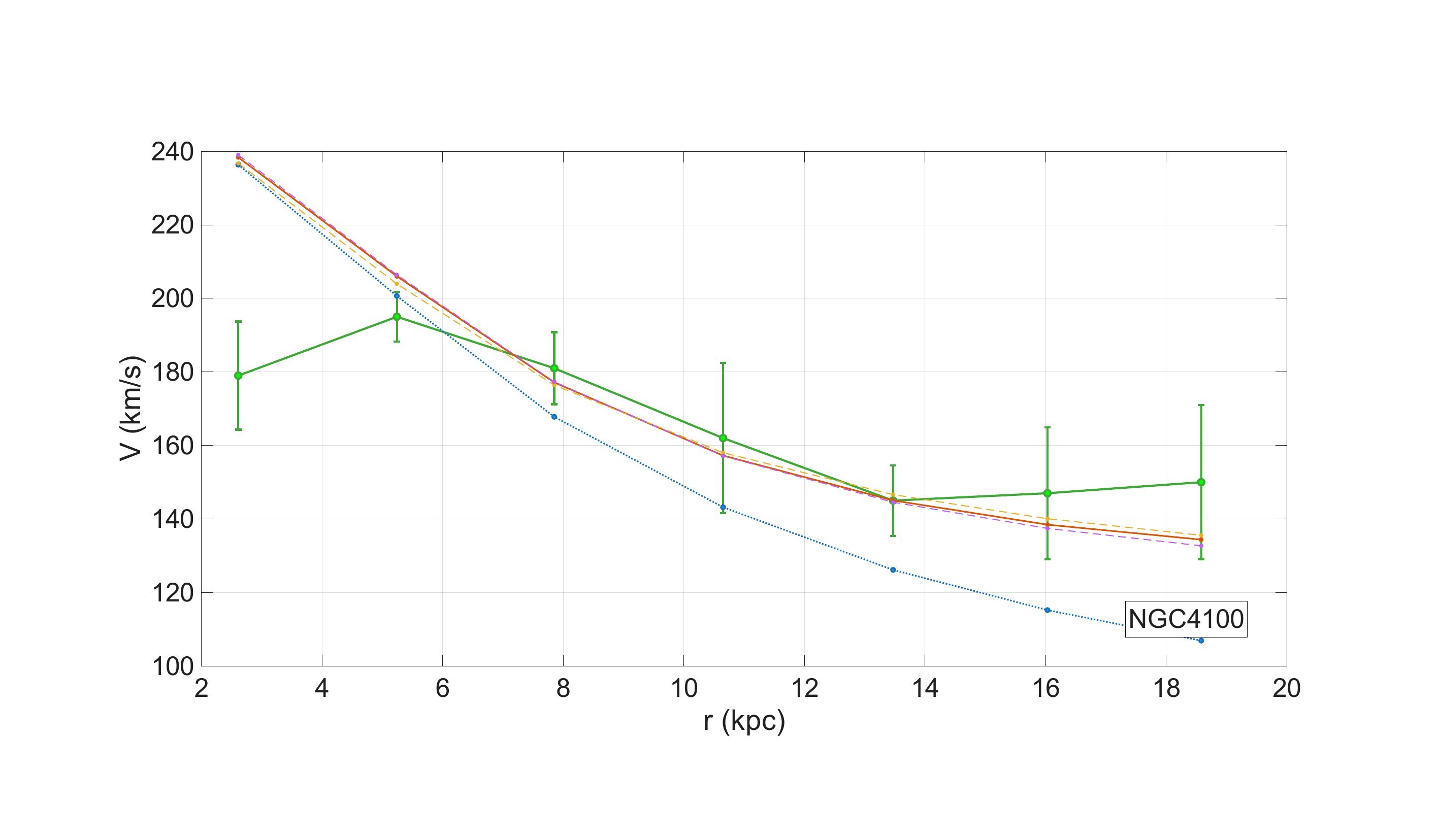}
\includegraphics[trim=4cm 3cm 5cm 4cm, clip=true, width=0.325\columnwidth]{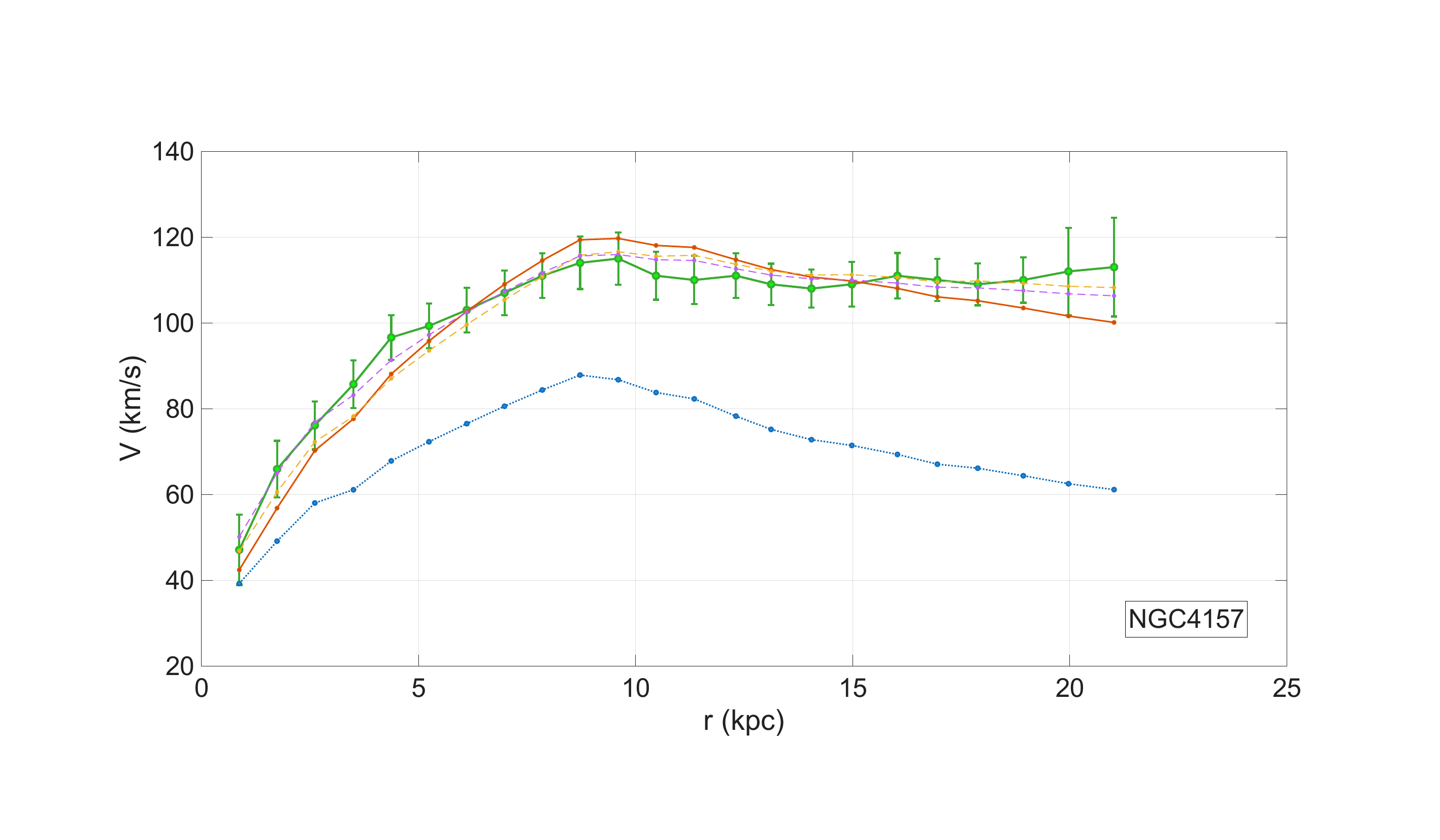}
\includegraphics[trim=4cm 3cm 5cm 4cm, clip=true, width=0.325\columnwidth]{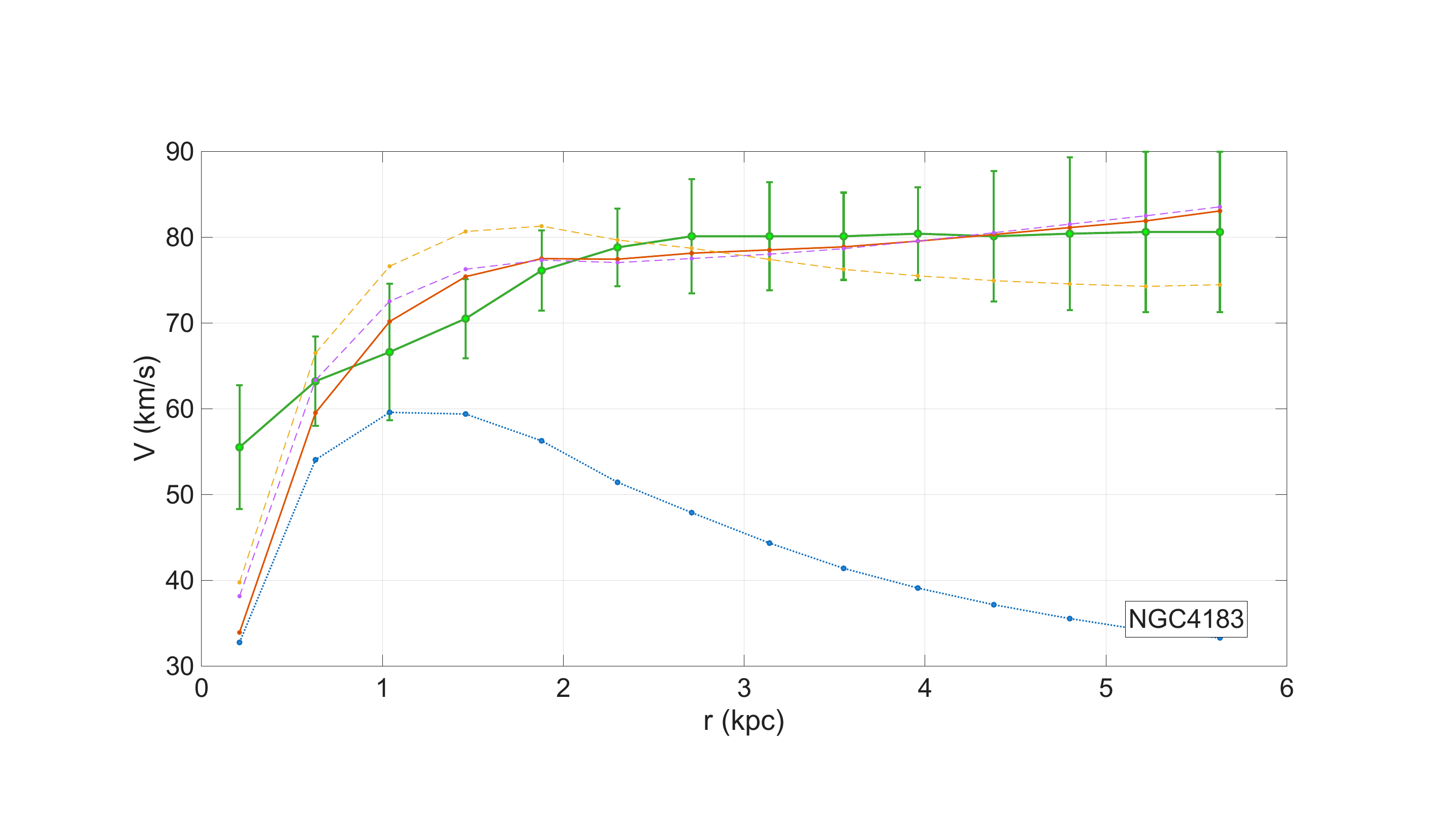}
\includegraphics[trim=4cm 3cm 5cm 4cm, clip=true, width=0.325\columnwidth]{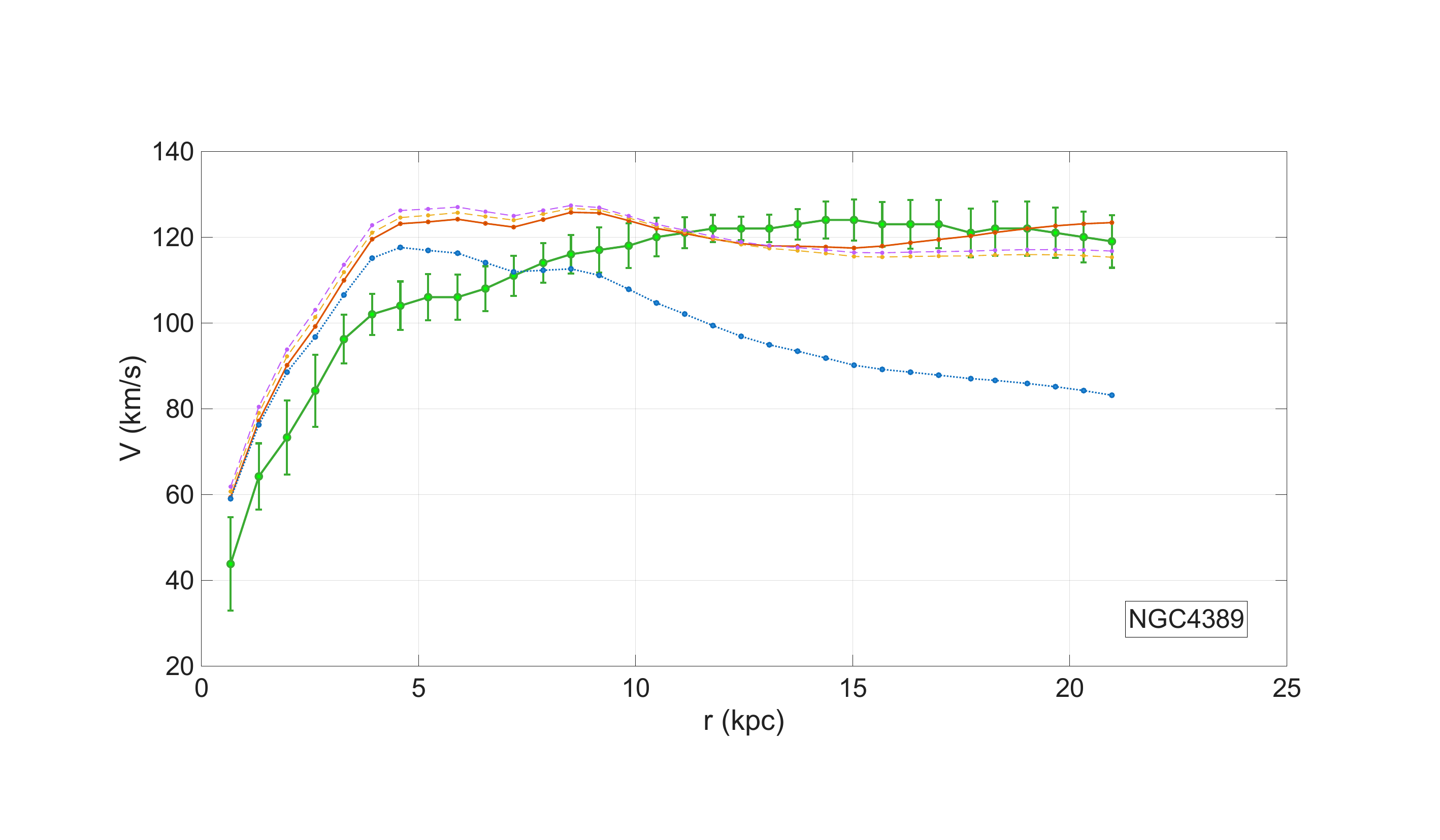}
\includegraphics[trim=4cm 3cm 5cm 4cm, clip=true, width=0.325\columnwidth]{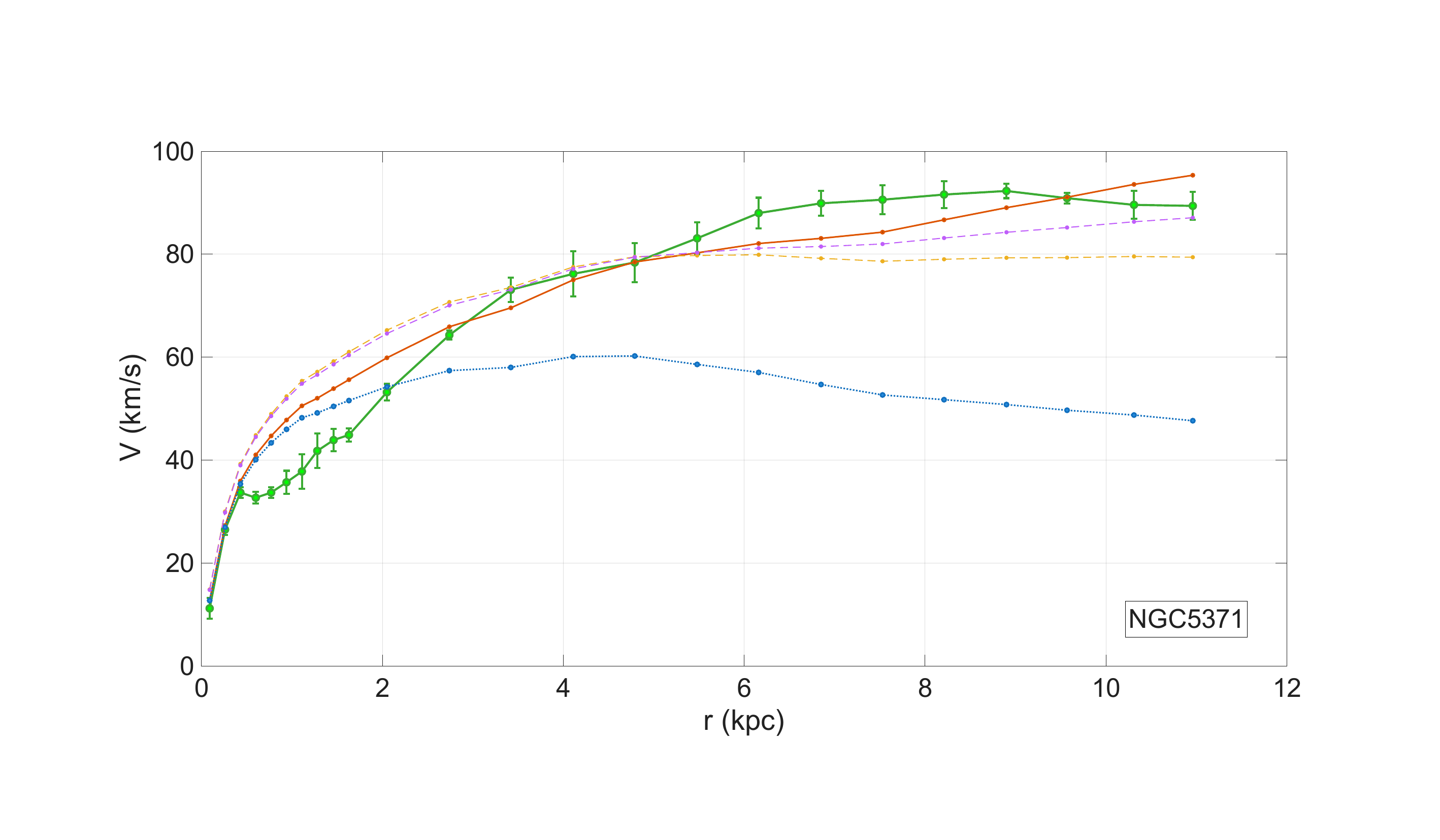}
\includegraphics[trim=4cm 3cm 5cm 4cm, clip=true, width=0.325\columnwidth]{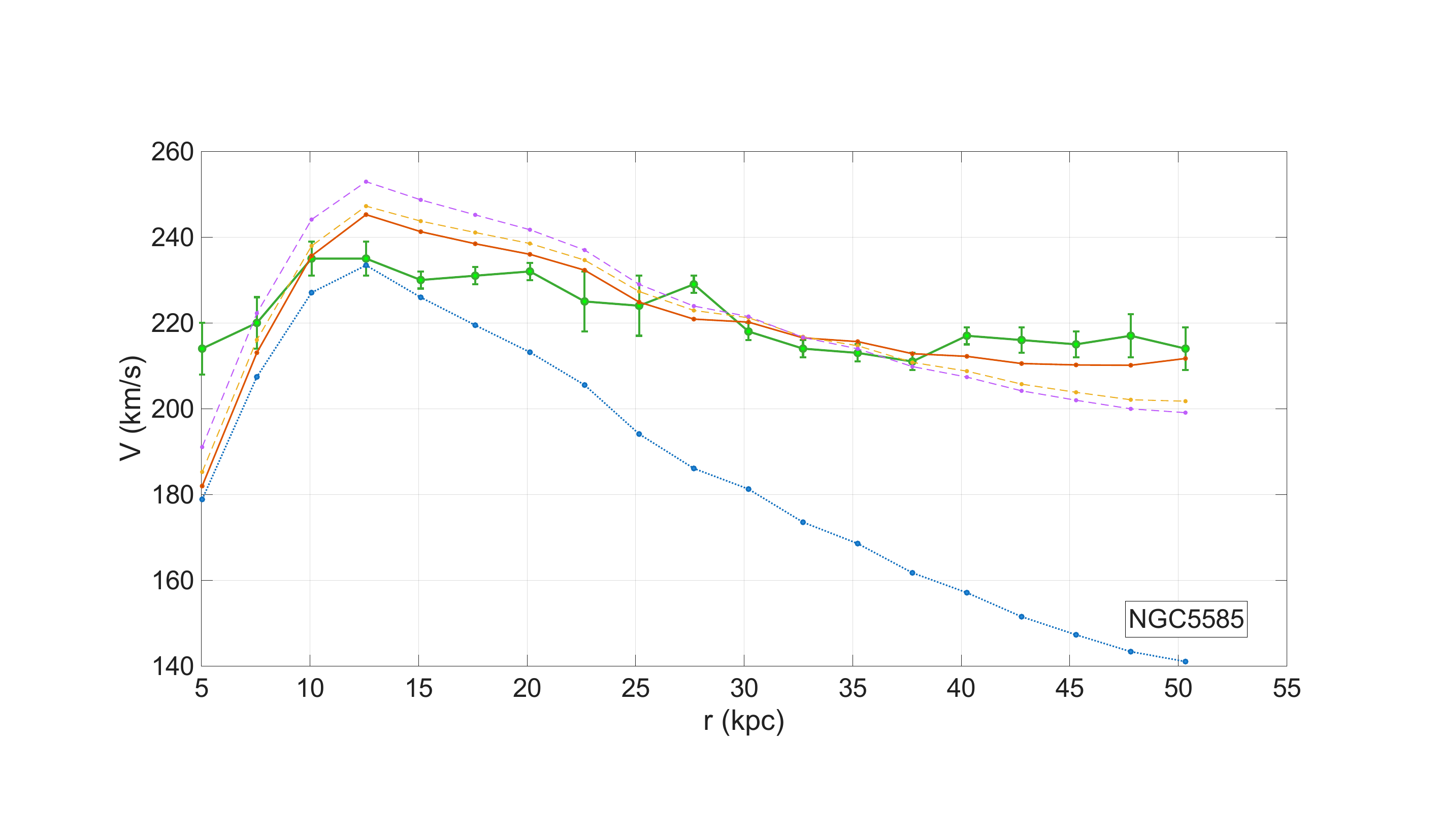}
\includegraphics[trim=4cm 3cm 5cm 4cm, clip=true, width=0.325\columnwidth]{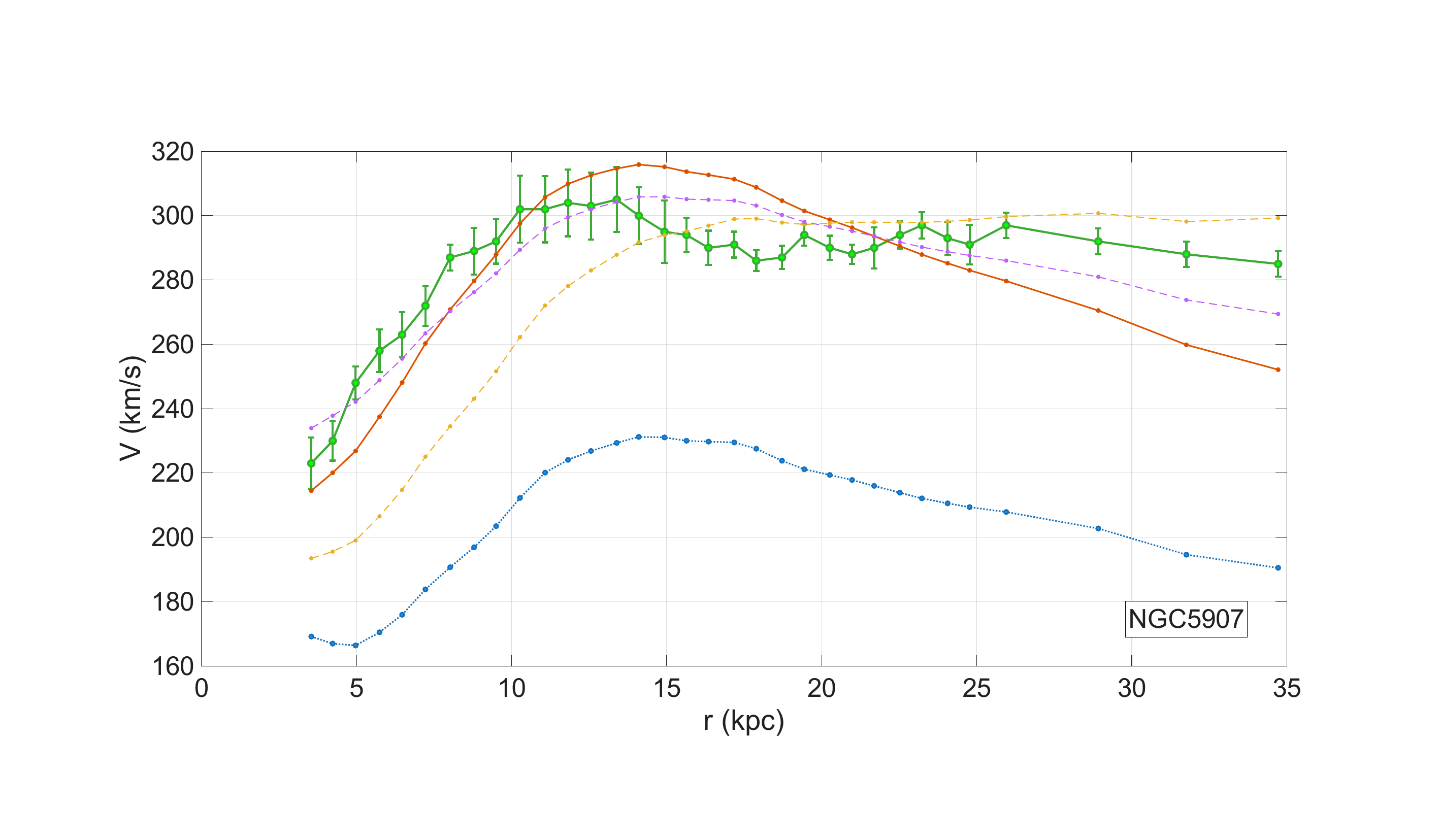}
\includegraphics[trim=4cm 3cm 5cm 4cm, clip=true, width=0.325\columnwidth]{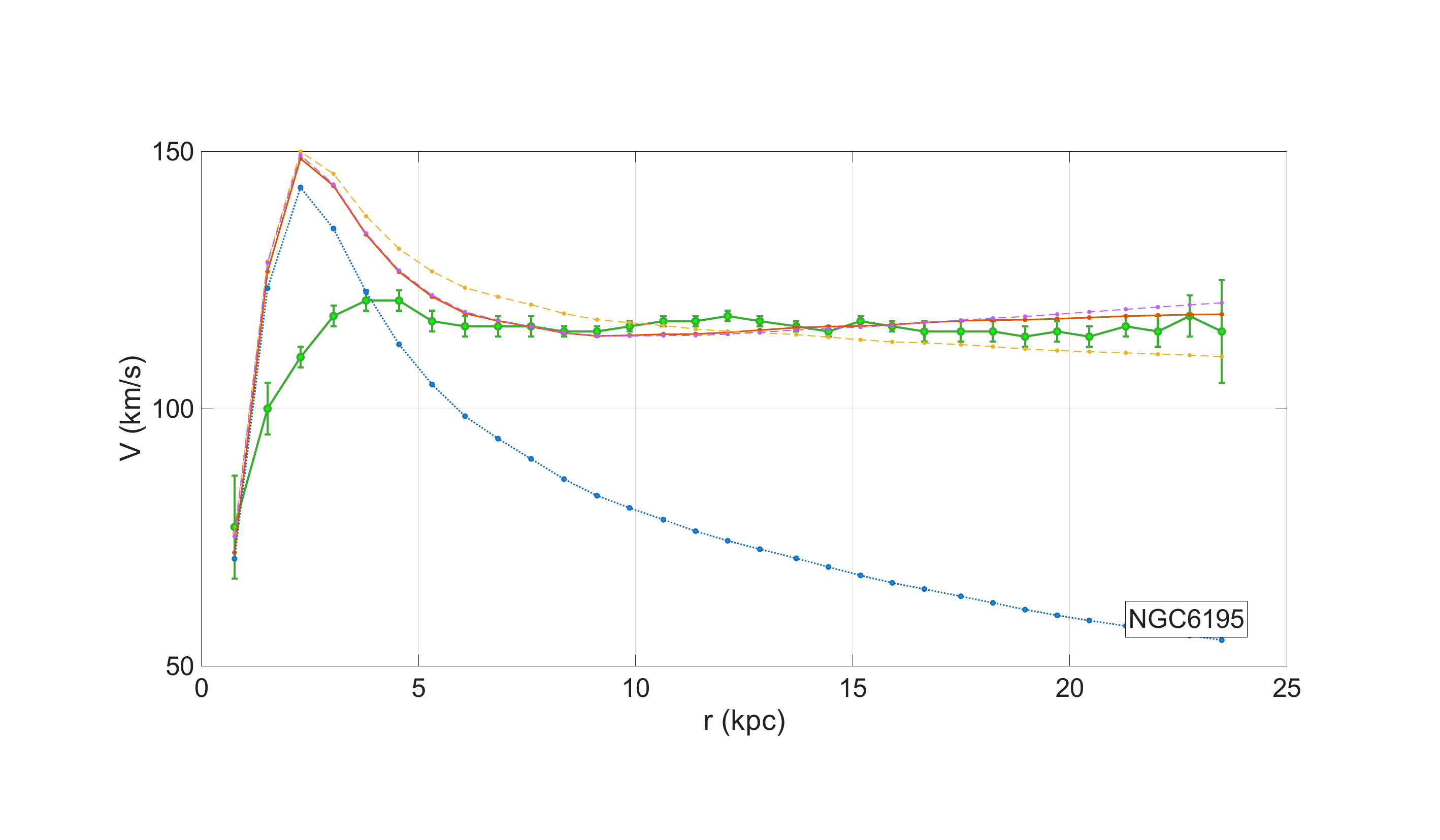}
\includegraphics[trim=4cm 3cm 5cm 4cm, clip=true, width=0.325\columnwidth]{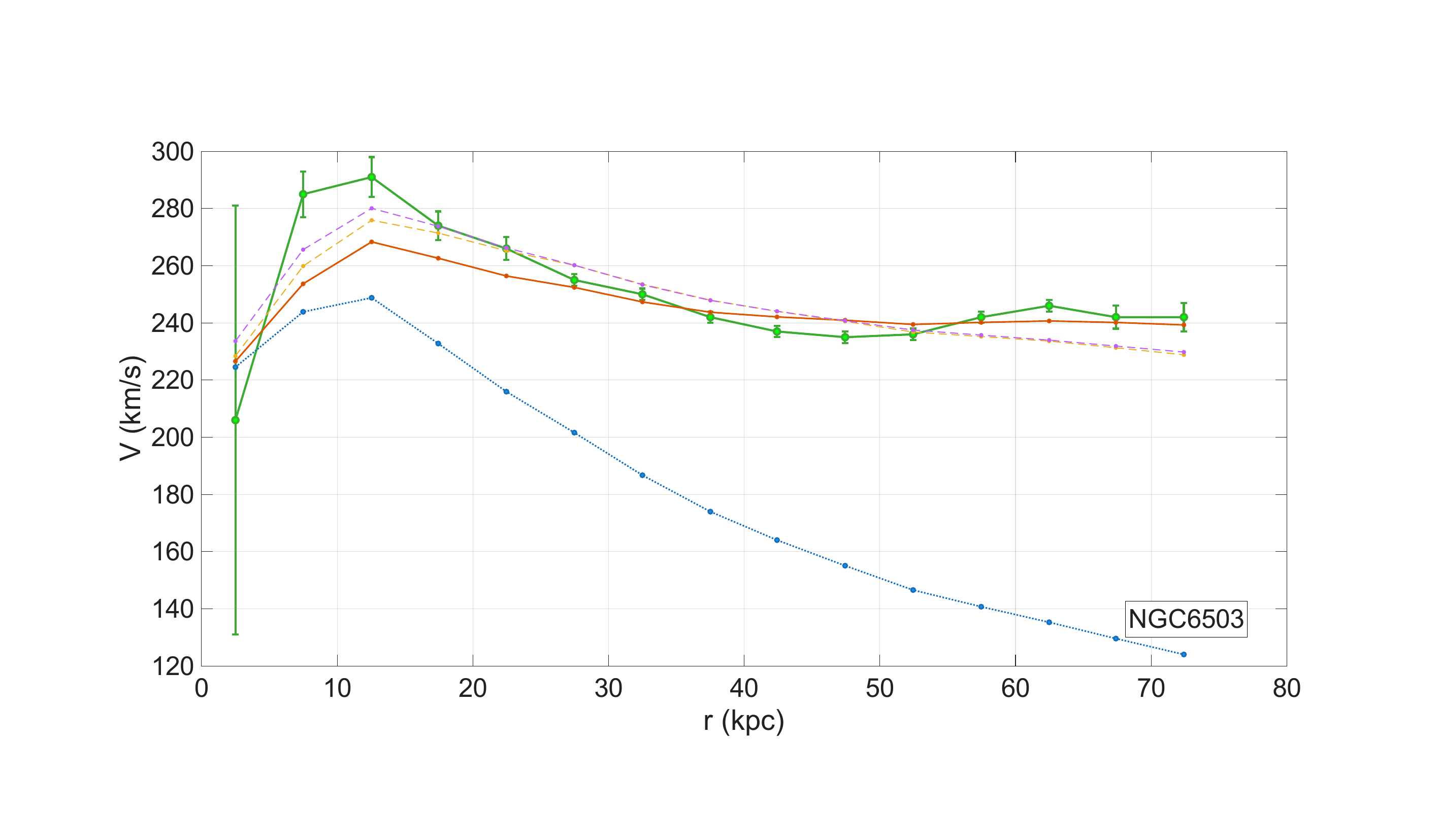}
\includegraphics[trim=4cm 3cm 5cm 4cm, clip=true, width=0.325\columnwidth]{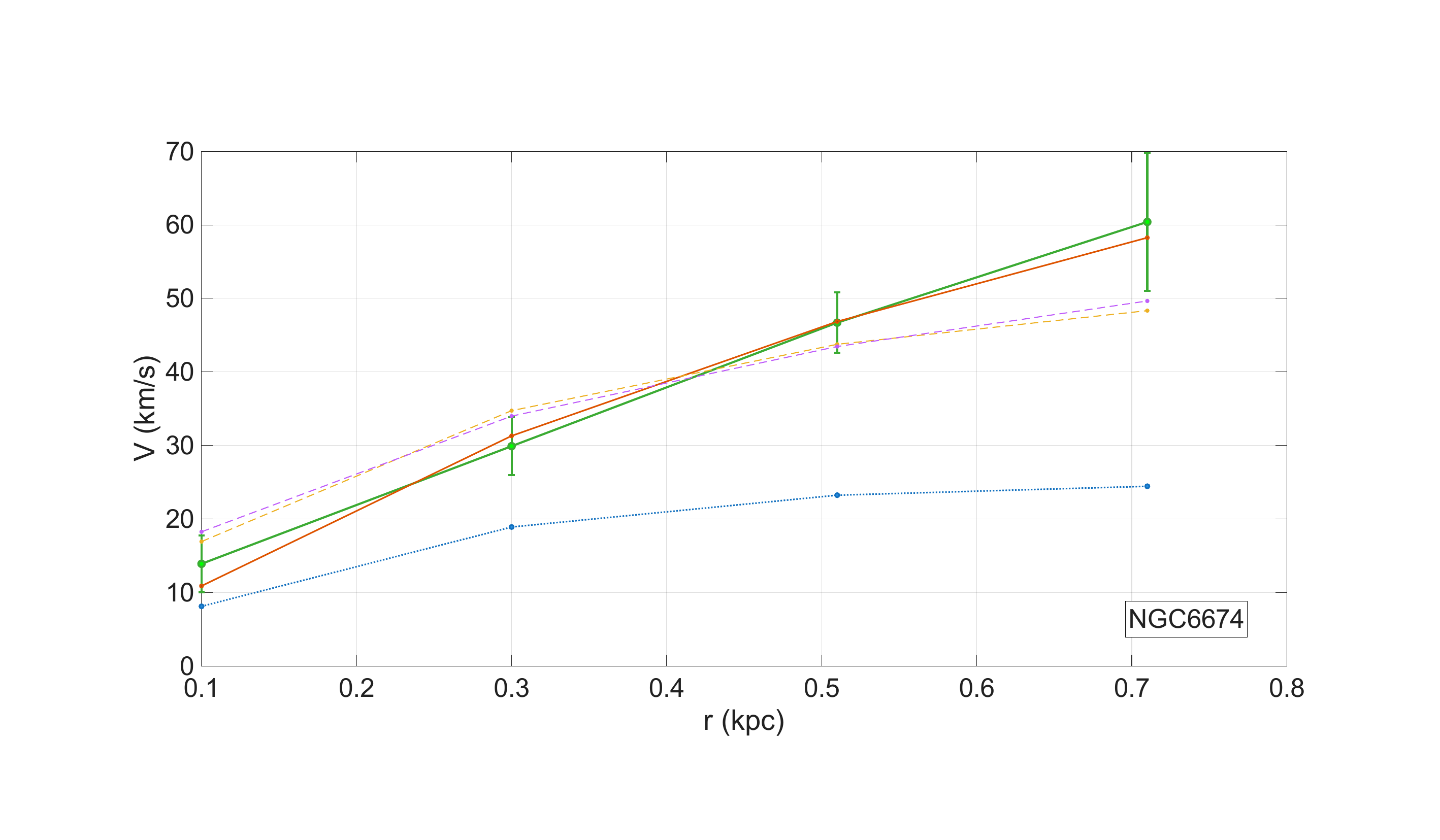}
\end{figure}

\begin{figure}
\centering
\includegraphics[trim=4cm 3cm 5cm 4cm, clip=true, width=0.325\columnwidth]{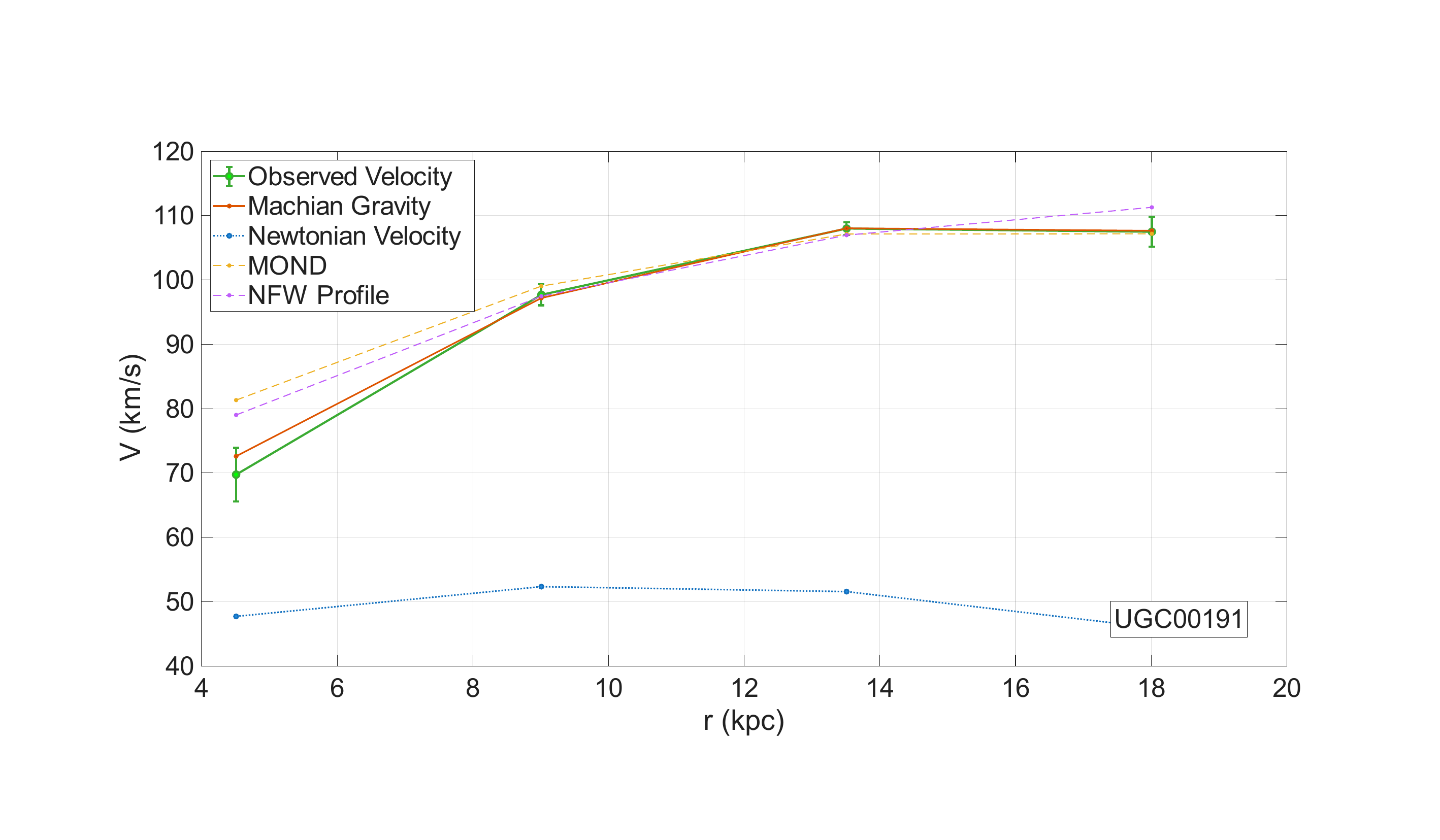}
\includegraphics[trim=4cm 3cm 5cm 4cm, clip=true, width=0.325\columnwidth]{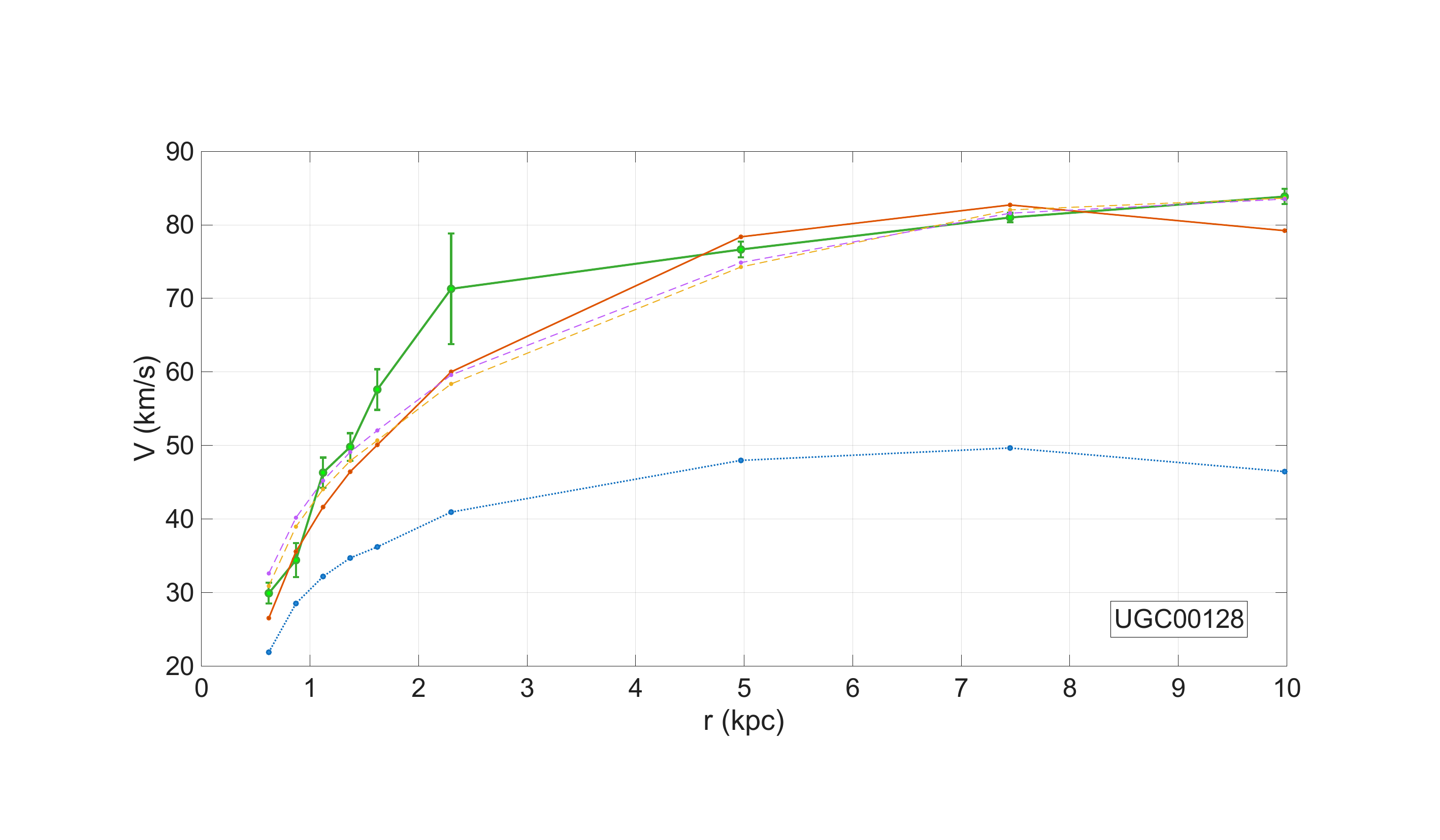}
s\includegraphics[trim=4cm 3cm 5cm 4cm, clip=true, width=0.325\columnwidth]{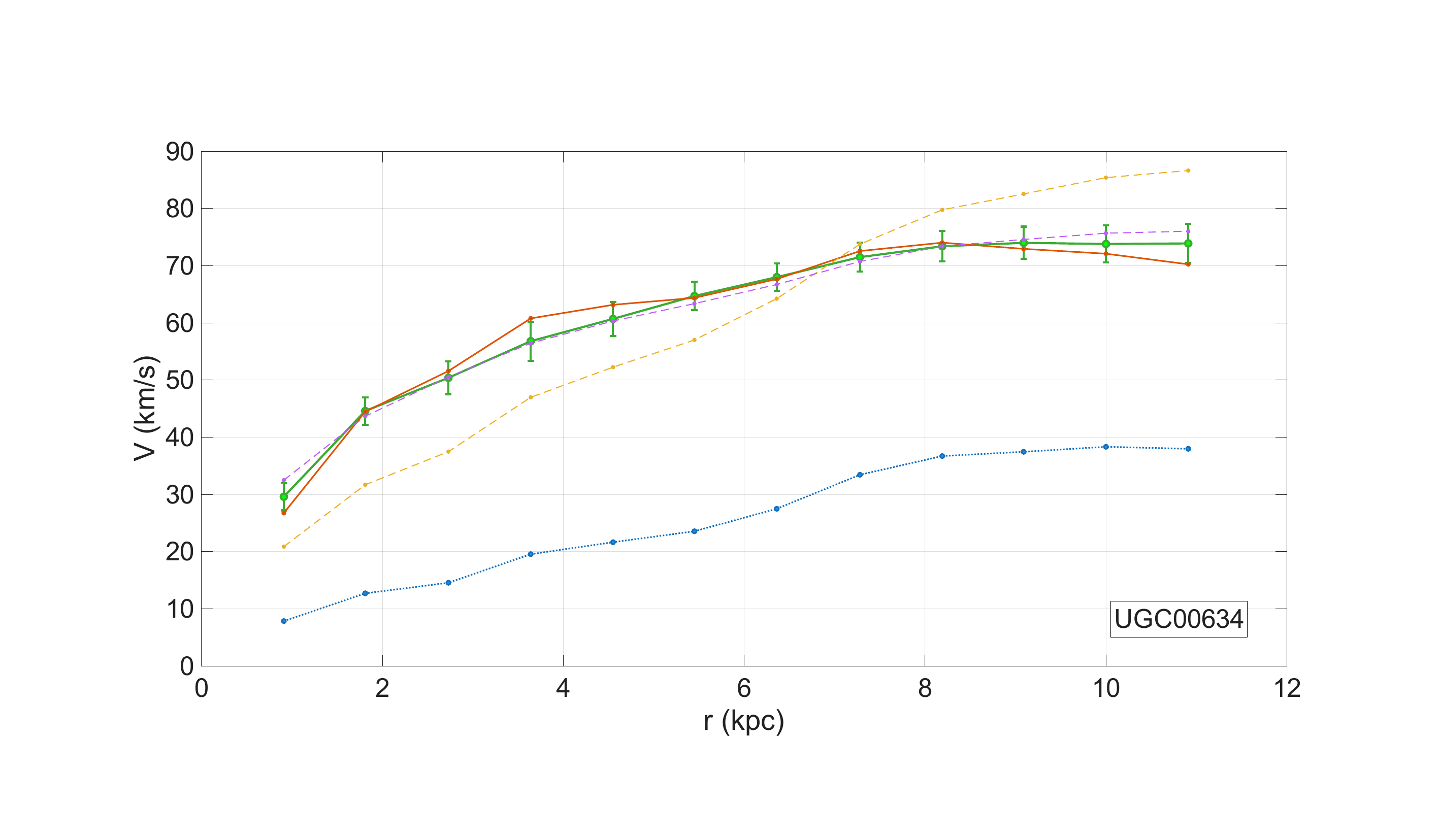}
\includegraphics[trim=4cm 3cm 5cm 4cm, clip=true, width=0.325\columnwidth]{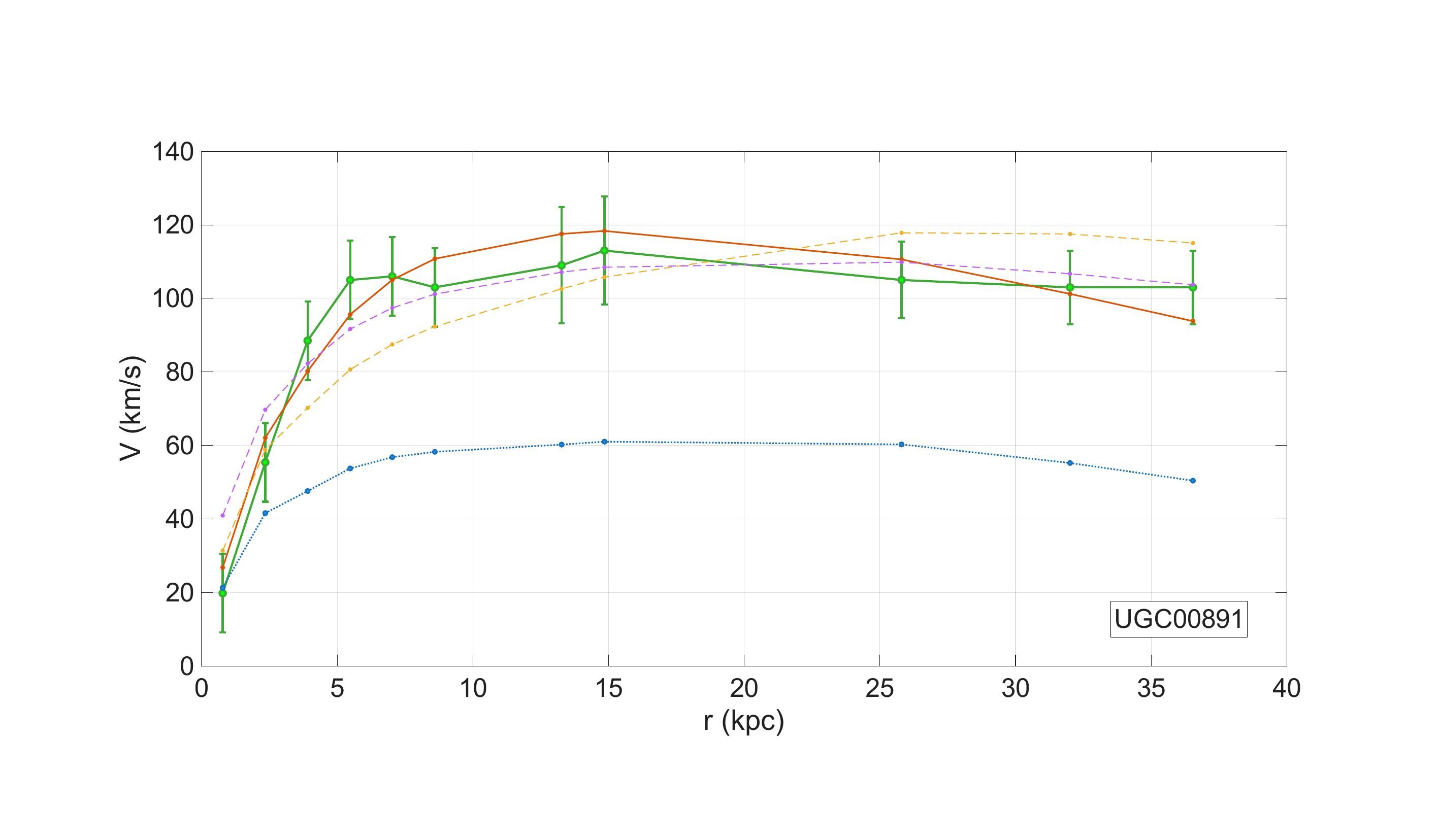}
\includegraphics[trim=4cm 3cm 5cm 4cm, clip=true, width=0.325\columnwidth]{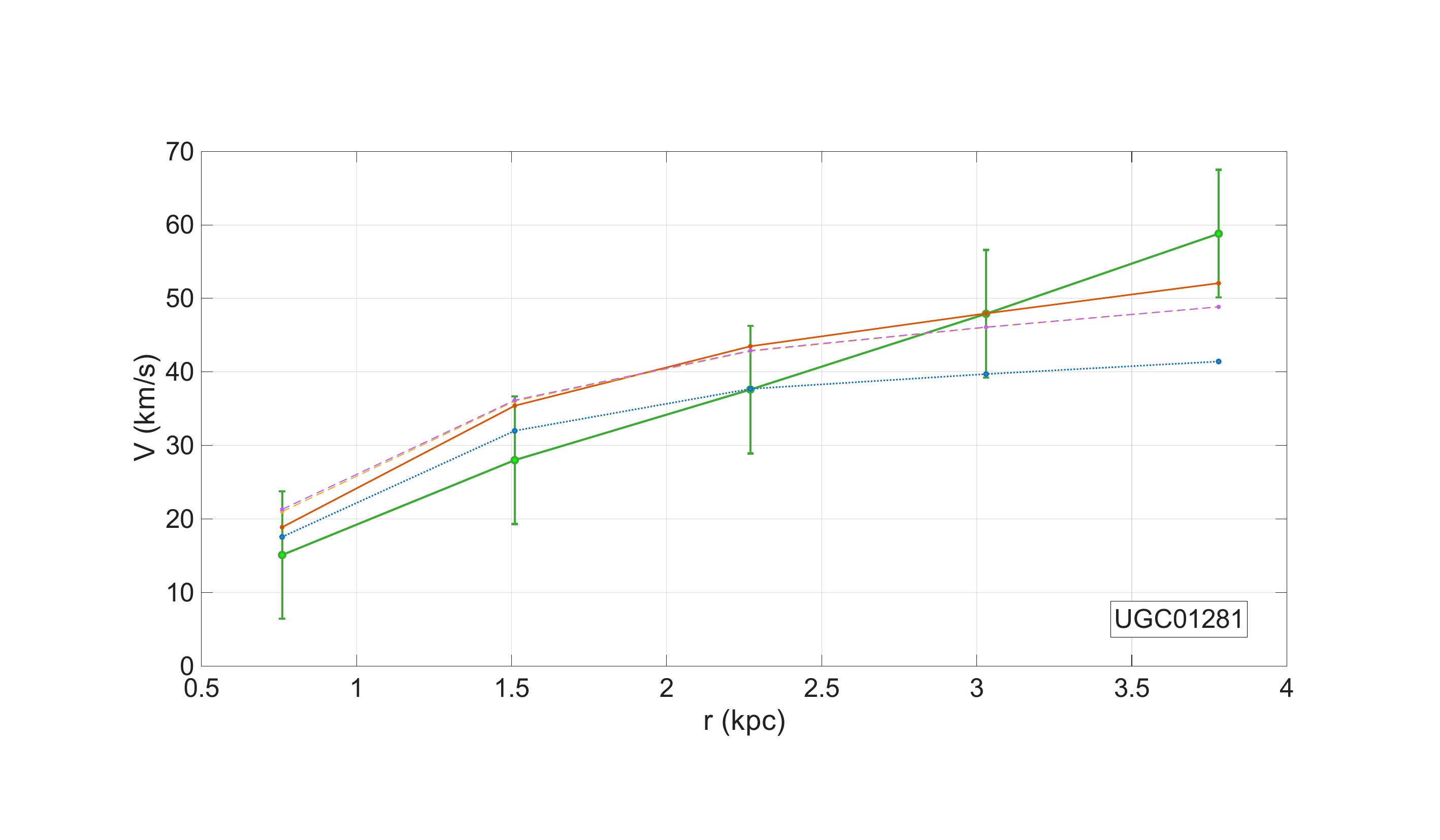}
\includegraphics[trim=4cm 3cm 5cm 4cm, clip=true, width=0.325\columnwidth]{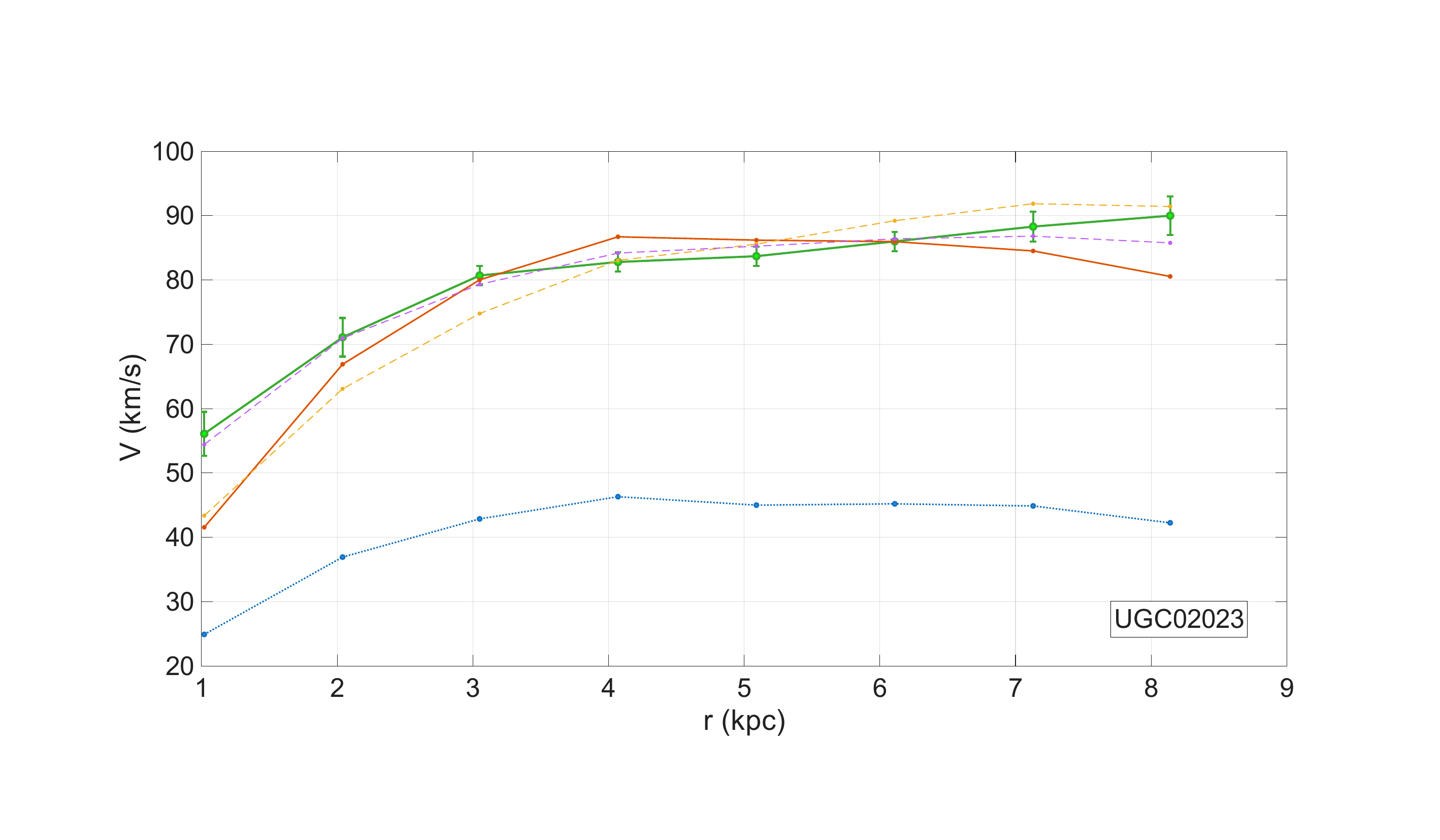}
\includegraphics[trim=4cm 3cm 5cm 4cm, clip=true, width=0.325\columnwidth]{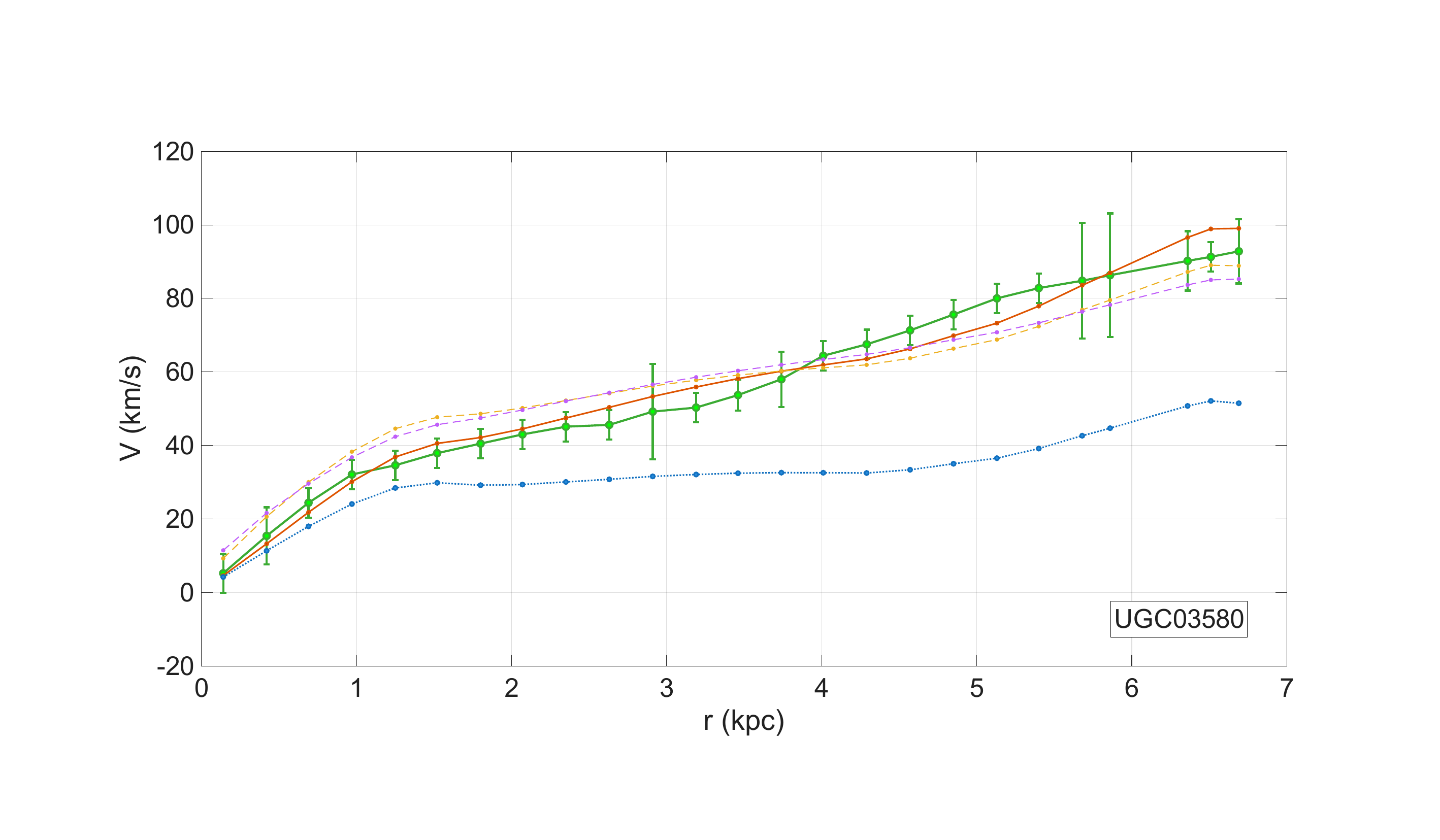}
\includegraphics[trim=4cm 3cm 5cm 4cm, clip=true, width=0.325\columnwidth]{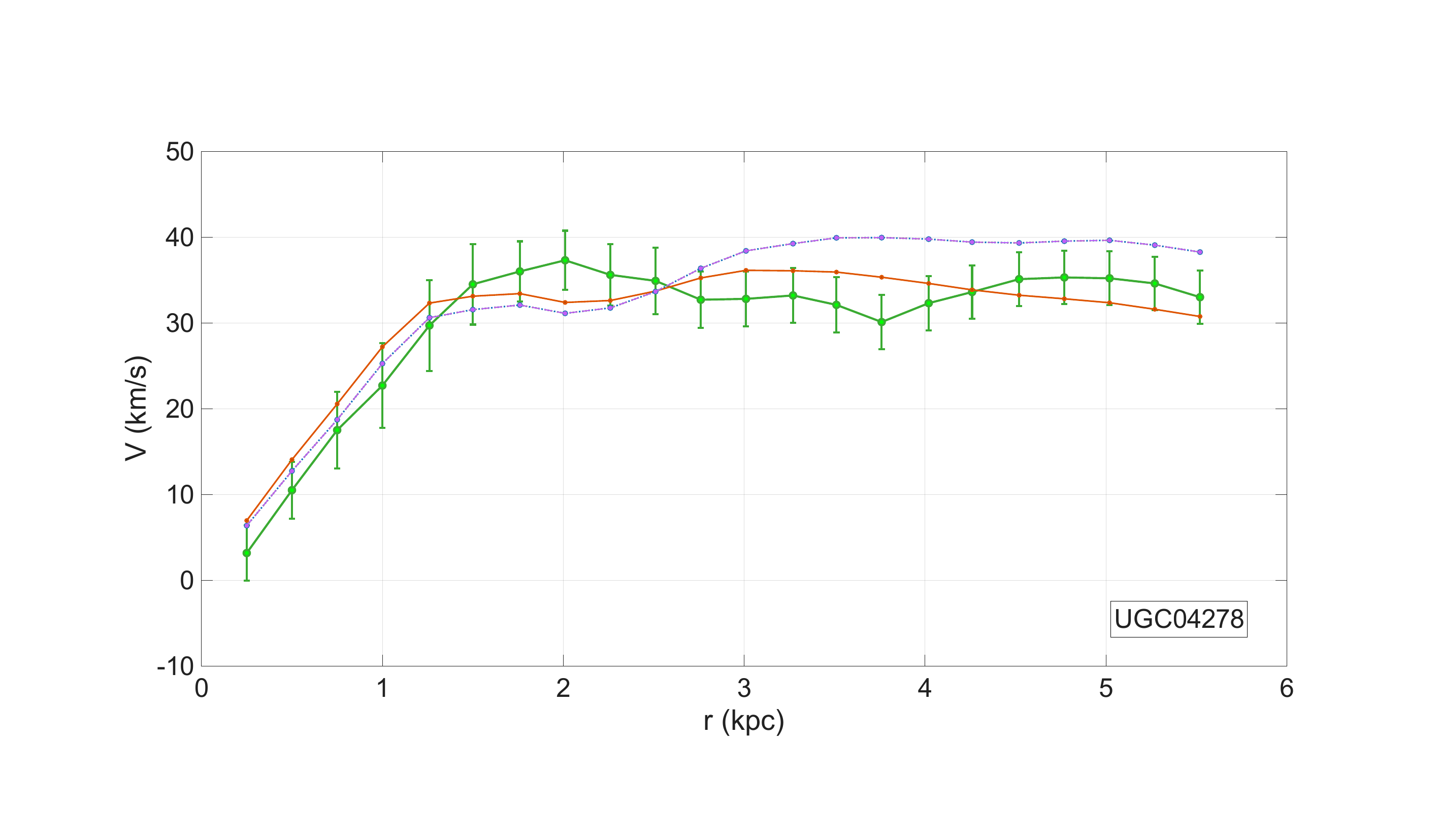}
\includegraphics[trim=4cm 3cm 5cm 4cm, clip=true, width=0.325\columnwidth]{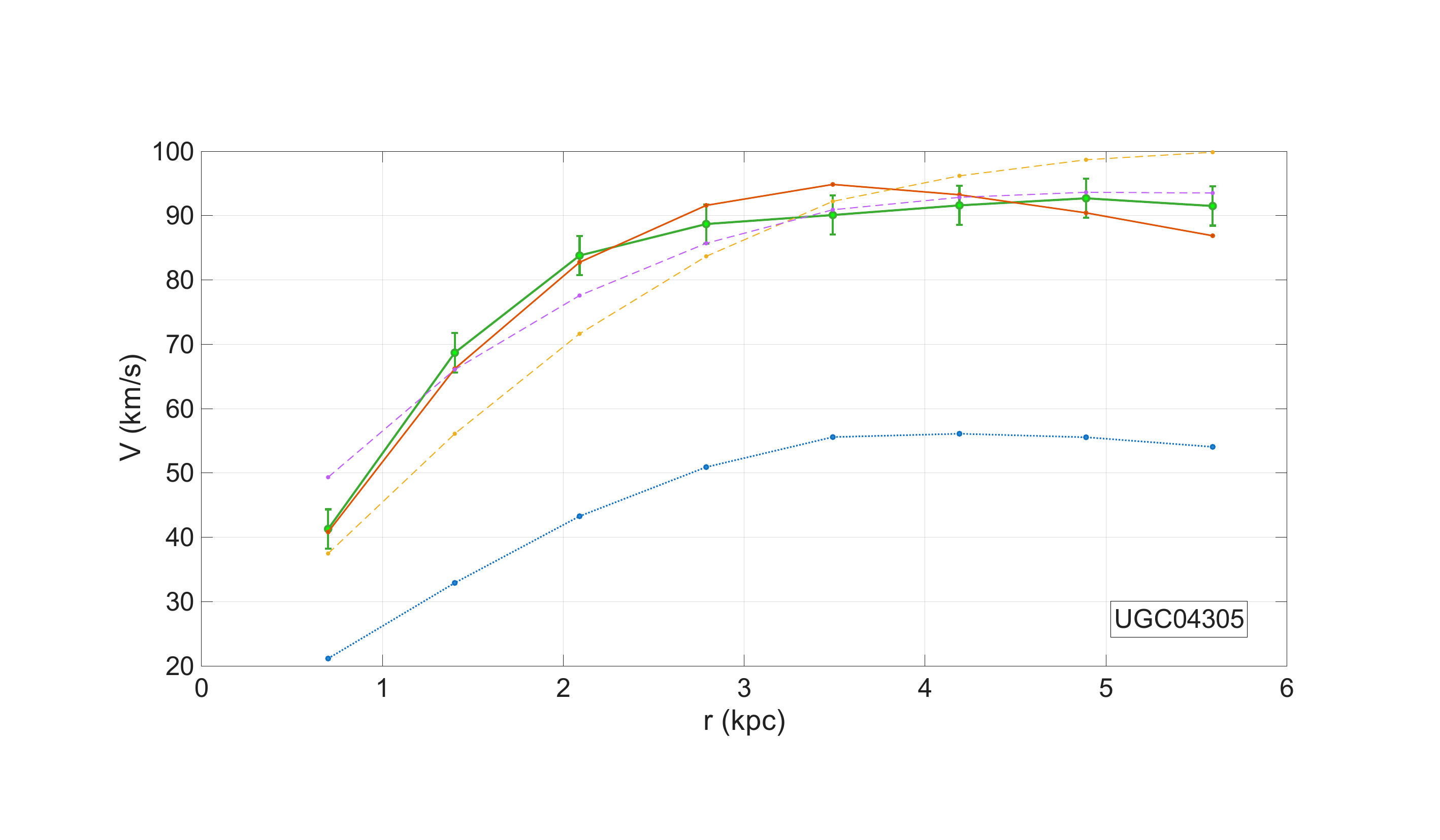}
\includegraphics[trim=4cm 3cm 5cm 4cm, clip=true, width=0.325\columnwidth]{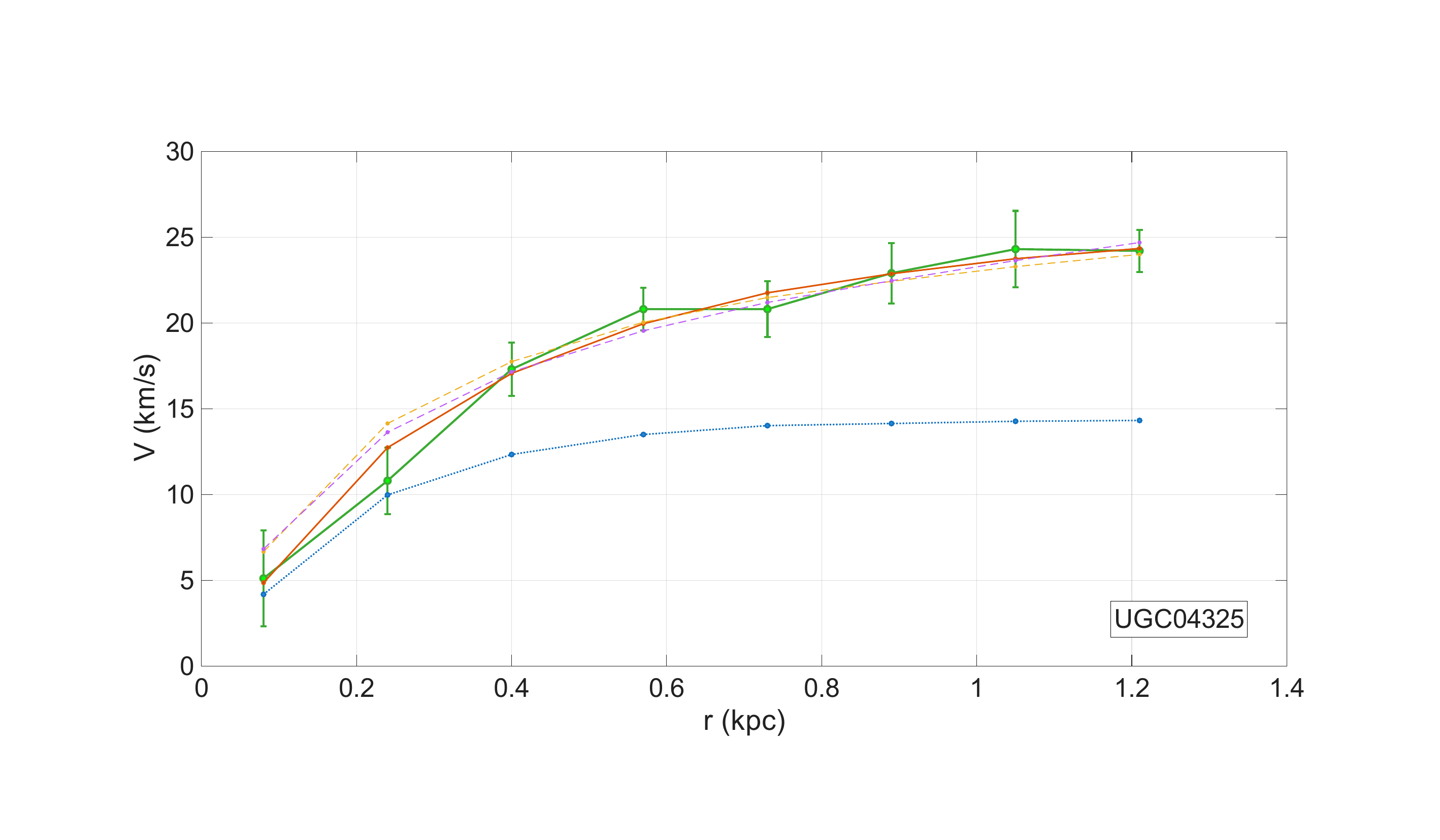}
\includegraphics[trim=4cm 3cm 5cm 4cm, clip=true, width=0.325\columnwidth]{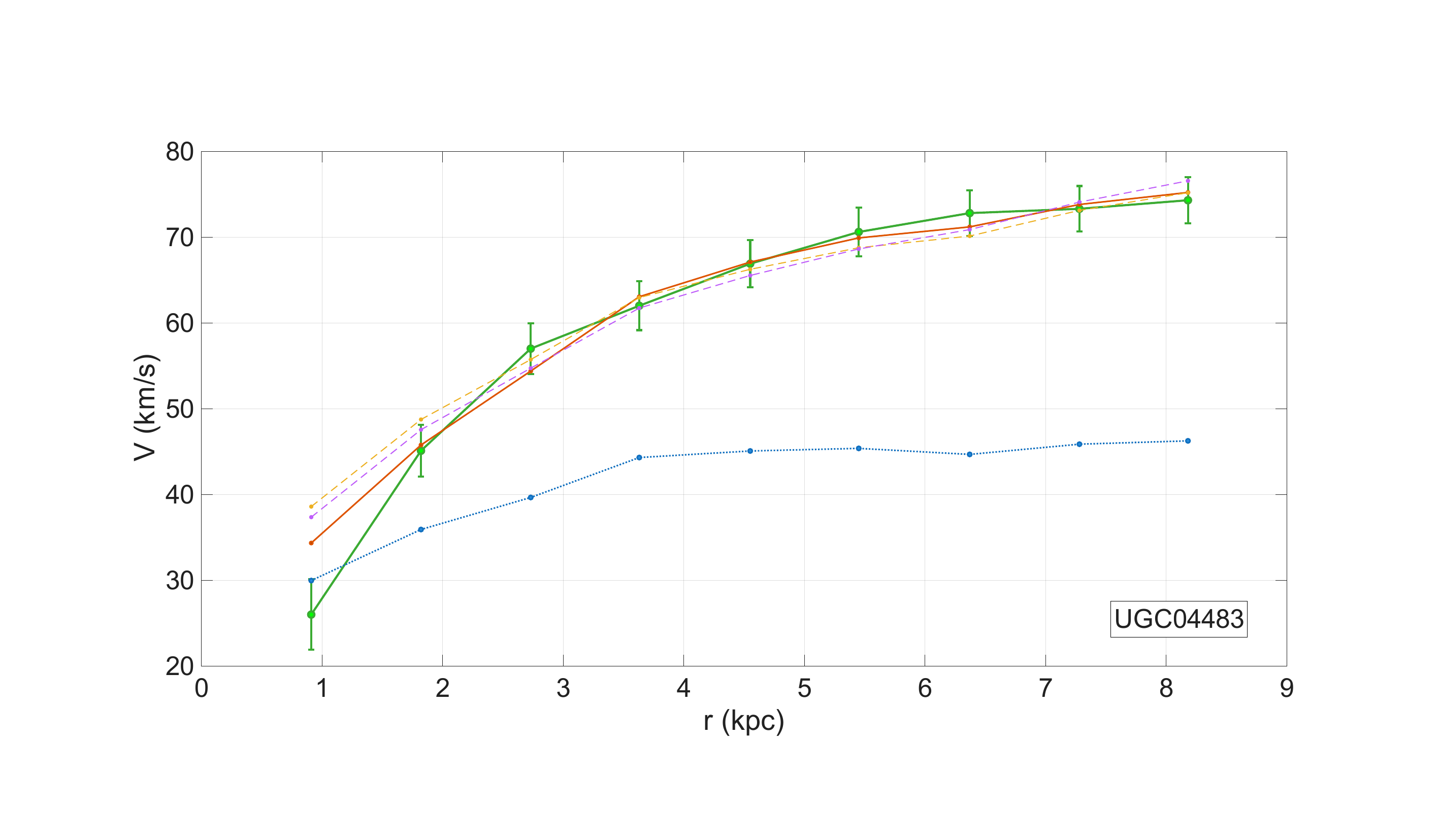}
\includegraphics[trim=4cm 3cm 5cm 4cm, clip=true, width=0.325\columnwidth]{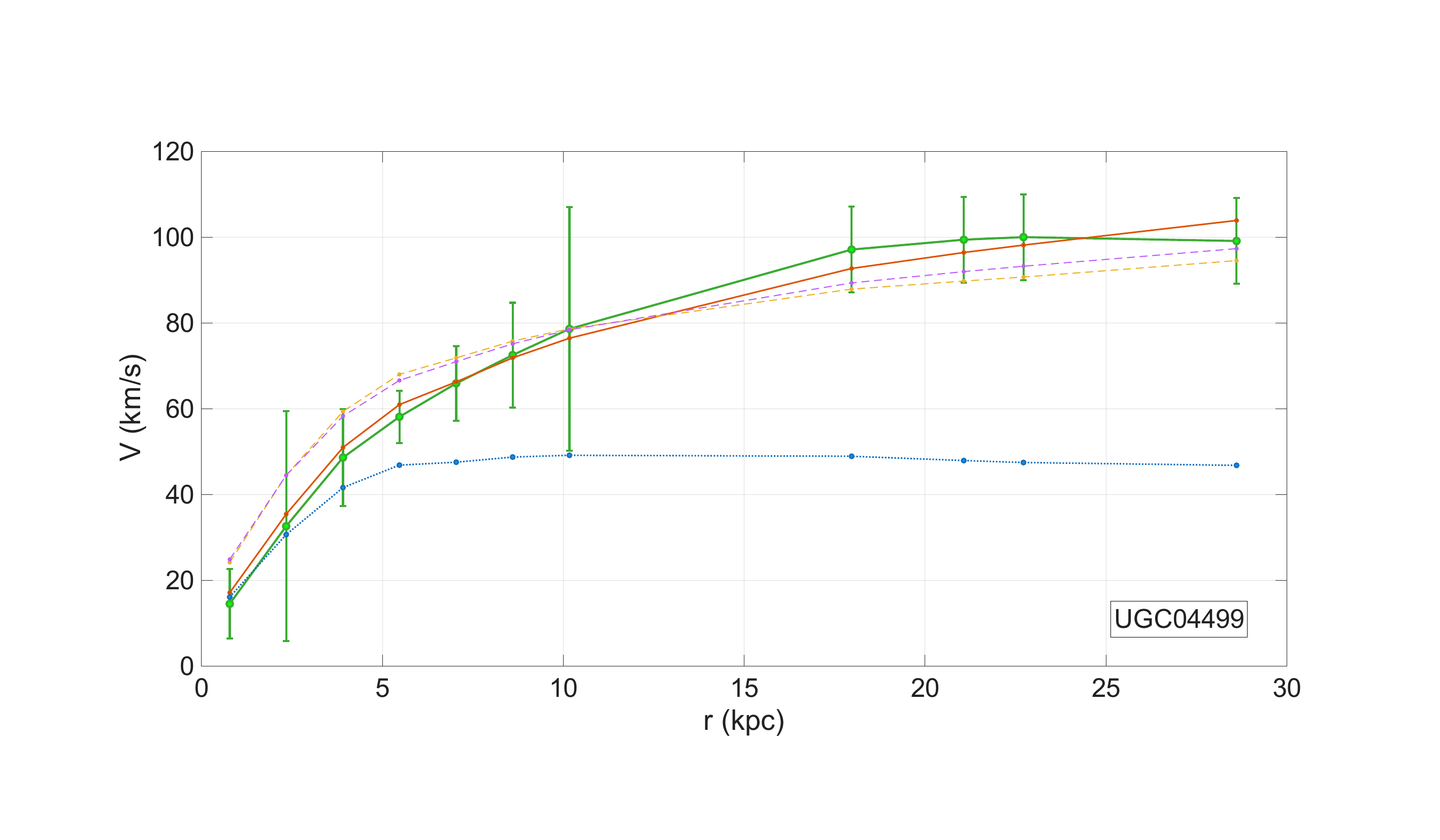}
\includegraphics[trim=4cm 3cm 5cm 4cm, clip=true, width=0.325\columnwidth]{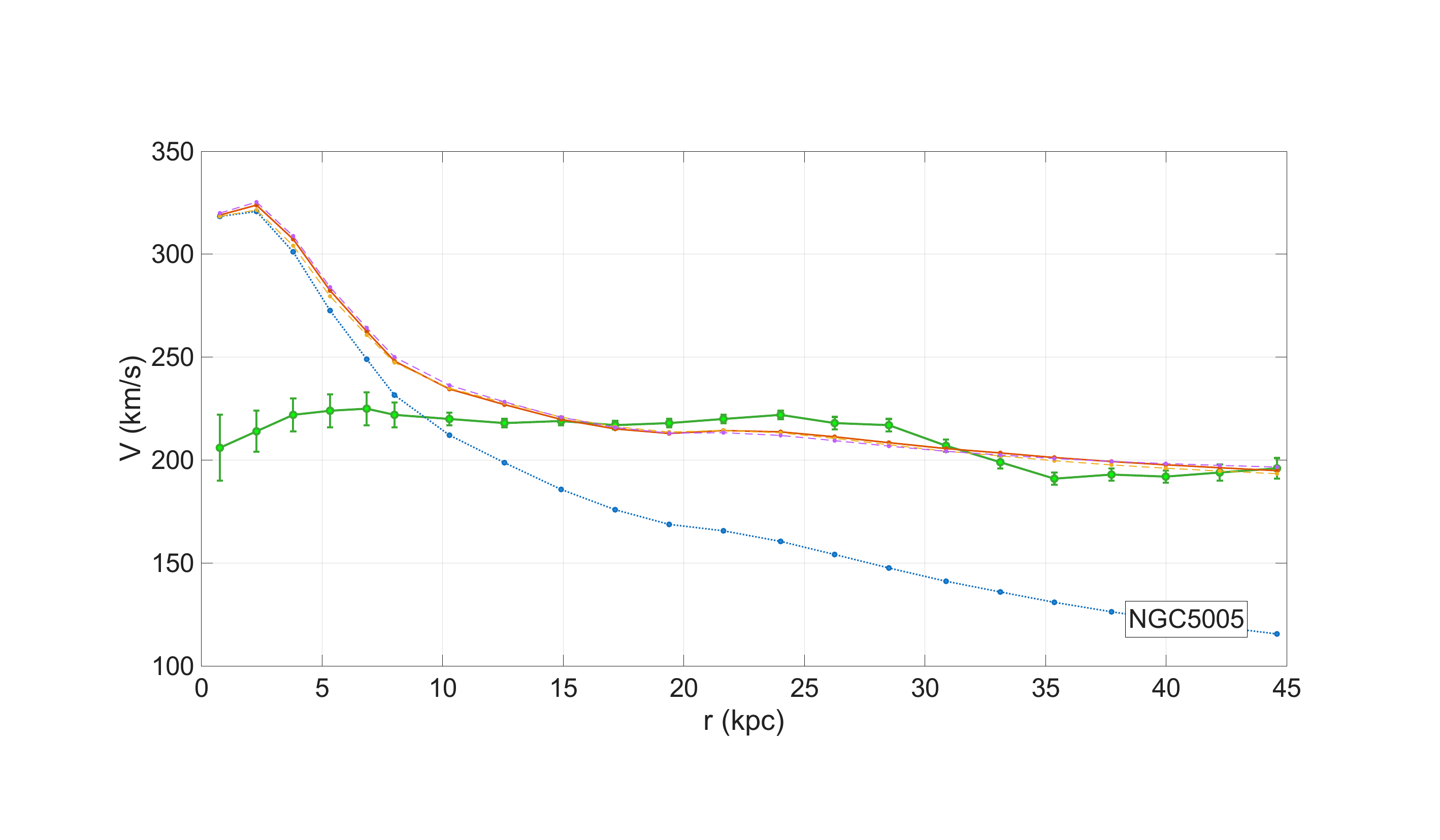}
\includegraphics[trim=4cm 3cm 5cm 4cm, clip=true, width=0.325\columnwidth]{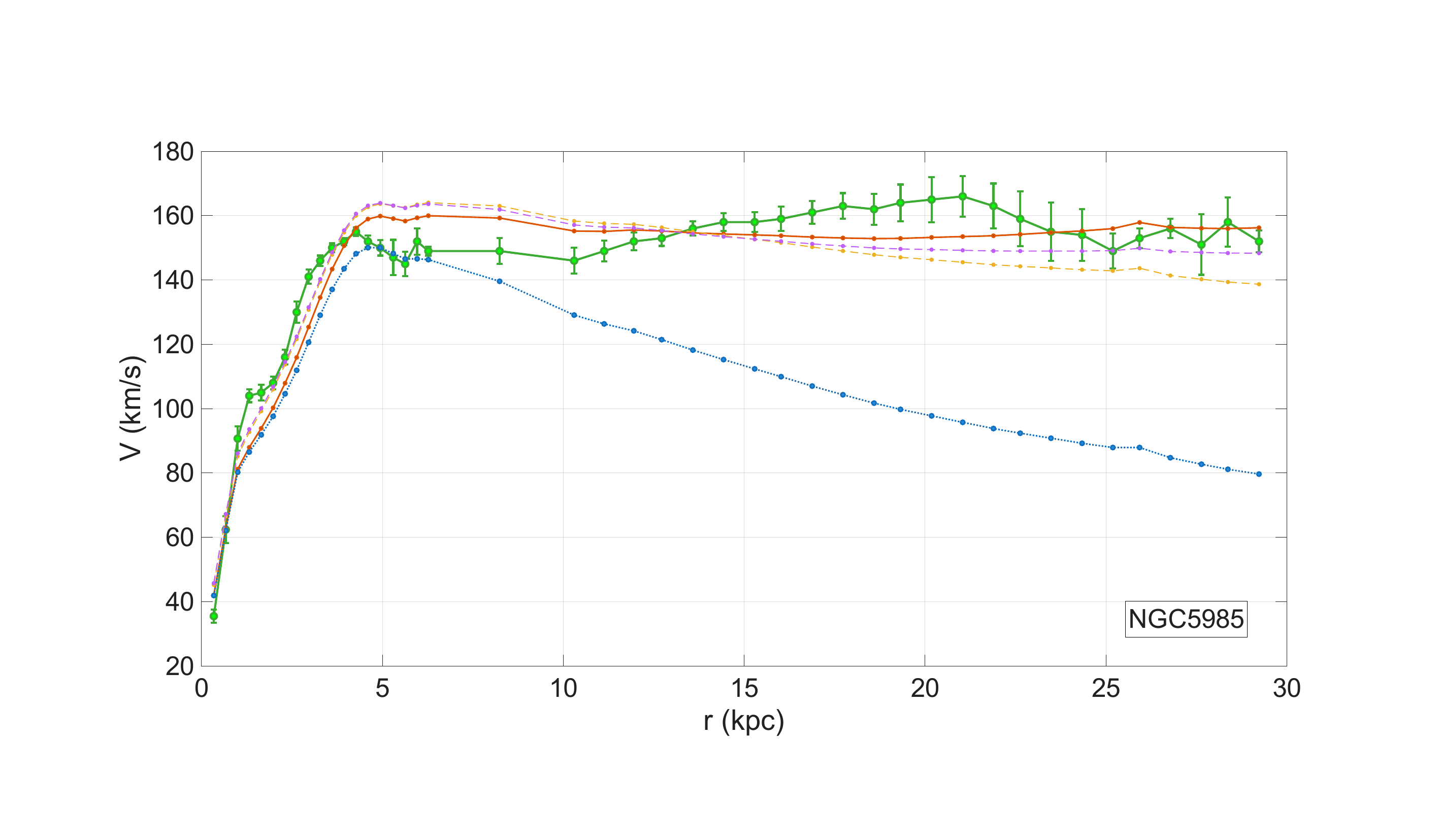}
\includegraphics[trim=4cm 3cm 5cm 4cm, clip=true, width=0.325\columnwidth]{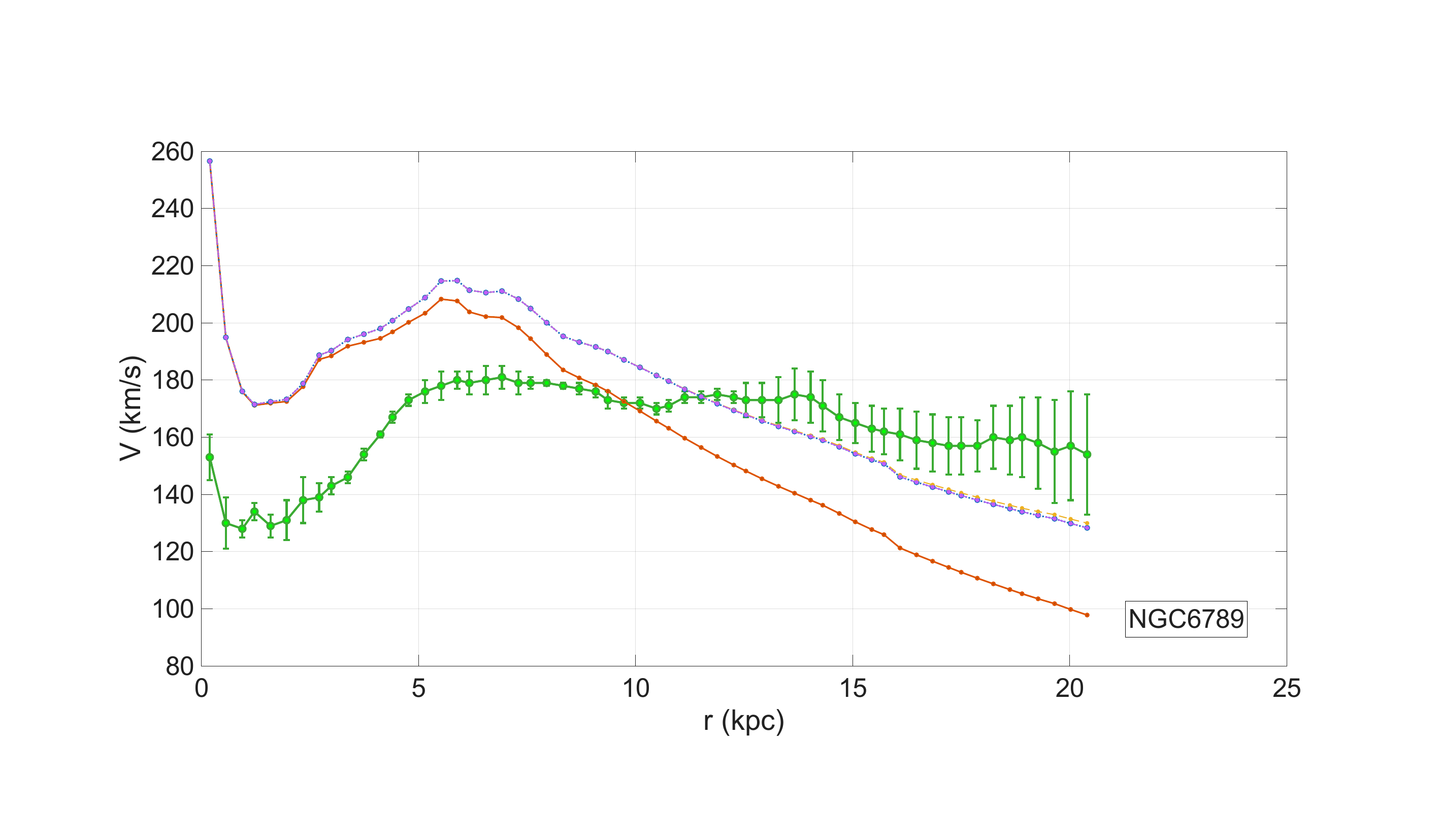}
\includegraphics[trim=4cm 3cm 5cm 4cm, clip=true, width=0.325\columnwidth]{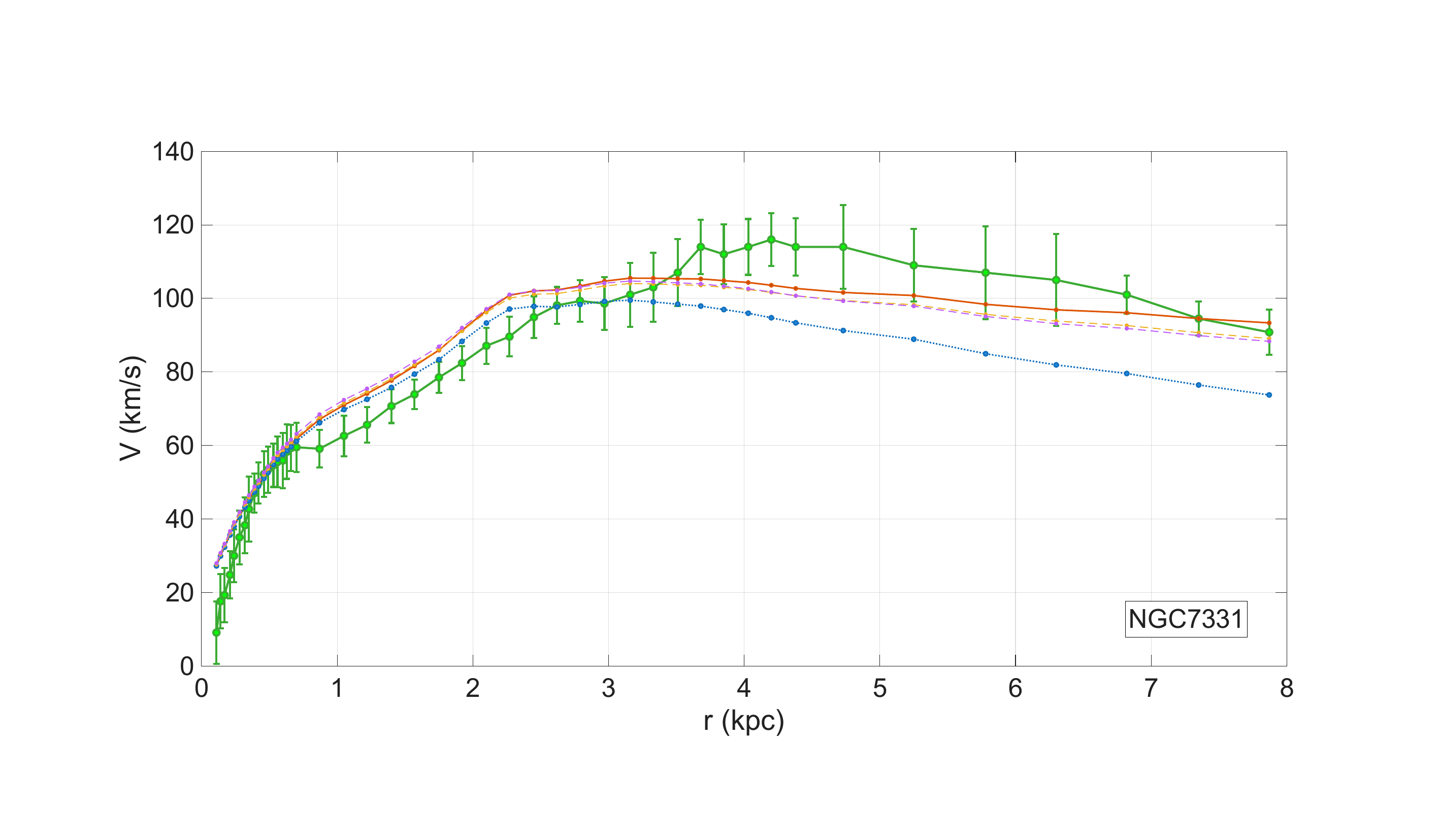}
\includegraphics[trim=4cm 3cm 5cm 4cm, clip=true, width=0.325\columnwidth]{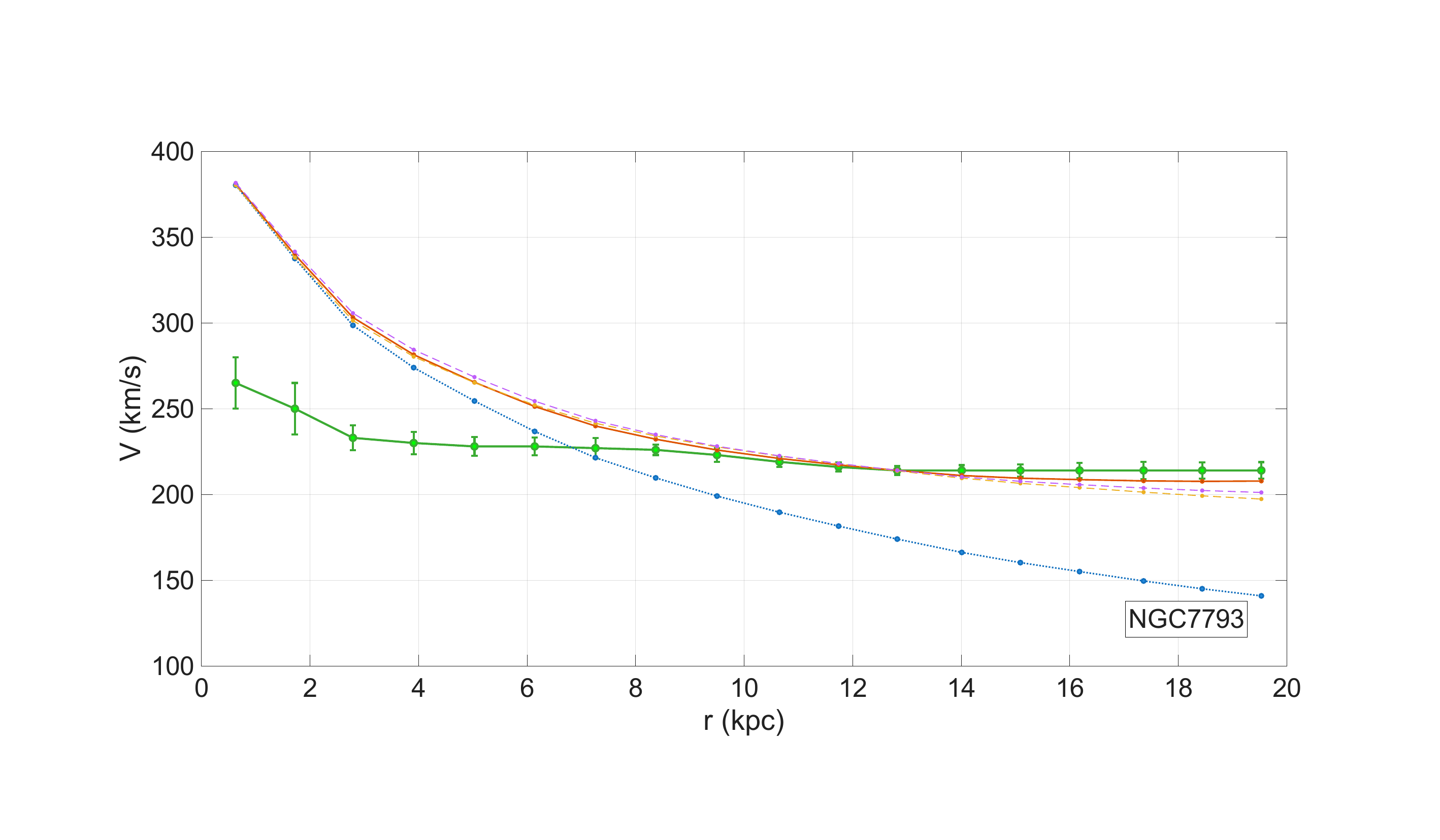}
\includegraphics[trim=4cm 3cm 5cm 4cm, clip=true, width=0.325\columnwidth]{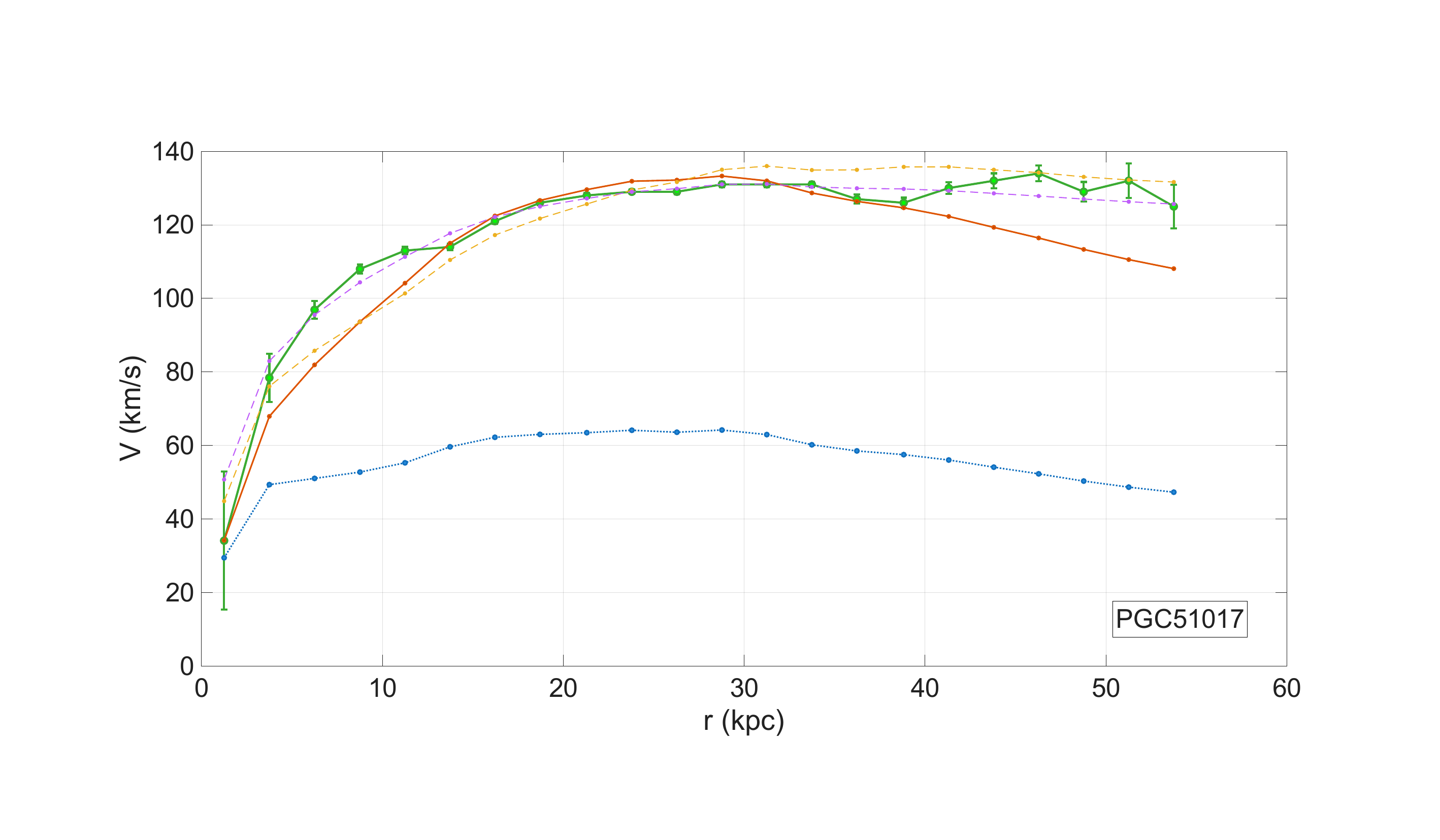}
\includegraphics[trim=4cm 3cm 5cm 4cm, clip=true, width=0.325\columnwidth]{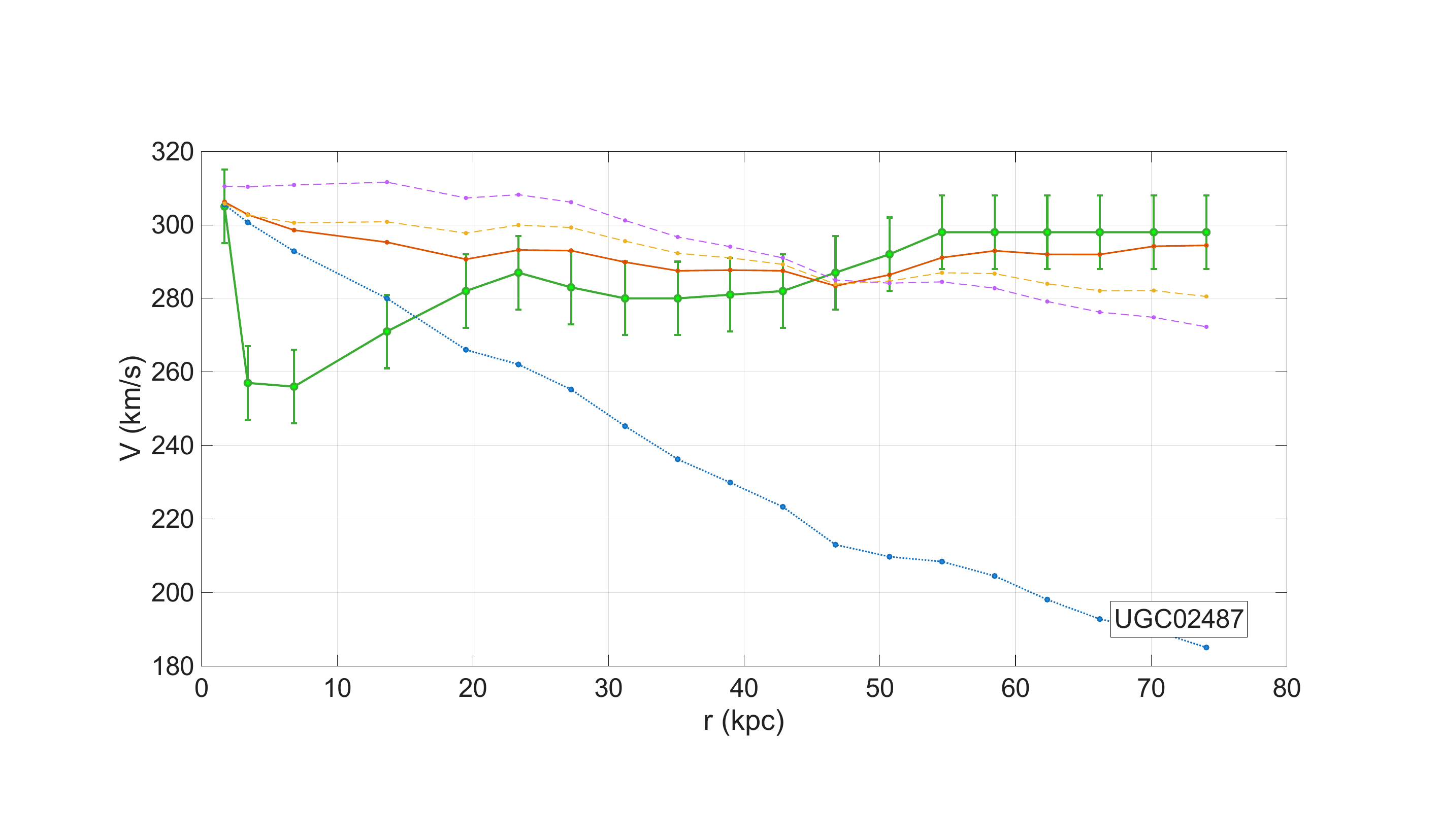}
\includegraphics[trim=4cm 3cm 5cm 4cm, clip=true, width=0.325\columnwidth]{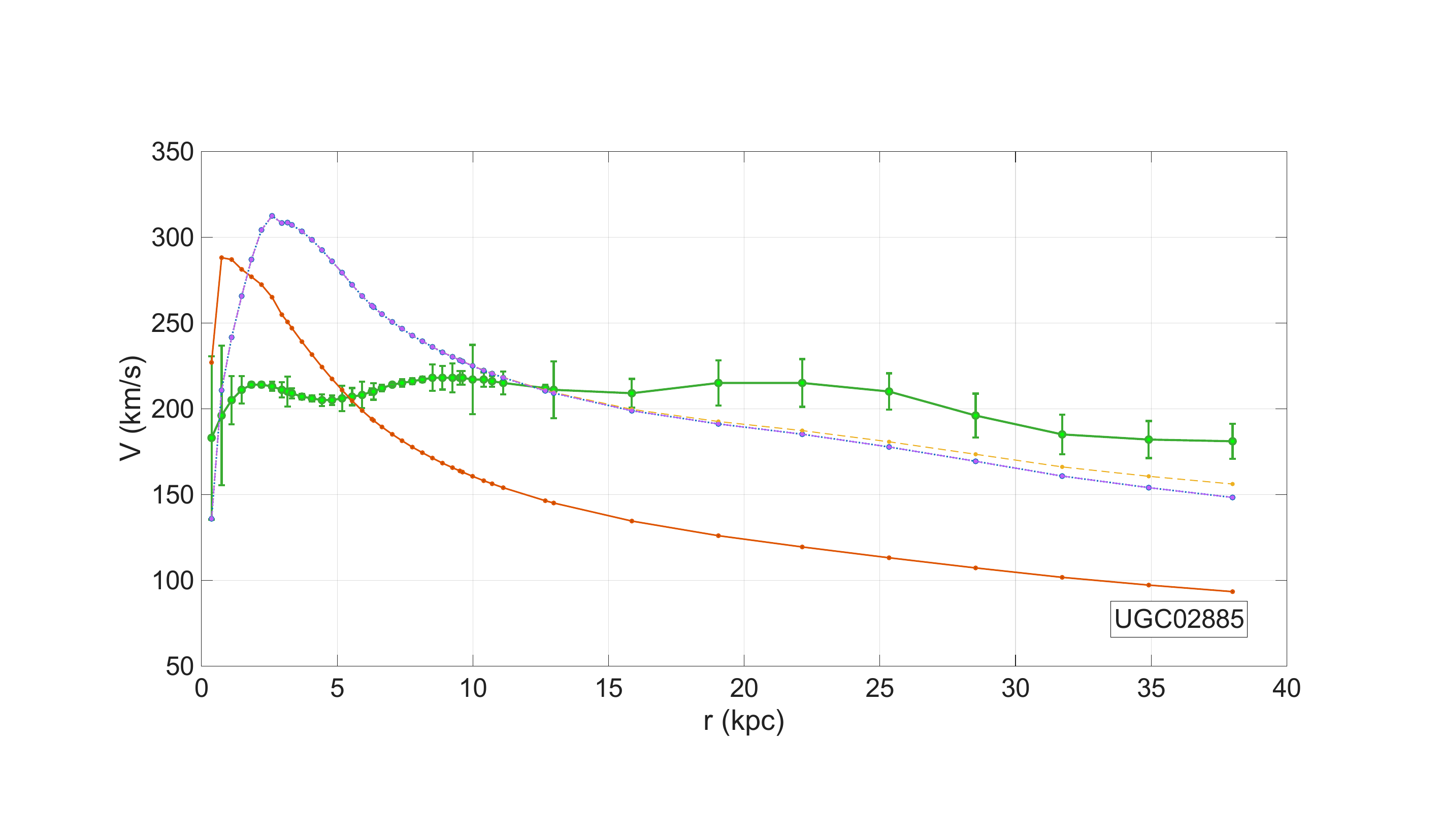}
\includegraphics[trim=4cm 3cm 5cm 4cm, clip=true, width=0.325\columnwidth]{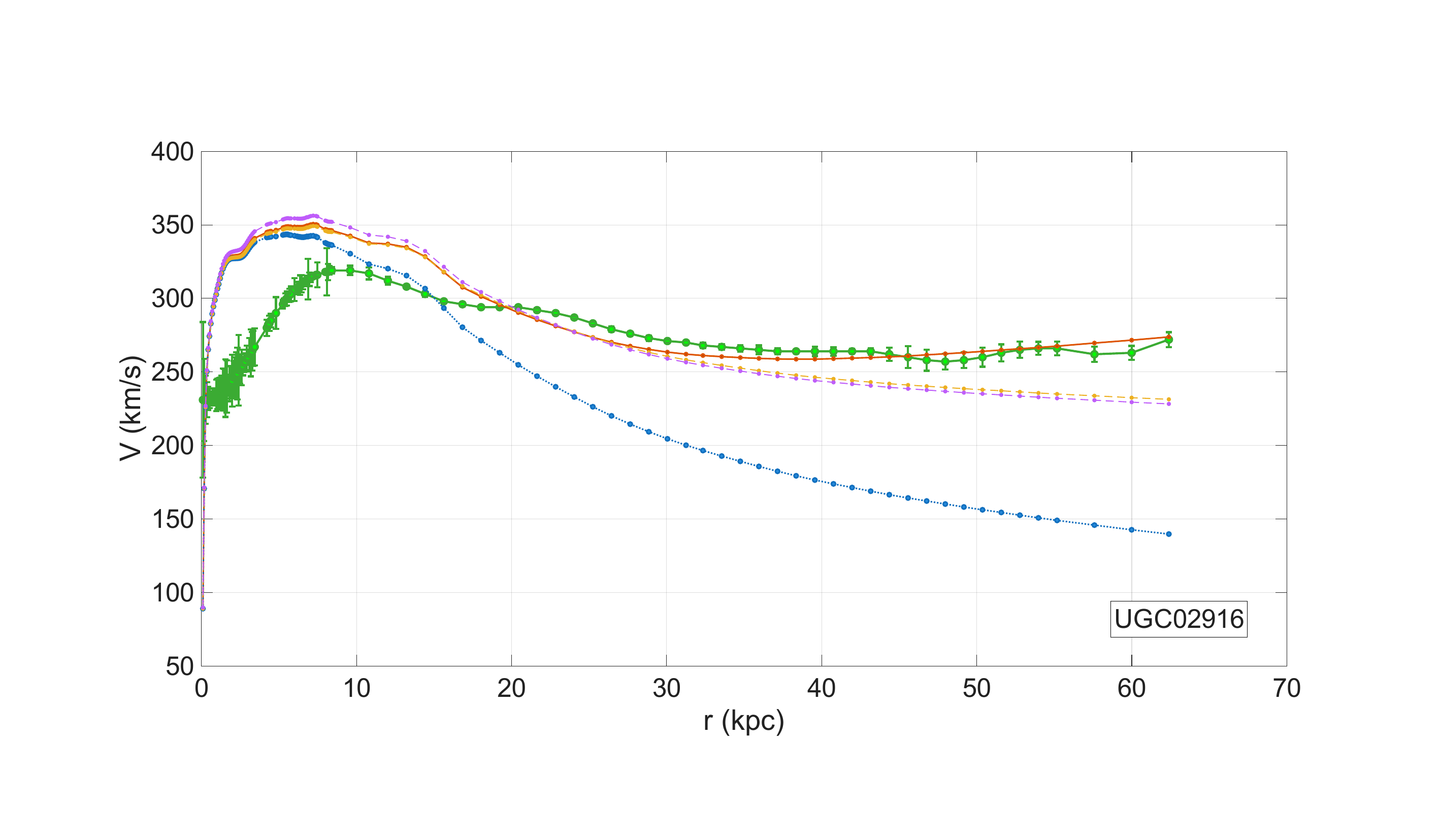}
\includegraphics[trim=4cm 3cm 5cm 4cm, clip=true, width=0.325\columnwidth]{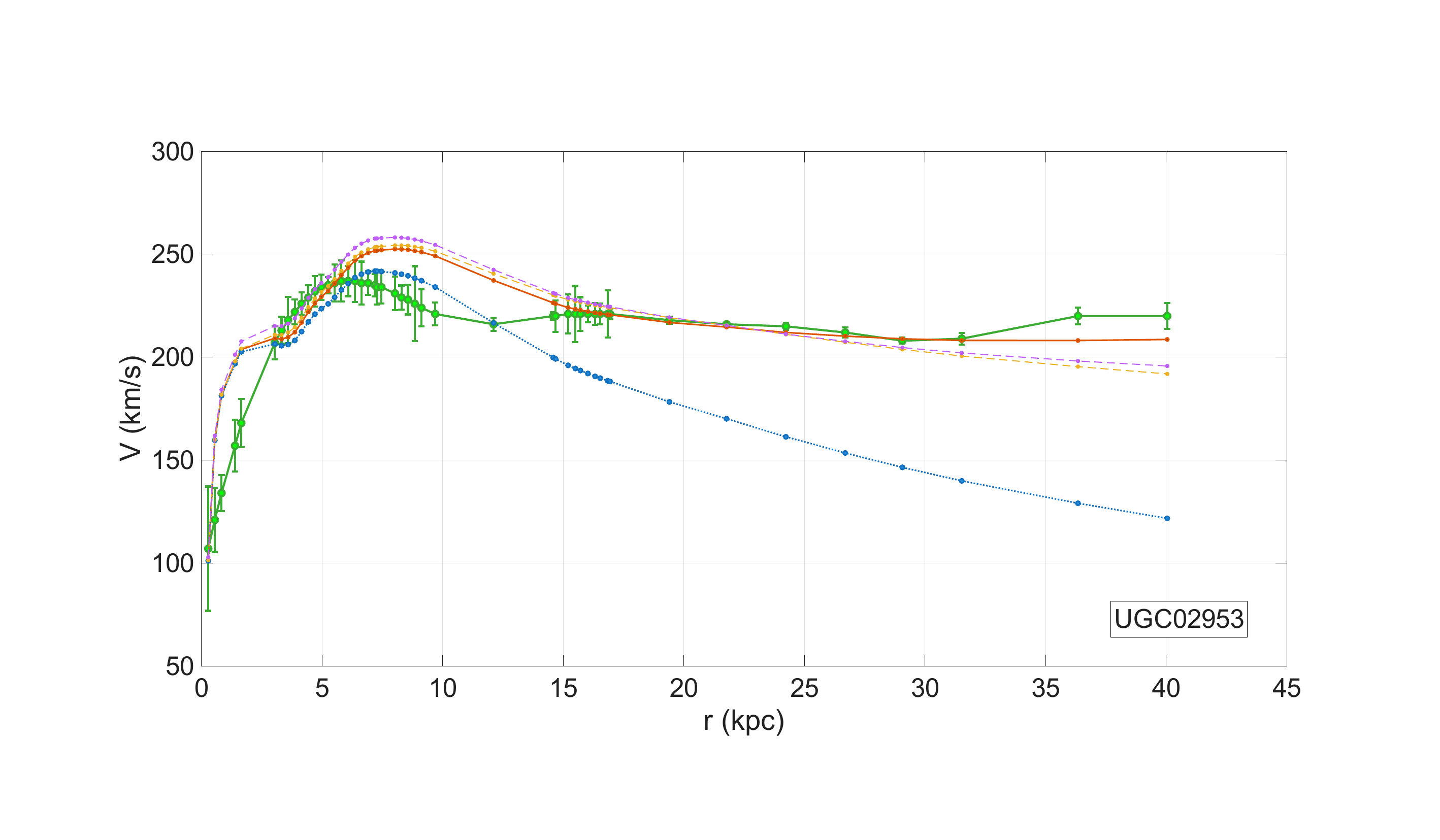}
\includegraphics[trim=4cm 3cm 5cm 4cm, clip=true, width=0.325\columnwidth]{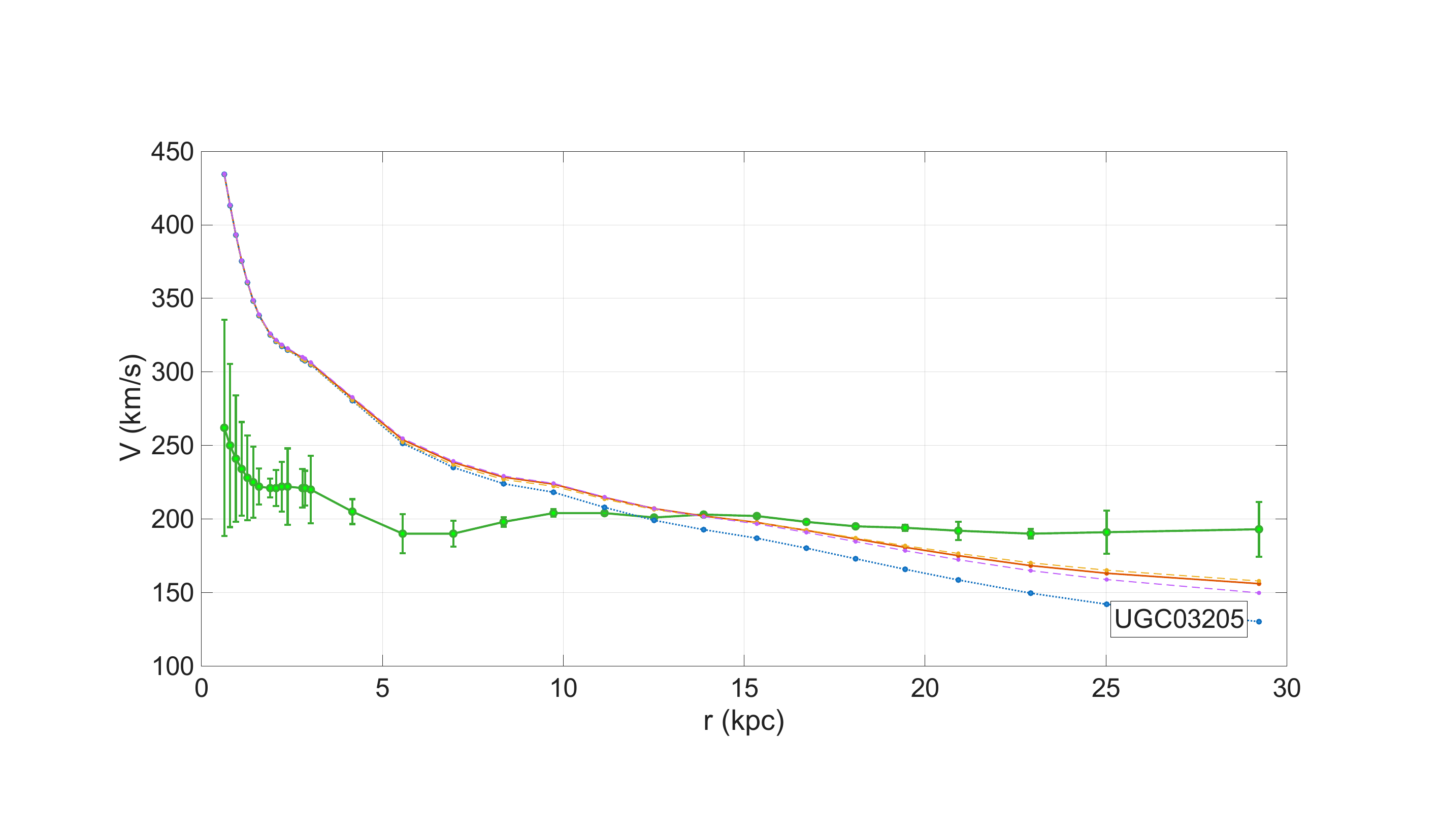}
\includegraphics[trim=4cm 3cm 5cm 4cm, clip=true, width=0.325\columnwidth]{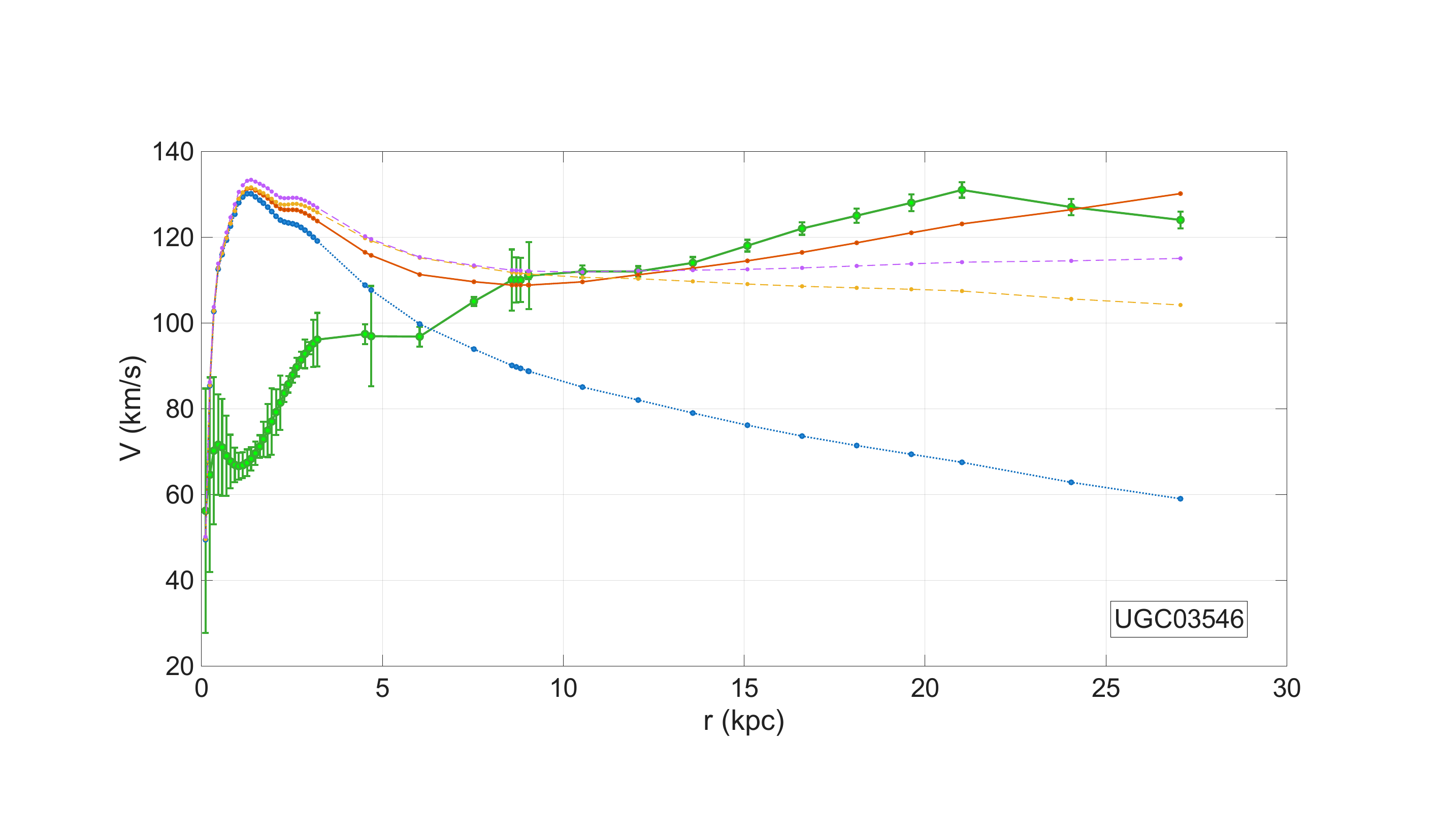}
\end{figure}

\begin{figure}
\centering
\includegraphics[trim=4cm 3cm 5cm 4cm, clip=true, width=0.325\columnwidth]{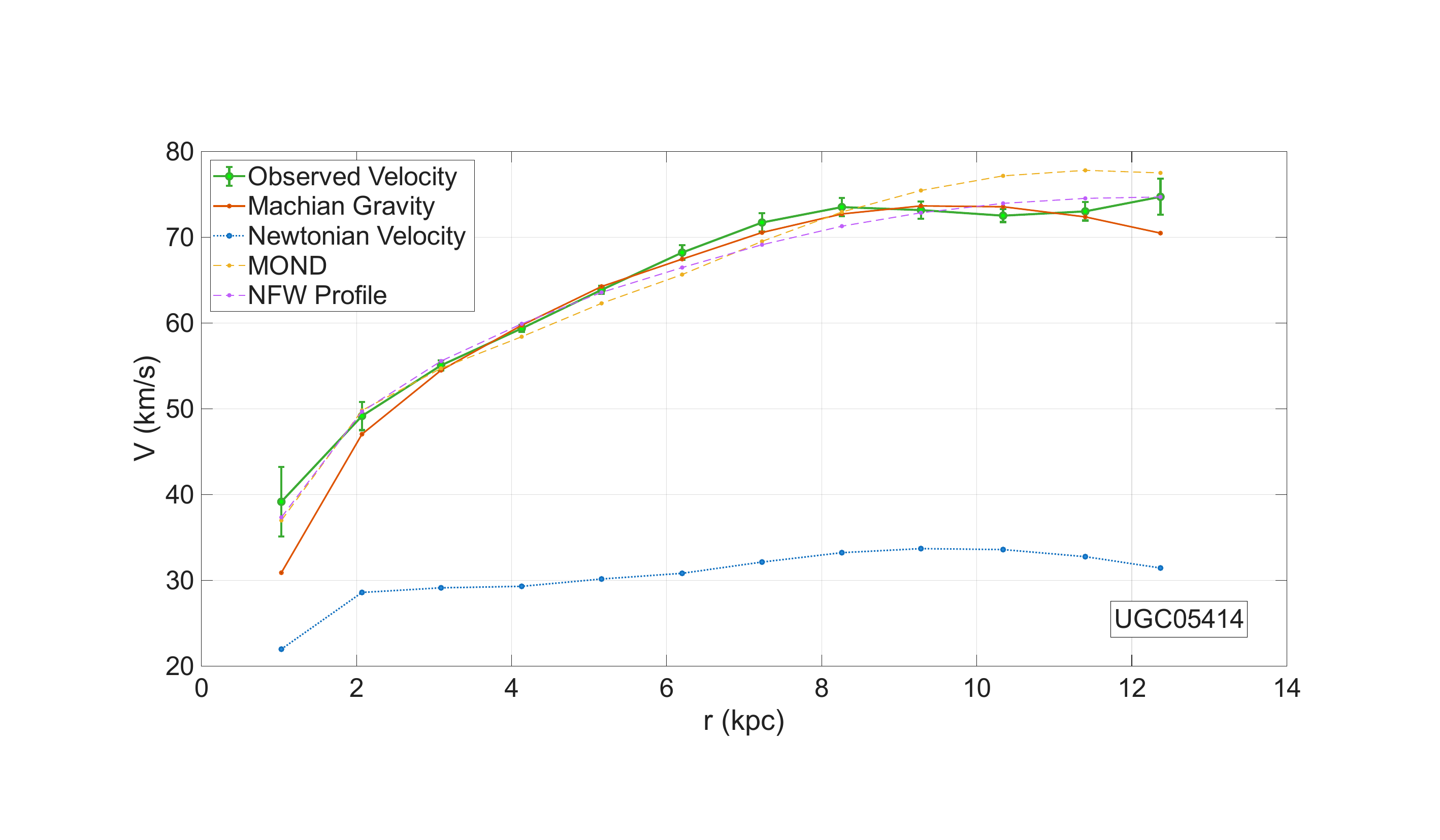}
\includegraphics[trim=4cm 3cm 5cm 4cm, clip=true, width=0.325\columnwidth]{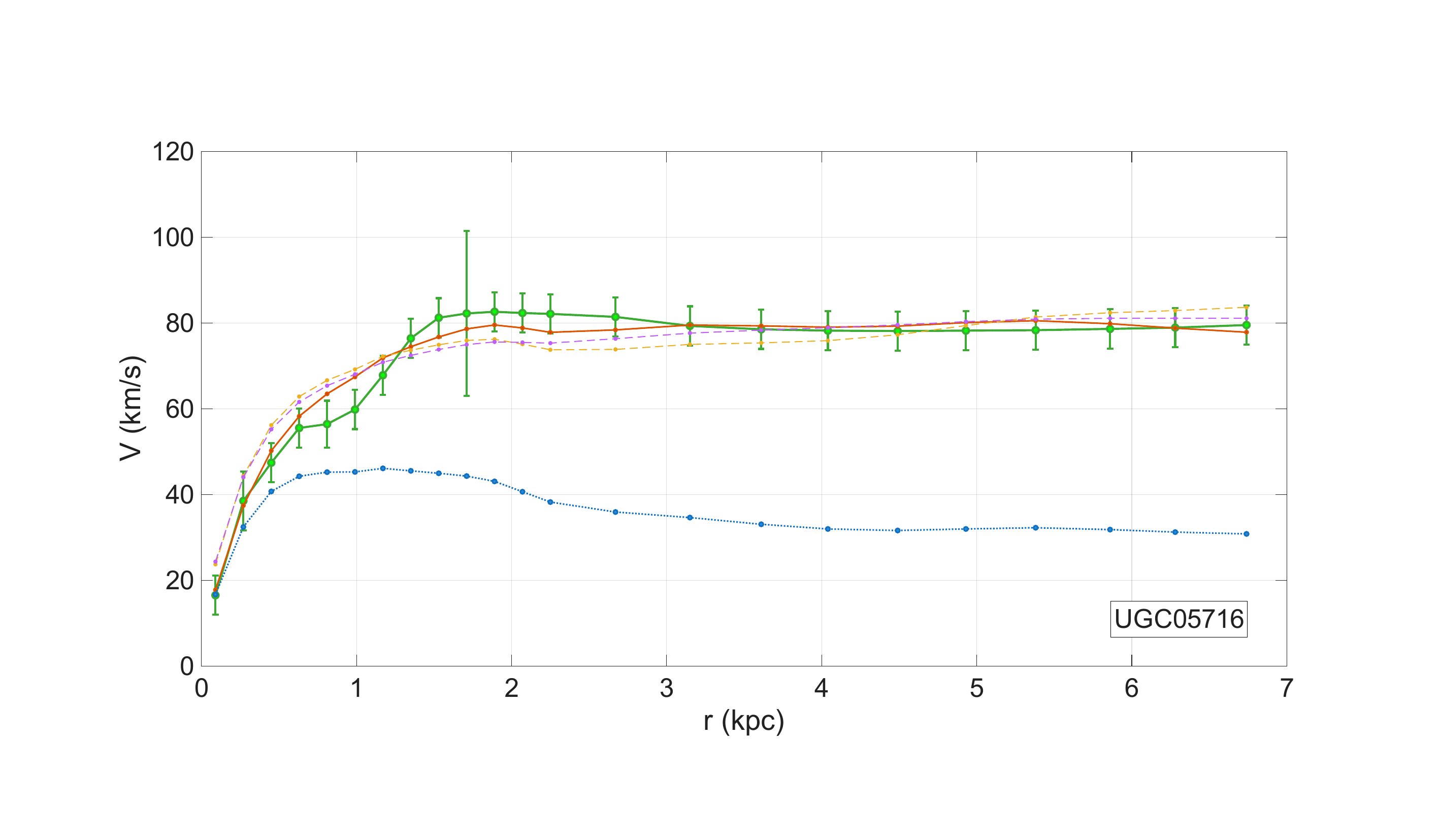}
\includegraphics[trim=4cm 3cm 5cm 4cm, clip=true, width=0.325\columnwidth]{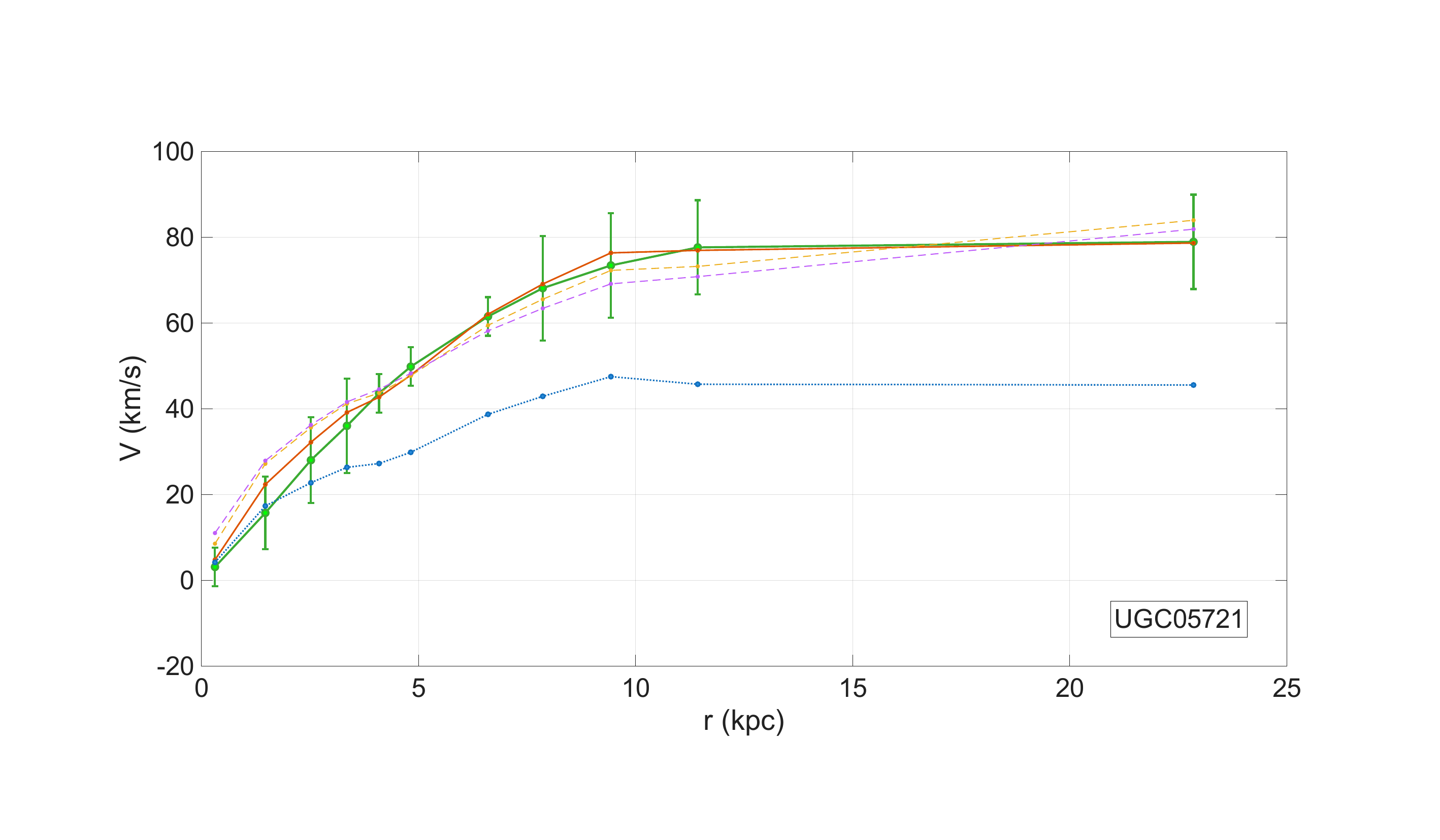}
\includegraphics[trim=4cm 3cm 5cm 4cm, clip=true, width=0.325\columnwidth]{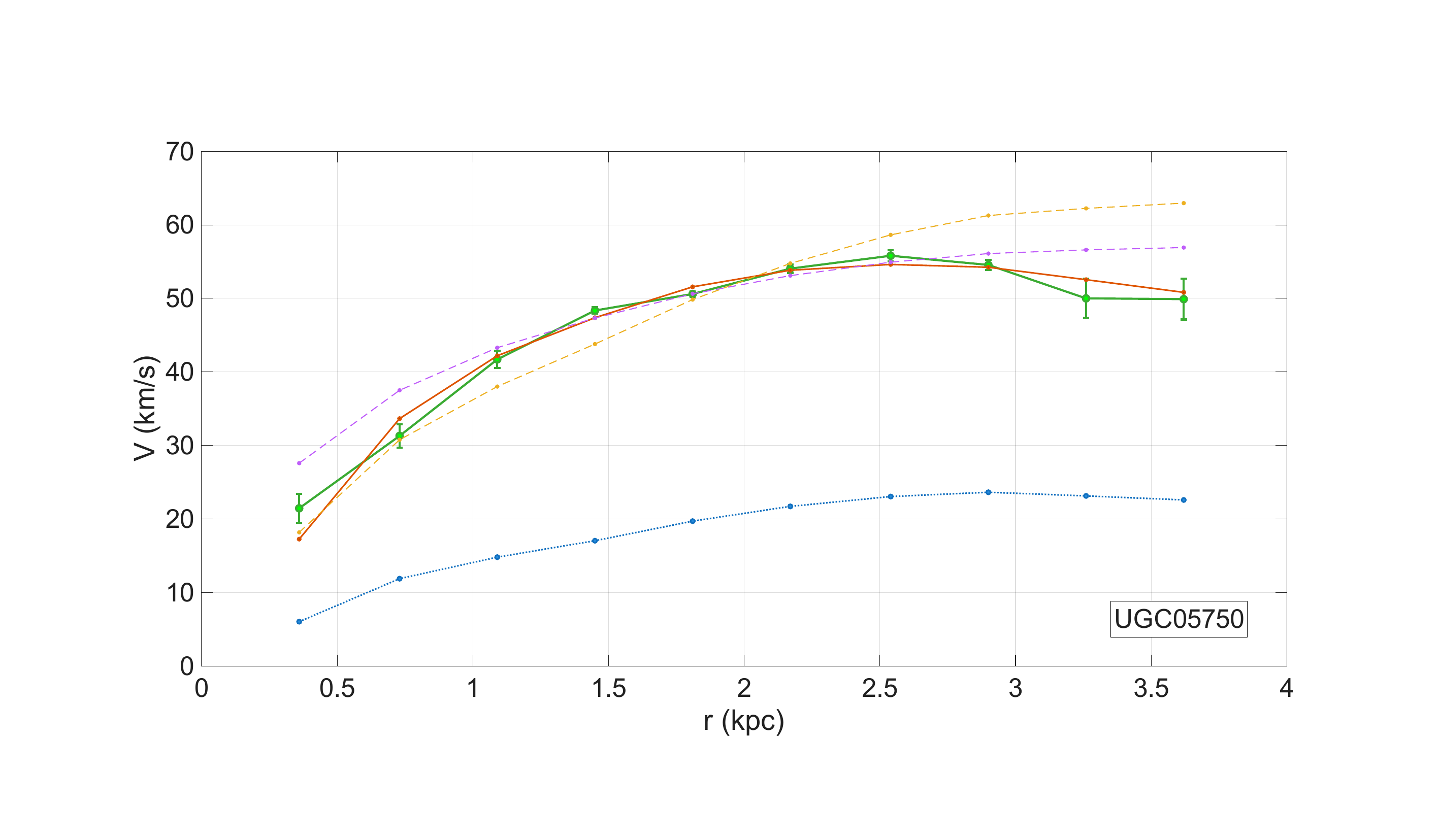}
\includegraphics[trim=4cm 3cm 5cm 4cm, clip=true, width=0.325\columnwidth]{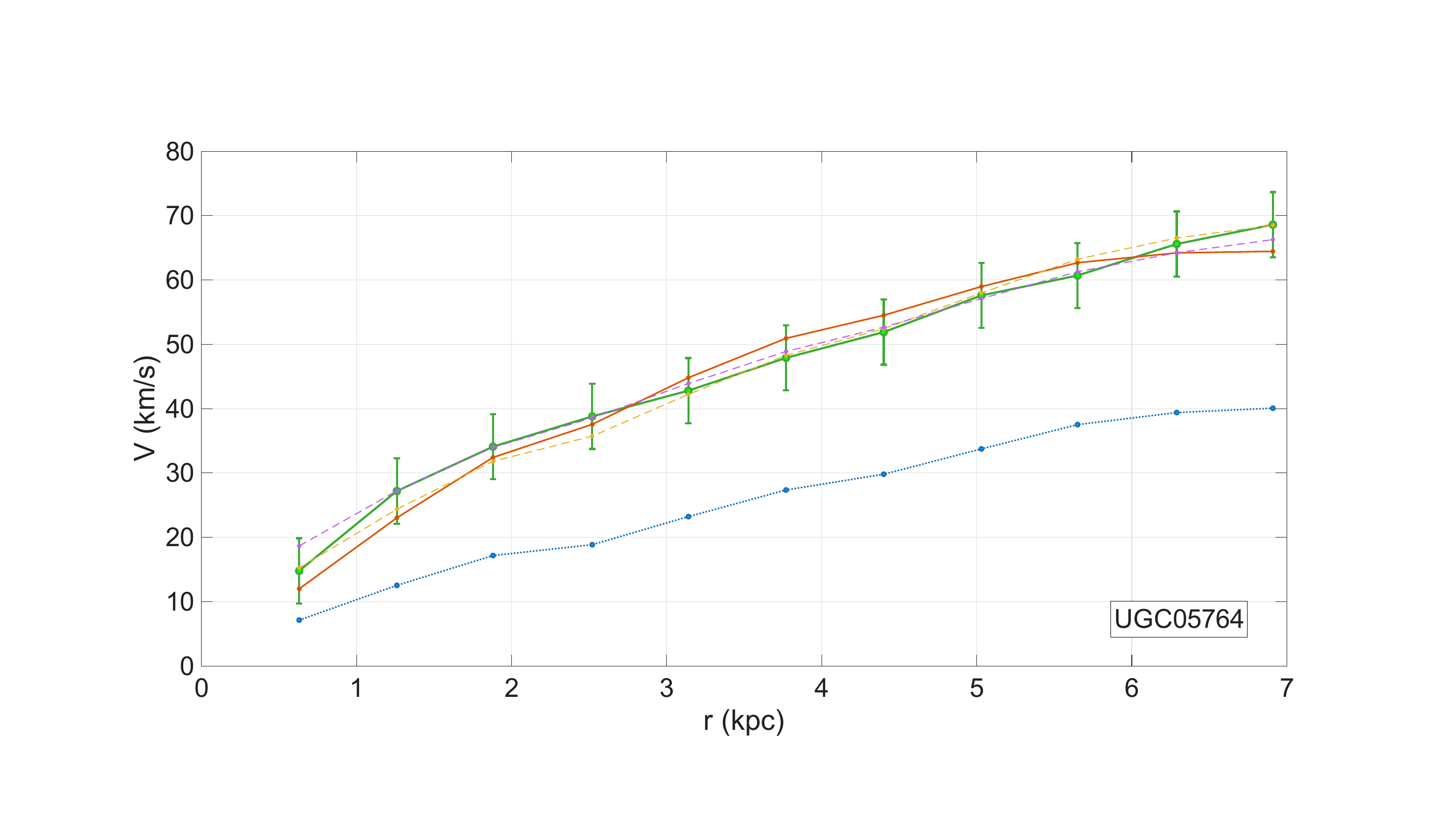}
\includegraphics[trim=4cm 3cm 5cm 4cm, clip=true, width=0.325\columnwidth]{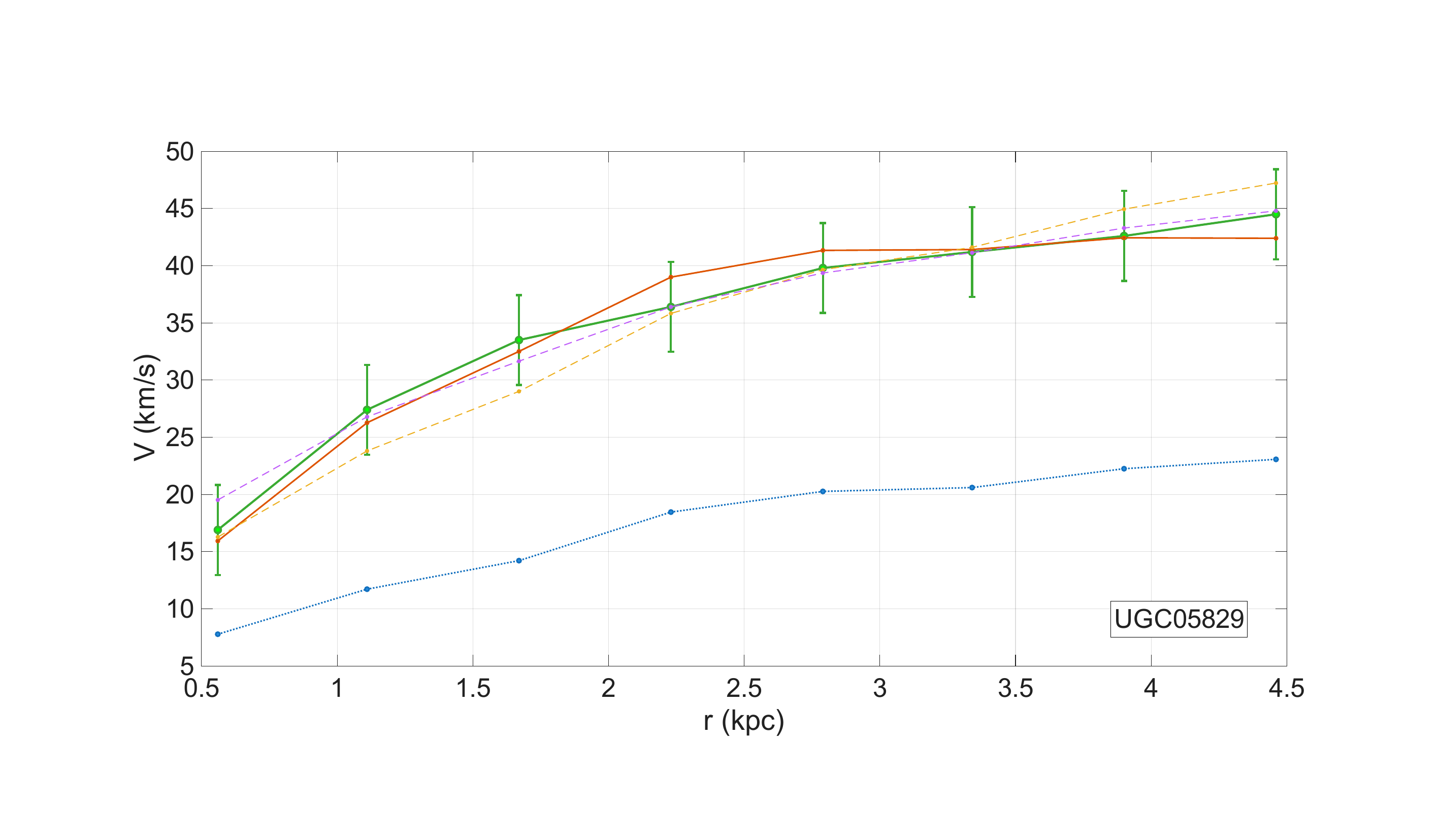}
\includegraphics[trim=4cm 3cm 5cm 4cm, clip=true, width=0.325\columnwidth]{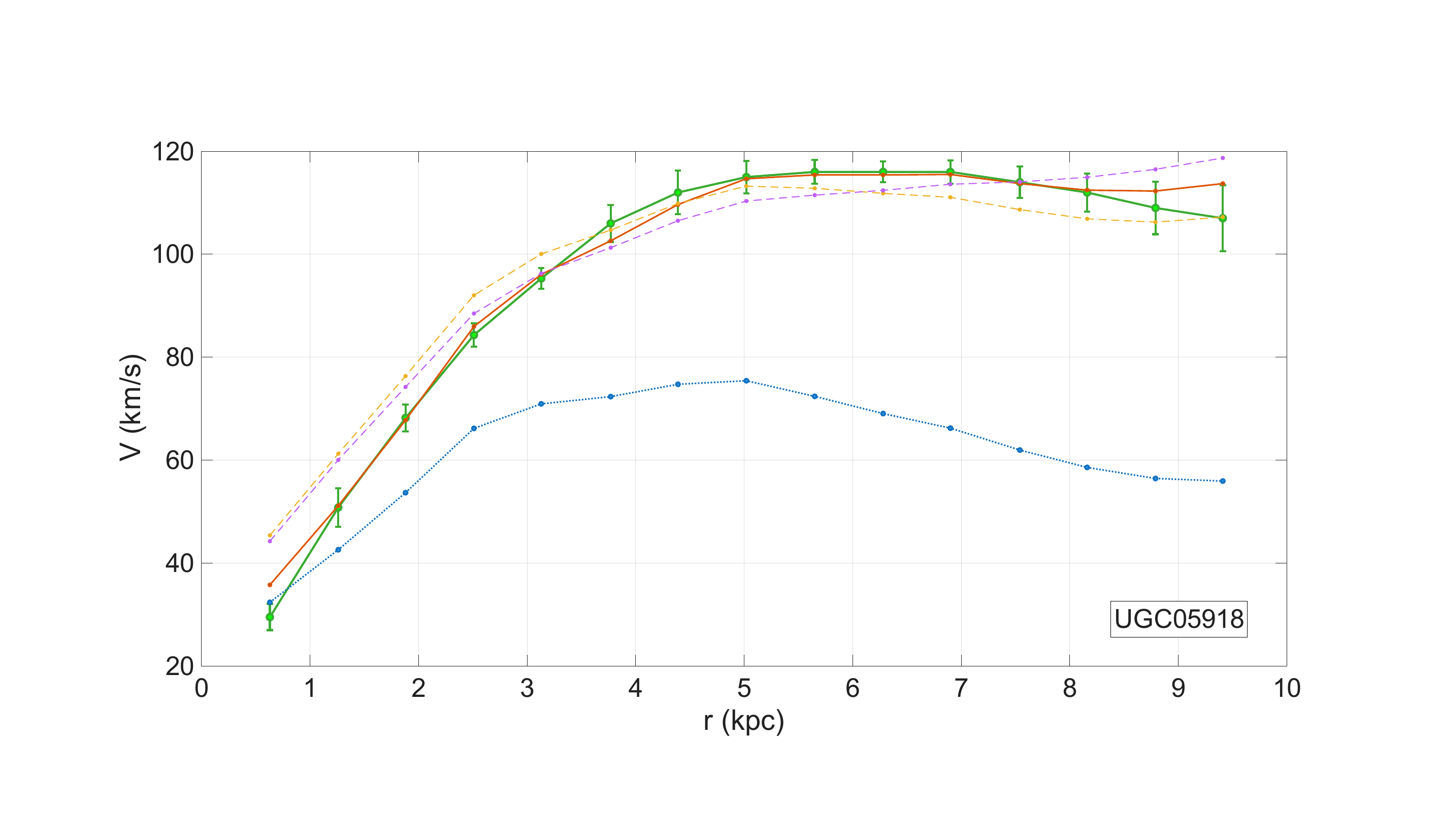}
\includegraphics[trim=4cm 3cm 5cm 4cm, clip=true, width=0.325\columnwidth]{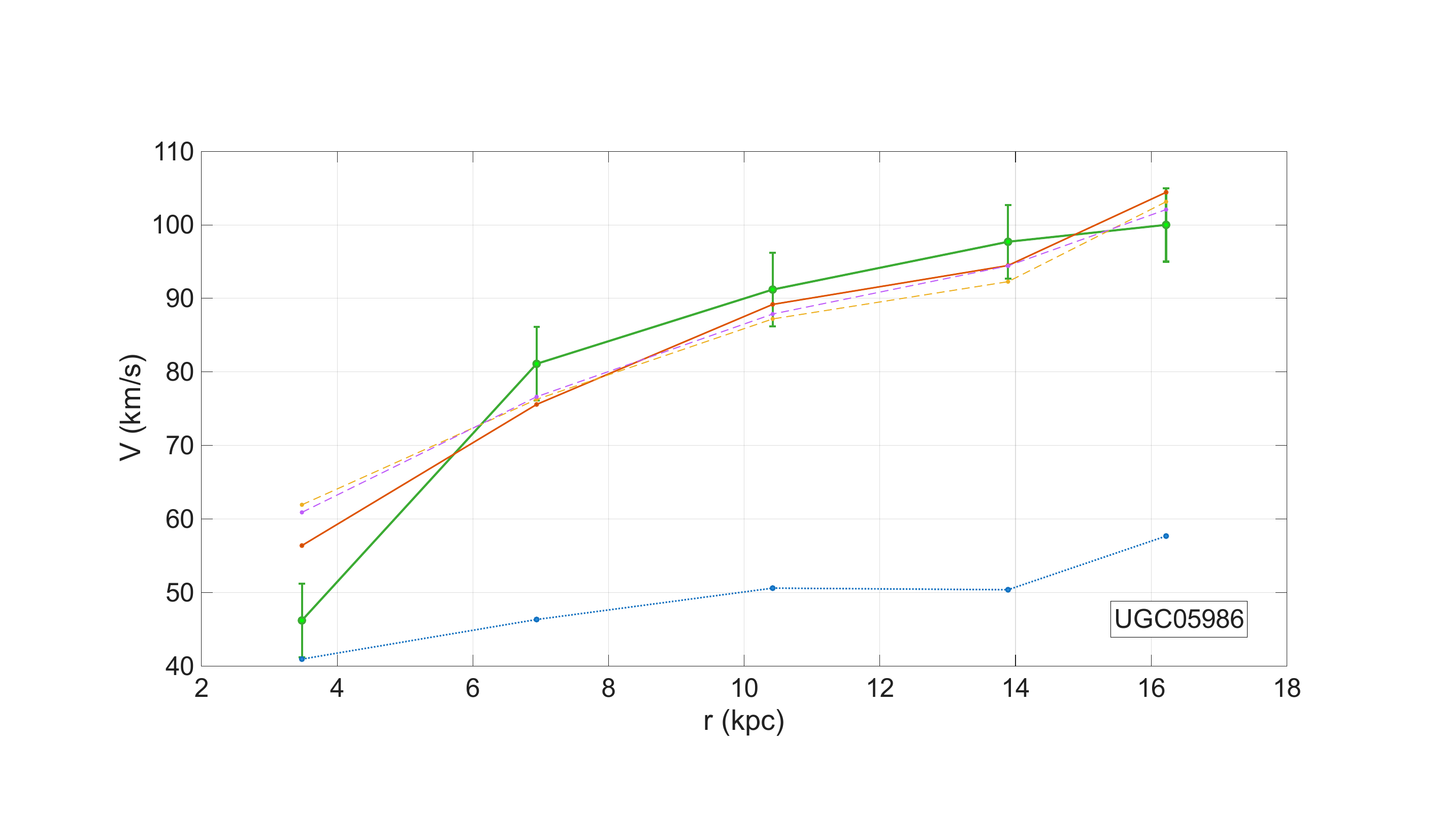}
\includegraphics[trim=4cm 3cm 5cm 4cm, clip=true, width=0.325\columnwidth]{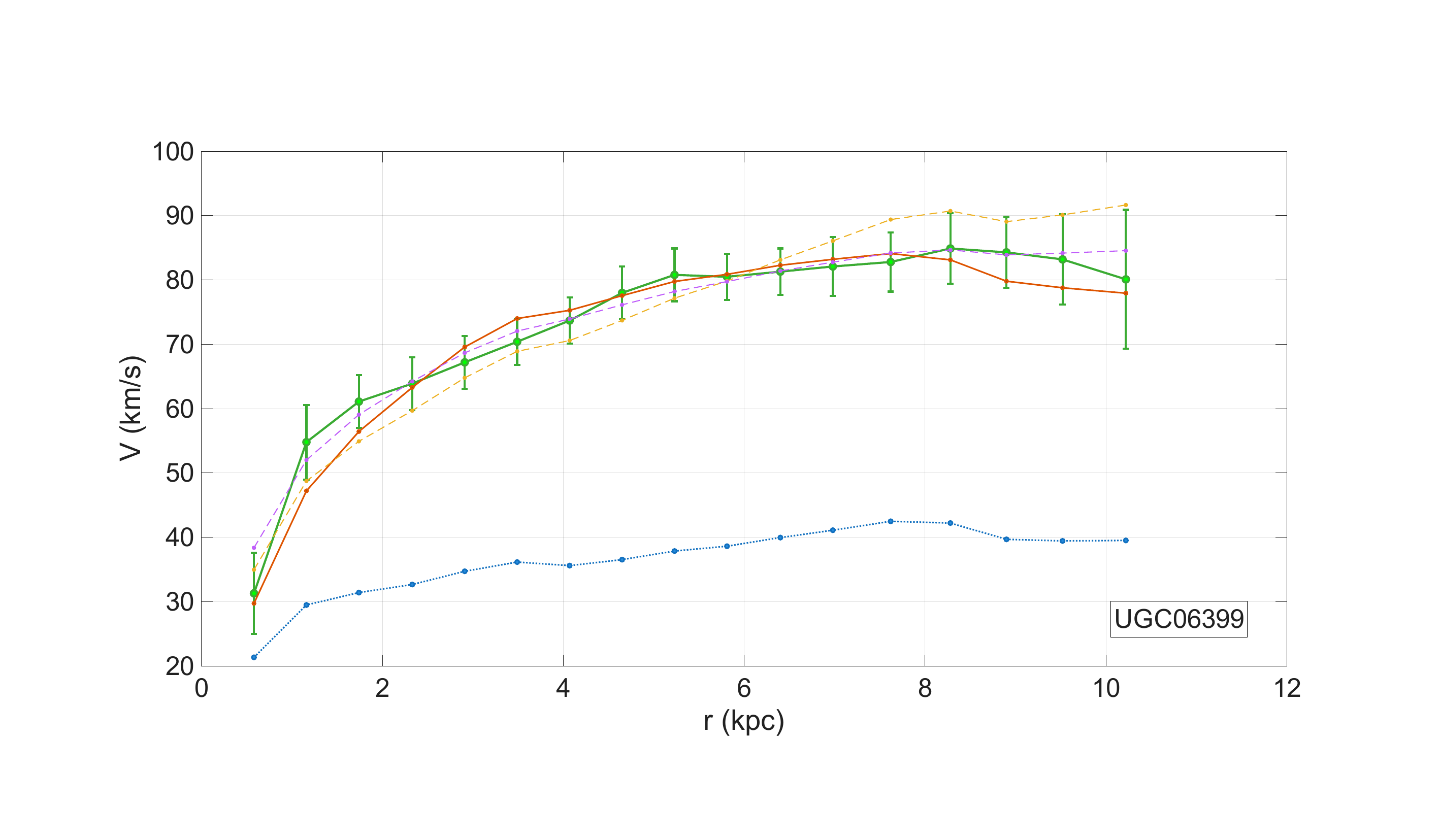}
\includegraphics[trim=4cm 3cm 5cm 4cm, clip=true, width=0.325\columnwidth]{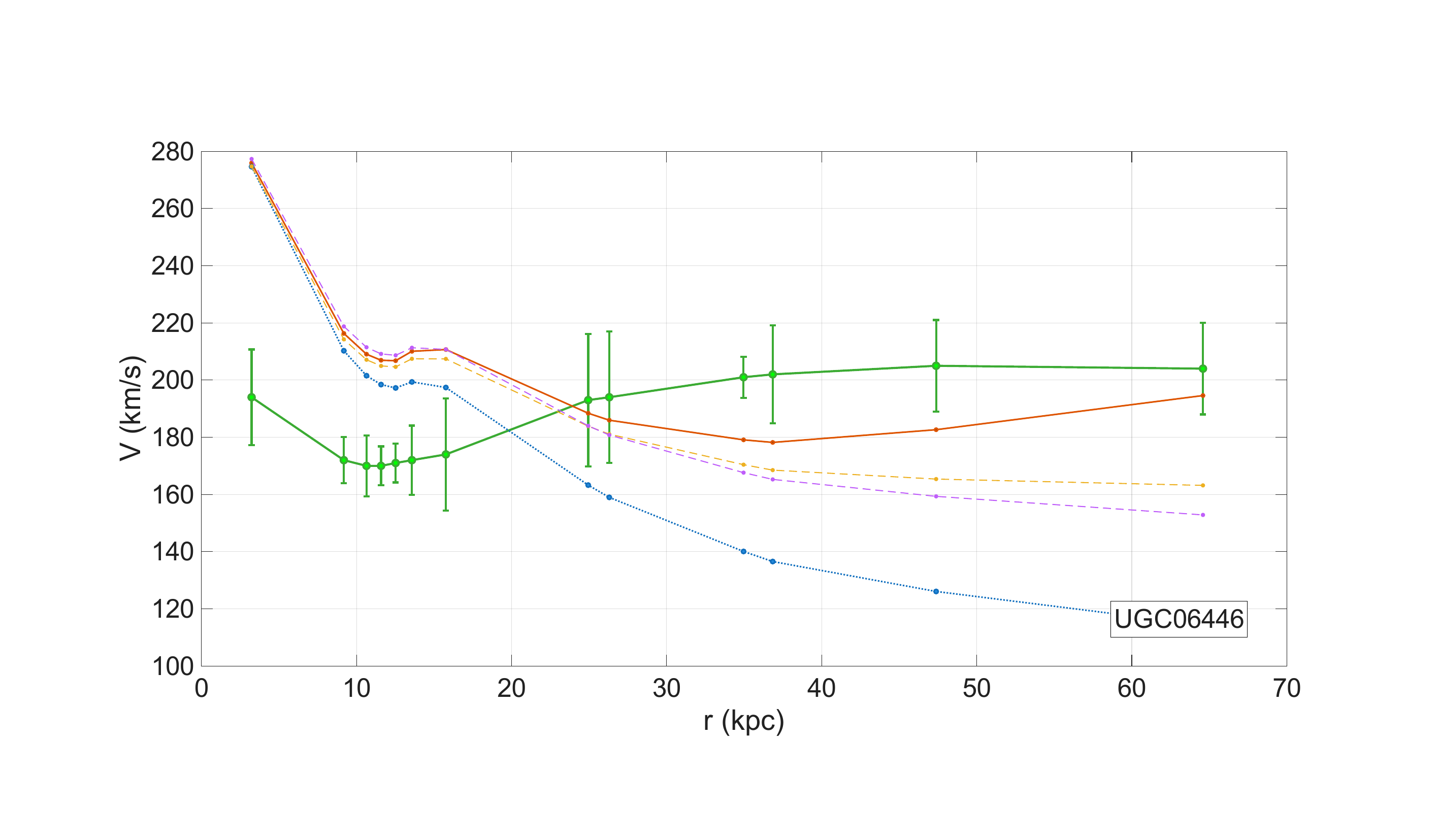}
\includegraphics[trim=4cm 3cm 5cm 4cm, clip=true, width=0.325\columnwidth]{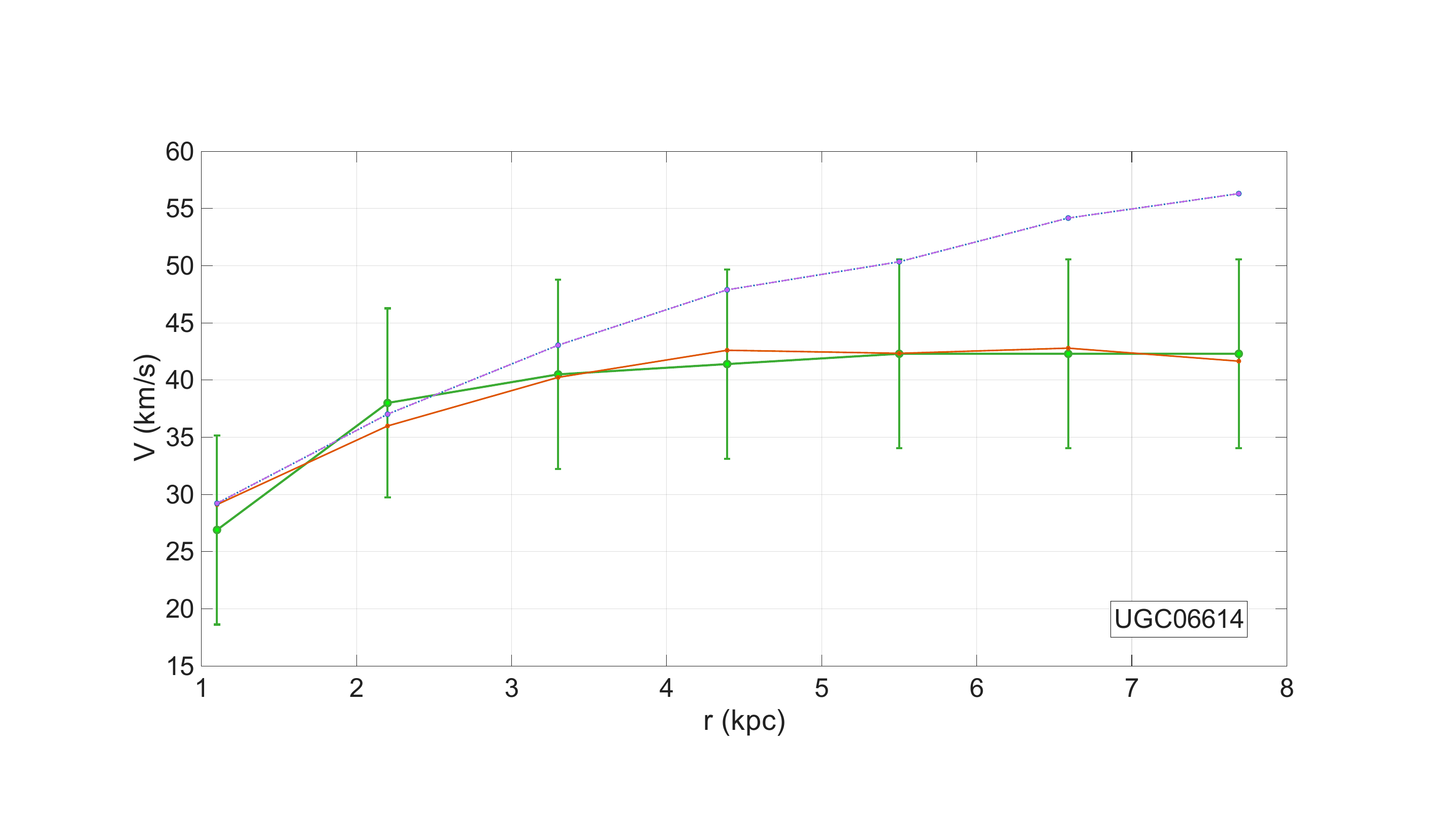}
\includegraphics[trim=4cm 3cm 5cm 4cm, clip=true, width=0.325\columnwidth]{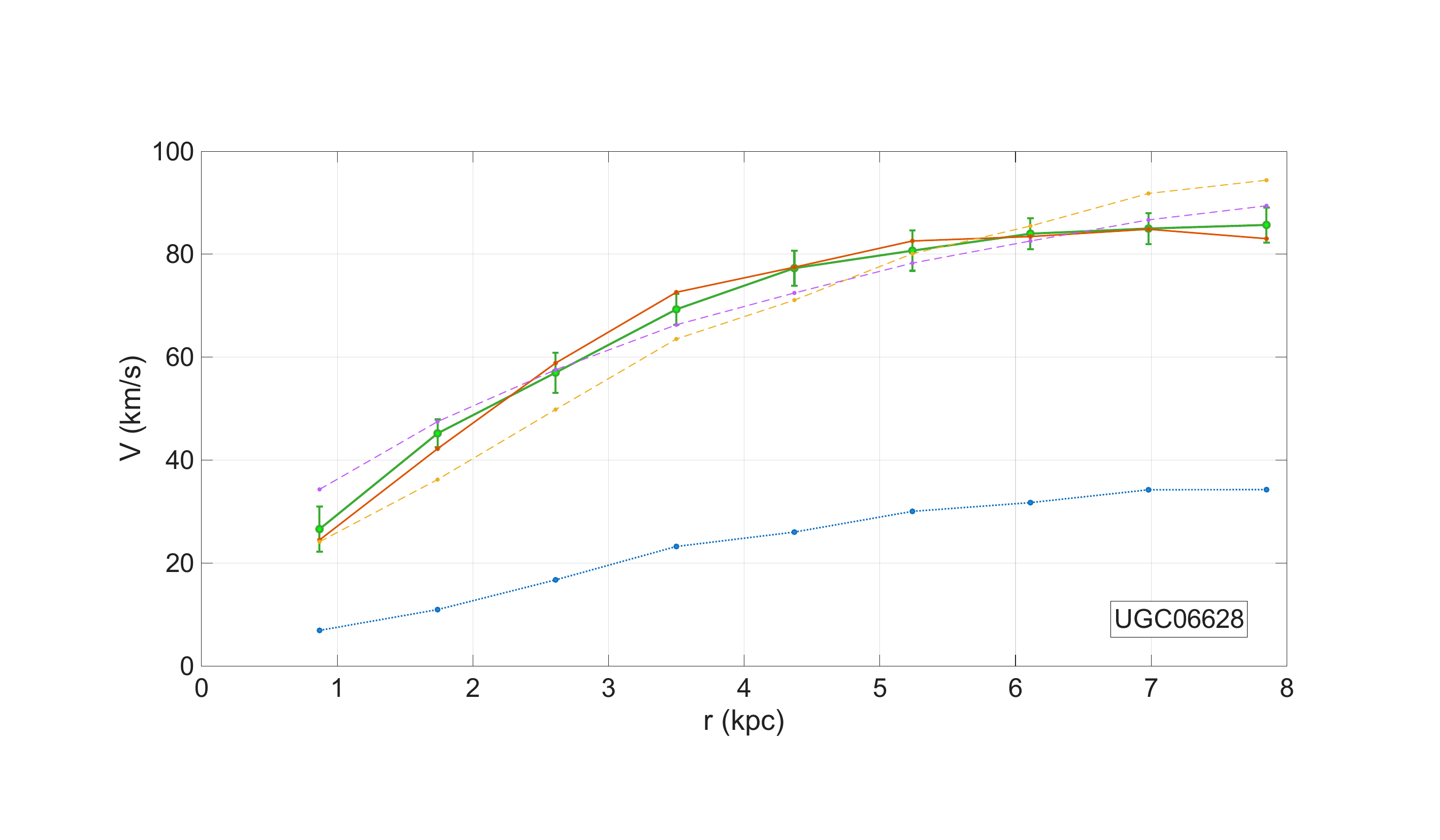}
\includegraphics[trim=4cm 3cm 5cm 4cm, clip=true, width=0.325\columnwidth]{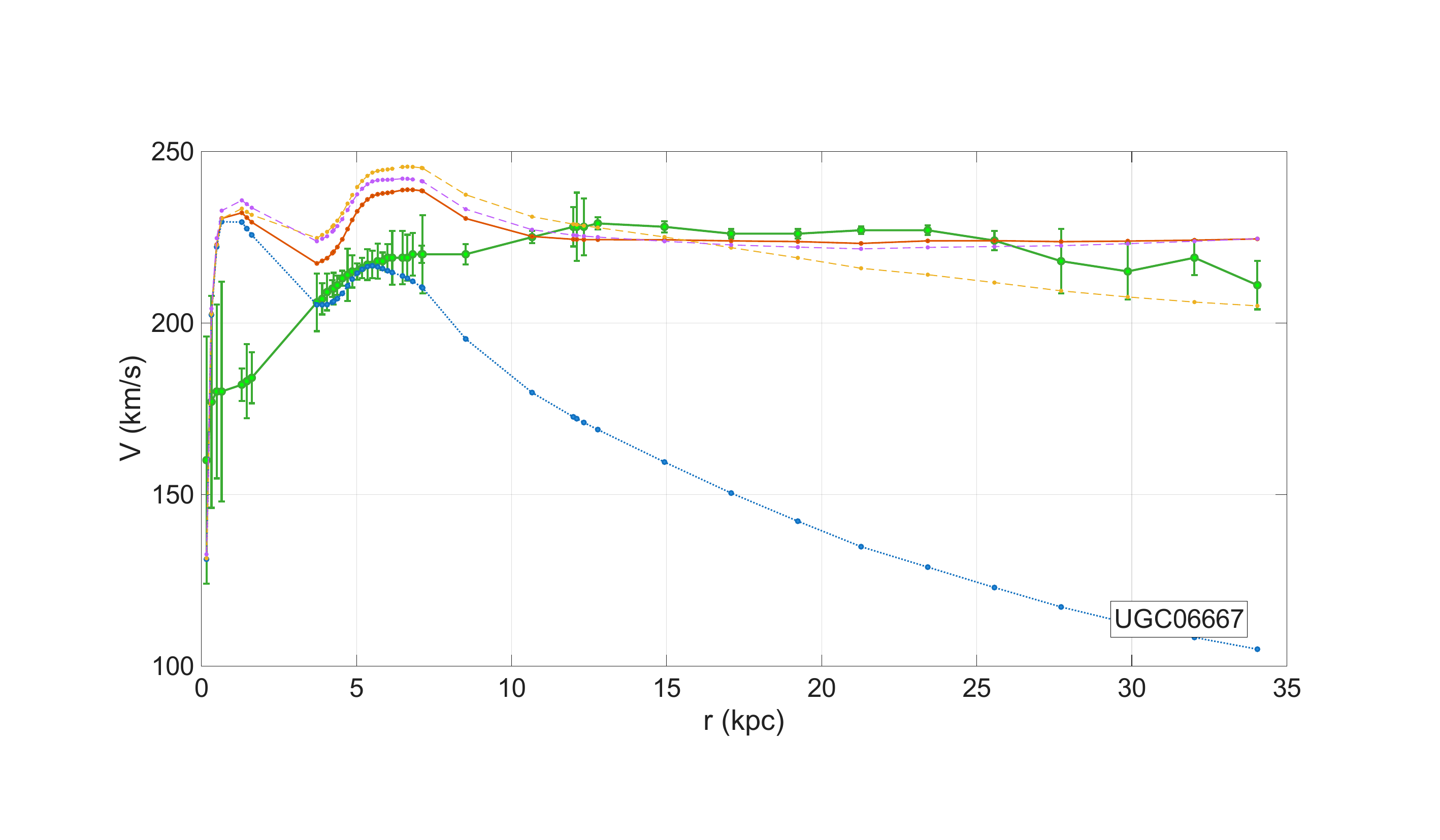}
\includegraphics[trim=4cm 3cm 5cm 4cm, clip=true, width=0.325\columnwidth]{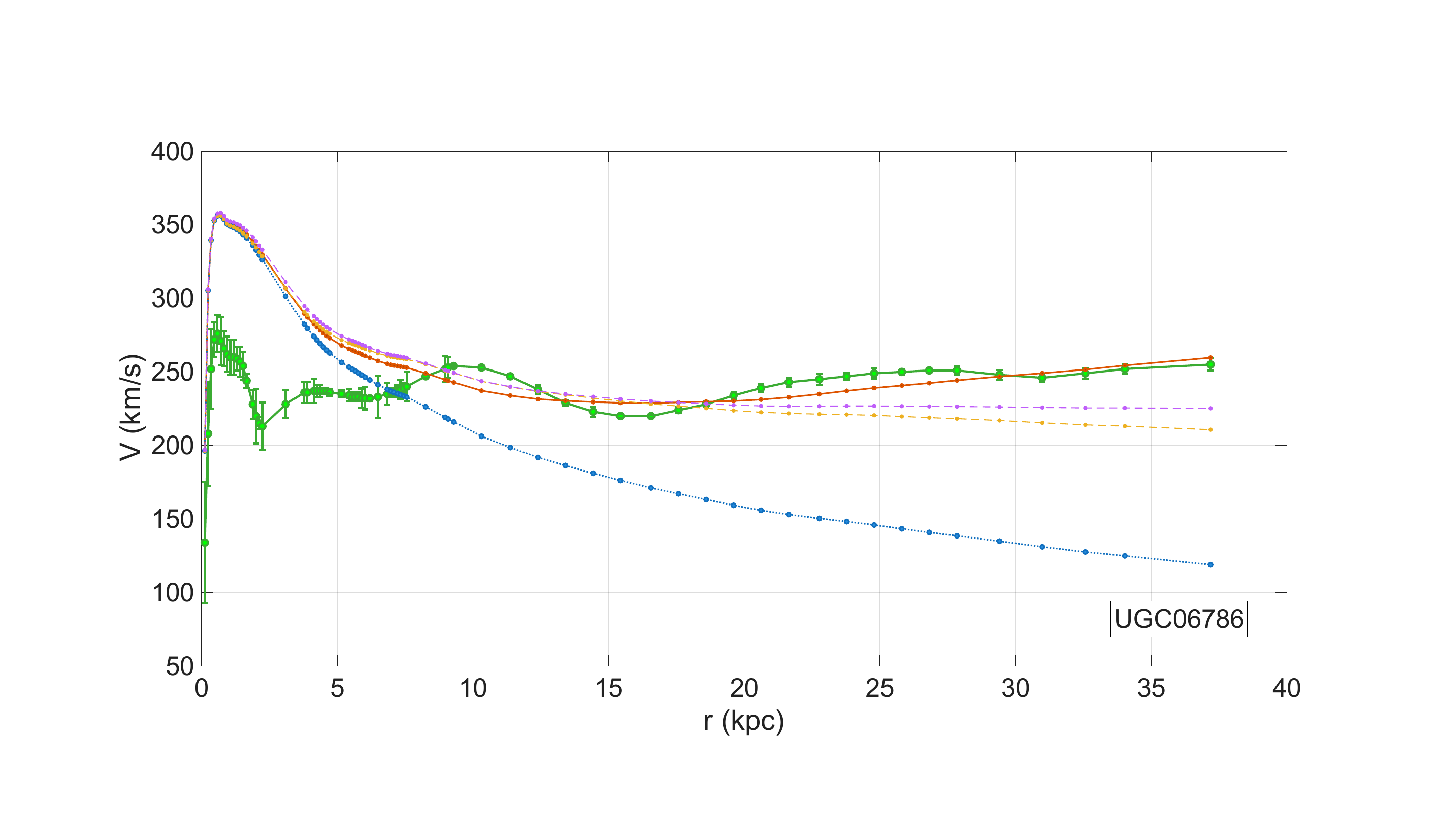}
\includegraphics[trim=4cm 3cm 5cm 4cm, clip=true, width=0.325\columnwidth]{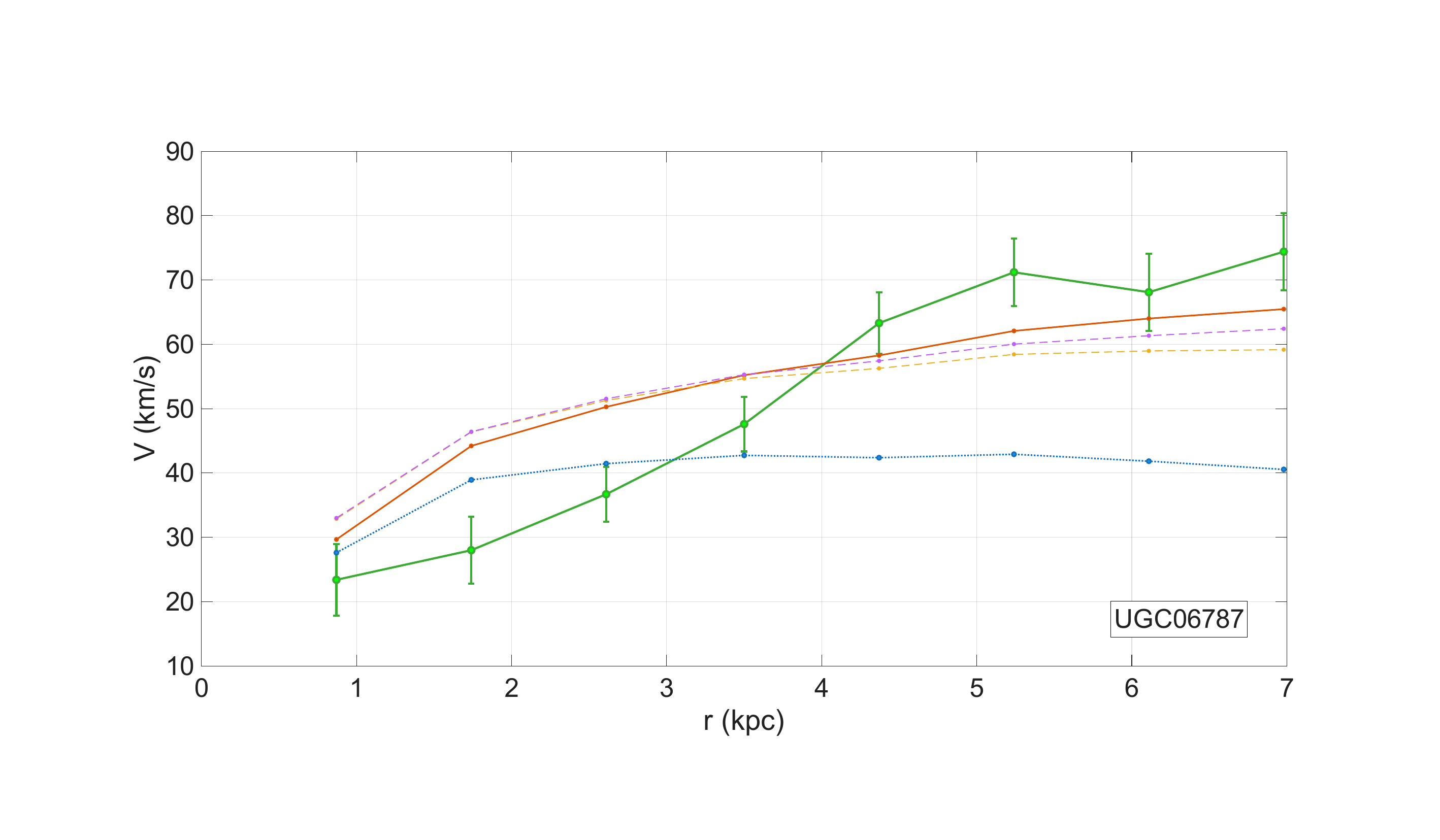}
\includegraphics[trim=4cm 3cm 5cm 4cm, clip=true, width=0.325\columnwidth]{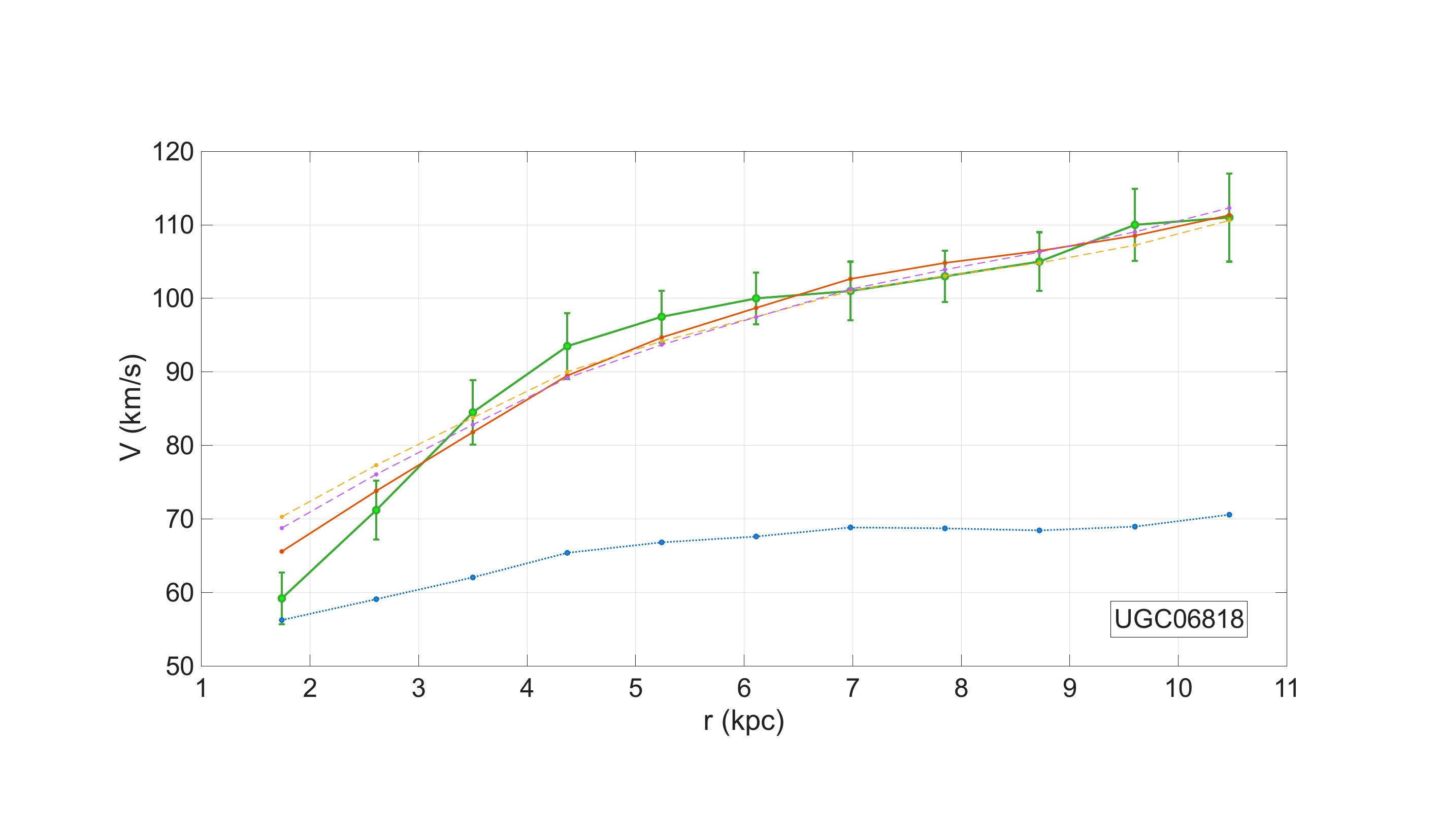}
\includegraphics[trim=4cm 3cm 5cm 4cm, clip=true, width=0.325\columnwidth]{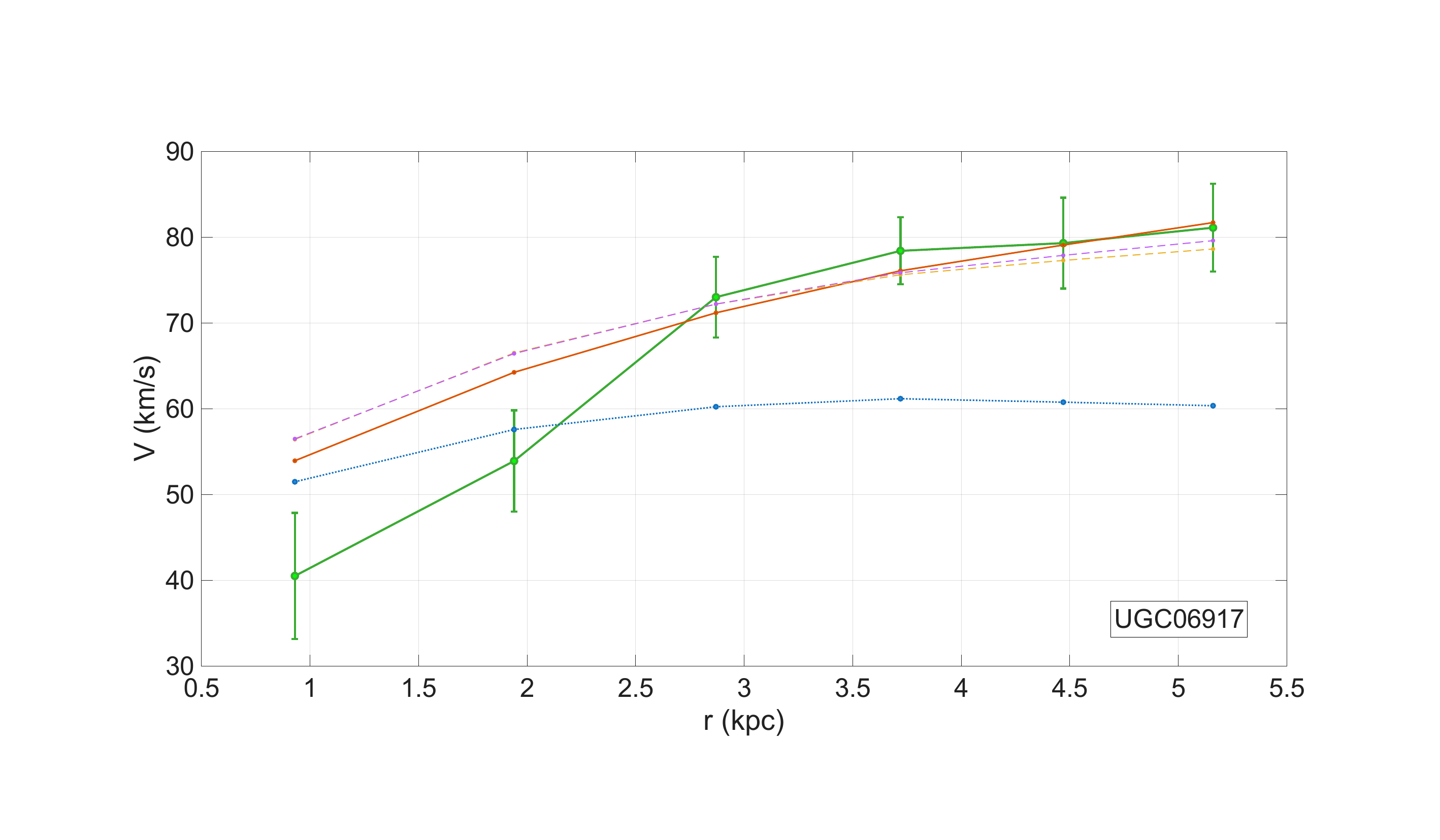}
\includegraphics[trim=4cm 3cm 5cm 4cm, clip=true, width=0.325\columnwidth]{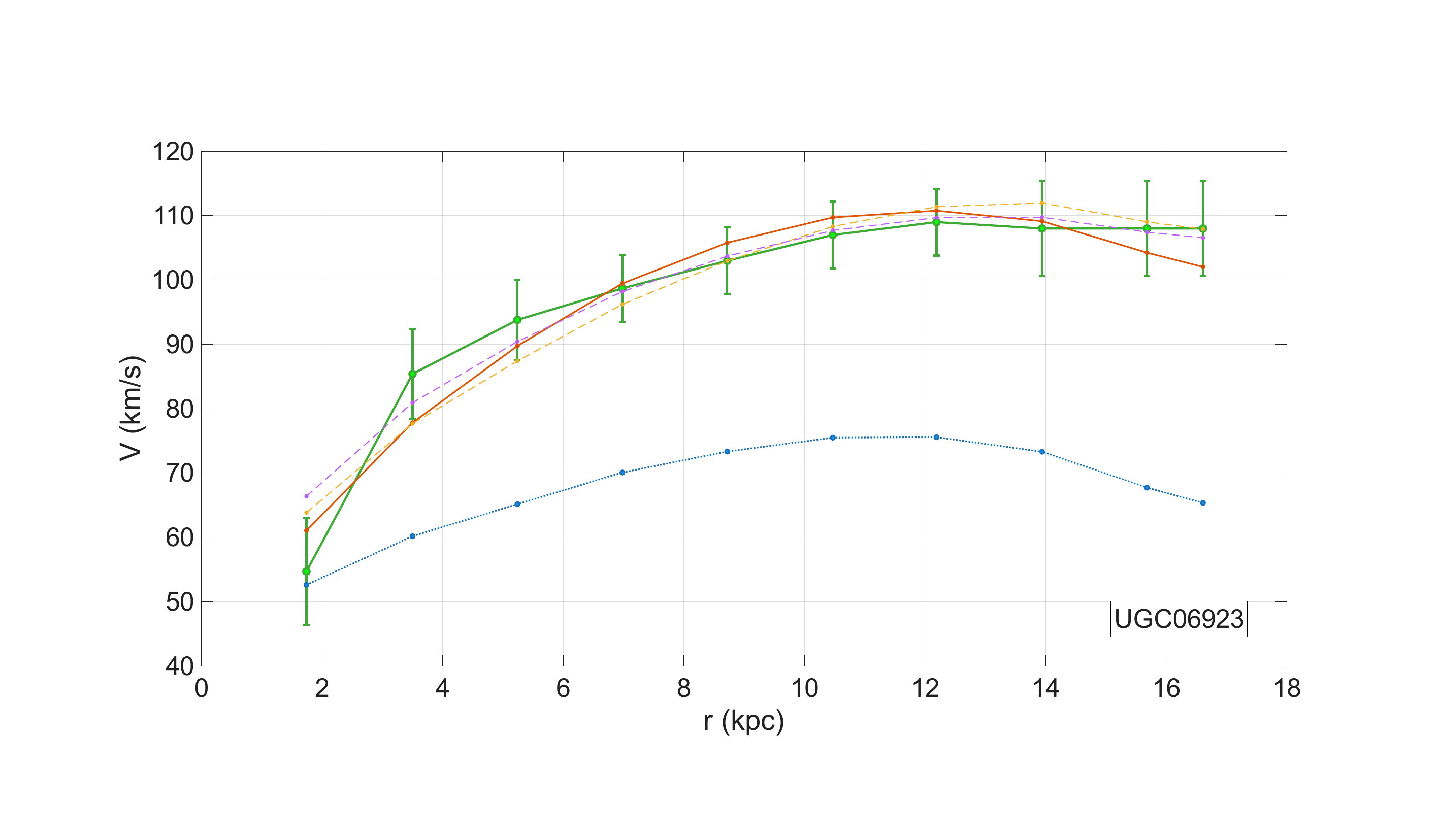}
\includegraphics[trim=4cm 3cm 5cm 4cm, clip=true, width=0.325\columnwidth]{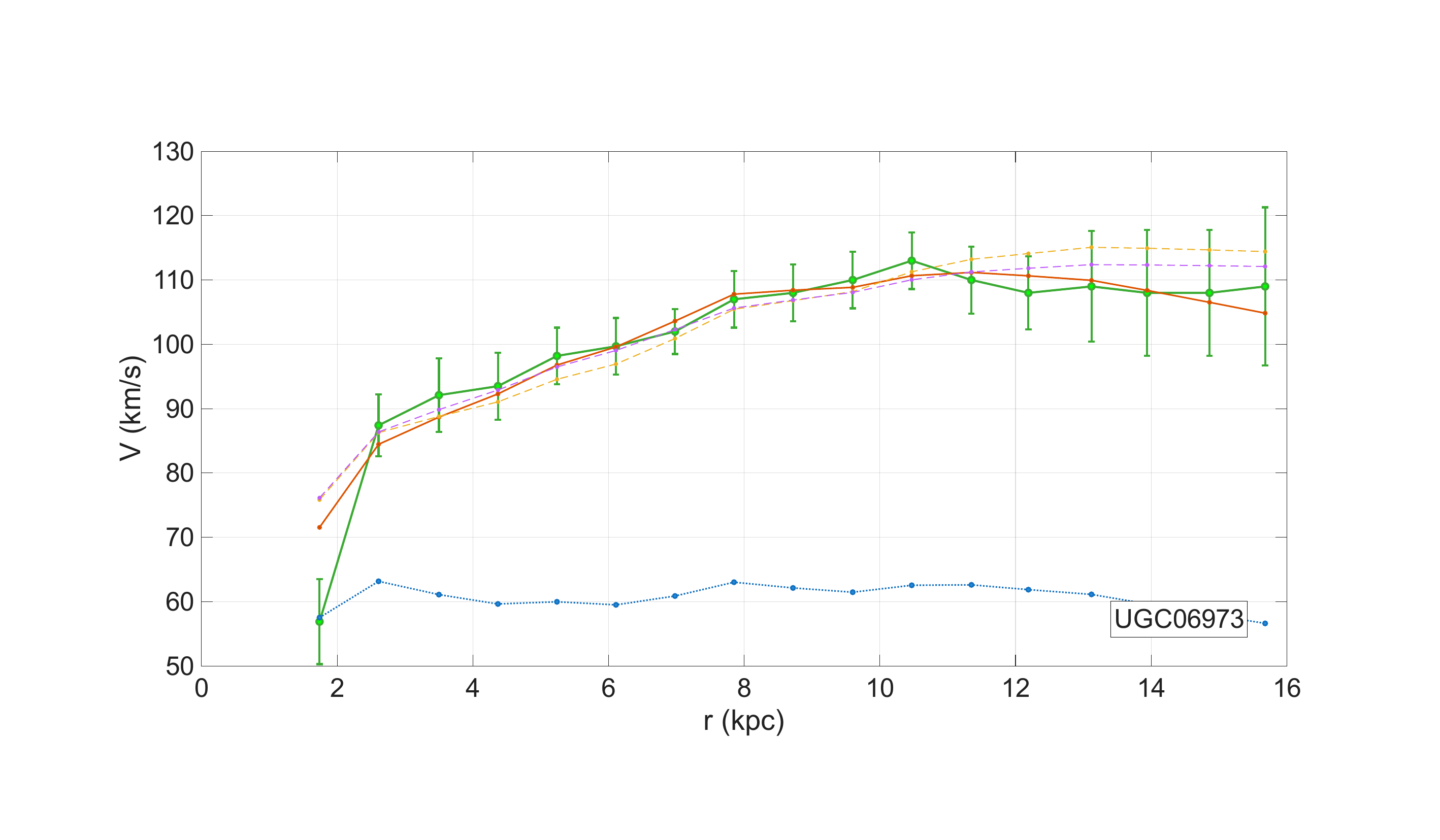}
\includegraphics[trim=4cm 3cm 5cm 4cm, clip=true, width=0.325\columnwidth]{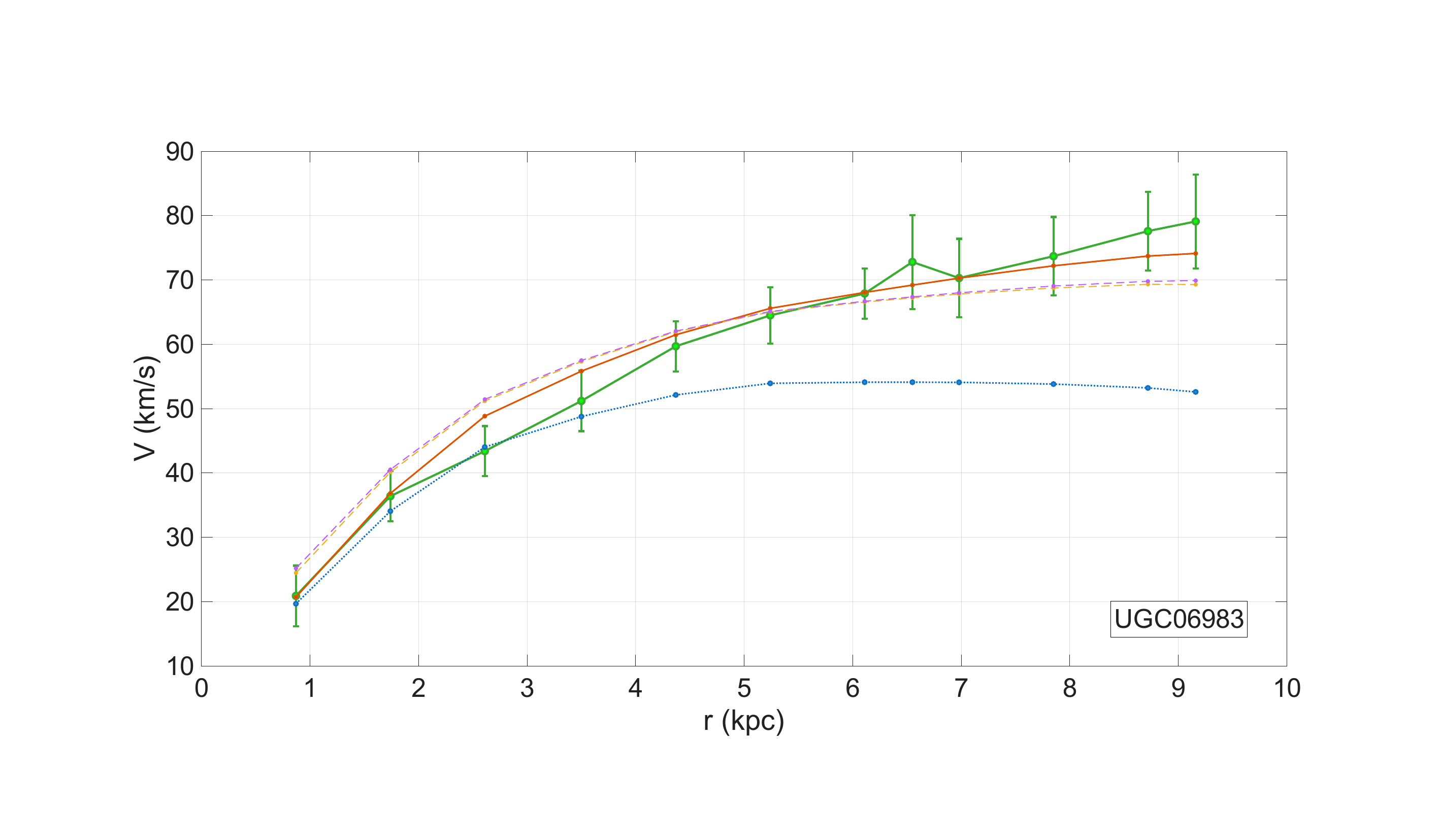}
\includegraphics[trim=4cm 3cm 5cm 4cm, clip=true, width=0.325\columnwidth]{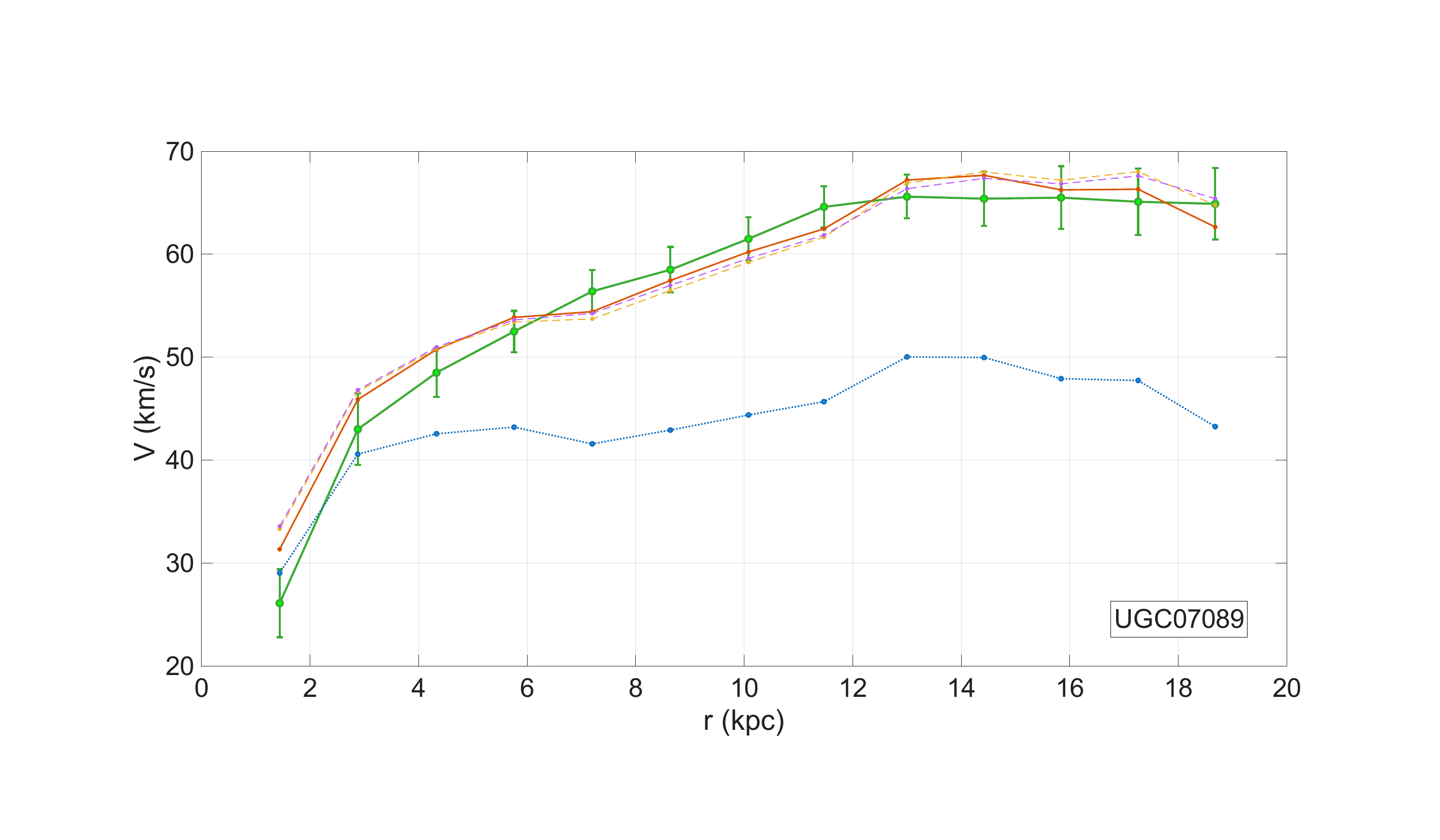}
\includegraphics[trim=4cm 3cm 5cm 4cm, clip=true, width=0.325\columnwidth]{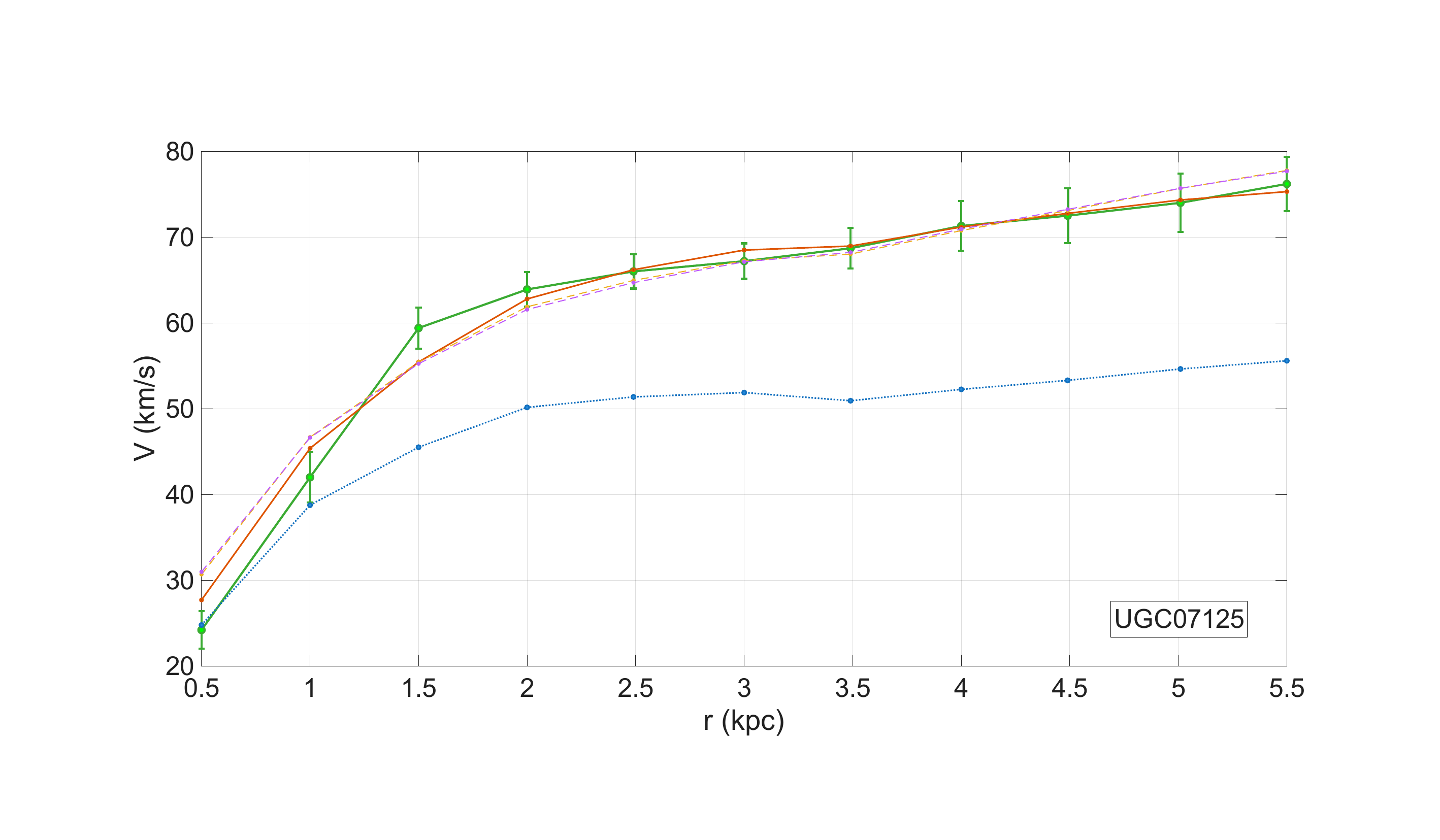}
\includegraphics[trim=4cm 3cm 5cm 4cm, clip=true, width=0.325\columnwidth]{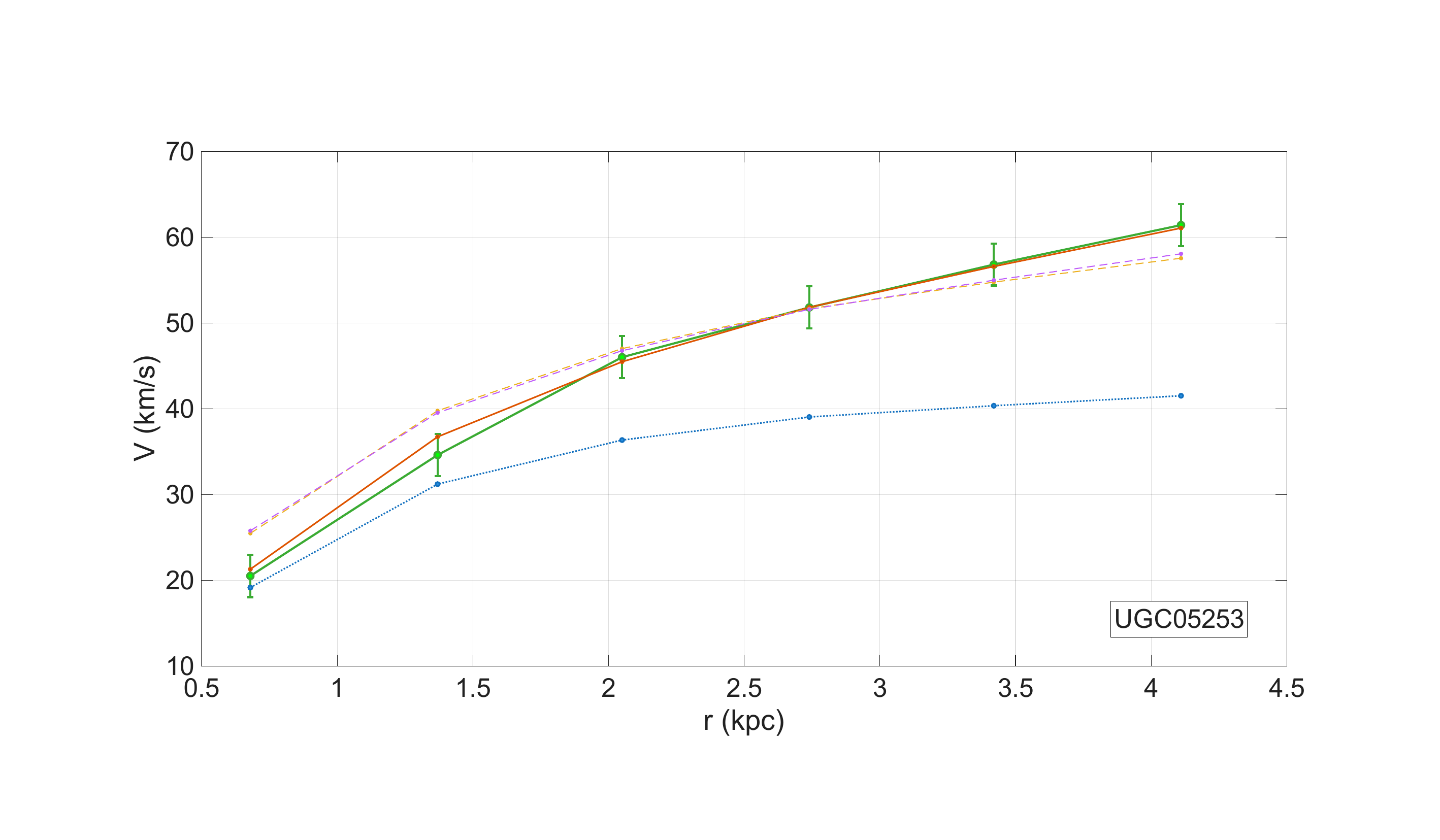}
\includegraphics[trim=4cm 3cm 5cm 4cm, clip=true, width=0.325\columnwidth]{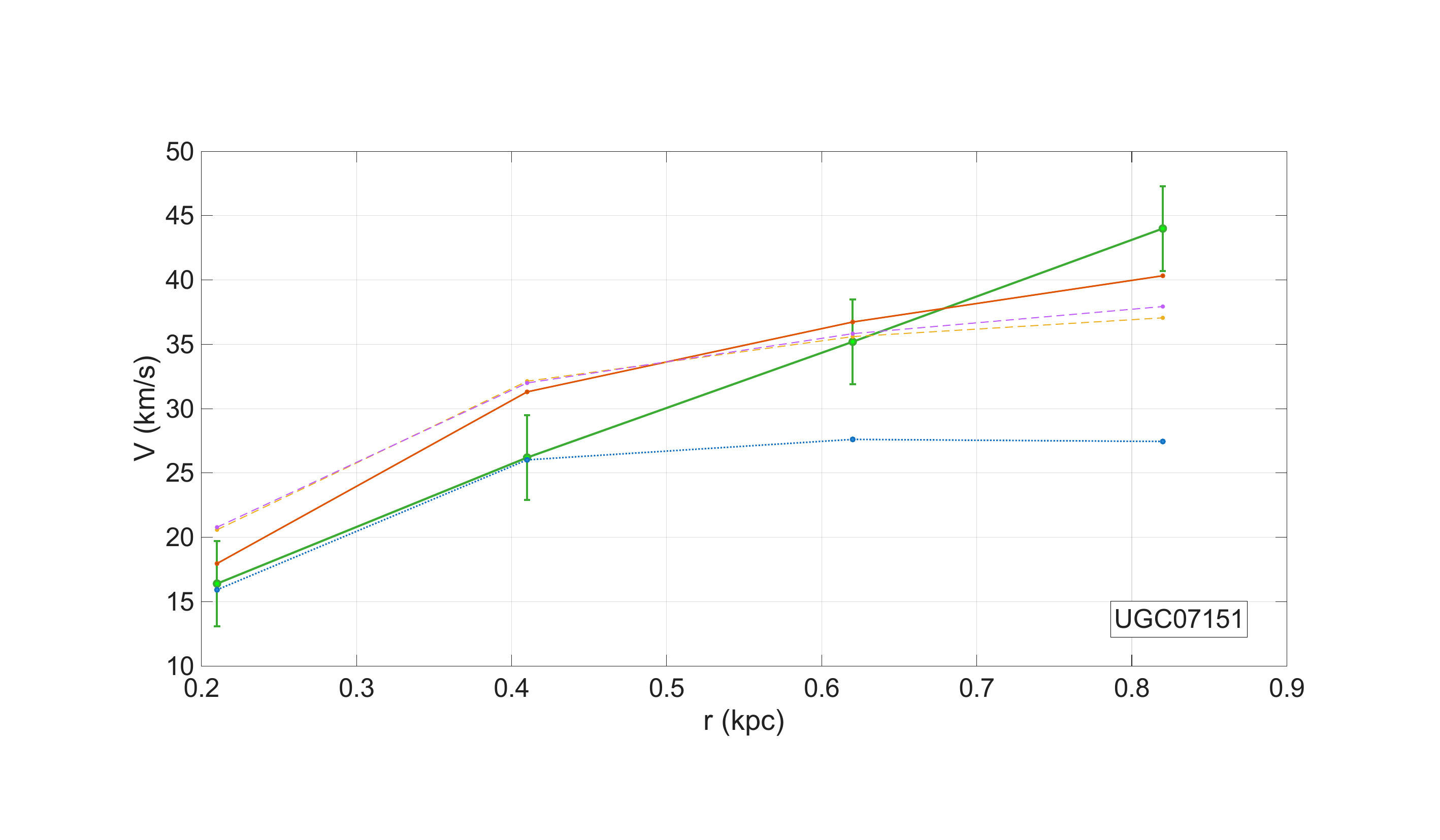}
\end{figure}

\begin{figure}
\centering
\includegraphics[trim=4cm 3cm 5cm 4cm, clip=true, width=0.325\columnwidth]{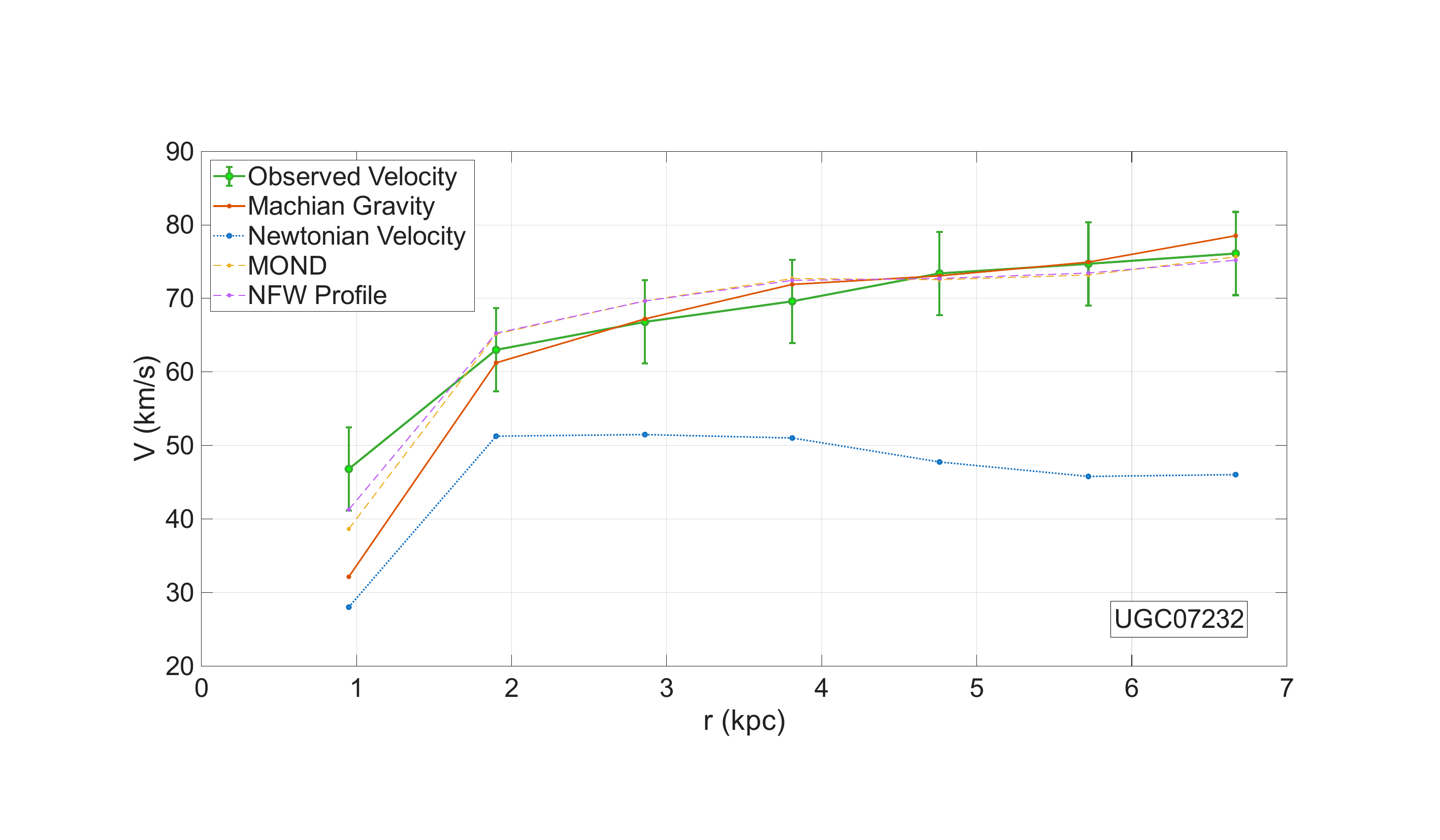}
\includegraphics[trim=4cm 3cm 5cm 4cm, clip=true, width=0.325\columnwidth]{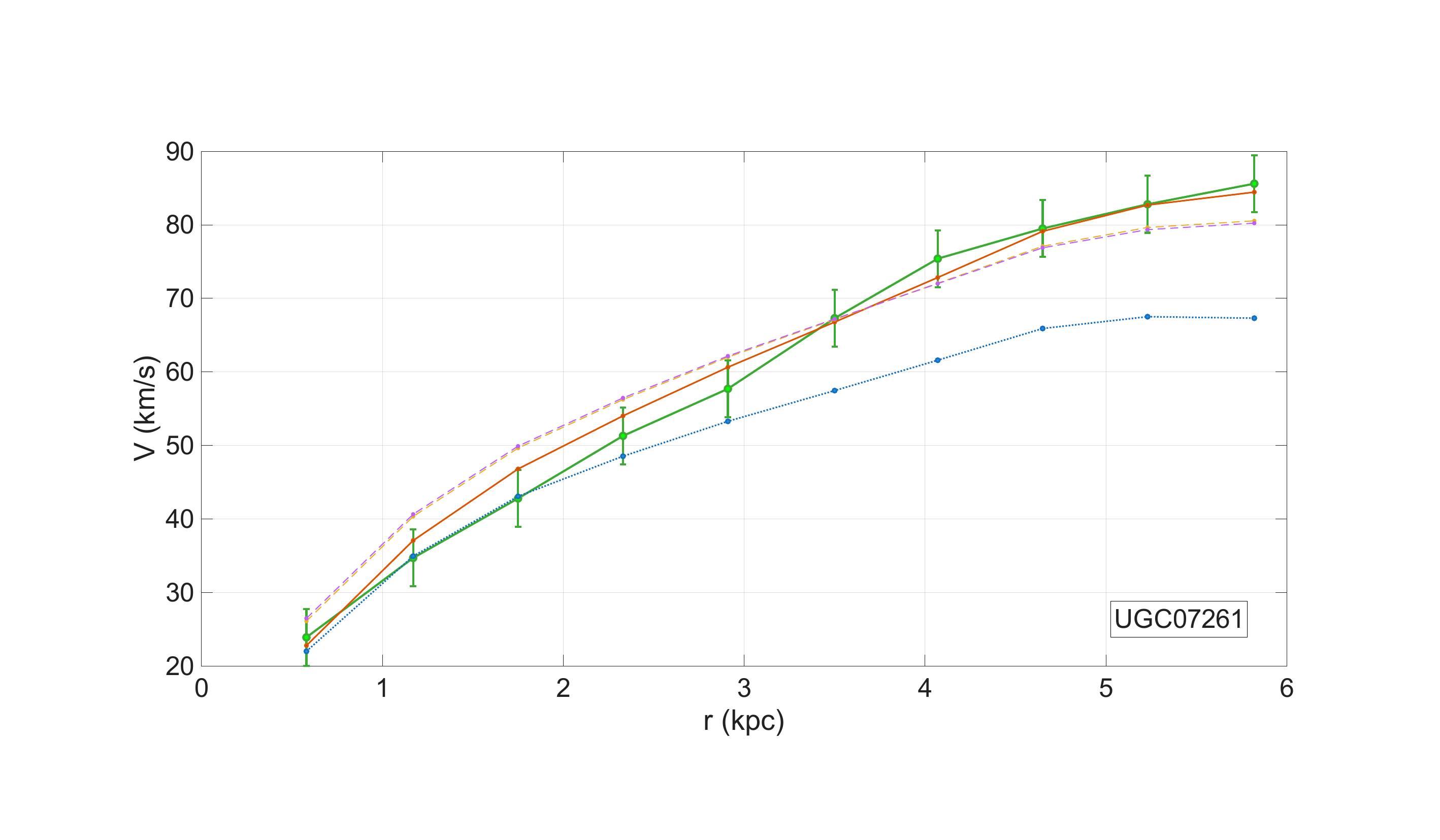}
\includegraphics[trim=4cm 3cm 5cm 4cm, clip=true, width=0.325\columnwidth]{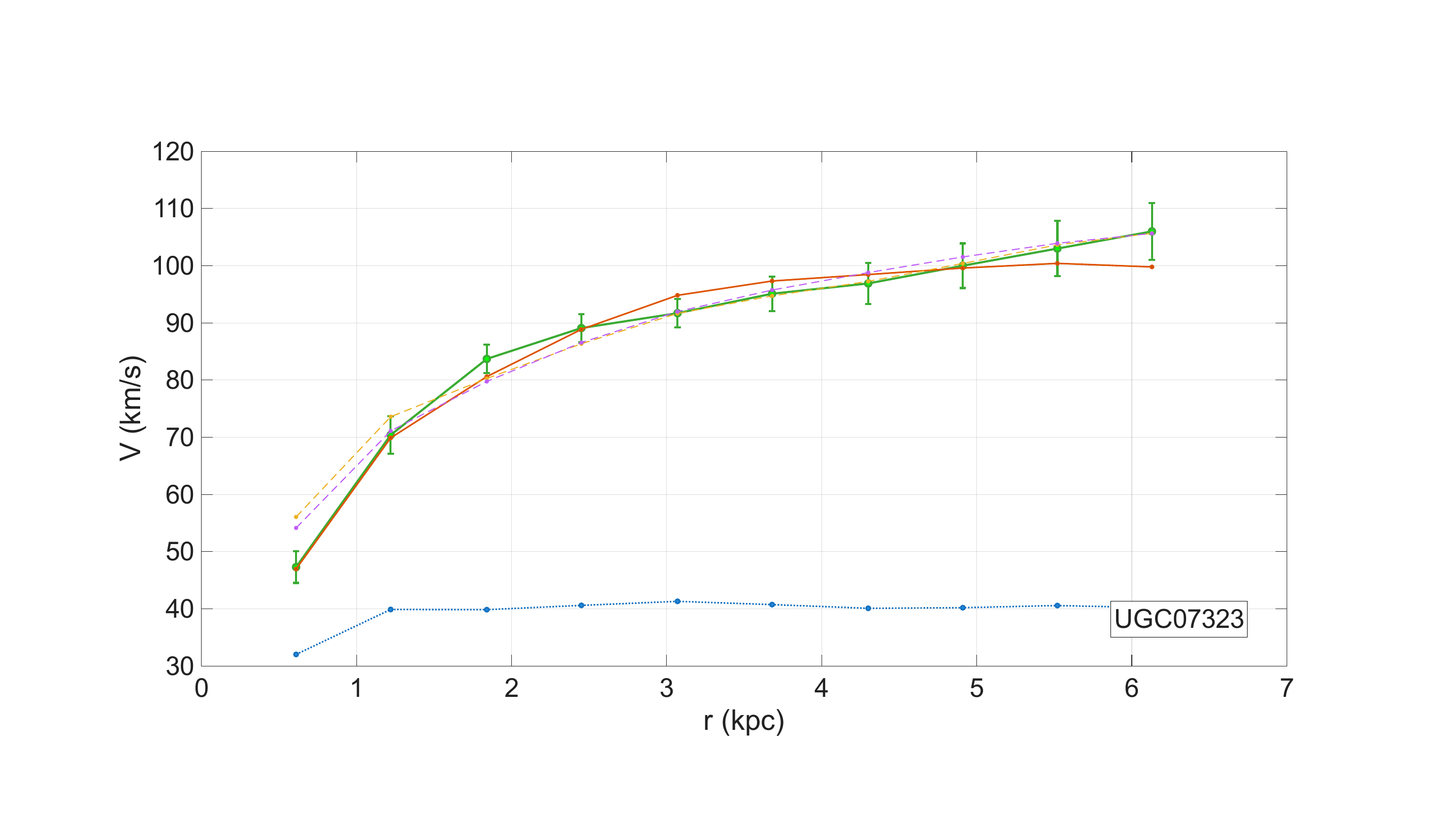}
\includegraphics[trim=4cm 3cm 5cm 4cm, clip=true, width=0.325\columnwidth]{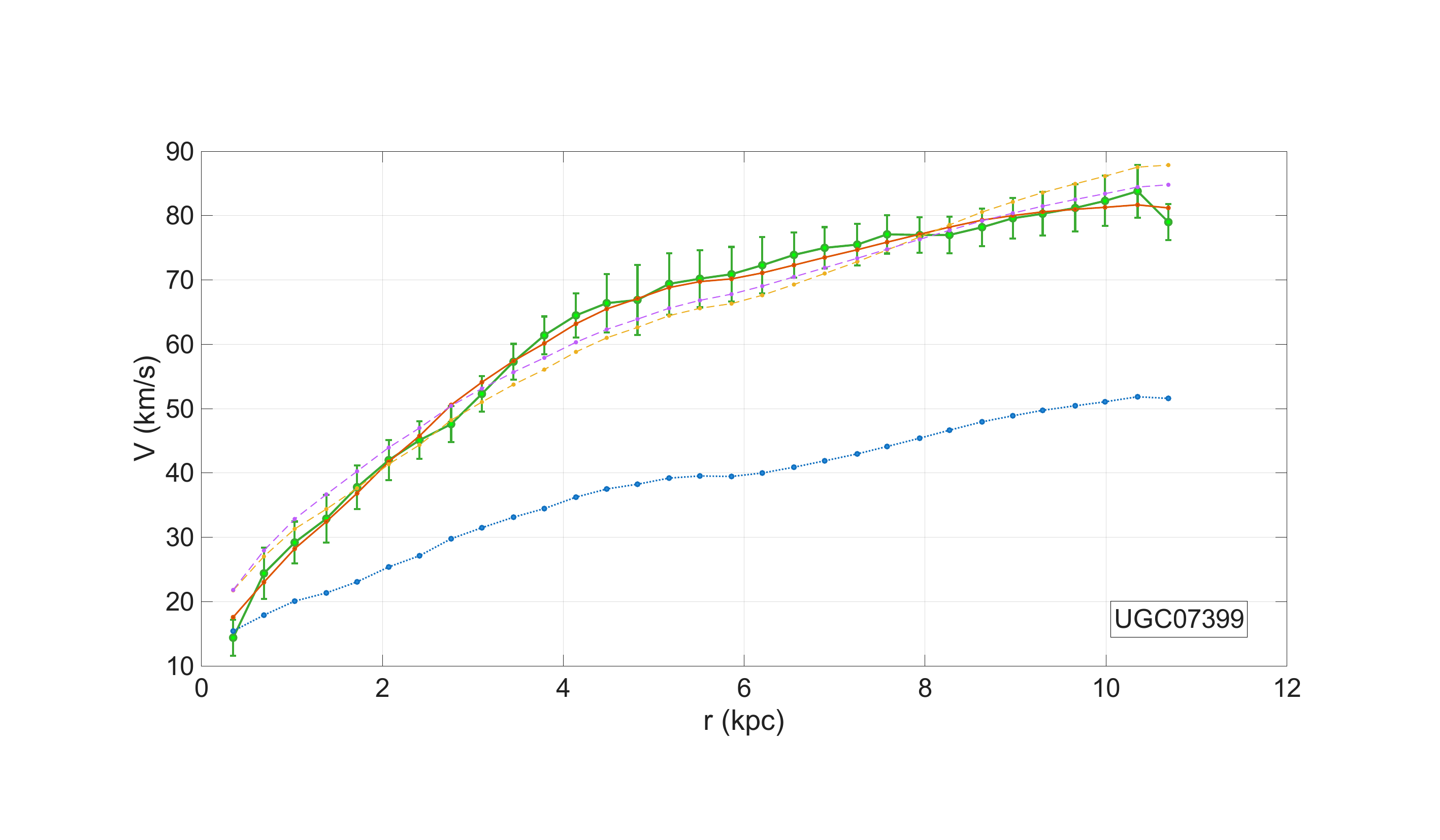}
\includegraphics[trim=4cm 3cm 5cm 4cm, clip=true, width=0.325\columnwidth]{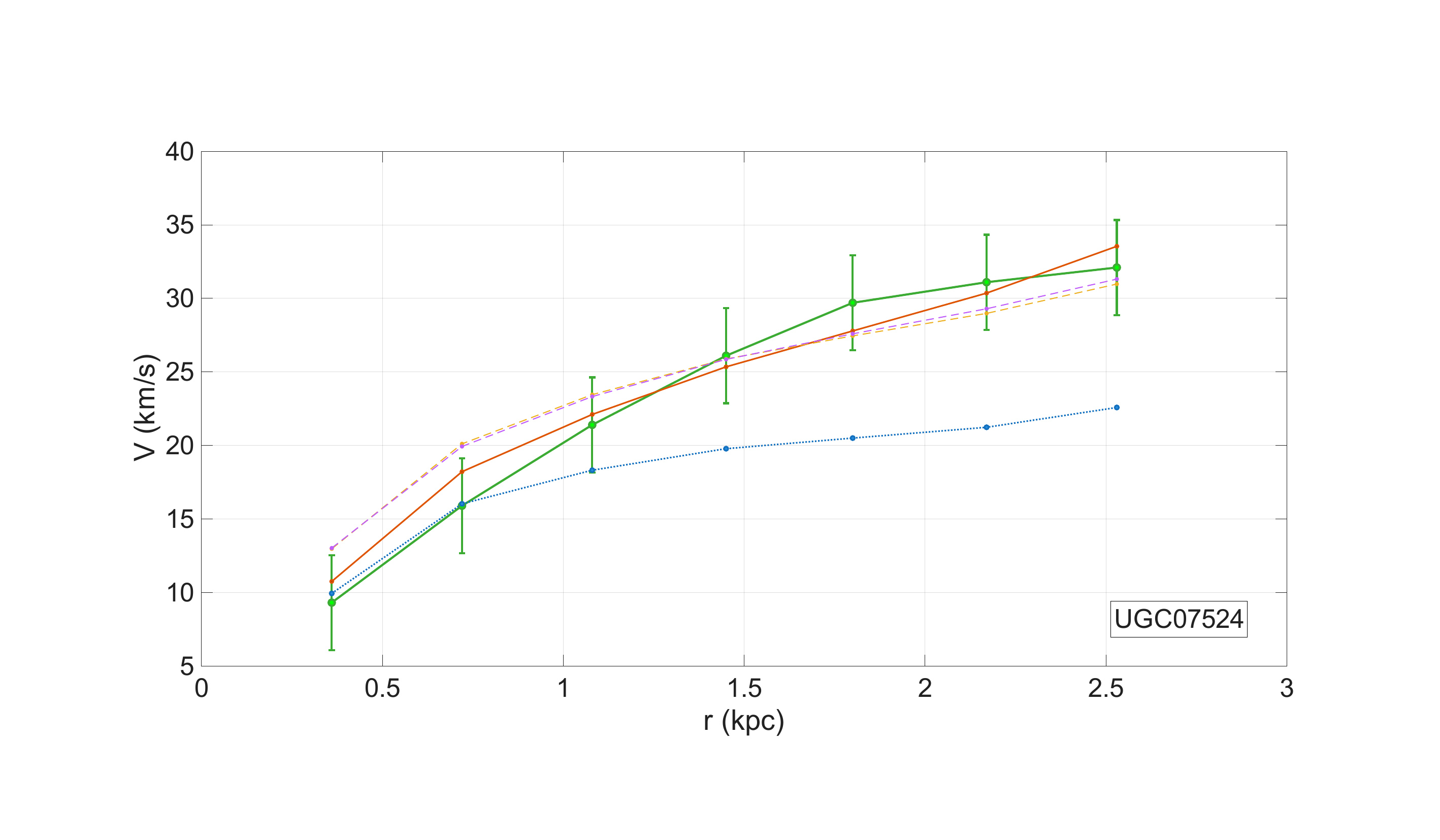}
\includegraphics[trim=4cm 3cm 5cm 4cm, clip=true, width=0.325\columnwidth]{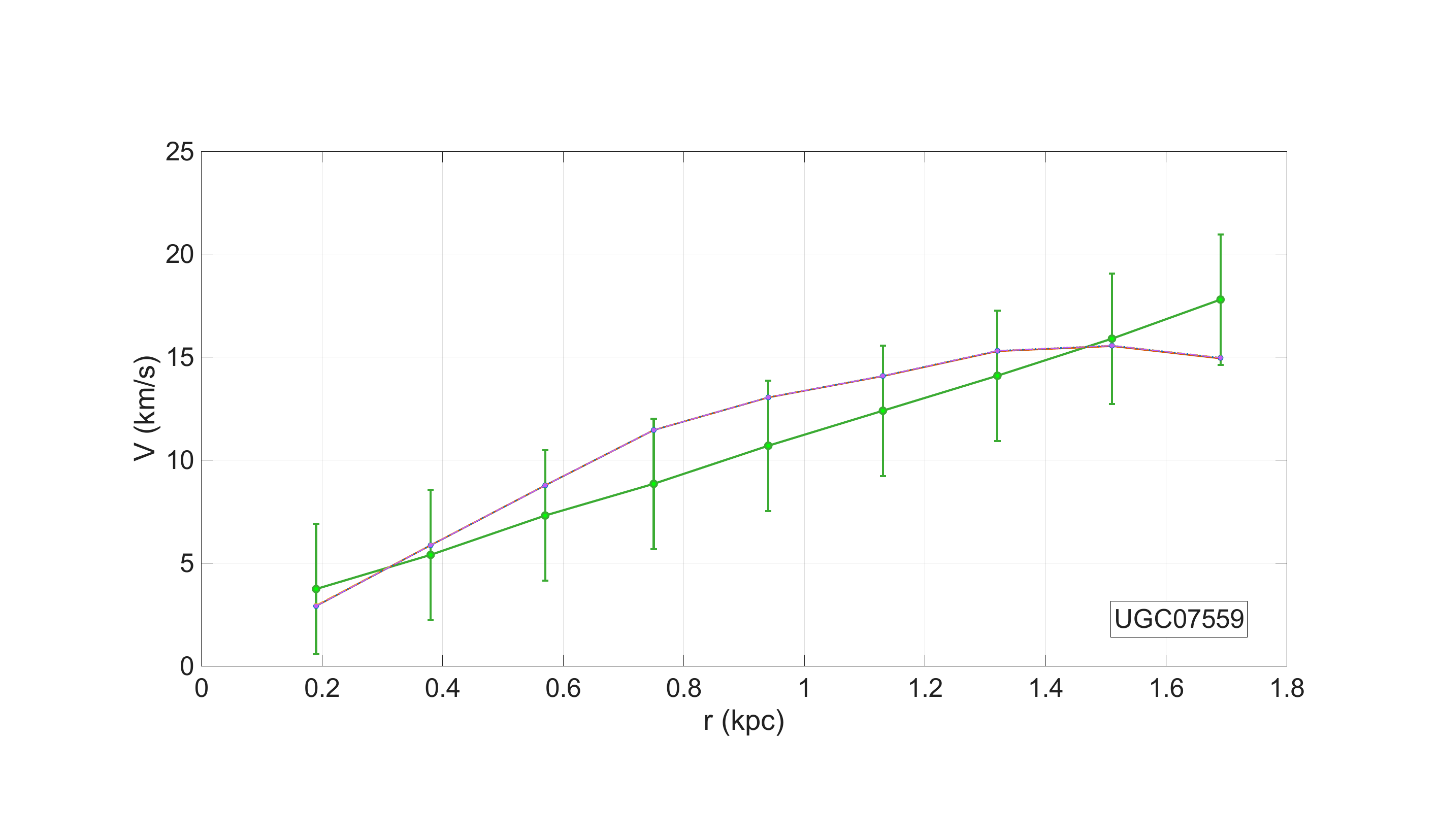}
\includegraphics[trim=4cm 3cm 5cm 4cm, clip=true, width=0.325\columnwidth]{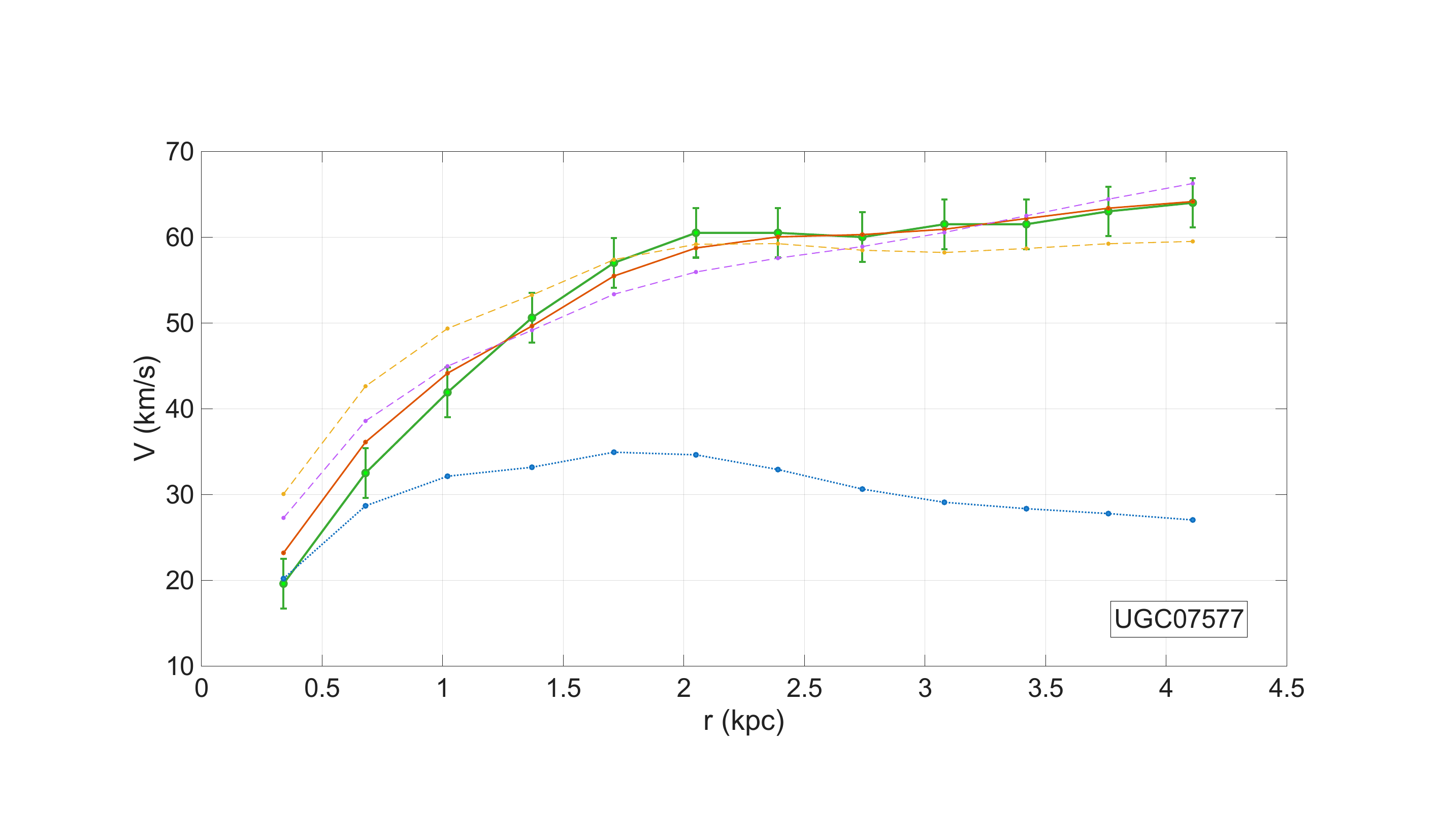}
\includegraphics[trim=4cm 3cm 5cm 4cm, clip=true, width=0.325\columnwidth]{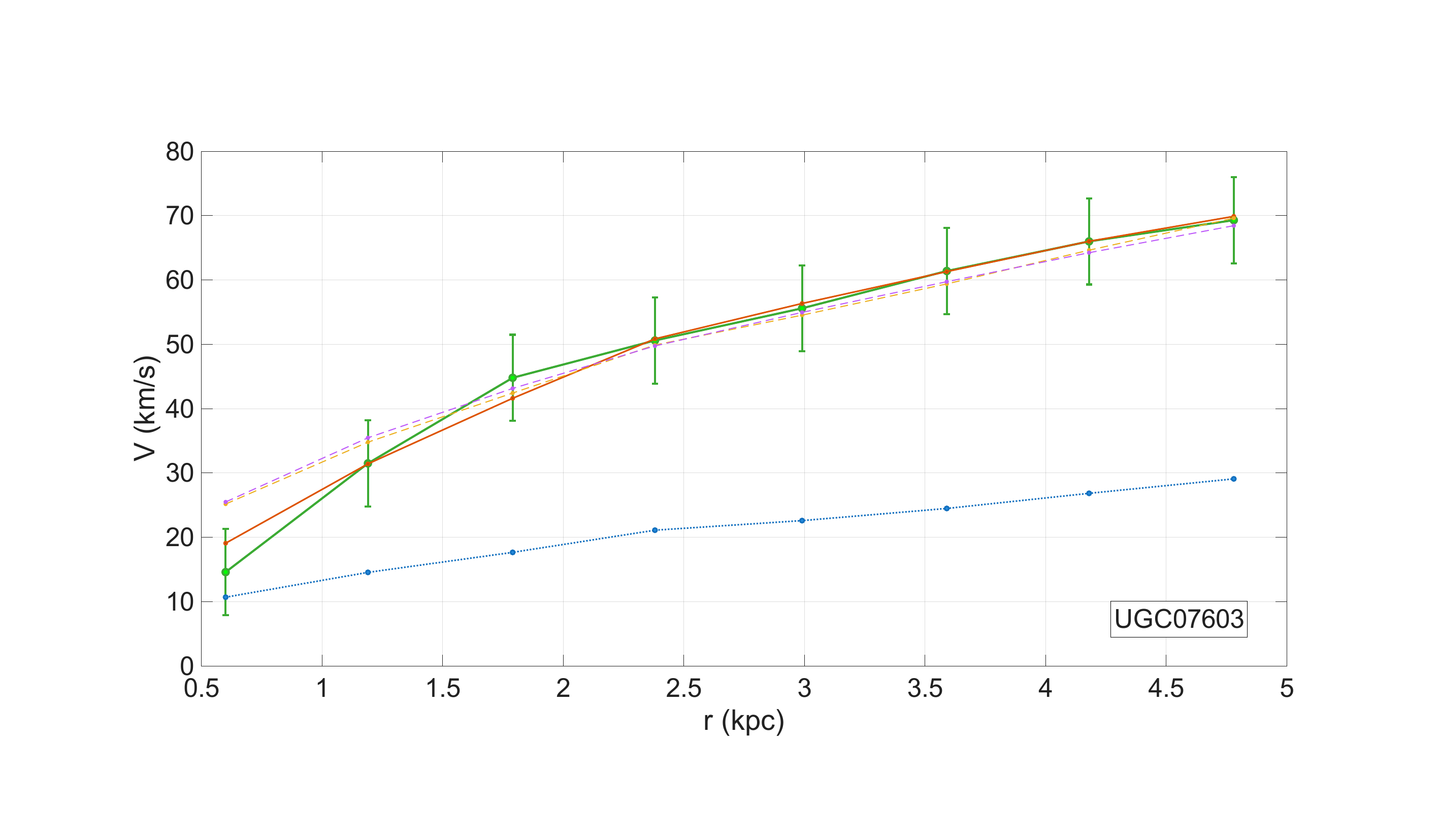}
\includegraphics[trim=4cm 3cm 5cm 4cm, clip=true, width=0.325\columnwidth]{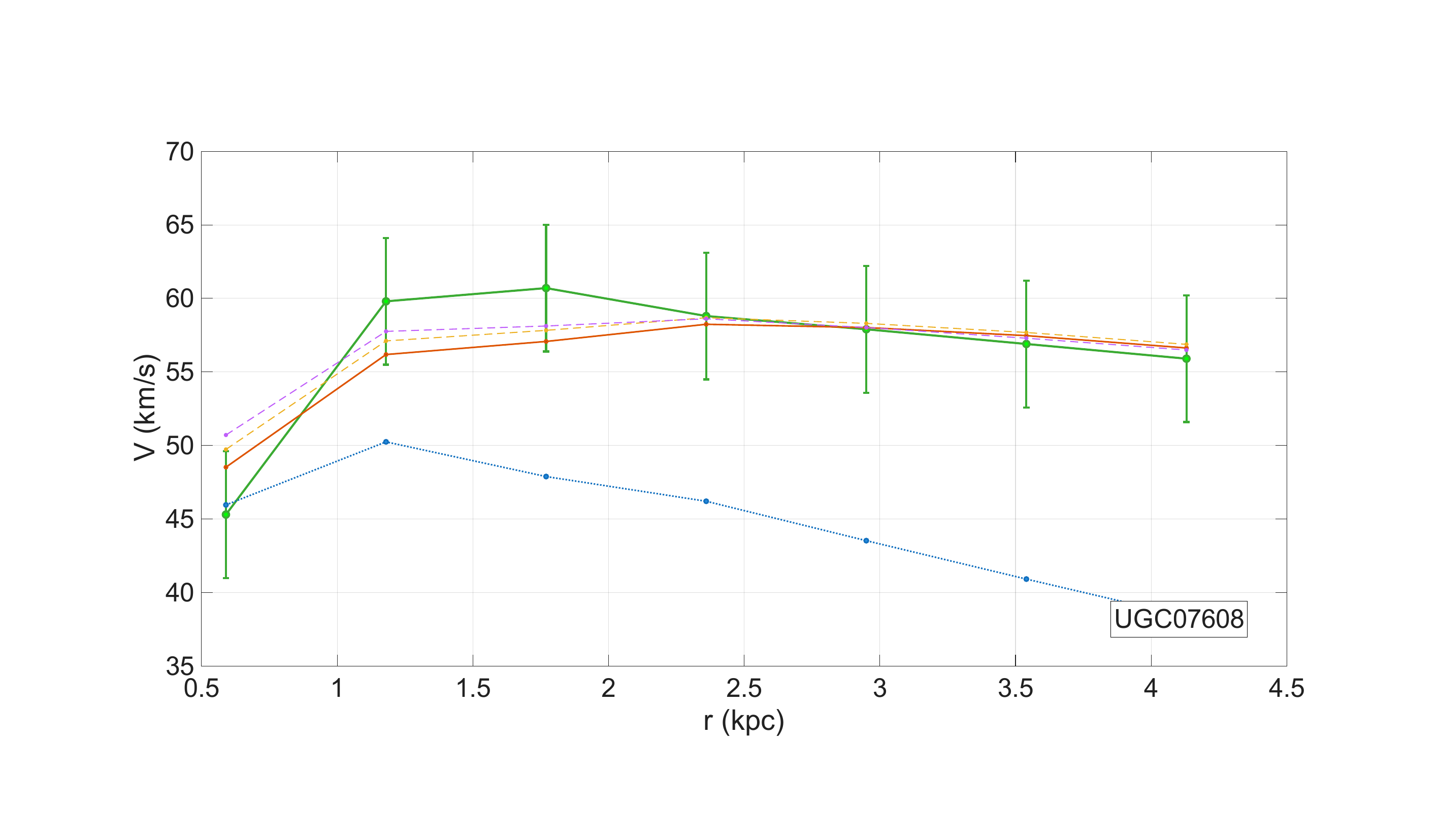}
\includegraphics[trim=4cm 3cm 5cm 4cm, clip=true, width=0.325\columnwidth]{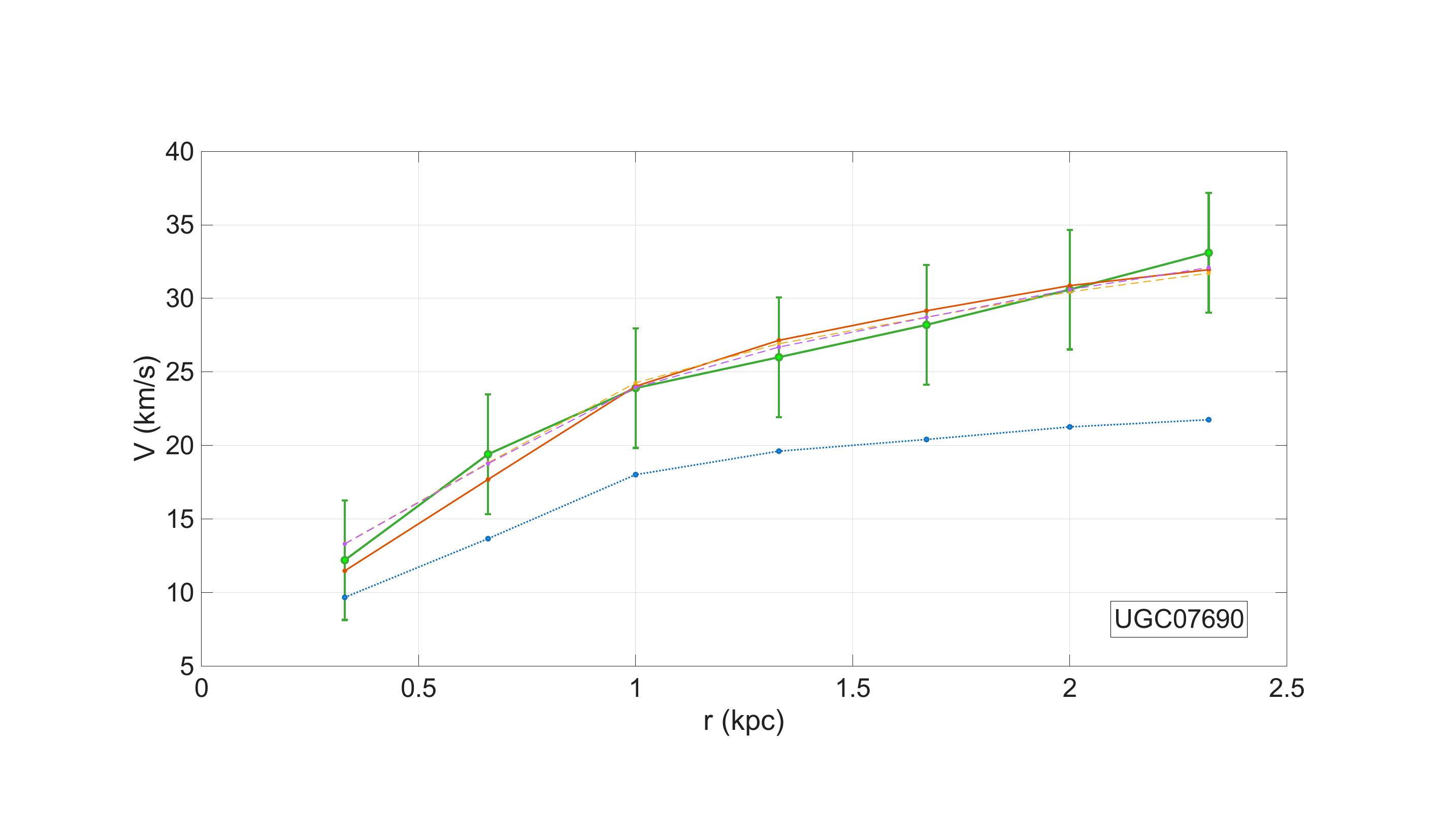}
\includegraphics[trim=4cm 3cm 5cm 4cm, clip=true, width=0.325\columnwidth]{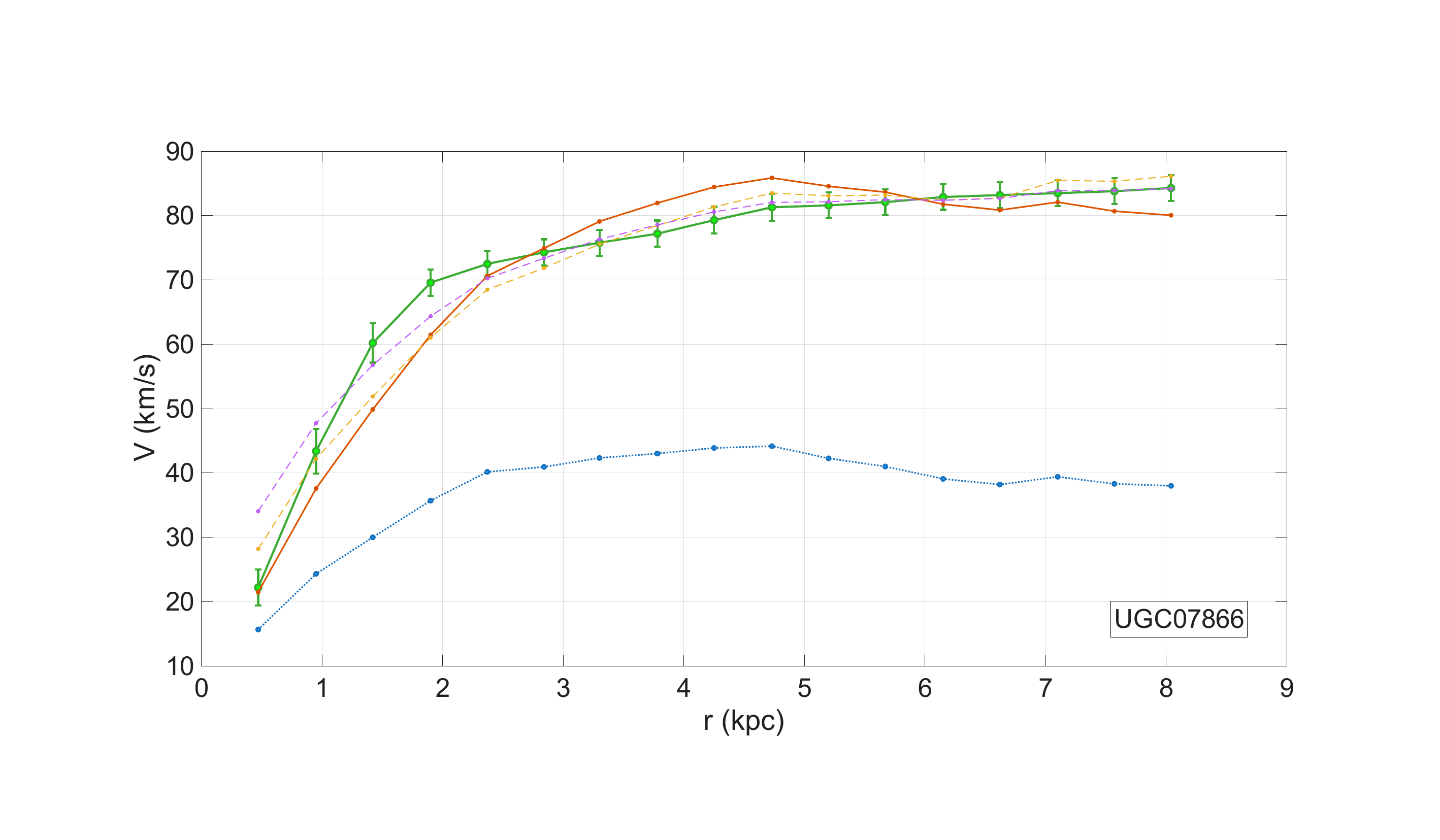}
\includegraphics[trim=4cm 3cm 5cm 4cm, clip=true, width=0.325\columnwidth]{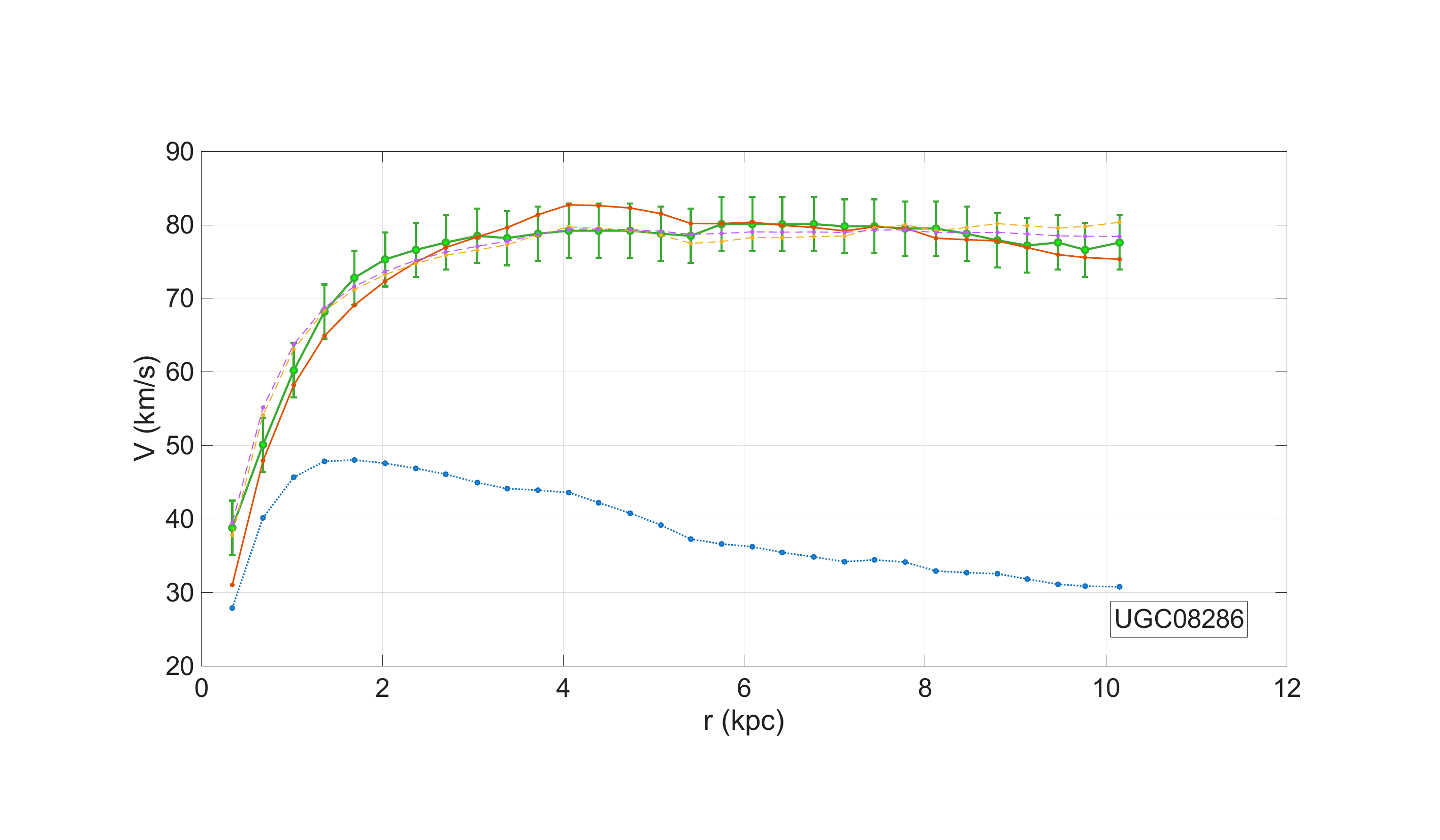}
\includegraphics[trim=4cm 3cm 5cm 4cm, clip=true, width=0.325\columnwidth]{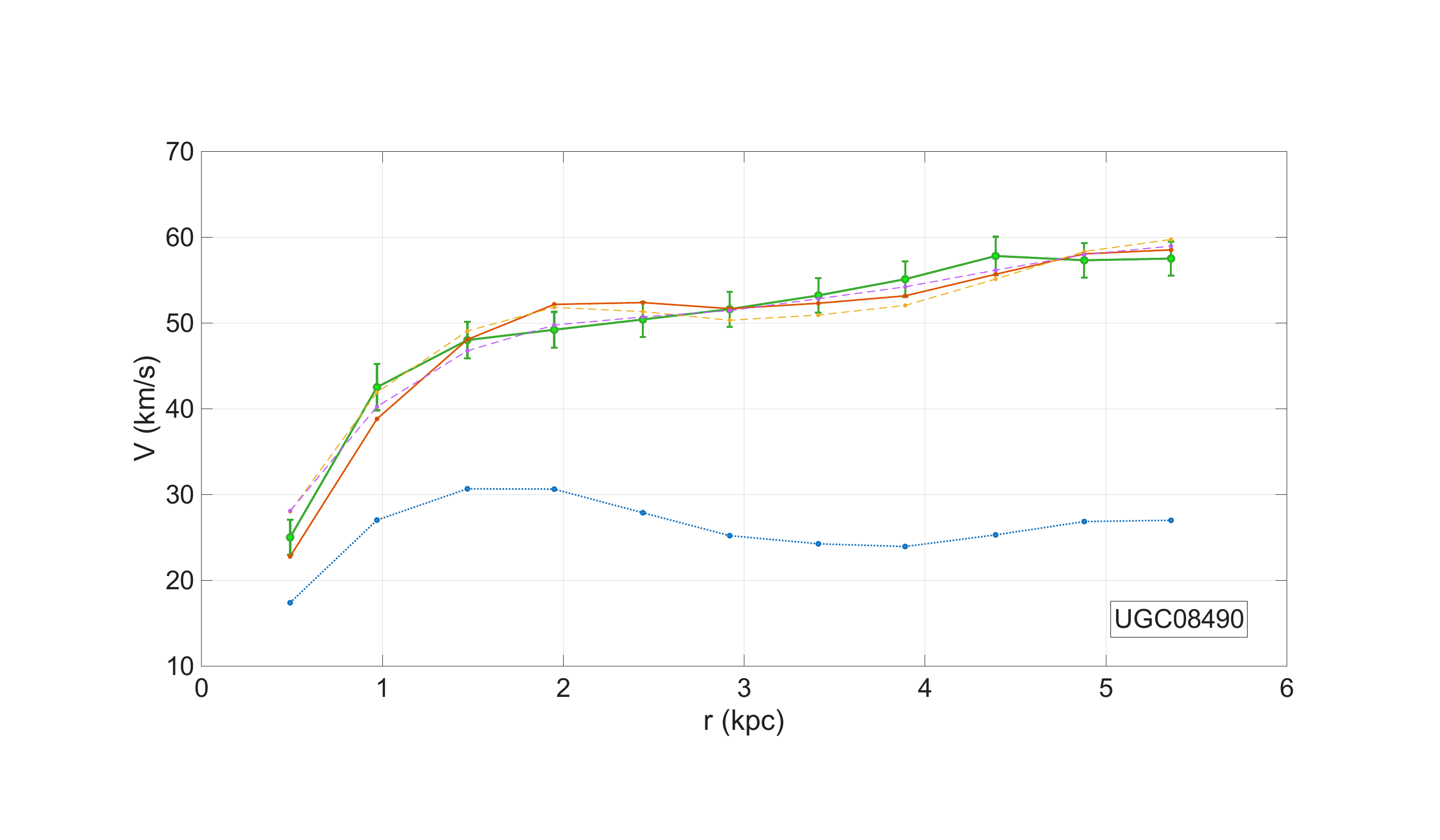}
\includegraphics[trim=4cm 3cm 5cm 4cm, clip=true, width=0.325\columnwidth]{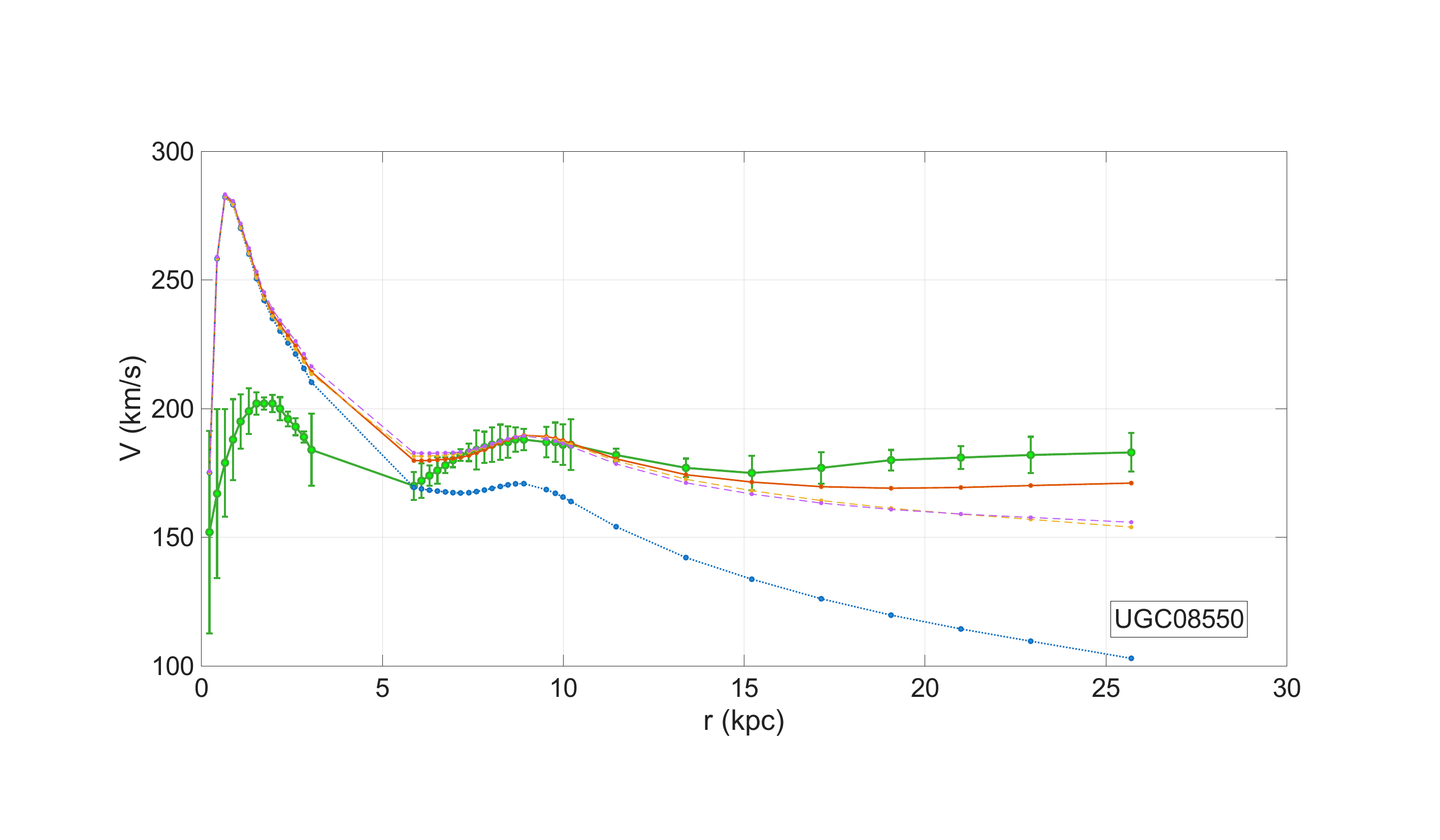}
\includegraphics[trim=4cm 3cm 5cm 4cm, clip=true, width=0.325\columnwidth]{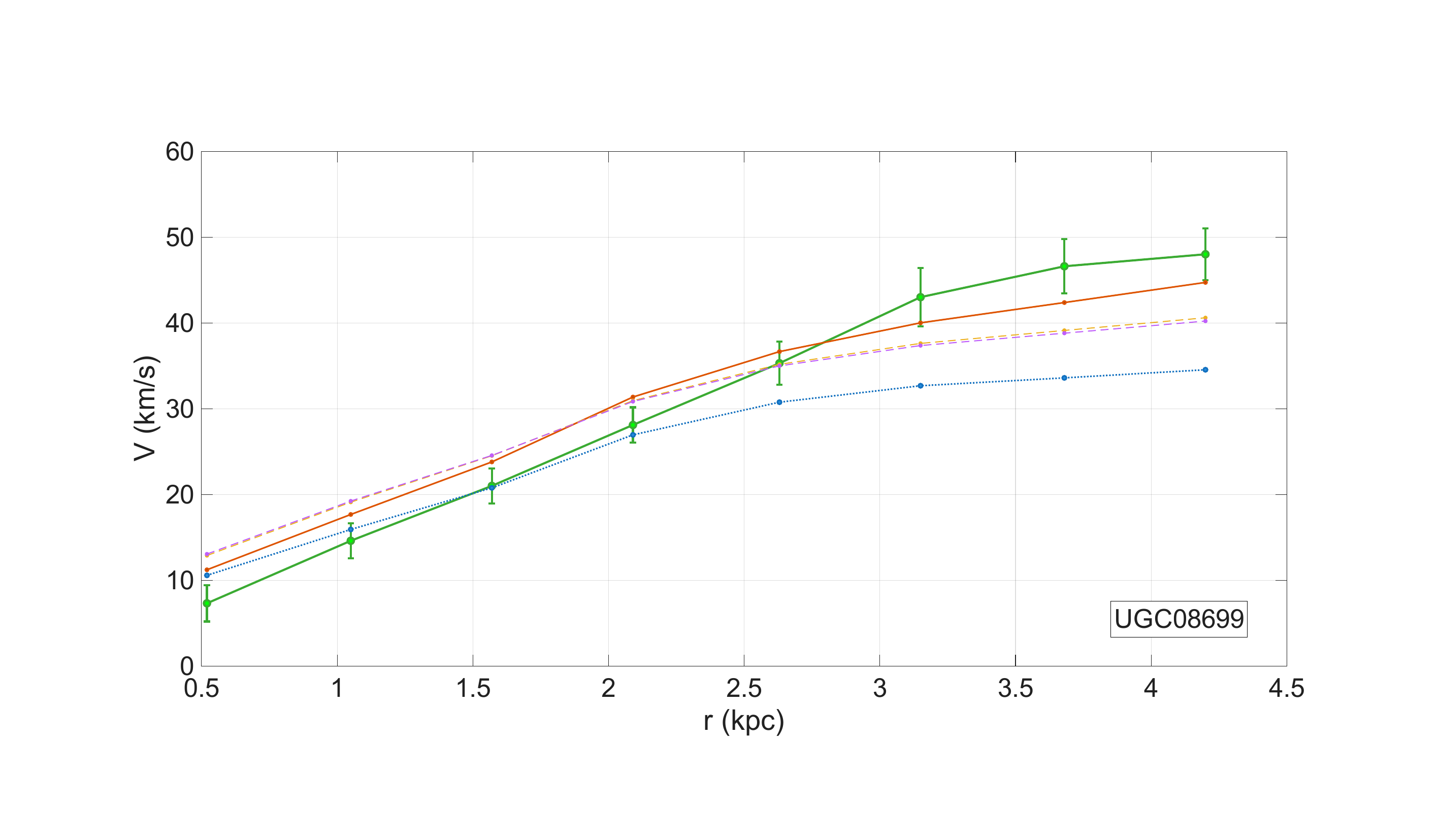}
\includegraphics[trim=4cm 3cm 5cm 4cm, clip=true, width=0.325\columnwidth]{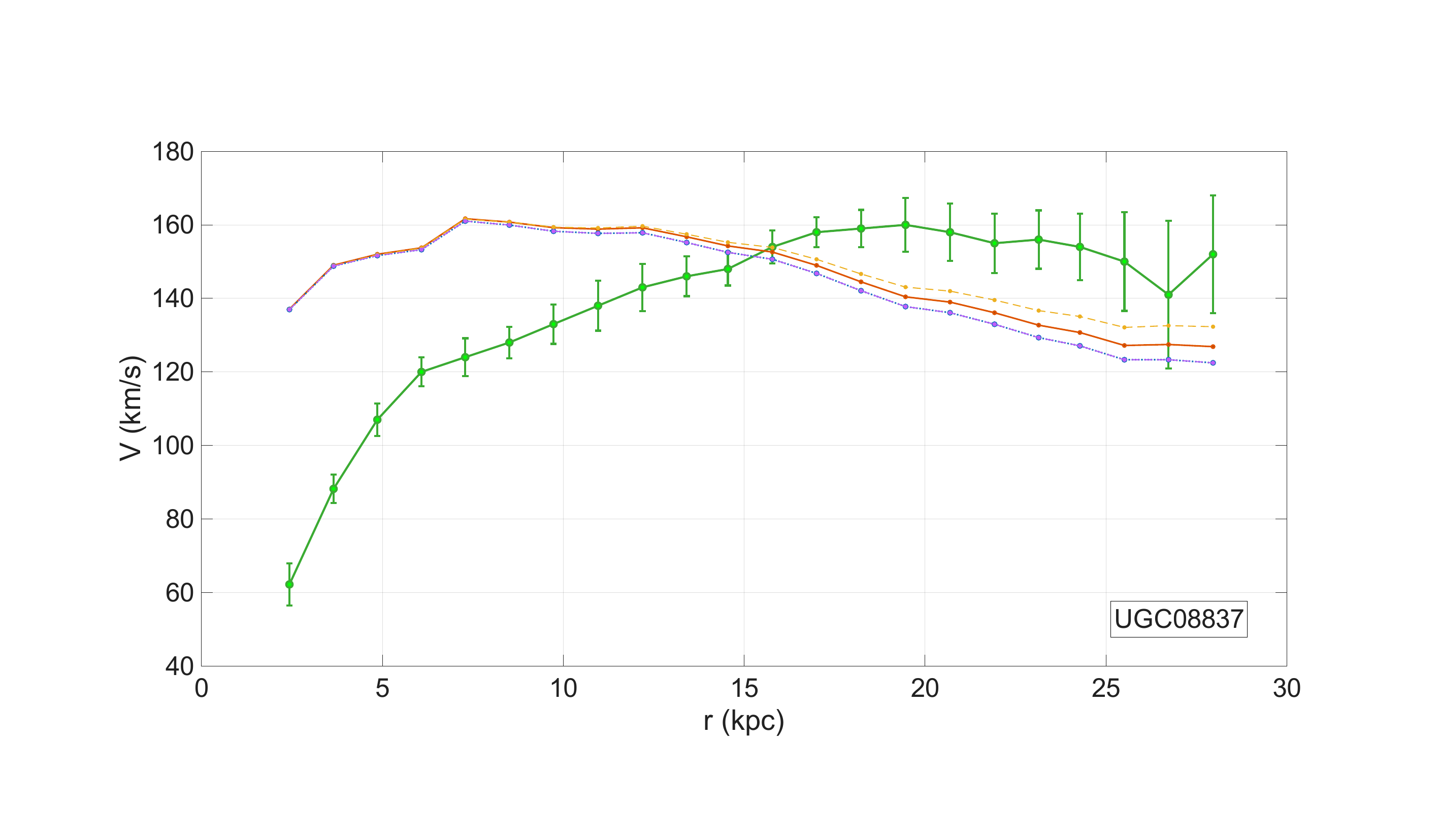}
\includegraphics[trim=4cm 3cm 5cm 4cm, clip=true, width=0.325\columnwidth]{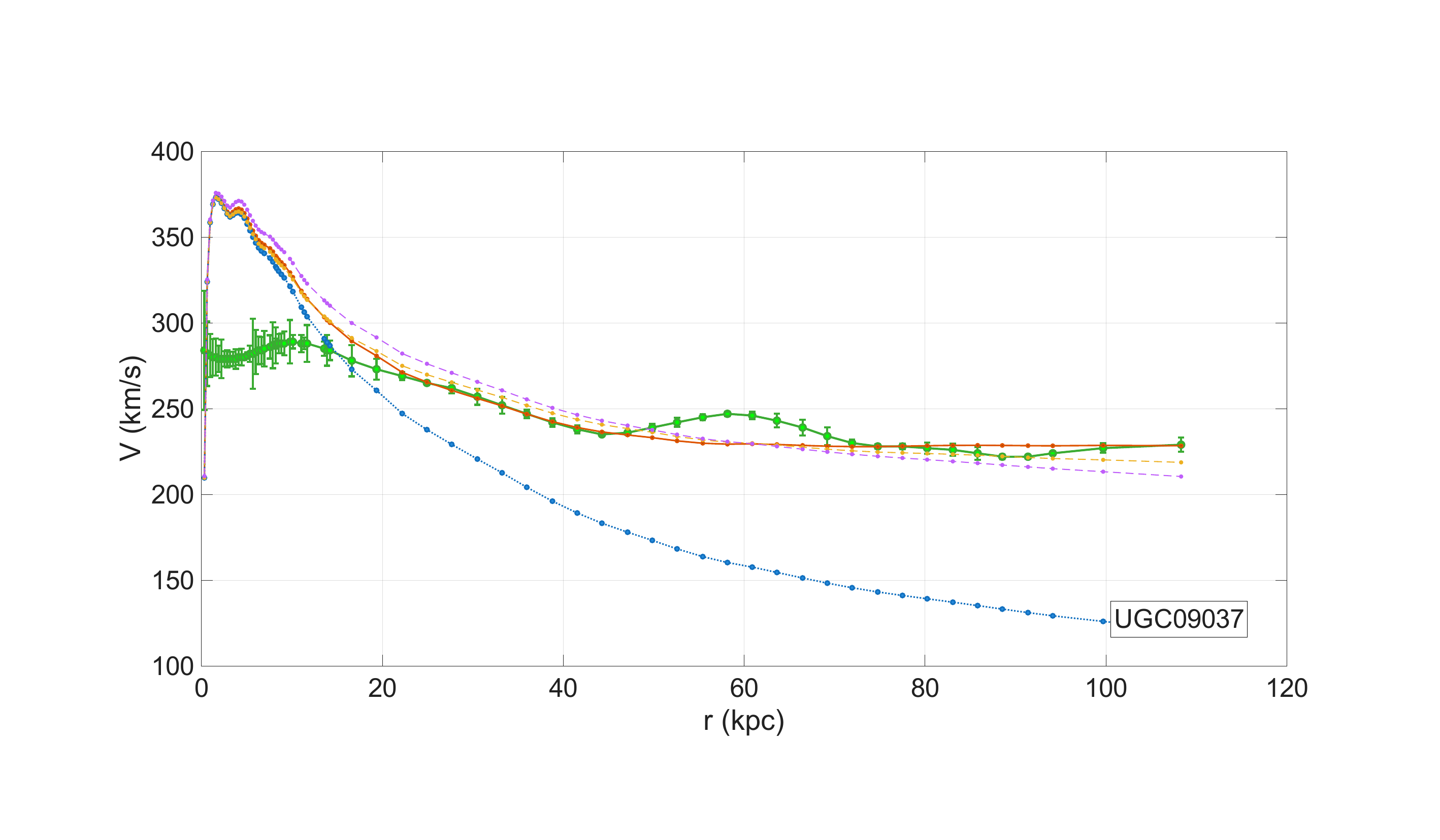}
\includegraphics[trim=4cm 3cm 5cm 4cm, clip=true, width=0.325\columnwidth]{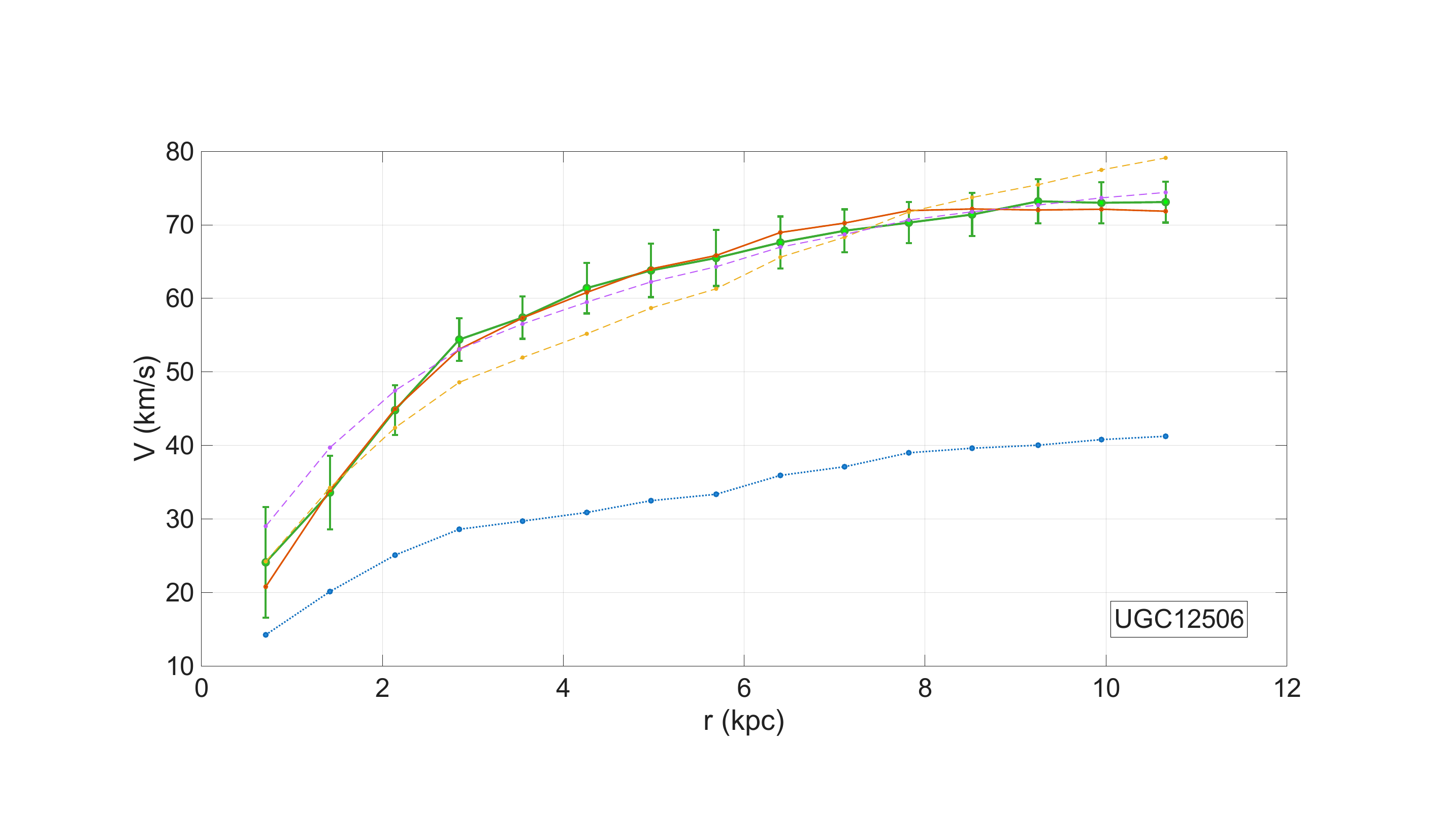}
\includegraphics[trim=4cm 3cm 5cm 4cm, clip=true, width=0.325\columnwidth]{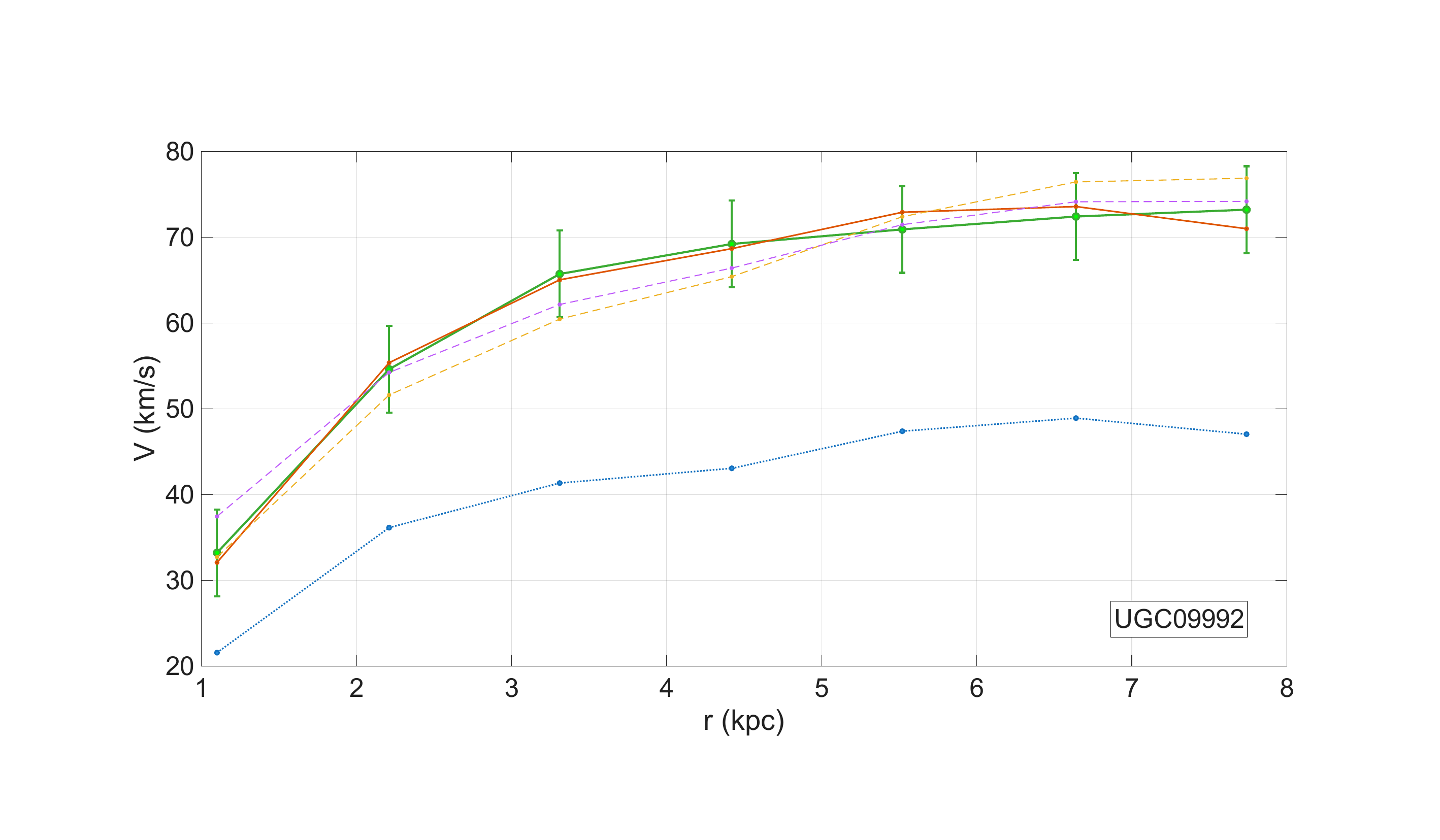}
\includegraphics[trim=4cm 3cm 5cm 4cm, clip=true, width=0.325\columnwidth]{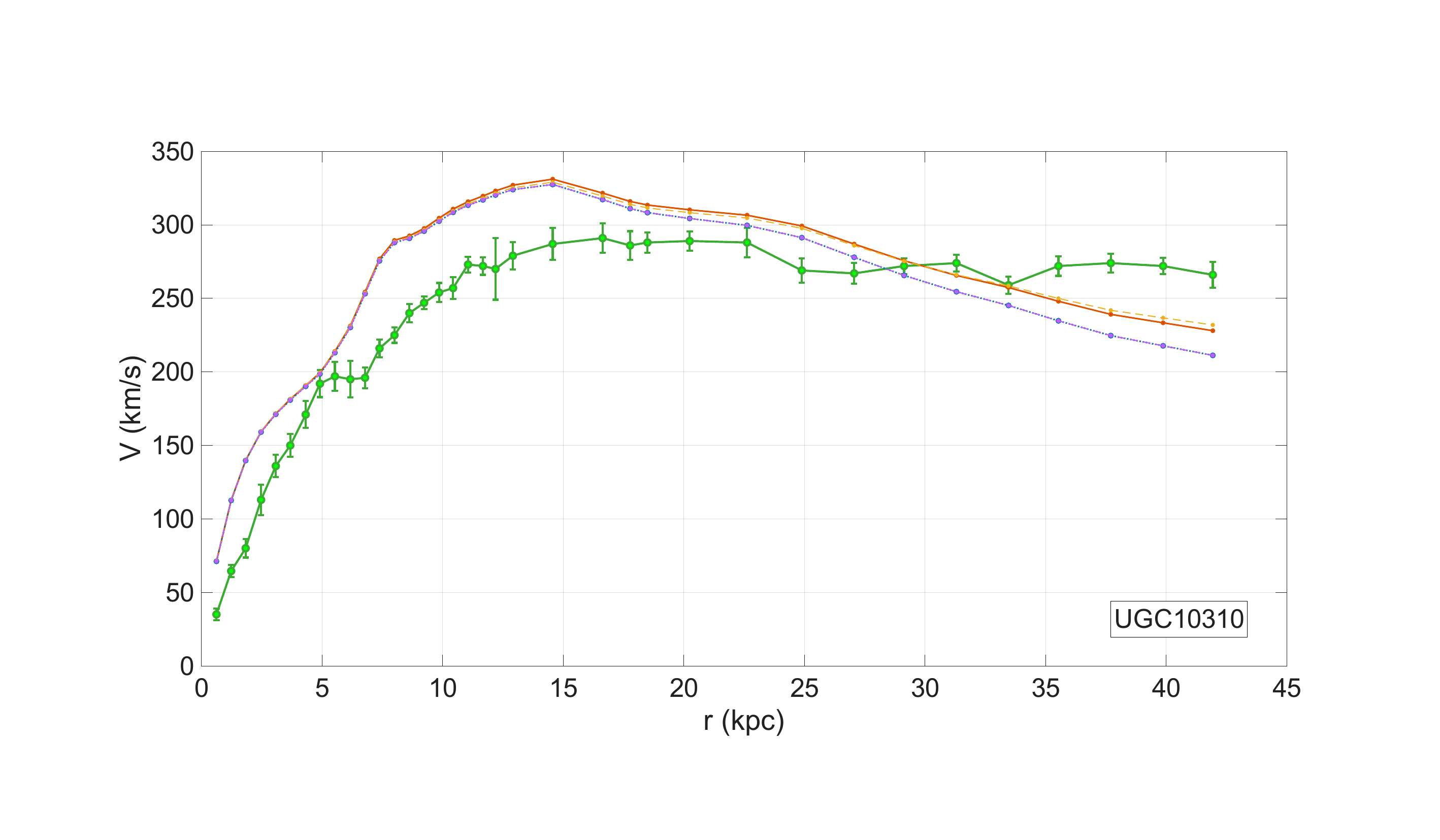}
\includegraphics[trim=4cm 3cm 5cm 4cm, clip=true, width=0.325\columnwidth]{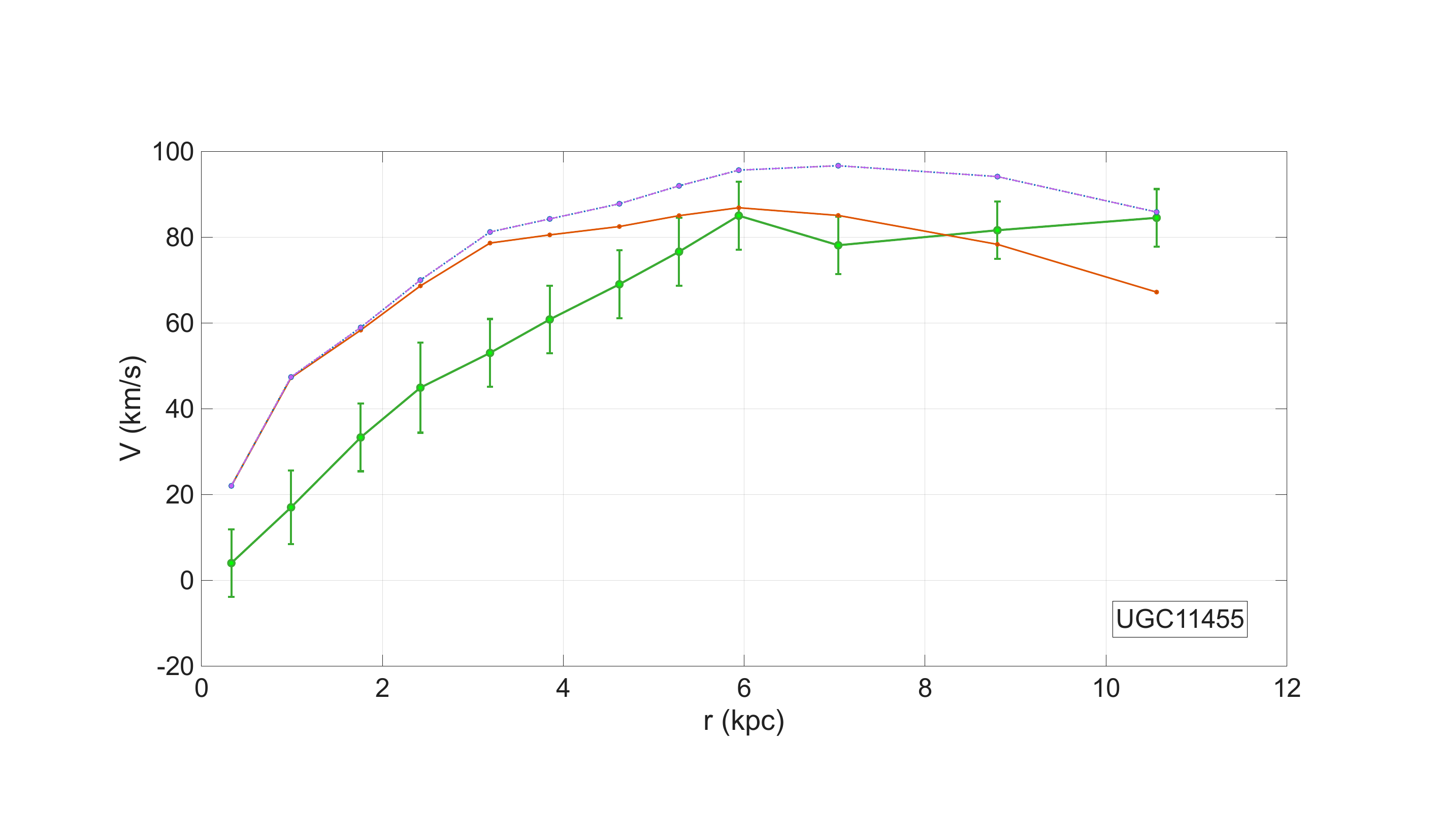}
\includegraphics[trim=4cm 3cm 5cm 4cm, clip=true, width=0.325\columnwidth]{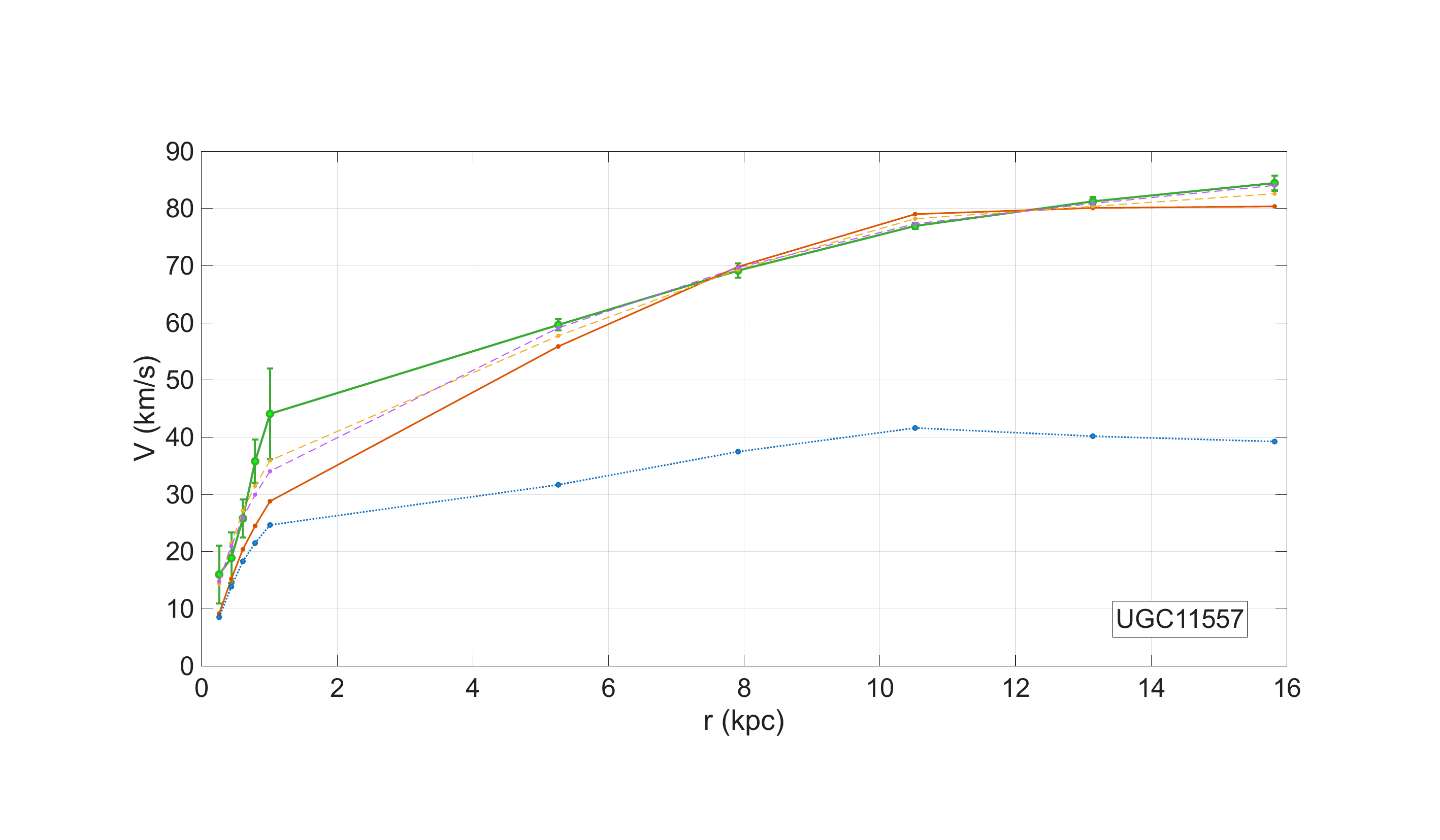}
\includegraphics[trim=4cm 3cm 5cm 4cm, clip=true, width=0.325\columnwidth]{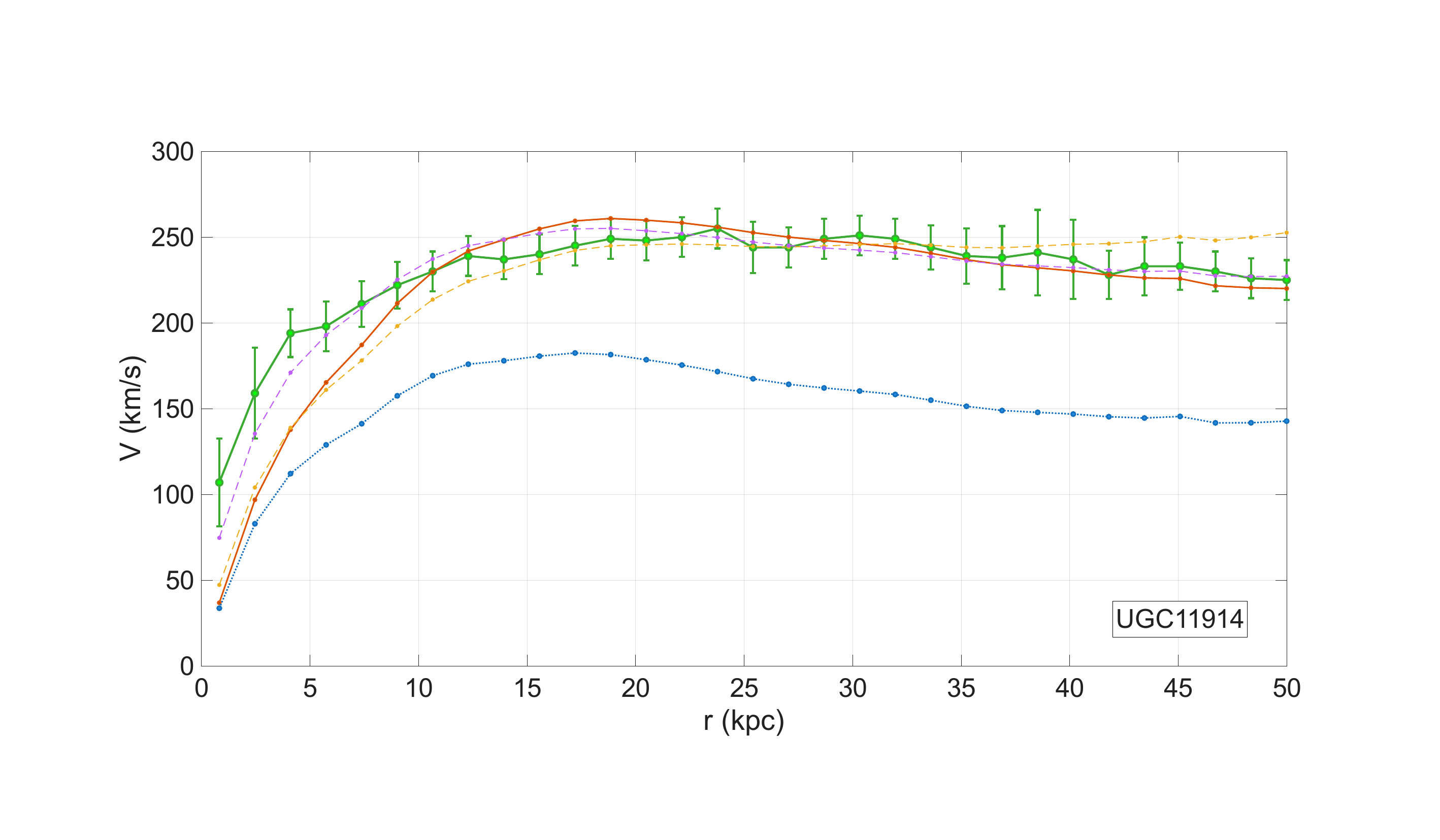}
\includegraphics[trim=4cm 3cm 5cm 4cm, clip=true, width=0.325\columnwidth]{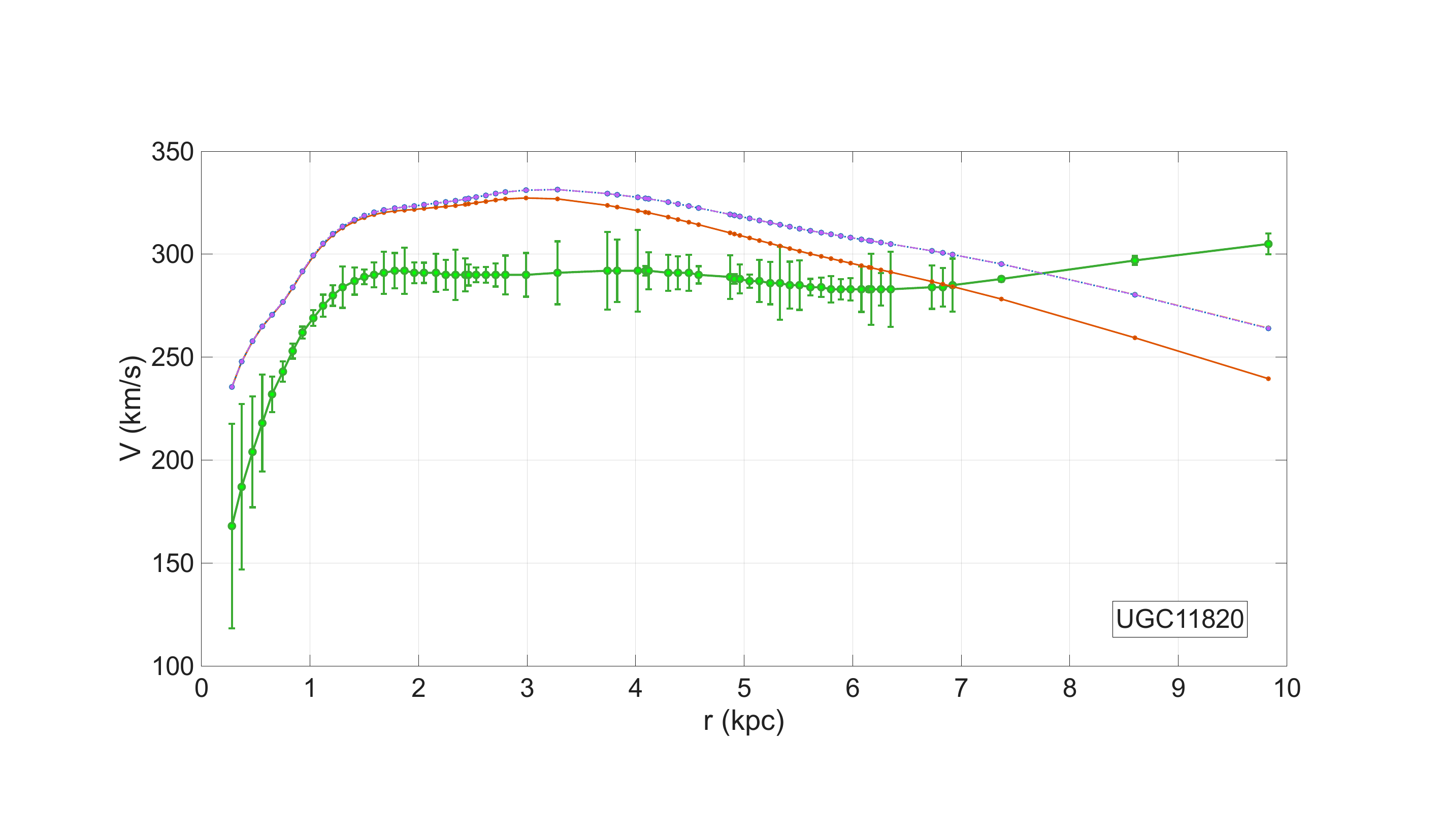}
\end{figure}

\begin{figure}
\centering
\includegraphics[trim=4cm 3cm 5cm 4cm, clip=true, width=0.325\columnwidth]{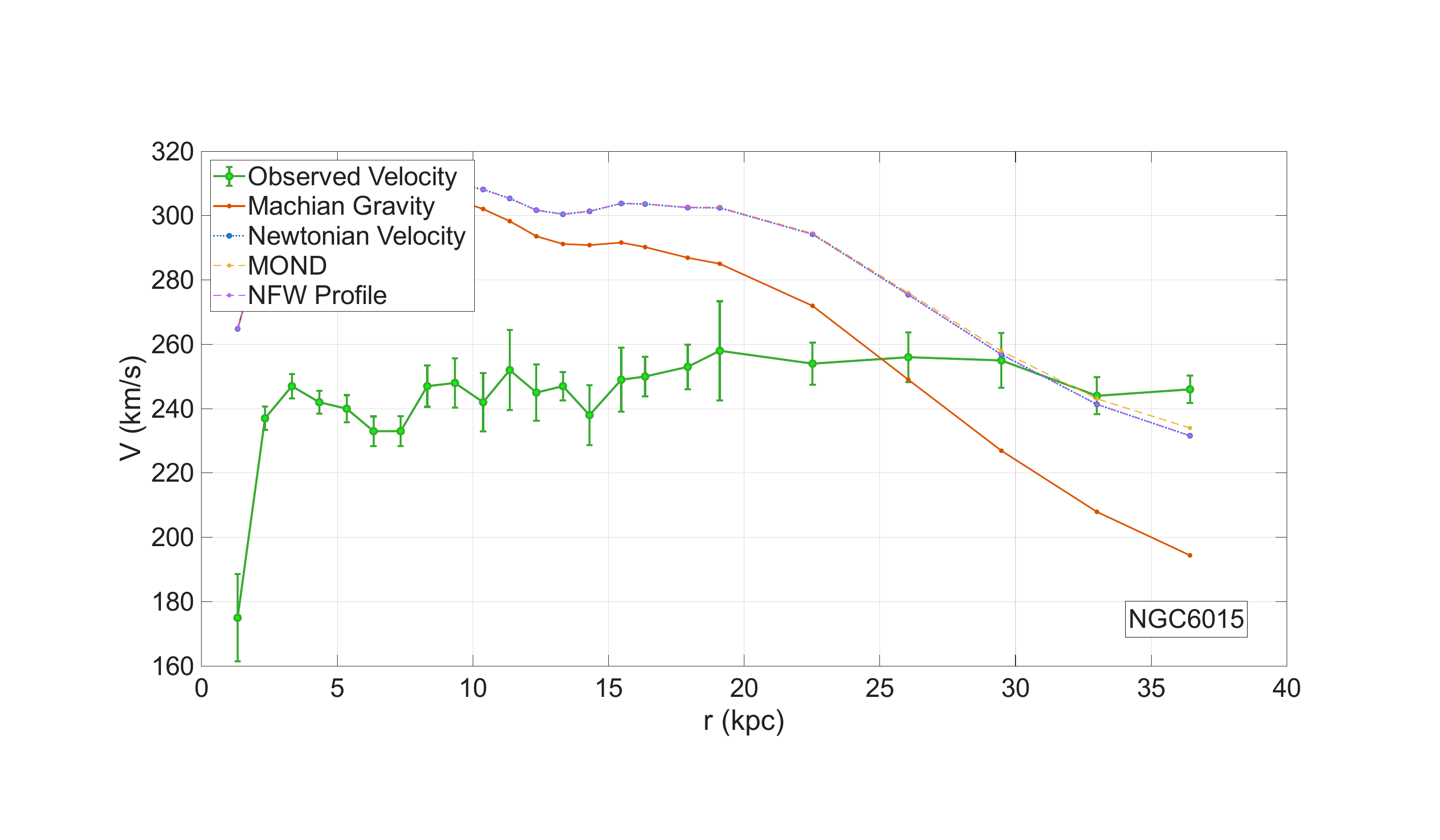}
\includegraphics[trim=4cm 3cm 5cm 4cm, clip=true, width=0.325\columnwidth]{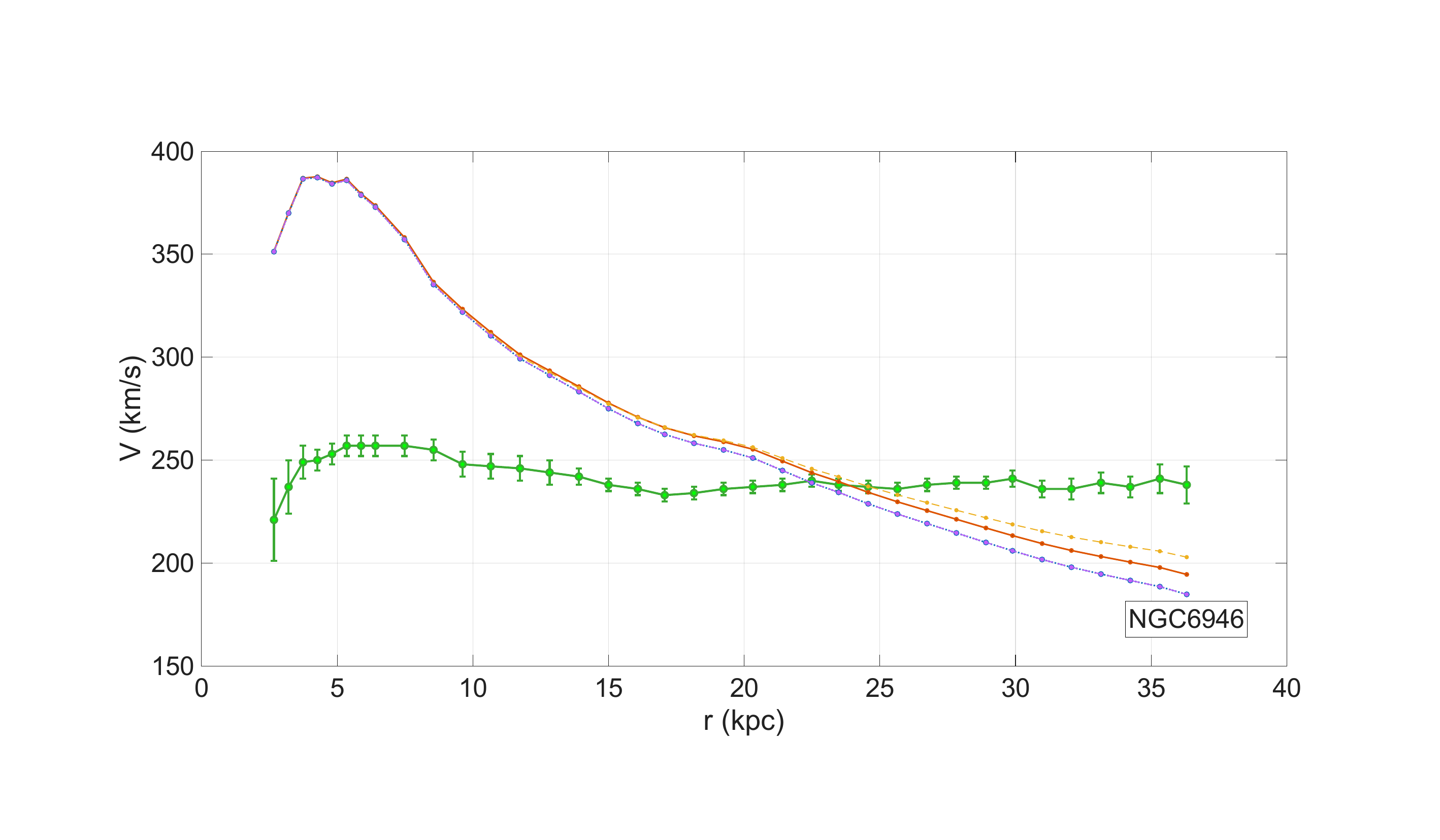}
\includegraphics[trim=4cm 3cm 5cm 4cm, clip=true, width=0.325\columnwidth]{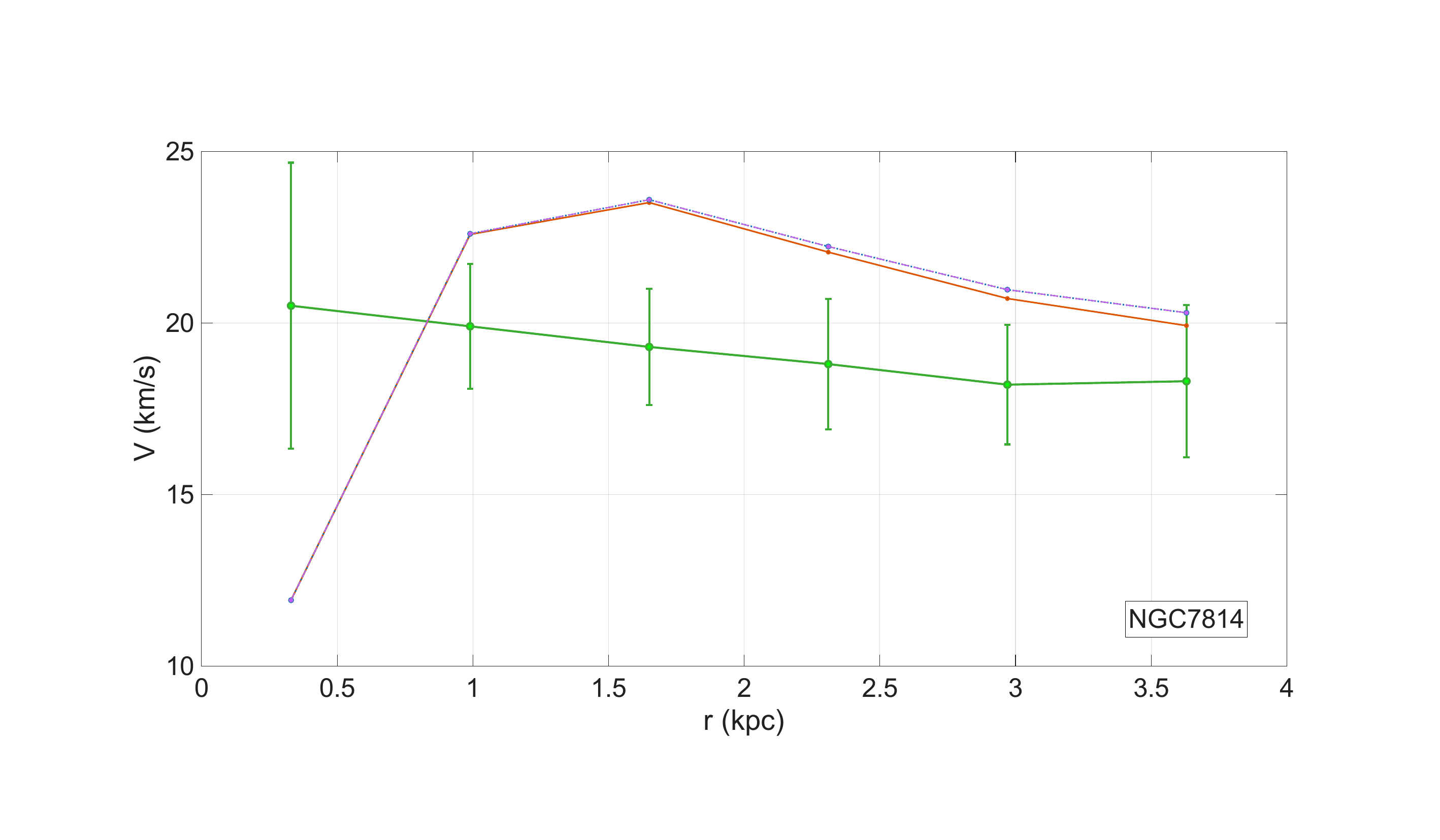}
\includegraphics[trim=4cm 3cm 5cm 4cm, clip=true, width=0.325\columnwidth]{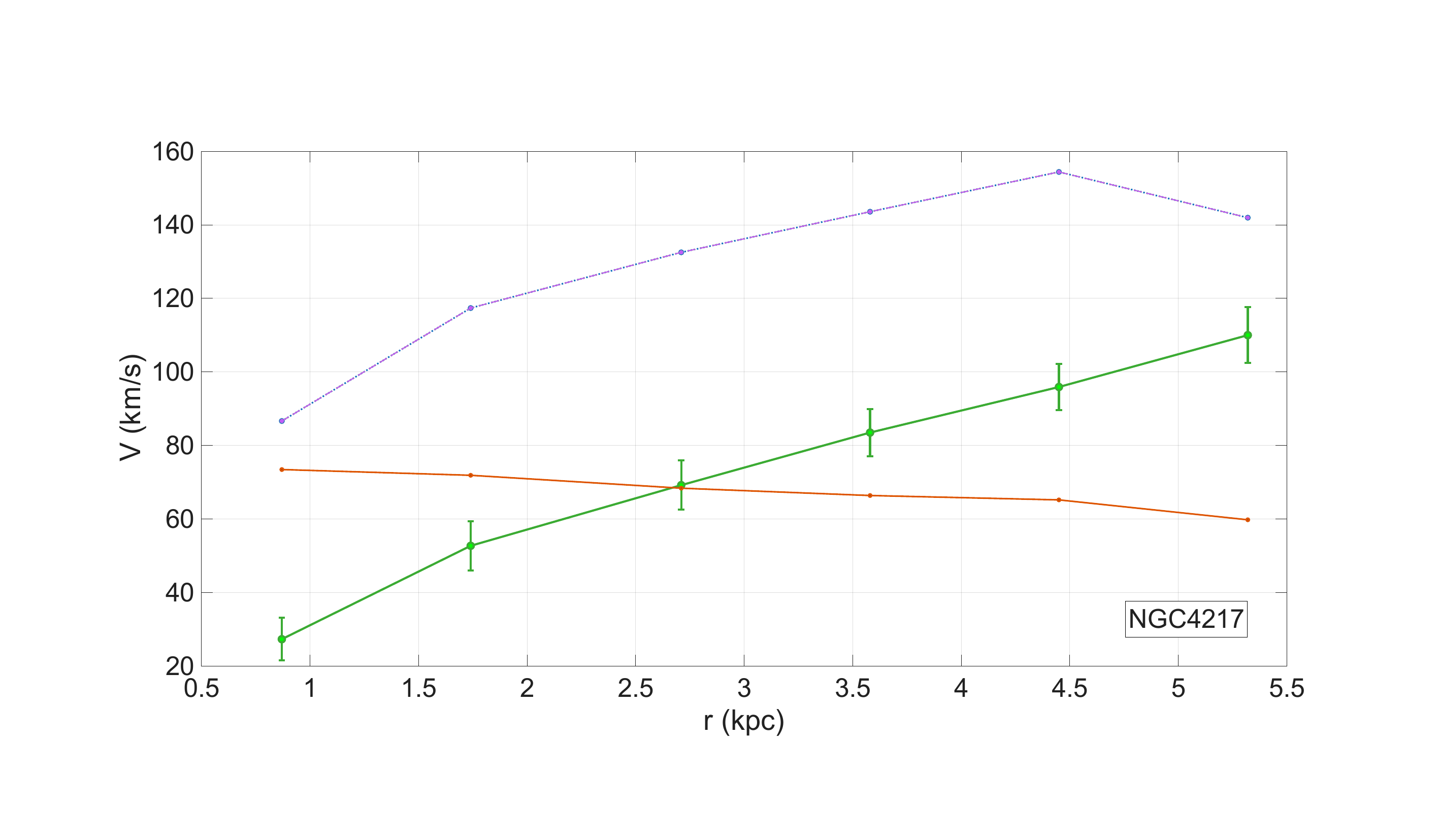}
\includegraphics[trim=4cm 3cm 5cm 4cm, clip=true, width=0.325\columnwidth]{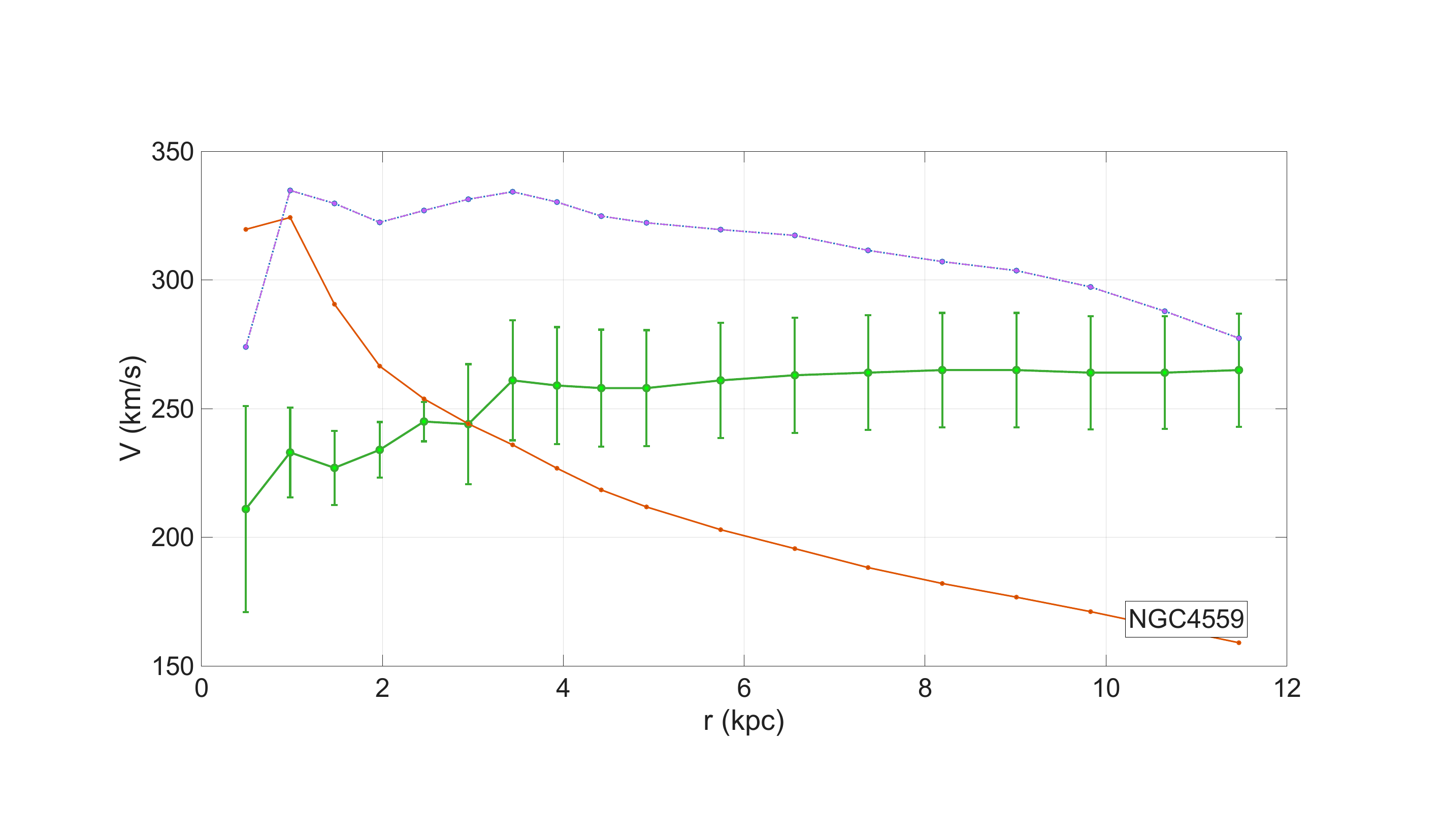}
\includegraphics[trim=4cm 3cm 5cm 4cm, clip=true, width=0.325\columnwidth]{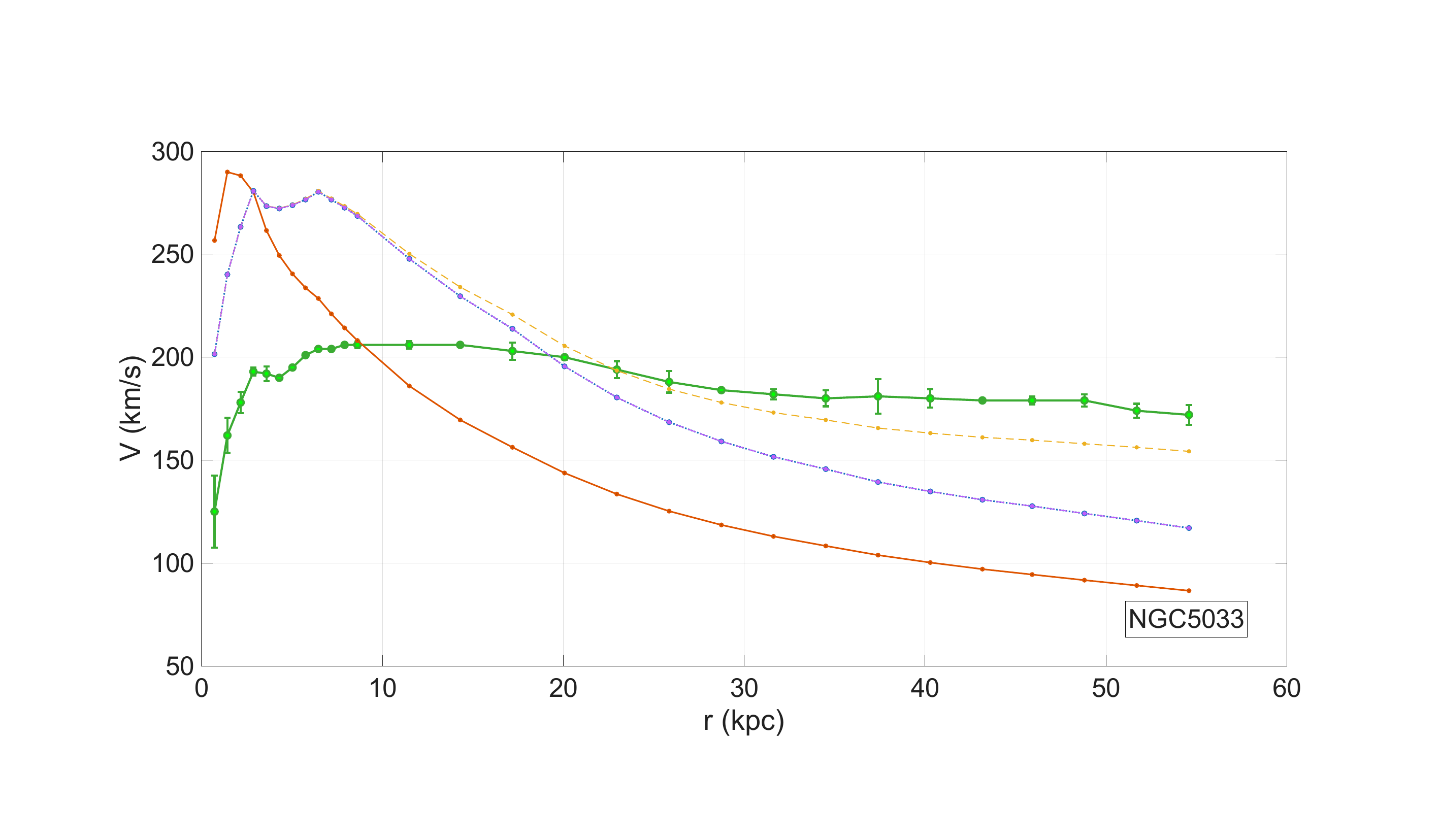}
\includegraphics[trim=4cm 3cm 5cm 4cm, clip=true, width=0.325\columnwidth]{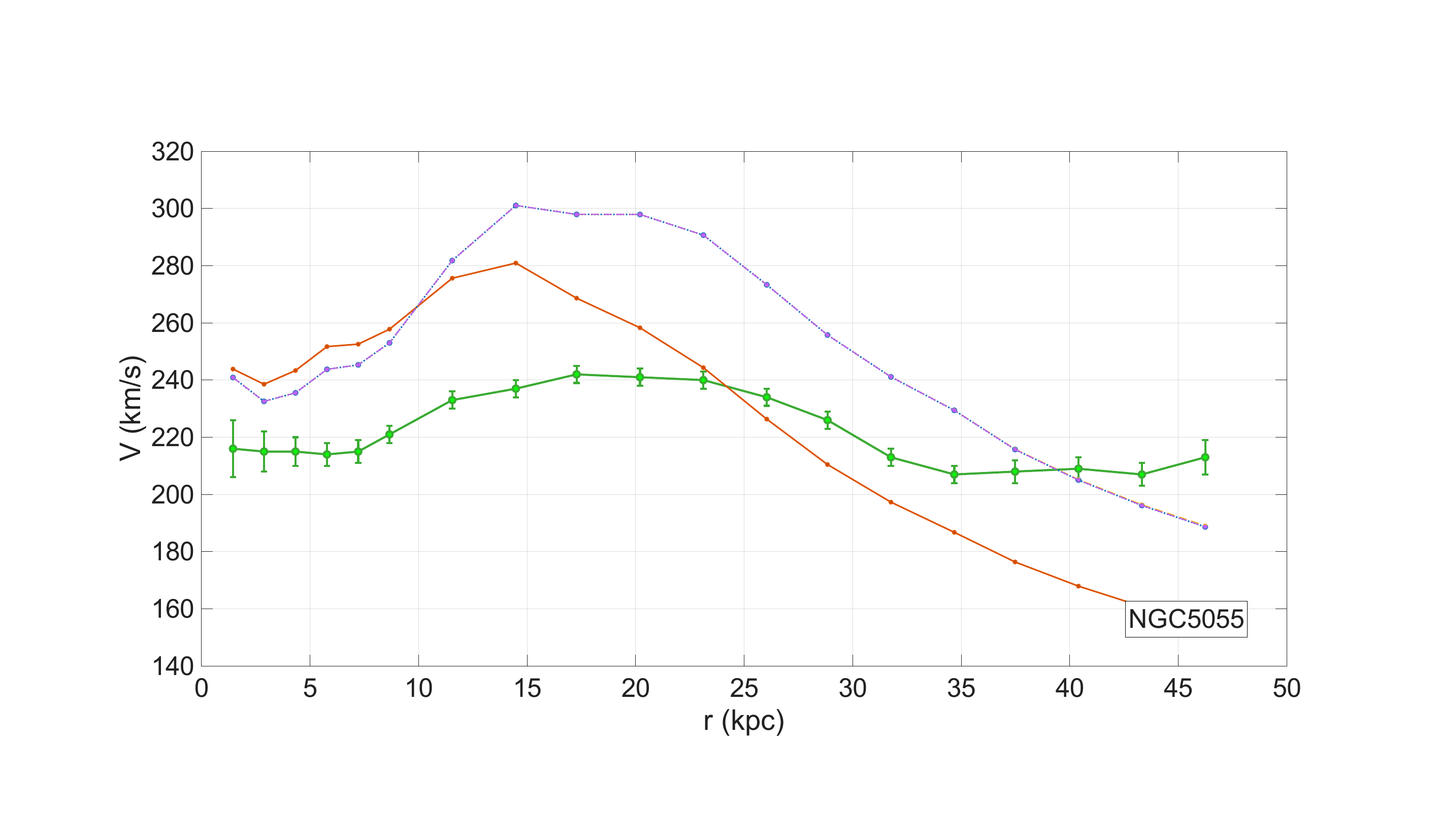}
\includegraphics[trim=4cm 3cm 5cm 4cm, clip=true, width=0.325\columnwidth]{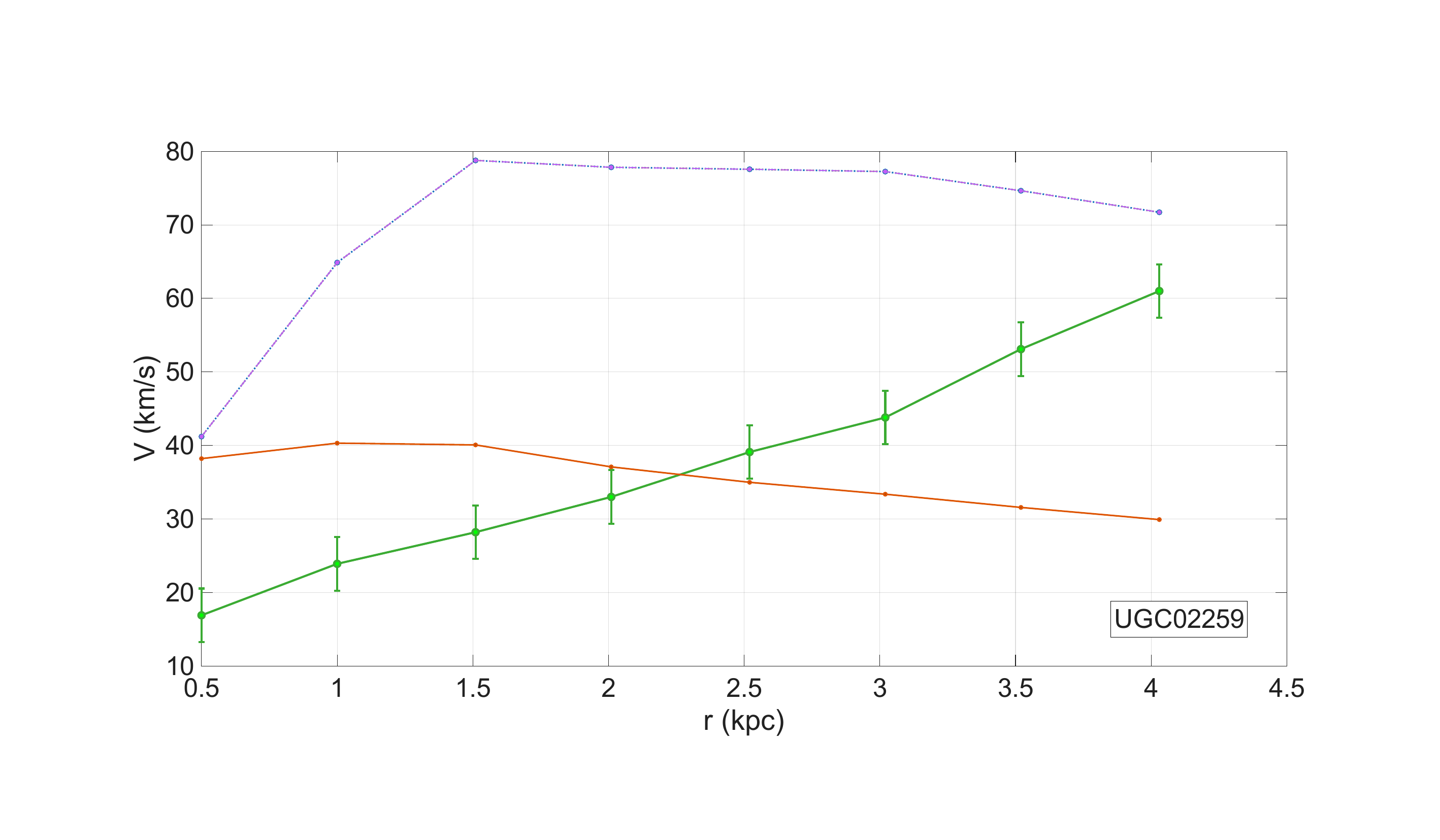}
\includegraphics[trim=4cm 3cm 5cm 4cm, clip=true, width=0.325\columnwidth]{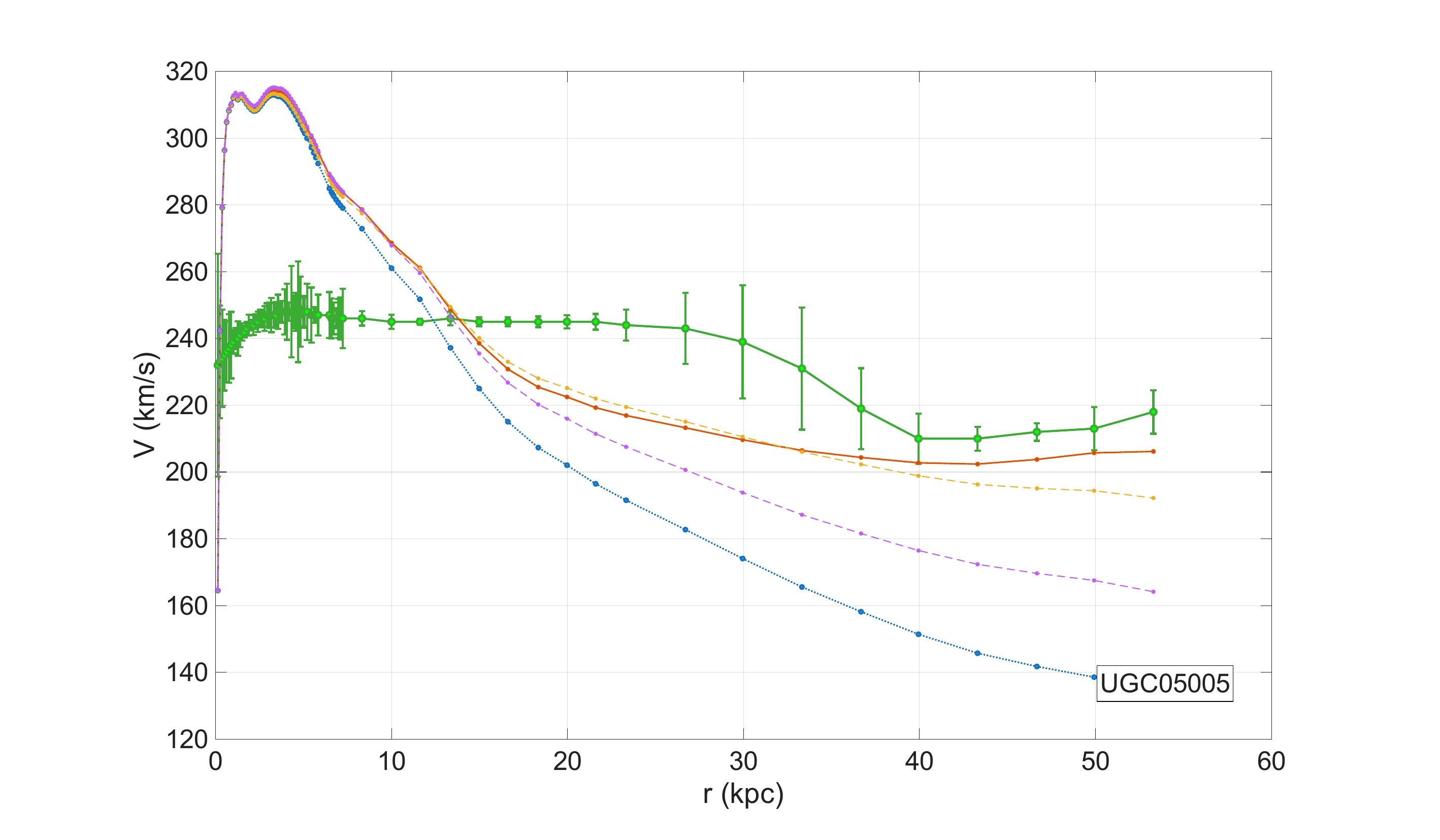}
\includegraphics[trim=4cm 3cm 5cm 4cm, clip=true, width=0.325\columnwidth]{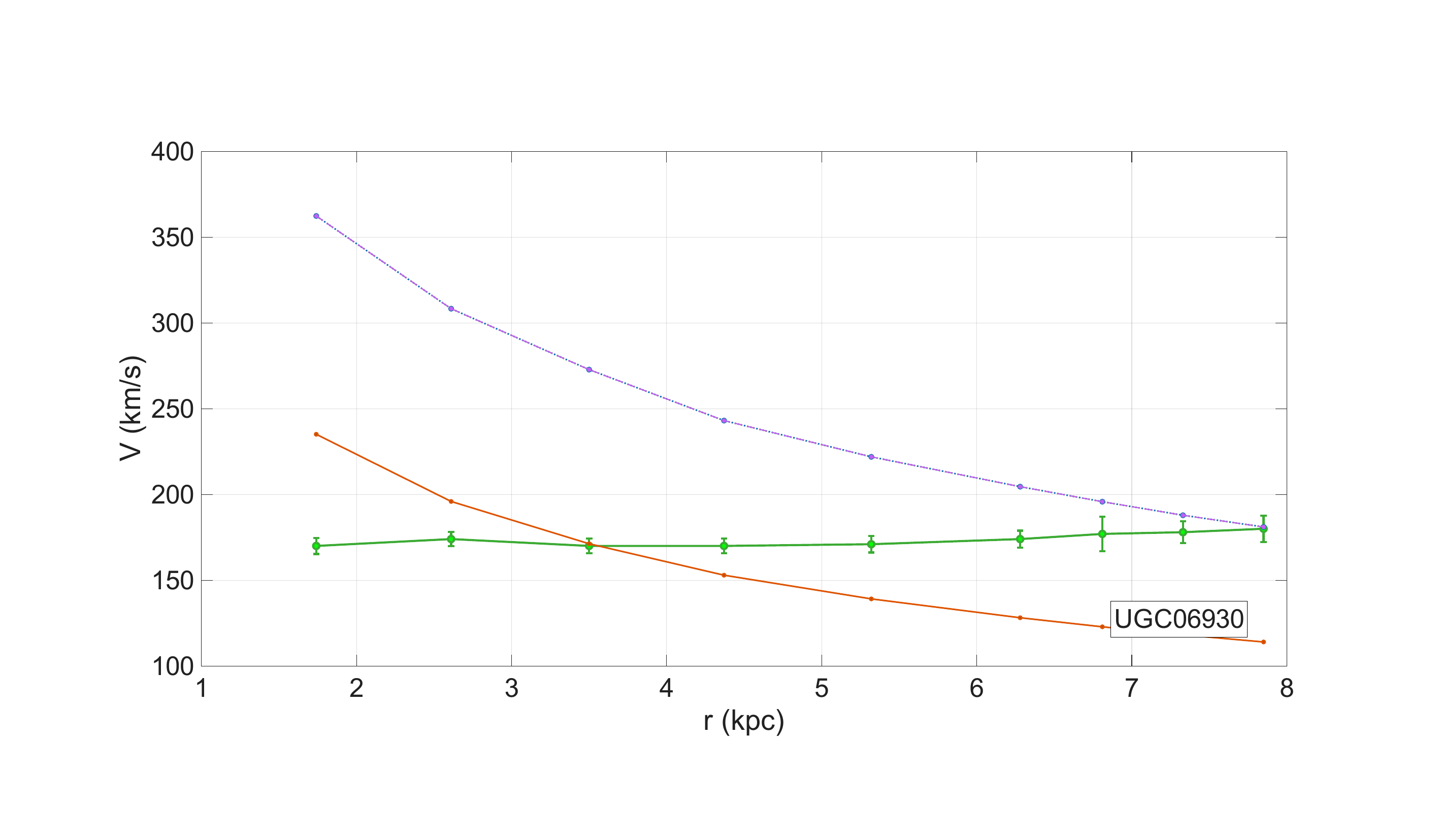}
\includegraphics[trim=4cm 3cm 5cm 4cm, clip=true, width=0.325\columnwidth]{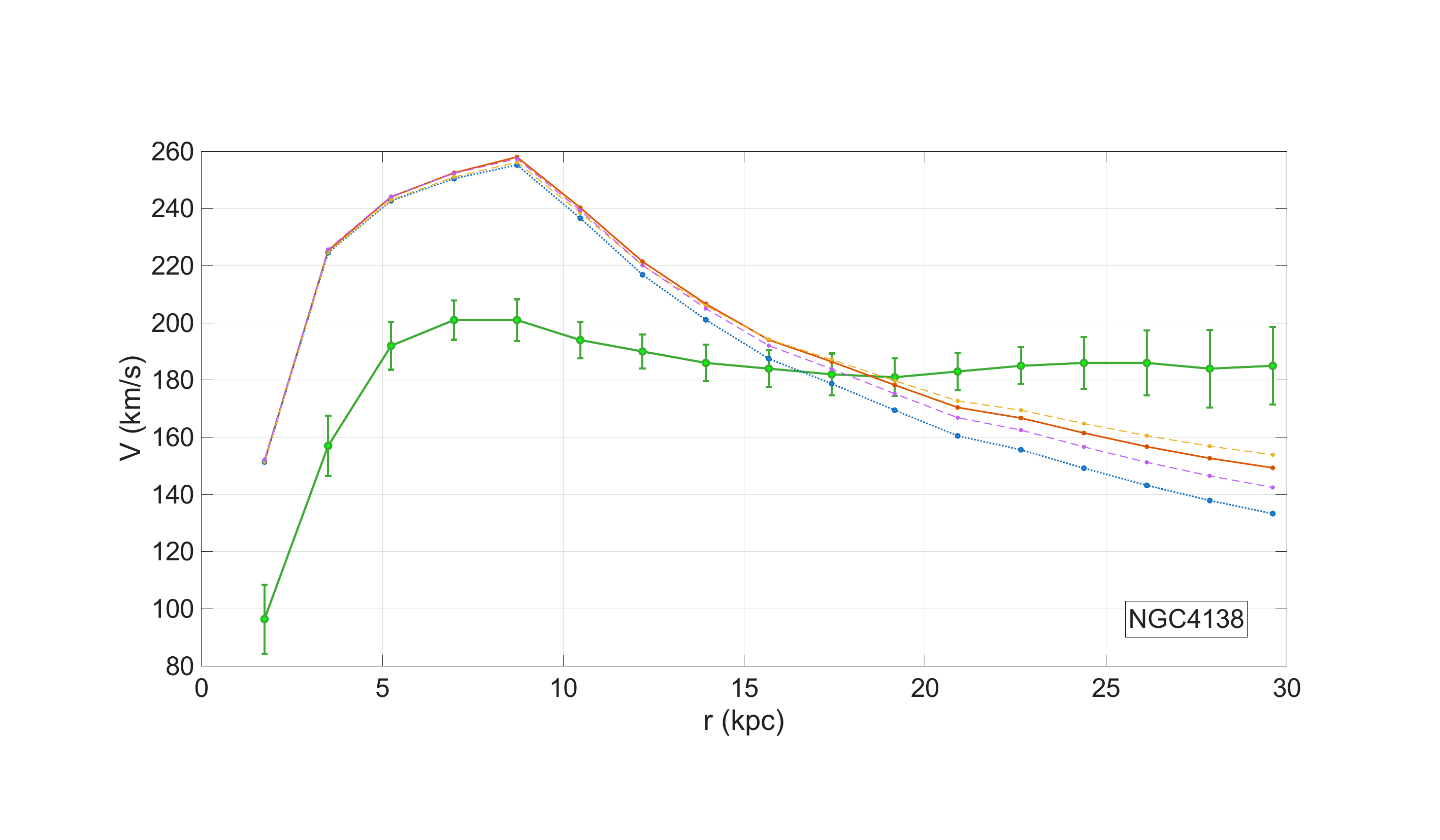}
\includegraphics[trim=4cm 3cm 5cm 4cm, clip=true, width=0.325\columnwidth]{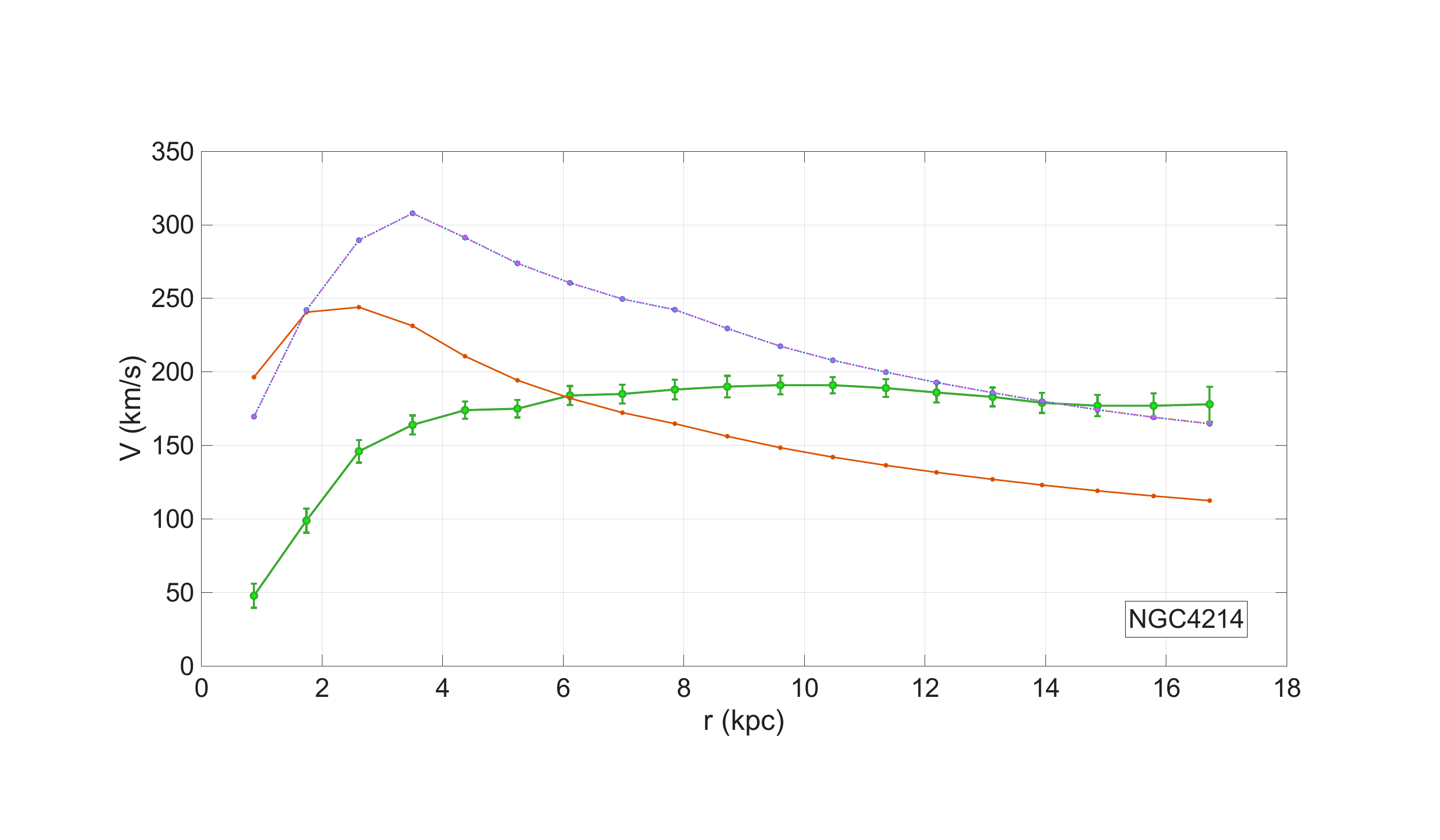}
\includegraphics[trim=4cm 3cm 5cm 4cm, clip=true, width=0.325\columnwidth]{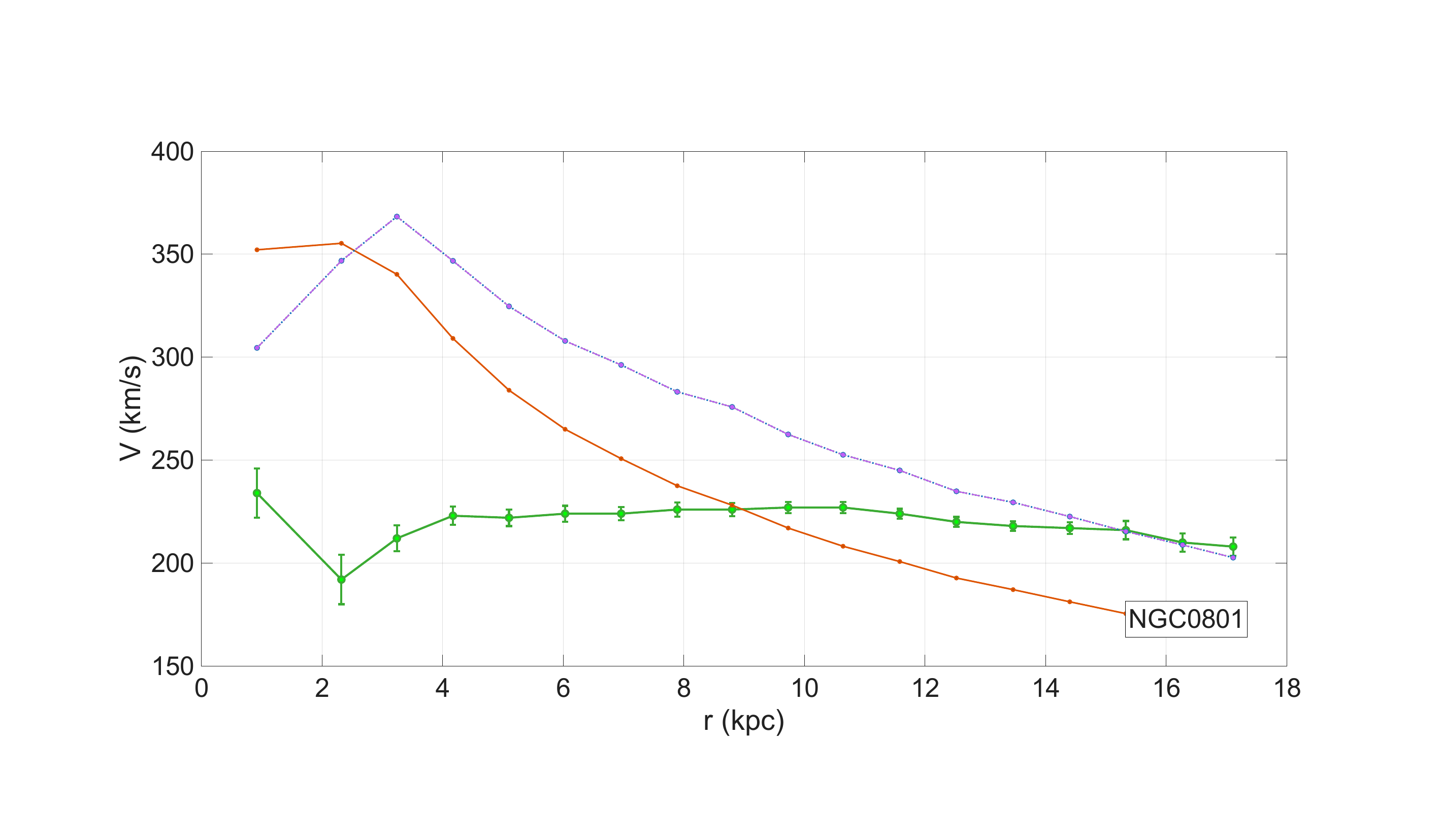}
\includegraphics[trim=4cm 3cm 5cm 4cm, clip=true, width=0.325\columnwidth]{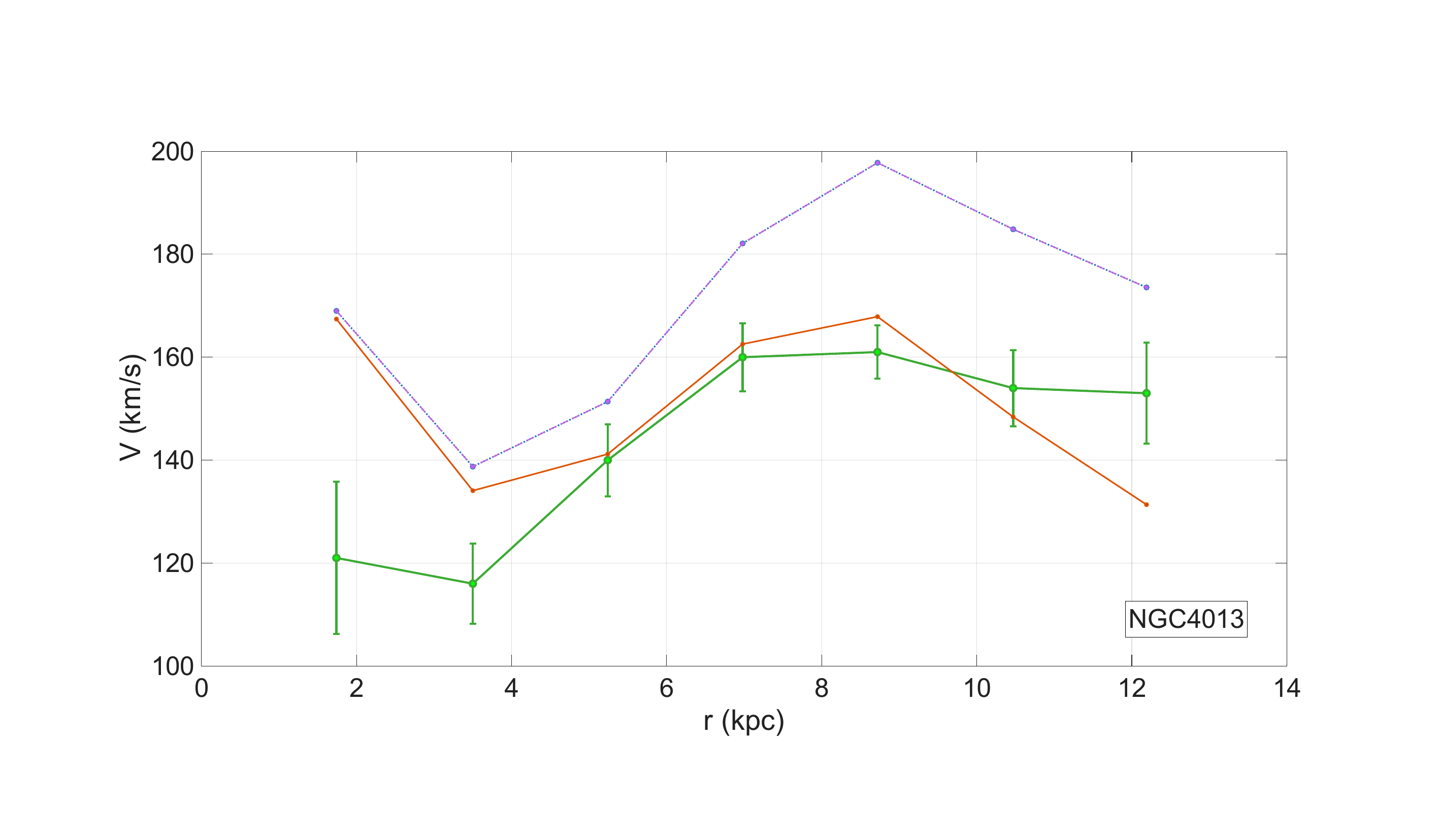}
\includegraphics[trim=4cm 3cm 5cm 4cm, clip=true, width=0.325\columnwidth]{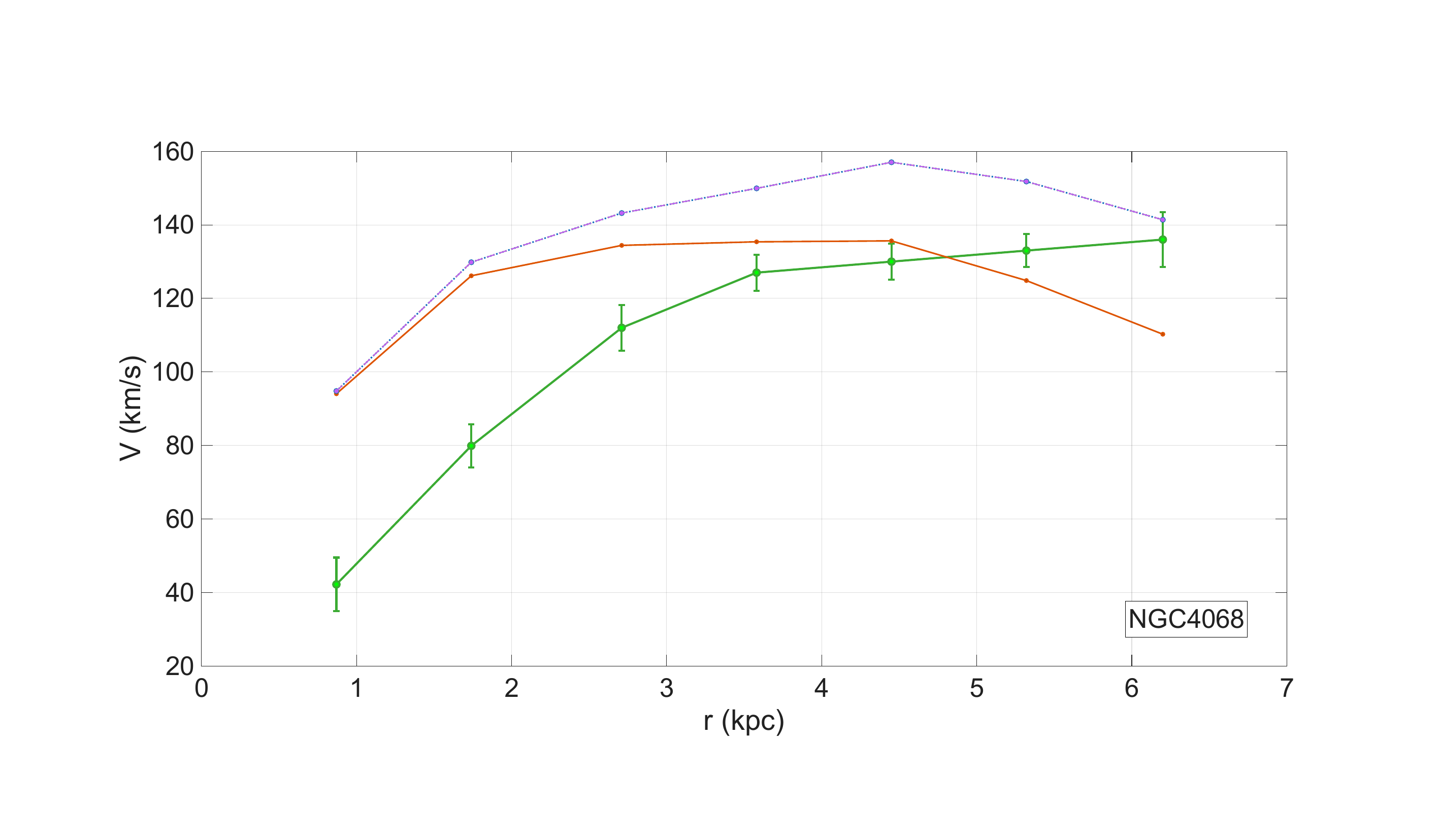}
\includegraphics[trim=4cm 3cm 5cm 4cm, clip=true, width=0.325\columnwidth]{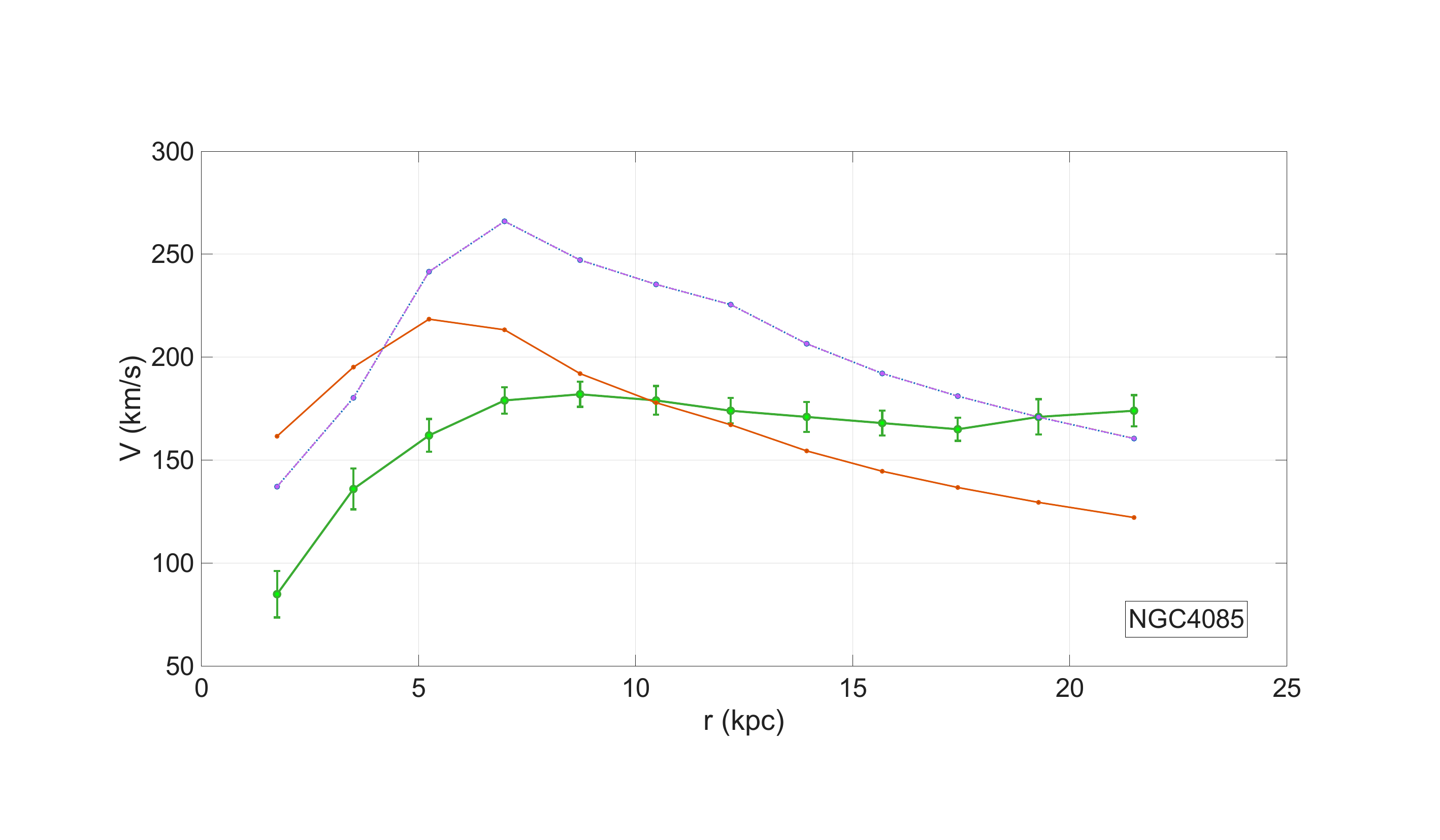}
\includegraphics[trim=4cm 3cm 5cm 4cm, clip=true, width=0.325\columnwidth]{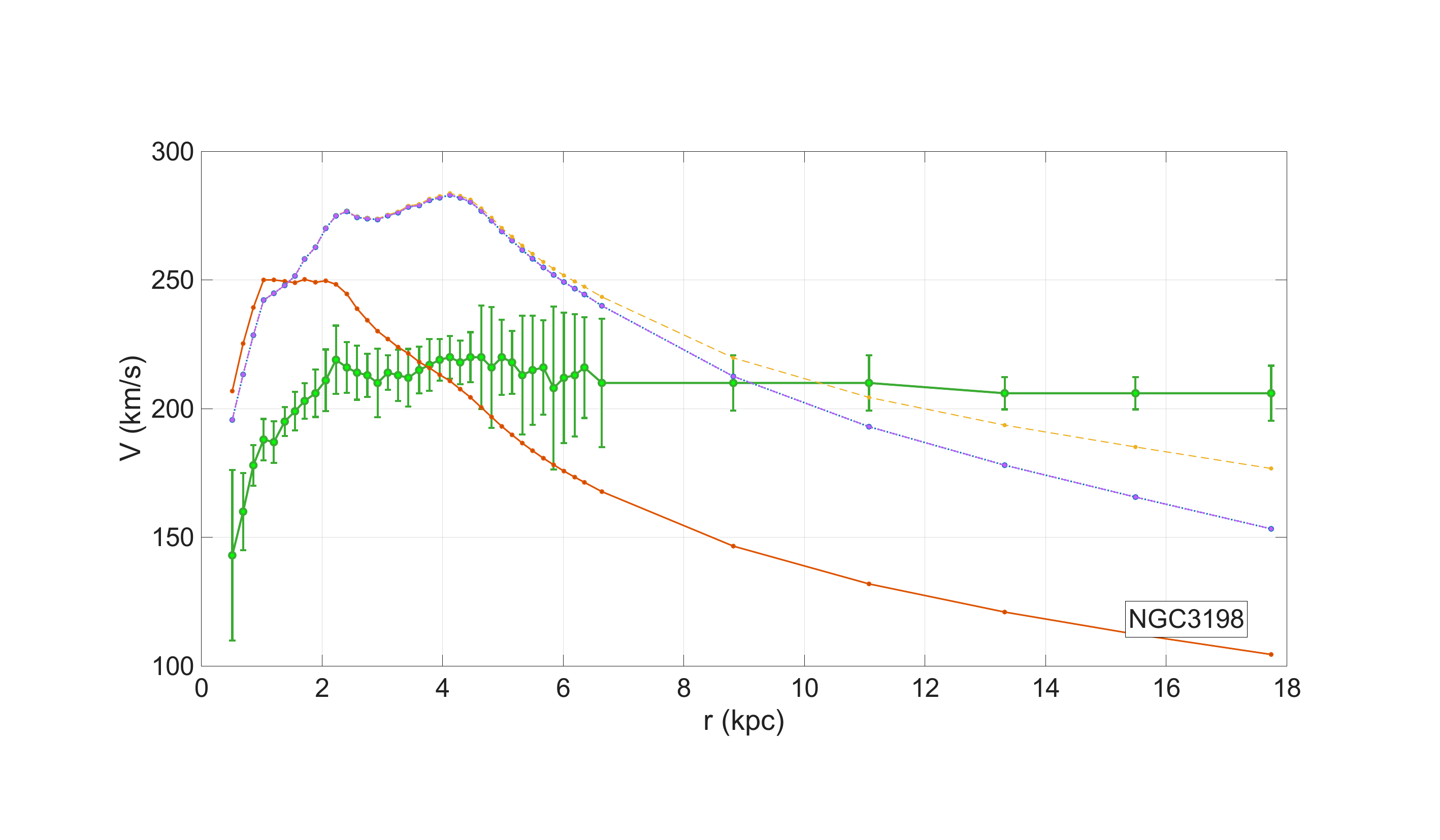}
\includegraphics[trim=4cm 3cm 5cm 4cm, clip=true, width=0.325\columnwidth]{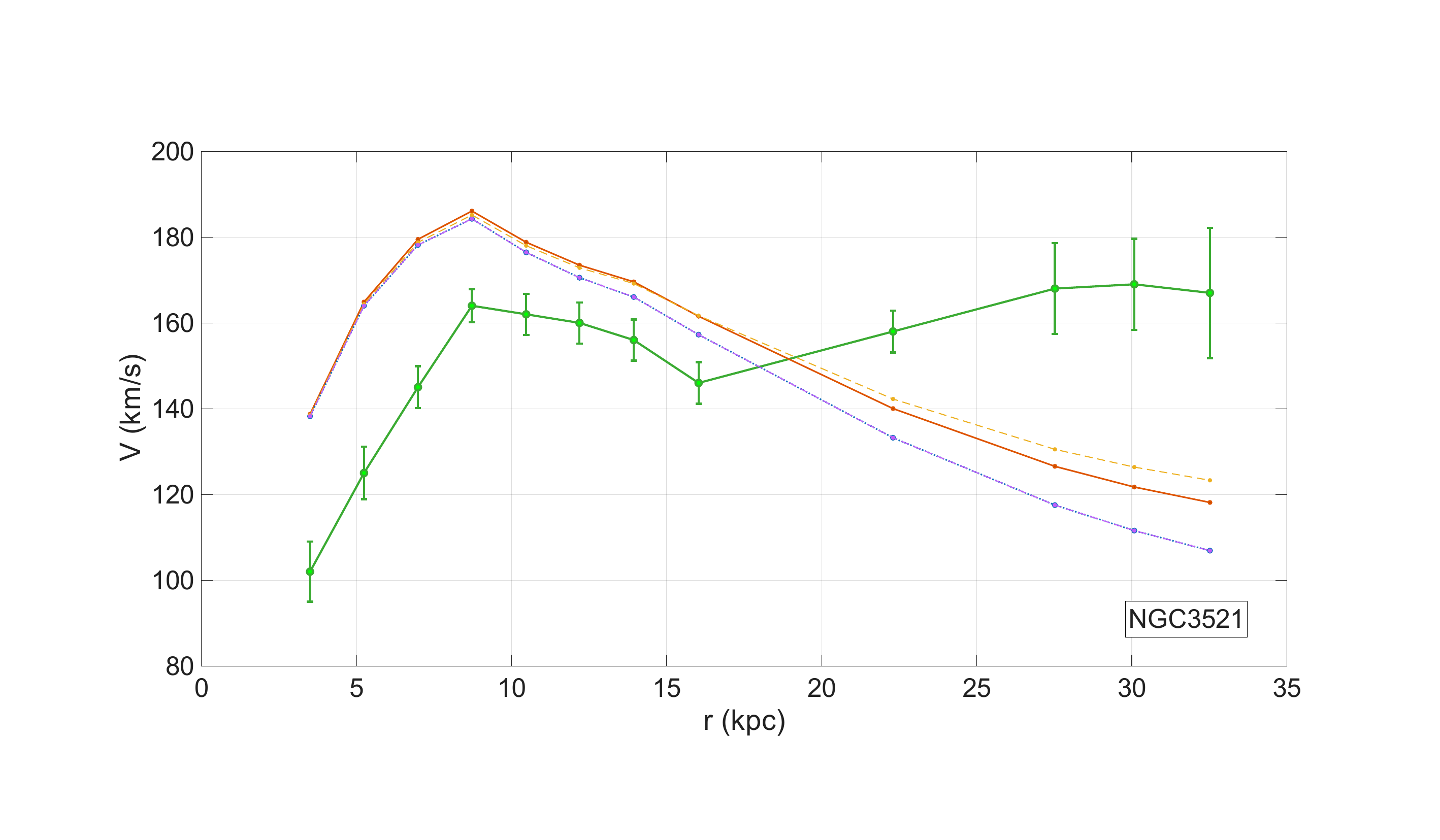}
\includegraphics[trim=4cm 3cm 5cm 4cm, clip=true, width=0.325\columnwidth]{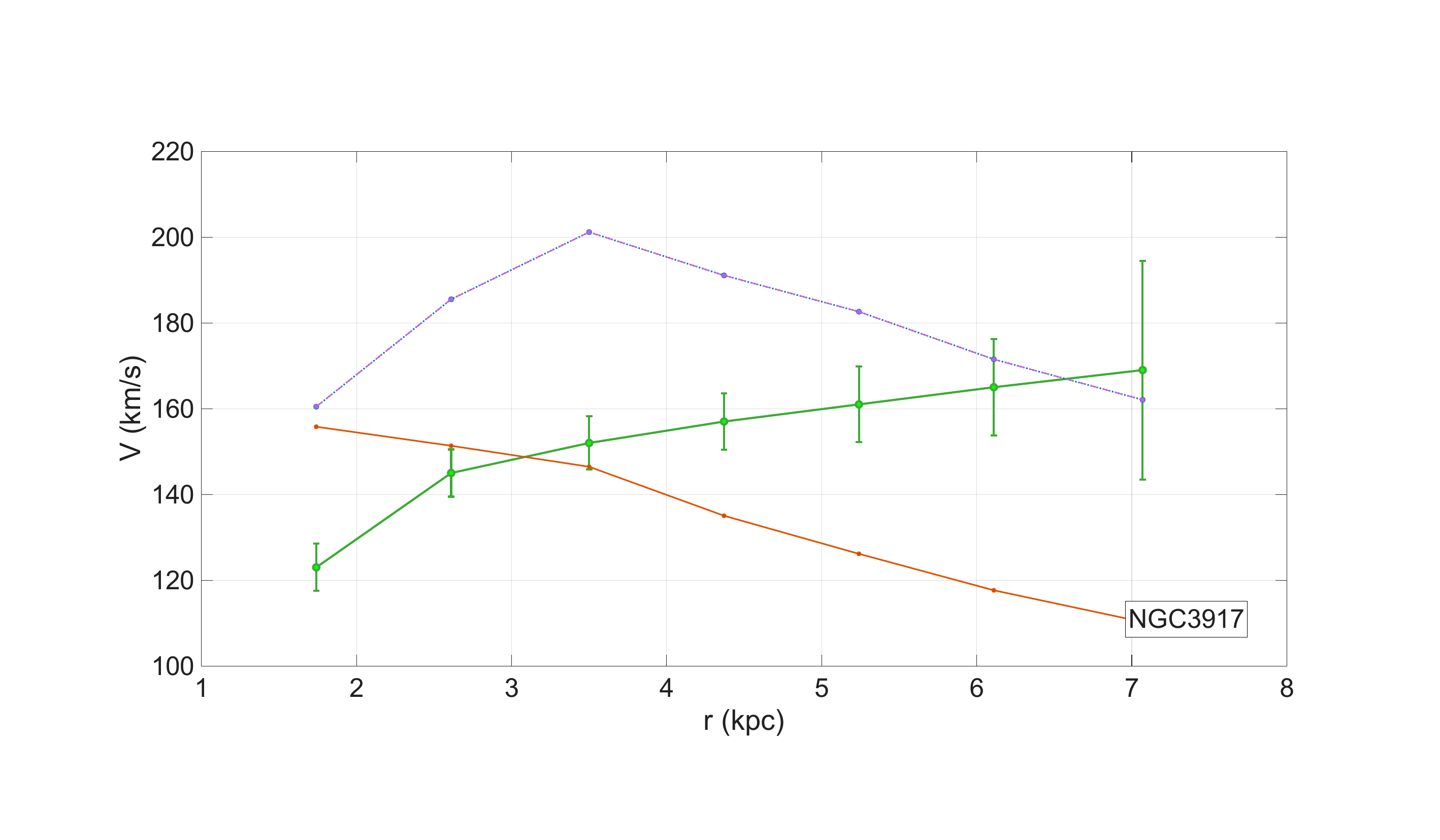}
\includegraphics[trim=4cm 3cm 5cm 4cm, clip=true, width=0.325\columnwidth]{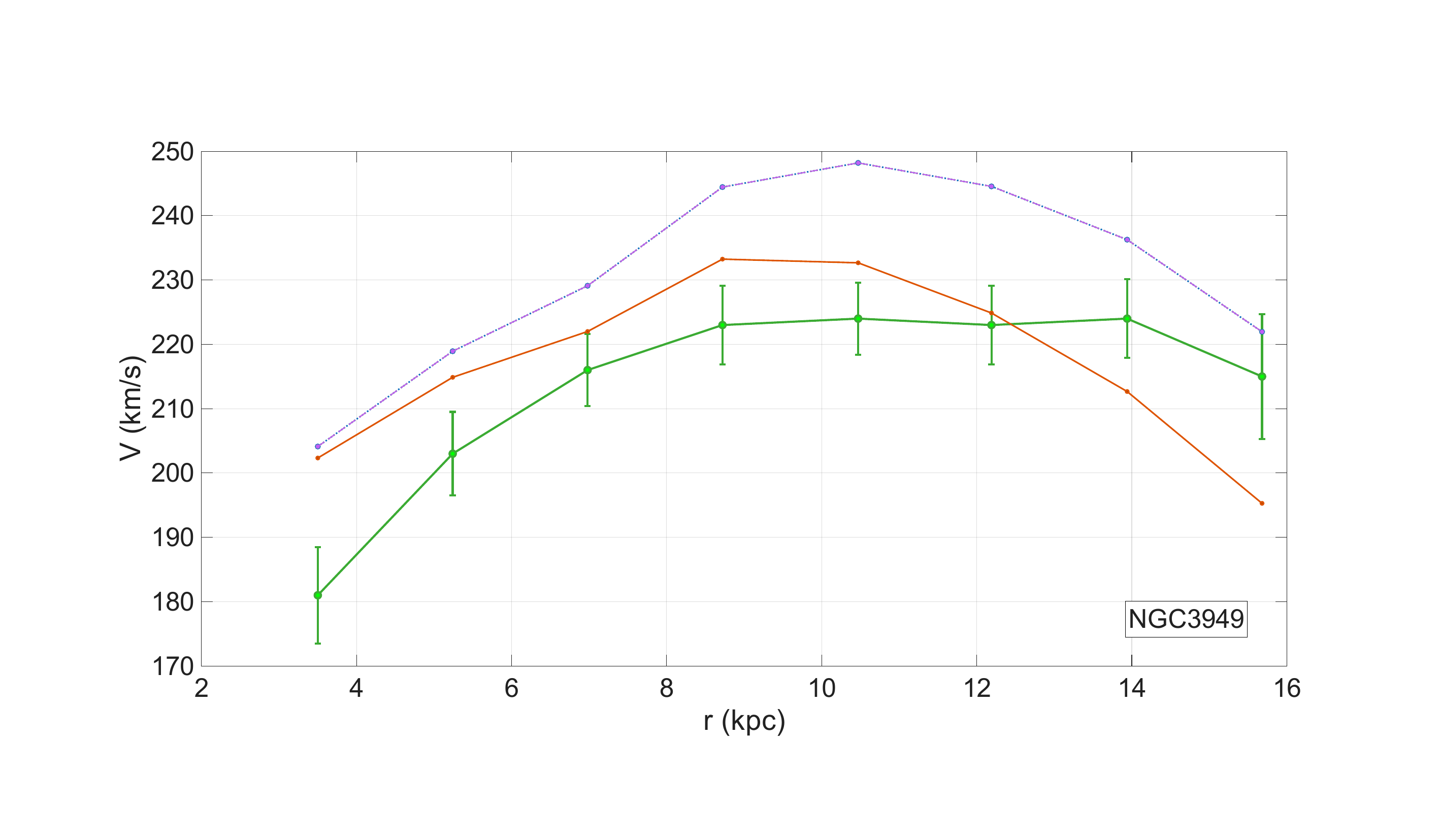}
\includegraphics[trim=4cm 3cm 5cm 4cm, clip=true, width=0.325\columnwidth]{Fig2/galaxy/69}
\end{figure}
\begin{figure}
\centering

\caption{\label{SPARC26}The graph displays the radial velocity profiles of 21 SPARC galaxies, wherein the fit using MG is not adequate. It's evident that for all these galaxies, the Keplerian velocity calculated from visible matter significantly exceeds the observed velocities, particularly at smaller radii. This indicates a potential issue in the mass modeling for these galaxies. }
\end{figure}

\newpage

\begin{table}
\begin{tabular}{|>{\rule[0.3cm]{0pt}{0.05cm}}p{0.14\columnwidth}|p{0.07\columnwidth}|p{0.08\columnwidth}|p{0.11\columnwidth}|p{0.11\columnwidth}|p{0.06\columnwidth}|p{0.09\columnwidth}|p{0.11\columnwidth}|}
\hline
{Galaxy} & {$R_{\rm max}$} & {$M$} & {$a$} & {$a_N$} & {$\lambda^{-1}$} & {$M_c$} & {$a_0$}\\
& {[kPc]} & {[$10^{9} M_\odot$]} & {[${\rm 10^{-9}cm/s^2}$]} & {[$10^{-9} {\rm cm/s^2}$]} & {[kpc]} & {[$10^{9} M_{\odot}$]} & {[${\rm 10^{-8}cm/s^2}$]}  \\[0.1cm]
\hline\hline
         D512-2 &  3.83  &  0.42  &  1.09  &  0.40  &  1.58  &  5.53 &  3.10 \\ 
         D564-8 &  3.07  &  0.08  &  0.66  &  0.11  &  2.57  &  15.42 &  3.25 \\ 
         D631-7 &  7.19  &  0.66  &  1.48  &  0.18  &  5.28  &  131.71 &  6.59 \\ 
         DDO064 &  2.98  &  0.55  &  2.39  &  0.86  &  1.46  &  10.97 &  7.14 \\ 
         DDO154 &  5.92  &  0.39  &  1.13  &  0.16  &  2.98  &  61.78 &  9.70 \\ 
         DDO161 &  13.37  &  2.98  &  1.06  &  0.23  &  13.26  &  595.51 &  4.72 \\ 
         DDO168 &  4.12  &  0.75  &  2.13  &  0.62  &  4.48  &  150.35 &  10.45 \\ 
         DDO170 &  12.33  &  2.42  &  1.02  &  0.22  &  4.75  &  68.11 &  4.21 \\ 
    ESO079-G014 &  16.67  &  66.17  &  6.16  &  3.32  &  13.70  &  753.17 &  5.59 \\ 
    ESO116-G012 &  9.86  &  7.53  &  4.12  &  1.08  &  9.16  &  872.42 &  14.50 \\ 
    ESO444-G084 &  4.44  &  0.29  &  2.87  &  0.21  &  1.61  &  58.11 &  31.28 \\ 
    ESO563-G021 &  42.41  &  382.57  &  7.44  &  2.96  &  33.40  &  8204.61 &  10.25 \\ 
         F561-1 &  9.66  &  7.41  &  0.85  &  1.11  &  4.60  &  3.19 &  0.21 \\ 
         F563-1 &  20.10  &  8.11  &  1.81  &  0.28  &  3.80  &  362.36 &  34.99 \\ 
        F563-V1 &  7.87  &  1.74  &  0.31  &  0.39  &  3.65  &  1.04 &  0.11 \\ 
        F563-V2 &  10.47  &  7.29  &  4.31  &  0.93  &  1.99  &  141.67 &  49.82 \\ 
        F565-V2 &  8.80  &  2.22  &  2.54  &  0.40  &  3.78  &  171.20 &  16.73 \\ 
         F567-2 &  9.59  &  3.74  &  0.92  &  0.57  &  2.41  &  10.55 &  2.54 \\ 
         F568-1 &  13.23  &  14.59  &  4.94  &  1.16  &  2.86  &  235.09 &  39.95 \\ 
         F568-3 &  17.98  &  16.38  &  2.60  &  0.71  &  7.10  &  290.83 &  8.03 \\ 
        F568-V1 &  17.63  &  9.82  &  2.56  &  0.44  &  2.65  &  198.94 &  39.44 \\ 
         F571-8 &  15.55  &  11.57  &  4.32  &  0.67  &  14.18  &  2313.13 &  16.04 \\ 
        F571-V1 &  13.59  &  4.85  &  1.68  &  0.37  &  6.72  &  227.82 &  7.02 \\ 
         F574-1 &  12.60  &  10.02  &  2.56  &  0.88  &  3.54  &  103.07 &  11.44 \\ 
         F574-2 &  10.83  &  5.32  &  0.48  &  0.63  &  7.53  &  0.52 &  0.01 \\ 
        F579-V1 &  15.16  &  18.09  &  2.78  &  1.10  &  1.54  &  60.60 &  35.55 \\ 
         F583-1 &  16.26  &  7.02  &  1.47  &  0.37  &  3.39  &  130.73 &  15.83 \\ 
         F583-4 &  7.29  &  2.71  &  2.17  &  0.71  &  4.66  &  59.85 &  3.84 \\ 
         IC2574 &  10.23  &  3.12  &  1.44  &  0.42  &  13.98  &  624.26 &  4.45 \\ 
         IC4202 &  25.90  &  234.65  &  7.63  &  4.88  &  130.72  &  46898.50 &  3.83 \\ 
       KK98-251 &  3.13  &  0.33  &  1.21  &  0.47  &  3.34  &  21.05 &  2.63 \\ 
        NGC0024 &  11.27  &  5.31  &  3.48  &  0.58  &  2.84  &  143.87 &  24.89 \\ 
        NGC0055 &  13.50  &  10.78  &  1.80  &  0.82  &  11.15  &  313.43 &  3.52 \\ 
        NGC0100 &  9.62  &  4.65  &  2.80  &  0.70  &  11.95  &  928.87 &  9.07 \\ 
        NGC0247 &  14.54  &  12.12  &  2.55  &  0.80  &  3.15  &  51.92 &  7.29 \\ 
        NGC0289 &  71.12  &  119.64  &  1.24  &  0.33  &  39.85  &  4725.59 &  4.15 \\ 
        NGC0300 &  11.80  &  4.69  &  2.40  &  0.47  &  8.06  &  521.33 &  11.19 \\ 
        NGC0801 &  59.82  &  410.36  &  2.53  &  1.60  &  256.86  &  82053.40 &  1.73 \\ 
        NGC0891 &  17.11  &  163.36  &  8.19  &  7.78  &  0.72  &  72.81 &  198.36 \\ 
        NGC1003 &  30.24  &  16.76  &  1.42  &  0.26  &  23.36  &  3351.63 &  8.56 \\ 
        NGC1090 &  30.09  &  98.06  &  2.76  &  1.51  &  76.59  &  19604.74 &  4.66 \\ 
        NGC1705 &  6.00  &  0.81  &  2.76  &  0.31  &  1.97  &  76.08 &  27.20 \\ 
        NGC2366 &  6.06  &  1.22  &  1.30  &  0.46  &  4.08  &  54.90 &  4.60 \\         
\hline
\end{tabular}
\end{table}
\begin{table}
\begin{tabular}{|>{\rule[0.3cm]{0pt}{0.05cm}}p{0.13\columnwidth}|p{0.07\columnwidth}|p{0.08\columnwidth}|p{0.11\columnwidth}|p{0.11\columnwidth}|p{0.06\columnwidth}|p{0.09\columnwidth}|p{0.11\columnwidth}|}
\hline
{Galaxy} & {$R_{\rm max}$} & {$M$} & {$a$} & {$a_N$} & {$\lambda^{-1}$} & {$M_c$} & {$a_0$}\\
& {[kPc]} & {[$10^{9} M_\odot$]} &{[${\rm 10^{-9}cm/s^2}$]} & {[$10^{-9} {\rm cm/s^2}$]} & {[kpc]} & {[$10^{9} M_{\odot}$]} & {[${\rm 10^{-8}cm/s^2}$]}  \\[0.1cm]
\hline\hline      
        NGC2403 &  20.87  &  17.97  &  2.79  &  0.57  &  18.63  &  3547.42 &  14.24 \\ 
        NGC2683 &  34.62  &  84.96  &  2.13  &  0.99  &  77.75  &  16973.95 &  3.91 \\ 
        NGC2841 &  63.64  &  269.62  &  4.40  &  0.93  &  36.33  &  14462.51 &  15.27 \\ 
        NGC2903 &  24.96  &  88.50  &  4.21  &  1.98  &  86.84  &  17694.20 &  3.27 \\ 
        NGC2915 &  10.04  &  1.57  &  2.41  &  0.22  &  3.33  &  259.62 &  32.56 \\ 
        NGC2955 &  35.43  &  391.83  &  4.71  &  4.35  &  1.32  &  46.46 &  36.94 \\ 
        NGC2976 &  2.27  &  3.33  &  10.39  &  9.02  &  8.46  &  0.00 &  0.00 \\ 
        NGC2998 &  42.28  &  207.16  &  3.16  &  1.62  &  91.76  &  41412.75 &  6.85 \\ 
        NGC3109 &  6.45  &  0.80  &  2.28  &  0.27  &  3.61  &  160.70 &  17.15 \\ 
        NGC3198 &  44.08  &  62.38  &  1.63  &  0.45  &  49.70  &  12473.53 &  7.04 \\ 
        NGC3521 &  17.74  &  96.95  &  7.75  &  4.29  &  0.94  &  20.91 &  32.80 \\ 
        NGC3726 &  32.52  &  86.41  &  2.78  &  1.14  &  166.05  &  17246.44 &  0.87 \\ 
        NGC3741 &  7.00  &  0.32  &  1.23  &  0.09  &  2.38  &  63.48 &  15.65 \\ 
        NGC3769 &  37.16  &  27.49  &  1.11  &  0.28  &  37.24  &  5495.73 &  5.52 \\ 
        NGC3877 &  11.35  &  97.43  &  8.15  &  10.54  &  6.54  &  8.84 &  0.29 \\ 
        NGC3893 &  19.05  &  71.81  &  4.74  &  2.76  &  62.51  &  14355.69 &  5.12 \\ 
        NGC3917 &  14.86  &  30.48  &  4.09  &  1.92  &  8.54  &  217.86 &  4.16 \\ 
        NGC3949 &  7.07  &  43.18  &  13.09  &  12.04  &  0.05  &  9.25 &  5303.97 \\ 
        NGC3953 &  15.68  &  179.59  &  9.55  &  10.18  &  17.46  &  0.00 &  0.00 \\
        NGC3972 &  8.72  &  19.15  &  6.67  &  3.51  &  11.19  &  475.55 &  5.30 \\         
        NGC3992 &  46.02  &  266.08  &  3.96  &  1.75  &  74.07  &  29852.53 &  7.58 \\ 
        NGC4010 &  10.47  &  22.58  &  4.61  &  2.87  &  31.09  &  4510.52 &  6.50 \\ 
        NGC4013 &  31.01  &  88.74  &  3.02  &  1.29  &  69.58  &  17742.05 &  5.11 \\ 
        NGC4051 &  12.19  &  85.38  &  6.22  &  8.01  &  8.35  &  0.00 &  0.00 \\ 
        NGC4068 &  2.33  &  0.49  &  2.44  &  1.26  &  7.31  &  98.08 &  2.56 \\ 
        NGC4085 &  6.20  &  28.82  &  9.67  &  10.45  &  4.56  &  0.00 &  0.00 \\ 
        NGC4088 &  21.48  &  128.65  &  4.57  &  3.89  &  1.66  &  43.12 &  21.77 \\ 
        NGC4100 &  22.76  &  69.03  &  3.60  &  1.86  &  51.38  &  13799.18 &  7.29 \\ 
        NGC4138 &  18.58  &  49.38  &  3.92  &  1.99  &  56.05  &  9849.35 &  4.37 \\ 
        NGC4157 &  29.61  &  122.32  &  3.75  &  1.94  &  147.53  &  24433.07 &  1.56 \\ 
        NGC4183 &  21.02  &  18.26  &  1.97  &  0.58  &  6.82  &  172.00 &  5.15 \\ 
        NGC4214 &  5.63  &  1.45  &  3.74  &  0.64  &  4.36  &  289.82 &  21.21 \\ 
        NGC4217 &  16.72  &  105.49  &  6.14  &  5.26  &  0.71  &  22.92 &  63.48 \\ 
        NGC4389 &  5.32  &  24.92  &  7.37  &  12.27  &  0.04  &  0.78 &  682.67 \\ 
        NGC4559 &  20.97  &  33.70  &  2.19  &  1.07  &  43.36  &  6733.50 &  4.99 \\ 
        NGC5005 &  11.47  &  205.21  &  19.84  &  21.74  &  0.25  &  22.17 &  510.66 \\ 
        NGC5033 &  44.59  &  138.51  &  2.79  &  0.97  &  31.64  &  4158.18 &  5.79 \\ 
        NGC5055 &  54.59  &  173.80  &  1.76  &  0.81  &  0.58  &  52.09 &  215.97 \\ 
        NGC5371 &  46.24  &  382.69  &  3.18  &  2.49  &  8.58  &  168.31 &  3.18 \\ 
        NGC5585 &  10.96  &  5.79  &  2.36  &  0.67  &  13.37  &  1157.39 &  9.03 \\ 
        NGC5907 &  50.33  &  232.78  &  2.95  &  1.28  &  101.11  &  46544.66 &  6.34 \\ 
        NGC5985 &  34.72  &  293.02  &  7.58  &  3.39  &  6.08  &  916.09 &  34.53 \\ 
        NGC6015 &  29.23  &  43.14  &  2.56  &  0.70  &  35.62  &  8624.61 &  9.48 \\   
\hline
\end{tabular}
\end{table}
\begin{table}
\begin{tabular}{|>{\rule[0.3cm]{0pt}{0.05cm}}p{0.13\columnwidth}|p{0.07\columnwidth}|p{0.08\columnwidth}|p{0.11\columnwidth}|p{0.11\columnwidth}|p{0.07\columnwidth}|p{0.09\columnwidth}|p{0.11\columnwidth}|}
\hline
{Galaxy} & {$R_{\rm max}$} & {$M$} & {$a$} & {$a_N$} & {$\lambda^{-1}$} & {$M_c$} & {$a_0$}\\
& {[kPc]} & {[$10^{9} M_\odot$]} &{[${\rm 10^{-9}cm/s^2}$]} & {[$10^{-9} {\rm cm/s^2}$]} & {[kpc]} & {[$10^{9} M_{\odot}$]} & {[${\rm 10^{-8}cm/s^2}$]}  \\[0.1cm]
\hline\hline          
        NGC6195 &  36.43  &  454.27  &  5.38  &  4.77  &  33.66  &  0.00 &  0.00 \\ 
        NGC6503 &  23.50  &  16.55  &  1.82  &  0.42  &  21.32  &  2799.00 &  8.58 \\ 
        NGC6674 &  72.41  &  259.05  &  2.62  &  0.69  &  76.15  &  37632.00 &  9.05 \\ 
        NGC6789 &  0.71  &  0.10  &  16.65  &  2.73  &  0.67  &  19.72 &  61.78 \\       
        NGC6946 &  20.40  &  78.10  &  3.77  &  2.62  &  14.22  &  0.00 &  0.00 \\ 
        NGC7331 &  36.31  &  288.33  &  5.06  &  3.05  &  208.40  &  18463.89 &  0.59 \\ 
        NGC7793 &  7.87  &  9.95  &  3.39  &  2.24  &  23.59  &  1984.28 &  4.97 \\ 
        NGC7814 &  19.53  &  90.13  &  7.60  &  3.29  &  41.34  &  18021.00 &  14.69 \\ 
       PGC51017 &  3.63  &  0.35  &  0.30  &  0.37  &  0.15  &  0.11 &  6.73 \\ 
       UGC00128 &  53.75  &  27.94  &  0.94  &  0.13  &  10.79  &  816.53 &  9.77 \\ 
       UGC00191 &  9.98  &  5.01  &  2.28  &  0.70  &  3.12  &  54.62 &  7.81 \\ 
       UGC00634 &  18.01  &  8.86  &  2.08  &  0.38  &  9.71  &  732.70 &  10.83 \\ 
       UGC00731 &  10.91  &  3.66  &  1.62  &  0.43  &  1.15  &  42.85 &  45.15 \\ 
       UGC00891 &  7.39  &  1.23  &  1.78  &  0.31  &  6.22  &  245.66 &  8.85 \\ 
       UGC01230 &  36.54  &  21.60  &  0.94  &  0.23  &  5.10  &  261.11 &  13.99 \\ 
       UGC02023 &  3.78  &  1.51  &  2.96  &  1.47  &  11.41  &  301.30 &  3.23 \\ 
       UGC02259 &  8.14  &  3.38  &  3.22  &  0.71  &  1.89  &  49.72 &  19.43 \\ 
       UGC02455 &  4.03  &  4.82  &  2.99  &  4.14  &  0.04  &  0.15 &  136.45 \\ 
       UGC02487 &  80.38  &  567.89  &  4.47  &  1.23  &  30.89  &  8618.41 &  12.58 \\ 
       UGC02885 &  74.07  &  589.68  &  3.89  &  1.50  &  126.90  &  117864.77 &  10.20 \\ 
       UGC02916 &  38.00  &  194.18  &  2.79  &  1.87  &  0.21  &  30.52 &  977.33 \\ 
       UGC02953 &  62.39  &  283.26  &  3.84  &  1.01  &  84.81  &  56638.51 &  10.98 \\ 
       UGC03205 &  40.04  &  137.92  &  3.92  &  1.20  &  63.17  &  27577.96 &  9.63 \\ 
       UGC03546 &  29.23  &  115.13  &  4.13  &  1.88  &  108.22  &  23018.33 &  2.74 \\ 
       UGC03580 &  27.06  &  21.90  &  1.84  &  0.42  &  30.07  &  4379.50 &  6.75 \\ 
       UGC04278 &  6.69  &  4.12  &  4.17  &  1.28  &  7.19  &  625.24 &  16.86 \\ 
       UGC04305 &  5.52  &  1.88  &  0.64  &  0.86  &  1.51  &  0.67 &  0.41 \\ 
       UGC04325 &  5.59  &  3.80  &  4.85  &  1.69  &  1.02  &  26.19 &  35.26 \\ 
       UGC04483 &  1.21  &  0.06  &  1.57  &  0.55  &  0.74  &  1.37 &  3.51 \\ 
       UGC04499 &  8.18  &  4.07  &  2.19  &  0.85  &  4.52  &  66.63 &  4.54 \\ 
       UGC05005 &  28.61  &  14.56  &  1.11  &  0.25  &  21.20  &  1808.47 &  5.61 \\ 
       UGC05253 &  53.29  &  221.94  &  2.89  &  1.09  &  110.38  &  44376.56 &  5.08 \\ 
       UGC05414 &  4.11  &  1.65  &  2.97  &  1.36  &  8.30  &  328.50 &  6.64 \\ 
       UGC05716 &  12.37  &  2.84  &  1.46  &  0.26  &  3.47  &  89.78 &  10.42 \\ 
       UGC05721 &  6.74  &  1.49  &  3.04  &  0.46  &  1.80  &  74.24 &  31.86 \\ 
       UGC05750 &  22.85  &  11.01  &  0.88  &  0.29  &  6.92  &  124.04 &  3.61 \\ 
       UGC05764 &  3.62  &  0.43  &  2.23  &  0.46  &  0.61  &  11.33 &  42.90 \\ 
       UGC05829 &  6.91  &  2.58  &  2.21  &  0.75  &  2.19  &  22.14 &  6.41 \\ 
       UGC05918 &  4.46  &  0.55  &  1.44  &  0.39  &  0.98  &  6.87 &  9.90 \\   
       UGC05986 &  9.41  &  6.85  &  3.94  &  1.08  &  7.24  &  604.94 &  16.07 \\ 
       UGC05999 &  16.22  &  12.55  &  2.00  &  0.66  &  8.66  &  323.52 &  6.01 \\ 
       UGC06399 &  7.85  &  4.23  &  3.17  &  0.96  &  4.01  &  99.27 &  8.60 \\ 
       UGC06446 &  10.22  &  3.71  &  2.03  &  0.50  &  2.13  &  60.44 &  18.55 \\ 
       UGC06614 &  64.59  &  196.51  &  2.09  &  0.66  &  111.37  &  39284.25 &  4.41 \\ 
\hline
\end{tabular}
\end{table}
\begin{table}
\begin{tabular}{|>{\rule[0.3cm]{0pt}{0.05cm}}p{0.13\columnwidth}|p{0.07\columnwidth}|p{0.08\columnwidth}|p{0.11\columnwidth}|p{0.11\columnwidth}|p{0.07\columnwidth}|p{0.09\columnwidth}|p{0.11\columnwidth}|}
\hline
{Galaxy} & {$R_{\rm max}$} & {$M$} & {$a$} & {$a_N$} & {$\lambda^{-1}$} & {$M_c$} & {$a_0$}\\
& {[kPc]} & {[$10^{9} M_\odot$]} &{[${\rm 10^{-9}cm/s^2}$]} & {[$10^{-9} {\rm cm/s^2}$]} & {[kpc]} & {[$10^{9} M_{\odot}$]} & {[${\rm 10^{-8}cm/s^2}$]}  \\[0.1cm]
\hline\hline  
       UGC06628 &  7.69  &  5.67  &  0.75  &  1.34  &  4.35  &  0.11 &  0.01 \\   
       UGC06667 &  7.85  &  2.14  &  3.03  &  0.48  &  1.39  &  76.81 &  55.65 \\   
       UGC06786 &  34.05  &  87.12  &  4.24  &  1.05  &  33.32  &  17420.10 &  21.87 \\ 
       UGC06787 &  37.19  &  122.13  &  5.67  &  1.23  &  42.35  &  24420.42 &  18.97 \\ 
       UGC06818 &  6.98  &  2.67  &  2.57  &  0.76  &  11.61  &  533.43 &  5.51 \\ 
       UGC06917 &  10.47  &  12.13  &  3.81  &  1.54  &  6.32  &  195.35 &  6.82 \\    
       UGC06923 &  5.16  &  4.37  &  4.13  &  2.29  &  12.67  &  870.12 &  7.55 \\ 
       UGC06930 &  16.61  &  16.50  &  2.28  &  0.83  &  6.04  &  137.85 &  5.26 \\ 
       UGC06973 &  7.85  &  59.84  &  13.37  &  13.53  &  0.05  &  9.41 &  5023.91 \\ 
       UGC06983 &  15.68  &  11.69  &  2.46  &  0.66  &  4.79  &  177.50 &  10.79 \\ 
       UGC07089 &  9.16  &  5.89  &  2.21  &  0.98  &  20.40  &  1174.06 &  3.93 \\ 
       UGC07125 &  18.68  &  8.13  &  0.73  &  0.32  &  8.81  &  61.64 &  1.11 \\ 
       UGC07151 &  5.50  &  3.95  &  3.42  &  1.82  &  2.74  &  22.79 &  4.23 \\ 
       UGC07232 &  0.82  &  0.14  &  7.65  &  2.98  &  1.75  &  28.68 &  13.08 \\ 
       UGC07261 &  6.67  &  3.29  &  2.81  &  1.03  &  3.34  &  45.09 &  5.63 \\ 
       UGC07323 &  5.82  &  6.13  &  4.08  &  2.52  &  17.65  &  1221.89 &  5.47 \\ 
       UGC07399 &  6.13  &  2.30  &  5.94  &  0.85  &  1.90  &  119.62 &  45.98 \\ 
       UGC07524 &  10.69  &  6.62  &  1.89  &  0.81  &  3.06  &  48.60 &  7.24 \\ 
       UGC07559 &  2.53  &  0.30  &  1.32  &  0.65  &  5.13  &  59.86 &  3.17 \\ 
       UGC07577 &  1.69  &  0.09  &  0.61  &  0.43  &  14.16  &  0.01 &  0.00 \\ 
       UGC07603 &  4.11  &  0.70  &  3.23  &  0.58  &  3.31  &  139.55 &  17.76 \\ 
       UGC07608 &  4.78  &  0.94  &  3.26  &  0.57  &  2.00  &  58.38 &  20.34 \\ 
       UGC07690 &  4.13  &  1.41  &  2.45  &  1.16  &  2.19  &  12.45 &  3.60 \\ 
       UGC07866 &  2.32  &  0.26  &  1.53  &  0.66  &  1.54  &  3.33 &  1.95 \\ 
       UGC08286 &  8.04  &  2.70  &  2.86  &  0.58  &  2.33  &  67.46 &  17.39 \\ 
       UGC08490 &  10.15  &  2.24  &  1.92  &  0.30  &  3.01  &  105.57 &  16.21 \\ 
       UGC08550 &  5.36  &  0.91  &  2.00  &  0.44  &  2.08  &  33.62 &  10.81 \\ 
       UGC08699 &  25.70  &  63.39  &  4.22  &  1.34  &  44.11  &  12673.65 &  9.08 \\ 
       UGC08837 &  4.20  &  1.16  &  1.78  &  0.92  &  11.63  &  232.90 &  2.40 \\ 
       UGC09037 &  27.96  &  97.50  &  2.68  &  1.74  &  255.33  &  19482.38 &  0.42 \\ 
       UGC09133 &  108.31  &  368.98  &  1.57  &  0.44  &  98.21  &  33142.66 &  4.79 \\ 
       UGC09992 &  3.89  &  0.72  &  0.98  &  0.66  &  1.80  &  2.25 &  0.97 \\ 
       UGC10310 &  7.74  &  3.98  &  2.24  &  0.93  &  2.19  &  24.31 &  7.09 \\ 
       UGC11455 &  41.93  &  435.07  &  5.47  &  3.45  &  268.62  &  86981.64 &  1.68 \\  
\hline
\end{tabular}
\end{table}
\begin{table}
\begin{tabular}{|>{\rule[0.3cm]{0pt}{0.05cm}}p{0.13\columnwidth}|p{0.07\columnwidth}|p{0.08\columnwidth}|p{0.11\columnwidth}|p{0.11\columnwidth}|p{0.07\columnwidth}|p{0.09\columnwidth}|p{0.11\columnwidth}|}
\hline
{Galaxy} & {$R_{\rm max}$} & {$M$} & {$a$} & {$a_N$} & {$\lambda^{-1}$} & {$M_c$} & {$a_0$}\\
& {[kPc]} & {[$10^{9} M_\odot$]} &{[${\rm 10^{-9}cm/s^2}$]} & {[$10^{-9} {\rm cm/s^2}$]} & {[kpc]} & {[$10^{9} M_{\odot}$]} & {[${\rm 10^{-8}cm/s^2}$]}  \\[0.1cm]
\hline\hline  
       UGC11557 &  10.56  &  18.09  &  2.19  &  2.26  &  7.81  &  0.00 &  0.00 \\ 
       UGC11820 &  15.82  &  5.67  &  1.46  &  0.32  &  6.58  &  178.47 &  5.74 \\ 
       UGC11914 &  9.83  &  159.33  &  30.67  &  22.98  &  12.92  &  0.00 &  0.00 \\ 
       UGC12506 &  49.99  &  236.81  &  3.28  &  1.32  &  11.43  &  1452.11 &  15.48 \\ 
       UGC12632 &  10.66  &  4.22  &  1.62  &  0.52  &  2.55  &  43.43 &  9.31 \\ 
       UGC12732 &  15.40  &  8.60  &  2.02  &  0.51  &  4.44  &  147.59 &  10.45 \\ 
        UGCA281 &  1.08  &  0.09  &  2.61  &  1.13  &  0.92  &  2.35 &  3.85 \\ 
        UGCA442 &  6.33  &  0.58  &  1.63  &  0.20  &  3.60  &  115.75 &  12.45 \\ 
        UGCA444 &  2.62  &  0.13  &  1.81  &  0.27  &  1.02  &  4.74 &  6.32 \\        
\hline
\end{tabular}
\caption{\label{clusterProperties}The table provides information on various parameters for each SPARC galaxy, including the radius of the last velocity data point ($R_{\rm max}$), mass ($M$), centripetal acceleration ($a$) at the last data point, centripetal acceleration calculated using Newtonian mechanics at the last velocity data point, best fit value of $\lambda^{-1}$, $M_c$, and the derived value of $a_0 = \frac{GM_c}{\lambda^{-2}}$. It's evident that the value of $a_0$ varies significantly among different galaxies, indicating that a uniform $a_0$ may not be sufficient to adequately explain the velocity profiles of all galaxies.}
\end{table}

\end{document}